\documentclass[a4paper,12pt,onecolumn,oneside]{report}
\pdfoutput=1

\usepackage{amsfonts}
\usepackage{amsmath}
\usepackage{amssymb}
\usepackage{mathrsfs}
\usepackage[english]{babel}
\usepackage[T1]{fontenc}
\usepackage[latin1]{inputenc}
\usepackage{fancyheadings}
\usepackage{textcomp}
\usepackage[pdftex]{graphicx}
\usepackage{pgfplots}
\pgfplotsset{compat=newest}
\usepackage{booktabs}

\newcommand{\clearemptydoublepage}{\newpage{\pagestyle{empty}\cleardoublepage}}

\def\a{\alpha}
\def\b{\beta}
\def\g{\gamma}
\def\G{\Gamma}
\def\d{\delta}
\def\D{\Delta}
\def\ep{\epsilon}
\def\vep{\varepsilon}
\def\z{\zeta}
\def\e{\eta}
\def\th{\theta}
\def\vth{\vartheta}

\def\la{\lambda}
\def\La{\Lambda}
\def\m{\mu}
\def\n{\nu}

\def\vp{\varpi}
\def\r{\rho}

\def\s{\sigma}
\def\S{\Sigma}
\def\t{\tau}
\def\f{\phi}

\def\vf{\varphi}
\def\ch{\chi}
\def\ps{\psi}
\def\o{\omega}

\def\tG{\tilde{G}}
\def\tR{\tilde{R}}
\def\ta{\tilde{\alpha}} 
\def\tr{\tilde{\rho}}
\def\to{\tilde{\omega}}

\def\tp{\tilde{\pi}}
\def\tvp{\tilde{\varpi}}

\def\tps{\tilde{\psi}}

\newcommand{\ti}[1]{\tilde{#1}}

\newcommand{\Xd}{X^{\cdot}}
\newcommand{\xid}{\xi^{\cdot}}
\newcommand{\chd}{\chi^{\cdot}}
\newcommand{\zd}{\zeta^{\cdot}}
\newcommand{\xd}{x^{\cdot}}
\newcommand{\vfd}{\varphi^{\cdot}}

\newcommand{\bn}{\bar{n}}

\newcommand{\bvf}{\bar{\varphi}}
\newcommand{\bvfd}{\bar{\varphi}^{\cdot}}

\newcommand{\hxi}{\hat{\xi}}
\newcommand{\hxid}{\hat{\xi}^{\cdot}}

\newcommand{\h}[1]{\hat{#1}}
\newcommand{\hp}{\hat{\pi}}

\providecommand{\abs}[1]{\lvert#1\rvert}
\providecommand{\norm}[1]{\lVert#1\rVert}
\newcommand{\emn}{\eta_{\mu \nu}}
\newcommand{\de}{\partial}
\newcommand{\half}{\frac{1}{2}}
\newcommand{\third}{\frac{1}{3}}
\newcommand{\fourth}{\frac{1}{4}}
\newcommand{\p}{\prime}
\newcommand{\pr}[1]{{#1}^{\prime}}
\newcommand{\beq}{\begin{equation}}
\newcommand{\eeq}{\end{equation}}
\newcommand{\beqnn}{\begin{equation*}}
\newcommand{\eeqnn}{\end{equation*}}
\newcommand{\bea}{\begin{eqnarray}}
\newcommand{\eea}{\end{eqnarray}}
\newcommand{\nn}{\nonumber}
\newcommand{\noi}{\noindent}
\newcommand{\ie}{\textit{i.e.}\,}
\newcommand{\via}{\textit{via}\,}

\newcommand{\xia}{\xi^{a}}

\newcommand{\xim}{\xi^{\mu}}
\newcommand{\chm}{\chi^{\mu}}

\newcommand{\xm}{x^{\mu}}

\newcommand{\vecx}{\vec{x}}
\newcommand{\veck}{\vec{k}}
\newcommand{\vecp}{\vec{p}}

\newcommand{\vecP}{\vec{P}}

\newcommand{\mbf}[1]{\mathbf{#1}}
\newcommand{\mbfG}{\mathbf{G}}
\newcommand{\mbfT}{\mathbf{T}}
\newcommand{\mbfg}{\mathbf{g}}
\newcommand{\mbfh}{\mathbf{h}}

\newcommand{\mbfn}{\mathbf{n}}

\newcommand{\tmbfg}{\tilde{\mathbf{g}}}
\newcommand{\tmbfK}{\tilde{\mathbf{K}}}
\newcommand{\tmbfG}{\tilde{\mathbf{G}}}
\newcommand{\tmbfT}{\tilde{\mathbf{T}}}
\newcommand{\mcal}[1]{\mathcal{#1}}
\newcommand{\mcalT}{\mathcal{T}}
\newcommand{\mcalI}{\mathcal{I}}
\newcommand{\mcalL}{\mathcal{L}}
\newcommand{\mcalH}{\mathcal{H}}
\newcommand{\mcalK}{\mathcal{K}}

\newcommand{\mscr}[1]{\mathscr{#1}}
\newcommand{\mscrT}{\mathscr{T}}
\newcommand{\mscrM}{\mathscr{M}}
\newcommand{\mscrB}{\mathscr{B}}
\newcommand{\bmscrM}{\bar{\mathscr{M}}}
\newcommand{\bmbfg}{\bar{\mathbf{g}}}
\newcommand{\mscrI}{\mathscr{I}}

\def\deum{\partial^{\mu}}

\newcommand{\dez}{\de_{z}}
\newcommand{\dey}{\de_{y}}

\newcommand{\deu}[1]{\de^{#1}}

\newcommand{\ded}[1]{\de_{#1}}
\newcommand{\dedq}[1]{\de_{#1}^{2}}
\newcommand{\dem}{\de_{\m}}
\newcommand{\den}{\de_{\n}}
\newcommand{\demden}{\de_{\m} \de_{\n}}
\newcommand{\boxf}{\Box_{4}}

\newcommand{\boxfi}{\Box_{5}}
\newcommand{\boxs}{\Box_{6}}

\newcommand{\dexim}{\de_{\xi^{\m}}}

\newcommand{\deximxin}{\de_{\xi^{\m}} \de_{\xi^{\n}}}

\newcommand{\dechm}{\de_{\chi^{\m}}}

\newcommand{\dechmchn}{\de_{\chi^{\m}} \de_{\chi^{\n}}}

\newcommand{\evb}{\Big\rvert_{X^{\cdot} = \bar{\varphi}^{\cdot}(\xi^{\cdot})}}
\newcommand{\evbh}{\Big\rvert_{X^{\cdot} = \bar{\varphi}^{\cdot}(\hat{\xi}^{\cdot})}}

\newcommand{\evbbh}{\bigg\rvert_{X^{\cdot} = \bar{\varphi}^{\cdot}(\hat{\xi}^{\cdot})}}

\newcommand{\Msf}{M_{6}^{4}}
\newcommand{\Mft}{M_{5}^{3}}
\newcommand{\Mfs}{M_{4}^{2}}
\newcommand{\MPq}{M_{P}^{2}}

\newcommand{\mg}[1]{g_{#1}}
\newcommand{\tmg}[1]{\tilde{g}_{#1}}
\newcommand{\umg}[1]{g^{#1}}

\newcommand{\Zp}{Z^{\prime}}
\newcommand{\Zpq}{{Z^{\prime}}^2}
\newcommand{\Yp}{Y^{\prime}}
\newcommand{\Ypq}{{Y^{\prime}}^2}
\newcommand{\Zpp}{Z^{\prime \prime}}
\newcommand{\Ypp}{Y^{\prime \prime}}

\newcommand{\Pp}{P^{\prime}}

\newcommand{\mZ}{\mathscr{Z}}
\newcommand{\mY}{\mathscr{Y}}

\newcommand{\mcalZpq}{{\mathcal{Z}^{\prime}}^2}
\newcommand{\mcalYp}{\mathcal{Y}^{\prime}}
\newcommand{\mcalYpq}{{\mathcal{Y}^{\prime}}^2}

\newcommand{\bla}{\bar{\la}}
\newcommand{\blac}{\bar{\la}_{c}}
\newcommand{\dla}{\d \! \la}

\newcommand{\ltp}{l_2^{\shortparallel}}
\newcommand{\ltn}{l_2^{\perp}}
\newcommand{\lo}{l_1}

\newcommand{\dvf}{\delta \! \varphi}
\newcommand{\hdvf}{\delta \! \hat{\varphi}}

\newcommand{\dvfn}{\delta \! \varphi_{\perp}}
\newcommand{\hdvfn}{\delta \! \hat{\varphi}_{\perp}}
\newcommand{\dvfp}{\delta \! \varphi_{\shortparallel}}
\newcommand{\hdvfp}{\delta \! \hat{\varphi}_{\shortparallel}}
\newcommand{\dvfz}{\delta \! \varphi^{z}}
\newcommand{\dvfy}{\delta \! \varphi^{y}}
\newcommand{\dvfzgi}{\delta \! \varphi^{z}_{gi}}
\newcommand{\dvfygi}{\delta \! \varphi^{y}_{gi}}

\newcommand{\dvfzp}{{\delta \! \varphi^{z}}^{\prime}}
\newcommand{\dvfyp}{{\delta \! \varphi^{y}}^{\prime}}

\newcommand{\bv}{\bar{v}}
\newcommand{\dv}{\delta v}

\newcommand{\dvn}{\delta v_{\perp}}
\newcommand{\dhvn}{\delta \hat{v}_{\perp}}
\newcommand{\dvsp}{\delta v_{\shortparallel}}
\newcommand{\dhvsp}{\delta \hat{v}_{\shortparallel}}

\newcommand{\aq}{\alpha^{2}}
\newcommand{\bq}{\beta^{2}}
\newcommand{\ha}[1]{h^{#1}}
\newcommand{\va}[1]{v^{#1}}

\begin{document}

\begin{titlepage}

\begin{center}
\large{UNIVERSIT\`A DEGLI STUDI DI MILANO}

\vspace{0.3cm}

\large{Scuola di Dottorato in Fisica, Astrofisica e Fisica Applicata}

\vspace{0.5cm}

\large{and}

\vspace{0.4cm}

\large{UNIVERSITY OF PORTSMOUTH}

\vspace{0.3cm}

\large{Institute of Cosmology and Gravitation}
\end{center}

\vspace{1.6cm}

\begin{center}
\huge{{\bfseries Modif\mbox{}ied Theories of Gravity}}

\vspace{0.3cm}

\large{s.s.d. FIS/02}
\end{center}

\vspace{16mm}

\begin{flushright}
\large{PhD Thesis of:} \large{\bfseries{Fulvio Sbis\`a}}

\vspace{0.2cm}
\large{Ciclo} \large{\bfseries{XXIII}} \phantom{vediamo quantoi}
\end{flushright}

\vspace{12mm}

\begin{flushleft}
\large{Thesis Director: \bfseries{Prof. Dietmar Klemm}}

\large{Thesis Director: \bfseries{Dr. Kazuya Koyama}}

\vspace{4mm}

\large{Director of the Doctoral School: \bfseries{Prof. Marco Bersanelli}}

\large{Director of the Doctoral School: \bfseries{Prof. David Wands}}
\end{flushleft}

\vspace{1cm}

\begin{center}
\large{A.A. 2012/2013}
\end{center}
\end{titlepage}

\clearemptydoublepage

\pagestyle{empty}

\begin{tabular}{rrl}
\qquad & \qquad &            \\
\qquad & \qquad &            \\
\qquad & \qquad &            \\
\qquad & \qquad &            \\
\qquad & \qquad &            \\
\qquad & \qquad &            \\
\qquad & \qquad &            \\
\qquad & \qquad &            \\
\qquad & \hspace{8cm} &  {\itshape  To Dietmar } \\
\end{tabular}

\clearemptydoublepage

\chapter*{Abstract}

\addcontentsline{toc}{chapter}{Abstract}

\pagestyle{fancyplain}
\lhead[\fancyplain{}{\rm\thepage} \mbox{  } \mbox{  } \mbox{  }  \sl{Abstract}]{\fancyplain{}{}}
\lfoot[\fancyplain{}{}]{\fancyplain{}{}}
\chead[\fancyplain{}{}]{\fancyplain{}{}}
\cfoot[\fancyplain{\rm\thepage}{}]{\fancyplain{\rm\thepage}{}}
\rhead[\fancyplain{}{}]{\fancyplain{}{\sl{Abstract} \mbox{  } \mbox{  } \mbox{  } \rm\thepage}}
\rfoot[\fancyplain{}{}]{\fancyplain{}{}}

The recent observational data in cosmology seem to indicate that the universe is currently expanding in an accelerated way. This unexpected conclusion can be explained assuming the presence of a non-vanishing yet extremely f\mbox{}ine tuned cosmological constant, or invoking the existence of an exotic source of energy, dark energy, which is not observed in laboratory experiments yet seems to dominate the energy budget of the Universe. On the other hand, it may be that these observations are just signalling the fact that Einstein's General Relativity is not the correct description of gravity when we consider distances of the order of the present horizon of the universe.

In order to study if the latter explanation is correct, we have to formulate new theories of the gravitational interaction, and see if they admit cosmological solutions which f\mbox{}it the observational data in a satisfactory way. Quite generally, modifying General Relativity introduces new degrees of freedom, which are responsible for the dif\mbox{}ferent large distance behaviour. On one hand, often these new degrees of freedom have negative kinetic energy, which implies that the theory is plagued by ghost instabilities. On the other hand, for a modif\mbox{}ied gravity theory to be phenomenologically viable it is necessary that the extra degrees of freedom are ef\mbox{}f\mbox{}iciently screened on terrestrial and astrophysical scales. One of the known mechanisms which can screen the extra degrees of freedom is the Vainshtein mechanism, which involves derivative self-interaction terms for these degrees of freedom.

In this thesis, we consider two dif\mbox{}ferent models, the Cascading DGP and the dRGT massive gravity, which are candidates for viable models to modify gravity at very large distances. Regarding the Cascading DGP model, we consider the minimal (6D) set-up and we perform a perturbative analysis at f\mbox{}irst order of the behaviour of the gravitational f\mbox{}ield and of the branes position around background solutions where pure tension is localized on the 4D brane. We consider a specif\mbox{}ic realization of this set-up where the 5D brane can be considered thin with respect to the 4D one.

We show that the thin limit of the 4D brane inside the (already thin) 5D brane is well def\mbox{}ined, at least for the conf\mbox{}igurations that we consider, and conf\mbox{}irm that the gravitational f\mbox{}ield on the 4D brane is f\mbox{}inite for a general choice of the energy-momentum tensor. We also conf\mbox{}irm that there exists a critical tension which separates background conf\mbox{}igurations which possess a ghost among the perturbation modes, and background conf\mbox{}igurations which are ghost-free. We f\mbox{}ind a value for the critical tension which is dif\mbox{}ferent from the value which has been obtained in the literature; we comment on the dif\mbox{}ference between these two results, and perform a numeric calculation in a particular case where the exact solution is known to support the validity of our analysis.

Regarding the dRGT massive gravity, we consider the static and spherically symmetric solutions of these theories, and we investigate the ef\mbox{}fectiveness of the Vainshtein screening mechanism. We focus on the branch of solutions in which the Vainshtein mechanism can occur, and we truncate the analysis to scales below the gravitational Compton wavelength. We consider the weak f\mbox{}ield limit for the gravitational potentials, while keeping all non-linearities of the mode which is involved in the screening.

We determine analytically the number and properties of local solutions which exist asymptotically on large scales, and of local (inner) solutions which exist on small scales. Moreover, we analyze in detail in which cases the solutions match in an intermediate region. We show that asymptotically f\mbox{}lat solutions connect only to inner conf\mbox{}igurations displaying the Vainshtein mechanism, while non asymptotically f\mbox{}lat solutions can connect both with inner solutions which display the Vainshtein mechanism, or with solutions which display a self-shielding behaviour of the gravitational f\mbox{}ield. We show furthermore that there are some regions in the parameter space of the theory where global solutions do not exist, and characterize precisely in which regions the Vainshtein mechanism takes place.

\clearemptydoublepage

\clearemptydoublepage

\chapter*{Author's note}

\pagestyle{fancyplain}
\lhead[\fancyplain{}{\rm\thepage} \mbox{  } \mbox{  } \mbox{  }  \sl{Note from the author}]{\fancyplain{}{}}
\lfoot[\fancyplain{}{}]{\fancyplain{}{}}
\chead[\fancyplain{}{}]{\fancyplain{}{}}
\cfoot[\fancyplain{\rm\thepage}{}]{\fancyplain{\rm\thepage}{}}
\rhead[\fancyplain{}{}]{\fancyplain{}{\sl{Note from the author} \mbox{  } \mbox{  } \mbox{  } \rm\thepage}}
\rfoot[\fancyplain{}{}]{\fancyplain{}{}}

This thesis is the result of the work I did during my Ph.D.~, which was a joint (co-tutoring) Ph.D.~between the University of Milan, Italy and the University of Portsmouth, UK. This version of the thesis is a slightly revised version of the thesis I submitted to the University of Portsmouth in December 2013. \\

\noi The chapter \ref{Nested branes with induced gravity} is based on the papers
\begin{itemize}
 \item F.~Sbis\`a and K.~Koyama, \emph{The critical tension in the Cascading DGP model}, submitted to JCAP (2014); ArXiv: 1405.7617 [hep-th]
 \item F.~Sbis\`a and K.~Koyama, \emph{Perturbations of Nested Branes With Induced Gravity}, JCAP \textbf{06}, 029 (2014); ArXiv: 1404.0712 [hep-th].
\end{itemize}
while the chapter \ref{Vainshtein Mechanism in Massive Gravity} is based on the paper
\begin{itemize}
 \item F.~Sbis\`a, G.~Niz, K.~Koyama and G.~Tasinato, \emph{Characterizing Vainshtein solutions in massive gravity}, Phys.~Rev.~D \textbf{86}, 024033 (2012); ArXiv: 1204.1193 [hep-th].
\end{itemize}

\clearemptydoublepage

\clearemptydoublepage

\pagestyle{fancyplain}
\lhead[\fancyplain{}{\rm\thepage} \mbox{  } \mbox{  } \mbox{  }  \sl{Contents}]{\fancyplain{}{}}
\lfoot[\fancyplain{}{}]{\fancyplain{}{}}
\chead[\fancyplain{}{}]{\fancyplain{}{}}
\cfoot[\fancyplain{\rm\thepage}{}]{\fancyplain{\rm\thepage}{}}
\rhead[\fancyplain{}{}]{\fancyplain{}{\sl{Contents} \mbox{  } \mbox{  } \mbox{  } \rm\thepage}}
\rfoot[\fancyplain{}{}]{\fancyplain{}{}}

\tableofcontents

\clearemptydoublepage

\listoffigures

\clearemptydoublepage

\chapter*{Conventions}

\addcontentsline{toc}{chapter}{Conventions}

\pagestyle{fancyplain}
\lhead[\fancyplain{}{\rm\thepage} \mbox{  } \mbox{  } \mbox{  }  \sl{Conventions}]{\fancyplain{}{}}
\lfoot[\fancyplain{}{}]{\fancyplain{}{}}
\chead[\fancyplain{}{}]{\fancyplain{}{}}
\cfoot[\fancyplain{\rm\thepage}{}]{\fancyplain{\rm\thepage}{}}
\rhead[\fancyplain{}{}]{\fancyplain{}{\sl{Conventions} \mbox{  } \mbox{  } \mbox{  } \rm\thepage}}
\rfoot[\fancyplain{}{}]{\fancyplain{}{}}

Unless explicitly said otherwise, throughout this thesis we use the following conventions:\\

For metric signature, connection, covariant derivative, curvature tensors and Lie derivative we follow the conventions of Misner, Thorne and Wheeler \cite{MisnerThorneWheeler}. Explicitly, the metric signature is the ``mostly plus'' one
\beq
\tag{0.1}
\eta_{AB} = diag ( -1, +1, \dots , +1 ) \quad ,
\eeq
so for example a spacelike unit vector $\mbf{n}$ has positive norm ($n_A n^A = +1$). In a metric manifold with metric $\mbfg$ we will always use the unique symmetric connection compatible with the metric (Levi-Civita connection). The sign convention for the covariant derivative associated to the connection is
\beq
\tag{0.2}
\nabla_{\!\! A} \, V^{B} = \de_{A} V^{B} + \G^{B}_{AL} V^{L} \qquad \qquad \nabla_{\!\! A} \, \omega_{B} = \de_{A} \, \omega_{B} - \G^{M}_{AB} \, \omega_{M} \quad ,
\eeq
and the Riemann curvature tensor is defined as
\beq
\tag{0.3}
R^{A}_{\,\,\, BMN} = \de_M \G^{A}_{NB} - \de_N \G^{A}_{MB} + \G^{A}_{ML} \G^{L}_{NB} - \G^{A}_{NL}  \G^{L}_{MB} \quad ,
\eeq
while the Ricci curvature tensor is defined as
\beq
\tag{0.4}
R_{MN} = R^{L}_{\,\,\, MLN} = \de_L \G^{L}_{MN} - \de_N \G^{L}_{ML} + \G^{S}_{SL} \G^{L}_{MN} - \G^{S}_{NL}  \G^{L}_{SM} \quad .
\eeq
The sign convention for the Einstein equation is
\beq
\tag{0.5}
R_{MN} - \half \, R \, g_{MN} = + \, \frac{8 \pi G}{c^4} \,\, T_{MN} \quad .
\eeq
The convention for the Lie derivative of a tensor $T_{\,\,\,\, AB}^{M}$ along a vector field $V^N$ is
\beq
\tag{0.6}
\label{Liepartial}
\big( \mathcal{L}_{\mbf{V}} \mbf{T} \big)_{\,\,\,\, AB}^{M} = V^L \de_L \, T_{\,\,\,\, AB}^{M} - ( \de_L V^M ) \, T_{\,\,\,\, AB}^{L} + ( \de_A V^L ) \, T_{\,\,\,\, LB}^{M} + ( \de_B V^L ) \, T_{\,\,\,\, AL}^{M} \quad .
\eeq

When dealing with models with one or two spatial extra dimensions, 6D indices are denoted by capital letters, so run from 0 to 5; 5D indices are denoted by latin letters, and run from 0 to 4, while 4D indices are denoted by greek letters and run from 0 to 3.

We def\mbox{}ine symmetrization and antisymmetrization without normalization
\beq
\tag{0.7}
A_{(M| \cdots |N)} \equiv A_{M \cdots N} + A_{N \cdots M} \qquad \qquad A_{[M| \cdots |N]} \equiv A_{M \cdots N} - A_{N \cdots M} \quad ,
\eeq
and we indicate the trace of a rank (1,1) or (0,2) tensor by $\textrm{tr}$, so
\beq
\tag{0.8}
\textrm{tr} D^{M}_{\,\, N} = D^{L}_{\,\, L} \qquad \qquad \textrm{tr} A_{MN} = \umg{MN} \, A_{MN} \quad .
\eeq

As for notation, abstract tensors are indicated with bold-face letters, while quantities which have more than one component but are not tensors (such as coordinates for example) are expressed in an abstract way replacing every index with a dot. For example, the sextet of coordinates $X^A$ are indicated in abstract form as $\Xd$, the quintet of coordinates $\xi^a$ are indicated in abstract form as $\xid$, and the quartet of coordinates $x^\m$ are indicated in abstract form as $\xd$.

When studying perturbations, the symbol $\simeq$ indicates usually that an equality holds at linear order.

We use throughout the text the (Einstein) convention of implicit summation on repeated indices, and we use units of measure where the speed of light has unitary value $c=1$. The reduced 4D Planck mass is defined as $M_P = (8 \pi G)^{-1/2} \sim 2.43 \times 10^{18} \, \textrm{GeV}$.

\clearemptydoublepage

\clearemptydoublepage

\chapter*{Abbreviations}

\addcontentsline{toc}{chapter}{Abbreviations}

\pagestyle{fancyplain}
\lhead[\fancyplain{}{\rm\thepage} \mbox{  } \mbox{  } \mbox{  }  \sl{Abbreviations}]{\fancyplain{}{}}
\lfoot[\fancyplain{}{}]{\fancyplain{}{}}
\chead[\fancyplain{}{}]{\fancyplain{}{}}
\cfoot[\fancyplain{\rm\thepage}{}]{\fancyplain{\rm\thepage}{}}
\rhead[\fancyplain{}{}]{\fancyplain{}{\sl{Abbreviations} \mbox{  } \mbox{  } \mbox{  } \rm\thepage}}
\rfoot[\fancyplain{}{}]{\fancyplain{}{}}

Throughout this thesis we use the following abbreviations:\\

\noindent GR: General Relativity\\
FLRW: Friedmann-Lema\^{\i}tre-Robertson-Walker\\
QFT: Quantum Field Theory\\
SM: Standard Model (of particle physics)\\
DM: Dark Matter\\
CDM: Cold Dark Matter\\
4D, 5D, 6D, \ldots : four dimensional, five dimensional, six dimensional, \ldots\\
Cod-1, cod-2, cod-3, \ldots : codimension-1, codimension-2, codimension-3, \ldots\\
KK: Kaluza-Klein\\
AHDD: Arkani-Hamed-Dimopolous-Dvali\\
RS: Randall-Sundrum\\
AdS: Anti-de Sitter\\
GRS: Gregory-Rubakov-Sibiryakov\\
GN: Gaussian Normal\\
GNC: Gaussian Normal Coordinates\\
FP: Fierz-Pauli\\
vDVZ: van Dam-Veltman-Zakharov\\
BD: Boulware-Deser\\
dRGT: de Rham-Gabadadze-Tolley\\
KNT: Koyama-Niz-Tasinato\\
GLM: G\"umr\"uk\c{c}\"uo\u{g}lu-Lin-Mukohyama .

\clearemptydoublepage

\clearemptydoublepage

\pagenumbering{arabic}

\pagestyle{fancyplain}
\renewcommand{\chaptermark}[1]{\markboth{#1}{}}
\renewcommand{\sectionmark}[1]{\markright{\thesection\ #1}}
\lhead[\fancyplain{}{\rm\thepage} \mbox{  } \mbox{  } \mbox{  } \sl\leftmark ]{\fancyplain{}{}}
\lfoot[\fancyplain{}{}]{\fancyplain{}{}}
\chead[\fancyplain{}{}]{\fancyplain{}{}}
\cfoot[\fancyplain{\rm\thepage}{}]{\fancyplain{\rm\thepage}{}}
\rhead[\fancyplain{}{}]{\fancyplain{}{\sl\rightmark \mbox{  } \mbox{  } \mbox{  } \rm\thepage}}
\rfoot[\fancyplain{}{}]{\fancyplain{}{}}

\chapter{Introduction}
\label{Introduction}

The universe displays a stunning variety of physical objects and phenomena. The (almost) empty and cold intergalactic space, the region around a black hole and a planet placed in one of the arms of a spiral galaxy are very different for average density, temperature and strength of the gravitational field, and bear little resemblance one to the other. The study of these objects and their properties is without doubt very interesting and important. However, from the point of view of a cosmologist, the questions that one would like to answer are more related to how these objects formed, how long ago this happened and what will happen to them in the future. More generally, one would like to understand if the universe itself, seen as a whole physical system, has always existed or not, how old it is in the latter case, and what will its final fate be. To be able to answer these questions, one should know what are the laws that govern its evolution and be able to solve the equations of motion. However, since we are not able to handle the complexity of a system as big and complicated as the universe, we are almost forced to tackle the problem trying to find a very simplified model, which grasps the essence of the phenomena under study but is simple enough to be handled mathematically. As we shall see, this is made possible by the assumption (corroborated by the observations) that the universe is homogeneous and isotropic on very large scales. This approach has proved to be very fruitful, and has led to the so called standard cosmological model, where many observed phenomena like the redshift of distant objects, the existence and spectrum of the Cosmic Microwave Background radiation (CMB) and the relative abundance of light elements find a natural explanation.

\section{The Homogeneous and Isotropic Universe}

Despite the huge variety of physical configurations mentioned above (even if we concentrate just on mass, the average density within a galaxy is typically $10^5$ larger that the average density of the universe \cite{KolbTurner}, and in turn galaxies contain objects which are much more dense than the galactic average, such as neutron stars), observing the universe at various length scales suggests that an averaged description on very large scales may be the simplified description we are looking for. In fact, once chosen a direction in the sky and averaged the observations over a solid angle of fixed opening $\vartheta$, it can be seen that progressively increasing the value of $\vartheta$ leads to a result which is independent of the direction we choose. In other words, on large scales the observable universe seems to be (spatially) highly isotropic around us. This is suggested by the number count of galaxies we see in the sky, but is also confirmed by the counting of radio sources we can detect, by the observations of X- and $\gamma$-ray backgrounds, and expecially by the striking smoothness ($\delta T/T \lesssim 10^{-5}$) of the Cosmic Microwave Background \cite{KolbTurner}.

To be able to build a model of the universe, however, it is not enough to know how it looks like from our planet: we need more information, namely we need to know how the universe would look like from other positions as well. Since we cannot achieve that in practice, we have to make some assumptions: it is natural to assume that we don't occupy a special position in the universe (Copernican Principle), and therefore that the universe itself would look isotropic (in an averaged sense as previously mentioned) also when seen from every other point. This condition implies that, on large scales, we can describe the observable universe as being spatially homogeneous\footnote{It can be seen that isotropy from every point implies homogeneity \cite{Wald84}.} and isotropic. Being impossible to prove it directly, this assumption has to be verified \emph{a posteriori} comparing the predictions of the model we would obtain with the observations: it is indeed very well confirmed by several different kinds of observations.

In describing the dynamics of the universe as a whole, we rely heavily on the knowledge we have of physical phenomena on earth and in the solar system. It is in fact natural to start from the laws which we know describe well physics on energies/length scales we can study on and around our planet (in a lab, or with high precision measurements in the solar system), and extrapolate their validity to arbitrary large scales. We are of course not granted that this is the correct thing to do, since new degrees of freedom or even new dynamical laws may show up as we increase the length scales and the complexity of the system under study. On the other hand, it is a very reasonable guess to start with. We will therefore assume that the correct framework to use to model the universe is the one offered by Einstein's General Relativity (GR) \cite{GR}, which is currently thought to describe correctly the gravitational interaction (up to very high energies), and that gravity is the only interaction responsible for the large scale structure of the universe. To be precise, we will consider an extension of the original theory, proposed by Einstein \cite{EinsteinCosmo}, where the cosmological constant is explicitly present in the equations of motion.

In this framework, gravity is seen as a geometrical effect, and the geometrical properties of the universe are encoded in the metric tensor $\mbf{g}$. The curvature of the universe is sourced by the energy-momentum tensor of matter fields $\mbf{T}$, and is determined by the Einstein equations\footnote{We use units of measure where the speed of light $c$ is one.}
\beq
\label{Einsteineq}
\mbf{G} + \La \mbf{g} = 8 \pi G \, \mbf{T}
\eeq
where $G$ is the Newton constant, $\La$ is the so-called cosmological constant and $\mbf{G}$ is the Einstein tensor. The large scale homogeneity and isotropy suggests to ``approximate'' the exact manifold $(\mscr{M}, \mbf{g})$ which describes our universe with a homogeneous and isotropic manifold. We suppose then that $(\mscr{M}, \mbf{g})$ is locally diffeomorphic to a homogeneous and isotropic manifold $(\bar{\mscr{M}},\bar{\mbf{g}})$, where $\bar{\mbf{g}}$ is the metric on $\bar{\mscr{M}}$, and that (in a sense to be formalized later) they are very similar when we focus only on very large scales. We indicate with $\phi$ the diffeomorphism which relate the two manifolds
\beq
\phi : \bar{\mscr{M}} \rightarrow \mscr{M} \quad .
\eeq
We expect that the homogeneous and isotropic metric $\bar{\mbf{g}}$ encodes the fundamental information on the large scale geometry of the real universe, despite having (due to the high symmetry) fewer degrees of freedom compared to $\mbf{g}$. The idea is to start from the Einstein equations for $\mbf{g}$, and obtain a set of equations for $\bar{\mbf{g}}$ which can be thought of describing the large scale dynamics of the real universe. This description turns out to be mathematically tractable, and very insightful. Furthermore, this approach allows us to approximately disentangle the large scale behavior of the universe from the dynamics of small scale structures which form inside it.

\subsection{The Robertson-Walker metric}
\label{The Robertson-Walker metric}

The condition of spatial homogeneity and isotropy is in fact highly stringent, and amounts to ask that there exist a class of observers (\emph{isotropic observers}) whose trajectories fill the universe, and to each of whom the universe appears spatially isotropic at every time. This implies that there is a natural 3+1 splitting of the spacetime $\bar{\mathscr{M}}$, and more precisely that $\bar{\mathscr{M}}$ can be foliated in three-dimensional spatial hypersurfaces $\Sigma_t$, parametrized by a timelike coordinate $t$, which have constant three-dimensional curvature \cite{Wald84}. Furthermore, it implies that each spatial hypersurface $\Sigma_t$ is locally isomorphic either to a 3-sphere (positive curvature), or to 3D flat Euclidean space (zero curvature), or to a 3-hyperboloid (negative curvature): it follows that on each $\Sigma_t$ we can choose a reference system such that locally the three-dimensional metric can be written as
\beq
\label{3DmetricmathcalSKR}
ds_{(3)}^2 = dR^2 + \mathcal{S}^{2}(K_t , R) \, \big( d\theta^2 + \sin^2 \!\theta \, d\phi^2 \big) \quad ,
\eeq
where the function $\mathcal{S}$ is defined as
\begin{equation}
\label{mathcalSKfunction}
 \mathcal{S}(K , R) \quad \left\{
  \begin{aligned}
   \phantom{i}&= \sqrt{K}^{\, -1} \, \sin \Big( R \, \sqrt{K} \Big) & K&>0 \\
   \phantom{i}&=R & K&=0 \\
   \phantom{i}&= \sqrt{\abs{K}}^{\, -1} \, \sinh \Big( R \, \sqrt{\abs{K}} \Big) & K&<0
  \end{aligned}
\right.
\end{equation}
and $K_t$ is a dimensionful\footnote{For the sake of precision, $K_t$ has inverse length squared dimension.} quantity which is proportional to the 3-dimensional curvature of the hypersurface $\Sigma_t$. Note that, despite in (\ref{mathcalSKfunction}) the function $\mathcal{S}$ is defined piecewise, it is actually smooth both in $R$ and in $K$.

To choose a reference system on $\bar{\mathscr{M}}$, we can start from a hypersurface $\bar{\S} = \S_{\bar{t}}$ of three-dimensional curvature $\bar{K}$, and follow the trajectories of the isotropic observers, assigning a fixed spatial coordinate label to each observer. Labelling each hypersurface by the proper time of a clock carried by any of the isotropic observers (by homogeneity, they all observe the same proper time difference) and assuming that the sign of $K_t$ does not change with time, we arrive at the following line element
\beq
\label{RWmetric0}
ds^2 = -dt^2 + A^2(t) \, \bigg[ dR^2 + \mathcal{S}^{2}(\bar{K} , R) \, \big( d\theta^2 + \sin^2 \!\theta \, d\phi^2  \big) \bigg]
\eeq
where $\theta$ and $\phi$ are angular coordinates (therefore dimensionless), $R$ is a (dimensionful) radial coordinate, and $A$ is a dimensionless function of $t$. In this system of reference the isotropic observers are by construction at rest, and therefore the reference system is called the \emph{comoving reference}. Note that it is possible to take into account all the cases corrispondent to $\bar{K} > 0$ in a unified way, and the same is true for the cases corrispondent to $\bar{K} < 0$. In fact, the 3D curvature of the hypersurface $\bar{\S}$ defines a characteristic length scale
\beq
\bar{R} \equiv \sqrt{\abs{\bar{K}}}^{\, -1}
\eeq
which can be used as a ``ruler'' for spatial measurements: we can decide to use as radial coordinate the adimensional ratio
\beq
\ch = \frac{R}{\bar{R}} = \sqrt{\abs{\bar{K}}} \, R \quad ,
\eeq
and absorb in $A(t)$ the multiplicative constant which is produced in the 4D line element defining
\beq
a(t) \equiv \frac{A(t)}{\sqrt{\abs{\bar{K}}}} \quad .
\eeq
With this choice, and defining in the spatially flat case $\ch \equiv R$ and $a(t) \equiv A(t)$, we arrive at the line element
\beq
\label{RWmetric1}
ds^2 = -dt^2 + a^2(t) \bigg[ d \ch^2 + S_{k}^{\, 2}(\ch) \, \big( d\theta^2 + \sin^2 \!\theta \, d\phi^2  \big) \bigg]
\eeq
where $k$ can take on only the values $+1$, $0$ and $-1$, and the function $S_{k}$ is defined as
\begin{equation}
\label{Skfunction}
 S_{k}(\chi) \quad \left\{
  \begin{aligned}
   \phantom{i}&=\sin \chi & k&=+1 \\
   \phantom{i}&=\chi & k&=0 \\
   \phantom{i}&=\sinh \chi & k&=-1 \quad .
  \end{aligned}
\right.
\end{equation}
The metric associated to the line elements (\ref{RWmetric0}) and (\ref{RWmetric1}) is called the \emph{Robertson-Walker metric}: as we mentioned previously, in the case $k = +1$ the 3D spatial hypersurfaces $\Sigma_t$ are locally isomorphic to 3-spheres, while in the case $k = 0$ they are locally isomorphic to a 3D flat Euclidean space and in the case $k = -1$ they are locally isomorphic to 3D hyperboloids. If we assume that the isomorphism is global, then the universe is called closed in the case $k = +1$ (and $\ch$ is defined for $0 \leq \ch \leq \pi$), flat in the case $k = 0$ ($0 \leq \ch < +\infty$) and open in the case $k = -1$ ($0 \leq \ch < +\infty$). Note that in the cases $k = \pm 1$ the coordinate $\ch$ is dimensionless while $a(t)$ is dimensionful, while the opposite happens in the $k = 0$ case.

It is useful sometimes to single out the part of the metric which is independent of the timelike coordinate $t$ (usually termed \emph{cosmic time}) and define \emph{spatial metric} the three-dimensional metric $\g_{ij}$ such that the Robertson-Walker line element takes the form
\begin{equation*}
ds^{2}= - dt^{2} + a^{2}(t) \, \gamma_{ij}(x) \,dx^{i}dx^{j} \quad .
\end{equation*} 
This metric defines a notion of distance on the three-dimensional hypersurfaces: taken any two points $P_1$ and $P_2$ on the same $\S_t$, the distance calculated using $\gamma_{ij}$ is called \emph{comoving distance} of the two points, and is indicated with $d_{C}(P_1,P_2)$. The spatial distance between $P_1$ and $P_2$ which is effectively measured is the one calculated using the full metric $g_{ij}$: it is called (instantaneous) \emph{physical distance} and is related to the comoving distance via the relation $d_{F}(P_1,P_2)= a(t)\,\, d_{C}(P_1,P_2)$. Note furthermore that redefining the time coordinate in the following way
\beq
\eta(t) \equiv \int^{t} \frac{d\xi}{a(\xi)}
\eeq
it is possible to factorize the dependence on the function $a$ and put the metric above in the form
\begin{equation*}
ds^{2}= a^{2}(\eta) \, \big( -d\eta^{2} + \gamma_{ij}(x) \,dx^{i}dx^{j} \big) \quad .
\end{equation*}
The time coordinate $\eta$ defined in this way is called \emph{conformal time}. A yet different way to write the line element (\ref{RWmetric1}) is obtained redefining the radial coordinate $\ch \rightarrow r$ in order to have the angular part of the metric independent from $k$: the line element in this coordinate system reads
\beq
\label{RWmetric2}
ds^2 = -dt^2 + a^2(t) \bigg[ \frac{dr^2}{1 - k r^2} + r^2 \big( d\theta^2 + \sin^2 \!\theta \, d\phi^2  \big) \bigg]
\eeq
where the radial coordinate $r$ is defined on $0 \leq r < 1$ in the case of positive spatial curvature ($k = + 1$), while it is defined on $0 \leq r < +\infty$ in the case of vanishing or negative spatial curvature ($k = 0, -1$). Note that, in the case of positive spatial curvature, this reference system covers only half of the space (it covers half of the three-dimensional spheres).

We can see that the requirement of homogeneity and isotropy drastically reduces the number of degrees of freedom: once specified the geometry of the spatial hypersurfaces (\ie specified if $k= 0$, $k=1$ or $k = -1$), the metric has just one degree of freedom, the \emph{scale factor} $a(t)$, which depends on just one of the four spacetime coordinates. The evolution of the universe is then constrained by the condition of homogeneity and isotropy to be just a uniform expansion/contraction of the three-dimensional spacelike hypersurfaces, encoded in the evolution of the scale factor. Its dynamics is determined by appropriate equations that are to be derived from the exact Einstein equations using the hypothesis of large scale homogeneity and isotropy.

\subsection{Perfect f\mbox{}luids}

The source term of the dynamical equations for the scale factor will involve (as we will see later) a spatial averaging procedure on the exact energy-momentum tensor of the universe. It is therefore important to understand what are the implications of spatial homogeneity and isotropy for the source term of Einstein equations.

Let us consider in general a tensor field $\bar{\mbf{T}}$ of type $(1,1)$ defined on $\bar{\mscr{M}}$ and let's impose the condition of homogeneity and isotropy on $\bar{\mbf{T}}$. This implies that, in the comoving reference, the components of the tensor depend just on the time coordinate; furthermore, the tensor is diagonal and its spatial components satify
\beq
\bar{T}_{i}^{\,\,j}(t) \propto \d_{i}^{\,\,j} \quad ,
\eeq
or equivalently, lowering one index,
\begin{align}
\label{homisofluid}
\bar{T}_{0 0} &= \mcal{F}(t) & \bar{T}_{0 i} &= 0 & \bar{T}_{ij} &= \mcal{G}(t) \, g_{ij}(t, \vecx)
\end{align}
where $\mcal{F}$ and $\mcal{G}$ are completely generic real functions. Note that the condition of homogeneity and isotropy does not tell anything about the time evolution of $\mcal{F}$ and $\mcal{G}$ and if they are independent one from the other or not. If we identify $\bar{\mbf{T}}$ with the stress energy tensor, then the interdependence between $\mcal{F}$ and $\mcal{G}$ is encoded in the continuity equation (which is implied by the equations of motion), and in the microscopic description of the system.

There is a well known class of physical systems which is described by an energy-momentum tensor of this form: \emph{perfect fluids}. A fluid living in a Minkowski spacetime is said to be perfect if, whatever its four-velocity profile $u^{\m}(x)$, the heat conduction is always absent and there are no shear stresses (\ie its viscosity is zero). Therefore (apart from its velocity profile) a perfect fluid is characterized by only two macroscopic quantities, its rest frame energy density $\rho(x)$ and pressure $p(x)$: this implies that its energy momentum tensor is of the form
\beq
\label{fluperMink}
T^{\mu\nu}= \big( \rho +p \big) u^{\mu} u^{\nu} + p\,\eta^{\mu\nu} \quad ,
\eeq
where $\rho$, $p$ and $u^{\mu}$ generally depend on all the four coordinates $x^{\mu}$. It follows that a perfect fluid living in a curved spacetime has a (lowered indices) energy-momentum tensor of the form
\beq
\label{flupercurved}
T_{\mu\nu}= \big( \rho +p  \big) u_{\mu} u_{\nu} + p \, g_{\mu\nu} \quad ,
\eeq
where $\r$, $p$ and $u_{\mu}$ depend on all the four coordinates $(t, \vecx)$. Considering now the equation (\ref{homisofluid}), we can see that an homogeneous and isotropic fluid always behaves as a perfect fluid which is at rest in the comoving reference and whose energy density and pressure are constant on the spatial hypersurfaces $\S_t$. If we relax the assumption of homogeneity and isotropy, it is not necessarily true that we can describe the matter-energy content of the universe as a (inhomogeneous and anisotropic) perfect fluid with nontrivial velocity profile, because heat conduction and viscosity may play a role. However, it turns out to be very fruitful to model the energy-matter content of the universe as a \emph{collection} of perfect fluids, so it is worthwhile to spend some more words on it. 

\subsection{Matter, radiation and vacuum}

There are many physical systems that can be macroscopically described as fluids. Their (different) microscopic structure shows up at macroscopic level via relations, which are called \emph{equations of state}, that link together the thermodynamical parameters of the fluid. A particular importance in cosmology is given to perfect fluids which are characterized by the very simple equation of state $p = w \rho$, where $w$ is a constant. Among this class of fluids, there are three special cases which deserve a more detailed discussion: the cases $w=0$, $w=1/3$ and $w= -1\,$.

The case $w=0$ is suitable to describe a gas of nonrelativistic particles, in other words particles whose kinetic energy is negligible compared to their rest energy, and can be used for example to describe the matter which constitutes galaxies. The case $w=1/3$ instead is suitable to describe a gas of ultrarelativistic particles, that is particles whose rest energy is negligible with respect to their kinetic energy, such as neutrinos. Note that also a system like the electromagnetic field can be described as a perfect fluid with the equation of state $p = (1/3) \rho$: this follows from the well known fact that the energy-momentum tensor of the electromagnetic field is traceless, and is consistent with the idea that we may see the electromagnetic field as a collection of photons (which are by definition ultrarelativistic being massless). Finally, the case $w=-1$ can be used to describe the so-called \emph{vacuum energy}. Quantum Field Theory suggests that also the vacuum state (that is, a configuration devoid of particles) possesses a non-zero energy (which is actually divergent unless we put a cutoff to the theory): the contribution of a quantum field to the classical energy-momentum tensor is expected to be the expectation value $\langle 0 \! \mid \! \hat{T}^{\mu\nu} \! \mid\! 0 \rangle$ on the vacuum state $\mid\! 0 \rangle$. In flat space, the requirement that the quantum theory and likewise the vacuum state are invariant with respect to Lorentz transformations imply that the above mentioned expectation value has the form $\langle 0 \!\mid \hat{T}_{\mu\nu}\mid \! 0 \rangle \propto \,\e_{\mu\nu}\,$; it follows that in a curved spacetime
\beq
\langle 0 \!\mid \hat{T}_{\mu\nu}\mid \! 0 \rangle \propto \, g_{\mu\nu} \quad .
\eeq
We can conclude that vacuum energy can be treated at semiclassical level as a perfect fluid with the equation of state $p(x) = - \rho(x)\,$. Note that the cosmological constant term in equation (\ref{Einsteineq}) is precisely of the form above. The cosmological constant in fact can be alternatively thought of as a second characteristic energy/length scale of the gravitational field (beside $G$) which shows up only at ultra large scales, or from another point of view can be thought of as describing the semiclassical effect of vacuum energy of quantum fields in the cosmological context. From the latter point of view, it is more logical to consider it as a source term, and move the cosmological constant term to the right hand side of the Einstein equations defining
\beq
T^{(\La)}_{\m\n} = - \frac{\La}{8 \pi G} g_{\m\n} \quad .
\eeq
This energy-momentum tensor is characterized by a pressure $p = - \La/8 \pi G$ and an energy density $\rho = \La/8 \pi G$. In the following we adopt this point of view and include the contribution of a (possibly non-zero) cosmological constant in the total energy-momentum tensor: we don't constrain \emph{a priori} the sign of $\La$ and allow it to have positive or negative value.

\section{The Friedmann-Lema\^{\i}tre-Robertson-Walker model}
\label{The Friedmann-Robertson-Walker model}

Before deriving the equations that govern the evolution of the scale factor, it is useful to specify how the large scale spatial homogeneity and isotropy is expressed in our formalism. Using the diffeomorphism $\phi$ which maps the reference manifold $(\bmscrM , \bmbfg )$ into the manifold $(\mscrM , \mbfg )$ which describes the ``real'' universe (or at least its observable part), we can pull-back the exact metric $\mbfg$ obtaining the metric $\phi_{\star} (\mbfg)$ which is defined on $\bmscrM$. We can define now the deviation from spatial homogeneity and isotropy as the difference of the two metrics on $\bmscrM$, which in comoving coordinates reads as
\beq
h_{\m \n}(t, \vec{x}) = \big( \phi_{\star} (\mbfg) \big)_{\m \n}(t, \vec{x}) - \bar{g}_{\m \n}(t, \vec{x}) 
\eeq 
where $t$ is the cosmic time and $\vec{x}$ indicates the spatial coordinates on the spacelike hypersurfaces $\S_t$. Note that since homogeneity and isotropy provide a natural way of splitting space and time on $\bmscrM$ (which is explicitly realized in the comoving reference), it makes sense to talk about operations which involve just the spatial coordinates. The tensor $h_{\m \n}(x)$ is not a perturbation and does not need to be small, actually it can be huge: the condition of large scale spatial homogeneity and isotropy is translated in the fact that $h_{\m \n}(x)$ gives approximately a vanishing contribution to the Einstein tensor when the latter is averaged on spatial volumes $\mathscr{V}$ large enough to render the homogeneity apparent (to be quantitative, spheres with diameter bigger than\footnote{One megaparsec (Mpc) is approximately $3.1 \times 10^{19} \, \textrm{km}$.} $100$ Mpc \cite{KolbTurner}). To be more precise, let's indicate with $\hat{\mbfG}$ the operator which associates to any metric the Einstein tensor built with the metric itself, and for every point $\vecx$ on $\Sigma_t$ let's consider a large enough volume $\mathscr{V}(\vecx)$ centered around it. The large scale spatial homogeneity and isotropy at a fixed time $t$ is expressed by the fact that, performing some spatial average over $\mathscr{V}(\vecx)$ of the pull-back of the ``real'' Einstein tensor, one gets approximately the Einstein tensor built with the homogeneous and isotropic metric
\begin{align}
\label{buh1}
\langle \big[ \phi_{\star} \big( \hat{\mbfG}(\mbfg) \big) \big]_{00} \rangle_{\mathscr{V}(\vecx)} &\simeq  \big( \hat{\mbfG} (\bmbfg) \big)_{00}(t, \vecx) \\[2mm]
\label{buh2}
\langle \textrm{tr} \big[ \phi_{\star} \big( \hat{\mbfG}(\mbfg) \big) \big]_{ij} \rangle_{\mathscr{V}(\vecx)} &\simeq  \textrm{tr} \big( \hat{\mbfG} (\bmbfg) \big)_{ij}(t, \vecx) \quad .
\end{align}
Here $\textrm{tr} [ \phantom{b} ]_{ij}$ stands for the trace over spatial components. Imposing that the large scale homogeneity and isotropy holds at every $t$, amounts to asking that the equations above hold at every $t$. This implicitly defines the time evolution of the scale factor: to obtain it, we should calculate the evolution of the full metric and then take the spatial average at every time. However, this is not doable in practice, and we would like to obtain some dynamical (differential) equations for the scale factor itself. Therefore, we consider the equations
\begin{align}
\label{Friedmann complicated1}
\big( \hat{\mbfG} (\bmbfg) \big)_{00}(t, \vecx) &= 8 \pi G \, \langle \big( \phi_{\star} (\mbfT) \big)_{00} \rangle_{\mathscr{V}(\vecx)} \\[2mm]
\label{Friedmann complicated2}
\textrm{tr} \big( \hat{\mbfG} (\bmbfg) \big)_{ij}(t, \vecx) &= 8 \pi G \, \langle \textrm{tr} \big( \phi_{\star} (\mbfT) \big)_{ij} \rangle_{\mathscr{V}(\vecx)}
\end{align}
which are written in terms of the scale factor, its derivatives and the averaged energy-momentum tensor. Note that these equations are not exactly compatible with the validity of (\ref{buh1})-(\ref{buh2}) at every time: if we start at time $t_i$ with a scale factor which satisfies (\ref{buh1})-(\ref{buh2}), its time evolution according to (\ref{Friedmann complicated1})-(\ref{Friedmann complicated2}) will not exactly satisfy (\ref{buh1})-(\ref{buh2}) at subsequent times. In other words, the time evolution of the complete metric (including deviations from from homogeneity and isotropy) does not commute with the operation of spatial averaging. The actual difference depends on the explicit form of the real metric as well as the details of the spatial averaging procedure. We decide to neglect this difference for the moment, therefore studying the evolution of the scale factor according to (\ref{Friedmann complicated1})-(\ref{Friedmann complicated2}), leaving the possibility to study the effect of this approximation later.

\subsection{The Friedmann equations}

We define $\bar{T}_{\m\n}(t, \vecx)$ as the homogeneous and isotropic tensor (therefore of the form (\ref{homisofluid})) whose non-zero components are obtained by spatial averaging the pullback of the real energy-momentum tensor
\begin{align}
\label{defbarT00}
\bar{T}_{00}(t, \vecx) &\equiv \langle \big( \phi_{\star} (\mbfT) \big)_{00} \rangle_{\mathscr{V}(\vecx)} \\[2mm]
\label{defbarTij}
\textrm{tr}\, \bar{T}_{ij}(t, \vecx) &\equiv \langle \textrm{tr} \big( \phi_{\star} (\mbfT) \big)_{ij} \rangle_{\mathscr{V}(\vecx)} \quad .
\end{align}
Note that we can then write the equations (\ref{Friedmann complicated1})-(\ref{Friedmann complicated2}) in a more familiar way as
\beq
\big( \hat{\mbfG}(\bmbfg) \big)_{\m \n} =  8 \pi G \, \bar{T}_{\m \n}
\eeq
since, out of the 10 components of this equation, just two of them are linearly independent due to the high symmetry of the system. Taking a suitable linear combination of these two equations one gets the \emph{Friedmann equations}
\bigskip
\begin{align}
\label{Fried}
\Big( \dfrac{\dot{a}}{a} \Big)^{\!2} \! & = \, \dfrac{8 \pi G}{3}\, \rho -\dfrac{k}{a^{2}}\\[5mm]
\label{Friedacc}
\dfrac{\ddot{a}}{a}\,\, & = \, - \dfrac{4 \pi G}{3} \, \left( \rho+3 p\right) \quad ,
\end{align}
and it is customary to refer to the first one simply as the Friedmann equation, and to the second one as the acceleration equation\footnote{We indicate derivatives with respect to the cosmic time with an overdot $\dot{a} \equiv da/dt$.}. Note that these two equations imply the \emph{continuity equation} 
\beq
\label{continuity}
\dot{\rho}\,=-3\,\dfrac{\dot{a}}{a}\,(\rho+p) \quad ,
\eeq
which actually expresses the fact that energy is conserved and can be obtained from $\nabla_{\!\mu} \bar{T}^{\mu\nu} = 0\,$. It is customary to define the \emph{Hubble parameter}
\beq
H(t) \equiv \frac{\dot{a}(t)}{a(t)}
\eeq
and the \emph{deceleration parameter}
\beq
q(t) \equiv - \frac{a(t) \ddot{a}(t)}{\dot{a}^2(t)}
\eeq
which are independent of the overall normalization of the scale factor. The Hubble parameter has the dimension of inverse time, and its value today $H_0$ can be taken to be a rough measure of the inverse of the age of the universe, as we shall see. It is also useful to define the \emph{critical density} of the universe $\:\rho_{crit}\equiv 3H^{2}\!/\,8 \pi G$ (which is a time dependent quantity) and the \emph{density parameter} $\:\Omega \equiv \rho / \rho_{crit}\,$: using these two quantities, the Friedmann equation reads
\beq
\label{denspar}
\Omega(t)\,-1=\dfrac{k}{a^{2}(t) H^{2}(t)} \quad ,
\eeq
and it is easy to see that the sign of $k$ is determined by the fact that $\rho$ is larger, smaller or equal to the critical density. In fact we have
\begin{alignat*}{2}
 \rho &<\rho_{crit} & \quad \Leftrightarrow \quad &k<0\\
 \rho &=\rho_{crit} & \quad \Leftrightarrow \quad &k=0\\
 \rho &>\rho_{crit} & \quad \Leftrightarrow \quad &k>0 \quad ,
\end{alignat*}
and this implies that the spatial geometry of the universe is directly linked to the total value of the density of energy (relatively to the the square of the Hubble parameter). 

To study the evolution of the scale factor, we should solve equations (\ref{Fried})-(\ref{Friedacc}) with appropriate initial conditions. This system of differential equations is however not closed, since there are two equations and three unknowns ($a$, $\rho$ and $p$): to be able to solve it, we need an additional equation, such as one which tells us how the average pressure $p$ of the universe is related to the average energy density $\r$ and to the scale factor $a$. If we knew the precise distribution and thermodynamic properties of all matter in the universe, we may construct an equation of state $p = p (\rho, a)$ which expresses the ``global'' thermodynamic properties of the universe. In practice, we model the matter/energy content of the universe as the sum of few contributions whose thermodynamic properties are simple and easy to handle. In fact, we consider a model in which the universe is filled with three components, which are nonrelativistic matter (which from now on will be simply called ``matter''), radiation (which comprises also ultrarelativistic matter) and vacuum energy. As said previously, all these components are perfect fluids which obey the simple equation of state $p = w \rho$ with $w$ respectively equal to $0$, $1/3$ and $-1$. Note that the evolution of the scale factor influences differently the energy density of every component since the continuity equation implies that
\beq
\rho (t) \propto a^{-3(1+w)}(t) \quad .
\eeq
In particular, for matter the energy density scales as $a^{-3}$, \ie inversely proportional to the spatial volume, while for radiation we have $\rho \propto a^{-4}$, which is consistent with idea that a dilatation/contraction of the spatial volume influences both the number density and the wavelength of photons. Instead, the dilatation/contraction of the spatial volume does not influence the energy density of the vacuum. It follows that, in order to determine the evolution of scale factor and therefore the history of the universe, it is essential to know not only the overall energy density, but also the relative abundances of the three different components.

Note that, once we specify the composition of the universe thereby fixing its equations of state, in principle to solve the system (\ref{Fried})-(\ref{Friedacc}) we need the initial conditions\footnote{We indicate with the pedix $\phantom{i}_0$ the quantities evaluated today.} $a_0$, $\dot{a}_0$, $k$, $\rho_0^{M}$, $\rho_0^{R}$, $\rho_0^{\La}$. However, the overall value of the scale factor is not physically observable, so to find $H(t)$, $\rho^{M}(t)$, $\rho^{R}(t)$ and $\rho^{\La}(t)$ it is enough to know $H_0$, $k$, $\rho_0^{M}$, $\rho_0^{R}$, $\rho_0^{\La}$. A nice way to parametrize the initial conditions for the Friedmann equations, and therefore to parametrize the cosmological models, is to introduce separate density parameters for every component type of perfect fluid which composes the energy-momentum tensor: we define
\beq
\label{densparameters}
\Omega_{M}(t) \equiv \frac{8 \pi G}{3} \, \frac{\rho^{M}}{H^{2}} \quad , \quad \Omega_{R}(t) \equiv \frac{8 \pi G}{3} \, \frac{\rho^{R}}{H^{2}} \quad , \quad \Omega_{\La}(t) \equiv \frac{8 \pi G}{3} \, \frac{\rho^{\La}}{H^{2}} \quad .
\eeq
It is also useful to incorporate the dependence on the sign of the spatial curvature in another density parameter, which however does \emph{not} come from an energy density and is therefore only a way of keep track of the spatial curvature: we define
\beq
\label{curvdensparameter}
\Omega_{K}(t) \equiv - \frac{k}{a^{2} H^{2}} \quad .
\eeq
In term of these cosmological parameters the Friedmann equations take the suggestive form
\begin{align}
\label{Friedmanneqsdensityparameters}
1 & = \Omega_{M}(t) + \Omega_{R}(t) + \Omega_{\La}(t) +\Omega_{K}(t) \\[2mm]
q(t) & = \frac{1}{2} \, \Omega_{M}(t) + \Omega_{R}(t) - \Omega_{\La}(t) \quad .
\end{align}

\subsection{The expanding universe}

Note that, even if we knew exactly the spatial geometry of the universe and the average energy density and pressure at a fixed time (for example at the present time), we couldn't infer from the Friedmann equations if the universe is expanding or contracting. To obtain this information, we should observe the universe at different times, and for example study if the average energy density has increased or decreased. However, there is a much more straightforward way to infer if the universe is expanding or contracting: the evolution of the scale factor influences the propagation of particles and electromagnetic radiation, since the scale factor is present in the geodesic equation which descibe their propagation. If we know the properties of the radiation when it was emitted from a distant body, we can obtain information on how the universe evolved during the propagation of the radiation by studying how its properties changed when we receive it. In this case, we don't need to observe the universe at different times, but it is enough to observe at the present time the radiation coming from distant objects. Furthermore, since the Friedmann-Lema\^{\i}tre-Robertson-Walker model (provided with information about the composition of the universe) gives very distinctive observational features in the propagation of matter and radiation, observing the radiation from far away objects provides a strong test of the validity of the assumptions we made, and on the validity of the model itself.

\subsubsection{The cosmological redshift}

Consider a photon (a light ray in practice) which is emitted in the comoving reference at cosmic time $t_e$ with frequency $\omega_e$: its propagation is described by null geodesics, and solving the geodesic equation it is easy to see \cite{CarrollBook} that the frequency for the same photon observed in the same reference at time $t_{r} > t_e$ is
\begin{equation}
\label{cosmoredshift}
\omega(t_{r}) = \dfrac{a(t_e)}{a(t_{r})} \,\, \omega(t_e) \quad .
\end{equation}
The expansion/contraction of the universe therefore determines a shift in the frequency of the electromagnetic radiation between its emission (for example by a galaxy) and its detection (for example by a telescope): the received frequency is lower than the emitted one if the universe expands, while it is higher if the universe contracts. Note that, crucially, this shift is frequency-independent, so photons of different initial frequency are shifted by the same amount during the propagation. The quantity used to express a generic frequency shift is the \emph{redshift} $z$ defined as $z \equiv \frac{\lambda_{r}-\lambda_e}{\lambda_e}$, where $\lambda$ is the wavelength of the radiation: the redshift due to the cosmological expansion is called \emph{cosmological redshift} and reads
\begin{equation*}
z = \dfrac{a(t_{r})}{a(t_e)} -1 \quad .
\end{equation*}
This property is very important because we know that, if we send polycromatic radiation to an atom/molecule on Earth, it absorbs/emits a very specific pattern of radiation. Indeed, the type of atom/molecule can be identified unambiguously observing the frequency pattern, and the identification is possible even studying just the relative frequency intervals in the pattern. Assuming that the laws of physics (and the constants of nature) are the same in every galaxy, we can identify the atom/molecule which emitted/absorbed a pattern of radiation in a distant galaxy studying the relative frequency intervals: from the overall shift between the received pattern and the pattern which the same atom/molecule would emit on Earth, we can obtain the redshift.

In general a frequency shift can be due to different effects, for example it can be due to the relative motion between emitter and observer (Doppler effect): when we study the radiation coming from a distant body, we expect the total redshift to include also a Doppler component due to peculiar velocities. Therefore, the Friedmann-Lema\^{\i}tre-Robertson-Walker model implies that if the universe is expanding we should observe that the radiation coming from most of the celestial bodies is redshifted, and going to higher redshifts we should observe less or none contributions from the (conventional) Doppler effect. We instead expect to observe to opposite if the universe is contracting. Experimentally, the observations are in extremely good agreement with the predictions of an \emph{expanding} Robertson-Walker universe.

\subsubsection{The Hubble's law}
\label{Hubbleslaw}

Another distinctive feature of the Friedmann-Lema\^{\i}tre-Robertson-Walker model is that one expects that the further away from us an object is, the more redshifted it appears to us. Roughly speaking, this is due to the fact that the more distant an object is, the more time it takes for its radiation to reach us: therefore the universe expands more between emission and absorption, and the cosmological redshift is bigger. However, if we want to translate this reasoning in a precise way, we have to define what we mean by ``distance'' in a cosmological context: in fact, to measure the physical distance $d_F$ defined above we should perform an instantaneous measurement, which is impossible to do in practice, while the only thing we can do in cosmology is to study the light signals which reach us after travelling throughout the universe. Therefore we define the \emph{luminosity distance} of a light source
\beq
d^2_L \equiv \frac{L}{4 \pi F} \quad ,
\eeq
where $L$ is the absolute luminosity of the source and $F$ is the energy flux measured by the observer. This definition is motivated by the fact that, in a Minkowski spacetime, the flux of incoming light is the ratio between the intrinsic luminosity and the surface area of a sphere of radius $d_F$, where $d_F$ is the (instantaneous) spatial distance between the emitter and the observer: this is just a consequence of energy conservation. Therefore in a Minkowski spacetime the luminosity distance and the instantaneous spatial distance are coincident. While in the Minkowski spacetime the cosmological redshift is by definition vanishing, we expect that in an expanding universe there is a relation between the luminosity distance of an object $d_L$ and its (cosmological) redshift $z$, and we expect that the bigger the distance the bigger the redshift.

In fact, considering for simplicity the case of a spatially flat universe, it can be shown \cite{KolbTurner} that the luminosity distance of an object is related to its cosmological redshift by
\beq
\label{hoho}
d_L(z) = H_0^{-1} \Big( z + \half (1 - q_0) z^2 + \mathcal{O}(z^3) \Big) \quad ,
\eeq
and we notice that, when the redshift is small, the luminosity distance-redshift relationship is \emph{linear}
\beq
\label{HumbleHubble}
d_L(z) \simeq H_0^{-1} z \quad .
\eeq
This relation is known as \emph{Hubble's Law}, and is indeed confirmed by observations: the geometrical explanation of the distance-redshift relation is one of the major successes of the Friedmann-Lema\^{\i}tre-Robertson-Walker model. 
Note that measurements of luminosity distances and redshifts of many objects in a suitable range of redshifts allows us to estimate both the present value of the Hubble parameter and the present value of the acceleration parameter.

\subsubsection{The expansion of the universe and thermodynamics}

Considering now the (free) motion of massive particles in the Robertson-Walker spacetime, we call the three-dimensional velocity $v^{i}$ of the particle expressed in the comoving reference \emph{peculiar velocity}: this name is motivated by the fact that $v^{i}$ is the ``excess'' (spatial) velocity of the test particle compared to the isotropic observers' one (which is zero in the comoving reference). Indicating $\abs{\vec{v}}^{2} \equiv g_{ij} v^{i} v^{j}$, it can be shown \cite{CarrollBook} that the geodesic equation implies
\begin{equation*}
\abs{\vec{v}}(t) \propto \dfrac{1}{a(t)} \quad .
\end{equation*}

This implies that, if the scale factor is increasing (and so the universe is expanding), the peculiar velocity of a particle kinematically decreases and eventually dies off, while the opposite would happen if the universe is contracting. In particular, a perfect gas of particles in thermal equilibrium in an expanding universe will get cooler and cooler, since its temperature is proportional to the average energy for degree of freedom. The cooling due to the expansion is true more generally for a collection of interacting gases of particles, included gas of massless particles (photons and neutrinos), apart possibly during phase transitions when energy and entropy are released into the system.

Note that, strictly speaking, a Friedmann-Lema\^{\i}tre-Robertson-Walker universe can never be in thermal equilibrium, since it is not stationary (in other words, the metric does not possess a time-like Killing vector field). However, if the interactions between different species occurr rapidly enough compared to the timescale of the expansion, the universe will to a good approximation evolve through a succession of nearly thermal states, with the temperature decreasing as the scale factor increases: naively, a reaction is occurring rapidly enough to mantain the thermal equilibrium if its interaction rate $\G$ satisfies \cite{KolbTurner}
\beq
\label{Desiree}
\G \gtrsim H \quad ,
\eeq
where $H$ is the Hubble parameter. If a type of particles is in thermal equilibrium with the other species (the \emph{thermal bath}) and at a certain point its interaction rates decrease so that (\ref{Desiree}) is not satisfied anymore, we say that it \emph{decouples} from the thermal bath. In particular, as can be deduced by (\ref{cosmoredshift}), a gas of non-interacting massless particles after decoupling mantains a thermal spectrum forever with temperature kinematically decreasing as $T \propto 1/a$.

\subsubsection{The universe in thermal equilibrium}

Despite the fact that the universe is markedly not in thermal equilibrium today, we have a very good reason to believe that it was in thermal equilibrium in the past. In fact, the universe is filled by an (almost perfectly) isotropic background radiation (first detected by Penzias and Wilson in 1967 \cite{PenziasWilson}, and successively studied in detail by several missions including COBE, WMAP and the recent mission PLANCK) which has an almost perfect blackbody spectrum at the temperature of $T \simeq 2.73$ K . The existence and spectrum of this Cosmic Microwave Background radiation (CMB) fits naturally in the Friedmann-Lema\^{\i}tre-Robertson-Walker cosmological model, as we shall see, and is thought to be the decoupled remnant of the radiation which was emitted when the universe was in thermal equilibrium.

The fact that the universe was in thermal equilibrium in the past, taking into account the thermodynamic considerations above, opens a exciting possibility. If we suppose that we know with precision the laws which describe particle interactions up to a given energy, then the precise knowledge of the expansion history of the universe permits us to trace back its thermal history, so long as the temperature of the universe corresponds to energy scales where we can trust our particles physics theories (to be very conservative, we can trust the Standard Model of particle physics at least till energies $\sim 1$ TeV $\simeq 1.602 \times 10^{-7}$ J). However, extreme care has to be taken when we try to extrapolate the physics we know on Earth and in the solar system to very different regimes (much bigger length-scales, very small accelerations, higher complexity of the system). It is well known that the observed (luminous) mass in galaxies is largely insufficient to explain the dynamics of stars in galaxies using Newton's laws of gravitation and inertia. A very natural assumption is to postulate that there is a substantial amount of matter which we don't observe since it does not emit light: however, cosmological constraints on the formation of light nuclear elements (together with the recent estimations of of the matter density parameter) imply \cite{KolbTurner} that most of this matter has to be non-baryonic in nature. Therefore, we are led to assume that there is a type of matter, which is termed Dark Matter (DM), which interacts very weakly with SM particles via the electromagnetic, weak and strong nuclear interactions, and makes its presence felt only via gravitational effects. Despite indirect evidence of its existence through dynamical properties of galaxies and clusters, weak gravitational lensing and cosmological structure formation, a definitive direct detection of dark matter is still missing.

\subsection{The expansion history of the universe}

In light of the considerations above, and unless we find strong evidence of the contrary, it seems reasonable to consider a cosmological model where the description of matter and radiation in cosmological context is based on the Standard Model of particle physics with the only addition of a weakly interacting, non-baryonic dark matter. We also assume that the dark matter is \emph{cold}, which means that it decoupled when it was non-relativistic. With these assumptions, since we know precisely the interaction rates of the Standard Model particles and by hypothesis the Cold Dark Matter (CDM) interacts with the SM particles only gravitationally, we can precisely reconstruct the history of the universe (and of the structures it contains) once we observationally determine the cosmological parameters. Note that, since the interaction properties of the Cold Dark Matter are very different from the ordinary baryonic matter, the details of the physical processes happening in the universe (and more generally the history of its composition) are sensitive to the relative abundance of the CDM with respect to the baryonic matter. Therefore, it is useful to divide the matter density $\r_M$ into the baryonic and the CDM contributions $\r_M = \r_{B} + \r_{C}$, and write the total matter density parameter $\Omega_{M}$ as the sum of the density parameter $\Omega_{B}$ for the baryonic matter and of the density parameter $\Omega_{C}$ for the CDM.

As we shall discuss in detail later on, the cosmological observations have reached a degree of precision which enables us to characterize precisely the values of the cosmological parameters. We now want to use the estimates (\ref{cosmoparametersobserved}) to describe qualitatively the past evolution of the universe and point out the main predictions and successes of the Standard Cosmological Model, while we will discuss in the next section the implications of these results in relation to our understanding of the universe.

\subsubsection{The evolution of the scale factor}

The information about the composition of the universe given by the estimates (\ref{cosmoparametersobserved}) implies that the energy density is positive definite: this means that the scale factor is a monotonically increasing function of the cosmic time. Note that, as already mentioned, the densities of the different components of our universe scale differently with the scale factor, and more precisely we have
\beq
\frac{\Omega_{\La}}{\Omega_{M}} \propto \Big( \frac{a}{a_0} \Big)^3 \qquad , \qquad \frac{\Omega_{M}}{\Omega_{R}} \propto \frac{a}{a_0} \quad .
\eeq
Therefore, apart from the transition periods when the energy density of two (or in principle several) components are comparable, one of the components is always much bigger than the others, and so effectively dominating the total energy density. It is then useful to solve approximately the equations for the scale factor neglecting the energy density of the components which are not dominating, and to patch together these solutions at the transition times. We will say that the universe is \emph{matter dominated} when $\Omega_{\La}$, $\Omega_{R}$ and $\Omega_{K}$ are negligible with respect to $\Omega_{M}$, and analogous definitions hold for \emph{radiation dominated}, \emph{curvature dominated} and \emph{vacuum dominated} universe. Under this approximation, we can explicitly solve the Friedmann and continuity equations for the different domination cases.

The observations then tell us that the universe was radiation dominated in the past, then at redshift $z = z_{eq} \sim 3 \times 10^3$ it became matter dominated, and it has (just) passed the transition between matter and vacuum domination, which happened at $z \sim 0.3$. Note furthermore that the pressure of matter and radiation is non-negative, while a positive vacuum energy has negative pressure: the second Friedmann equation tells us that the second derivative of the scale factor has been negative in the past till $z \sim 0.6$, and is now positive (equivalently, the deceleration parameter $q$ was positive in the past and is now negative). Therefore, the universe has been expanding in a decelerated way until very recently, and is now expanding in an accelerated way.

Following the evolution of the scale factor backwards in time, the universe seems to approach a singular state, since $a \rightarrow 0$, $\rho \rightarrow + \infty$ and the curvature of spacetime diverges: this singularity is usually called \emph{Big Bang}. Note that the Big Bang is a fictitious singularity, in the sense that we do not expect General Relativity to be a reliable description of gravity and of the geometry of spacetime when curvature and energy are so high. We expect in fact GR to be the effective theory of a quantum theory of gravity, whose details are not clear yet, and that at least at energies higher than the \emph{Planck energy} $E_{\textup{pl}} \simeq 1.2 \times 10^{19} \,\textrm{GeV}$ we cannot make reliable calculations without taking into account the quantum aspects of gravity. Nonetheless, it is useful to fix the origin of time assigning the value $t = 0$ to the fictitious singularity: with this convention, if we assume that quantum gravity effects are under control for energies below the Planck energy, then the Standard Cosmological Model describes our universe for $t \geq t_{\textup{pl}}\,$, where $t_{\textup{pl}} = 10^{-43} \, \textrm{s}$ is the \emph{Planck time}. Even if we don't know what happens before the Planck time, we may think that in some sense the Big Bang actually marks the birth of our universe. From this point of view, we can use the Friedmann equations to estimate the age of our universe. We can get an upper limit to this value extrapolating linearly the evolution of the scale factor back in time (since $\dot{a}(t)$ is negative for most of time in the past, the actual age will be lower): this procedure gives the value $H_0^{-1}$, which corresponds roughly to $10^{10}$ years. A more careful treatment using the actual solutions of the Friedmann equations shows that this rough estimate gives the correct timescale for the age of the universe.

\subsubsection{Successes of the Standard Cosmological Model}

The Friedmann-Lema\^{\i}tre-Robertson-Walker cosmological model, despite its conceptual simplicitly, is very successful in predicting several aspects of the universe which are indeed observed. As we already mentioned, the existence of cosmological redshift and the Hubble's law find a natural explanation in the FLRW model: moreover, the thermodynamic study of the matter and the radiation which fill the universe permits to draw precise conclusions concerning the abundance of the light nuclear elements, the existence and properties of the CMB and the mechanism of structure formation. We give below a brief account of these successes following \cite{KolbTurner}, where a much more detailed exposition can be found.

As the universe cools down, when the temperature\footnote{We use here units of measure where the Boltzmann costant $k_B$ has unitary value, so that we can express energy and temperature in the same unity of measure.} of the thermal bath becomes comparable to the nuclear binding energy per nucleon (1 - 8 MeV), it starts to become possible for protons and neutrons to combine into nuclei, and the nucleosynthesis of light elements slowly begins. It can be shown \cite{KolbTurner} that the formation of elements heavier than ${}^{4}_{}\textrm{He}$ is suppressed, and, since ${}^{4}_{}\textrm{He}$ is more tightly bound than D, ${}^{3}_{}\textrm{H}$ and ${}^{3}_{}\textrm{He}$, all the neutrons end up being secluded into the former: on the other hand, the weak interaction rate fixes the neutron to proton ratio to be $\sim 1/7$ when ${}^{4}_{}\textrm{He}$ is syntetized. Therefore, the FLRW model predicts that approximately the 75\% of the baryonic mass in the universe is made of Hydrogen and the 25\% of ${}^{4}_{}\textrm{He}$, while D, ${}^{3}_{}\textrm{H}$, ${}^{3}_{}\textrm{He}$ and ${}^{7}_{}\textrm{Li}$ are present just in traces: these predictions about the relative abundance of light elements are indeed confirmed by the observations.

When the temperature of the thermal bath becomes comparable to the ionization energy of the Hydrogen atom ($\sim 10$ eV), the energy of the photons starts to be not sufficient to excite the bound state nuclei-electrons, and eventually atoms start to be formed (electrons and protons \emph{recombine}). The fact that the density of free electrons drops, in turn implies that the interaction rate between photons and matter drops, and soon after the recombination the photons decouple from the baryons. Since before the decoupling the photons and the matter were in thermodynamic equilibrium, after the decoupling the photons travel freely throughout the universe mantaining a blackbody spectrum whose temperature decays like $a^{-1}$. These photons constitute the radiation which we now detect at the redshifted temperature of $T_0 \simeq 2,73$ K, that is the CMB. The CMB can therefore be considered an istantaneous pictures of how the universe was at $t \simeq 300.000 \, \textrm{yrs}$.

The CMB is strikingly smooth, in the sense that relative variations of temperature at different directions in the sky are as small as $\delta T/T \simeq 10^{-5}$. On the other hand, the presence of any structure at the decoupling should have been reflected in local variations of the temperature of the Planckian spectrum, which should have been remained imprinted in the CMB we see today: therefore, the smoothness of the CMB tells us that the structures we see nowadays were not present at decoupling, but there were very small density perturbations. In other words, the deviations from homogeneity and isotropy were small at \emph{all} scales at the decoupling. This suggests the following general picture: small density perturbations which were already present at decoupling grew because of their self-gravity and eventually formed the huge inhomogeneities we observe nowadays via the gravitational instability. A careful study of the evolution of coupled matter and gravity perturbations in the expanding universe confirms that this idea provides a consistent explanation of the mechanism of cosmological structure formation: it is worthwhile to notice that the presence of the CDM is crucial in this picture, because without the CDM the growth of baryon perturbations wouldn't have been rapid enough to be able to form structures at the present time.

\section{The late time acceleration problem}
\label{late time acceleration problem}

\subsubsection{The $\La$CDM model}

In the previous sections, we motivated the fact that a natural way to study the evolution of the universe is to consider a description averaged on very large scales, and to assume that the particle content of the universe is made up of Standard Model particles with the addition of Cold Dark Matter (CDM). Furthermore, we included in the equations of gravitation a cosmological constant term, which can be seen as the minimal modification to the original equations proposed by Einstein since it respects the structure of the original theory. It is customary to refer to this cosmological model as to the $\La$CDM model. This model is theoretically attractive because it is conceptually simple, mathematically tractable and it is testable against observations. Furthermore, it is highly successful because, apart from preserving the standard successes of the Friedmann-Lema\^{\i}tre-Robertson-Walker cosmological model, it provides a self-consistent fit to all the observational data available so far.

However, the success of the $\La$CDM model at the same time raises deep theoretical problems. In fact, despite the ability to fit the data, the values of the cosmological parameters inferred by the observations (and the picture of the universe which they suggest) are extremely puzzling from a fundamental point of view. To elucidate the nature of the problem and the possible ways out, we now first introduce the topic of the extimation of cosmological parameters, and then turn to discuss the problem of the late time acceleration of the universe.

\subsection{The composition of our universe}

The estimation of the cosmological parameters
\beq
H_{0} \qquad \Omega_{B_0} \qquad \Omega_{C_0} \qquad \Omega_{R_0} \qquad \Omega_{\La_0} \qquad \Omega_{K_0}
\eeq
is something that has to be done observationally, comparing the theoretical predictions with the observational data. The observational estimation of these parameters has recently become a very active field of research: on one hand this is due to the fact that the theoretical framework just described is flexible enough to account for different kinds of observations, but at the same time simple enough to permit its predictions to be tested with precision. On the other hand, it is due to the fact that the amount and precision of observational data has recently reached an unprecedented level. It is also a quite technical field, therefore we give in following just the basic underlying ideas.

\subsubsection{Standard candles and standard rulers}
\label{Standard candles and standard rulers}

One of the most important concepts in modern observational cosmology is the notion of \emph{standard candle} and \emph{standard ruler}. A standard candle is an (astrophysical) object whose absolute luminosity is precisely known, while a standard ruler is an absolute length scale which is accurately known and which is imprinted in one or several cosmological features. By absolute luminosity we mean the flux of energy (in form of light) per unit time across a sphere which closely surrounds the emitting object. The importance of standard candles in cosmology lies in the fact that the observed luminosity of a source is influenced both by its absolute luminosity and by the evolution history of the Hubble parameter, so if we know the absolute luminosity we can gain informations on the evolution history. Likewise, the observed length scale corresponding to the absolute length of a standard ruler is influenced by the evolution history of the Hubble parameter, and therefore a precise knowledge of the absolute length enables us to characterize the evolution history.

The astrophysical objects which come closer to be standard candles are Type Ia supernovae. They are quite rare objects, since we expect to see few of them per century in a Milky-Way-sized galaxy, but have the advantage to be very bright (their brightness is comparable to their host galaxy's one) and so potentially observable at high redshift ($z \sim 1$). This is important to test the evolution history of the Hubble parameter, as can be seen looking at (\ref{hoho}): low redshift supernovae ($z \ll 1$) enables to estimate just the Hubble parameter today, while observing also high redshift ones enables to estimate also the deceleration parameter. They are however not perfect standard candles, since nearby type Ia supernovae display a scatter of about 40\% in their peak brightness \cite{CarrollTroddenTASIlectures}. However, the observed differences in their peak luminosities turns out to be very closely correlated with observed differences in the shapes of their light curves: type Ia supernovae explosions can then be considered a one-parameter family of events, and observing both the peak brightnesses and the light curves enables to compensate for the difference and standardize their peak brightness, significantly reducing the scatter. In this sense, type Ia supernovae are ``standardizable candles''.

The standard ruler in cosmology is instead provided by the characteristic scale of acoustic oscillations in the photon-baryon fluid. As we already mentioned, before decoupling the nuclei and electrons were tightly coupled with photons: in this regime, baryons and photons moved in unison and can be treated as a single fluid \cite{Dodelson}. Since the perturbations from homogeneity and isotropy were small, it is sufficient to work at first order in perturbations, and it is useful to decompose the relative perturbation $\d$ of the density of the baryon-photon fluid in Fourier modes
\beq
\d(\eta,\vecx) = \int \! dk^3 \, \d(\eta,\veck) \, \, e^{i \veck \cdot \vecx}
\eeq
where $\eta$ indicates the conformal time. For modes inside the horizon, a Newtonian analysis suffices and it can be shown that every mode $\d(\eta,\veck)$ obeys a forced and damped harmonic oscillator equation, where the damping is due to the expansion of the universe, the forcing to the gravitational potential, and the harmonic force to the pressure exerted by the photons. Neglecting the damping term, the solution to the associated homogeneous equation is approximately given by
\beq
\label{baryonphoton1}
\d(\eta,\veck) \supset A_{k} \sin(k \, c_s \eta) + B_{k} \cos(k \, c_s \eta) \quad ,
\eeq
where $c_s$ is the sound speed of the baryon-photon fluid, while for modes inside the horizon the damping term introduces only a smooth modulation which does not significantly distort the oscillating pattern of the solution (\ref{baryonphoton1}). The coefficients $A_{k}$ and $B_{k}$ are to be determined by the initial conditions, and comparison with the CMB anisotropy spectrum tells that $A_{k} \ll B_{k}$ and $B_{k}$ is nearly independent of $k$. Therefore we approximately have a pure oscillating contribution in the density perturbations
\beq
\label{baryonphoton2}
\d(\eta,\veck) \supset B_{k} \cos(k \, c_s \eta) \quad .
\eeq
Focusing on a fixed mode $k$, this tells us that the amplitude of every mode oscillates periodically in time. Focusing on a fixed time, on the other hand, this contribution to the amplitudes of the modes displays a periodic oscillation in $k$. The acoustic oscillations of the baryon-photon fluid therefore fix a characteristic scale in Fourier space when the density perturbations are studied at a fixed time: this scale is set by the physics of a tightly coupled baryon-photon plasma, which is quite well understood, and therefore we can predict this scale with great accuracy. The periodicity scale set by the acoustic oscillations remains imprinted in both the CMB anisotropies spectrum and in the large scale distribution of galaxies.

\subsubsection{Observations and cosmological parameters}
\label{Observations and cosmological parameters}

To understand why standard candles and standard rulers can allow us to determine observationally the cosmological parameters, suppose that we know that an object we observe is a standard candle of absolute luminosity $L$. From Earth, we can experimentally determine both its redshift $z$ and the flux of light $F$ received from it. On the other hand (neglecting peculiar velocities) if we know how the scale factor evolves during the propagation of the light signal, we can determine the comoving distance $\ch$ of the object from us as a function of the redshift $\ch(z)$, using the geodesic equation for light rays
\beq
\label{lafavorita}
\chi = \int_{t_e}^{t_r} \frac{dt}{a(t)} = \int_{a_e}^{a_r} \frac{da}{a^2 H(a)} = a_{0}^{-1} \int_{0}^{z} \frac{d\z}{H(\z)} \quad .
\eeq
Note that the latter relation really depends on the redshift through the expansion history of the universe and not directly on $z$, so it depends on the cosmological parameters of the model. The received flux $F$ is generally uniquely determined by the redshift, the comoving distance between the object and us and the absolute luminosity $L$: therefore, the experimental determination of the redshift and of the flux of energy from a standard candle allows us to probe the expansion history and therefore the value of the cosmological parameters.

To be quantitative, it is customary to consider the ratio between the absolute luminosity and the received flux and more precisely the quantity
\beq
d_L = \sqrt{\frac{L}{4 \pi F}} \quad ,
\eeq
which we've already encountered in section (\ref{Hubbleslaw}) and is called the \emph{luminosity distance} of the source, since in flat space is exactly equal to the physical distance. To derive the dependence of the received flux $F$ on the redshift (where $F$ is defined as the flux of energy  per unit time through a detector divided by the detector's surface area), we can use the definition of flux of energy carried by a collection of individual particles (in this case photons) which pass isotropically through a spherical surface of area $A$
\beq
F = \frac{E N}{\D t \, A}
\eeq
where $N$ is the number of photons (which for simplicity we assume to have the same energy) which pass across the surface in a time $\D t$ and $E$ is the energy of every photon. If we consider two spherical surfaces centered around a body which emits radiation, one ($\mcal{S}_0$) which closely surrounds it and one ($\mcal{S}_{\ch}$) of comoving radius $\ch$, the number of photons which pass through $\mcal{S}_0$ is equal to the number of photons which pass through $\mcal{S}_\ch$ since the number of photons is conserved during the propagation. However, the ratio between the flux of energy $F_0$ through $\mcal{S}_0$ ($F_0 = L/A_0$ where $A_0$ is the area of $\mcal{S}_0$) and the flux of energy $F$ through $\mcal{S}_\ch$ is influenced by three factors: 1. the energy gets redshifted of a factor $1 + z$ during the propagation; 2. the time it takes for $N$ photons to pass across the surface $\mcal{S}_{\chi}$ is higher of a factor $1 + z$ compared to the time it takes for them to pass across $\mcal{S}_0$; 3. the physical areas of the surfaces which has fixed comoving radiuses increase as the universe expands. Therefore, the ratio between the absolute luminosity of a source and the flux of energy detected by an observer whose comoving distance from the source is $\chi$ reads
\beq
\frac{L}{F_\ch}  = \frac{E_{0}}{E_{\chi}} \, \frac{\D t_{\chi}}{\D t_{0}} \, \frac{N}{N} \, A_{\chi} = (1 + z)^2 \, A_{\chi} \quad ,
\eeq
where $E_{0}$ is the energy of the photons when emitted while $\D t_{0}$ is the time interval needed for the $N$ photons to pass across $\mcal{S}_0$. Since the area of a surface of comoving radius $\ch$ in the system of coordinates (\ref{RWmetric1}) is $A_{\chi} = 4 \pi \, a^2_0 \, S^2_k(\chi)$, we get
\beq
d_L (z) = (1 + z) \, a_0 \, S_k(\chi)
\eeq
and using (\ref{lafavorita}) we then have
\beq
d_L (z) = (1 + z) \, a_0 \, S_k \bigg( \frac{1}{a_0} \int_{0}^{z} \frac{d\z}{H(\z)} \bigg) \quad .
\eeq
In the spatially flat case the $a_0$ factors cancel out, while in the spatially curved cases we can use the definition of curvature density parameter $\Omega_{K} = -k/ a^2 H^2$ to get
\begin{equation}
\label{luminositydistancez}
 d_L (z) \quad \left\{
  \begin{aligned}
   \phantom{i}&=(1 + z) \, \int_{0}^{z} \frac{d\z}{H(\z)} & k&=0 \\[1mm]
   \phantom{i}&=(1 + z) \, \frac{H_{0}^{-1}}{\sqrt{\abs{\Omega_{K0}}}} \, S_k \bigg( \sqrt{\abs{\Omega_{K0}}} \int_{0}^{z} \! \frac{H_{0}}{H(\z)} \, d\z \bigg) & k&=\pm 1  \quad .\\
  \end{aligned}
\right.
\end{equation}
Each choice of cosmological parameters gives a unique evolution history $H(z)$, and therefore determine uniquely $d_L (z)$: detecting the redshift $z_{\star}$ and the received flux $F_{\star}$ of a standard candle we obtain experimentally the value $d_L^{\star} = d_L(z_{\star})$, which then constrains the evolution history and the values of the cosmological parameters.

For standard rulers, the situation is very similar. Considering an astrophysical object, we can never measure its real physical diameter just observing the light which comes from it, but we can measure the angle $\vartheta$ subtended by the object. The angle $\vth$ is generally uniquely determined by the redshift of the object, the comoving distance between the object and us and by its absolute (physical) diameter $D$: therefore, taking into account the relation (\ref{lafavorita}), the experimental determination of the redshift and of the apparent angular scale of a standard ruler allows us to probe the expansion history and therefore the value of the cosmological parameters. Again, to be quantitative it is useful to consider the ratio between the absolute diameter of the object and the angle $\vartheta$ which subtends it, defining the quantity
\beq
d_A = \frac{D}{\vartheta}
\eeq
which is called the \emph{angular diameter distance}, since in flat space a source of length $l$ whose distance from us is $d$ subtends an angle $\vartheta = l/d$. Suppose that an astrophysical object has a known physical diameter and is placed at comoving distance $\ch$ from us: calling $t_e$ the cosmic time when radiation is emitted from the object and $t_r$ the the cosmic time when the radiation is detected on Earth, the comoving diameter of the object at emission is
\beq
D_C = \frac{D}{a_e}  \quad ,
\eeq
where $a_e = a(t_e)$. Since in the reference system (\ref{RWmetric1}) the comoving diameter of the object is related to the angle which subtends it when seen from Earth by the relation $D_C = \vth S_k(\ch)$, we have that the angular distance reads
\beq
d_A = a_e \, S_{k}(\ch) = \frac{a_0 S_{k}(\ch)}{1 + z} \quad .
\eeq
Therefore, the angular diameter distance and the luminosity distance are related by
\beq
\label{dLanddA}
d_L(z) = (1 + z)^2 d_A(z) \quad ,
\eeq
so relations very similar to (\ref{luminositydistancez}) hold also for $d_A(z)$.

Note that, concerning standard candles, we can determine the distance of low redshift supernovae by astrophysical means, and therefore obtain their absolute luminosity. This implies that, using the standardizing procedure, we can obtain the absolute luminosity of higher redshift supernovae as well, once we observe their light curves. Concerning standard rulers, on the other hand, we know to a very high precision the absolute physical scale corresponding to acoustic oscillations in the baryon-photon plasma. As we mentioned above, this scale is imprinted both in the CMB (where it is linked to the angular scale of the first acoustic peak) and in the large scale structure of galaxies (where it is linked to the position of the BAO peak): therefore, we can use independently these two determinations at different redshifts of the angular scale corresponding to the acoustic oscillations to estimate the cosmological parameters. In particular, the parameter which relates the angular scale of the first acoustic peak of the CMB to the angular diameter distance of the physical scale of the sound horizon on the last scattering surface is called the \emph{CMB shift parameter}.

As we already mentioned, the field of observational cosmology is at present very active. A real breakthrough came at the end of last century, when the Supernova Search Team \cite{Riess98} and the Supernova Cosmology Project \cite{Perlmutter98} using data on the luminosity distance-redshift relation for type Ia supernovae indipendently provided evidence for a non-zero cosmological constant and a negative value of $q_0$. For this very surprising and important result the Nobel Prize in Physics 2011 was awarded to S. Perlmutter, B.~P. Schmidt, and A.~G. Riess. Using data from luminosity distance of type Ia supernovae \cite{Riess:2009pu}, from the large scale distribution of galaxies \cite{Percival09} and from the angular spectrum of anisotropies of the CMB from the satellite PLANCK it is possible to rigorously test the $\La$CDM cosmological model, and the model shows to provide a consistent fit to the data. Recently a general agreement in the community has been reached on the values of the cosmological parameters: the data on the CMB collected by the recent mission PLANCK \cite{Ade:2013zuv}, together with the data on the CMB polarization at low multipoles collected by the mission WMAP \cite{Komatsu10} provide the values 
\begin{align}
\label{cosmoparametersobserved}
h &= 0.673 \pm 0.012 & \Omega_{B_0} h^2 &= 0.02205 \pm 0.00028 \nn \\[2mm]
\Omega_{C_0} h^2 &= 0.1199 \pm 0.0027 & \Omega_{\La_0} &= 0.685^{+0.018}_{-0.016} \nn \\[2mm]
\Omega_{K_0} &= -0.037^{+0.043}_{-0.049} & \phantom{K_0}
\end{align}
where we have defined $H_{0} = 100 \, h$ km/s/Mpc.

\subsection{The acceleration problem}
\label{The acceleration problem}

We may conclude that the $\La$CDM model is very satisfactory since it gives a consistent description of all the cosmological observations up to date. Note that, as we already mentioned, the observed values of the cosmological parameters (\ref{cosmoparametersobserved}) imply that the universe is at present vacuum dominated, and it is expanding in an \emph{accelerated} way $q_0 < 0$. A closer look to (\ref{cosmoparametersobserved}), on the other hand, gives a somewhat strange feeling. It seems in fact that 70\% of the energy density in the universe is in the form of a mysterious component with negative pressure, a property which we never observe in particle colliders and in earth-based labs experiments. Also, the elusive dark matter hasn't been observed in colliders yet, but nevertheless seems to be the dominant component of nonrelativistic matter and in fact significantly more abundant that the ``normal'' baryonic matter ($\Omega_{C} \sim 6.5 \, \Omega_{B}$). Instead of confirming the picture we had about how nature works, and enriching it with new details, the recent cosmological observations suggest a radically different picture. This, although unexpected, is not \emph{a priori} wrong or worrying, and we may just accept it as an observational evidence.

However, if we are to accept a radically new picture of how nature works, we would like to understand it both from the phenomenological and the fundamental point of view. The problem is that we don't understand at a fundamental level why the $\La$CDM model should be correct. As we said, we haven't yet observed directly the particles which should constitute the dark matter. More importantly, the observed value of the cosmological constant $\Omega_{\La} \neq 0$, $\Omega_{\La} \sim \Omega_{M}$ is actually very puzzling and difficult to understand, as we will see soon. It is therefore reasonable to wonder if instead some of the assumptions at the core of the $\La$CDM model are maybe not correct, and if we are maybe misinterpreting the observational data. It is in fact possible that gravity is not described by GR at very large scales, or that there exist new degrees of freedom (or even new laws of nature!) which show up only when we increase enormously the length scales and the complexity of the system under study. Or it may be that the Copernican principle is not really valid (which however would be puzzling from a philosophical point of view). If one or several of this things are true, then the conclusion that $\La$ is non-zero may be ill based. It seems indeed worth exploring these other routes, before concluding that the picture of the universe drawn by the $\La$CDM model is reliable.

\subsubsection{The cosmological constant problem}

The invariance with respect to general coordinate transformations and the energy conservation, which are at the heart of the formulation of GR, allow the addition of a term $\La \, g_{\m\n}$ to the (1915) Einstein equations \cite{GR} which does not alter the structure of the theory, as first recognized by Einstein himself \cite{EinsteinCosmo}. Although we are not forced to keep such a term, since we don't observe its effects in the solar system or on earth, it is not obvious that we should set it to zero either: it may in fact describe a second characteristic constant of the gravitational force \cite{Padmanabhan}. A non-zero value of $\La$ introduces into the theory a length scale
\beq
r_{\La} \sim \sqrt{\frac{1}{\abs{\La}}}
\eeq
above which the cosmological constant term would strongly affect the spacetime: the gravitational interaction would then be characterized by two parameters, one which describes the strength of the interaction (Newton's constant $G$) and one which describe its large scale behavior ($\La$). There is however a problem, coming from the fact that cosmological observations imply that today $\Omega_{\La_0} \sim \Omega_{M_0}$. The energy density of matter and vacuum scale very differently with the scale factor $\rho_{\La}/\rho_{M} = a^3$, so the time when these densities are comparable is a very special and rare one in the history of the universe: for most of the time, vacuum energy is either dominating or negligible compared to matter. On the other hand, the time when astrophysical structures form is another very special moment is the cosmic history, and is correlated with the time of matter-radiation equality. The fact that $\Omega_{\La_0} \sim \Omega_{M_0}$ today means that matter-vacuum equality and the formation of structures happen roughly at the same time: however this is a priori highly unlikely to happen, since we don't expect correlations between $\Omega_{\La} / \Omega_{M}$ and $\Omega_{R} / \Omega_{M}$. To say the same thing differently, an extreme fine tuning in initial conditions would be necessary for this to happen: this problem is known as the \emph{coincidence problem} (or also as the ``new'' cosmological constant problem). It is fair to say that, in this approach, the small and fine tuned value of $\La$ is no more a mystery than the fine tuning in other constants of nature \cite{DurrerMaartens}. Furthermore, anthropic arguments may provide a way out of this problem \cite{WeinbergCC,Garriga:2002tq}. In fact, in some cosmological models the effective cosmological constant takes a wide variety of values in different parts of the universe/cosmological eras; most of these parts/eras would however be extremely ``inhospitable'', since they would not allow the formation of stars and planets and the development of an ``intelligent'' form of life (observers). In particular, if the effective cosmological constant is positive and too big, structure formation by gravitational instability is impossible; if instead it is negative and too big in absolute value, the universe re-collapses before intelligent life had the time to develop. The very fact that the human race is able to formulate the coincidence problem implies that we should observe a value for the effective cosmological constant which lies inside a tiny ``anthropic range''\cite{Garriga:2002tq}. The appeal of this approach to the coincidence problem has been strengthened by the discovery that string theory possesses a huge number of low-energy vacua (the string theory landscape) \cite{SusskindLandscape}.

The situation is in any case deeply worsened by the fact that we expect a contribution of exactly the same form as $\La \, g_{\m\n}$ coming from the source term of the Einstein equations. As we already mentioned, in Quantum Field Theory the vacuum state $\mid \!\! 0 \rangle$ seems to possess a non-zero energy and pressure, and if the field theory is Lorentz invariant it should produce a contribution to the energy momentum tensor of the form
\beq
T_{\m\n}^{(\textup{vac})} = \langle 0 \! \mid \! \hat{T}_{\m\n} \! \mid \! 0 \rangle = - \r_{\textup{vac}} \, g_{\m\n} \quad .
\eeq
Despite the fact that this is an expectation value in a quantum theory, while GR is a classical theory, we expect that such a term should be included as a source in the Einstein equations, since vacuum energy has shown to have measurable effects at classical level (consider for example the Casimir effect). To understand what may be a reasonable value for $\r_{\textup{vac}}$, let's consider as an example a free (\ie non-interacting) scalar field in a Minkowski spacetime. In a canonical quantization approach, every Fourier mode $\veck$ of the field is equivalent to a quantum harmonic oscillator, which is known to possess a non-zero vacuum energy $E_0(\veck) = \hbar \, \o(\veck)/2$ where $\o(\veck) = \sqrt{m^2 + k^2}$. Therefore, summing up the contributions of every single mode, we find that the total vacuum energy of the field diverges. However, we may assume that the quantum field theory description is reliable only below a momentum cut-off scale $k_{\textup{cut}}$: we definitely expect the description not to be adequate for energies above the Planck energy $E_{\textup{pl}} = \sqrt{\hbar c^5/G} \sim 10^{19}$ GeV, but to be conservative we may lower the cutoff at the TeV energy scale $\sim 10^{-16} E_{\textup{pl}}$. Summing the vacuum energy of the modes up to the cutoff, we have that the vacuum energy scales as the cutoff energy scale at the fourth power \cite{WeinbergCC}
\beq
\label{vacuumenergyscalewithcutoffenergy}
\r_{\textup{vac}} \sim \frac{E^{4}_{\textup{cut}}}{\hbar^3 c^3}
\eeq
where we have explicitly shown the $c$ and $\hbar$ coefficients for dimensional clarity. Note that if we assume that the value of $\La$ estimated by the cosmological observations is due to vacuum energy, we have
\beq
\label{vacuumenergyobservationallyinferred}
\r^{(\textup{obs})}_{\textup{vac}} \sim 10^{-8} \,\,\, \textrm{erg}/\textrm{cm}^3
\eeq
while using (\ref{vacuumenergyscalewithcutoffenergy}) we get the theoretical estimates
\beq
\r^{(\textup{th})}_{\textup{vac}} \sim 10^{112} \,\,\, \textrm{erg}/\textrm{cm}^3 \quad (\textrm{Planck}) \qquad \qquad \r^{(\textup{th})}_{\textup{vac}} \sim 10^{48} \,\,\, \textrm{erg}/\textrm{cm}^3 \quad (\textrm{TeV}) \quad .
\eeq
We can see that, if we take the cutoff to the Planck scale, there is a difference of about 120 orders of magnitude between the observed value and the theoretical expectation, and even in the case of the TeV cutoff scale the difference is nearly 60 orders of magnitude. This extreme clash between predictions and observations is sometimes called the ``old'' cosmological problem: it can be expressed as the fact that vacuum energy seems to be much smaller than predicted, but it can also be restated as the fact that vacuum energy seems to gravitate much less then expected. 

In general, we expect that the only observable signature of both vacuum energy and a ``true'' cosmological constant is its effect on spacetime, and therefore the two in principle very different contributions cannot be distinguished by observations \cite{DurrerMaartens}. Therefore, we should write the cosmological constant present in the Einstein equations as an ``effective'' constant which is the sum of a ``bare'' cosmological constant and of a vacuum energy contribution
\beq
\La_{\textup{eff}} = \La + 8 \pi G \r_{\textup{vac}} \quad .
\eeq
To match the observed value, it is necessary that the two term cancel with a relative precision which is almost incredible: $(\La-\La_{\textup{vac}})/\La \sim 10^{-56}$ in the TeV scale cutoff case, and even more so in the Planck scale cutoff case. Therefore an extreme fine-tuning between the two contributions is needed to be consistent with the observations. Note that we may interpret this tuning as a renormalization of the energy of the vacuum, in which case we may choose $\La$ in such a way that the renormalized value of the vacuum energy is equal to the observationally inferred value (\ref{vacuumenergyobservationallyinferred}). From this point of view, the fine tuning is not a problem, as long as it is radiatively stable. However, it can be shown \cite{Kaloper:2013zca} that this is not the case: the addition of loop corrections to the vacuum energy dramatically shifts the value of the vacuum energy, by $\mcal{O}(1)$ in the units of the ultraviolet cutoff. To neutralize it one must retune the classical term by hand order by order in perturbation theory, so the fine tuning is indeed a problem.

It is natural to wonder whether the two problems we have highlighted above are two faces of the same problem or are two different problems. It may well be that the reason why vacuum energy is very small (or gravitates very little), and the reason why cosmological observations suggest a non-zero $\La$, are in some sense independent. It is in fact reasonable to expect that, since vacuum energy is so smaller than predicted (gravitates so much less than expected), it may actually be exactly zero (not gravitate at all). This may be due to a symmetry which prevents a non-zero vacuum energy or to a completely different reason: understanding this seems one of the most difficult problems in contemporary physics. Nevertheless, unless we find a incontrovertible indication of the contrary, we may take the point of view that, however difficult to solve, this problem is disentangled from the implications of cosmological observations. This is the point of view we take in this thesis: without addressing the problem of why vacuum energy is so small (does not almost gravitate), we try to understand why in cosmology we observe a non-zero and fine tuned $\La$.

\subsubsection{Backreaction, dark energy and modif\mbox{}ied gravity}
\label{Backreaction, dark energy and modified gravity}

If we want to explain the cosmological observations without resorting to a non-zero cosmological constant, some of the hypothesis which underlie the $\La$CDM model have to be relaxed. Despite the fact that all of them may be not correct, for simplicity we can study what happens if we relax in turn just one of these assumptions, namely the large scale homogeneity and isotropy, the assumption that the universe is filled only with CDM and standard model particles, and the fact that gravity is described by GR at all scales. In the following, we describe briefly the main advantages/disadvantages of the different cases.

As we said previously, while large scale isotropy is very well tested observationally, homogeneity is not. It is usually assumed that we don't occupy a special place in the universe (the Copernican principle), which implies homogeneity, but since this is a philosophical assumption, it may be wrong after all. In fact, if the Earth was situated near the center of a huge, nearly spherical structure, the supernovae observations may be explained as due to the inhomogeneity, without having a non-zero $\La$ \cite{Enqvist06, Vanderveld06}. However, apart from being philosophically puzzling, this scenario poses another fine tuning problem, regarding the characteristic of the spherical structure and our position inside it. Moreover, it is not so clear whether it is consistent with all the cosmological observations, not just supernovae \cite{DurrerMaartens}. A different possibility is that the fact that inhomogeneities go non-linear produce a sizable effect on the evolution of the scale factor. As we said in section (\ref{The Friedmann-Robertson-Walker model}), the time evolution does not commute with the averaging procedure on the Einstein equations. Therefore, the ``real'' scale factor that describes our universe is different from the one we get by solving the Friedmann equations, and it may be that this difference is crucial in judging if $\La$ is zero or not: the universe may seem to accelerate at late times just because we don't take into account properly this effect. The influence of inhomogeneities on the evolution of the scale factor is known as \emph{backreaction} (see for example \cite{EllisNicolaiDurrerMaartens} and references in \cite{DurrerMaartens}): this would provide a dramatic resolution of the coincidence problem, since in this case the formation of structures and the apparent acceleration are correlated since they are both a consequence of the fact that inhomogeneities go non-linear. However, there is no convincing demonstration that the backreaction is indeed able to explain the apparent acceleration. It should be noted anyway that it may significantly affect the estimation of cosmological parameters, even if it does not lead to acceleration \cite{DurrerMaartens}.

Alternatively, if we take $\La = 0$, neglect backreaction and assume that large scale homogeneity and isotropy hold, we are forced to admit that either gravity is not described exactly by GR, or that there are new degrees of freedom whose contribution to the energy-momentum tensor is responsible for the acceleration of the universe. The situation is somewhat similar to what happened when deviations from the predicted orbits were observed for some planets in the solar system: in the case of the anomalies of the orbits of Uranus and Neptune, the existence of a new, unobserved planet was postulated. Pluto was indeed discovered later on. On the other hand, the anomalous precession of the perihelion of Mercury could not be explained as the effect of a yet unobserved object (originally called Vulcan): the discrepancy was shown to be due to the inadequacy of the Newtonian theory of gravity, and the resolution of the problem was the result of the development of a new theory of gravity, General Relativity. If we consider GR to be the correct theory of gravitational interaction, even at extremely large scales, then the cosmological observations can be explained by adding a source term in the Einstein equations, which by equation (\ref{Friedacc}) have to satisfy $\rho + 3 p < 0$. This is a very unusual property, since at the classical level the matter we observe in Earth-based experiments has positive energy and non-negative pressure. Therefore, not only we have to introduce an ad-hoc matter which we don't observe on Earth and in the solar system, but this matter has to have very exotic properties. On the other hand, at quantum level such a property is not so strange, and can be enjoyed also by a very simple system such as a (classical) scalar field. This new component of the energy-momentum tensor is usually termed \emph{dark energy}, and there are several different models/scenarios (such as for example quintessence models, K-essence and others, see \cite{EllisNicolaiDurrerMaartens}) which address the late time acceleration problem following this idea. However, most of them are not well motivated (so far) from the point of view of fundamental physics, and generally do not solve the coincidence problem, since some sort of fine tuning seems to be required anyway \cite{DurrerMaartens}.

Finally, we may assume that there is not such a thing as dark energy, but the observations just signal the breakdown of the validity of GR at ultra large scales. From this point of view, the explanation of the apparent acceleration is to be found in formulating a new theory of gravity, which should reproduce very well the results of GR at scales from a micron up to astrophysical scales, but should deviate from it at ultra large scales. This approach is usually called \emph{modified gravity}: for an extensive review, see \cite{ModifiedGravityAndCosmologyHugeReview}. There are several modified gravity scenarios which have been studied, among which $f(R)$ gravity, braneworld models and massive gravity. Braneworld models have the appealing feature to be in a loose sense motivated by fundamental physics, since the existence of extra dimensions and ``branes'' where matter is localised is a important ingredient in string theory. However, quite generally, braneworld models which modify gravity at large distances are mainly phenomenological, in the sense that there are usually no precise indications about how to embed them into string theory. Overall, one of the crucial points is that it is very difficult to modify gravity at large distances, without introducing changes at intermediate and small distances: typically, the modifications can be traced back to the presence of new (gravitational) degrees of freedom, which however seems to contribute also at small scales. In order this not to happen, it is necessary that there is a ``screening'' mechanism which efficiently suppresses the contributions of the new degrees of freedom in the contexts where GR results have to be reproduced. Another problem is that modifying gravity at large scales quite often produces new degrees of freedom which have (at least in some configurations) negative kinetic energy (in which cases they are called \emph{ghosts}). This is usually regarded as unacceptable, since at quantum level the vacuum would be unstable.

\section{Thesis summary}

In this thesis, we explore the possibility of formulating a consistent theory of modified gravity, motivated by the problem of the late time acceleration of the universe. For definiteness, we consider two different models of modified gravity: a braneworld model, the Cascading DGP model, of which we study the minimal setup (6D), and a class of massive gravity models known as dRGT massive gravity. Rather than on their cosmological aspects, we focus on the theoretical consistency of these two models, namely the absence of ghost instabilities (in the Cascading DGP case) and the ability of reproducing the predictions of Einstein gravity at terrestrial and astrophysical scales (in the dRGT massive gravity case). In particular, in the first case we investigate the presence of a critical value for the tension of the 4D brane which separates configurations which are plagued by ghost instabilities from configurations which are stable, while in the second case we establish if the model exibits an efficient screening mechanism which permits to recover GR results at small and intermediate scales.

The thesis is therefore structured as follows: in chapter \ref{Braneworlds and the DGP Model} we introduce braneworld models and the DGP model, which, although not providing itself a modified gravity solution to the late time acceleration problem, have been studied extensively and provided ideas and tools which turned out to be useful to propose new models. We also discuss the problems of the DGP model, and suggest that it is worthwhile to consider generalization of this model which may be free from its shortcomings.

In chapter \ref{Nested branes with induced gravity} we introduce the 6D Cascading DGP, and consider a realization of this scenario where the codimension-1 brane can be considered thin with respect to the codimension-2 brane. We study first order perturbations around background configurations where the codimension-2 brane is equipped with positive tension, and confirm that gravity on the codimension-2 brane is regularized by the presence of the codimension-1 brane. We also confirm the existence of the critical tension, and find that, differently from the claims in the literature, it is possible to avoid the ghost for every value of the free parameters of the model, provided we put enough tension on the codimension-2 brane. We comment on this difference and support the validity of our result with a numerical check in a case where the exact solution can be found explicitly.
 
We then consider the case of massive gravity, which we discuss in detail in chapter \ref{Ghost Free Massive Gravity}, and concentrate on the recently proposed class of models known as dRGT massive gravity, which apart from the mass is characterized by two adimensional free parameters. The models which belong to this class are free of the Boulware-Deser ghost and propagate exactly five degrees of freedom, so they are consistent interacting theories of a massive spin-2 field.

Finally, in chapter \ref{Vainshtein Mechanism in Massive Gravity} we study in detail the efficiency of the screening mechanism known as ``Vainshtein mechanism'' in this class of models. We consider spherically symmetric solutions, and select one of the two branches of solutions which have been found, which is the only branch where the Vainshtein mechanism can work. We characterize completely the number and properties of solutions which exist asymptotically on large scales, and of solutions which exist around the origin at small scales. We provide a complete characterization of the phase space of these theories in relation to the existence of global solutions and to the way the Vainshtein mechanism works, which is an important step in establishing the viability of these theories.

\clearemptydoublepage
\chapter{Braneworlds and the DGP model}
\label{Braneworlds and the DGP Model}

In the framework of modified gravity, theories with extra spatial dimensions and in particular the so-called braneworld models have attracted a lot of attention. Apart from providing a geometrical mechanism of modifying gravity at large distances, they have played a crucial role in the recent construction of a class of ghost-free massive gravity theories. Therefore, we dedicate this chapter to a general introduction to braneworld theories and in particular to the DGP model.

\section{Introduction to braneworlds}

\subsection{Historical introduction}

\subsubsection*{Kaluza-Klein theories}

The idea that there may be some spatial dimensions in addition to the three we have experience of is in fact not a recent one. Already in 1921, Theodor Kaluza \cite{Kaluza1921} (reprinted with English translation in \cite{ModernKaluzaKleinTheories}) studied a five dimensional extension of General Relativity, and noticed that the degrees of freedom of the metric associated with the extra dimension could be interpreted as a vector field in our four dimensional world (plus an additional scalar). Recognizing in this vector the 4-potential of electromagnetic theory, the Einstein equations for the 5D metric would produce respectively the Einstein equations and the Maxwell equations for gravity coupled to the electromagnetic field, thereby geometrically giving a unified description of these two forces. Oskar Klein in 1926 \cite{Klein1926} (also reprinted with English translation in \cite{ModernKaluzaKleinTheories}) proposed that, if the extra dimension is compact and of radius $r$, deviations to the known laws would not show up for length scales larger than $r$, or for energies less than $1/r$, thereby we wouldn't be able to observe them if $r$ is small enough (say $r < 10^{-19} \textrm{m}$, corresponding to an energy $E \sim 1 \textrm{TeV}$). This idea of the extra dimensions being rolled up and small is usually referred to as the Kaluza-Klein (KK) scenario: it has been almost universally adopted for a long time to explain why we don't observe the extra dimensions, despite their existence, and typically the characteristic radius of the extra dimensions was assumed to be incredibly small, of the order of the Planck length $l_{pl} = \sqrt{\hbar G/c^3} \sim 10^{-35} \,\textrm{m}$. The very idea of the existence of extra dimensions had a big push by the discovery in the 1970's that string theory, one of the most promising candidates for unifying general relativity and quantum mechanics as well as providing a unification of all the forces, is only consistent if there is a suitable number of extra dimensions (6 for superstring theory).

\subsubsection*{Braneworlds and large extra dimensions}

A conceptual revolution began around 1960 \cite{Fronsdal59,Joseph62} when the idea that matter and force fields, instead of propagating in all the space, could be confined to a surface in a higher dimensional space started being discussed. At the beginning of the 1980's, Akama \cite{Akama82} and independently Rubakov and Shaposhnikov \cite{RubakovShaposhnikov83} proposed an explicit particle physics realisation of the localization phenomenon, while Visser \cite{Visser85} and Squires \cite{Squires85} proposed a gravitational realization of the same phenomenon. The idea of matter being localized on a surface, or on a ``brane'', became much more popular with the discovery in the 1990's that extended objects, called p-branes, are of fundamental importance in string theory. In particular there are objects called D-branes to which the ends of open strings are attached, while closed strings can propagate in the bulk. The idea that gravity could propagate in the extra dimensions (in string theory it is described by closed strings) while matter and Standard Model interactions could be confined to a brane, led Arkani-Hamed, Dimopolous and Dvali \cite{ArkaniHamed:1998rs, ArkaniHamed:1998nn} (AHDD) to propose that the characteristic length of compact extra dimensions could be much bigger than the Planck length, and in fact macroscopic (even at sub-millimeter scale). The crucial observation is that, while particle interactions are probed by high energy colliders at energies up to the TeV scale and therefore for length scales down to $10^{-19}$ m, gravity is tested only for length scales down to $5 \times 10^{-5} \, \textrm{m} = 0.05 \, \textrm{mm}$ \cite{Kapner:2006si}. This idea led to the proposal that the observed Newton constant $G$ may be not the fundamental strength of gravity, but it is an effective strength related to the fundamental strength $G_{\star}$ \via the relation $G \propto G_{\star}/V$ where $V$ is the volume of the compact extra dimensions. This idea opened up the fascinating possibility of having a fundamental (Planck) scale for gravity as low as 1 TeV (with the possibility of realistically observing quantum gravity effects in particle colliders) \cite{Antoniadis:1998ig}, and from another point of view of explaining the observed weakness of gravity compared to the other interactions as an effect of the ability of gravity to propagate in all the spatial dimensions.

\subsubsection*{Non factorizable geometry and localization of gravity}

In the braneworld picture, more often than not it is assumed that some mechanism (the presence of a bulk soliton in QFT, or the very existence of D-branes in string theory) localizes matter and the standard model interactions. Once assumed the existence of such a mechanism, explaining why the extra dimensions are not observed reduces to explain why gravity behaves as in the (4D) GR despite propagating in more than four dimensions. Despite the widespread belief that compact (although not necessarily extremely small) extra dimensions are needed to reproduce 4D gravity in a suitable distance range, it was shown by Randall and Sundrum in a famous series of two papers \cite{Randall:1999ee, Randall:1999vf} that, if the bulk metric is not factorizable, this is not the case. In particular, a flat 4D brane with non-zero tension $T$ in a 5D bulk with negative cosmological constant $\La$ causes the bulk to become a 5D Anti-deSitter space $AdS_5$ (if $T$ and $\La$ are appropriately tuned), with warped metric $ds^2 = e^{-\abs{y}/L} \e_{\m\n} dx^{\m}dx^{\n} + dy^2$ where $y$ is the extra dimension and $L \propto \sqrt{1/\La}$. In particular, they showed how the warping in the bulk metric between two flat branes could be used to explain the hierarchy between the electroweak mass scale and the gravitational Planck scale \cite{Randall:1999ee}, and how the warping could effectively localize gravity on one brane even if the extra dimension is not compact \cite{Randall:1999vf}. However, in the Randall-Sundrum model the extra dimension is not truly infinite since its \emph{volume} is still finite due to the warping. As a result, the relation between the fundamental Planck scale of gravity and the 4D effective one is very similar to the one which holds in the AHDD model, with the radius of the extra dimension replaced by the characteristic length $L$ of $AdS_{5}$. Likewise, in both AHDD and RS models the modifications to the Newton law happen at \emph{small} distances, where the critical length is set by the characteristic length of the extra dimensions: below that scale, gravity is mediated by all the KK tower of graviton modes, while above that scale gravity is mediated only by the zero mode and Einstein gravity is reproduced.

\subsubsection*{Multi-branes models and inf\mbox{}inite volume extra dimensions}

In 2000, Kogan and collaborators (the ``Millennium'' model) \cite{Kogan:1999wc} and independently Gregory, Rubakov and Sibiryakov (GRS) \cite{Gregory:2000jc} instead showed that it is possible to construct a braneworld model where gravity looks like GR at observable scales, but behaves differently both at smaller \emph{and larger} scales. In the Millennium model, two positive tension branes (one of which is supposed to describe our universe) are positioned at the fixed points of an $S_1/\mathbb{Z}_2$ orbifold, and between them there is a negative tension brane which moves freely in a 5D bulk equipped with a negative cosmological constant. Surprisingly, for some configurations of the model the spectrum of 4D graviton modes contains a ultralight massive state (beside a massless mode) which couples to matter much more strongly than the other massive states in the tower of Kaluza-Klein (KK) states. Therefore, we can have configurations where gravity at intermediate distances is mediated by \emph{two} 4D graviton modes, the massles zero mode and the ultralight KK mode: instead, at short distances (which can be tuned to be smaller than one micron) also the other states in the KK tower become important, while for large distances (which can be tuned to be of the order of the present Hubble radius) the ultralight massive state does not contribute appreciably and gravity is mediated only by the massless zero mode.

In the GRS model, instead, a brane of positive tension $\s > 0$ (where matter is confined) is flanked in the fifth dimension by two branes of negative tension $-\s/2$ (see also \cite{Charmousis:1999rg}). The bulk cosmological constant $\La$ is negative between the negative tension branes, while it is zero outside: tuning $\La$ appropriately, there exists a configuration where the bulk is $AdS_5$ between the negative tension branes and flat outside. Interestingly, for a range of choices of the parameters of the model, the gravitational interaction behaves as four dimensional at intermediate distances, while it behaves as five dimensional both at small distances and at large distances (however, the 5D gravitational constant has different values in the two regimes). The situation is different from the Millennium model, where gravity is 4D both at intermediate and at large distances, but gets weaker passing from intermediate to large distances. More importantly, in the GRS model the extra dimension is not only non-compact, but in fact truly infinite, since the geometry is flat ouside the negative tension branes and therefore the volume of the extra dimension is infinite indeed.

In theories with one infinite volume extra dimension, gravity is mediated (at all distances) by an infinite number of graviton modes. It may seem unlikely that such theories can reproduce Einstein gravity on the brane in a suitable range of distances, since this property was historically associated with the fact that gravity is effectively mediated only by one massless graviton. However, the answer lies in the fact that the wave function of the massive graviton modes in the extra dimension, when evaluated on the brane, has a non-trivial profile as a function of the mass. In fact, soon after the introduction of the GRS model it was proposed \cite{Dvali:2000rv, Csaki:2000pp} that the ability of theories with one infinite volume extra dimension to reproduce 4D gravity can be understood as if gravity were mediated by a \emph{metastable} 4D graviton, or in other words by a continuous superposition of 4D massive gravitons peaked around $m = 0$ with a finite width. The GRS model in fact was shown to belong to this class of models. 

\subsubsection*{The DGP model}

Later in the same year, Dvali, Gabadadze and Porrati (DGP) showed that it is not even necessary to consider multi-brane models to have 4D gravity in a infinite volume extra dimensional space. In the DGP model \cite{DGP00}, there is just one 4D brane in an infinite-volume 5D bulk, but crucially the action contains an induced gravity term localized on the brane, which is responsible for the peaked profile in the mass space. The gravitational field therefore looks five dimensional for very large distances, while looks four dimensional at small distances: however, at linearized level, the theory does not reproduce Einstein gravity at small scales but is similar to a scalar-tensor theory of gravitation. On the other hand, it was soon recognised that the small scales phenomenology of the model may be richer than what the linearized analysis suggests, due to the fact that non-linearities may become important even at astrophysical scales. 

The DGP model inspired a lot of activity, both to establish its phenomenological viability \cite{Deffayet:2001uk,Tanaka:2003zb,Gruzinov:2001hp,PorratiVDVZ, LutyPorratiRattazzi,NicolisRattazzi} and to explore its potential ability to address long standing theoretical problems like the cosmological constant problem and more recent ones as the late time acceleration problem of cosmology (see section \ref{late time acceleration problem}). In the cosmological context, a breakthrough came when it was shown \cite{Deffayet00} that the DGP model admits ``self-accelerating'' solutions, opening the door to the idea of explaining the late time acceleration as a purely geometric and ``modified gravity'' phenomenon \cite{Deffayet:2001pu}, without resorting to the idea of dark energy. Concerning the cosmological constant problem, it has been shown \cite{Dvali:2002pe} that infinite volume extra dimensions provide a way to bypass the no-go theorem formulated by Weinberg \cite{WeinbergCC}, and therefore are extremely appealing from that point of view. 

However, the attempts were not crowned by success. It has been shown that the self accelerating cosmological solution contains a ghost \cite{Gorbunov:2005zk, LutyPorratiRattazzi, NicolisRattazzi, DGPspectereoscopy} and therefore cannot be quantum mechanically stable. Furthermore, a careful analysis has shown that there is strong tension between the theoretical predictions and the cosmological data, which in practice rule out the DGP self-accelerating solution as an explanation for the late time acceleration \cite{MaartensMajerotto06}. From another point of view, it has been shown that the DGP model cannot solve the Cosmological Constant problem by ``degravitating'' sources with very large characteristic length scales, since its gravitational potential does not decay fast enough at large distances \cite{Dvali:2007kt,deRham:2007rw}.

Nevertheless, the richness of ideas and approaches to several problems of modern physics which were conceived by studying the DGP model, even if it is not successful itself, suggest that it may be worth trying to find generalizations of the DGP model which may be similar enough to its original formulation to preserve the good features, and different enough to be free of its shortcomings. Therefore, we dedicate this chapter to the presentation of the DGP model, and we will consider its generalizations in the next chapter.

\subsection{Mathematical preliminaries}
\label{Mathematical preliminaries on braneworlds}

Let $\mscrM$ be a $N$-dimensional ($N \geq 4$) manifold. We call a $D$-dimensional brane (or a $(D-1)$-brane for short) a $D$-dimensional submanifold $\S$ of $\mscrM$. We define \emph{codimension} of the brane the number $N-D$. Despite being a subset of $\mscrM$, we can equivalently consider $\S$ to be a separate manifold equipped with an \emph{embedding function}
\beq
\vf : \S \rightarrow \mscrM
\eeq
which specifies the ``position'' of $\S$ inside $\mscrM$ when seen as a subset. Being the dimensionalities of $\mscrM$ and $\S$ different, $\vf$ is not invertible, and can be used to pull-back to $\S$ tensors of type $(0,k)$ defined on $\mscrM$ and push-forward to $\mscrM$ tensors of type $(n,0)$ defined on $\S$. In particular, for every $p \in \S$, if $\big\{\mbf{w}_{(j)}\big\}_{j}$ ($j = 1, \ldots, D$) is a basis of tangent vectors in $T_{p}\S$, then $\big\{\vf^{\star}(\mbf{w}_{(j)})\big\}_{j}$ is a linearly independent set of vectors in $T_{\vf(p)}\mscrM$, where $\vf^{\star}$ indicates the push-forward with respect to the embedding function. We define the $D$-dimensional subset of $T_{\varphi(p)}\mscrM$ spanned by this set of vectors to be the tangent space to $\S$ (seen as a subset of $\mscrM$) and we will denote it as $T_{\vf(p)}\S$.

We will in general consider two different atlases of maps, one which defines coordinates on $\mscrM$ and another one which defines coordinates on $\S$. Indicating with $X^M$ the coordinates on $\mscrM$ and with $\xi^m$ the coordinates on $\S$, the embedding function reads in coordinates $\vf^M(\xi^m)$: if we work with the coordinate expression of tangent vectors, a basis of $T_{\vf(p)}\S \subset T_{\varphi(p)}\mscrM$ is given by the directional derivatives of the embedding function
\beq
\label{parallelvectorsgeneral}
v_{(a)}^A(p) \equiv \bigg\{ \frac{\de}{\de \xi^a}\Big\rvert_{p} \vf^A \bigg\}_{\!a} \qquad \qquad a = 0, \ldots, D-1
\eeq
(where the derivative is evaluated in the coordinate expression of the point $p$), and this relation in turn defines the abstract tangent vectors $\mathbf{v}_{(a)}$. If the ambient manifold $\mscrM$ is a metric manifold $(\mscrM, \mbfg)$, the embedding induces a metric structure on the brane $\S$ as well: we define the \emph{induced metric} $\tmbfg$
\beq
\tmbfg: T\S \times T\S \rightarrow \mathbb{R} \qquad \qquad \tmbfg \equiv \vf_{\star} (\mbfg) 
\eeq
where $\vf_{\star}$ indicates the pullback with respect to the embedding function. In coordinates the previous relation reads
\beq
\tilde{g}_{ab}(\xi^{\cdot}) = \mbfg \big( \mathbf{v}_{(a)} , \mathbf{v}_{(b)} \big)(\xi^{\cdot}) \quad ,
\eeq
and explicitly
\beq
\label{Induced metric general}
\tilde{g}_{ab}(\xi^{\cdot}) = \frac{\de \vf^{A}(\xi^{\cdot})}{\de \xi^{a}}
\frac{\de \vf^{B}(\xi^{\cdot})}{\de \xi^{b}} \, \, g_{AB}(X^{\cdot}) \Big\rvert_{X^{\cdot}
= \vf^{\cdot}(\xi^{\cdot})}
\eeq
where we used the notational convention of indicating the set of coordinates $X^M$ and $\xi^m$ respectively with $\Xd$ and $\xid$, while the embedding function $\vf^a$ is indicated with $\vfd$. We assume here that the metric $\mbfg$ is nondegenerate and pseudo-Riemannian.

In the following we will be mostly interested in codimension-1 brane, for which there is a fair amount of dedicated terminology and geometrical concepts to which we now turn.

\subsubsection{Codimension-1 braneworlds}

We denote in general with a tilde $\tilde{\phantom{a}}$ quantities pertaining to the codimension-1 brane. Taken a basis $\big\{\mbf{b}_{(a)}\big\}_{a}$ ($a = 1, \ldots, D$) of $T\S \subset T\mscrM$, we can define the vector $\mathbf{n}(\xi^{\cdot})$ normal to the cod-1 brane in the following way
\begin{equation*}
 \mathbf{n}(\xi^{\cdot}) : \quad \left\{
  \begin{aligned}
   \langle \mathbf{n}(\xi^{\cdot}) \vert \mathbf{b}_{(a)}(\xi^{\cdot}) \rangle_{\mathbf{g}} &= 0\\[1mm]
   \abs{\langle \mathbf{n}(\xi^{\cdot}) \vert \mathbf{n}(\xi^{\cdot}) \rangle_{\mathbf{g}}} &= 1
  \end{aligned}
\right.
\end{equation*}
where $\langle \phantom{n} \vert \phantom{n} \rangle_{\mathbf{g}}$ indicate the scalar product associated to the metric $\mathbf{g}$. There are two possibilities, depending on the sign of the squared modulus of $\mbfn$: if the normal vector is spacelike $\norm{\mbfn} > 0$, the brane is said to be timelike, while if the normal vector is timelike $\norm{\mbfn} < 0$, the brane is said to be spacelike. We will consider only the case of a spacelike normal vector, which corresponds to having a ``spatial'' extra dimension. Even fixing the sign of $\norm{\mbfn}$, the system above does not define uniquely the normal vector since there are two possible choices which define the local orientation of the brane. Note that we can uniquely decompose a vector $\mbf{w}$ into an orthogonal component $\mbf{w}_{\perp} = w_{\perp} \mathbf{n}$ and a parallel component $\mbf{w}_{\shortparallel}$ such that $\langle \mbf{w}_{\shortparallel} \vert \mathbf{n} \rangle_{\mathbf{g}} = 0$.

Using the normal vector we can define the first fundamental form of the cod-1 brane\footnote{The notation $\mbf{g}(\mathbf{n},\_)$ indicates the 1-form which, to every vector $\mathbf{r}$, associates the number $\mbf{g}(\mathbf{n},\mathbf{r})$.}
\beq
\mbf{P}(\xid) \equiv \mbf{g} - \mbf{g}(\mathbf{n},\_) \otimes \mbf{g}(\mathbf{n},\_) \quad ,
\eeq
where $\mbf{g}$ is evaluated in $\Xd = \vfd(\xid)$, and $\mbf{n}$ is evaluated in $\xid$. Acting on two vectors $\mbf{c}$ and $\mbf{d}$, the first fundamental form give as a result the scalar product computed with $\mbf{g}$ between the parallel components of the two vectors
\beq
\mbf{P} \big( \mbf{c}, \mbf{d}\big) = \mbf{P} \big( \mbf{c}_{\shortparallel}, \mbf{d}_{\shortparallel} \big) = \mbf{g} \big( \mbf{c}_{\shortparallel}, \mbf{d}_{\shortparallel} \big) \quad ,
\eeq
and therefore extracts the notion of metric on the brane from the bulk metric $\mbf{g}$. To get an intrinsic object which defines metric concepts on the brane we can pull-back the first fundamental form to the brane using the embedding function, obtaining the (already introduced) induced metric
\beq
\tmbfg \equiv \vf_{\star} (\mbfg) = \vf_{\star} (\mbf{P}) \quad .
\eeq
From the induced metric we can construct the associated symmetric and metric compatible connection, and the curvature tensors and scalar, which characterise the intrinsic geometry of the brane.

The second fundamental form of the cod-1 brane is defined as
\beq
\mbf{K} \equiv - \half \, \mcal{L}_{\mbf{n}} \, \mbf{P} \quad ,
\eeq
and instead characterizes the extrinsic geometry of the brane. Like the first fundamental form, it is a brane parallel object in the sense that it acts only on the parallel components of the vectors
\beq
\mbf{K} \big( \mbf{c}, \mbf{d}\big) = \mbf{K} \big( \mbf{c}_{\shortparallel}, \mbf{d}_{\shortparallel} \big) \quad .
\eeq
To obtain from the second fundamental form an intrinsic object which describes the extrinsic geometry we can pull-back $\mbf{K}$ to the brane, obtaining the extrinsic curvature $\tilde{\mathbf{K}}(\xi^{\cdot})$
\beq
\tilde{\mathbf{K}} \equiv \vf_{\star} \big( \mbf{K} \big) = - \half \, \vf_{\star} \big( \mcal{L}_{\mbf{n}} \, \mbf{g} \big) \quad .
\eeq
Using the expression (\ref{Liepartial}) for the Lie derivative, and taking advantage of the fact that $\mathbf{n}$ and $\mathbf{v}_{(a)}$ are orthogonal for every $a$, we can express it as
\beq
\tmbfK(\xi^{\cdot}) = \tmbfK^{[og]}(\xi^{\cdot}) + \tmbfK^{[pg]}(\xi^{\cdot}) + \tmbfK^{[b]}(\xi^{\cdot}) \quad ,
\eeq
where we defined
\begin{align}
\tilde{K}^{[og]}_{ab}(\xi^{\cdot}) &\equiv - \half \, \bigg( \big( \de_{\mathbf{n}} \, \mathbf{g} \big) \big( \mathbf{v}_{(a)} , \mathbf{v}_{(b)} \big) \bigg) \\[2mm]
\tilde{K}^{[pg]}_{ab}(\xi^{\cdot}) &\equiv \half \, \bigg( \big( \de_{\mathbf{v}_{(a)}} \, \mathbf{g} \big) \big( \mathbf{n} , \mathbf{v}_{(b)} \big) + \big( \de_{\mathbf{v}_{(b)}} \, \mathbf{g} \big) \big( \mathbf{n} , \mathbf{v}_{(a)} \big) \bigg) \\[2mm]
\tilde{K}^{[b]}_{ab}(\xi^{\cdot}) &\equiv \half \, \bigg( \mathbf{g} \big( \mathbf{n} , \de_{\xi^{a}} \mathbf{v}_{(b)} \big) + \mathbf{g} \big( \mathbf{n} , \de_{\xi^{b}} \mathbf{v}_{(a)} \big) \bigg)
\end{align}
where $\mbfg$, $\de_{\mathbf{v}_{(a)}} \, \mathbf{g}$ and $\de_{\mathbf{n}} \, \mathbf{g}$ are evaluated in $\Xd = \vfd(\xid)$. The first two pieces are named ``orthogonal gradient'' and ``parallel gradient'' as they are non-zero when the bulk metric has non-zero derivative respectively in the directions orthogonal and parallel to the cod-1 brane, even when the cod-1 brane is not bent. The third piece is instead due to the bending, since it is non-zero when the brane is bent even if the bulk metric is constant. The three contributions read in coordinates
\begin{align}
\label{Camilla}
\tilde{K}^{[og]}_{ab}(\xi^{\cdot}) &\equiv - \half \, \frac{\de \varphi^{A}(\xid)}{\de \xi^{a}}
\frac{\de \varphi^{B}(\xid)}{\de \xi^{b}} \, n^L(\xid) \frac{\de \, g_{AB}}{\de X^L}\Big\rvert_{X^{\cdot} = \vf^{\cdot}(\xid)} \\[2mm]
\label{Pip}
\tilde{K}^{[pg]}_{ab}(\xi^{\cdot}) &\equiv \half \, n^A(\xid) \, \frac{\de \varphi^{B}(\xid)}{\de \xi^{(a}} \, \frac{\de \varphi^{L}(\xid)}{\de \xi^{b)}} \frac{\de \, g_{AB}}{\de X^L}\Big\rvert_{X^{\cdot} = \vf^{\cdot}(\xid)}\\[2mm]
\label{Pop}
\tilde{K}^{[b]}_{ab}(\xi^{\cdot}) &\equiv n_{L}(\xid) \, \frac{\de^2 \varphi^{L}(\xid)}{\de \xi^{a} \de \xi^{b}}
\end{align}
where $n_M(\xid) = g_{LM}(\vfd(\xid)) \, n^M(\xid)$.

Note that it is always possible, at least locally, to use (N-1) of the N bulk coordinates to parametrize the brane: for definiteness we can indicate the coordinates on the brane with $\xid$, and the bulk coordinates as $\Xd = (\xid, z)$, so essentially we recognize $z$ as the extra dimension. In this case all the components of the embedding function are trivial but $\vf^z$, and (with a little abuse of notation) we call $\vf$ the nontrivial component
\beq
\vfd(\xid) = (\xid, \vf(\xid)) \quad .
\eeq
Using this gauge fixing between the bulk coordinates and the brane coordinates, the system is now characterised by the bulk metric $g_{AB}(\Xd)$ and by one scalar function, the nontrivial component of the embedding $\vf$. We can express the objects which define the geometrical properties of the brane using these quantities: the induced metric takes the simplified form
\beq
\label{indmetbic}
\tmg{ab}(\xid) = \frac{\de \varphi (\xid)}{\de \xi^a} \frac{\de \varphi (\xid)}{\de \xi^b} \, \mg{zz}\big( \vfd(\xid) \big) + \frac{\de \varphi (\xid)}{\de \xi^{(a}} \, \mg{z|b)}\big( \vfd(\xid) \big) + \mg{ab}\big( \vfd(\xid) \big) \quad ,
\eeq
and a 1-form orthogonal to the brane can be found as
\beq
N_A (\xid) \equiv \Big( - \frac{\de \vf}{\de \xia} (\xid), 1 \Big) \quad .
\eeq
Normalizing $N$ we obtain the normal form to the cod-1 brane
\beq
n_A (\xid) \equiv \frac{\vep}{\sqrt{g^{LM} \, N_L \, N_M }} \, \, \, \Big( - \frac{\de \vf}{\de \xia} (\xid) , 1 \Big)
\eeq
where $g^{LM}$ is evaluated in $\Xd = (\xid,\vf(\xid))$, $N_L$ is evaluated in $\xid$ and $\vep = \pm 1$ distinguishes between the two possible orientation choices. Using the results above we can express the extrinsic curvature in a simplified way as well, and for example the bending contribution to the extrinsic curvature reads
\beq
\tilde{K}^{[b]}_{ab}(\xid) = \frac{1}{\sqrt{N_L \, N^L }} \, \, \, \frac{\de^2 \varphi(\xid)}{\de \xi^{a} \de \xi^{b}} \quad .
\eeq

\section{The DGP model}
\label{The DGP model}

The DGP model \cite{DGP00}, in its original formulation, is a codimension-1 braneworld model in five dimensions. The complete spacetime $\mscrM = \mscr{B} \cup \Sigma$ is made up of a five dimensional bulk $\mscr{B} = \mscr{B}_{-} \cup \mscr{B}_{+}$ constituted by the two disjoint pieces $\mscr{B}_{-}$ and $\mscr{B}_{+}$, which have in common a four dimensional boundary $\Sigma = \partial \mscr{B}_{-} = \partial \mscr{B}_{+}$. We assume that the topology of $\mscr{B}_{-}$ and $\mscr{B}_{+}$ is the same as $\mathbb{R}^4 \times \mathbb{R}$. The action of the model is 
\begin{multline}
\label{DGPaction}
S = 2 \Mft \int_{\mscrB} \!\! d^5 X \, \sqrt{-g} \, R + 2 \Mfs \int_{\S} \!\! d^4 x \, \sqrt{-\ti{g}} \, \ti{R} + \int_{\S} \!\! d^4 x \, \sqrt{-\ti{g}} \, \mscr{L}_{M} +\\
+ S_{GH}(\S_{-}) + S_{GH}(\S_{+}) 
\end{multline}
where $S_{GH}(\S_{-})$ and $S_{GH}(\S_{+})$ are the Gibbons-Hawking terms\footnote{$S_{GH} = -4 \Mft \int \!\! d^4 x \, \sqrt{-\ti{g}} \, \ti{K}$, where $K$ is the trace of the extrinsic curvature of the brane.} \cite{Boundaryterm1,Boundaryterm2} on the two sides of the brane, and $\mscr{L}_{M}$ is the matter Lagrangian. Here $g$ is the determinant of the bulk metric and $R$ is the Ricci scalar constructed from it, while $\ti{g}$ is the determinant of the induced metric on the brane and $\ti{R}$ is the Ricci scalar constructed from it. We assume that the mass scales $\Mft$ and $\Mfs$ obey the hierarchy $\Mfs/\Mft \gg 1$. The distinctive feature of this action is the \emph{induced gravity term} 
\beq
2 \Mfs \int_{\S} \!\! d^4 x \, \sqrt{-\ti{g}} \, \ti{R}
\eeq
which as we shall see is responsible for the recovery of the correct 4D Newtonian behavior of gravity on the brane, for small and intermediate distances. This piece of the action can be introduced at classical level purely on phenomenological grounds, but can be also understood as contribution coming from loop corrections in the low energy effective action of a quantum description where matter is confined on the brane \cite{DGP00}.

The equations of motion for this system are
\begin{align}
\label{Bulkeqbfnonz2}
\mbfG =& \,\, 0 \qquad \,\, (\textrm{bulk})\\[2mm]
\label{jcbfnonz2}
\Mft \big[ \tmbfK - \tmbfg \, \textrm{tr} \, \tmbfK \big]_{\pm} + \Mfs \, \tmbfG =& \,\, \tmbfT \qquad (\textrm{brane})
\end{align}
where $\mbfG$ and $\tmbfG$ are the Einstein tensors constructed respectively from $\mbfg$ and $\tmbfg$, $\tmbfK$ is the extrinsic curvature of the brane and $\tmbfT$ is the energy momentum tensor of the matter localized on the brane. Equation (\ref{Bulkeqbfnonz2}) is simply the vacuum Einstein equation in the bulk, while (\ref{jcbfnonz2}) is the Israel junction condition \cite{Israeljc} on the brane. The notation $[\phantom{a}]_{\pm}$ indicates the jump across the brane of the quantity in square parenthesis, or equivalently $[\phantom{a}]_{\pm} = [\phantom{a}]_{\S_{+}} - [\phantom{a}]_{\S_{-}}$.

It is customary to assume that $\mscr{B}_{-}$ and $\mscr{B}_{+}$ are diffeomorphic and to impose a reflection symmetry across the brane ($\mathbb{Z}_2$ symmetry). In this case it is enough to solve the equations of motion in one of the two pieces to know the solution in all the bulk. Assuming that the $\mathbb{Z}_2$ symmetry holds, the equations of motion become
\begin{align}
\label{Bulkeqbfz2}
\mbfG =& \,\, 0 \qquad \,\, (\textrm{bulk})\\[2mm]
\label{jcbfz2}
2 \Mft \big( \tmbfK - \tmbfg \, \textrm{tr} \, \tmbfK \big) + \Mfs \, \tmbfG =& \,\, \tmbfT \qquad (\textrm{brane})
\end{align}
where for definiteness the bulk equation is considered in $\mscrB_{+}$ and the extrinsic curvature is evaluated in $\S_{+}$ with the orientation choice corresponding to the normal vector that points inward $\mscrB_{+}$. Note that assigning the energy-momentum tensor on the brane is not enough to fix univocally the solution of the system above, so an additional condition is needed to render the model self-consistent. This is typical of codimension-1 braneworld models: apart from the junction conditions, a condition on the behavior of the bulk metric at spatial infinity is to be imposed (where by spatial infinity we mean infinity in the extra dimension sense, that is in the direction which is normal to the brane). If we restrict ourselves to solutions where the gravitational field becomes weak at spatial infinity, we can always work at first order in perturbations and write the complete solution as the sum of a particular sourced solution and the general solutions of the homogeneous system of equations. It is standard then to impose as boundary condition at spatial infinity the requirement that the sourced solution decays asymptotically to zero, and that the homogeneous solution is a superposition of outgoing waves only, formalizing the idea that nothing can enter our universe from the extra dimension.

We will call $\Xd = (\xd,y)$ the coordinates in the bulk. Although we could use a generic coordinate system on the brane, we will use four of the five bulk coordinates to parametrize the brane (for the sake of precision $\xd$), which (as we mentioned in the previous section) we can always do at least locally. Following the terminology of subsection (\ref{Mathematical preliminaries on braneworlds}), the embedding function reads
\beq
\label{fouroffive}
\vf^{\cdot}(\xd) = \big( x^{\cdot}, \vf(\xd) \big) \quad .
\eeq
The system is completely determined once we know the bulk metric $\mbfg(\Xd)$ and the brane embedding function $\vf(\xd)$. Using the relations (\ref{Bulkeqbfz2})-(\ref{jcbfz2}), it is straightforward to see that if the brane is empty ($\tmbfT = 0$), the configuration
\begin{align}
\mbfg(\Xd) &= \bar{\mbfg}(\Xd) = \boldsymbol{\e} \nn \\[2mm]
\vf(\xd) &= \bar{\vf}(\xd) = 0 \label{straightflatDGPsolution}
\end{align}
is a solution of the equations of motion, since all the curvature tensors (constructed from the bulk and from the induced metric) vanish. In fact using (\ref{indmetbic}) it is easy to see that the induced metric is flat as well
\beq
\bar{\tilde{\mbfg}}(\xd) = \boldsymbol{\e} \quad .
\eeq
Therefore, a straight brane in a flat bulk is a vacuum solution of the theory. This vacuum solution is quite different from the warped solution of an empty (but of course tensionful) Randall-Sundrum brane, and it may seem surprising that gravity on a DGP brane can be very similar to 4D GR. We turn now to the analysis of weak gravity in the DGP model.

\subsection{Weak gravity in the DGP model}

Let's study perturbations around the flat-Minkowski solution, still using four of the five bulk coordinates to parametrize the brane so that (\ref{fouroffive}) holds. We indicate with $\pi(\xd)$ the perturbation of the embedding function $\vf$ and with $h_{ab}(\xd,y)$ the perturbation of the bulk metric, explicitly
\begin{align}
\vf(\xd) &= \pi(\xd)\\[2mm]
g_{ab}(\xd,y) &= \e_{ab} + h_{ab}(\xd,y) \quad .
\end{align}
We define the perturbation of the induced metric as
\beq
\tilde{h}_{\m\n}(\xd) \equiv \tmg{\m\n}(\xd) - \e_{\m\n} \quad ,
\eeq
and we indicate with $\tilde{\mathcal{T}}_{\m\n}$ the perturbation of the energy-momentum tensor localized on the brane. Note that we can always choose (at full non-linear level) the reference system in the bulk in such a way that $h_{55}(\Xd) = h_{5\m}(\Xd) = 0$, in which case the perturbation of the induced metric takes the form
\beq
\ti{h}_{\m\n} = h_{\m\n}\big\rvert_{y = 0^+} + \dem \pi \, \den \pi \quad .
\eeq

While the above definitions do not assume that $\pi$, $h_{ab}$ and $\mathcal{T}_{\m\n}$ are small, we now focus on studying perturbative solutions to (\ref{Bulkeqbfz2})-(\ref{jcbfz2}) at first order. It is very useful to choose a gauge which simplifies the expressions as much as we can: a common choice is to use Gaussian Normal Coordinates (GNC), where the brane is placed at $y = 0$ and the only non-zero components of the bulk metric perturbations are the 4D ones. This reference system is therefore defined by
\beq
\pi^{(GN)}(\xd) = 0 \qquad \qquad h^{(GN)}_{5a}(\Xd)= 0 \quad ,
\eeq
and have the good property that the induced metric is exactly the bulk metric computed in $y = 0^{+}$
\beq
\tilde{g}^{(GN)}_{\m\n}(\xd) = g^{(GN)}_{\m\n}\big\rvert_{y=0^{+}}(\xd) \quad .
\eeq
We will use instead a different gauge choice, introduced by \cite{Garriga:1999yh}, where we do not fix the position of the brane, and so the bending becomes a physical perturbation mode. On one hand, this is mathematically useful since it permits to simplify the bulk equations. On the other hand, it is also physically useful because the bending mode has a direct geometrical interpretation and its dynamics turn out to be characterised by a different length scale compared to the bulk perturbations, which is important at non-linear level. Without fixing the bending, it is possible to impose more gauge conditions on the bulk metric, and in fact it is possible to impose
\beq
h_{55}(\Xd) = h_{5\n}(\Xd)= 0 \qquad \qquad \eta^{\m\n} h_{\m\n}(\Xd) = \de_{\m} \, h^{\m}_{\,\,\,\n}(\Xd) = 0
\eeq
where indices are raised with the background inverse metric $\e^{\m\n}$. In this gauge, the only non-zero components of the bulk metric perturbations are the 4D ones and the bulk metric is transverse-traceless (TT-gauge), which is the 5D equivalent of what is usually done in GR to study gravitational waves \cite{Wald84}. Note that these gauge conditions can be imposed only in source-free regions, which is always true in our case since we consider an empty bulk.

We can now derive the dynamical equations for the relevant degrees of freedom in this gauge. First, note that the trace of the junction conditions (\ref{jcbfz2}) gives
\beq
\label{bendingeq}
\Box_4 \pi = - \frac{1}{6 \Mft} \, \tilde{\mathcal{T}}
\eeq
where $\tilde{\mathcal{T}} = \eta^{\m\n} \tilde{\mathcal{T}}_{\m\n}$, and we use the notations $\de_{\m} = \frac{\de}{\de x^{\m}}$, $\Box_4 = \e^{\m\n} \demden$ and $\Box_5 = \Box_4 +\de^{2}_{y}$. The latter equation confirms that $\pi$ is not a gauge mode but instead a physical perturbation mode which is sourced by the trace of the energy momentum tensor. The junction condition reads

\beq
\label{jcfirstorderwithpi}
- \half \Big(2 \Mft \, \de_y + \Mfs \, \Box_4 \Big)\Big\rvert_{y = 0^{+}} h_{\m\n} = \tilde{\mathcal{T}}_{\m\n} + 2 \Mft \, \e_{\m\n}  \, \Box_4 \pi  - 2 \Mft \demden \pi \quad ,
\eeq
and we see that the bending mode acts as a source for the bulk metric $h_{\m\n}$ along with the energy momentum tensor. Using the trace equation (\ref{bendingeq}) we can write the equations of motion for the bulk metric in a suggestive way: the bulk equation (\ref{Bulkeqbfz2}) reads
\beq
\label{bulkeqfirstorder}
\Box_5 h_{\m\n} = 0 \quad ,
\eeq
while the junction condition becomes

\beq
\label{jcfirstorder}
- \half \Big(2 \Mft \, \de_y + \Mfs \, \Box_4 \Big)\Big\rvert_{y = 0^{+}} h_{\m\n} = \tilde{\mathcal{T}}_{\m\n} - \third \e_{\m\n}  \, \tilde{\mathcal{T}} - 2 \Mft \demden \pi \quad .
\eeq

\subsubsection{The DGP propagator}

A powerful way to study solutions to linear differential equations in presence of sources is to derive the \emph{propagator}, which roughly speaking is the solution corresponding to a perfectly localized source (it is the Green's function of the differential equation). More precisely, it can be defined as the object
\beq
\mathcal{D}_{\m\n}^{\,\,\,\,\,\,\a\b}(x,y; x^{\prime})
\eeq
such that the solution to the linear differential equation corresponding to a source configuration $\tilde{\mathcal{T}}_{\m\n}(x)$ is
\beq
\label{propagatordef}
h_{\m\n}(x,y) = \int d^4 x^{\p} \, \mathcal{D}_{\m\n}^{\,\,\,\,\,\,\a\b}(x,y;x^{\p}) \, \tilde{\mathcal{T}}_{\a\b}(x^{\p}) \quad .
\eeq
To calculate the DGP propagator, note that we can neglect the term $2 \Mft \demden \pi$ in equation (\ref{jcfirstorder}) since it produces in momentum space a contribution $\sim p_{\m} p_{\n}$, which has no effect at first order if we consider (as we do) test bodies whose energy-momentum tensor is conserved. Therefore the propagator for our system obeys
\begin{align}
\label{propagatorbulkeqfirstorder}
\Box_5 \, \mathcal{D}_{\m\n}^{\,\,\,\,\,\,\a\b}(x,y;x^{\p}) &= 0 \\[2mm]
 \label{propagatorjcfirstorder}
- \half \Big(2 \Mft \, \de_y + \Mfs \, \Box_4 \Big)\Big\rvert_{y = 0^{+}} \mathcal{D}_{\m\n}^{\,\,\,\,\,\,\a\b}(x,y;x^{\p}) &= \Big[ \half \Big( \d_{\m}^{\,\,\a} \d_{\n}^{\,\,\b} + \d_{\n}^{\,\,\a} \d_{\m}^{\,\,\b} \Big) - \third \,\e_{\m\n} \e^{\a\b} \Big] \, \d^{(4)}(x-x^{\p}) \quad .
\end{align}
To find a solution to this system, we can factorize a scalar part $\mathcal{D}_{S}(x-x^{\p},y)$ which depends on the coordinates (where we have made manifest that the propagator can depend only on the difference of the coordinates, due to the 4D translational inveriance of the model) and a purely numerical part which carries the tensor structure $\mathcal{S}_{\m\n}^{\,\,\,\,\,\,\a\b}$
\beq
\mathcal{D}_{\m\n}^{\,\,\,\,\,\,\a\b}(x-x^{\p},y) = \mathcal{S}_{\m\n}^{\,\,\,\,\,\,\a\b} \, \mathcal{D}_{S}(x-x^{\p},y) \quad .
\eeq
Roughly speaking, the tensor structure gives the relative weight between the different components of the resulting metric $h_{\m\n}$, while the scalar part fixes the dependence of the components from the coordinates. Substituting this expression into (\ref{propagatorbulkeqfirstorder})-(\ref{propagatorjcfirstorder}) one gets that the tensor structure is
\beq
\label{tensorstructure}
\mathcal{S}_{\m\n}^{\,\,\,\,\,\,\a\b} = \half \Big( \d_{\m}^{\,\,\a} \d_{\n}^{\,\,\b} + \d_{\n}^{\,\,\a} \d_{\m}^{\,\,\b} \Big) - \third \,\e_{\m\n} \e^{\a\b} \quad ,
\eeq
while the scalar propagator obeys
\begin{align}
\label{scalarpropagatorbulkeqfirstorder}
\Box_5 \, \mathcal{D}_{S}(x-x^{\p},y) &= 0 \\[2mm]
 \label{scalarpropagatorjcfirstorder}
- \half \Big(2 \Mft \, \de_y + \Mfs \, \Box_4 \Big)\Big\rvert_{y = 0^{+}} \mathcal{D}_{S}(x-x^{\p},y) &= \d^{(4)}(x-x^{\p}) \quad .
\end{align}

In the case where the source is static $\tilde{\mathcal{T}}_{\a\b}(x^{\p}) = \tilde{\mathcal{T}}_{\a\b}(\vec{x}^{\, \p})$, the metric $h_{\m\n}$ evaluated on the brane (from equation (\ref{propagatordef})) takes the form
\beq
h_{\m\n}(\vec{x},0) = \mathcal{S}_{\m\n}^{\,\,\,\,\,\,\a\b} \int d^3 \vec{x}^{\, \p} \, \, \tilde{\mathcal{T}}_{\a\b}(\vec{x}^{\, \p}) \, V(\vec{x} - \vec{x}^{\, \p}) \quad ,
\eeq
where $V(\vec{x} - \vec{x}^{\, \p})$ is the (static) potential
\beq
V(\vec{x} - \vec{x}^{\, \p}) = \int dt^{\p} \, \, \mathcal{D}_S (\vec{x} - \vec{x}^{\, \p}, t^{\p}, 0) \quad .
\eeq
Note that the potential actually depends only on the relative distance $r = \norm{\vec{x} - \vec{x}^{\, \p}}$, due to the rotational symmetry of the system. The potential for the DGP model can be found exactly, and reads \cite{DGP00} 
\beq
V(r) = \frac{1}{\pi^2 M_4^2} \, \frac{1}{r} \, \bigg[ \sin \Big( \frac{r}{r_c} \Big) \textrm{Ci} \Big( \frac{r}{r_c} \Big) + \half \cos \Big( \frac{r}{r_c} \Big) \Big( \pi - 2 \, \textrm{Si} \Big( \frac{r}{r_c} \Big) \Big) \bigg]
\label{V} 
\eeq
where ${\rm Ci}(z) \equiv - \int_{z}^{+ \infty} {\rm cos}(t) \, dt/t$ and ${\rm Si}(z)\equiv \int_0^z {\rm sin}(t) \, dt/t$ are respectively the Cosine integral function and the Sine integral function, and the distance scale $r_c$ is defined as follows
\beq
r_c \equiv \frac{M_4^2}{2 M_5^3} \quad .
\label{rc}
\eeq
It can be seen that $r_c$ is a ``crossover'' scale where the behavior of the gravitational potential changes from $4D$ to $5D$. In fact, at short distances $r \ll r_c$ the potential behaves as
\beq
V(r) \simeq \frac{1}{\pi^2 M_4^2} \, \frac{1}{r} \, \bigg[ \frac{\pi}{2} + \bigg( \gamma - 1 + \ln \Big( \frac{r}{r_c} \Big) \bigg) \, \frac{r}{r_c} + \mathcal{O}(r^2) \bigg]
\label{short}
\eeq
and at leading order it has the 4D Newtonian $1/r$ scaling (here $\gamma \simeq 0.577$ is the Euler-Mascheroni constant), while at large distances $r \gg r_c$ we obtain
\beq
V(r) \simeq \frac{1}{\pi^2 M_4^2} \, \frac{1}{r} \, \bigg[ \frac{r_c}{r} + \mathcal{O} \Big( \frac{r_c^3}{r^3} \Big) \bigg]
\label{long}
\eeq
so at leading order it has now the 5D behavior $1/r^2$. This results suggests that we may hope to reproduce GR results using the DGP model as long as we set $r_c$ to be much bigger than the length scales we are interested in, and tune
\beq
\frac{1}{\Mfs} \sim G \quad .
\eeq
Note that this implies the following hierarchy of scales
\beq
r_g \equiv \frac{M}{\Mfs} \lll r_c \quad .
\eeq

\subsubsection{Weak GR gravity \textit{vs.} weak DGP gravity}
\label{Weak GR gravity vs weak DGP gravity}

The story is however more complex than that. Let's consider for definiteness a static and spherically symmetric point source of mass $M$: $\tilde{\mathcal{T}}_{\a\b}(\vec{x}^{\, \p}) = M \, \d_{\a}^{\,\, 0} \d_{\b}^{\,\, 0} \, \d^{(3)}(\vec{x}^{\, \p})$ (which may model a star or a planet). In this case the metric on the brane reads
\beq
h_{\m\n}(\norm{\vecx},0) = \mathcal{S}_{\m\n}^{\,\,\,\,\,\,00} \, M \, V(\norm{\vecx}) \quad ,
\eeq
and one can easily see from (\ref{tensorstructure}) that the off-diagonal components of $\mathcal{S}_{\m\n}^{\,\,\,\,\,\,00}$ are zero while $\mathcal{S}_{00}^{\,\,\,\,\,\,00} = 2/3 = 2 \, \mathcal{S}_{ii}^{\,\,\,\,\,\,00}$. Note furthermore that at first order we have for the induced metric
\beq
\ti{h}_{\m\n}(t,\vecx) \simeq h_{\m\n}(t,\vecx) \quad .
\eeq
Therefore, indicating $r = \norm{\vecx}$ and writing the induced metric in terms of the gravitational potentials $\Psi$ and $\Phi$
\begin{align}
\label{gravitationalpotentials1}
\ti{h}_{00}(r) &= - 2 \, \Phi(r) \\[1mm]
\label{gravitationalpotentials2}
\ti{h}_{0i}(r) &= 0 \\[1mm]
\label{gravitationalpotentials3}
\ti{h}_{ij}(r) &= - 2 \, \Psi(r) \, \d_{ij} \quad ,
\end{align}
we have that for $r \ll r_{c}$
\beq
\label{gravpotDGPlin}
\Phi(r) = 2 \, \Psi(r) \qquad \qquad \Psi(r) = - \frac{M}{\Mfs} \, \frac{1}{12 \, \pi \, r} \quad .
\eeq
The situation is quite different from GR, where one has \cite{MisnerThorneWheeler}
\beq
\label{gravitationalpotentialsGR}
\Phi(r) = \Psi(r) \qquad \qquad \Psi(r) = -GM/r \quad .
\eeq
Despite the fact that (for $r \ll r_{c}$) the two potentials in the DGP model scale as $1/r$, it is apparent that in DGP we can never reproduce the complete GR line element. In fact, suitably tuning the value of $\Mfs$ we can reproduce one of the two potentials, but never both of them. The fact is that, experimentally, we can test both the potentials independently: non-relativistic test bodies (for example a planet orbiting around a star) are in fact influenced only by $\Phi(r)$, while the propagation of light is influenced by both of the potentials. Therefore if we put right the orbits of planets then the light deflection comes out wrong, and conversely if we reproduce the correct light deflection then the orbit of planets does not agree with observations anymore: the relative error we get is as big as 25\% (see \emph{e.g.} \cite{Lue:2005ya}). It seems then that the weak field gravity in the DGP model is irreparably different from the weak field gravity in GR. This difference can be traced back to the fact that in GR the tensor structure is
\beq
\label{GRtensorstructure}
S_{\m\n}^{\,\,\,\,\,\,\a\b} = \half \Big( \d_{\m}^{\,\,\a} \d_{\n}^{\,\,\b} + \d_{\n}^{\,\,\a} \d_{\m}^{\,\,\b} \Big) - \half \,\e_{\m\n} \e^{\a\b} \quad ,
\eeq
and, as a consequence of the coefficient of the last term being $\half$ instead of $\third$, one has
\beq
S_{00}^{\,\,\,\,\,\,00} = S_{ii}^{\,\,\,\,\,\,00} \quad .
\eeq

Regarding the bending mode, as we already saw in the linear approximation it obeys equation (\ref{bendingeq}). Considering the same form for the source term $\tilde{\mathcal{T}}_{\a\b}(\vec{x}^{\, \p}) = M \, \d_{\a}^{\,\, 0} \d_{\b}^{\,\, 0} \, \d^{(3)}(\vec{x}^{\, \p})$ we used to find the gravitational potentials, we find the following profile for the bending mode in presence of a static, spherically symmetric and point-like source
\beq
\label{bendinglinsol}
\pi(r) = - \frac{M}{6 \Mft} \, \frac{1}{4 \pi r}
\eeq
where (the notation is not a happy one in this case) the $\pi$ in the denominator of the right hand side is the number $3.14159 26\ldots$, while the $\pi$ in the left hand side is the bending mode. 

\subsection{Nonlinearities and the Vainshtein mechanism}
\label{Nonlinearities and the Vainshtein mechanism}

From what we said above, it may seem that solar system observations rule out the DGP model for every choice of parameters. However, this conclusion relies on the implicit assumption that, since the motion of planets and light in the solar system is described by weak field (\ie linearized) GR, in the DGP model it should be described by the weak field approximation of DGP. In GR, the scale at which non-linearities become important around a spherically symmetric source is $r_s = GM$: we are then implicitly assuming that the scale at which non-linearities become important in DGP is the scale $r_g \equiv M/\Mfs \sim r_s$ corresponding to the scale at which non-linearities become important in GR, or at least much smaller than the length scales we can probe in earth-solar system measurements. This is however not obvious.

To verify this, we should evaluate all the non-linear terms when the dynamical variables take on their weak field value, and recognize at which length scales such non-linear terms become comparable to the linear ones. Naively, we may in fact expect the presence of a different scale where non-linearities become important in the DGP model: following \cite{PorratiVDVZ}, we notice that the profile for the bending mode in the linear approximation (\ref{bendinglinsol}) becomes very large even for $r \gg r_g $, since
\beq
\pi(r) = - r_c \, r_g  \, \frac{1}{12 \pi r} \quad .
\eeq
This can be traced back to the fact that, at linear level, $h_{\m\n}$ receives contributions both from the extrinsic curvature term (multiplied by $\Mft$) and from the induced gravity term (multiplied by $\Mfs$): as a result of the competition between these two terms, there is a crossover scale $r_c$ above which $h_{\m\n}$ couples to $\ti{\mathcal{T}}_{\m\n}$ with effective strength $G_5= 1/\Mft$, while below $r_c$ it couples with effective strength $G_4 = 1/\Mfs$. At ``small'' scales the behavior of $h_{\m\n}$ is then dictated by $\Mfs \tmbfG$, which sets the scale $r_g = M/\Mfs$ where non-linear terms in $h_{\m\n}$ become important. The bending mode $\pi$, instead, at linear order receives contributions only from the extrinsic curvature term, and therefore couples to $\ti{\mathcal{T}}$ with effective strength $G_5= 1/\Mft$ at all scales: as a result, the solution (\ref{bendinglinsol}) contains only $M/\Mft \sim r_c r_g$. However, at quadratic order we have
\beq
\tilde{h}_{\m\n} = h_{\m\n}\Big\rvert_{y=0^+} + \de_{\m} \pi \de_{\n} \pi \quad ,
\eeq
so the equation of motion for the bending mode acquires a contribution from the induced gravity term as well: the competition between the linear term controlled by $\Mft$ and the quadratic one controlled by $\Mfs$ may introduce a new scale where non-linearities become important.

\subsubsection{The Vainshtein radius}
\label{The Vainshtein radius}

It is actually not difficult to see that, for a static, spherically symmetric point-like source of mass $M$, the term $\de_{\m} \pi \de_{\n} \pi$ (evaluated with the linear profile (\ref{bendinglinsol})) becomes of the same order of $h_{\m\n}(y=0^+)$ at the \emph{Vainshtein radius}
\beq
\label{VainshteinRadius3}
r_V = \sqrt[3]{\frac{M \Mfs}{M_5^6}} \sim \sqrt[3]{r_g \, r_c^2} \quad ,
\eeq
and therefore below this radius the linear approximation cannot be trusted. The hierarchy between $r_g$ and $r_c$ implies that $r_g \ll r_V \ll r_c$: we conclude that the linear approximation for the DGP model breaks down at distances which are much bigger than the distance where the linear approximation breaks down in GR. To be quantitative, using $H_0 \sim 70 \, \textrm{km/s/Mpc}$ and $r_c \sim c/H_0$ we get\footnote{1 MegaParsec (Mpc) is approximately $1 \, \textrm{Mpc} \simeq 3.09 \times 10^{19} \, \textrm{km}$} $r_c \sim 4.3 \times 10^3 \, \textrm{Mpc}$ and for the sun\footnote{$M_{sun} \sim 2 \times 10^{30}$ kg} we get $r_g^{sun} \sim 1.5 \, \textrm{km}$ and finally $r_V^{sun} \sim 3 \times 10^{15} \, \textrm{km} \sim 10^2 \, \textrm{pc}$. Note that the average distance between Pluto and the sun is $\sim 6 \times 10^{9} \, \textrm{km} \sim 10^{-6} \,r_V$: in practice, the light deflection experiments and the orbits of planet and satellites take place in the range $r_g < r < r_V$, so the analysis of the previous section does not apply. Note that we have not shown that above $r_V$ the linear approximation holds: in the complete perturbative expansion there will be interaction terms containing all powers of $\pi$, $h$ and mixed terms $\pi^{n} h^{m}$, each of which, when evaluated on the linear solutions, may become important at a different scale. In principle some non-linear terms may become of the same order of the linear ones at scales which are even higher then $r_V$.

However, it has been shown \cite{Deffayet:2001uk,Tanaka:2003zb,Gruzinov:2001hp,PorratiVDVZ, LutyPorratiRattazzi,NicolisRattazzi} that the approximation where $h_{\m\n}$ is treated at first order while we keep non-linear terms in $\pi$ is consistent, and $r_V$ is indeed the highest of the scales where non-linearities become important. To find out what happens below $r_V$ (\ie for radii smaller than $r_V$ but bigger than the scales where other non-linear terms become important), we can consider the approximated equations of motion where we keep the linear terms in $h$ and the \emph{quadratic} terms in $\pi$. This is equivalent to postulate the following ordering of amplitudes
\beq
\label{amplitudesorderingbending}
h_{\m\n} \sim \epsilon^2 \qquad\qquad \pi \sim \epsilon 
\eeq
and truncate the equations at the $\epsilon^2$ level. This does not change the extrinsic curvature part since corrections start at $\epsilon^3$ level ($h \, \pi$ terms), and changes just the induced gravity term which becomes
\begin{multline}
\ti{G}_{\m\n} = - \half \, \Box_4 h_{\m\n}\Big\rvert_{y = 0^+} + \Box_4 \pi \, \demden \pi - \dem \de^{\la} \pi \, \den \de_{\la} \pi -\\
- \half \, \e_{\m\n} \Big( ( \Box_4 \pi )^2 - \de_{\a} \de^{\la} \pi \, \de^{\a} \de_{\la} \pi \Big) + \mcal{O}(\epsilon^3) \quad .
\end{multline}
Note that, despite the fact that calculating the Einstein tensor from $\dem \pi \den \pi$ one would expect terms with three derivatives, all these terms cancel leaving out an expression which is of second order in derivatives. This property is highly nontrivial and very restrictive, and defines a very interesting class of Lagrangians (the \emph{Galileon} Lagrangians) of which the Lagrangian for the bending mode in the DGP model is just a particular case, as we will see in section \ref{Demixing in the decoupling limit and galileons}. Taking the trace of the junction conditions, we obtain the non-linear equation for the bending mode
\beq
\label{bendingeqIIorder}
\Box_4 \pi + \frac{r_c}{3} \, \Big( (\Box_4 \pi)^2 - \de_{\a} \de^{\la} \pi \, \de^{\a} \de_{\la} \pi \Big) = - \frac{1}{6 \Mft} \, \tilde{\mathcal{T}}
\eeq
which for a static, spherically symmetric, point-like source of mass $M$ can be exactly integrated \cite{NicolisRattazziTrincherini} to give
\beq
\label{pino}
\frac{\pi^{\p}}{r} + \frac{2 r_c}{3} \, \Big( \frac{\pi^{\p}}{r} \Big)^{\!2} = \frac{M}{6 \Mft} \, \frac{1}{4 \pi r^3}
\eeq
where we indicate derivatives with respect to $r$ with a prime. Inserting the linear profile (\ref{bendinglinsol}) in the previous equation one recognizes that the non-linear term becomes comparable to the linear term at the radius $r = r_V$: the Vainshtein radius is therefore not only the radius where non-linearities in $\pi$ become comparable to $h_{\m\n}$ in the induced metric, but also the radius where non-linearities become important in the equation of motion for $\pi$ itself. Equation (\ref{pino}) is an algebraic equation in $\pi^{\p}/r$, in fact a quadratic equation at fixed $r$: we can then solve it exactly obtaining
\beq
\label{bendingnonlinearsolution}
\Big[ \frac{\pi^{\p}(r)}{r} \Big]_{\pm} = - \frac{3}{4 r_c} \bigg( 1 \pm \sqrt{1 + \frac{2}{9 \pi} \, \frac{r^2_c r_g}{r^3}} \bigg) \quad .
\eeq
There are two branches of solutions, characterised by the sign $+$ or $-$: the $-$ solutions is decaying at infinity, while the $+$ one is not (we have $\pi \propto r^2$ for very large radii). The solution we are interested in here is the decaying one, since it has to reduce to (\ref{bendinglinsol}) when $r \gg r_V$: from the previous equation we can obtain the asymptotic behaviors (note that $r^2_c r_g = r^3_V/4$)
\begin{equation}
\label{bendingnonlinearasymptotic}
 \pi^{\p}(r) \quad \left\{
  \begin{aligned}
   \phantom{i} &= \frac{1}{12 \pi} \, \frac{r_c r_g}{r^2} \qquad \qquad \textrm{for} \, r \gg r_V \\[2mm]
   \phantom{i} &= \sqrt{\frac{1}{8 \pi}} \, \sqrt{\frac{r_g}{r}} \qquad \qquad \!\!\!\! \textrm{for} \, r \ll r_V \quad .
  \end{aligned}
\right.
\end{equation}

\subsubsection{The Vainshtein mechanism}

We can pictorially sum up the situation in the following way. The presence of a static point source on the brane has (in our language/gauge choice) two separate effects: it creates a nontrivial profile for the embedding $\pi$ of the brane, and it creates a nontrivial metric $h_{\m\n}$ in the 5D spacetime. The latter effect can in turn be split in the presence of a significant leaking of the gravitational force into the bulk (encoded in $\de_y h_{\m\n}$ in the junction conditions) and the presence of a significant gravitational force on the brane (encoded in $\Box_4 h_{\m\n}$ in the junction conditions). The situation we described so far is then the following: there are two relevant length scales, the crossover radius $r_c$ and the Vainshtein radius $r_V \ll r_c$. Above the crossover scale, the leaking of the gravitational force into the bulk is non-negligible (the extra dimension ``opens up'') so gravity on the brane has a 5D behavior. Below $r_c$ the gravitational leaking is instead negligible, and gravity on the brane is essentially 4D. Above the Vainshtein radius, the bending does not contribute appreciably to the induced gravity term, but acts as a source for $h_{\m\n}$ in such a way that the tensor structure of gravity on the brane is different from the one characteristic of GR. When we approach $r_V$, instead, non-linearities in $\pi$ start becoming important and $\pi$ starts contributing significantly to the induced gravity term.

Nonlinearities in $\pi$ change the bending profile (as we saw) with respect to the linear case, and influence the induced metric since at order $\epsilon^2$ we have
\beq
\label{inducedmetricquadraticpi}
\tilde{h}_{\m\n} = h_{\m\n}\Big\rvert_{y=0^+} + \de_{\m} \pi \de_{\n} \pi \quad .
\eeq
Furthermore, quadratic terms in $\pi$ are likely to modify the way the bending sources the metric $h_{\m\n}$, and so the behavior of the gravitational potentials may be significantly different from what we found in the context of the linear approximation. To study that, we focus on length scales smaller than the crossover scale and larger than the scales where other non-linear terms become important. This implies that we can work at order $\epsilon^2$ (in the sense of (\ref{amplitudesorderingbending})), and at the same time safely neglect the $\de_y h_{\m\n}$ term in the junction conditions. Therefore the junction conditions give
\beq
\Mfs \tG_{\m\n} =  \tilde{\mathcal{T}}_{\m\n} - \third \e_{\m\n}  \, \tilde{\mathcal{T}} - 2 \Mft \demden \pi - \frac{\Mfs}{3} \, \e_{\m\n}  \, \tilde{R}[\pi] \quad ,
\eeq
where
\beq
\label{giov}
\ti{R}[\pi] = (\Box_4 \pi)^2 - \de_{\a} \de^{\la} \pi \, \de^{\a} \de_{\la} \pi \quad .
\eeq

Let's consider as we did before a static, spherically symmetric source of mass $M$. Let's suppose that its radius is smaller than $r_V$, but bigger than any scale where other non-linear terms (other than quadratic in $\pi$) become important: we may schematically model this configuration considering a point-like source of mass $M$ and assuming that the theory at order $\ep^2$ holds down to $r = 0$. The spherical symmetry allows us to write the induced metric in the same form (\ref{gravitationalpotentials1})-(\ref{gravitationalpotentials3}) used at linear level, where now the gravitational potentials $\Phi$ and $\Psi$ contain a contribution from the bending mode as well as a contribution from $h_{\m\n}$, according to (\ref{inducedmetricquadraticpi}). Using the fact that $\tR_{00} = \triangle_3 \Phi$ and $\tG_{00} = 2 \triangle_3 \Psi$, where $\triangle_3$ is the Laplacian operator, we have
\begin{align}
\Mfs \triangle_{3} \Phi &=  \tilde{\mathcal{T}}_{00} + \frac{1}{6} \, \tilde{\mathcal{T}} - \Mft \triangle_{3} \pi - \frac{1}{3} \, \Mfs \, \tilde{R}[\pi] \\[3mm]
\Mfs \triangle_{3} \Psi &=  \half \Big( \tilde{\mathcal{T}}_{00} + \third \, \tilde{\mathcal{T}} \Big) + \frac{1}{6} \, \Mfs \, \tilde{R}[\pi]
\end{align}
where the induced curvature scalar takes the form
\beq
\ti{R}[\pi] = \frac{2}{r^2} \frac{d}{dr} (r \pi'^2) \quad .
\eeq
Note first of all that, if we neglect the quadratic terms $\tilde{R}[\pi]$, integrating the equations above we get exactly the solutions (\ref{gravpotDGPlin}) for the gravitational potentials (in the regime $r \ll r_c$) which we obtained using the propagator tecnique, so the analysis is consistent. Secondly, keeping the quadratic terms in $\pi$, we see that the two gravitational potentials couple differently with the energy-momentum tensor and the linear contribution in $\pi$ (which is the origin of the factor of two difference between the potentials in (\ref{gravpotDGPlin})), but at the same time the non-linear contributions from the bending mode have opposite sign in the two cases. Integrating the equations above on a sphere of radius $r$ and centered on the point-like mass we get
\begin{align}
\frac{\Phi^{\p}}{r} &= \frac{5}{6}\, \frac{r_g}{4 \pi r^3} - \frac{1}{2 \, r_c} \, \frac{\pi^{\p}}{r} - \frac{2}{3}\, \Big( \frac{\pi^{\p}}{r} \Big)^2 \\[3mm]
\frac{\Psi^{\p}}{r} &= \frac{1}{3}\, \frac{r_g}{4 \pi r^3} + \frac{1}{3}\, \Big( \frac{\pi^{\p}}{r} \Big)^2 \quad ,
\end{align}
and using the $r \gg r_V$ and $r \ll r_V$ behaviors (\ref{bendingnonlinearasymptotic}) we arrive at
\begin{equation}
\label{gravpotentialsDGPoutsideVainshtein}
 r \gg r_V \quad \left\{
  \begin{aligned}
   \quad \frac{\Phi^{\p}}{r} &= \frac{r_g}{6 \, \pi r^3} + \mcal{O} \Big( \frac{1}{r^6} \Big) \\[2mm]
   \quad \frac{\Psi^{\p}}{r} &= \frac{r_g}{12 \, \pi r^3} + \mcal{O} \Big( \frac{1}{r^6} \Big)
  \end{aligned}
\right.
\end{equation}
and
\begin{equation}
\label{gravpotentialsDGPinsideVainshtein}
 r \ll r_V \quad \left\{
  \begin{aligned}
   \quad \frac{\Phi^{\p}}{r} &= \frac{r_g}{8 \pi r^3} + \mcal{O} \Big( \frac{1}{r^{3/2}} \Big) \\[2mm]
   \quad \frac{\Psi^{\p}}{r} &= \frac{r_g}{8 \pi r^3} \quad .
  \end{aligned}
\right.
\end{equation}
It is apparent that for $r \gg r_V$ the ``linear'' DGP behavior is reproduced, with the factor two difference between the potentials, while well inside the Vainshtein radius the potentials are equal one to the other and therefore linear GR is reproduced. This is due to the fact that the quadratic contributions in $\pi^{\p}$ have opposite signs for the two potentials, and counterbalance the different way the two potentials couple to the energy-momentum tensor and to the linear terms in $\pi$.

Note that, at order $\ep^2$, the function $\ti{R}[\pi]$ defined in \eqref{giov} is exactly the (4D) scalar curvature on the brane constructed with the induced metric $\ti{\textbf{g}}$, and using the asymptotic behaviors (\ref{bendingnonlinearasymptotic}) it is easy to see that $\ti{R}$ vanishes both for $r \ll r_V$ and for $r \gg r_V$. Therefore the situation on the brane is the following: a static, spherically symmetric massive body is surrounded by a thick spherical shell inside which the (4D) scalar curvature is non-zero, and such that the Vainshtein radius $r_V$ is bigger than the inner radius $r_i$ and smaller than the outer radius $r_o$ of the thick shell. This spherical shell marks the transition from the inner volume $r < r_i$, where Einstein gravity is reproduced (at least at leading order), to the outer volume $r > r_o$, where gravity is still 4D but Einstein gravity is not reproduced. This outer volume in turn extends till a second thick spherical shell appears, this time such the crossover scale $r_c$ is bigger than the inner radius $R_i$ and smaller than the outer radius $R_o$ of the shell, which marks the transition from 4D gravity to 5D gravity, which then extends to spatial infinity.

We can conclude that (quadratic) non-linearities in the bending mode restore the agreement with GR on length scales where non-linearities in GR are still negligible. The fact that agreement with GR is restored via (derivative) self-coupling of a light degree of freedom is known as \emph{Vainshtein mechanism}, and has been proposed for the first time by A.~Vainshtein \cite{Vainshtein72} in the context of massive gravity (which will be treated in the chapters \ref{Ghost Free Massive Gravity} and \ref{Vainshtein Mechanism in Massive Gravity}).

\subsection{Cosmology in the DGP model}

Let's study now cosmological solutions in the DGP model. Following \cite{Deffayet00}, we consider configurations where the 5D metric in the Gaussian normal coordinates reads
\beq
\label{cosmometricDGP}
ds^2 = -N^2(\t, y) d\t^2 + A^2(\t, y) \g_{ij} dx^i dx^j + B^2(\t, y) dy^2
\eeq
where $\g_{ij}$ is a metric on a three dimensional space of constant curvature, and (as in section (\ref{The Robertson-Walker metric})) a parameter $k = +1, 0, -1$ identifies the three possible cases for the sign of the spatial curvature. The brane is located at $y = 0$, where $y$ is the extra dimension, and the induced metric reads
\beq
ds^2 = \tilde{g}_{\m\n} dx^\m dx^\n = - n^2(\t) d\t^2 + a^2(\t) \g_{ij} dx^i dx^j \quad ,
\eeq
where we denote with lower case letters the values of the bulk metric components evaluated on the brane 
\beq
n(\t) = N(\t, 0) \qquad a(\t) = A(\t, 0) \qquad b(\t) = B(\t, 0) \quad .
\eeq
We assume that the matter content of the brane have the usual cosmological form
\beq
\tilde{T}^{\,\,\n}_{\m}(\t) = \mathrm{diag} \big(-\r(\t),p(\t),p(\t),p(\t) \big) \quad .
\eeq
Note that it is always possible to set $n(\t) = 1$ using the gauge freedom and rescaling the time coordinate $\t \rightarrow t$. Using this freedom the Hubble parameter on the brane takes the usual form
\beq
H(t) = \frac{\dot{a}(t)}{a(t)} \quad .
\eeq
We will make the further assumption that the bulk is flat, or equivalently that the bulk metric \eqref{cosmometricDGP} can be transformed into the 5D Minkowski metric by a suitable change of coordinates.

\subsubsection{The modif\mbox{}ied Friedmann equations}

These assumptions imply that we can derive an evolution equation for $H(t)$ without solving the full equations and find the exact metric in the bulk. It can be shown that the Friedmann equation in this case takes the form \cite{Deffayet00}
\beq
\label{DGPFriedmannequations}
H^2 + \frac{k}{a^2} - \epsilon \, \frac{1}{r_c} \, \sqrt{H^2 + \frac{k}{a^2}} = \frac{1}{3 M_4^2} \, \r \quad ,
\eeq
where $r_c = \MPq / 2\Mft$ is the DGP crossover scale, and also that the usual conservation equation holds for matter on the brane
\beq
\dot{\r} + 3 H (p + \r) = 0 \quad .
\eeq
Note that there are two branches of solutions, identified by the the value $\epsilon = \pm 1$ of the parameter $\epsilon$ in \eqref{DGPFriedmannequations}, which corresponds to the sign of the jump of $\de_y A$ across the brane. 

Inspecting the Friedmann equation, we can see that the usual 4D Friedmann equation is reproduced whenever the square root term in \eqref{DGPFriedmannequations} is subdominant with respect to the other two terms. Explicitly this happens when
\beq
\sqrt{H^2 + \frac{k}{a^2}} \gg \frac{1}{r_c}
\eeq
and, considering the cases $k = 0$ or $k = -1$, we find
\beq
\label{convcosmocondition}
H^{-1} \ll r_c
\eeq
so the usual 4D cosmological evolution is reproduced when the Hubble radius is smaller than the crossover scale. Taking as initial condition at a certain $t = \bar{t}$ a configuration where the universe is expanding and satisfies \eqref{convcosmocondition}, we want to study how the late time cosmology predicted by this model looks like. We assume that the 4D universe is filled with matter whose energy density is non-negative and goes to zero when $a \rightarrow + \infty$ (if the equation of state of matter is $p = w \r$, this means that $w \geq -1$).  

\subsubsection{Late time cosmology}

It turns out that the late time cosmological evolution is quite different depending on which branch we consider. To see it more clearly, it is useful to recast the Friedmann equation in the following form:
\beq
\label{DGPFriedmannequations2}
\sqrt{H^2 + \frac{k}{a^2}} = \frac{1}{2 r_c} \Bigg( \epsilon + \sqrt{1 + \frac{4 r_c}{3 M_4^2} \, \r}\, \Bigg) \quad .
\eeq

Let's start by considering the branch of solutions defined by $\epsilon = -1$. Considering just the cases $k = 0$ and $k = -1$, where the universe expands forever (\emph{i.e.} $a(t) \rightarrow +\infty$ for $t \rightarrow +\infty$), we have that at late times the matter density goes to zero, so we can expand the square root in the right hand side of \eqref{DGPFriedmannequations2} to obtain
\beq
\sqrt{H^2 + \frac{k}{a^2}} = \frac{1}{6 M_5^3} \, \r \quad ,
\eeq
which is called the \emph{5D regime}. In this branch, the universe continues expanding with $H \rightarrow 0$ for $t \rightarrow +\infty$, but at late times the expansion rate changes from $(H^2 + \frac{k}{a^2}) \propto \r$ to $(H^2 + \frac{k}{a^2}) \propto \r^2$. In practice, when the Hubble radius reaches the crossover scale $r_c$ the universe starts feeling the extra dimension, and there is a transition in the expansion rate. This branch is usually called the \emph{normal branch}.

Now consider the branch defined by $\epsilon = +1$. Also in this case we restrict the analysis to the cases $k = 0$ and $k = -1$, where $a(t) \rightarrow +\infty$ for $t \rightarrow +\infty$ and we have that at late times the matter density goes to zero. Differently from the normal branch, in this case we have
\beq
\sqrt{H^2 + \frac{k}{a^2}} > H_{self} \equiv \frac{1}{r_c}
\eeq
and the Hubble parameter is bounded from below
\beq
H > H_{self} = \frac{1}{r_c} \quad .
\eeq
This means that, for $t \rightarrow +\infty$, the energy density goes to zero and the scale factor goes to infinity, but the Hubble parameter asymptote the finite and non-zero value $H_{self}$. Therefore, when the Hubble radius reaches the crossover scale $r_c$ and the universe starts feeling the extra dimension, the universe enters an \emph{accelerating} phase: note that this happens for geometrical reasons, without the need of a cosmological constant or of a source term which propels the accelerated expansion. This branch is usually called the \emph{self-accelerating branch}.

It can be shown explicitly \cite{Deffayet00} that these solutions can be embedded in the 5D Minkowski spacetime, and therefore the treatment is self-consistent.

\subsubsection{Estimation of cosmological parameters}

The existence of a self-accelerating cosmological solution is very interesting from the point of view of the late time acceleration problem, since it may explain this puzzling phenomenon by geometrical means \cite{Deffayet:2001pu}. However, a necessary condition for this picture to be feasible is that the cosmological solutions of the DGP model provide a consistent fit to the observational data.

To see if this is indeed the case or not, and to estimate the best fit cosmological parameters, it is useful to express the modified Friedmann equations in DGP cosmology using appropriate density parameters. Note that we can define $\Omega_{M}$, $\Omega_{R}$ and $\Omega_{K}$ in the same way as we did in (\ref{densparameters}) and (\ref{curvdensparameter}), while $\Omega_{\La}$ is absent in this case since $\La = 0$. However, we can take into account the fact that the Friedmann equation is modified by a term dependent by the parameter $r_c$ by introducing a new density parameter
\beq
\Omega_{r_c} \equiv \frac{1}{4 H_0^2 r_c^2} \quad ,
\eeq
in terms of which the (modified) Friedmann equation reads \cite{MaartensMajerotto06}
\beq
\frac{H^2(z)}{H^2_0} = \Big( \sqrt{\Omega_M (1 + z)^3 + \Omega_{r_c}} + \sqrt{\Omega_{r_c}} \, \Big)^2 + \Omega_K (1 + z)^2
\eeq
and the acceleration equation reads \cite{MaartensMajerotto06}
\beq
\frac{\ddot{a}}{a} \, \frac{1}{H_0^2} = \Big( \sqrt{\Omega_M (1 + z)^3 + \Omega_{r_c}} + \sqrt{\Omega_{r_c}} \, \Big) \, \bigg( \sqrt{\Omega_{r_c}} + \frac{2 \, \Omega_{r_c} - \Omega_M (1 + z)^3}{2 \sqrt{\Omega_M (1 + z)^3 + \Omega_{r_c}}} \bigg) \quad .
\eeq
Note that the equations above are written in a dimensionless form, since we used the redshift $z$ as the indipendent (evolution) variable instead of the cosmic time $t$, taking advantage of the fact that they are related in a biunivoque way (at least between the Big Bang and now).

\begin{figure*}
\begin{center}
\includegraphics[width=7cm]{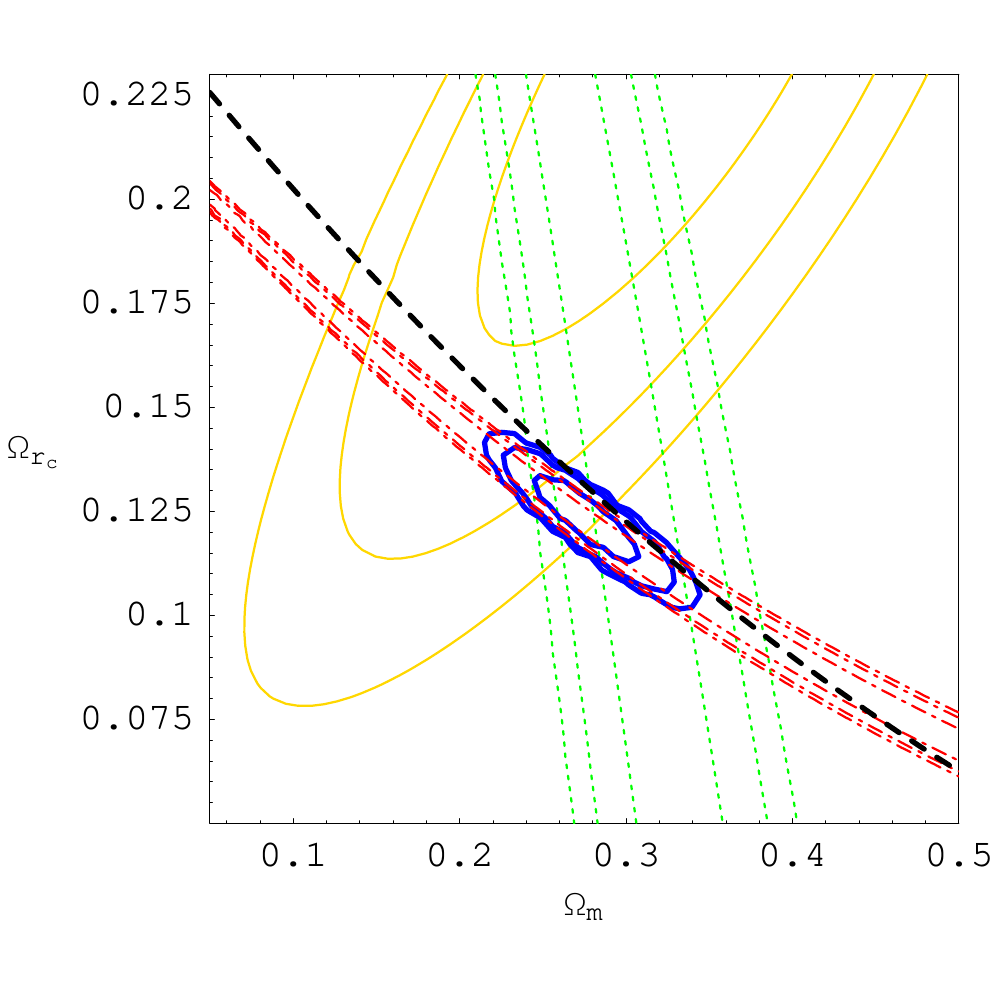}\quad
\includegraphics[width=7cm]{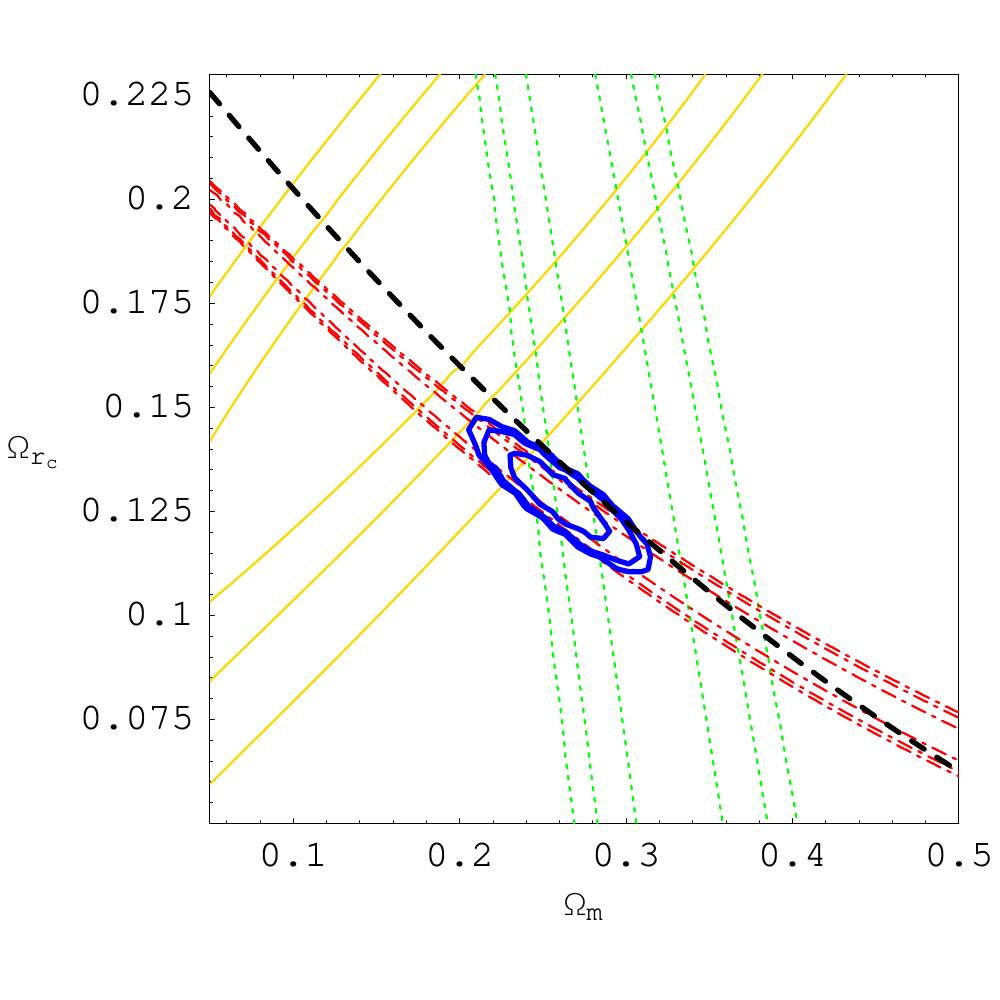}\\
\includegraphics[width=7cm]{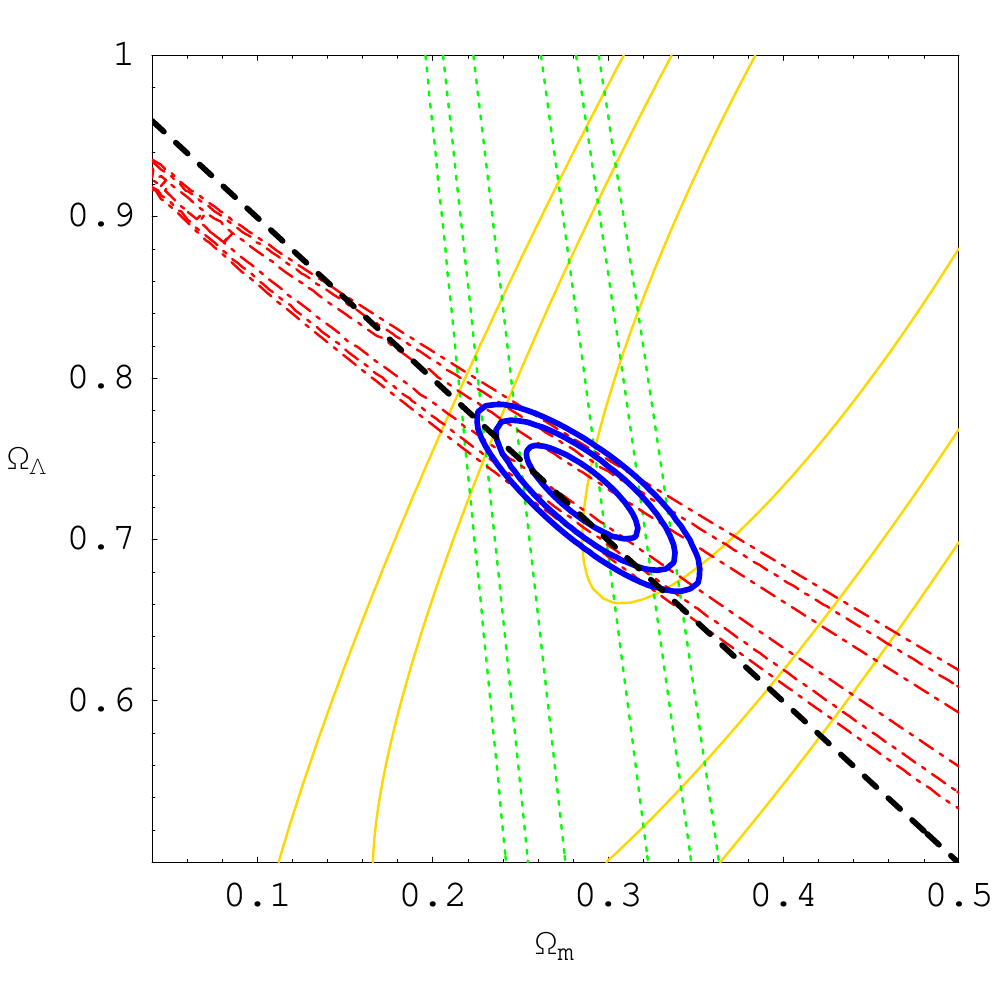}\quad
\includegraphics[width=7cm]{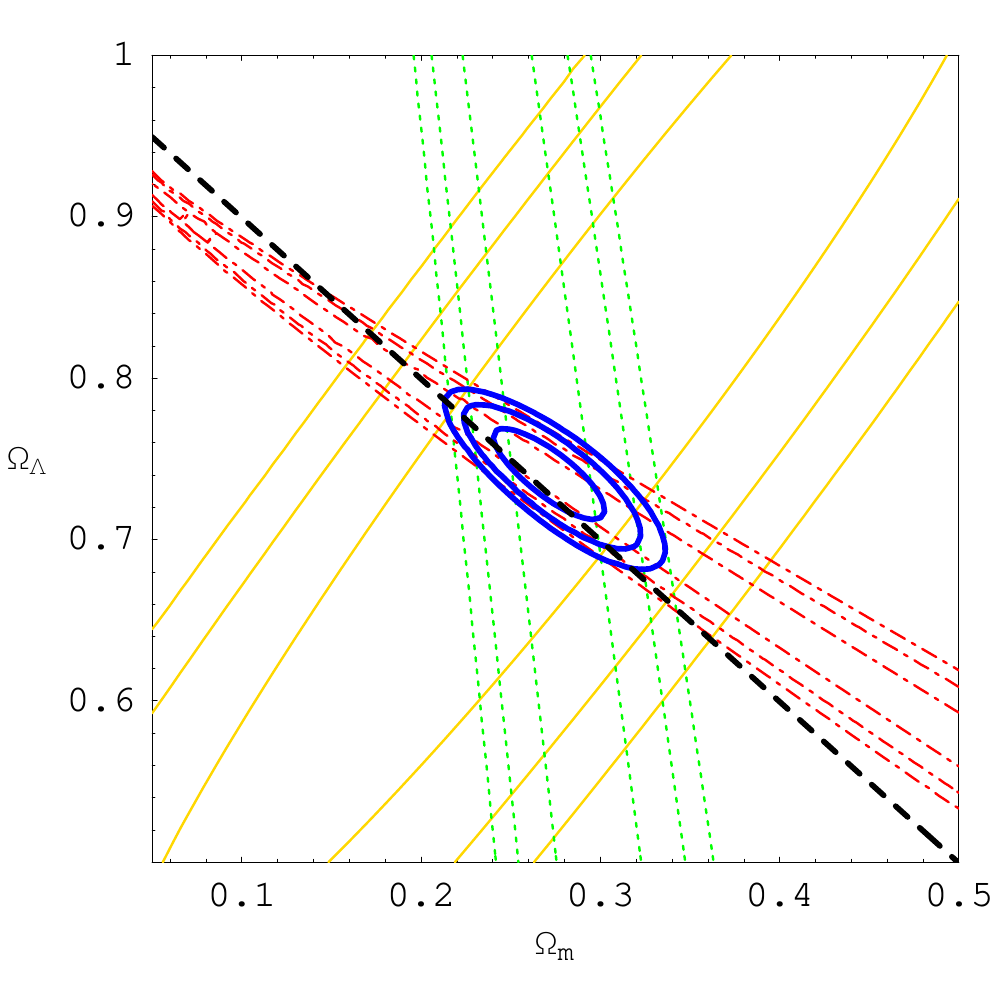}
\caption[Observational constraints for DGP and $\La$CDM]{Joint constraints [solid thick (blue)] on self-accelerating DGP models (above) and on $\La$CDM (below) from the SNe data [solid thin (yellow)], the BO measure $A$ [dotted (green)] and the CMB shift parameter $S$ [dot-dashed (red)]. The left plots and the right plots use different supernovae datasets. The thick dashed (black) line represents the flat models, $\Omega_K=0$. From \cite{MaartensMajerotto06}.} \label{sac}
\end{center}
\end{figure*}

Having expressed the expansion history $H(z)$ in terms of the cosmological parameters, it is possible to estimate how well the theory fits the relevant sets of observational data. Quite in general, every type of cosmological observation is nearly insensitive with respect to changes in the space of the cosmological parameters in some specific directions, while it is sensitive to changes in the other directions: this phenomenon is called \emph{degeneracy}. To constrain the values of the cosmological parameters in a satisfactory way, it is necessary to break the degeneracy by performing a joint fit to several types of observations whose degenerate directions are not parallel: a good choice is to consider the Type Ia supernovae magnitude-redshift relation, the CMB shift parameter $R$ and the position of the BAO peak. In figure \ref{sac}, the $68\%$, $95\%$ and $99\%$ confidence regions in the $\Omega_{r_c}$- $\Omega_M$ plane is displayed for these three types of observations, as well as the confidence levels for the joint fit: it is manifest that for the $\La$CDM model the three $68\%$ confidence regions have a non-empty intersection, while this does not happen for the DGP self-accelerating cosmological solutions.
This implies that there is tension between the DGP cosmology and the data, while the $\La$CDM model provides a significantly better fit to the data \cite{MaartensMajerotto06,Rydbeck:2007gy}. Since DGP and $\La$CDM have the same number of free parameters, and both of them need to be fine tuned, we can conclude that the observational data disfavor the DGP self-accelerating cosmology in comparison to the $\La$CDM model. 

To obtain a stronger constraint on the significance of the fit for the DGP self-accelerating cosmology, and a decisive sentence about its phenomenological feasibility, it is necessary to consider also cosmological observations which test other aspects of the DGP gravity. In fact, different descriptions of the gravitational interaction in general not only give different predictions for the ``background'' evolution of the spacetime, namely the evolution of the scale factor, but also give different predictions for the behaviour of metric perturbations around the background, both at linear level and at non-linear level. The behaviour of gravitational perturbations, in turn, influences structure formation and leaves a mark in several observable aspects of the universe, for example the Large Scale Structure, and influences the weak lensing properties as well. Therefore, to thoroughly test a modified gravity model, it is necessary to combine both distance measures, which probe the background evolution, and growth measures, which probe the evolution of perturbations. This is important also to distinguish different models, such as DGP and appropriately tuned quintessence models, which can produce the same expansion history but give different predictions for the growth of perturbations.

The study of perturbations in the DGP cosmology is notoriously a difficult task, due to the fact that one necessarily has to solve the time dependent five-dimensional equations of motion since the bulk gravitational field responds to, and backreacts on, matter density perturbations on the brane \cite{Koyama:2005kd}. Despite the fact that for linear perturbations on sub-horizon scales one can  analytically take into account 5D effects, and derive effective 4D equations for the matter perturbations and the gravitational potentials \cite{Koyama:2005kd}, for scales comparable to the horizon scale and above one has to resort to numerical computations \cite{Cardoso:2007xc}, which are computationally very demanding. However, it has been shown that it is possible to parametrize the modified gravity effects on all scales in a computationally efficient way, which allows to study in detail the tension of the DGP predictions with observational data, both regarding the background evolution and the growth properties. Such an approach permits to conclude that the DGP predictions show a statistical discrepancy of $\sim 5 \s$ with the observational data \cite{Fang:2008kc}: this result effectively rules out the DGP self-accelerating cosmological solutions from a phenomenological point of view.

\section{Theoretical problems of the DGP model}

We have seen that in the DGP model there is a branch of cosmological solutions which display a transition form the usual 4D cosmological evolution to an accelerated one. This happens without the need of introducing dark energy or a non-zero cosmological constant: it happens for geometric reasons. This result motivated the hope to explain the late time acceleration of the universe by geometrical means \cite{Deffayet:2001pu}, where the transition to the accelerated phase is a consequence of the fact that the correct theory of gravity is not GR, and the difference starts to be felt when the Hubble radius reaches the critical scale $r_c$. Despite being a very appealing possibility, this cannot solve the fine tuning problem which is present in the case of the cosmological constant, since to explain the cosmological observations we have to tune the 5D mass scale $M_5^3$ (and therefore $r_c$) such that the transition happens (in cosmological terms) very close to the matter-radiation equality. More importantly, as we mentioned above the cosmological observations effectively rule out the DGP self-accelerating cosmology from a phenomenological point of view.

Beside this aspect, the DGP model is problematic also from the point of view of theoretical consistency, since it is plagued by the presence of a ghost instability in the self-accelerating branch and by an unacceptably low strong coupling scale. We briefly discuss these two problems below. These issues (both the theoretical and phenomenological ones) are serious enough to force us to abandon the self-accelerating DGP cosmology. However, there is still the possibility that some generalizations of the DGP model may be ghost-free and fit the data significantly better than the original version, thereby providing a geometrical explanation for the late time cosmic acceleration.

\subsection{Ghost instabilities}

A ghost field is, by definition, a field who has negative kinetic energy. Considering for example the following free Lagrangian density for a relativistic scalar field $\f$ in a Minkowski spacetime (indices are raised/lowered with the flat metric $\e^{\m\n}$/$\e_{\m\n}$)
\beq
\label{ghostLagrangianfree}
\mscr{L} = - \frac{\ep}{2} \, \dem \f \, \de^{\m} \f - \frac{\vep}{2} \, m^2 \f^2 \quad ,
\eeq
where $\ep = \pm 1$ and $\vep = \pm 1$, and performing the Legendre transform with respect to $\dot{\f}$ (here an overdot indicates a time derivative), we obtain the Hamiltonian density
\beq
\label{ghostHamiltonianfree}
\mscr{H} = \ep \, \Big( \frac{1}{2} \, \dot{\f}^2 + \frac{1}{2} \, \big( \vec{\nabla} \f \big)^{\! 2} \Big) + \frac{\vep}{2} \, m^2 \f^2
\eeq
in terms of which the Hamiltonian is defined as
\beq
H \equiv \int_{\mathbb{R}^3} \! d^3 x \,\, \mscr{H}[\f,\dot{\f}] \quad .
\eeq
It is easy to see that, if $\ep =\vep = +1$, the Hamiltonian is positive semi-definite and therefore bounded from below, since its minimum value is $H = 0$ which corresponds to the trivial configuration $\f(t,\vecx) = 0$. However, if $\ep =\vep = -1$, the Hamiltonian is negative semi-definite and therefore bounded from above, since now the trivial configuration $\f = 0$ corresponds to its maximum value $H = 0$; finally, in the case $\ep = - \vep$, the Hamiltonian is indefinite and so it is not bounded either from below or from above. The field $\f$ is called a \emph{ghost field} if $\ep = -1$, while is called a \emph{tachyon field} if $\vep = - 1$. If the Lagrangian density is not Lorentz invariant, the part of the kinetic term which decides if the field is a ghost or not is the one which contains the time derivative of the field (the ``velocity'' of the field), or the conjugate momentum in the Hamiltonian formulation. This definitions extend in a straightforward way to more general cases than scalar fields.

As we explain in appendix \ref{Classical and quantum ghosts}, at classical level a free ghost field is a perfectly fine degree of freedom, but as soon as it interacts with a non-ghost field the system becomes unstable. Perturbing a given solution of the equations of motion by a small amount, the fact that the (classical) instability develops or not (and in the former case how fast this happens) depends on the properties of the interaction as well as on the form of the (small) initial perturbation. On the other hand, at quantum level the instability is more severe, since (under very reasonable assumptions) the existence of an interaction term between the ghost sector and the standard sector always produce a spontaneous decay of the initial quantum state with the emission of an infinite amount of radiation and particles. Note that a ghost always effectively interact with the Standard Model fields, because even if there is no direct coupling, both the ghost and the ordinary matter feel the gravitational force, which produce a (graviton mediated) effective interaction. As we show in appendix \ref{Classical and quantum ghosts}, if the ghost-standard fields quantum system is exactly described by a Lorentz-invariant action of the form
\beq
\label{quantumghostSM}
\mscr{L} = \mscr{L}_{\f}[\f,\de \f \,] + \mscr{L}_{SM}[\psi_{(j)}, \de \psi_{(j)}] + \mscr{L}_{int}[\f,\psi_{(j)}] \quad ,
\eeq
where $\mscr{L}_{int}$ is a local interaction term, there is no way to avoid the catastrophic instability.

The only possibility to accomodate ghosts in the theory is to admit that the Lagrangian density (\ref{quantumghostSM}) is just an effective action which describes the quantum dynamic below a cut-off momentum scale, and see if there exists a range of values for the cut-off which on one hand renders the decay rate into ghosts and SM particles acceptably small, and on the other hand preserves the successes of the Standard Model of particle physics. Note that we don't expect (\ref{quantumghostSM}) to be a good description at arbitrarily high energies anyway, nevertheless it is usually believed that the validity of such a description extends at least till the energies where the Standard Model of particle physics is probed, and maybe much further (even till the Planck energy, in the most optimistic case). However, as we explain in the appendix \ref{Classical and quantum ghosts}, to be consistent with the experimental bounds about gamma-rays and SM particles backgrounds (for example positrons), we need to impose that the validity of the description (\ref{quantumghostSM}) breaks down at energies significantly \emph{lower} than the scales where we probe the Standard Model in accelerators. If the theory which is valid above the cut-off is not ghost-free, the breakdown of the effective theory has to be associated to the breakdown of some assumptions which are at the basis of our current comprehension of nature, namely Lorentz-invariance or locality of the interactions. We conclude that, even if strictly speaking it is possible to include ghosts in a low energy effective theory without violating observational bounds, their presence requires a very unorthodox theoretical description.

Therefore, the presence of a ghost excitation in the self-accelerating branch of the DGP cosmology \cite{Koyama:2005tx, Koyama:2007za, Gorbunov:2005zk, LutyPorratiRattazzi, NicolisRattazzi, DGPspectereoscopy} implies that, even if these cosmological solutions were not ruled out by observations, the self-accelerating DGP cosmology would not provide a ``natural'' resolution of the late time acceleration problem.

\subsection{Strong coupling problem}
\label{Strong coupling problem}

So far we have considered the DGP model from a purely classical point of view, and therefore the action (\ref{DGPaction}) has been studied as a classical action. However, we may assume that the DGP model possesses an underlying (fundamental) quantum description, and that the action (\ref{DGPaction}) is just an effective description at classical level of the underlying quantum theory. It is natural from this point of view to ask which is the domain of validity of the effective classical description, or in other words what is the class of solutions of (\ref{DGPaction}) which provide a good approximation of the associated full quantum solutions. In particular, considering a classical solution $\bar{\vf}$ of the DGP model (where with $\vf$ we collectively indicate all the fields which appear in the action), we would like to understand at which scales (if any) quantum corrections to the classical solution start to be non-negligible. An obvious analogue is to be found in condensed matter systems: even if their behaviour is described by quantum mechanics, to some extent we can describe vibrations in solids as classical waves in a continuous medium. However, when the wavelength of the vibrations become comparable to the inter-molecular distance, then quantum effects start to be important and the classical solutions are no more a good effective description of the system.

Since we don't know the details of the fundamental quantum description of the DGP model, it is useful to study the problem semiclassically: we start writing the general configuration as the sum of a background part and a perturbation part
\beq
\vf = \bar{\vf} + \d \vf \quad ,
\eeq
and perform a semi-classical quantization, which means that we quantize only the perturbation $\d \vf$ around the classical background $\bar{\vf}$. Considering the straight-flat solution (\ref{straightflatDGPsolution}) of the DGP model as the background, we can expand the classical action in powers of the perturbation: we find an infinite sequence of interaction terms, each of which is suppressed (once we canonically normalize the kinetic terms) by a characteristic scale $\La_{(i)}$. The smallest of these scales is called the strong coupling scale $\La$. To estimate the quantum corrections, we can calculate the 1-loop effective action, using the strong coupling scale as the cut-off: this generates an infinite sequence of operators (since the DGP action is non-renormalizable) built from $\d \vf$ and its derivatives, each of which is suppressed by powers of the strong coupling scale or higher energy scales. In the DGP case, the operator which fixes the strong coupling scale is the cubic self-interaction term for the bending perturbation $\pi$ \cite{LutyPorratiRattazzi}, and $\La$ reads
\beq
\La = \sqrt[3]{m^2 M_4}
\eeq
where $m \equiv 1/r_c$. It has been shown \cite{NicolisRattazzi} that, to study quantum corrections in the DGP model, it is not necessary to consider the complete DGP action but it is sufficient to consider only the behaviour of the perturbation of the bending mode $\pi$, described by the Lagrangian density
\beq
\mathscr{L}_{\pi} = - \half \, \dem \pi \deum \pi + \frac{1}{\La^3} \, \dem \pi \deum \pi \, \Box \pi \quad .
\eeq
This action has the notable property that, when we calculate the 1-loop corrections, the coefficient in front of the interaction term does not get renormalized \cite{LutyPorratiRattazzi} (a more general non-renormalization theorem can be proved \cite{Hinterbichler:2010xn,deRham:2012ew} for all the Galileon Lagrangians, see section \ref{Demixing in the decoupling limit and galileons}): therefore, the classical action we started with can be trusted as long as we consider energies below the strong coupling scale or length scales bigger than the inverse of the strong coupling scale $r_{\star} = 1/\La$. However, setting $M_4^2 \sim 1/G$ (to reproduce Newtonian gravity) and $m \sim H_0^{-1}$ (which means that the crossover scale is of the same order of the Hubble radius), we obtain that for the perturbations of the DGP model around the straight-flat solution we have \cite{LutyPorratiRattazzi}
\beq
r_{\star} \simeq 1000 \,\, \textrm{km} \quad .
\eeq
We may conclude that the classical solutions of the DGP model lose predictivity for distances below $1000$ km, since at these scales quantum corrections become important and so, to be able to calculate the gravitational field and the bending of the brane in that range of distances, we should know the fundamental quantum description. This result would imply that we cannot calculate the gravitational force between two masses if they are closer than 1000 km, and so the DGP model would be phenomenologically useless. Note that the interaction term which fixes the strong coupling scale is the same term which is responsible for the effectiveness of the Vainshtein mechanism: on one hand, it is responsible for the fact that classical non-linearities become important at anomalously large length scales (compared to the GR case), and on the other hand it is responsible for the fact that the strong coupling scale is anomalously small.

This conclusion is however premature. In fact, if we want for example to compute the gravitational attraction between the Earth and a test body, the straight-flat solution (\ref{straightflatDGPsolution}) is not the correct background solution around which to perform the semi-classical quantization. As a matter of fact, not even the ``linear'' profile for the bending (\ref{bendinglinsol}) around a massive object provides a good background, since as we already mentioned non-linearities become important at astrophysical scales. Considering the background which takes into account also the self-shielding of the bending mode, it is possible to show that the length scale at which quantum corrections become important is severely suppressed, and on the Earth surface is approximately 1 cm \cite{NicolisRattazzi}. Note that this is still above (of about two orders of magnitude) the minimum distance at which GR has been tested: there is therefore a possible danger, since we do not control the quantum corrections in a range of length scales where experimentally we find that they have to be still small. On the other hand, it is reasonable to expect that, since they are small at distances of 1 cm, they don't increase steeply below that distance, so the situation is not as worrying as it were if we had $r_{\star} \sim$ 1000 km. The fact that quantum corrections which we don't know how to calculate may become important at length scales where we can test experimentally the validity of the classical solutions is known as \emph{strong coupling problem}.

\clearemptydoublepage
\chapter{Nested branes with induced gravity}
\label{Nested branes with induced gravity}

In the previous chapter we motivated that it is worthwhile to study generalizations of the DGP model, with the hope of finding new models which are similar enough to the original formulation to preserve its good features, but different enough to be free of its shortcomings. In particular, for what concerns the problem of the late time acceleration of the universe, it is reasonable to expect that (at least some) generalizations of the DGP model still admit cosmological self-accelerated solutions: furthermore, the effective Friedmann equations in the new model will be necessarily modified with respect to the original DGP ones, and so they may fit the data better. Moreover, different geometrical set-ups and/or more sophisticated constructions may provide a mechanism to get rid of the ghost and help with the strong coupling problem, hopefully leading to a phenomenologically acceptable theory.

\subsubsection{Higher dimensional generalizations of the DGP model}

A quite natural way to generalize the DGP model is to consider a higher codimension setup. Higher codimension branes are very interesting in their own rights, and have been extensively studied (see for example the references in \cite{Cline:2003ak,Vinet:2004bk} and, for a more general review, \cite{Burgess:2004ib,Burgess:2013ara}). In particular, codimension-2 branes have attracted a lot of attention since they enjoy the property that pure vacuum energy does not produce 4D curvature on the brane, but merely curve the extra dimensions. This is very interesting from the point of view of the cosmological constant problem, because it may explain by geometrical means why the value of the cosmological constant inferred by cosmological observations is strikingly smaller than the value predicted theoretically. On the other hand, higher codimension branes are notoriously very delicate to deal with, since the thin limit of a brane is not well defined when its codimension is $\geq 2$ if gravity is described by (the higher dimensional generalization of) GR \cite{GerochTraschen}. Moreover, if we put energy-momentum other than pure tension on a brane with codimension $\geq 2$, the gravitational field on the brane diverges in the thin limit. These two results imply that one has necessarily to model the internal structure of the brane. Even if one is interested in an effective description valid on scales much larger than the brane thickness, one has first to solve the coupled evolution of the internal structure and of the external fields, and only afterwards one can obtain from the exact solution the information relevant on scales larger than the brane thickness.

Despite this, generalizations of the DGP model obtained by including an induced gravity term in the action for a brane of codimension $\geq 2$ have themselves been extensively studied (for the earliest works see  \cite{Dvali:2000xg,Dvali:2001gm,Dvali:2001gx,Dvali:2001ae,Gabadadze:2003ck}). However, it is not clear if going to higher codimensions may help with the problem of the ghosts in the self-accelerating cosmology, since in some realizations of the codimension-2 DGP model we find ghosts even among the perturbations around the straight-flat solution \cite{Dubovsky:2002jm} (see however \cite{Berkhahn:2012wg}). Anyway, this seems to be a regularization dependent property, since in another realization of the codimension-2 DGP model (which differs from the previous one in the way it is regularized on very small scales) perturbations around the straight-flat solution have been shown to be ghost free \cite{Kolanovic:2003am}. From the point of view of the cosmological constant problem, instead, considering DGP branes of codimension two is promising because they may provide a realization of the degravitation mechanism \cite{Dvali:2002pe,Dvali:2002fz,ArkaniHamed:2002fu,Dvali:2007kt,deRham:2007rw}.

\subsubsection{Multi-branes models and the Cascading DGP model}

Beside pure codimension-$n$ set-ups, it is possible to generalize the DGP model by considering more elaborate braneworld constructions. For example, we can consider intersecting branes scenarios, where a 4D brane lies at the intersection of higher dimensional branes; a generalization of the DGP model is then obtained by equipping the branes with an induced gravity term (see \cite{Corradini:2007cz,Corradini:2008tu} for an analysis of maximally symmetric solutions and cosmological solutions in this set-up). Another interesting possibility is to consider nested brane set-ups, where a 4D brane is embedded inside higher dimensional branes (see \cite{Morris:1997hj,Edelstein:1997ej} for a field theory realization), and add induced gravity terms to the branes. In particular, a realization of the latter idea (the \emph{Cascading DGP} model \cite{deRham:2007xp}) has been claimed to have very interesting properties, such as the fact that matter with a generic equation of state can be localized on the thin 4D brane, and that in the minimal (6D) set-up there is a critical value for the tension of the 4D brane above which perturbations around the pure tension solutions are ghost-free. Furthermore, it has been shown that this model admits self-accelerating solutions \cite{Minamitsuji:2008fz}, and that it seems to provide a promising setup for the degravitation mechanism \cite{deRham:2007rw,Moyassari:2011nb}.

In this chapter, we study the Cascading DGP model with the aim of verifying some of these claims by explicitly solving the Einstein equations at first order in perturbations. We consider the minimal set-up of the model (6D), and choose a particular regularization of the model (which we call the nested branes realization of the Cascading DGP model). We study the scalar sector of perturbations around the pure tension solutions, confirming that there is no restriction on the equation of state of the matter which can be put on the thin 4D brane. Furthermore, we confirm the existence of a critical tension, which however we find to have a different value from the one that was obtained in the literature, and find a geometrical interpretation for its existence. To perform our analysis, we develop an approach to study perturbations on nested branes with induced gravity which can be generalized to other background configurations. Our results provide a solid basis for further studies of the Cascading DGP model.

\section{Branes of codimension 2 and higher}
\label{Branes of codimension 2 and higher}

In this section, we explain in detail some of the interesting features which characterize branes with codimension larger than one. We first show explicitly that putting a source term which has the form of pure tension on a codimension-2 brane leaves the induced metric flat, then discuss the issue of the thin limit, and the fact that only pure tension can be put on a thin brane of codimension $\geq 2$.

\subsection{Codimension-2 branes and conical spaces}
\label{Codimension-2 branes and conical spaces}

Let's consider a 6D manifold $\mscr{M}$ which is the product of a 2 dimensional Riemannian space $\mscr{C}_2$ and of the 4D Minkowski space
\beq
\label{Jana}
\mscr{M} = \mscr{C}_2 \times Mink_4 \quad .
\eeq
Due to the product structure of the spacetime, it is possible to define a reference system on $\mscr{M}$ by separately introducing a reference system $\zd = (\z^1,\z^2)$ on the extra dimensional manifold $\mscr{C}_2$, and a reference system $\xd = (x^0, x^1, x^2, x^3)$ on the 4D space $Mink_4$. Furthermore, the metric on $\mscr{M}$ can be written as follows
\beq
\label{Bersch}
ds^2 = \g_{ij} \, d\z^i d\z^j + \eta_{\m\n} \, dx^\m dx^\n \quad ,
\eeq
where $\boldsymbol{\g}$ is the 2D Riemannian metric on $\mscr{C}_2$ and is independent of the 4D coordinates $\xd$. It is easy to see that all the connection coefficients which contain 4D indices are identically zero, and the same is true for the components of the Riemann and Ricci tensor. Therefore, the 6D Einstein tensor is of the form
\begin{align}
G_{ij} &= G^{(2)}_{ij} \\[1mm]
G_{i\m} &= 0 \\[1mm]
G_{\m\n} &= - \half \, R^{(2)} \, \eta_{\m\n} \quad ,
\end{align}
where $\mathbf{R}^{(2)}$, $\mathbf{G}^{(2)}$ and $R^{(2)}$ are respectively the Ricci tensor, Einstein tensor and Ricci scalar of the 2D space $\mscr{C}_2$. Furthermore, since the Einstein tensor of a two dimensional Riemannian manifold vanishes identically, the only non trivial components of the 6D Einstein tensor are the $\m\n$ ones.

We would like to use the ansatz (\ref{Bersch}) to construct a solution of the 6D Einstein equations
\beq
\label{Einsteineq6D}
\Msf \, \mathbf{G} = \mathbf{T}
\eeq
when the energy-momentum tensor $\mathbf{T}$ is localized in the extra dimensions. Note first of all that, since as we mentioned above the hypothesis (\ref{Jana}) implies $G_{ij} = G_{i\m} = 0$, the only structure for the 6D energy-momentum tensor which is compatible with the ansatz for the geometry is
\beq
\label{regularizedenmom6D}
T_{AB}(\zd,\xd) = T^{(loc)}_{\m\n}(\zd,\xd) \, \d_{A}^{\,\,\,\m} \, \d_{B}^{\,\,\,\n} \quad .
\eeq
A configuration of this kind corresponds to a situation where the momentum has zero components in the extra dimensions, and it can flow only along the 4D directions. Furthermore, the energy-momentum is separately conserved on each $\zd$-constant 4D slice, and there is no pressure between different slices. Inserting this expression as well as the ansatz (\ref{Bersch}) into the Einstein equations (\ref{Einsteineq6D}), we obtain the following system
\beq
\label{tiritiriti}
- \frac{\Msf}{2} \, R^{(2)} \, \e_{\m\n} = T^{(loc)}_{\m\n} \quad .
\eeq
This implies that the localized energy-momentum tensor, to be compatible with the ansatz (\ref{Jana}), has to be independent of the 4D coordinates $\xd$ and has to be proportional to the 4D metric $\e_{\m\n}$; without loss of generality, we can write it in the form
\beq
\label{regularizedenmompure6D}
T^{(loc)}_{\m\n}(\zd) = - \la \, f(\zd) \, \e_{\m\n}
\eeq
where
\beq
\label{ClaireClaire}
\int_{\mscr{C}_2} \!\! d^2\z \, \sqrt{\g} \, f(\zd) = 1 \quad .
\eeq
We will furthermore ask that $f(\zd)$ is circularly symmetric, and localized around $\r = 0$. By parametrizing the extra dimensional manifold with polar coordinates $(\r , \vartheta)$, this implies that $f(\zd) = f(\r , \vartheta) = f(\r)$, and that there exists a radius $R >0$ such that $f(\r)$ vanishes for $\r \geq R$. Under these hypothesis, to find a solution to the system (\ref{tiritiriti}) we have to find a two dimensional manifold which is circularly symmetric and whose Ricci curvature is non-zero only inside a circle of (coordinate) radius $R$. As shown in appendix \ref{Conical geometry}, a regularized cone has all the requested properties, and provides a solution to the problem. In particular, the deficit angle of the outer part of the cone is determined uniquely by $\la$, and explicitly given by
\beq
\label{deficitanglepuretension6D}
\a = \frac{\la}{\Msf} \quad .
\eeq 
Note that this implies a higher bound on the value of $\la$, since for $\a = 2 \pi$ the cone becomes degenerate: to obtain a well-defined solution, we have to impose $\la < 2 \pi \Msf$.

Considering now a mathematical (thin) codimension-2 brane $\mcal{B}$ positioned at $\r = 0$, we can parametrize it with 4D coordinates $\chd$ such that the embedding function reads
\beq
\phi^{A}(\chd) = \big(0, 0, \chd \big) \quad .
\eeq
Using the general relation (\ref{Induced metric general}), the metric induced on the brane reads
\beq
\label{Jure}
\ti{g}_{\m\n}(\chd) = \e_{\m\n} \quad .
\eeq
We can think of $T^{(loc)}_{\m\n}$ as the energy-momentum tensor present inside a physical (thick) codimension-2 brane, centered around the thin brane $\mcal{B}$. If we are interested in an effective description valid on length scales much larger than the brane thickness, we may consider a limiting description in which the integrated energy-momentum tensor
\beq
\mscr{T}_{\m\n} \equiv \int_{\mscr{C}_2} \!\! d^2\z \, \sqrt{\g} \,\, T^{(loc)}_{\m\n}
\eeq
is exactly localized on $\mcal{B}$ (with infinite density). Using (\ref{tiritiriti}) and (\ref{Jure}), we get
\beq
\label{CandyCandy}
\mscr{T}_{\m\n} = - \la \, \ti{g}_{\m\n} \quad ,
\eeq
which implies that $\mscr{T}_{\m\n}$ describes a situation where there is pure tension on the brane $\mcal{B}$. More in general, we can think of the energy-momentum defined by (\ref{regularizedenmom6D}) and (\ref{regularizedenmompure6D}) as describing a thick cod-2 brane which contains just vacuum energy, whose distribution inside the brane is determined by $f$. Since we showed that this source configuration corresponds to the geometry (\ref{Jana}), where $\mathscr{C}_2$ is a regularized cone and the 4D slices are flat, we conclude that vacuum energy on a codimension-2 brane does not produce curvature on the 4D dimensions, but merely curves the extra dimensions (it creates the tip of the cone). Note that the geometry of the external part of the cone does not depend on how the vacuum energy is distributed inside the thick brane, but depends only on its total amount $\la$. If we interpret the localized source configuration (\ref{CandyCandy}) as the limit of a sequence of configurations where $f$ becomes more and more peaked and $\la$ is held fixed, we can associate it to a spacetime of the form (\ref{Jana}) where $\mathscr{C}_2$ is a ``sharp'' cone with deficit angle (\ref{deficitanglepuretension6D}). In this sense we can say that pure tension on a (thin) cod-2 brane generates a conical singularity.

This result suggests a striking way to look at the cosmological constant problem. The standard approach based on a 4D description of nature is that, if GR is valid, then a non-zero $\La$ produces curvature in the observable universe and strongly affects the spacetime at distances larger than $r_{\La} \sim 1/\sqrt{\abs{\La}}$; the theoretical expectations for $\La$ give a value for $r_{\La}$ which is largely incompatible with the astrophysical and cosmological observational data. From this point of view, the only way to solve the problem still assuming the validity of GR is to admit that, for some reason, the theoretical expectations are vastly wrong and the cosmological constant is much smaller than naively expected. On the other hand, if we relax the assumption that nature is four dimensional at a fundamental level, a new possibility opens up, namely the fact that the cosmological constant may be not at all small, but merely produce little or no effects on the 4D universe we have experience of. The 6D toy model we just studied may suggest that, instead of trying to explain why the vacuum energy is so small, we may try to explain why it gravitates so little. However, for this picture to work it is necessary that the curvature of the 4D universe remains small even when, together with tension, also matter and radiation are present on the codimension-2 brane. Following this approach, we could look for models where, if we add matter and radiation to the brane or if a phase transition happens and the vacuum energy changes abruptly, the geometry induced on our 4D universe dynamically relaxes towards a nearly flat configuration. This mechanism is usually called \emph{self-tuning}; see for example \cite{Koyama:2007rx} and references therein for a review of this idea. 

Unfortunately, the step from putting pure tension on a codimension-2 brane to putting a generic energy-momentum tensor is anything but straightforward. We now turn to the discussion of this issue.

\subsection{Thin limit of branes of codimension-2 and higher}
\label{Thin limit of branes of codimension-2 and higher}

In the previous subsection, we considered a special class of configurations for the geometry (\ref{Jana}) and for the energy-momentum tensor (\ref{regularizedenmom6D}) of a thick cod-2 brane. We saw that, for this class of configurations, the geometry of the spacetime outside the thick brane depends only on the ``total'' energy, namely the integral of $T_{00}$ in the extra dimensions, while it is independent of how the energy-momentum is distributed inside the brane; the details of the internal distribution of the energy-momentum just change the internal structure. Therefore, if we are interested in a description valid on scales much larger than the brane thickness and not in its internal structure, we may take into account all these configurations in a unified way thinking that the source is concentrated (with infinite density) on a thin brane $\mcal{B}$. In this way, we avoid introducing the internal structure which is to be ignored anyway, since we know how to relate the concentrated source to the external geometry. The procedure of effectively describing a concentrated source via a perfectly localized source is known as a \emph{thin limit} description.

In general, considering a specific thick brane set-up, a thin brane configuration is defined by considering a set-up where all the branes has zero thickness and no internal structure, and by equipping them with physical properties using functions which vanish everywhere but on the thin branes. The feasibility of giving a thin limit \emph{description} to the original theory depends on the possibility to define physical properties on the thin branes in such a way that the thin brane configurations have a one-to-one correspondence with the external fields solutions. If this happens, as long as we are interested in a description which is valid at length scales much larger than the (real) branes thickness, we can effectively work only at the thin level and assign directly a thin limit source configuration and obtain the external field configuration and viceversa.

\subsubsection{The case of branes of codimension-2 and higher}
\label{The case of branes of codimension-2 and higher}

The problem of establishing which localized source set-ups in GR admit a thin limit description has been thoroughly studied by Geroch and Traschen in the seminal paper \cite{GerochTraschen}. Starting from the observation that, from a mathematical point of view, a thin source configuration is correctly described by a distribution and not by a function of the spacetime coordinates, they developed a general framework to define a thin limit description based on the fact that, for a specific class of localized set-ups, the Einstein equations can be made sense as an equation between distributions. However, they proved that this framework can be applied only to shell configurations in GR, and not to string or point-like configurations. Furthermore, they proved that, for the case of string defects, it is impossible to define a general law which associates thin source configurations to thick source configurations in such a way that the thin source configuration and the external field configuration are related in a one-to-one way. This results implies that the thin limit is not well defined for strings sources in 4D GR, and this conclusion can be extended to general codimension-2 defects in higher dimensional generalizations of GR.

This conclusion may look surprising from the point of view of the results we obtained in section \ref{Codimension-2 branes and conical spaces}, where we constructed explicitly thin solutions for a codimension-2 brane containing pure tension, using the fact that the external configuration does not depend on how vacuum energy is distributed inside the brane. The results of \cite{GerochTraschen}, however, does not imply that it is impossible to construct specific thin solutions: they merely imply that it is in general impossible to give a general thin limit \emph{description} of cod-2 branes. In fact, the results of section \ref{Codimension-2 branes and conical spaces} heavily depend on the assumption (\ref{Jana}) we made on the geometry of the 6D manifold, which implies the condition (\ref{regularizedenmom6D}) on the energy-momentum tensor. It is in fact possible to see that, if we consider geometrical configurations where the manifold $\mscr{M}$ inside the brane is \emph{not} factorizable, then we can generate the same external solution (a cone with deficit angle given by (\ref{deficitanglepuretension6D}) and flat 4D slices) with an internal source different from vacuum energy (\ref{regularizedenmompure6D})-(\ref{ClaireClaire}), and with a total energy which does not satisfy (\ref{deficitanglepuretension6D}). An explicit example in this sense is given in \cite{GerochTraschen}, which however violates (\ref{regularizedenmom6D}) as well, since it has non-zero pressure in the extra dimensions (in our 6D set-up this corresponds to $T_{\r\r} \neq 0$). Therefore \emph{in general} there is no unique correspondence between internal source configuration and external fields configuration, even though such a unique correspondence can be defined if we restrict to certain subclasses of configurations.

\subsubsection{Generic sources on thin branes}

A even more compelling evidence for the impossibility of giving a thin limit description to cod-2 branes is given by considering sources which do not have the structure of pure tension. It has in fact been shown \cite{Cline:2003ak, Vinet:2004bk} that, if we put on a codimension-2 brane energy-momentum different from pure tension, the external field configuration on the brane diverges when we send the brane thickness to zero (unless we allow for Gauss-Bonnet terms in the bulk action \cite{Bostock:2003cv}). This is in some sense not surprising, since also in Maxwell electrostatics the electric field of a static and string-like (or point-like) electric charge configuration diverges on the source: this behavior can be related to the fact that the Green's function of the Laplace operator is finite only in one dimension, while diverges in the other cases. Therefore, we actually \emph{expect} this divergence to appear when we consider a linearized analysis of the gravitational field of cod-2 branes; from this point of view, it is the pure tension case that is to be considered exceptional, since only in this particular case the exact non-linear solution happens to be finite. The crucial difference is that Maxwell's electrostatics is a linear theory, and so this divergent behavior does not forbid to give a thin limit description to string-like and point-like electric charge configurations \cite{GerochTraschen}. This is however not true in GR, which is a non-linear theory.

These results imply that to describe codimension-2 branes we forcefully have to model the internal structure of the brane, and to define the law which describes how this structure reacts to changes in the external fields and in the energy-momentum content of the brane.

\section{The Cascading DGP model}
\label{The Cascading DGP model}

The fact that (for branes of codimension higher than one) it is necessary to take explicitly into account the internal structure of the brane and its dynamics, is very inconvenient although theoretically sensible. On one hand, quite often braneworld models are used to study at a phenomenological level the effect of matter localization, without having in mind a precise realization of the localization mechanism and therefore of the internal structure. On the other hand, even if we indeed have in mind a concrete realization of the braneworld set-up, it is generically very difficult to solve explicitly the coupled equations of motion for the evolution of the internal structure and of the external fields, especially if we don't consider only highly symmetric configurations.

Another troubling feature of higher codimension braneworlds is the possible presence of ghost instabilities, which casts serious doubts on the phenomenological validity of these models and has to be avoided. Quite recently, a model has been proposed which seems to be able to solve at once both the problem of the divergence in the thin limit and the ghosts problem, the Cascading DGP model \cite{deRham:2007xp}. This model, which we present below, seems to be promising also for the late time acceleration problem, since it has been shown to admit self-accelerating solutions \cite{Minamitsuji:2008fz}, and for the cosmological constant problem, since it is a candidate for an explicit realization of the degravitation mechanism \cite{deRham:2007rw,Moyassari:2011nb}. Other interesting results related to the Cascading DGP model can be found in \cite{Afshordi:2008rd,deRham:2009wb,Khoury:2009tk,Agarwal:2009gy,Wyman:2010jp,Agarwal:2011mg}.

\subsection{The general formulation and the minimal set-up}

In the general formulation of the Cascading DGP model, a $D$-dimensional bulk $\mcal{B}$ ($D\geq 6$) contains a hierarchical sequence of branes of increasing dimensionality $dim = 4,5, \ldots, D-1$ recursively embedded one into the other (the $4$D brane is embedded into the $5$D one, which is embedded into the $6$D one and so on). Each of the branes is equipped with an induced gravity term, which implies that the model can be considered as a higher dimensional generalization of the DGP model. The minimal set-up is the six-dimensional one, which is described by the action
\begin{multline}
\label{CascadingDGP6D}
S = 2 \Msf \int_{\mcal{B}} \!\! d^6 X \, \sqrt{-g} \, R + 2 \Mft \int_{\mcal{C}_1} \!\! d^5 \xi \, \sqrt{-\ti{g}} \, \ti{R} + \\
+ \int_{\mcal{C}_2} \!\! d^4 \ch \, \sqrt{-g^{(4)}} \, \Big( 2 \Mfs R^{(4)} + \mscr{L}_{M} \Big)
\end{multline}
where $\mcal{C}_1$ indicates the cod-1 brane and $\mcal{C}_2$ indicates the cod-2 brane. Here $\ti{\mathbf{g}}$ indicates the metric induced on the cod-1 brane, while $\mathbf{g}^{(4)}$ indicates the metric induced on the cod-2 brane and the Lagrangian $\mscr{L}_{M}$ describes the matter localized on the cod-2 brane: the presence of the Gibbons-Hawking terms (see section \ref{The DGP model}) for each brane is implicitly understood. Similarly to the DGP model, where a $Z_2$ reflection symmetry is enforced across the brane, also in the Cascading DGP model a reflection symmetry is enforced: in the minimal set-up, a $Z_2 \times Z_2$ (double) reflection symmetry is imposed in the bulk (and therefore, by continuity, a $Z_2$ symmetry is imposed on the cod-1 brane). The induced gravity term on the cod-2 brane is in particular necessary for the ability of the theory to reproduce Newtonian gravity on small scales, and this condition fixes the parameter $\Mfs$ to be equal to the Planck mass $M^2_P$: therefore the theory described by the action (\ref{CascadingDGP6D}) has two free parameters, and it is convenient to use the mass scales
\begin{align}
\label{6DCascadingMassScales}
m_5 &\equiv \frac{\Mft}{\Mfs} & m_6 &\equiv \frac{\Msf}{\Mft}
\end{align}
and the associated length scales
\begin{align}
\label{6DCascadingLengthScales}
l_5 &\equiv \frac{1}{m_5} & l_6 &\equiv \frac{1}{m_6} \quad .
\end{align}

\subsubsection{The codimension-1 brane as a regulator}

The study of weak gravitational fields in the Cascading DGP model has first been performed in \cite{deRham:2007xp, deRham:2007rw} where, keeping rigidly fixed the position of the branes, the propagator for weak perturbations of the metric field around the configuration where both the branes and the bulk are flat has been derived. Most interestingly, the cod-2 brane-to-brane propagator (which is the evaluation on the cod-2 brane of the propagator corresponding to a source positioned on the cod-2 brane itself) turns out to be finite, while in the limit $M_5 \rightarrow 0$ one recovers the logaritmically divergent propagator characteristic of pure codimension-2 branes. This result has been confirmed by a more detailed analysis \cite{deRham:2010rw} which studied weak perturbations of the gravitational field around background configurations where pure tension is localized on the cod-2 brane. This implies that the presence of the cod-1 brane with its induced gravity term regularizes gravity, and therefore we can localize on the cod-2 brane an energy-momentum tensor of a generic form; we can say that, concerning the problem of the divergence of the propagator in pure cod-2 branes, the cod-1 brane acts as a ``regulator''.

Furthermore, it has been shown that, if $m_6 \ll m_5$, the static and spherically symmetric gravitational potential has the $6D$ behavior $\propto 1/r^3$ at very large scales ($r \gg l_6$), while it has the $5D$ behavior $\propto 1/r^2$ at intermediate scales ($l_5 \ll r \ll l_6$) and finally it has the $4D$ behavior $\propto 1/r$ at small scales ($r \ll l_5$). In the case $m_6 \gg m_5$, instead, there is a direct transition from the $6D$ behavior to the $4D$ behavior at the scale $r \sim l_{56}$, where $l_{56} \equiv 1/\sqrt{m_5 m_6}$. In the former case, gravity ``cascades'' step by step from $6D$ to $5D$ to $4D$ coming from large to small distances, which justifies a posteriori the name of the model. It is important to note that, despite the gravitational potentials display the $4D$ behavior at small scales, the tensor structure is different from the one of GR \cite{deRham:2007xp,deRham:2007rw}, and so the predictions of the theory are not compatible with the solar system observations in the range of length scales where an analysis at linear order in perturbations is valid. This is analogous to the situation in the DGP model, in which case the agreement with observation is however restored at non-linear order (see section \ref{Nonlinearities and the Vainshtein mechanism}). It has in fact been suggested \cite{deRham:2007xp,deRham:2007rw} that the Vainshtein mechanism is effective also in the Cascading DGP model, and that, in the case $m_6 \ll m_5$, a double version of this mechanism is responsible first for the recovery of the $5D$ tensor structure at intermediate scales, and finally for the recovery of the $4D$ tensor structure at small scales.

\subsubsection{Ghosts in the 6D set-up}

Along with these promising properties of the 6D Cascading DGP model, it has been noted \cite{deRham:2007xp,deRham:2007rw} that, considering small gravitational perturbations around the Minkowski bulk when the cod-2 brane is tensionless, one of the perturbation fields is a ghost. More specifically, if we consider a scalar-vector-tensor decomposition of the metric perturbations with respect to the 4D Lorentz group, the ghost mode has been shown to belong to the scalar sector, which is expected to be the most subtle sector since it contains the fields which describe the fluctuation of the position of the branes. It has however been noted that, if we instead consider a background configuration where the cod-2 brane is tensionful, this conclusion can change; in fact, it has been proposed \cite{deRham:2007xp,deRham:2007rw,deRham:2010rw} that, if the background (cod-2) tension $\bla$ is larger than the critical tension
\beq
\label{critical tension intro}
\blac^{\textup{dRKT}} \equiv \frac{2}{3} \, m_6^2 \, \Mfs \quad ,
\eeq
then the perturbations at linear order are ghost-free, while the ghost appears only when the (cod-2) background tension is smaller than the critical tension $\blac^{\textup{dRKT}}$. This implies that there is an interval of values for the vacuum energy in the 4D observable universe such that the 6D Cascading DGP model is perturbatively ghost free; to confirm the phenomenological viability of the model, the absence of the ghost should be confirmed also at full non-linear level. Remarkably, the fact that the cod-2 tension is bounded from above by the value $\bla_M = 2 \pi \Msf$ (as we shall see below) constrains the value of the free parameters $m_5$ and $m_6$ which corresponds to phenomenologically viable realizations of the model, since only if $\blac < \bla_M$ there exists a window of values for the cod-2 tension where the theory is ghost free. Indeed, as a consequence of the result (\ref{critical tension intro}), it has been claimed \cite{deRham:2007xp,deRham:2010rw} that only the realizations of the 6D Cascading DGP model where gravity cascades ``step by step'' ($m_6 < m_5$) are physically acceptable.

Interestingly, two set-ups which are connected to the Cascading DGP idea have been studied in \cite{deRham:2007rw}, and it has been shown that in these set-ups it is not even necessary to put tension on the cod-2 brane to avoid the ghost. These set-ups are not characterized by the recursive embedding of a 4D brane into a 5D brane, but are constructed by promoting a 4D brane to a \emph{compact} 5D or 6D object (see \cite{Kaloper:2007ap} and \cite{Kolanovic:2003am} for similar constructions). In the first case, the cod-2 brane becomes a compact cod-1 brane, more precisely a 5D spherical surface. In the second case, the cod-2 brane becomes a 6D sphere, which is however characterized by a mass scale different from the bulk one, so that the cod-2 brane can be understood as a medium with non-zero gravitational permeability \cite{Kolanovic:2003am}. The divergence of the gravitational field typical of thin cod-2 branes is regularized by the fact that the coordinate radius $\Delta$ of the 5D spherical surface (respectively of the 6D sphere) is different from zero, and $\Delta$ acts as the regularization parameter. Since the brane is now fully 5D (respectively, 6D), it has to be equipped with a 5D (respectively, 6D) induced gravity term (and not a 4D one as in the action (\ref{CascadingDGP6D})). It turns out that the two set-ups produce the same 4D low energy effective action, which however differs from the 4D effective action derived from (\ref{CascadingDGP6D}) since there is a coupling between the metric on the cod-2 brane and the field which describes the fluctuations of the physical radius of the same brane. The presence of this coupling is in fact crucial, since it is responsible for the sign flip in the kinetic term of the field which is a ghost in the 6D Cascading DGP model, and which is healthy in these set-ups. 

\subsection{The Cascading DGP as a scenario}
\label{The Cascading DGP as a scenario}

Quite in general, it is possible to equip the cod-2 brane with a true 4D induced gravity term only if it is possible to define a thin limit for the cod-2 brane; however, for the reasons explained in section \ref{Thin limit of branes of codimension-2 and higher}, it seems unlikely that the thin limit for the Cascading DGP model can be defined in full generality. In fact, we know that changing the internal structure of a pure cod-2 brane we obtain different phenomenologies even when the brane becomes thin; in the Cascading DGP case, beside the freedom to choose the cod-2 internal structure, we have the additional freedom to choose how the internal structures of the two (cod-1 and cod-2) branes are related one to the other. Therefore, the analysis of Geroch and Traschen \cite{GerochTraschen} imply that, unless we are able to show that the Cascading DGP admits a description in terms of distributions, we need to view the Cascading DGP not as a model but more precisely as a \emph{scenario}. We in fact expect that different ways to specify the internal structures of the branes (both singularly and in relation one to the other) in the Cascading DGP set-ups may lead to truly different models, with different phenomenologies.

\subsubsection{Thin limit and hierarchy between branes}

This implies that, if we want to study the phenomenology of the 6D Cascading DGP scenario in a clean way, we should consider configurations where both of the branes are thick, and model the internal structures of the branes, with particular attention to their mutual relation. This is however extremely complicated, and probably not doable in practice. To facilitate the analysis, however, we could consider particular cases in which there is a hierarchy of scales between the two branes, with the hope that this permits to describe the system with a good approximation by considering one of the branes thin (relatively to the other). 
\begin{figure}[htb!]
\centering
\begin{tikzpicture}
\begin{scope}[scale=.9,rotate=30,>=stealth]
\fill[color=green!20!white] (-6,-1.5) rectangle (6,1.5);
\filldraw[fill=red!20!white,draw=black] (0,0) ellipse (2cm and 1cm);
\draw[very thick] (-6,-1.5) -- (6,-1.5);
\draw[very thick] (-6,1.5) -- (6,1.5);
\draw[gray,thin,<->] (0,0) -- node[color=black,anchor=west]{$l_2^{\perp}$} (0,1);
\draw[gray,thin,<->] (0,0) -- node[color=black,anchor=north]{$l_2^{\shortparallel}$} (2,0);
\draw[gray,thin,<->] (5,-1.5) -- node[color=black,anchor=east]{$l_1$} (5,1.5);
\end{scope}
\end{tikzpicture}
\caption[Characteristic scales for the cod-1 and cod-2 branes]{Characteristic scales for the cod-1 brane (green) and the cod-2 brane (ellipse, violet)}
\label{characteristic nested scales}
\end{figure}
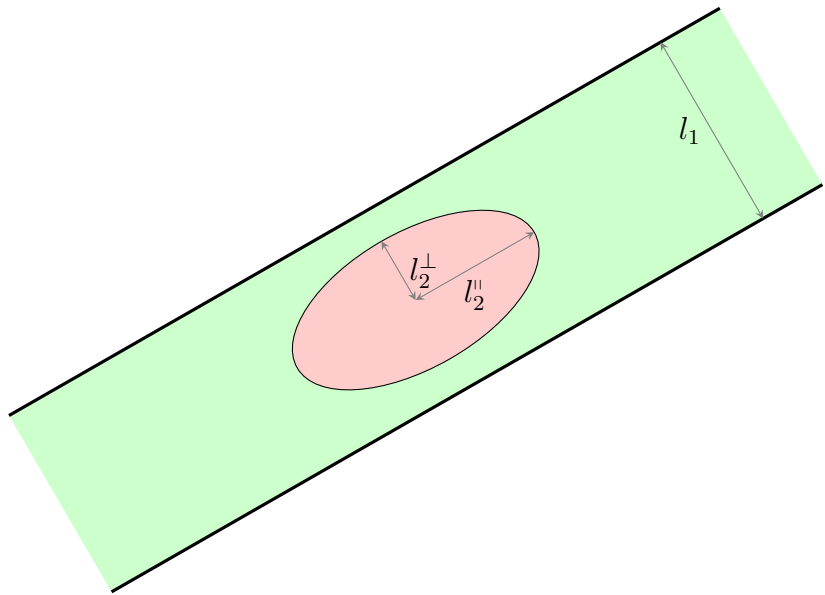
To clarify this point, let's consider in fact the simple schematic description of figure \ref{characteristic nested scales} where the 2D sections of the branes in the extra dimensions are plotted (each point in the figure represents a 4D spacetime): we indicate with $l_1$ the thickness of the cod-1 brane, with $l_2^{\shortparallel}$ the thickness of the cod-2 brane in the parallel directions (from the point of view of the cod-1 brane) and with $l_2^{\perp}$ the thickness of the cod-2 brane in the normal direction.

Among the infinite possible choices for the three representative thicknesses, we could consider cases where there exist definite hierarchic relations between them: for example, we can consider a case (case A) where we have $\ltp \sim \ltn$, $\ltn \ll \lo$, and a case (case B) where we have $\ltp \gg \ltn$, $\ltn \sim \lo$. In the first case, the cod-2 brane is nearly circular in section and its radius is much smaller than the cod-1 thickness: we may consider the cod-2 brane to be thin relatively to the cod-1 brane, and expect that in this case the 6D configurations are very much similar to ``conical'' configurations with a small regularized tip (such as in a pure cod-2 case), where the effect of the broad cod-1 brane is only to distort the 5D sections of the cone from the circular shape. In the second case, instead, we may consider the cod-1 brane to be thin with respect to the cod-2 brane: we could take advantage of this hierarchy and describe this situation by considering a perfectly thin cod-1 brane, and ask that the matter in the cod-1 brane is distributed only inside a ribbon of width $\sim \ltp$. 

\subsubsection{The nested branes realization of the Cascading DGP scenario}

The situation described by the second case discussed above has definitely some good aspects: on one hand, it permits us to study the problem using the formalism of cod-1 branes, which in particular implies that the internal structure of the cod-1 brane does not play a role, and has a clear connection with the (cod-1) DGP model. On the other hand, it is possible to show (as we shall see) that the analysis of \cite{deRham:2010rw} implicitly makes use of this assumption, which implies that some important properties, such as the fact that the presence of the cod-1 brane regularizes gravity and the existence of the critical tension, should be enjoyed by this class of configurations. Therefore, in the following we will consider only realizations of the Cascading DGP scenario where the characteristic length scales of the branes satisfy the hierarchy $\ltp \gg \ltn$, $\ltn \sim \lo$, and which can be described to a good approximation by assuming that the cod-1 brane is perfectly thin, while energy and momentum are distributed inside a ``ribbon'', which constitutes the cod-2 brane. Henceforth, we refer to this realization of the 6D Cascading DGP scenario as the \emph{nested branes realization of the 6D Cascading DGP scenario}.

Note that, despite for the selected class of configurations the internal structure of the cod-1 brane does not play a role, a priori we don't know if the thin limit of a ``ribbon'' cod-2 brane inside an already thin cod-1 brane is well defined or not. This is in fact a very important point to establish; if this (second) thin limit is well defined, it is possible to work with a thin cod-2 brane and forget the internal structure of the cod-2 brane as well, thereby simplifying further the analysis. Furthermore, if this is the case, it is straightforward to equip the cod-2 brane with a truly 4D induced gravity term.

\section{Nested branes with induced gravity}

Regarding the nested branes realization of the 6D Cascading DGP scenario, it is reasonable to expect that whether the thin limit of the cod-2 brane is well defined or not does not depend on the presence of the induced gravity term on the cod-2 brane. Somewhat similarly, the analysis of \cite{deRham:2007xp} seems to indicate that gravity is regularized due to the presence of the cod-1 induced gravity term, not due to the cod-2 induced gravity term. On the other hand, the existence of a critical tension above which the theory is ghost-free and the recovery of Newton gravity on small scales seem to be tightly linked to the presence of the induced gravity term on the cod-2 brane.

These considerations suggest that it may be convenient to split the analysis of the 6D Cascading DGP model in two parts. Following this suggestion, in this and the following two sections (\ref{Perturbations of nested branes with induced gravity} and \ref{Thin limit of nested branes with induced gravity}) we develop a framework to study perturbations in a general set-up where a thin codimension-1 brane equipped with an induced gravity term is embedded in a 6D ambient space (although the extension to a general $N$-dimensional case is straightforward), and contains an energy-momentum tensor localized inside a ``ribbon'' codimension-2 brane (without making any hypothesis on the form of the latter source term). We refer to this set-up as the \emph{nested branes with induced gravity} set-up, in contrast with the nested branes realization of the 6D Cascading DGP scenario where the hypothesis of the presence of the cod-2 induced gravity terms is made. This analysis permits us to show that the thin limit of the cod-2 brane inside the thin cod-1 brane is well defined (at least when considering first order perturbations around pure tension backgrounds), which means that we can equip the (thin) cod-2 brane with a 4D induced gravity term. In section \ref{Ghosts in the 6D Cascading DGP model}, instead, we specialize this framework to study the properties of the nested brane realization of the 6D Cascading DGP scenario: notably, we will concentrate on the existence of ghost modes in relation to the tension of the 4D brane.

\subsection{The set-up}
\label{The set-up}

Before turning to the study of the solutions of the equations of motion, it is useful to introduce the formalism we use to study the nested branes with induced gravity set-up: this also allows us to set the notation and the conventions.

In the following, we consider a 6D spacetime with a 5D submanifold (codimension-1 brane) $\mcal{C}_1$ embedded in it; we suppose that the brane is (globally) orientable, and divides the spacetime in two disconnected pieces which are diffeomorphic. The complete spacetime is then divided in two patches with a common boundary: we assume that there is a 4D submanifold (codimension-2 brane) $\mcal{C}_2$ embedded in the codimension-1 brane, which likewise divides the codimension-1 brane in two patches whose common boundary is the codimension-2 brane. Note that the branes $\mcal{C}_1$ and $\mcal{C}_2$ are mathematical branes, \emph{i.e.} thin branes; we assume that the codimension-1 brane is equipped with a (5D) induced gravity term, and that matter and tension are confined inside the codimension-1 brane and localized around the codimension-2 brane $\mcal{C}_2$. More specifically, we distinguish between a physical (thick) cod-2 brane, inside which matter and tension are confined (the ``ribbon'' cod-2 brane), and a mathematical (thin) cod-2 brane, with respect to which the $Z_2$ symmetry is imposed (see below). When the thin limit of the cod-2 brane is performed, the physical brane coincides with the mathematical one.

We assume that a $Z_{2}$ symmetry holds across the codimension 1 brane, so that the geometry of the whole spacetime can be obtained once we know the geometry of one of the two 6D patches, which we will call the bulk. The fact that the codimension 1 brane is thin implies that solving the equations of motion for the whole system reduces to solving the Einstein equation in the bulk
\beq
\label{Amsterdam}
G_{AB} = 0 \quad ,
\eeq
and imposing that the Israel junction conditions hold at the boundary (\emph{i.e.} at the cod-1 brane)
\begin{equation}
\label{junction conditionsch3}
2 M_6^4 \, \Big( \ti{K}_{ab} - \ti{g}_{ab} \, \big( \tilde{g}^{cd} \tilde{K}_{cd}  \big) \Big) + \Mft \, \ti{G}_{ab} = \ti{T}_{ab} \quad .
\end{equation}
Here $\ti{K}_{ab}$ is the extrinsic curvature of the codimension-1 brane, $\ti{G}_{ab}$ is the Einstein tensor constructed from the metric induced on the codimension-1 brane and $\ti{T}_{ab}$ is the 5D energy-momentum tensor present on the codimension-1 brane. The choice of orientation of the cod-1 brane which fixes the extrinsic curvature is the one defined by the normal vector which points inward the bulk. Furthermore, a $Z_{2}$ symmetry with respect to the cod-2 brane is assumed to hold inside the cod-1 brane: note that we do \emph{not} impose a $Z_{2} \times Z_{2}$ symmetry as in the original formulation of the Cascading DGP model, since the fact that a $Z_{2}$ symmetry holds inside the cod-1 brane does not imply that a double $Z_{2}$ symmetry holds in the bulk.

\subsubsection{The geometrical set-up}

Since the cod-1 and cod-2 branes are submanifolds of the 6D ambient space, they may be considered as separate manifolds, each one equipped with its atlas of reference systems plus an embedding function which describes how they are embedded in the ambient space. We therefore consider a 6D reference system $X^{A}= (z,y,x^{\m})$ in the bulk, while a 5D reference system $\xi^{a} = (\xi, \xi^{\m})$ is introduced on the cod-1 brane and a 4D reference system $\chi^{\m}$ is introduced on the codimension-2 brane. The position of the cod-1 brane in the bulk is described by the embedding function $\vfd$ whose component expression is $\varphi^{A}(\xi^{a})$, while the position of the cod-2 brane inside the cod-1 brane is described by the embedding function $\ti{\a}^{\cdot}$ whose expression in coordinates is $\ti{\a}^{a}(\chi^{\m})\,$. The bulk metric $\mathbf{g}$ induces on the codimension-1 brane the metric $\ti{\mathbf{g}} \equiv \varphi_{\star} \big( \mathbf{g} \big)$, where $\varphi_{\star}$ indicates the pullback with respect to the embedding function $\vf^{\cdot}$, and in turn the metric $\ti{\mathbf{g}}$ induces on the codimension-2 brane a metric $\mathbf{g}^{(4)} \equiv \ti{\a}_{\star} \big( \mathbf{\ti{g}} \big)$, where $\ti{\a}_{\star}$ indicates the pullback with respect to the embedding function $\ti{\a}^{\cdot}$. In general, quantities pertaining to the cod-1 brane are denoted by a tilde $\tilde{\phantom{a}}$, while quantities pertaining to the cod-2 brane are denoted by a superscript $\phantom{a}^{(4)}$. In coordinate representation, the metric $\ti{\mathbf{g}}$ induced on the cod-1 brane is given by the expression (\ref{Induced metric general}) while the metric $\mathbf{g}^{(4)}$ induced on the cod-2 brane reads
\beq
g^{(4)}_{\m \n}(\chi^{\cdot}) = \frac{\de \tilde{\a}^{a}(\chi^{\cdot})}{\de \chi^{\m}}
\frac{\de \tilde{\a}^{b}(\chi^{\cdot})}{\de \chi^{\n}} \, \, \tilde{g}_{ab}(\xi^{\cdot})
\Big\rvert_{\xi^{\cdot} = \tilde{\a}^{\cdot}(\chi^{\cdot})} \quad .
\eeq
There is a particular class of reference systems for the codimension-1 brane which has an important role in the following: the cod-1 reference systems which are Gaussian Normal with respect to the cod-2 brane. Given a reference system $\chd$ on the cod-2 brane, every point of the cod-1 brane which lies in a neighborhood of the cod-2 brane can be reached from a unique point $\chd$ of the cod-2 brane following a geodesic of the induced metric $\ti{\mathbf{g}}$ which is normal to the cod-2 brane. We can then define a reference system on the cod-1 brane (at least in the neighborhood of the cod-2 brane) by assigning to each point the coordinates of the starting point on the cod-2 brane, and the value of the affine parameter $\hxi$ of the normal geodesic, with the convention that $\hxi$ is zero for the point belonging to the cod-2 brane. We refer to this class of reference systems as codimension-1 Gaussian Normal reference systems, and we indicate quantities evaluated in this coordinate systems with an overhat $\h{\phantom{a}}$. By construction, we have that \cite{CarrollBook}
\begin{align}
\h{g}_{\xi\xi}(\hxi,\chd) &= 1 & \h{g}_{\xi\m}(\hxi,\chd) &= 0
\end{align}
and moreover we have that, choosing a fixed $\hxi$, the 4D tensor $\h{g}_{\m\n}(\hxi,\chd)$ (seen as a function of $\chd$) is the induced metric on the 4D slice characterized by that particular $\hxi$. From the point of view of the notation, we synthetically indicate the cod-1 GN coordinates as $\hxid \equiv ( \hxi, \chd )$. The requirement that a $Z_{2}$ symmetry with respect to the cod-2 brane is assumed to hold inside the cod-1 brane is formalized asking that, when expressed in cod-1 GNC, the $\m\n$ and $\xi\xi$ components of the induced metric $\h{g}_{ab}$ and of the extrinsic curvature $\h{K}_{ab}$ are symmetric with respect to the reflection $\hxi \rightarrow - \hxi$, while the $\xi\m$ components are antisymmetric.

\subsubsection{The source set-up}

We assume that the (5D) energy-momentum tensor present on the cod-1 brane, which sources the junction conditions (\ref{junction conditionsch3}), is localized around the (mathematical) cod-2 brane or in other words is localized inside the (physical) cod-2 brane. By ``localized'' we mean that, first of all, momentum, momentum flux and pressure vanish in the direction orthogonal to the cod-2 brane, which means that the cod-1 energy-momentum tensor in cod-1 GN coordinates reads
\beq
\label{angelica}
\hat{T}_{ab}(\hxi, \chd) = \d_{a}^{\, \, \m} \, \d_{b}^{\, \, \n} \, \hat{T}^{(loc)}_{\m\n}(\hxi, \chd) \quad .
\eeq
Secondly, we ask that there exists a (finite) localization length $l_2$ such that $\h{T}^{(loc)}_{\m\n}$ vanishes when it is evaluated at a distance $\hxi$ from the cod-2 brane which is bigger than $l_2$ (the length $l_2$ corresponds, in the language of section \ref{The Cascading DGP as a scenario}, to the ``parallel'' thickness $\ltp$). More precisely, considered a cod-1 GN reference system $( \hxi , \chd )$, to each point $\chd$ of the cod-2 brane we can associate the positive number $l_2(\chd)$ which is the local \emph{coordinate} thickness (in the $\hxi$ direction) of the cod-2 brane. We define $l_2$ to be the upper bound of these local thicknesses
\beq
l_2 \equiv \textrm{sup} \{ l_2(\chd) \}_{\chd \in \, \mcal{C}_2} \quad .
\eeq
We say that the cod-2 brane has a finite thickness if and only if the superior extreme of the local thicknesses is finite, in which case we simply call $l_2$ the thickness of the (physical) cod-2 brane. Note that the definition is independent of the choice of the coordinate system on the cod-2 brane. We furthermore define the cod-2 energy-momentum tensor as the 4D tensor $T^{(4)}_{\m\n}(\chd)$ obtained by the pillbox integration of $\h{T}_{ab}$ across the cod-2 brane, so that we have
\beq
\label{cristina}
\int_{-l_2}^{+l_2} d \hxi \,\, \hat{T}_{ab}(\hxi, \chd) = \d_{a}^{\, \, \m} \, \d_{b}^{\, \, \n} \, T^{(4)}_{\m\n}(\chd) \quad .
\eeq
We can consider the latter tensor as the ``would be'' thin limit source configuration if the thin limit description were well-defined. From this point of view, we can consider different configurations $\ti{T}_{ab}$ which correspond to the same $T^{(4)}_{\m\n}$ as different regularizations of the perfectly localized source $T^{(4)}_{\m\n}$. Note that, since as we already mentioned we don't know if the thin limit is well-defined for nested braneworld set-ups, we don't assume \emph{a priori} that the cod-2 energy-momentum tensor fixes uniquely the solution for the geometry outside the thick cod-2 brane.

In the following, we perform the pillbox integration of the junction conditions across the cod-2 brane; we will use the notation
\beq
\label{strange equation1}
\int_{-}^{+} \! d\hxi \equiv \int_{-l_2}^{+l_2} \! d\hxi
\eeq
and also, given a quantity $\mscr{Q}(\xi)$ defined on the cod-1 brane, we indicate
\begin{align}
\mscr{Q}\Big\rvert_{+} &\equiv \mscr{Q}\Big\rvert_{\xi = +l_2} & \mscr{Q}\Big\rvert_{-} &\equiv \mscr{Q}\Big\rvert_{\xi = -l_2}
\end{align}
and finally
\beq
\label{strange equation2}
\Big[ \mscr{Q} \Big]_{\pm} \equiv \mscr{Q}\Big\rvert_{\xi = +l_2} - \mscr{Q}\Big\rvert_{\xi = -l_2} \quad .
\eeq

\subsection{Pure tension solutions}
\label{Pure tension solutions}

Pure tension solutions in nested branes set-ups have been studied already in \cite{Gregory:2001xu,Gregory:2001dn} (without induced gravity terms) and \cite{Dvali:2006if} (with induced gravity terms) in the context of 5D braneworld models, where extended sources inside the 4D brane were used to investigate the non-perturbative properties of these theories. In this subsection, we study the solutions of the 6D nested branes with induced gravity set-up which correspond to pure tension source configurations. These solutions will be used as background solutions for the perturbative study of the next subsections. For this reasons, we indicate with an overbar the quantities, such as the energy-momentum tensor and the metrics induced on the cod-1 and cod-2 branes, which correspond to these background configurations.

Let's consider localized source configurations of the form (\ref{angelica}), in which the tensor $\h{T}_{ab}$ does not depend on the 4D coordinates $\chd$ and where the localized cod-1 energy-momentum tensor in cod-1 Gaussian Normal Coordinates reads
\beq
\label{marina}
\bar{T}^{(loc)}_{\m\n}(\hxi) = - f(\hxi) \, \bla \,\, \h{g}_{\m\n}(\hxi) \quad ,
\eeq
where the function $f$ vanishes for $\abs{\hxi} > l_2$ and satisfies
\beq
\label{PimPumPam}
\int_{-}^{+} \! d\hxi \, f(\hxi) = 1 \quad .
\eeq
This function can be considered to be a regularized version of the Dirac delta function, and describes the internal structure of the thick cod-2 brane; to be compatible with the $Z_2$ symmetry present inside the cod-1 brane, it has to be even with respect to the reflection $\hxi \rightarrow - \hxi$. Note that when $f$ tends to a Dirac delta, the cod-2 energy momentum tensor tends to
\beq
\label{TakeOnMe}
\bar{T}^{(4)}_{\m\n}(\chd) \rightarrow - \bla \,\, \bar{g}^{(4)}_{\m\n}(\chd) \quad ,
\eeq
which is the energy-momentum tensor corresponding to a (thin) pure tension source (note that the minus sign is due to the fact that we use the mostly plus signature, and is responsible for the fact that positive tension corresponds to positive energy density $\bar{T}^{(4)}_{00}$). To find a solution to the equations of motion, we consider a geometrical ansatz which respects the translational invariance in the 4D directions parallel to the cod-2 brane: we assume that the cod-2 brane is placed at $\xi = 0$
\beq
\label{Robert}
\bar{\a}^{a}(\chi^{\cdot}) = \big( 0, \chi^{\m} \big) \quad ,
\eeq
while the cod-1 brane has the following embedding
\beq
\label{Patty}
\bar{\varphi}^{A}(\xi^{\cdot}) = \big( Z(\xi), Y(\xi), \xi^{\m} \big)
\eeq
and the bulk metric is the 6D Minkowski metric
\beq
\label{George}
\bar{g}_{AB}(X^{\cdot}) = \eta_{AB} \quad .
\eeq
We assume furthermore that the function $Y(\xi)$ is a diffeomorphism, which in particular means that $dY/d\xi$ never vanishes. It is not difficult to see that the $\xi\xi$ component of the cod-1 induced metric reads
\beq
\bar{g}_{\xi\xi} (\xi) = \Zpq(\xi) + \Ypq(\xi) \quad ;
\eeq
we can then use the gauge freedom to rescale the coordinate $\xi \rightarrow \hxi$ in such a way that $\bar{g}_{\xi\xi} (\hxi) = 1$, which implies that the non trivial components of the embedding function in the new coordinate system satisfy
\beq
\label{Kelly}
{Z^{\prime}}^{2}(\hxi) + {Y^{\prime}}^{2}(\hxi) = 1 \quad .
\eeq
Furthermore, since $Y$ is a diffeomorphism, we can always choose the new coordinate $\hxi$ such that
\beq
\label{Aha}
Y^{\prime}(\hxi) = \sqrt{1 - {Z^{\prime}}^{2}(\hxi)} \quad .
\eeq
Note that in the $(\hxi,\xim)$ reference system the metric induced on the cod-1 brane $\bar{g}_{ab}$ is the 5D Minkowski metric (which implies that $(\hxi,\xim)$ is a cod-1 GN reference system) and that the metric induced on the cod-2 brane $\bar{g}^{(4)}_{\m \n}$ is the 4D Minkowski metric. Using (\ref{Kelly}), the 6D 1-form normal to the cod-1 brane reads
\beq
\label{Hanayo}
\bar{n}_{M}(\hxi) = \varepsilon \, \big( Y^{\prime}(\hxi), - Z^{\prime}(\hxi), 0, 0, 0, 0 \big) \quad ,
\eeq
and the only non-vanishing component of the extrinsic curvature of the cod-1 brane is
\beq
\bar{K}_{\xi \xi}(\hxi) = \varepsilon \, \frac{Z^{\prime \prime}(\hxi)}{\sqrt{1 - {Z^{\prime}}^{2}(\hxi)}} \quad ,
\eeq
where $\varepsilon = \pm 1$ encodes the choice of the orientation of the cod-1 brane. In particular, the choice $\vep = + 1$ corresponds to the normal vector which points in the direction of increasing $z$, which means that we construct the full $Z_2$ symmetric 6D space using the part of the 6D Minkowski space which stays on the positive $z$ side with respect to the cod-1 brane. The opposite is true for the choice $\vep = - 1$.

\subsubsection{Conical space and nested branes}

It is easy to see that the bulk equations of motion are identically satisfied, while the only components of the junction conditions which are not trivially satisfied are the $\m\n$ ones
\beq
\label{Roy}
- 2 M_6^4 \, \bar{K}_{\xi \xi} \, \eta_{\m \n} = \bar{T}^{(loc)}_{\m \n} \quad ,
\eeq
which can be rewritten as
\beq
\label{Botswana}
\varepsilon \, \frac{Z^{\prime \prime}(\hxi)}{\sqrt{1 - {Z^{\prime}}^{2}(\hxi)}} = \frac{\bla}{2 M_6^4} \, f(\hxi) \quad .
\eeq
Note that, since the function $f(\hxi)$ is even, $Z(\hxi)$ has to be even as well, which in particular implies that $\Zp(0) = 0$. Furthermore, since the system is invariant with respect to rigid translations of the branes in the bulk, we can choose the bulk coordinates in such a way that we have $Z(0) = 0$ while the bulk metric remains the 6D Minkowski metric. With this choice, the function $Z(\hxi)$ is determined by the Cauchy problem
\beq
\label{Marina Rei}
\left\{
\begin{aligned}
\varepsilon \, \frac{Z^{\prime \prime}(\hxi)}{\sqrt{1 - {Z^{\prime}}^{2}(\hxi)}} &= \frac{\bla}{2 M_6^4} \, f(\hxi) \\[2mm]
Z(0) = \Zp(0) &= 0 \quad .
\end{aligned}
\right.
\eeq
To find the solution $Z(\hxi)$, it is useful to introduce the function $P(\hxi) \equiv \Zp(\hxi)$ and to consider the associated Cauchy problem
\beq
\label{summersummer}
\left\{
\begin{aligned}
\Pp(\hxi) &= \mcal{D} \big( \hxi, P(\hxi) \big) \\[2mm]
P(0) &= 0 \quad ,
\end{aligned}
\right.
\eeq
where
\beq
\mcal{D} \big( \hxi, P \big) = \varepsilon \,\frac{\bla}{2 M_6^4} \, f(\hxi) \,\, \sqrt{1 - P^2} \quad :
\eeq
in fact, the latter Cauchy problem is now of first order and, if we find the solution $P(\hxi)$ of (\ref{summersummer}), the solution of (\ref{Marina Rei}) is found by taking the primitive of $P$ which vanishes in $\hxi = 0$
\beq
\label{AllezAllez}
Z(\hxi) = \int_0^{\hxi} \! d\z \, P(\z) \quad .
\eeq
The real function $\mcal{D}$ is defined on the domain $A = \mathbb{R} \times [-1, 1]$ , and is continuous in $\hxi$ and locally Lipschitzian with respect to $P$ in the open domain $\mathring{A} = \mathbb{R} \times (-1, 1)$: therefore, by the Picard-Lindel\"of theorem (see for example \cite{Hale} for the definition of locally Lipschitzian and for the formulation of this theorem) the Cauchy problem (\ref{summersummer}) admits a unique solution $P(\hxi)$ in a neighborhood of $\hxi = 0$. Moreover, this local solution can be extended (at least) as long as $\abs{P(\hxi)}$ remains smaller than 1. The local solution can actually be derived explicitly, as a consequence of the fact that $\arcsin P$ is an explicit primitive of the function $\Pp/\sqrt{1 - P^2}$: we have in fact that, in the domain where the solution $P(\hxi)$ exists, it reads explicitly
\beq
\label{satisfaction}
P(\hxi) = \sin \bigg( \vep \, \frac{\bla}{2 \Msf} \, \int_0^{\hxi} \! d\z \, f(\z) \bigg) \quad .
\eeq
Note that, if $\bla/4 \Msf < \pi/2$, the function on the right hand side of (\ref{satisfaction}) is smaller than one in absolute value for every value of $\hxi$, since the relation (\ref{PimPumPam}) implies that
\begin{align}
\int_0^{\hxi} \! d\z \, f(\z) &\leq \half & \int_0^{l_2} \! d\z \, f(\z) &= \half \quad .
\end{align}
This implies that the solution $P(\hxi)$ can be extended to all the real axis, and therefore the Cauchy problem (\ref{summersummer}) possesses a (unique) global solution; in particular, $P(\hxi)$ is constant for $\abs{\hxi} \geq l_2$, and explicitly
\beq
P(\hxi) = \pm \sin \bigg( \vep \, \frac{\bla}{4 \Msf} \bigg) \qquad \qquad \textrm{for} \qquad \qquad \hxi \gtrless \pm l_2 \quad .
\eeq
This implies that, if $\bla/4 \Msf < \pi/2$, the Cauchy problem (\ref{Marina Rei}) admits a unique global solution which is given by (\ref{AllezAllez}) and (\ref{satisfaction}), and for $\abs{\hxi} \geq l_2$ reads
\beq
\label{Youpiyou}
Z(\hxi) = \Zp_+ \,\, \abs{\hxi} + Z_0 \qquad \qquad \textrm{for} \qquad \qquad \abs{\hxi} \geq l_2
\eeq
where, indicating $\Zp_{+} \equiv \Zp\big\rvert_{+}$, we have
\beq
\label{Dom}
\arcsin Z^{\p}_{+} = \varepsilon \, \frac{\bla}{4 M_6^4}
\eeq
and $Z_0$ is given by
\beq
Z_0 = \int_0^{l_2} P(\z) \, d\z - l_2 \, P(l_2) \quad .
\eeq
Having found a global solution for $Z(\hxi)$, we can obtain a global solution for $Y(\hxi)$ using (\ref{Aha}); note that we can chose freely the initial condition for $Y$, and we decide to impose $Y(0) = 0$ which in particular implies that $Y(\hxi)$ is odd. Therefore, the global solution for $Y$ reads
\beq
Y(\hxi) = \int_0^{\hxi} \! d\z \, \sqrt{1 - P^2(\z)} \quad ,
\eeq
and in particular for $\abs{\hxi} \geq l_2$ it takes the form
\beq
\label{Annabelle}
Y(\hxi) = \Yp_+ \,\, \hxi \pm Y_0 \qquad  \qquad \textrm{for} \qquad  \qquad \hxi \gtrless \pm l_2
\eeq
where
\beq
\Yp_+ = \sqrt{1 - \Zpq_+} = \cos \bigg( \varepsilon \, \frac{\bla}{4 M_6^4} \bigg)
\eeq
and $Y_0$ reads
\beq
Y_0 = \int_0^{l_2} \sqrt{1 - P^2(\z)} \,\, d\z - l_2 \, \sqrt{1 - P^2(l_2)} \quad .
\eeq
Note that the slopes of $Z(\hxi)$ and of $Y(\hxi)$ outside the thick cod-2 brane depend only on the total amount of tension $\bla$, and are independent of the internal structure of the thick cod-2 brane. Furthermore, if we keep $\bla$ constant and perform a limit in which $l_2 \rightarrow 0^+$, $P(l_2)$ remains constant and $P(\hxi)$ remains bounded, so $\Zp_+$ and $\Yp_+$ remain constant while $Z_0$ and $Y_0$ tend to zero. Therefore, if we restrict ourselves to configurations of the type (\ref{TakeOnMe})-(\ref{George}), we can give a thin limit description to this set-up where the tension $\bla$ is perfectly localized on the mathematical cod-2 brane $\mcal{C}_2$ and the components of the embedding function read
\beq
\label{CestLaVie}
\left\{
\begin{aligned}
Z(\hxi) &= \varepsilon \, \sin \Big( \frac{\bla}{4 M_6^4} \Big) \,\, \abs{\hxi} \\[2mm]
Y(\hxi) &= \cos \Big( \frac{\bla}{4 M_6^4} \Big) \,\, \hxi
\end{aligned}
\right.
\eeq
for every value of $\hxi$.

To understand the geometrical meaning of the solution defined by (\ref{Robert})-(\ref{George}) and (\ref{CestLaVie}), let's consider the $\bla > 0$ case. For definiteness, we can take $\vep = +1$ and $Z^{\p}_{+} > 0$ (the other possible choice $\vep = -1$ and $Z^{\p}_{+} < 0$ gives the same spacetime): the bulk is then a slice of flat space of opening
\beq
\label{Asus}
\vth = 2 \arctan \bigg( \frac{\Yp_+}{\Zp_+} \bigg) = 2 \arctan \Bigg( \frac{\sqrt{1 - \Zpq_+}}{\Zp_+} \, \Bigg) \quad .
\eeq
The whole 6D spacetime is reconstructed considering two copies of the bulk, which then can be seen as a flat 6D spacetime with total deficit angle $\a = 2 \pi - 2 \vth$, and gluing them together. The thin limit solution (\ref{Robert})-(\ref{George}) and (\ref{CestLaVie}) then corresponds to a geometric configuration which is the product of the 4D Minkowski space and a two dimensional cone of deficit angle $\a$, and using (\ref{Dom}) and (\ref{Asus}) we have
\beq
\a = \frac{\bla}{M_6^4} \quad .
\eeq
This in particular implies that it is not possible to localize on the cod-2 brane an arbitrary amount of tension: $\bla$ is in fact bounded from above by the value $\bla_M = 2 \pi \Msf$ which corresponds to a deficit angle of $2 \pi$, in which case the 2D cone becomes degenerate (a half-line).

We have then obtained that, also in the nested brane set-up, pure tension $\bla$ on a cod-2 brane produces a conical defect of deficit angle $\bla/M_6^4$: this is a non-trivial result, since the nested brane set-up violates the condition of circular symmetry (matter is confined inside the thin cod-1 brane) that we used in section \ref{Codimension-2 branes and conical spaces} for the case of pure cod-2 branes.

\subsubsection{Bulk-based and brane-based approaches}

In the analysis above, to find the geometric configuration corresponding to a pure tension source, we started from a (known) solution of the equation of motion in the bulk, and considered a general embedding (trajectory) for the cod-1 brane. To find a complete solution of the equations of motion, we wrote the junction conditions as equations for the (unknown) trajectory of the brane, and solved them to find a specific profile. This kind of approach to find a solution of the equations of motion is known in general as a \emph{bulk-based} approach. It is very useful when we know that the bulk have some symmetries, or we can guess the form of a sensible solution, so that we don't have to work with the most general form of the bulk metric but we can characterize it in terms of few parameters (or even no parameters as in our case).

An alternative approach which is widely used in codimension-1 braneworld theories is to fix instead the embedding of the brane (which is always possible provided the source on the brane is smooth), and solve for the bulk metric using the junction conditions and the bulk equations. This approach is usually called a \emph{brane-based} approach. An useful choice in this case is to use Gaussian Normal Coordinates with respect to the cod-1 brane. Besides fixing the position of the cod-1 brane, this also reduces the number of unknown bulk metric components and makes very transparent the connection between the bulk geometry and the induced geometry, since the induced metric is just the bulk metric evaluated on the brane. On the other hand, the global geometry of the bulk and branes configuration is usually more clear in the bulk-based approach.

An important question to ask is which of the two approaches is more convenient in the case of the nested brane scenario. Note first of all that the thin solution (\ref{Robert})-(\ref{George}) and (\ref{CestLaVie}) does not have a corresponding solution in the brane based approach. In fact, it is possible to set up a coordinates system where the brane is straight if and only if the normal vector field is smooth, while in the case (\ref{Robert})-(\ref{George}) and (\ref{CestLaVie}) the normal vector is discontinuous at the cod-2 brane (as can be seen from (\ref{Hanayo})). However, a thick solution of the form (\ref{Robert})-(\ref{George}), (\ref{Youpiyou}) and (\ref{Annabelle}) can indeed be expressed in a brane-based way, and this is always true apart from the limit configuration. The brane-based configurations corresponding to thick pure tension sources in the 6D nested brane set-up have in fact been worked out in \cite{deRham:2010rw}, and used as background configurations for a perturbative study at first order. In these configurations, the cod-1 and cod-2 branes are respectively positioned at $z = 0$ and $y = z = 0$, where $(z,y,\xd)$ are the bulk coordinates, while the bulk corresponds to the $z \geq 0$ domain and its metric reads
\beq
\label{backgroundthemline}
ds^2 = \big( 1 + \b^2 \big) \, dz^2 + 2 \b \, \ep_{n}(y) \, dz dy + dy^2 + \e_{\m \n} dx^{\m} dx^{\n} \quad ,
\eeq 
where $\b$ is a real parameter and $\ep_{n}(y)$ is a family of smooth and odd functions which in the $n \rightarrow \infty$ limit tend to the symmetric step function\footnote{The symmetric step function is defined as $\s(y) = 2 \, \theta (y) -1$, where $\theta$ is the Heavyside theta function.}. The parameter $\b$ is related to the total tension $\bla$ by \cite{deRham:2010rw}
\beq
\arctan \b = \frac{\bla}{4 \Msf}
\eeq
and it is in fact possible to show that these configurations are equivalent to the bulk-brane configurations (\ref{Robert})-(\ref{George}), (\ref{Youpiyou}) and (\ref{Annabelle}), since there exists a change of coordinates which links the two descriptions (see appendix \ref{Conical space in brane-based coordinates}).

Note that, in the limit $n \rightarrow +\infty$, the metric (\ref{backgroundthemline}) is not defined on the $z$ axis, and is in fact discontinuous across $y = 0$: a (coordinate) singularity in the bulk appears in the limit, and the bulk ``splits'' in two pieces. This is just the reflection of the fact that the normal vector changes rapidly across the cod-2 brane, and in the $n \rightarrow +\infty$ limit it becomes discontinuous: since the brane sits at the fixed position $z = 0$, the non-trivial behavior of the normal vector has to be encoded in the bulk metric. This is true also more in general: looking at the expressions (\ref{Camilla})-(\ref{Pop}) for the components of the extrinsic curvature tensor of a cod-1 brane, we note that $\ti{\mbf{K}}$ is built from the first derivatives of the bulk metric and the second derivatives of the embedding function. This implies that, to have an extrinsic curvature tensor which diverges at one point (as is required by the junction conditions if the energy momentum tensor is perfectly localized), either the embedding function is (at least) cuspy, or the bulk metric is (at least) discontinuous, or both.

Therefore, the bulk based approach is in general the only one in which it is possible to obtain a \emph{continuous} expression for the thin limit configuration in a nested brane set-up. In the case of pure tension, all the singularity is carried by the embedding, which converges to a cuspy configuration, while the bulk metric and all its derivatives remain smooth also in the limit and converge uniformly to their limiting configuration. Furthermore, the connection between these configurations and the conical geometry is much more clear in the bulk-based than in the brane-based approach.

\section{Perturbations of nested branes with induced gravity}
\label{Perturbations of nested branes with induced gravity}

We want now to study perturbations at first order in the nested branes with induced gravity set-up around the pure tension configurations which we derived in the previous section. One of our main aims is to verify if the presence of the cod-1 induced gravity term regularizes gravity. If this happens, despite the extrinsic geometry of the cod-1 brane diverges in the thin limit (as is implied by the junction conditions), the intrinsic geometry remains regular (or at least continuous). In our perturbative study, we would like to use an approach in which the properties of the bulk metric and cod-1 embedding configuration reflects most clearly the fact that the intrinsic geometry diverges or not when the thin limit on the cod-2 brane is taken. This suggests that the bulk-based approach is perhaps better suited than the brane-based one: in fact, at least in the pure tension case, the fact that the gravitational field is finite is expressed by the continuity of the bulk metric/embedding configuration, while in the latter case a singularity in the bulk appears in the thin limit even if gravity is regularized. Of course, as we have already mentioned, the latter singularity is purely a coordinate singularity, and in principle the two approaches are equivalent. In general, we feel that the bulk-based approach is geometrically more suited to the study of the nested branes set-up; therefore, we decide to follow this approach to study perturbations around the pure tension configurations.

\subsection{Perturbations in the bulk-based approach}

\subsubsection{General considerations}

In the following, we perturb both the bulk metric and the embedding of the cod-1 brane
\begin{align}
g_{AB}(X^{\cdot}) &= \bar{g}_{AB}(X^{\cdot}) + h_{AB}(X^{\cdot}) \\[2mm]
\varphi^{A}(\xi^{\cdot}) &= \bar{\varphi}^{A}(\xi^{\cdot}) + \d \varphi^{A}(\xi^{\cdot}) \quad ,
\end{align}
while we decide to keep fixed the position of the cod-2 brane in the cod-1 coordinate system (\emph{i.e.} it is still located at $\xi = 0$); in particular, we still use the 4D coordinates of the cod-1 brane to parametrize the cod-2 brane, so the embedding of the cod-2 brane reads
\beq
\label{Joan}
\tilde{\a}^{a}(\chi^{\cdot}) = \bar{\a}^{a}(\chi^{\cdot}) = \big( 0, \chi^{\m} \big)
\eeq 
also at perturbative level. We def\mbox{}ine the perturbation of the metric induced on the cod-1 brane as follows
\beq
\tilde{h}_{ab}(\xi^{\cdot}) \equiv  \tilde{g}_{ab}(\xi^{\cdot}) - \bar{g}_{ab}(\xi^{\cdot})  \quad ,
\eeq
and we note that, using the general definition (\ref{Induced metric general}), we can write it as the sum of a ``metric perturbation'' and of a ``bending perturbation'' part
\beq
\tilde{h}_{ab}(\xi^{\cdot}) = \tilde{h}^{[mp]}_{ab}(\xi^{\cdot}) + \tilde{h}^{[bp]}_{ab}(\xi^{\cdot}) \quad ,
\eeq
where
\begin{align}
\tilde{h}^{[mp]}_{ab} & \equiv \frac{\de \bvf^{A}(\xi^{\cdot})}{\de \xi^{a}} \, 
\frac{\de \bvf^{B}(\xi^{\cdot})}{\de \xi^{b}} \,\, h_{AB}(X^{\cdot}) \Big\rvert_{X^{\cdot} = \bvf^{\cdot}(\xi^{\cdot})} \\[2mm]
\tilde{h}^{[bp]}_{ab} & \equiv 
\frac{\de \dvf^{A}(\xi^{\cdot})}{\de \xi^{a}} \, \frac{\de \bvf^{B}(\xi^{\cdot})}{\de \xi^{b}} \,\, \e_{AB} + \frac{\de \bvf^{A}(\xi^{\cdot})}{\de \xi^{a}} \, \frac{\de \dvf^{B}(\xi^{\cdot})}{\de \xi^{b}} \,\, \e_{AB} \quad .
\end{align}
Similarly, we define the perturbation of the metric induced on the cod-2 brane as
\beq
h^{(4)}_{\m \n}(\chi^{\cdot}) \equiv g^{(4)}_{\m \n}(\chi^{\cdot}) - \bar{g}^{(4)}_{\m \n}(\chi^{\cdot}) \quad ,
\eeq
and in particular we have that, since the embedding of the cod-2 brane in the cod-1 brane is trivial, the metric induced on the cod-2 brane takes the form
\beq
g^{(4)}_{\m \n}(\chi^{\cdot}) = \tilde{g}_{\m \n}(0,\chd)
\eeq
and therefore
\beq
h^{(4)}_{\m \n}(\chi^{\cdot}) = \tilde{h}_{\m \n}(0,\chd) \quad .
\eeq

Every quantity defined in terms of the bulk metric and the cod-1 brane embedding can be written in general as the sum of a background part and a perturbation part; we use the convention that indices on the perturbation part of every quantity (and on the background part as well) are lowered/raised with the background metric. For example, remembering the definition (\ref{parallelvectorsgeneral}) of the parallel vectors $\mathbf{v}_{(a)}$, we consider the perturbative decomposition
\beq
v^{A}_{(a)} = \bv^{A}_{(a)} + \dv^{A}_{(a)} \quad ,
\eeq
where
\begin{align}
\label{Truman}
\bv^{A}_{(a)} &\equiv \frac{\de \bvf^A}{\de \xi^a} & \dv^{A}_{(a)} &\equiv \frac{\de \, \dvf^A}{\de \xi^a}
\end{align}
and the index-lowered background and perturbation parts read
\begin{align}
\bv_{A}^{(a)} &\equiv \e_{AB} \, \bv^{B}_{(a)} & \dv_{A}^{(a)} &\equiv \e_{AB} \, \dv^{B}_{(a)}
\end{align}
On the other hand, indices on the perturbed quantities (i.e. on the sum of the background part and the perturbation part) are lowered/raised with the full perturbed metric, and so we have
\beq
v_{A}^{(a)} \equiv g_{AB} \, v^{A}_{(a)} \simeq \bv_{A}^{(a)} + \dv_{A}^{(a)} + h_{AB} \, \bv^{B}_{(a)} \quad .
\eeq
In addition to the convention that latin letters $a$, $b$, $\ldots\,$ indicate 5D indices which run from 0 to 4, in this chapter we use also the convention that the letters $i$, $j$ and $k$ indicate 2D indices which run on the extra dimensions $z$ and $y$. These 2D indices are raised/lowered with the identity matrix, so we have for example
\begin{align}
\bvf_i^{\p} &\equiv \d_{ij} \, \bvf^{j \, \p} & \bn^{i} &\equiv \d^{ij} \, \bn_{j} \quad .
\end{align}

\subsubsection{Perturbation of the source}

Concerning the source term, we consider a perturbed energy-momentum tensor which in cod-1 GNC is of the form (\ref{angelica}), and such that the localized energy momentum tensor reads
\beq
\label{Gigio}
\hat{T}^{(loc)}_{\m\n}(\hxi, \chd) = - f(\hxi) \, \big( \bla + \dla \big) \,\h{g}_{\m\n}(\hxi, \chd) + \h{\mcalT}_{\m\n}(\hxi, \chd) \quad ,
\eeq
where $\h{\mcalT}_{\m\n}$ is the energy-momentum of the matter present inside the (thick) cod-2 brane. Note that we perturb both the matter content and the tension ($\dla$) of the cod-2 brane. At linear order, the equation (\ref{Gigio}) reads
\beq
\label{Gigia}
\hat{T}^{(loc)}_{\m\n} = \bar{T}^{(loc)}_{\m\n} - f(\hxi) \, \bla \, \h{h}_{\m\n} + \d \hat{T}^{(loc)}_{\m\n} \quad ,
\eeq
where $\bar{T}^{(loc)}_{\m\n}$ is the (thick) background pure tension source term
\beq
\label{marinaz}
\bar{T}^{(loc)}_{\m\n}(\hxi) = - f(\hxi) \,\, \bla \,\, \e_{\m\n} \quad ,
\eeq
$f(\hxi) \, \bla \, \h{h}_{\m\n}$ is a pertubation term coming from the background tension, and
\beq
\label{Gigetta}
\d \hat{T}^{(loc)}_{\m\n}(\hxi, \chd) = - f(\hxi) \,\, \dla \,\, \e_{\m\n} + \h{\mcalT}_{\m\n}(\hxi, \chd)
\eeq
is the perturbation term due to the tension perturbation and to the matter. Note that in principle $\bar{T}^{(loc)}_{\m\n}$ and $\d \hat{T}^{(loc)}_{\m\n}$ may be characterized by different localization lengths $l_2$ and $l^{\p}_2$: we ask that they are of the same order of magnitude $l_2 \sim l^{\p}_2$, and in the following for simplicity we will indicate with $l_2$ the biggest between $l_2$ and $l^{\p}_2$. In particular, in the following we implicitly assume this convention when we use the notation defined in the equations (\ref{strange equation1})-(\ref{strange equation2}). It is important to notice that, in principle, the presence of the matter may alter the distribution of the tension inside the thick cod-2 brane, as a consequence of generalized Casimir effects. We assume that such effects are not present, and therefore the form of the background solution and the form of $\h{\mcalT}_{\m\n}$ are independent in our analysis.

In analogy to what we did above, we define the matter cod-2 energy-momentum tensor as follows
\beq
\label{Kristin}
\mcalT^{(4)}_{\m\n}(\chd) \equiv \int_{-}^{+} \! d \hxi \,\, \h{\mcalT}_{\m\n}(\hxi, \chd) \quad .
\eeq

\subsubsection{The Scalar-Vector-Tensor decomposition}

Since at background level the 4D slices of the bulk at $z,y$ constant are isomorphic to the 4D Minkowski space, it is possible to consider the Scalar-Vector-Tensor decomposition of the perturbation field $h_{AB}(\Xd)$ with respect to the 4D coordinates $\xm$. The convenience of this type of decomposition is first of all that, at linear order, the three sectors (tensor, vector and scalar) decouple, and so the equations of motion in each sector may be simpler to solve than the complete equations. Secondly, it is possible that, concerning some specific property of the system, only one of the sectors is really relevant in that respect, and so the decomposition helps to focus on the property we may be interested in. We consider then the following decomposition of the bulk metric perturbation in TT-tensor, T-vector and scalar parts
\begin{align}
h_{\mu \nu} &= h^{TT}_{\mu \nu} + \partial_{(\mu} V_{\nu)} + \e_{\mu \nu} \,\pi + \partial_{\mu}\partial_{\nu} \vp \\[2mm]
h_{z \mu} &= D_{\mu} + \partial_{\mu} \s \\[2mm]
h_{y \mu} &= B_{\mu} + \partial_{\mu} \t \\[2mm]
h_{yy} &= \psi \\[2mm]
h_{zy} &= \r \\[2mm]
h_{zz} &= \o
\end{align}
where all the quantities are functions of the bulk coordinates $X^{\cdot}$. Above, $h^{TT}_{\mu \nu}$ is a transverse-traceless symmetric tensor, while $V_\mu$ , $D_\mu$ and $B_\mu$ are transverse 1-forms and $\o$, $\r$, $\psi$, $\s$, $\t$, $\pi$ and $\vp$ are scalars; in particular, concerning the scalar parts, we call \emph{trace part} the scalar field which multiplies the Minkowski metric (in this case $\pi$), while we call \emph{derivative part} the scalar field derivated twice with respect to the 4D coordinates (in this case $\vp$). Analogously to the bulk case, we can consider the Scalar-Vector-Tensor decomposition of quantities defined on the cod-1 brane with respect to the $\xim$ coordinates, and also of quantities defined on the cod-2 brane with respect to the $\chm$ coordinates. Note that at background level the $z$ and $y$ components of the cod-1 embedding function does not depend on the 4D coordinates $\xim$, while the 4D components of the cod-1 embedding function are related to the coordinates  $\xim$ by the identity map. As a consequence of this fact, we can perform the Scalar-Vector-Tensor decomposition of the perturbation of the cod-1 embedding function: we have that $\dvf^{z}$ and $\dvf^{y}$ are scalars, while we can decompose the (index lowered) 4D components as follows
\beq
\dvf_{\m} = \dvf^{T}_{\m} + \frac{\de}{\de \xim} \, \dvf_4
\eeq
where $\dvf^{T}_{\m}$ is a transverse 1-form (\emph{i.e.} $\e^{\m \a} \ded{\xi^\a} \dvf^{T}_{\m} = 0$) and $\dvf_4$ is a scalar. Furthermore, the structure of the background embedding function implies that the scalar sector of the metric induced on the cod-1 brane is constructed from the fields which belong to the scalar sector in the bulk (evaluated on the brane), and the same is separately true for the vector sector and the tensor sector. We have in fact that
\begin{align}
\tilde{h}^{[mp]}_{\xi \xi} &= {Z^{\prime}}^{2} \, \tilde{\omega} + 2 \, Z^{\prime} \, Y^{\prime} \, \tilde{\rho} + {Y^{\prime}}^{2} \, \tilde{\psi} \\[2mm]
\tilde{h}^{[mp]}_{\xi \m} &= \Big( \Zp \, \tilde{D}_{\m} + Y^{\prime} \, \tilde{B}_{\m} \Big) + \de_{\xi^{\m}} \Big( Z^{\prime} \, \tilde{\s} + Y^{\prime} \, \tilde{\t} \Big) \\[2mm]
\tilde{h}^{[mp]}_{\m \n} &= \ti{h}^{TT}_{\m\n} + \de_{\xi^{(\m}} \, \ti{V}_{\n)} + \ti{\pi} \, \eta_{\m\n} + \deximxin \, \tvp
\end{align}
and
\begin{align}
\tilde{h}^{[bp]}_{\xi \xi} &= 2 \, \Big( \Zp \, \dvfzp + \Yp \, \dvfyp \Big) \\[2mm]
\tilde{h}^{[bp]}_{\xi \m} &= \dvf^{T \, \p}_{\m} + \de_{\xi^{\m}} \Big( \Zp \, \dvfz + \Yp \, \dvfy + \dvf^{\p}_4 \Big) \\[2mm]
\tilde{h}^{[bp]}_{\m \n} &= \de_{\xi^{(\m}} \dvf_{\n)}^T + 2 \, \deximxin \dvf_4 \quad ,
\end{align}
where the tilded quantities $\ti{h}^{TT}_{\m\n}$, $\ti{\pi}$, \ldots are defined as the evaluation on the brane of their bulk counterparts
\begin{align}
\ti{h}^{TT}_{\m\n}(\xid) &\equiv h^{TT}_{\m\n}(\Xd) \evb & \ti{\pi}(\xid) &\equiv \pi(\Xd) \evb & &\dots \quad .
\end{align}
It is easy to see that $\Zp \, \tilde{D}_{\m} + Y^{\prime} \, \tilde{B}_{\m}$ and $\ti{V}_{\n}$ are 5D transverse vectors, while $\ti{h}^{TT}_{\m\n}$ is a 5D symmetric traceless and transverse tensor and $\big( Z^{\prime} \, \tilde{\s} + Y^{\prime} \, \tilde{\t} \big)$,  $\tp$, $\tvp$, $\to$, $\tr$ and $\tps$ are 5D scalars. Note that the TT tensor and the scalar-trace parts are missing in the brane perturbation contribution to the 5D metric perturbations. It is possible to see that, in turn, the scalar sector of the metric induced on the cod-2 brane is constructed from the fields which belong to the scalar sector on the cod-1 brane, and that the same is separately true for the vector sector and the tensor sector. We indicate with a superscript $\phantom{i}^{(4)}$ the bulk and cod-1 quantities evaluated on the cod-2 brane, for example
\beq
\pi^{(4)}(\chd) = \tp(\xid) \Big\rvert_{\xid = \bar{\a}^{\cdot}(\chd)} \quad .
\eeq

Considering the decomposition of the energy momentum tensor, for the matter cod-1 energy-momentum tensor we can write
\beq
\label{Tracy}
\hat{\mcalT}_{\m\n} = \hat{\mcalT}_{\m\n}^{TT} + \de_{\xi^{( \m}} \, \hat{\mcalT}^{V}_{\n)} + \deximxin \, \hat{\mcalT}^{(de)} + \eta_{\m\n} \, \hat{\mcalT}^{(tr)} \quad ,
\eeq
where the symmetric tensor $\hat{\mcalT}_{\m\n}^{TT}$ is transverse and traceless while $\hat{\mcalT}^{V}_{\n}$ is a transverse vector and $\hat{\mcalT}^{(tr)}$, $\hat{\mcalT}^{(de)}$ are scalars. Concerning the cod-2 matter energy-momentum tensor, also in this case we can perform the decomposition
\beq
\mcalT^{(4)}_{\m\n} = \mcalT_{(4) \, \m\n}^{TT} + \de_{\xi^{( \m}} \, \mcal{T}^{V}_{(4) \,\n)} + \deximxin \, \mcalT^{(4)}_{(de)} + \eta_{\m\n} \, \mcalT^{(4)}_{(tr)} \quad ,
\eeq
where the symmetric transverse-traceless tensor $\mcalT_{(4) \, \m\n}^{TT}$, the transverse vector $\mcal{T}^{V}_{(4) \,\n}$ and the scalar parts $\mcalT^{(4)}_{(tr)}$ and $\mcalT^{(4)}_{(de)}$ are linked to their cod-1 counterparts by the relations analogous to (\ref{Kristin}). Note that we can use the continuity equation to link the scalar parts to the trace of the cod-2 energy-momentum tensor
\begin{align}
\mcalT^{(4)}_{(tr)} &= \third \, \mcalT^{(4)} & \boxf \mcalT^{(4)}_{(de)} &= - \third \, \mcalT^{(4)} \quad ,
\end{align}
where $\mcalT^{(4)} \equiv \eta^{\m\n} \, \mcalT^{(4)}_{\m\n}$.

The fact that the scalar sectors on the cod-1 and on the cod-2 brane are constructed entirely from the fields which belongs to the scalar sector in the bulk (and that the same is separately true for the vector sector and the tensor sector) permits to consistently study the perturbations of the nested brane set-up by studying separately the three (decoupled) sectors. The Scalar-Vector-Tensor decomposition is especially useful in the nested branes realization of the Cascading DGP model. In fact, the scalar sector seems to be much more subtle than the other sectors, as a consequence of the fact that the perturbation of the bending of the cod-1 brane plays a role only in this sector (we will see in fact that the 4D components of perturbation of the cod-1 bending do not appear in the equations of motion). Furthermore, some properties of the model, such as the existence of a critical tension which divides background configurations with ghosts from ghost-free background configurations, are entirely due to the peculiarities of the scalar sector of the theory. Therefore, in the following we will concentrate only on the scalar sector of the perturbation fields.

\subsection{Gauge invariant variables}

When studying perturbations in GR, or more in general in theories which enjoy general covariance, a very important issue is that of gauge invariance. On one hand, one wants to make sure that he knows if the perturbation fields are physical or are pure gauge fields, since the latter ones have no physical relevance and so eventual pathologies in their behavior are harmless. On the other hand, an intelligent choice for the gauge fixing is sometimes crucial to be able to cast the equations in a form which is explicitly solvable. The gauge fixing procedure is however potentially dangerous in the nested branes with induced gravity set-up. As we already mentioned, some gauge conditions which are perfectly fine when the cod-2 brane is thick, are not compatible with the thin limit geometry (we mentioned in particular the case of the Gaussian Normal Coordinates in the bulk with respect to the cod-1 brane). As a consequence of this fact, if we impose some ``risky'' gauge condition we may end up seeing divergences appear in the thin limit, which are however consequence of the gauge choice and not of the fact that the theory has problems.

\subsubsection{Gauge invariant variables}

To circumvent this issue, we choose to work with gauge invariant variables. These are variables which are constructed (in each separate sector) from the perturbation fields in such a way that, as we change the gauge, their value remains invariant despite the value of the fields they are constructed with indeed varies. Considering an infinitesimal change of coordinates in the bulk (remember that we work at first order in perturbations)
\beq
X^{\p \, A} = X^A - \La^A(\Xd) \quad ,
\eeq
the metric tensor transforms as follows
\beq
h^{\p}_{MN}(\Xd) = h_{MN}(\Xd) + \de_{X^{(M}} \,\, \La_{N)}(\Xd)
\eeq
and the perturbation of the embedding transforms as
\beq
\dvf^{\p \, A}(\xid) = \dvf^{A}(\xid) - \ti{\La}^A(\xid) \quad ,
\eeq
where we defined $\ti{\La}^{L} \equiv \La^{L}(\bvf^{\cdot}(\xid))$ and it is intended that a prime here does not denote a derivative with respect to $\xi$ but just that the quantities are expressed in the new coordinate system. Considering only the scalar sector, the 4D components of the (index lowered) gauge parameter $\La_{N} \equiv \e_{NL} \La^L$ can be decomposed as
\beq
\La_{\m} = \frac{\de}{\de \xm} \, \La_{4} \quad :
\eeq
it is possible to see that (at least away from the thin limit) we can use the gauge freedom to set $\s^{\p} = \t^{\p} = \vp^{\p}$, and that this is a complete gauge fixing. Therefore, in the scalar sector there are four independent gauge invariant variables; we choose to work with the following variables
\begin{align}
\pi^{gi} &\equiv \pi \\[2mm]
\psi^{gi} &\equiv \psi - 2 \ded{y} \t + \dedq{y} \varpi \\[2mm]
\omega^{gi} &\equiv \omega - 2 \ded{z} \s + \dedq{z} \varpi \\[2mm]
\rho^{gi} &\equiv \rho - \ded{z} \t - \ded{y} \s + \ded{z} \ded{y} \varpi
\end{align}
which coincide with the fields $\pi$, $\psi$, $\o$ and $\r$ in the gauge $\s = \t = \vp = 0$. Note that indicating $i,j = z,y$ we have in compact form
\beq
h^{gi}_{ij} = h_{ij} - \frac{\de_{(i} h_{\m j)}}{\de_\m} + \de_i \de_j \, \varpi \quad ,
\eeq
where $h_{\m j}/\de_\m$ is a notation which means
\beq
\frac{h_{\m z}}{\de_\m} = \s \qquad \qquad \frac{h_{\m y}}{\de_\m} = \t \quad .
\eeq
Concerning the degrees of freedom which describe the position of the cod-1 brane, the components of the perturbation of the cod-1 embedding function are not gauge invariant, and in fact they can be gauged to zero at least away from the thin limit. However, we can give a gauge invariant description of the brane position by considering the following gauge invariant versions of the embedding perturbations
\begin{align}
\dvfz_{gi} &\equiv \dvfz + \Big[ \s - \half \, \dez \vp \Big]\evb \\[2mm]
\dvfy_{gi} &\equiv \dvfy + \Big[ \t - \half \, \dey \vp \Big]\evb \\[2mm]
\dvf_4^{gi} &\equiv \dvf_4 + \Big[ \half \, \vp \Big]\evb
\end{align}
which again coincide with $\dvfz$, $\dvfy$ and $\dvf_4$ in the gauge $\s = \t = \vp = 0$. We refer to this variables as the brane-gauge invariant variables. Note that they can be expressed in compact form as
\begin{align}
\dvf^{i}_{gi} &= \dvf^{i} + \Big[ \frac{h_{\m i}}{\de_\m} - \half \, \de_i \vp \Big]\evb \\[2mm]
\dvf^{gi}_{4} &= \dvf_{4} + \half \, \tvp \quad .
\end{align}
The perturbation of the metric induced on the cod-1 brane can be expressed in terms of the master variables (in the scalar sector) as follows
\begin{align}
\label{LeJeuxSonFait}
\ti{h}_{\xi\xi} &= \bvf^{i \, \p} \bvf^{j \, \p} \, h_{ij}^{gi}\evb + 2 \, \bvf_i^{\p} \, \dvf^{i \, \p}_{gi} \\[2mm]
\ti{h}_{\xi\m} &= \dexim \Big( \bvf_i^{\p} \, \dvf^{i}_{gi} + \dvf_{4}^{gi \, \p} \Big) \\[2mm]
\ti{h}_{\m\n} &= \tp \, \emn + 2 \, \deximxin \, \dvf_{4}^{gi} \quad .
\end{align}

The possibility to describe in a gauge invariant way both the perturbations of the bulk metric and the perturbations of the brane embedding, permits to study the perturbations around the pure tension solutions in the nested brane set-up in a purely gauge-invariant way. Note that this property is more general and holds also for perturbations around different background solutions.

\subsubsection{Bulk equations of motion}

We turn now to the study of the bulk equations of motion
\beq
G_{AB}(\Xd) = 0 \quad .
\eeq
Note that the Bianchi identity
\beq
\nabla_{M} \, G^{M}_{\,\, N} \simeq \de_{M} \, G^{M}_{\,\, N} = 0
\eeq
links together (in a differential way) the components of the Einstein tensor (independently of the fact that the metric solves the Einstein equations or not). We can exploit these relations to select a minimal subset of independent equations. Consider for example the Bianchi identity characterized by $N = z$
\beq
\de_{z} \, G^{z}_{\,\, z} + \de_{y} \, G^{y}_{\,\, z} + \de_{\m} \, G^{\m}_{\,\, z} = 0 \quad ;
\eeq
it is clear that, if we impose $G^{z}_{\,\, z} = G^{y}_{\,\, z} = 0$, then the 4-divergence of $G^{\m}_{\,\, z}$ automatically vanishes. On the other hand, in the scalar sector the mixed component $G^{\m}_{\,\, z}$ necessarily has the form $\de_{\m} [A]$, where $[A]$ is an appropriate functional of the metric components; therefore, $G^{z}_{\,\, z} = G^{y}_{\,\, z} = 0$ actually implies that $\boxf [A] = 0$. Since we assume boundary conditions which assure that the operator $\boxf$ is invertible, we have that $G^{z}_{\,\, z} = G^{y}_{\,\, z} = 0$ implies $[A] = 0$ which in turn implies $G^{\m}_{\,\, z} = 0$. For the same reason, we have that $G^{y}_{\,\, z} = G^{y}_{\,\, y} = 0$ implies $G^{\m}_{\,\, y} = 0$. Note furthermore that, in the scalar sector, the 4D components of the bulk Einstein tensor are of the form $G^{\m}_{\,\, \n} = \d_{\,\, \n}^{\m} [\textit{tr}] + \deu{\m} \den [\textit{de}]$ where $[tr]$ and $[de]$ are appropriate functionals of the metric components: taking the divergence $\deu{\n}$ of the $\n$ components of the Bianchi identities, one then gets that $G^{z}_{\,\, \n} = G^{y}_{\,\, \n} = 0$ implies $[tr] + \boxf [de] = 0$. We can conclude that solving $G^{z}_{\,\, z} = G^{y}_{\,\, z} = G^{y}_{\,\, y} = 0$ together with $[tr] = 0$ or $[de] = 0$ (or any set of four independent equations built from the former ones) is equivalent to solve the bulk Einstein equations $G_{AB} = 0$.

We choose to impose that $G^{z}_{\,\, z} = G^{y}_{\,\, z} = G^{y}_{\,\, y} = 0$ and that the trace part of the bulk Ricci tensor vanishes. Using the gauge invariant variables, these equations respectively read
\begin{align}
& \boxf \, \psi^{gi} + 3 \, \boxf \pi^{gi} + 4 \, \de_{y}^{2} \, \pi^{gi} = 0 \label{Clara 1} \\[4mm]
& \boxf \, \r^{gi} + \de_{z} \de_{y} \, 4 \pi^{gi} = 0 \label{Clara 2} \\[4mm]
& \boxf \, \o^{gi} + 3 \, \boxf \pi^{gi} + 4 \, \de_{z}^{2} \, \pi^{gi} = 0 \label{Clara 3} \\[4mm]
& \Box_6 \, \pi^{gi} = 0 \quad . \label{Clara}
\end{align}

\subsubsection{Codimension-1 brane curvature tensors}

For future reference, we give here the explicit form at first order of the Einstein tensor $\ti{G}_{ab}$ built from the metric induced on the cod-1 brane, and of the extrinsic curvature $\ti{K}_{ab}$ of the cod-1 brane, in terms of the gauge invariant variables. The perturbation of the cod-1 Einstein tensor $\d\tilde{G}_{ab} \equiv \tilde{G}_{ab} - \bar{G}_{ab} = \tilde{G}_{ab}$ explicitly reads
\begin{align}
\d \tilde{G}_{\xi \xi} &= \frac{3}{2} \, \boxf \, \tp \label{Lorette} \\[2mm]
\d \tilde{G}_{\m \xi} &= - \, \frac{3}{2} \, \dexim \, \tp^{\p} \label{Lorette2} \\[2mm]
\d \tilde{G}_{\m \n} &= \deximxin \, \Big[ \!- \frac{1}{2} \, \bvf^{i\, \p} \bvf^{j\, \p} \, \ti{h}^{gi}_{ij} - \tp + \bvf_{i}^{\p \p} \, \dvf^{i}_{gi} \Big] + \nn \\[2mm]
\phantom{a} &\hspace{3cm}+ \eta_{\m\n} \, \Big[ \frac{1}{2} \, \bvf^{i\, \p} \bvf^{j\, \p} \, \, \boxf \ti{h}^{gi}_{ij} + \frac{3}{2} \, \tp^{\p \p} + \boxf \, \tp - \bvf_{i}^{\p \p} \, \boxf \, \dvf^{i}_{gi} \, \Big] \quad , \label{Lorelei}
\end{align}

\noi while the perturbation of the cod-1 extrinsic curvature tensor $\d\tilde{K}_{ab} \equiv \tilde{K}_{ab} - \bar{K}_{ab}$ explicitly reads
\begin{align}
\d\tilde{K}_{\xi\xi}(\xi^{\cdot}) &= - \frac{1}{2} \, \bvf^{i\, \p} \bvf^{j\, \p} \, \bn^{k} \, \de_{k} \, h^{gi}_{ij} \, \evb + \bn^i \bvf^{j\, \p} \, \bvf^{k\, \p} \, \de_{k} \, h^{gi}_{ij} \, \evb + \label{MaMiFaccia}\\[1mm]
&\hspace{5.7cm}+ \half \, \bn^i \bn^j \, \ti{h}^{gi}_{ij} \, \big( \bn_k \bvf^{k \, \p \p} \big) + \bn_i \, \dvf^{i \, \p \p}_{gi} \nn \\[3mm]
\d\tilde{K}_{\xi\m}(\xi^{\cdot}) &=  \dexim \, \Big[ \, \frac{1}{2} \, \bn^i \bvf^{j\, \p} \, \ti{h}^{gi}_{ij} +  \bn_i \, \dvf^{i \, \p}_{gi} \, \Big] \\[3mm]
\d\tilde{K}_{\m\n}(\xi^{\cdot}) &= - \frac{1}{2} \, \bn^{k} \, \de_{k} \, \pi\evb \, \e_{\m\n} + \deximxin \, \bn_i \, \dvf^{i}_{gi} \quad . \label{IlFavore}
\end{align}

\noi Note that above we used again the convention that a prime means a derivative with respect to $\xi$.

\subsection{Master variables}

We have seen above that, considering only the scalar sector, the equations of motion can be written in terms of four ``metric'' variables, and three ``bending'' variables: to find a solution of the equations, we have to solve a system of coupled differential equations for these variables with boundary conditions on the brane and at infinity. In principle, therefore, we should work with seven variables (actually six because the 4D brane embedding perturbation $\dvf_4$ does not appear in the curvature tensors (\ref{Lorette})-(\ref{IlFavore}) as a consequence of the translational invariance of the background configuration in the 4D directions). The equations of motion have a two-fold role; on one hand they link the variables in a differential way, providing constraints, and on the other hand they provide evolution equations. In some cases, it is possible to separate two classes of variables (possibly by redefining fields) in such a way that one of the two classes contain variables which are completely determined by (differential) constraint equations in terms of the variables in the other class, while the latter variables obey decoupled evolution equations. If this happens, it is possible to formulate the dynamical problem purely in terms of the latter variables, which are therefore called \emph{master variables}.

\subsubsection{Metric master variables}

In our case, considering for the moment just the bulk equations, it is apparent that the field $\pi$ obeys a decoupled equation (eq.~(\ref{Clara})), while the other gauge invariant variables are completely determined in terms of the solution for $\pi$ by the equations (\ref{Clara 1})-(\ref{Clara 3}). Therefore, the field $\pi$ is the master variable (in the sense of Mukohyama \cite{Mukohyama:2000ui}) of the scalar sector in the bulk (see also \cite{Kodama:2000fa}). In fact we have
\begin{align}
\boxf \, \psi^{gi} &= - 3 \, \boxf \pi - 4 \, \de_{y}^{2} \, \pi \\[2mm]
\boxf \, \r^{gi} &= - \de_{z} \de_{y} \, 4 \pi \\[2mm]
\boxf \, \o^{gi} &= - 3 \, \boxf \pi - 4 \, \de_{z}^{2} \, \pi \quad ,
\end{align}
or in compact form
\beq
\label{desiree}
\boxf h^{gi}_{ij} = - 3 \, \d_{ij} \, \boxf \pi - 4 \, \de_{i} \de_{j} \, \pi \quad .
\eeq
If we now consider the branes as well, the use of the master variable introduces some subtleties. In fact, note that the perturbation of the cod-1 curvature tensors $\d \tilde{G}_{ab}$ and $\d \tilde{K}_{ab}$ contain the gauge invariant variables $h^{gi}_{ij}$ in the following forms (taking the 4D trace of the $\m\n$ components and the 4D divergence of the $\xi\m$ components of $\d \tilde{G}_{ab}$ and $\d \tilde{K}_{ab}$):
\begin{align}
&\boxf \ti{h}^{gi}_{ij} & &\ti{h}^{gi}_{ij} & &\de_k h^{gi}_{ij} \Big\rvert_{X^{\cdot} = \bar{\varphi}^{\cdot}(\xi^{\cdot})} \quad .
\end{align}
Therefore, concerning the metric perturbations, it is possible to express the cod-1 curvature tensors entirely in terms of the master variable $\pi$, at the price of working explicitly with the Green's function of the 4D D'Alembert operator $\boxf$ (we indicate the Green's function with the notation $1/\boxf$) and with derivatives of third order in the bulk coordinates $z$ and $y$ (note that both of these peculiarities appear just in the extrinsic curvature). The presence of the 4D Green's function is not necessarily a problem, since we can always perform the Fourier transform with respect to the 4D coordinates and work in the mixed representation $(z,y,k^\m)$; in this case the non-local relation in the coordinate space becomes a local relation in the momentum space. The presence of third derivatives is more subtle, however it does not imply that the system is unstable since it is just a consequence of the substitution (\ref{desiree}): the original system of differential equations is of second order and has a well-defined initial values formulation. Therefore, if we accept to live with these subtleties, we can describe the scalar sector of the metric perturbations in our set-up entirely in terms of one variable, $\pi$, and with the fields $\tp$ and $\pi^{(4)}$ obtained evaluating $\pi$ respectively on the cod-1 and on the cod-2 brane.

\subsubsection{Bending master variables}

Concerning the brane-gauge invariant variables, we have already mentioned that the translational invariance of the background configuration in the 4D directions implies that the 4D components of perturbation of the cod-1 bending does not appear in the equations of motion. Therefore, in principle we have two master variables for the bending mode, $\dvfzgi$ and $\dvfygi$. It is customary to describe the perturbations of the brane embedding by projecting the bending mode in the normal direction and in the parallel direction to the brane; we define therefore the normal component of the bending $\dvfn$ and the parallel component $\dvfp$
\begin{align}
\dvfn &\equiv \bn_i \, \dvf_{gi}^{i} & \dvfp &\equiv \bvf^{\p}_i \, \dvf_{gi}^i \quad ,
\end{align}
in terms of which the brane gauge invariant variables read
\beq
\label{Troppissimo}
\dvf^{i}_{gi} = \dvfn \, \bn^{i} + \dvfp \, \bvf^{i \, \p} \quad .
\eeq
The reason for considering this decomposition is that the normal and parallel components of the bending have a geometrical meaning which is more intuitive with respect to the separate $z$ and $y$ components. However, it is important to keep in mind that this is true only when the normal vector is smooth, since when it is discontinuous (as in the thin limit of the nested branes set-up) the normal and parallel components of the bending are not well defined on the cod-2 brane (while the $z$ and $y$ components of the bending are). If we insert the relation (\ref{Troppissimo}) in the expressions (\ref{Lorette})-(\ref{IlFavore}) for the cod-1 curvature tensors, and we use the relation
\beq
\bvf_i^{\p \p} \, \dvf^{i}_{gi} = \frac{\Zpp}{\sqrt{1 - \Zpq}} \,\, \dvfn \quad ,
\eeq
we find that the Einstein tensor $\ti{G}_{ab}$ depends only on the normal component of the bending $\dvfn$. This is consistent with the fact that, from the point of view of the intrinsic geometry, $\dvfp$ represents just a change of coordinates. On the other hand, as can be explicitly checked, the parallel component $\dvfp$ does \emph{not} disappear from the extrinsic curvature: in fact we have
\beq
\bn_i \, \dvf^{i \, \p}_{gi} = \dvfn^{\, \p} + \frac{\Zpp}{\sqrt{1 - \Zpq}} \, \dvfp \quad ,
\eeq
and for example
\beq
\d\tilde{K}_{\xi\m}(\xi^{\cdot}) \supset \dexim \, \bigg( \dvfn^{\, \p} + \frac{\Zpp}{\sqrt{1 - \Zpq}} \, \dvfp \bigg) \quad .
\eeq
It can be checked that $\dvfp$ appears also in $\d\tilde{K}_{\xi\xi}$ while it is absent from $\d\tilde{K}_{\m\n}$. This seems somehow in contrast with the observation of \cite{GarrigaVilenkin,Ishibashi:2002nn} that $\dvfn$ is the only physically observable fluctuation of the brane.

The fact is that, from the point of view of the bulk, the normal component of the bending is the only perturbation which changes the shape of the brane, while the parallel component doesn't. However, when working with thin cod-1 branes, the junction conditions are most naturally written in terms of the brane coordinates (since the energy-momentum tensor is defined only on the brane); therefore we have to pull-back the second fundamental form to the brane, to obtain the extrinsic curvature\footnote{Note that some authors call ``extrinsic curvature'' the object we call second fundamental form.}. The extrinsic curvature measures how fast the normal form changes when we move along a direction in the \emph{brane} coordinate system: a change of coordinates on the brane has the effect to make the normal form change more or less rapidly, and so has the same effect (from the brane point of view) as if we kept the brane coordinates unchanged and changed the shape of the brane (from the bulk point of view). Therefore, it is to be expected that $\dvfp$ appears in the extrinsic curvature (with our definition). The only exception to this argument is when the normal form is constant, such as when the brane is straight in a homogeneous bulk metric: in this case, a change of coordinates on the brane have no effect on the normal form. In fact, if we set $\Zpp = \Ypp = 0$ in our case, the parallel component of the bending $\dvfp$ indeed disappears from the extrinsic curvature.

We conclude that, if we study perturbations around a bent background embedding, then in the scalar sector there are two gauge invariant degrees of freedom which describe the fluctuations of the brane (or three if there is no invariance with respect to translations in the 4D directions). This in particular implies that, when we use the bulk based approach, we should in general consider both $\dvfn$ and $\dvfp$. However, in the background solution the embedding of the cod-1 brane is straight outside the cod-2 brane: this means that for $\abs{\hxi} > l_2$ the parallel component disappears from the junction conditions. In the thin limit, the brane is straight everywhere apart from $\hxi = 0$ where the embedding is not derivable; there is therefore the possibility that, if the thin limit of the perturbed configuration exists, then in this limit we could describe the bending of the brane only in terms of the normal component $\dvfn$. If this happens, the whole system (in the scalar sector) is described by two master variables: the ``metric'' master variable $\pi$, and the ``bending'' master variable $\dvfn$.

\subsection{The regularization of gravity}

We now turn to the analysis of the perturbed junction conditions. In particular, our main aim in this section is to verify that gravity on the cod-2 brane can be regularized by the induced gravity term on the cod-1 brane, and to understand this phenomenon from the point of view of the bulk based approach. In this section we don't perform the thin limit (yet), but still work with a thick cod-2 brane; we extract from the junction conditions the equations for the perturbations, and comment on the role of the induced gravity term regarding the singular structure of the equations around the cod-2 brane. These equations are the basis of the discussion of the thin limit which will be presented in the next section. The bulk-based approach permits to characterize in an intuitive way the singular structure of the solutions around the cod-2 brane, and to see quite clearly why the induced gravity term is needed to allow thin limit solutions where gravity does not diverge.

\subsubsection{The perturbed junction conditions}

To study the junction conditions at perturbative level, it is very convenient to use the cod-1 Gaussian Normal Coordinates. In fact, in this system of reference the localized energy-momentum tensor takes a particularly simple form, as the relation (\ref{angelica}) shows; furthermore, it simplifies the expression for the perturbed junction conditions since in this case $\h{h}_{\xi\xi} = \h{h}_{\xi\m} = 0$. However, it may seem strange that, having paid attention to work in a gauge-invariant way in the bulk, we choose a very specific reference system on the cod-1 brane; therefore it is worthwhile to spend some words to justify this choice.

The problem in the bulk, as we already mentioned, is that the profile of the normal vector tends to be not smooth in the thin limit, and so we cannot use the bulk Gaussian Normal Coordinates in this limit (equivalently, the latter choice of bulk reference system would produce coordinate singularities as we approach the thin limit). However, this does not happen inside the cod-1 brane: for example, considering the mathematical cod-2 brane $\mcal{C}_2$ which divides the cod-1 brane in two domains (which we may call the ``$\xi$ positive'' and the ``$\xi$ negative'' part), we can work separately in the two parts and impose in each of them the cod-1 Gaussian Normal Coordinates. We can then join in a continuous way the two ``partial'' reference systems, and obtain a cod-1 Gaussian Normal Coordinates system (at least) in a neighborhood of the cod-2 brane. This procedure work perfectly also in the thin limit, since the fact that the embedding becomes cuspy from the bulk point of view does not have any influence on our ability to impose the Gaussian Normal Coordinates in each of the two parts, and to join them continuously. Therefore, the use of Gaussian Normal Coordinates inside the cod-1 brane is justified in our set-up.

Perturbing the junction conditions (\ref{junction conditionsch3}) around the background (pure tension) solutions, we obtain the background relation (\ref{Roy}) (which we disregard in the following) plus a perturbation piece: the latter contains, in the left hand side, a term
\beq
- 2 M_6^4 \, \bar{K}_{\xi\xi} \, \h{h}_{\m\n}
\eeq
which cancels (again as a consequence of the background relation) the source term 
\beq
- f(\hxi) \, \bla \, \h{h}_{\m\n}
\eeq
which arises when we perturb the metric which multiplies the unperturbed tension. Disregarding the latter terms as well, the perturbation of the junction conditions reads
\begin{align}
- 2 \, M_6^4 \, \e^{\m\n} \d \hat{K}_{\m\n} + M_5^3 \, \d \hat{G}_{\xi\xi} &= 0 \label{Bronzi1} \\[3mm]
2 M_6^4 \, \d \hat{K}_{\m \xi} + M_5^3 \, \d \hat{G}_{\xi \m} &= 0 \label{Bronzi2} \\[3mm]
2 M_6^4 \Big( \d \hat{K}_{\m\n} - \e_{\m\n} \big( \e^{cd} \d \h{K}_{cd} \big) \Big) + M_5^3 \, \d \hat{G}_{\m\n} &= \d \hat{T}^{(loc)}_{\m\n} \quad , \label{Bronzi3}
\end{align}
where $\d \hat{T}^{(loc)}_{\m\n} = - f(\hxi) \, \dla \, \e_{\m\n} + \h{\mcalT}_{\m\n}$ and all the quantities are functions of $(\hat{\xi}, \chd)$.

Considering only the scalar sector, the $\xi\xi$ component of the perturbation of the junction conditions (eq.~(\ref{Bronzi1})) reads

\beq
\label{xixijunctioncondgi}
2 \, M_6^4 \, \Big( 2 \, \bn^{k} \, \de_{k} \, \pi\evbh - \bn_i \, \boxf \, \hdvf^{i}_{gi} \Big) + \frac{3}{2} \, M_5^3 \, \boxf \, \hp = 0 \quad ,
\eeq

\noi while the $\xi\m$ components of the perturbation of the junction conditions (eq.~(\ref{Bronzi2})) read

\beq
\label{ximujunctioncondgi}
2 \, M_6^4 \, \dechm \bigg( \frac{1}{2} \, \bn^i \bvf^{j\, \p} \, \h{h}^{gi}_{ij} + \bn_i \, \hdvf^{i \, \p}_{gi} \, \bigg) - \frac{3}{2} \, M_5^3 \, \dechm \, \hp^{\p} = 0 \quad .
\eeq 

\noi Regarding the $\m\n$ components of the perturbation of the junction conditions (eq.~(\ref{Bronzi3})), the derivative part reads

\beq
\label{dermunujunctioncondgi}
2 M_6^4 \, \dechmchn \, \bn_i \, \hdvf^{i}_{gi} + M_5^3 \, \dechmchn \bigg( \! - \frac{1}{2} \, \bvf^{i\, \p} \bvf^{j\, \p} \, \h{h}^{gi}_{ij} - \hp + \bvf_{i}^{\p \p} \, \hdvf^{i}_{gi} \bigg) = \dechmchn \hat{\mcal{T}}^{(de)} \quad ,
\eeq

\noi while the trace part reads

\begin{multline}
2 M_6^4 \, \bigg( \frac{3}{2} \, \bn^{k} \, \de_{k} \, \pi\evb - \bn_i \, \boxf \, \hdvf^{i}_{gi} + \frac{1}{2} \, \bvf^{i\, \p} \bvf^{j\, \p} \, \bn^{k} \, \de_{k} \, h^{gi}_{ij} \, \evb - \\[2mm]
- \bn^i \bvf^{j\, \p} \, \bvf^{k\, \p} \, \de_{k} \, h^{gi}_{ij} \, \evb - \half \, \bn^i \bn^j \, \big( \bn_k \bvf^{k \, \p \p} \big) \, \h{h}^{gi}_{ij} - \bn_i \, \hdvf^{i \, \p \p}_{gi} \bigg) \, \emn + \\[2mm]
+ M_5^3 \, \bigg( \frac{1}{2} \, \bvf^{i\, \p} \bvf^{j\, \p} \, \, \boxf \h{h}^{gi}_{ij} + \frac{3}{2} \, \hp^{\p \p} + \boxf \, \hp - \bvf_{i}^{\p \p} \, \boxf \, \hdvf^{i}_{gi} \bigg) \, \e_{\m\n} = \Big( \hat{\mcal{T}}^{(tr)} - f(\hxi) \, \dla \Big) \, \e_{\m\n} \quad . \label{gloria}
\end{multline}

\subsubsection{The pure tension case}
\label{The pure tension case}

To check the consistency of our analysis, we consider first the case of a pure tension perturbation: in this case, we know that the exact solution is given by a configuration of the type we found in section \ref{Pure tension solutions}, and more precisely it is the solution we obtain by performing the substitution $\bla \rightarrow \bla + \dla$ in the background configuration. We want to recover the same solution at first order in $\dla$ using our general perturbative analysis. In our formalism, a pure tension perturbation corresponds to the case $\h{\mcalT}_{\m\n} = 0$; since the perturbation of the source and the background solution are invariant with respect to translations in the 4D directions, we consider the following ansatz for the bending perturbation fields
\begin{align}
\label{Doma1}
\hdvf^{z}_{gi} &= \hdvf^{z}_{gi} \big( \hxi \big) & \hdvf^{y}_{gi} &= \hdvf^{y}_{gi} \big( \hxi \big) & \hdvf_{\m}^{gi} &= 0
\end{align}
and, since the bulk metric in the background solutions is the 6D Minkowski metric, we consider the following ansatz for the metric perturbation fields
\beq
\label{Doma2}
h_{AB}(\Xd) = 0 \quad .
\eeq
This in particular implies that $\h{h}_{\xi\xi} = 2 \, \bvf_i^{\p} \, \hdvf^{i \, \p}_{gi}$ and $\h{h}_{\xi\m} = 0$, and so the requirement that the coordinate system $(\hxi,\chd)$ is Gaussian Normal inside the cod-1 brane is equivalent to the condition
\beq
\label{Francesca}
\bvf_i^{\p} \, \hdvf^{i \, \p}_{gi} = 0 \quad .
\eeq

It is easy to see that the bulk equations (\ref{Clara 1})-(\ref{Clara}) are identically satisfied, and that the same is true for the $\xi\xi$, $\xi\m$ components and for the derivative part of the $\m\n$ components of the junction conditions (equations (\ref{xixijunctioncondgi}), (\ref{ximujunctioncondgi}) and (\ref{dermunujunctioncondgi})). The only equation which is not trivially satisfied is the trace part of the $\m\n$ components of the junction conditions (eq. (\ref{gloria})) which reads
\beq
\label{Eszter}
2 M_6^4 \,\, \bn_i \, \hdvf^{i \, \p \p}_{gi}  = f(\hxi) \, \dla \quad .
\eeq
To solve this equation, it is useful to recast it in terms of the parallel vectors $\mathbf{v}_{(a)}$ introduced in (\ref{parallelvectorsgeneral}), or better of their gauge invariant generalizations. More specifically, the perturbation of the parallel vectors $\dv^A_{(a)}$ (defined in (\ref{Truman})) can itself be decomposed into a parallel and a orthogonal component with respect to the background configuration of the cod-1 brane. In particular, we define the orthogonal and parallel component of the (gauge invariant) perturbation of the $(\xi)$ parallel vector $\dv^{A}_{(\xi)} = \dvf^{A\, \p}$ as follows
\begin{align}
\d v_{\perp}(\xi) &\equiv \bn_i \, \dvf_{gi}^{i \, \p} & \d v_{\shortparallel}(\xi) &\equiv \bvf_i^{\p} \, \dvf_{gi}^{i \, \p} \quad .
\end{align}
We can then express $\bn_i \, \dvf_{gi}^{i \, \p \p}$ using $\dvn$ and $\dvsp$: in fact, since
\beq
\bar{n}^{\p}_{i} = - \frac{\Zpp}{\sqrt{1 - \Zpq}} \, \bvf_i^{\p} \quad ,
\eeq
we have
\beq
\bn_i \, \dvf_{gi}^{i \, \p \p} = \d v^{\p}_{\perp} + \frac{\Zpp}{\sqrt{1 - \Zpq}} \, \d v_{\shortparallel} \quad .
\eeq
Since the cod-1 GNC condition (\ref{Francesca}) implies that $\d \h{v}_{\shortparallel}$ vanishes identically, we can rewrite the equation (\ref{Eszter}) as
\beq
\label{tumblr}
2 M_6^4 \,\, \dhvn^{\p}  = f(\hxi) \, \dla \quad ,
\eeq
which can be integrated to give
\beq
\dhvn(\hxi) = \frac{\dla}{2 M_6^4} \int_{0}^{\hxi} f(\z) \, d\z \quad .
\eeq
This in particular implies that
\beq
\label{Petra}
\dhvn(\hxi) = \pm \frac{\dla}{4 M_6^4} \qquad \textrm{for} \quad \hxi \gtrless \pm l_2 \quad ,
\eeq
and since we have in general
\beq
\dvfn^\p = - \frac{\Zpp}{\sqrt{1 - \Zpq}} \,\, \dvfp + \dvn \quad ,
\eeq
we conclude that for $\abs{\hxi} \geq l_2$ we have
\beq
\label{Hanna}
\dvfn (\hxi) = \frac{\dla}{4 M_6^4} \, \abs{\hxi} + \dvf_0 \quad ,
\eeq
where $\dvf_0$ is an integration constant. Note that the equations of motion does not fix $\dvf_0$, which is then arbitrary; this is consistent with the fact that, since the bulk is exactly Minkowsky, a rigid traslation of the brane is a symmetry of the system.

To understand the geometrical meaning of this configuration, we notice that the (total) embedding function is of the form
\beq
\vf^A(\hxi) = \Big( \mZ(\hxi), \mY(\hxi), 0,0,0,0 \Big)
\eeq
where $\mZ = Z + \dvf^{z}$ and $\mY = Y + \dvf^{y}$. The solution defined by (\ref{Doma1}), (\ref{Doma2}) and (\ref{Hanna}) corresponds to a configuration where the bulk is a ($Z_2$ symmetric) couple of slices of the 6D Minkowski spacetime, such that the total deficit angle is $\a = 4 \vartheta$ where
\beq
\tan \vartheta = \frac{d \mZ (\mY)}{d \mY}\bigg\rvert_{+} = \frac{d \mZ (\hxi)}{d \hxi}\bigg\rvert_{+} \, \bigg( \frac{d \mY (\hxi)}{d \hxi} \bigg)^{\!\! -1}\bigg\rvert_{+} \quad .
\eeq
Applying the $\arctan$ to both sides of the former equation, and expanding at first order in $\dvf^i$ the resulting relation we get
\beq
\vartheta = \arctan \, \frac{\Zp}{\sqrt{1 - \Zpq}}\Bigg\rvert_{+} + \Big( \Yp \dvf^{z \, \p} - \Zp \dvf^{y \, \p} \Big)\bigg\rvert_{+} \quad ,
\eeq
which can be rewritten as
\beq
\vartheta = \arcsin \, \Zp\big\rvert_{+} + \dvn\Big\rvert_{+} \quad :
\eeq
using the background relation (\ref{Dom}) and the relation (\ref{Petra}) we get
\beq
\vartheta = \frac{\bla + \dla}{4 M_6^4} \quad .
\eeq
Since the deficit angle is $\a = 4 \vartheta$, we conclude that a pure tension perturbation $\dla$ on the cod-2 brane produces a variation of the deficit angle
\beq
\d \a = \frac{\dla}{M_6^4} \quad ,
\eeq
while the bulk metric remains the Minkowski metric. This is the same result we get from the exact solutions we obtained in section \ref{Pure tension solutions}, and therefore suggests that our perturbative analysis is consistent.

\subsubsection{The induced gravity term and gravity regularization}

To obtain the equations (\ref{xixijunctioncondgi})-(\ref{gloria}) we did not assume that the matter is perfectly localized on the cod-2 brane; they are in fact just the junction conditions for a cod-1 brane with induced gravity, written in a general coordinate system, with the additional assumption that energy and momentum flow only parallel to the 4D directions (equation (\ref{angelica})). For this reason, we know that these equations admit smooth solutions and that the solution is unique, as in every smooth cod-1 brane set-up; the analysis is actually equivalent to a brane-based one, since we can change coordinates in the bulk and go to a Gaussian Normal Coordinate system. However, if we want to understand what happens when the cod-2 brane becomes thinner and thinner, the only thing we can do is to make some hypothesis on the behavior of the perturbation fields near the cod-2 brane when the thin limit is taken, and \emph{a posteriori} check that these assumptions are compatible with the structure of the equations of motion. This is where the bulk-based approach reveals to be very insightful.

In fact, we remember that in the background solutions the source of the divergence in the extrinsic curvature is easily recognizable: the divergence is entirely due to the embedding function, which in the thin limit is continuous but not derivable on the cod-2 brane, while the bulk metric is smooth also in the limit. This behavior assures that the gravitational field on the cod-2 brane is finite, since the induced metric is built from the first derivatives of the embedding (which is bounded), while the extrinsic curvature is built from its second derivatives, and therefore can diverge at the ``cuspy'' point. This leads us to consider the hypothesis that also for a generic (weak) source the singularity in the extrinsic curvature is carried completely by the embedding function, which converges to a cuspy profile, while the bulk metric and all its partial derivatives of every order remain smooth also in the limit. Of course, in the general case the bulk will not be flat, and the position of the cod-2 brane in the bulk coordinates as well as the local opening of the cod-1 brane ($\dvn\rvert_{+}$) will depend on the 4D coordinates $\chm$. 

Under this hypothesis, the only terms in the junction conditions which can diverge are those which contain second derivatives with respect to $\hxi$ of the embedding functions (background or perturbation part) and the second derivatives with respect to $\hxi$ of the bulk perturbations evaluated on the brane (e.g. $\hp^{\p\p}$), while the evaluation on the brane of the bulk perturbations and their derivatives with respect to the bulk coordinates do not diverge. Taking a look at the equations (\ref{Lorette})-(\ref{Lorelei}) and (\ref{MaMiFaccia})-(\ref{IlFavore}), it is apparent that the only components of the induced Einstein tensor and of the extrinsic curvature which can diverge are $\d \h{G}_{\m\n}$ and $\d \h{K}_{\xi\xi}$, and in particular we have
\begin{align}
\d \h{G}_{\m\n} &\approx \dechmchn \Big[ \, \bvf_i^{\p\p} \, \hdvf^{i}_{gi} \, \Big] + \Big[ \, \frac{3}{2} \, \hp^{\p\p} - \bvf_i^{\p\p} \, \boxf \hdvf^{i}_{gi} \, \Big] \emn \label{Zoi} \\[2mm]
\d \h{K}_{\xi\xi} &\approx \half \, \bn^i \bn^j \, \ti{h}^{gi}_{ij} \, \big( \bn_k \bvf^{k \, \p \p} \big) + \bn_i \, \dvf^{i \, \p \p}_{gi} 
\end{align}
where the symbol $\approx$ here means that the left hand side and the right hand side have the same diverging parts. Note that both $\d \h{G}_{\m\n}$ and $\d \h{K}_{\xi\xi}$ appear only in the $\m\n$ components of the junction conditions (\ref{Bronzi3}). This is compatible with the fact that these components are the only ones which are sourced by the energy-momentum tensor, and so in the thin limit $\d \h{G}_{\m\n}$ and $\d \h{K}_{\xi\xi}$ can hopefully balance the divergence on the right hand side of the equations coming from the source. Keeping only the diverging parts, the equation (\ref{Bronzi3}) reads
\beq
- 2 M_6^4 \, \d \h{K}_{\xi\xi} \, \e_{\m\n} + M_5^3 \, \d \hat{G}_{\m\n} = \Big( \h{\mcalT}^{(tr)} - f(\hxi) \, \dla \Big) \, \e_{\m\n} + \dechmchn \, \h{\mcalT}^{(de)}
\eeq
and, crucially, we note that the extrinsic curvature can contribute a diverging part only to the trace part of the equation (since $\d \h{K}_{\xi\xi}$ is multiplied by $\emn$), while the Einstein tensor $\d \h{G}_{\m\n}$ coming from the induced gravity term contributes also to the derivative part $\dechmchn$, as is evident looking at (\ref{Zoi}). Therefore, if there is no induced gravity term on the cod-1 brane (which is equivalent to say that $\Mft = 0$), our hypothesis on the singular behavior of the perturbation fields is compatible only with a matter source whose energy-momentum tensor has zero derivative part $\h{\mcalT}^{(de)} = 0$. However, the continuity equation implies that if $\h{\mcalT}^{(de)} = 0$ then also $\h{\mcalT}^{(tr)} = 0$; we conclude that, if our hypothesis is valid, then in absence of the induced gravity term on the cod-1 brane only pure tension can be put on the thin cod-2 brane.

This conclusion of course depends crucially on our hypothesis on the singular behavior of the perturbation fields: there may be other choices of the singular behavior that permits to localize matter different from pure tension on the thin cod-2 brane, even when $\Mft = 0$. However, a look at the equation (\ref{Bronzi3}) immediately shows that in this case the derivative part of the $\m\n$ components of the extrinsic curvature tensor has to diverge in the thin limit; this implies that $\bn_i \, \dvf^{i}_{gi}$ has to diverge, as is clear from the equation (\ref{IlFavore}). In this case, the $\xi\xi$ component of the junction conditions (\ref{xixijunctioncondgi}) implies that also $\pi$ in the thin limit has to diverge on the cod-2 brane; since the Ricci scalar built from the metric induced on the cod-2 brane is given by
\beq
R^{(4)} = - 3 \, \boxf \, \pi^{(4)} \quad ,
\eeq
we conclude that in this case the gravitational field diverges on the cod-2 brane, and therefore gravity is not regularized. This result suggests that the induced gravity term on the cod-1 brane is necessary for the regularization of gravity. However, when $\bn_i \, \dvf^{i}_{gi}$ and $\pi^{(4)}$ diverge our perturbative analysis at linear order breaks down, so a full non-perturbative analysis is needed to settle completely this point.

\section{Thin limit of nested branes with induced gravity}
\label{Thin limit of nested branes with induced gravity}

In the previous section, we proposed an ansatz for the behavior of the perturbation fields near the cod-2 brane, and motivated that, if the fields satisfy this ansatz, the quantities $\d \h{G}_{\m\n}$ and $\d \h{K}_{\xi\xi}$ can balance to divergence in the junction conditions coming from the localized energy-momentum tensor, still having a finite induced metric on the cod-2 brane. In this section our aim is two-fold: first of all, still considering the already mentioned ansatz, we want to understand if, taking the limit where the width $l_2$ of the cod-2 brane tends to zero (cod-2 thin limit), the equations of motion give rise to a consistent system of equations. Secondly, if these equations are indeed consistent, we want to understand if the internal structure of the cod-2 brane plays a role also when $l_2 \rightarrow 0$, and if this is not the case (in which case the cod-2 thin limit is well defined) we want to derive the thin limit equations of motion of the system.

\subsection{The thin limit of the background}

Before studying what happens to the equations for the perturbations when we send to zero the cod-2 brane thickness, it is useful to discuss the thin limit of the background configurations, since (as we shall see) they play a non-trivial role in the pillbox integration of the junction conditions across the cod-2 brane. To do that, it is useful to introduce first an alternative perspective on the thin limit, which in practice is more suited to obtain the thin configurations. 

\subsubsection{A new perspective on the thin limit}

In section \ref{Thin limit of branes of codimension-2 and higher}, we introduced the concept of the thin limit description of a theory with localized sources as a description which provides a very good approximation to the true theory when we focus on length scales which are much bigger than the typical localization scales of the sources. However, it is possible to see the thin limit from an alternative (but equivalent) perspective. Instead of considering a theory with sources of fixed thickness, and considering an effective description at larger and larger scales, we could consider a fixed theory, and consider source configurations whose thickness becomes smaller and smaller. This alternative way to see the thin limit is very useful to derive practically the thin limit description.

From this perspective, considering a thick source configuration $S_0$ of the theory and its corresponding field configuration $F_0$, the thin limit is performed by constructing a sequence of source configurations $\{S_n\}_n$ where at each $n \in \mathbb{N}$ the thickness and internal structure of all the branes are uniformly rescaled (contracted), and such that the density of energy momentum is rescaled (expanded) in such a way that the integrated amount of energy-momentum stay constant as $n$ varies. To this sequence of source configurations we can associate a sequence of external field configurations $\{F_n\}_n$, where $F_n$ is the solution of the equations of motion corresponding to the source configuration $S_n$. If the sequence of field configurations $\{F_n\}_n$ converges when $n \rightarrow + \infty$, and the limiting configuration $F_{\infty}$ depends only on the integrated amount of energy-momentum and not on the details of its internal distribution, then the $F_{\infty}$ is the field configuration in the thin limit description.

We follow the latter approach to study the thin limit of the nested branes with induced gravity set-up. Therefore, concerning the source configuration, in the following the localizing function $f(\hxi)$ is replaced by a sequence of functions $f_{[n]}(\hxi)$ which is a realization of the Dirac delta function. Moreover, the ``matter'' energy-momentum tensor $\h{\mcalT}_{\m\n}(\hxi, \chd)$ is replaced by a sequence of tensors $\h{\mcalT}^{[n]}_{\m\n}(\hxi, \chd)$ whose localization length tends to zero when $n \rightarrow + \infty$, but such that the cod-2 energy momentum tensor $\mcalT^{(4)}_{\m\n}(\chd)$ (defined in \ref{Kristin}) is independent from $n$. Likewise, the fields (for example $h_{AB}$, $\bar{\vf}^A$, $\dvf^A$, $\pi$ and so on) which describe the geometry of the set-up, both at background and perturbative level, are replaced by sequences of fields labelled by $n$.

\subsubsection{Thin limit of the background}
\label{Thin limit of the background}

Following the approach to the thin limit discussed above, the pure tension (background) source configurations become a sequence of tensors of the form
\beq
\label{angelicabackground}
\bar{T}^{[n]}_{ab}(\hxi, \chd) = - \d_{a}^{\, \, \m} \, \d_{b}^{\, \, \n} \, f_{[n]} \big( \hxi \big) \, \bla \,\, \bar{g}^{[n]}_{\m\n}(\hxi) \quad ,
\eeq
where $\bar{g}^{[n]}_{\m\n}$ is the induced metric on the cod-1 brane expressed in cod-1 GNC, $f_{[n]} \big( \hxi \big)$ is a sequence of even functions which satisfy
\begin{align}
\int_{-\infty}^{+\infty} f_{[n]} \big( \hxi \big) \, d \hxi &= 1 & f_{[n]} \big( \hxi \big) &= 0 \,\,\,\, \text{for} \,\,\,\, \abs{\hxi} \geq l^{[n]}_2 \quad ,
\end{align}
and $l^{[n]}_2$ is a sequence of positive numbers that converges to zero: $l^{[n]}_2 \rightarrow 0^+$ for $n \rightarrow + \infty$. The analysis of section \ref{Pure tension solutions} implies that there exist exact solutions for this class of sources such that the bulk, induced and double induced metrics are Minkowski (and in particular independent from $n$)
\begin{align}
g^{[n]}_{AB} &= \e_{AB} & \ti{g}^{[n]}_{ab} &= \e_{ab} & g^{(4) [n]}_{\m\n} &= \e_{\m\n} \quad ,
\end{align}
while the embedding of the cod-1 brane is $n$-dependent and non-trivial
\beq
\label{Pattysequence}
\bar{\varphi}_{[n]}^{A}(\hxi, \chd) = \big( Z_{[n]}(\hxi), Y_{[n]}(\hxi), \ch^{\m} \big) \quad ,
\eeq
and the cod-2 emdedding is trivial
\beq
\label{Robertsequence}
\bar{\a}_{[n]}^{a}(\chi^{\cdot}) = \big( 0, \chi^{\m} \big) \quad .
\eeq

\begin{figure}[htp!]
\begin{center}
\includegraphics{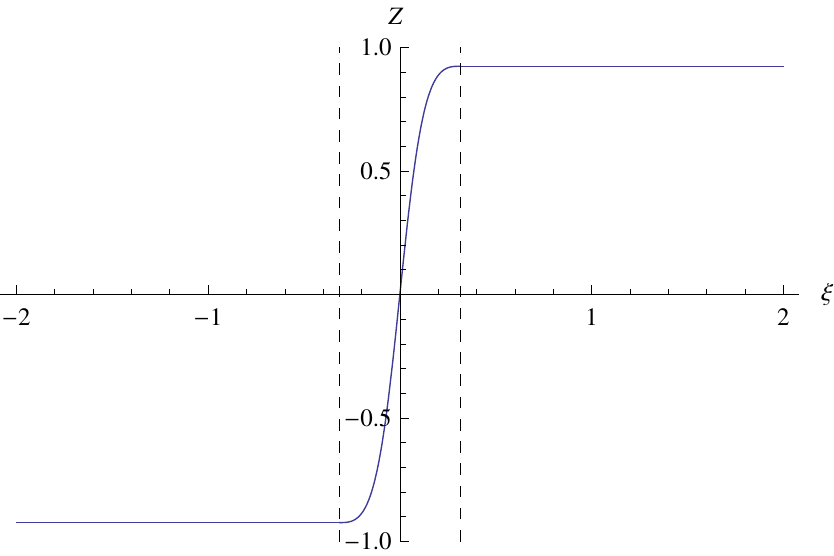}
\caption[The background embedding function $\Zp$]{Numerical plot of the background embedding function $\Zp$.}
\label{background embedding Z}
\end{center}
\end{figure}

\begin{figure}[htp!]
\begin{center}
\includegraphics{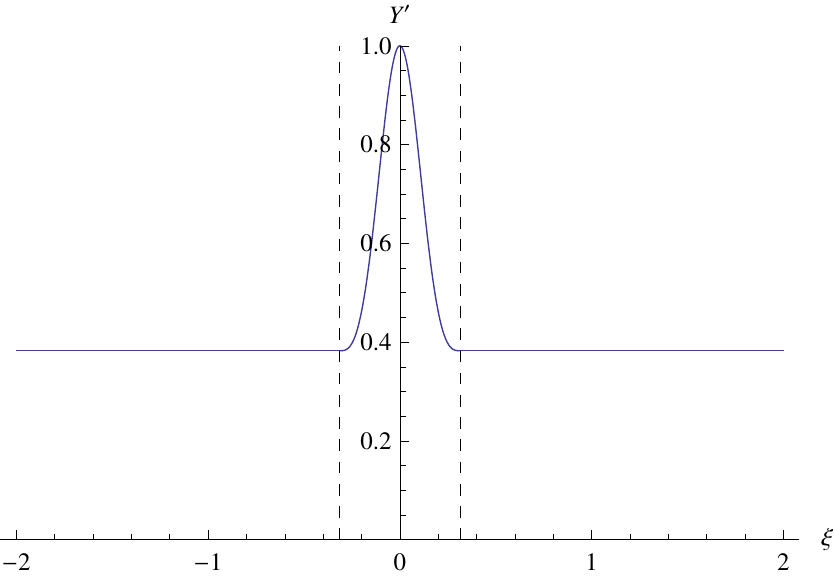}
\caption[The background embedding function $\Yp$]{Numerical plot of the background embedding function $\Yp$.}
\label{background embedding Y}
\end{center}
\end{figure}

It turns out to be important, for the pillbox integration of the equations for the perturbations, to understand carefully the behavior of $\Zp_{[n]}$ and $\Yp_{[n]}$ in the limit $n \rightarrow + \infty$. Introducing the regulating function $\ep_{[n]}(\hxi)$
\beq
\ep_{[n]}(\hxi) \equiv \int_{0}^{\hxi} f_{[n]} (\z) \, d \z \quad ,
\eeq
it is possible to express exactly the solution for the embedding function as follows
\begin{align}
\label{Balotelli}
\Zp_{[n]}(\hxi) &= \sin \bigg( \frac{\bla}{2 \Msf} \, \ep_{[n]}(\hxi) \bigg) \\[2mm]
\Yp_{[n]}(\hxi) &= \cos \bigg( \frac{\bla}{2 \Msf} \, \ep_{[n]}(\hxi) \bigg) \quad .
\end{align}
Note first of all that, consistently with the symmetry properties of $\Zp$ and $\Yp$, we have
\begin{align}
\Zp_{[n]}(0) &= 0 & \Yp_{[n]}(0) &= 1
\end{align}
independently of $n$. Secondly, consider a fixed value $\hxi$ different from zero (say positive, although the case $\hxi < 0$ is analogous): since $l^{[n]}_2 \rightarrow 0$, there exists a natural number $N$ such that, for $n \geq N$, we have $l^{[n]}_2 < \hxi$. Since by definition $\ep_{[n]}(\hxi) = 1/2$ for $\hxi \geq l^{[n]}_2$, we deduce that
\begin{align}
\Zp_{[n]}(\hxi > 0) &\xrightarrow[n \rightarrow + \infty]{} \sin \bigg( \frac{\bla}{4 \Msf} \bigg) & \Yp_{[n]}(\hxi > 0) &\xrightarrow[n \rightarrow + \infty]{} \cos \bigg( \frac{\bla}{4 \Msf} \bigg) \quad .
\end{align}
Putting together these results, we get that $\Zp_{[n]}$ and $\Yp_{[n]}$ converge respectively to the functions $\Zp_{\infty}$ and $\Yp_{\infty}$ which explicitly read
\beq
\Zp_{\infty} \big( \hxi \big) =
\begin{cases}
\sin \Big( \bla/4 \Msf \Big) & \text{for $\hxi > 0$} \\
0 & \text{for $\hxi = 0$} \\
- \sin \Big( \bla/4 \Msf \Big) & \text{for $\hxi < 0$}
\end{cases}
\eeq
and
\beq
\Yp_{\infty} \big( \hxi \big) =
\begin{cases}
\cos \Big( \bla/4 \Msf \Big) & \text{for $\hxi \neq 0$} \\
1 & \text{for $\hxi = 0$} \quad .
\end{cases}
\eeq
Note that, somewhat unexpectedly, the sequence of functions $\Yp_{[n]}$ converges to a \emph{discontinuous} function: this result is confirmed by the figures \ref{background embedding Z} and \ref{background embedding Y} where we plot numerically $\Zp_{[n]}$ and $\Yp_{[n]}$ for $n = 10$ and $\bla = (3/4) \bla_M$ using the explicit form for $\ep_{[n]}(\hxi)$ introduced in section \ref{Numerical check} (equation \ref{epnexplicit}).
The vertical dashed lines indicate the boundaries of the physical cod-2 brane.

\subsection{Pure cod-1 and cod-2 junctions conditions}

We now turn to the study of the junction conditions for the perturbations. To find the exact solution of the equations (\ref{xixijunctioncondgi})-(\ref{gloria}), we should in general solve the equations both for $\abs{\hxi} > l_2$, i.e. outside the physical cod-2 brane (external solution), and inside the cod-2 brane (internal solution), and join smoothly these solutions at the boundaries. However, if we are not interested in the field configuration inside the physical cod-2 brane, we can perform a pillbox integration across it and obtain conditions which relate the value of the fields on the two sides of the cod-2 brane to the cod-2 energy-momentum tensor. Following the approach to the thin limit explained above, this corresponds to solve the cod-1 junction conditions in the domain $\hxi \in \big( - \infty, -l_2^{[n]} \big) \cup \big( l_2^{[n]} , + \infty \big)$, perform the pillbox integration over the interval $\hxi \in \big( -l_2^{[n]},  l_2^{[n]} \big)$, and finally take the $n \rightarrow + \infty$ limit; we then obtain a set of equations for the ``external'' fields, which are valid for $\hxi \neq 0$, and a set of conditions which relate the value of the external fields at $\hxi = 0^-$ and $\hxi = 0^+$. We will refer to the former set of equations as \emph{pure codimension-1 junction conditions} and to the latter set of conditions as \emph{codimension-2 junction conditions}.

To perform the pillbox integration, it is necessary to make an ansatz on the behavior of the fields inside and around the cod-2 brane in the $n \rightarrow + \infty$ limit. Consistently with our analysis in the previous section, we assume that the perturbation of the bulk metric and its derivatives of every order converge uniformly to smooth limiting functions, and that the perturbation of the components of the embedding and all its 4D derivatives converge uniformly to continuous limiting functions. On the other hand, we assume that the first derivative with respect to $\xi$ of the components of the embedding converge (pointwise) to limiting functions which are not necessarily continuous in $\xi = 0$.

Note that, at linear order in perturbations, the effect on the metric and on the bending modes of a pure tension source perturbation and of a matter source perturbation is additive. Moreover, we already studied the effect of a pure tension source perturbation in our framework in section \ref{The pure tension case}, and will derive the exact solution in section \ref{The pure tension perturbation case}. Therefore, in the following we consider a pure matter source perturbation. This in particular allows to impose the boundary condition that the metric and bending perturbations decay at spatial infinity, which implies that the operator $\boxf$ is invertible.

\subsubsection{Pure cod-1 junctions conditions}

To obtain the pure cod-1 junction conditions, which as we mentioned are valid for $\hxi \neq 0$, it is sufficient to impose the conditions $\bvf_i^{\p\p} = 0$ and $\bn_i^{\p} = 0$ in the equations (\ref{xixijunctioncondgi})-(\ref{gloria}). Dropping temporarily the dependence on $n$ to avoid the notation to become too cumbersome, the $\xi\xi$ component of the junction conditions then becomes
\beq
\label{Gingerxixi}
2 \, M_6^4 \, \Big( 2 \, \de_{\bar{\textbf{n}}} \, \pi\evbh - \boxf \, \hdvfn \Big) + \frac{3}{2} \, M_5^3 \, \boxf \, \hp = 0 \quad ,
\eeq
while the $\xi\mu$ components become
\beq
\label{Gingerximu}
2 \, M_6^4 \, \dechm \bigg( \frac{1}{2} \, \bn^i \bvf^{j\, \p} \, \h{h}^{gi}_{ij} + \hdvfn^{\p} \bigg) - \frac{3}{2} \, M_5^3 \, \dechm \, \hp^{\p} = 0 \quad .
\eeq
Regarding the $\m\n$ components, the derivative part reads
\beq
\label{Gingerdermunu}
2 M_6^4 \, \dechmchn \, \hdvfn + M_5^3 \, \dechmchn \bigg( \! - \frac{1}{2} \, \bvf^{i\, \p} \bvf^{j\, \p} \, \h{h}^{gi}_{ij} - \hp \bigg) = 0 \quad ,
\eeq
while the trace part reads
\begin{multline}
2 M_6^4 \, \bigg( \frac{3}{2} \, \de_{\bar{\textbf{n}}} \, \pi\evbh + \frac{1}{2} \, \bvf^{i\, \p} \bvf^{j\, \p} \, \bn^{k} \, \de_{k} \, h^{gi}_{ij} \, \evbh - \bn^i \bvf^{j\, \p} \, \bvf^{k\, \p} \, \de_{k} \, h^{gi}_{ij} \, \evbh - \\[2mm]
- \boxfi \, \hdvfn \bigg) \, \emn + M_5^3 \, \bigg( \frac{1}{2} \, \bvf^{i\, \p} \bvf^{j\, \p} \, \, \boxf \h{h}^{gi}_{ij} + \frac{3}{2} \, \hp^{\p \p} + \boxf \, \hp \bigg) \, \e_{\m\n} = 0 \quad . \label{Gingertracemunu}
\end{multline}
These four equations are actually not independent, but are linked by differential relations if we take into account the bulk equations. To see this, it is useful to indicate the $\xi\xi$ equation with the symbol $(\xi\xi)$, the $\xi\m$ equation with $\dechm (\xi\m)$, the derivative part of the $\m\n$ equation with $\dechmchn (\textrm{de})$ and the trace part of the $\m\n$ equation with $(\textrm{tr}) \, \emn$. Expressing the equations in terms of the master variables (using the relation (\ref{desiree}), which encodes part of the bulk equations), and using the relations  
\begin{align}
\bvf^{i\, \p} \bvf^{j\, \p} \, \de_{i} \, \de_{j} \, \pi \evbh &= \hp^{\p \p} \\[2mm]
\bvf^{i\, \p} \bvf^{j\, \p} \, \bn^{k} \,\, \de_{i} \, \de_{j} \, \de_{k} \, \pi \evbh &= \de^2_{\hxi} \, \bigg( \de_{\bar{\textbf{n}}} \, \pi\evbh \bigg) \quad ,
\end{align}
it is possible to see that the equations above are linked by the relations
\begin{align}
\de_\xi \, (\xi\xi) + \boxf (\xi\m) &= 0 \\[2mm]
\de_{\chm} \de_\xi \, (\xi\m) + \de_{\chm} \big( \boxf (\textrm{de}) + (\textrm{tr}) \big) &= 0 \quad .
\end{align}
This implies that only two of the four equations (\ref{Gingerxixi})-(\ref{Gingertracemunu}) are independent: for example, the equations (\ref{Gingerxixi}) and (\ref{Gingertracemunu}) imply the equations (\ref{Gingerximu}) and (\ref{Gingerdermunu}). We choose to work with the equation $(\xi\xi) + \boxf (\textrm{de})$ and with the trace equation $(\xi\xi) + \boxf (\textrm{de}) + 4 \, (\textrm{tr})$, which in terms of master variables read respectively

\beq
\label{purecod1pi}
2 \, M_6^4 \, \de_{\bar{\textbf{n}}} \, \pi\evbh + M_5^3 \, \boxfi \, \hp = 0
\eeq
and

\beq
\label{purecod1bending}
\boxfi \, \hdvfn = \half \, \de_{\bar{\textbf{n}}} \, \pi\evbh + 2 \, \de^2_{\hxi} \, \bigg( \frac{\de_{\bar{\textbf{n}}}}{\boxf} \, \pi\evbh \bigg) \quad .
\eeq

\noi Note that the field $\pi$ obeys a decoupled equation also on the (pure) cod-1 brane.

\subsubsection{Cod-2 junctions conditions}

We now turn to the analysis of the junction conditions (\ref{xixijunctioncondgi})-(\ref{gloria}) across the cod-2 brane, where $\bvf_i^{\p\p} \neq 0$ and $\bn_i^{\p} \neq 0$. For the sake of clarity, we reinstate the explicit dependence on the index $n$ which controls the thin limit.

Note first of all that the $\xi\xi$ and $\xi\mu$ equations (\ref{xixijunctioncondgi}) and (\ref{ximujunctioncondgi}) are not sourced, and so a necessary condition for the existence of the thin limit is that the $\xi\xi$ and $\xi\mu$ components of the pure cod-1 junction conditions are continuous in $\hxi = 0$. This is in fact automatically true for the $\xi\xi$ equation (\ref{Gingerxixi}), due to the $Z_2$ symmetry present inside the cod-1 brane. In fact, considering our ansatz on the singular behavior of the fields, $\boxf \, \hp_{\infty}$ is continuous. Furthermore, we have that
\beq
\hdvfn^{[n]} = \Yp_{[n]} \hdvf^{z\,[n]}_{gi} - \Zp_{[n]} \hdvf^{y\,[n]}_{gi} \quad ,
\eeq
where $\hdvf^{z\,[n]}_{gi}$ and $\hdvf^{y\,[n]}_{gi}$ converge to the continuous functions $\hdvf^{z}_{\infty}$ and $\hdvf^{y}_{\infty}$, of which the former is even while the latter is odd. As we already mentioned, we have $\Yp_{[n]}\big\rvert_+ = \Yp_{[n]}\big\rvert_-$ and $\Zp_{[n]}\big\rvert_+ = - \Zp_{[n]}\big\rvert_-$, and therefore we obtain
\beq
\lim_{n \rightarrow + \infty} \Big[ \hdvfn^{[n]} \Big]_{\pm} = \hdvf^{z}_{\infty}(0) \lim_{n \rightarrow + \infty} \Big[ \Yp_{[n]} \Big]_{\pm} - \hdvf^{y}_{\infty}(0) \lim_{n \rightarrow + \infty} \Big[ \Zp_{[n]} \Big]_{\pm} \quad ,
\eeq
which vanishes since $\big[ \Yp_{\infty} \big]_{\pm} = 0$ and $\hdvf^{y}_{\infty}(\hxi = 0) = 0$. A completely analogue reasoning shows that
\beq
\lim_{n \rightarrow + \infty} \bigg[ \de_{\bar{\textbf{n}}} \, \pi_{[n]}\evbh \bigg]_{\pm} = 0 \quad ,
\eeq
since by symmetry we have that $\de_y \pi_{\infty}$ vanishes in $\Xd = \bvfd(0^{\cdot})$.

The continuity in $\hxi = 0$ of the $\xi\mu$ equation (\ref{Gingerximu}) is instead not automatically implied by the symmetry properties of the perturbation fields: since all the terms in equation (\ref{Gingerximu}) are odd, we get the consistency equation
\beq
\label{consistency}
\Bigg[ M_6^4 \, \sin \bigg( \frac{\bla}{2 \Msf} \bigg) \Big( \h{h}^{\infty}_{zz} - \h{h}^{\infty}_{yy} \Big) + 4 \, M_6^4 \, \hdvfn^{\infty \, \p} - 3 \, M_5^3 \, \hp_{\infty}^{\p} \Bigg]_{0^+} = 0 \quad ,
\eeq
where we used the fact that $\h{h}^{\infty}_{zy}(\hxi = 0) = 0$ by parity reasons, and we used the explicit expressions for $\Zp_{\infty}\big\rvert_{0^+}$ and $\Yp_{\infty}\big\rvert_{0^+}$. Furthermore, we used the notation $\big\rvert_{0^{+}}$ and $\big[ \phantom{a} \big]_{0^{+}}$ to indicate the evaluation in $\hxi = 0^+$.

On the other hand, the derivative and trace part of the $\mu\nu$ components of the junction conditions for the perturbations, equations (\ref{dermunujunctioncondgi}) and (\ref{gloria}), contain diverging pieces, and so it is necessary to perform a pillbox integration of these equations across the cod-2 brane. Note that, considering our ansatz on the behavior of the perturbations at the cod-2 brane, the only term which diverges in the left hand side of the equation (\ref{dermunujunctioncondgi}) is $\bvf_{i}^{\p \p} \, \hdvf^{i}_{gi}$; therefore, the derivative part of the $\m\n$ components of the junction conditions produces the condition (contracting with $\e^{\m\n}$)
\beq
\label{kindness der}
M_5^3 \lim_{n \rightarrow + \infty} \int_{-}^{+} \! d \hxi \,\, \bvf_{i \, [n]}^{\p \p} \, \boxf \, \hdvf^{i \, [n]}_{gi} = \boxf \, \mcal{T}^{(4)}_{(de)} \quad .
\eeq
Similarly, the only terms which diverge in the left hand side of the equation (\ref{gloria}) are the ones which are derivated twice with respect to $\hxi$; therefore, the trace part of the $\m\n$ components of the junction conditions produce the condition
\begin{multline}
\label{kindness trace}
- M_6^4 \lim_{n \rightarrow + \infty} \int_{-}^{+} \! d \hxi \, \bigg( \bn_{[n]}^i \bn_{[n]}^j \, \Big( \bn^{[n]}_k \bvf_{[n]}^{k \, \p \p} \Big) \, \h{h}^{gi \, [n]}_{ij} + 2 \, \bn^{[n]}_i \, \hdvf^{i \, [n] \, \p \p}_{gi} \bigg) + \\[2mm]
+ M_5^3 \lim_{n \rightarrow + \infty} \int_{-}^{+} \! d \hxi \, \bigg( \frac{3}{2} \, \hp_{[n]}^{\p \p} - \bvf_{i \, [n]}^{\p \p} \, \boxf \, \hdvf^{i \, [n]}_{gi} \bigg) = \mcal{T}^{(4)}_{(tr)} \quad .
\end{multline}
The continuity equation for the cod-2 energy-momentum tensor $\e^{\m\a} \de_{\a} \mcal{T}^{(4)}_{\m\n} = 0$ implies that it is equivalent to impose the conditions (\ref{kindness der}) and (\ref{kindness trace}), or the condition (\ref{kindness der}) and the following condition
\beq
\label{kindness trace2}
M_6^4 \lim_{n \rightarrow + \infty} \int_{-}^{+} \! d \hxi \, \bigg( \bn_{[n]}^i \bn_{[n]}^j \, \Big( \bn^{[n]}_k \bvf_{[n]}^{k \, \p \p} \Big) \, \h{h}^{gi \, [n]}_{ij} + 2 \, \bn^{[n]}_i \, \hdvf^{i \, [n] \, \p \p}_{gi} \bigg) - 3 \, M_5^3 \, \hp_{\infty}^{\p}\Big\rvert_{0^+} = 0 \quad .
\eeq
The integrations in the equations (\ref{kindness der}) and (\ref{kindness trace2}) are performed explicitly in the appendix \ref{Pillbox integration of nested branes}. It turns out that the equation (\ref{kindness trace2}) reproduces exactly the condition (\ref{consistency}), which is a confirmation of the consistency of our analysis, while the equation (\ref{kindness der}) produces the condition
\beq
\label{Exagerate1}
2 \, \Mft \, \tan \bigg( \frac{\bla}{4 \Msf} \bigg) \, \boxf \, \dvfn^{\infty}\Big\rvert_{0^+} = \boxf \, \mcal{T}^{(4)}_{(de)} \quad .
\eeq
The latter equation, together with (\ref{consistency}), constitutes the thin limit cod-2 junction conditions. Note that we can express the equation (\ref{consistency}) in terms of the master variables, obtaining
\beq
\label{Exagerate2}
\Bigg[ 4 \, M_6^4 \, \hdvfn^{\infty \, \p} - 4 \, M_6^4 \, \sin \bigg( \frac{\bla}{2 \Msf} \bigg) \frac{\big( \de_z^2 - \de_y^2 \big)}{\boxf} \, \pi_{\infty}\evbbh - 3 \, M_5^3 \, \hp_{\infty}^{\p} \Bigg]_{0^+} = 0 \quad .
\eeq

\subsection{Thin limit master equations}

In the previous section we showed that, in the limit where the thickness of the cod-2 brane tends to zero, the cod-1 junction conditions for the perturbations naturally split into two pure cod-1 junction conditions, equations (\ref{purecod1pi}) and (\ref{purecod1bending}), and two cod-2 junction conditions, equations (\ref{Exagerate1}) and (\ref{Exagerate2}). We want to discuss now the implications of these results for what concerns the thin limit of the cod-2 brane in the nested brane with induced gravity set-up.

Note first of all that the details of the internal structure of the cod-2 brane, encoded in the precise form of the realization of the Dirac delta $f_{[n]}$ and of the regulating function $\ep_{[n]}$, do not play a role in the limit $n \rightarrow + \infty$. Furthermore, the master variable $\hdvfp$ disappears from the thin limit equations, so in the thin limit the system is described purely in terms of two master variables, the field $\pi$ and the normal component of the bending $\hdvfn$. This implies that, if the system of thin limit equations for $\pi$ and $\hdvfn$ is consistent, the thin limit of the cod-2 brane in the nested brane with induced gravity set-up is well defined (at least when we consider first order perturbations around pure tension solutions). To address the latter point, consider the system of equations formed by the bulk and of the pure cod-1 junction conditions
\begin{gather}
\label{bulkeqdiscussion}
\boxs \, \pi = 0 \\[4mm]
\label{purecod1pidiscussion}
2 \, M_6^4 \, \de_{\bar{\textbf{n}}} \, \pi\evbh + M_5^3 \, \boxfi \, \hp = 0 \\[3mm]
\label{purecod1bendingdiscussion}
\boxfi \, \hdvfn = \half \, \de_{\bar{\textbf{n}}} \, \pi\evbh + 2 \, \de^2_{\hxi} \, \bigg( \frac{\de_{\bar{\textbf{n}}}}{\boxf} \, \pi\evbh \bigg) \quad ,
\end{gather}
where the fields $\pi$ and $\hdvfn$ are actually the thin limit fields $\pi_{\infty}$ and $\hdvfn^{\infty}$, but for the sake of simplicity we drop the $\infty$ symbol. This system of equations has the same structure of the equations for the metric perturbations and the bending in a 6D cod-1 DGP model; in particular, the equation (\ref{purecod1bendingdiscussion}) is the 6D equivalent of the bending equation $\boxf \hdvfn \propto \ti{T}$ which holds in the 5D DGP model, with the only difference that the energy-momentum tensor $\ti{T}$ vanishes (we are outside the cod-2 brane) and that we didn't impose the transverse-traceless condition on the metric perturbations. In the 6D cod-1 DGP model, the validity of the equations (\ref{purecod1pidiscussion}) and (\ref{purecod1bendingdiscussion}) for every value of $\hxi$ together with (\ref{bulkeqdiscussion}) and the boundary condition that the fields decay at spatial infinity, singles out a unique solution of the system of equations. For this to happen, it is crucial that the fields are continuous and derivable with continuous first partial derivatives.

Concerning the system of equations (\ref{bulkeqdiscussion})-(\ref{purecod1bendingdiscussion}), the main difference between the 6D nested branes with induced gravity set-up and the 6D cod-1 DGP model is that in the former set-up the fields $\hp$ and $\hdvfn$ are continuous but not derivable with respect to $\hxi$ in $\hxi = 0$. This implies that the system of equations does not single out a unique solution anymore, since there is freedom in choosing how to join the solutions for the fields in $\hxi = 0^+$ and in $\hxi = 0^-$. To render the solution unique again, we have to provide two relations which fix this freedom: this is precisely what the cod-2 junction conditions do. Considering the metric master variable, we can complement the bulk and the pure cod-1 equations with the condition (\ref{Exagerate2}), to obtain the system
\begin{gather}
\label{fantasy1}
\boxs \, \pi = 0 \\[4mm]
\label{fantasy2}
2 \, M_6^4 \, \de_{\bar{\textbf{n}}} \, \pi\evbh + M_5^3 \, \boxfi \, \hp = 0 \\[3mm]
\label{fantasy3}
3 \, M_5^3 \, \hp^{\p}\Big\rvert_{0^+} = \Bigg[ 4 \, M_6^4 \, \hdvfn^{\p} - 4 \, M_6^4 \, \sin \bigg( \frac{\bla}{2 \Msf} \bigg) \frac{\big( \de_z^2 - \de_y^2 \big)}{\boxf} \, \pi\evbbh \Bigg]_{0^+} \quad ,
\end{gather}
while considering the bending master variable, we can complement the pure cod-1 equation with the condition (\ref{Exagerate1}), to obtain the system
\begin{gather}
\label{fantasy4}
\boxfi \, \hdvfn = \half \, \de_{\bar{\textbf{n}}} \, \pi\evbh + 2 \, \de^2_{\hxi} \, \bigg( \frac{\de_{\bar{\textbf{n}}}}{\boxf} \, \pi\evbh \bigg) \\[3mm]
\label{fantasy5}
2 \, \Mft \, \tan \bigg( \frac{\bla}{4 \Msf} \bigg) \, \boxf \, \dvfn\Big\rvert_{0^+} =  \boxf \,\mcal{T}^{(4)}_{(de)} \quad .
\end{gather}
Therefore, the equation (\ref{Exagerate2}) provides a boundary condition of the Neumann type for $\hp$ on the side of the cod-2 brane, while the relation (\ref{Exagerate1}) provides a boundary condition of the Dirichlet type for $\hdvfn$ on the side of the cod-2 brane. The consistency of the system of coupled differential equations composed of the bulk equation, the pure cod-1 junction conditions and the cod-2 junction conditions for the master variables indicates that the thin limit of the cod-2 brane in the nested branes with induced gravity set-up is well defined, at least when we consider first order perturbations around the pure tension solutions. 

It is interesting to comment on how the presence of matter on the cod-2 brane sources the total field configuration according to the coupled system of equations for $\pi$ and $\hdvfn$. First of all, note that if there is no matter on the cod-2 brane then the configuration $\pi = 0$, $\hdvfn = 0$ is a solution of the system of equations (it is the background solution in fact). If we turn on the energy-momentum tensor on the cod-2 brane, $\h{\mcalT}$ forces the cod-2 brane to move (equation (\ref{fantasy5})); this movement acts as a boundary condition for the movement of the cod-1 brane (equation (\ref{fantasy4})), producing a non-trivial cod-1 bending profile. This profile necessarily has non-vanishing first $\hxi$-derivative on the side of the cod-2 brane, and this acts as a source for the metric master variable $\pi$, since it produces a non-trivial boundary condition for $\hp$ on the side of the cod-2 brane (equation (\ref{fantasy3})). As a consequence, a non-trivial profile for $\pi$ in the bulk and on the cod-1 brane is created (equations (\ref{fantasy1}) and (\ref{fantasy2})). The profile of $\hp$ on the cod-1 brane in turn acts as a source for $\hdvfn$ on the cod-1 brane (equation (\ref{fantasy4})), and so on.

\section{Ghosts in the 6D Cascading DGP model}
\label{Ghosts in the 6D Cascading DGP model}

In this section we want to use the results obtained in the previous sections regarding the nested branes with induced gravity set-up, to study the problem of ghosts in the nested branes realization of the 6D Cascading DGP model. As we already mentioned, it has been claimed \cite{deRham:2007xp, deRham:2010rw} that the perturbations around the pure tension solutions in the 6D Cascading DGP model are ghost-free if the background tension is bigger than a \emph{critical tension} $\blac^{\textup{dRKT}}$ which reads
\beq
\label{critical tension theirs}
\blac^{\textup{dRKT}} \equiv \frac{2}{3} \, m_6^2 \, \Mfs \quad ,
\eeq
while contain a ghost if the background tension is smaller than $\blac^{\textup{dRKT}}$. It has also been claimed that, performing a 4D scalar-vector-tensor decomposition of the perturbation modes, the ghost mode belongs to the scalar sector of the theory. Despite this very interesting result, a geometric interpretation of the existence of the critical tension and a clear understanding of the mechanism responsible for the presence/absence of the ghost are still missing. This may be related to the fact that the analysis of \cite{deRham:2007xp, deRham:2010rw} is performed with the brane-based approach: in fact, as we suggested above, the geometry of nested branes configurations is more transparent in the bulk-based approach. Therefore, it seems reasonable that performing the same analysis with the bulk-based approach may shed light on the interesting phenomenon which is responsible for the emergence of the critical tension.

\subsection{The critical tension}

\subsubsection{The thin limit of the 6D Cascading DGP model}

The nested branes realization of the Cascading DGP model is closely related to the nested branes with induced gravity set-up: in fact, the only difference is that in the former set-up there is a (4D) induced gravity term on the cod-2 brane which modifies the source term of the latter set-up. More specifically, in the nested branes with induced gravity set-up the equations of motion are (\ref{Amsterdam}) and (\ref{junction conditionsch3}), and the cod-1 source terms are of the form (\ref{angelica}), where
\beq
\h{T}^{(loc)}_{\m\n} (\hxi,\chd) = - f(\hxi) \,\, \la \,\, \h{g}_{\m\n} (\hxi,\chd) + \h{\mcalT}_{\m\n} (\hxi,\chd)
\eeq
and we assumed that the matter energy-momentum tensor $\h{\mcalT}_{\m\n}$ vanishes for $\abs{\hxi} > l_2$. In the nested branes realization of the 6D Cascading DGP scenario, the equations of motion are still (\ref{Amsterdam}) and (\ref{junction conditionsch3}) and the energy-momentum tensor still satisfies (\ref{angelica}), but the localized energy-momentum tensor $\h{T}^{(loc)}_{\m\n}$ takes the form 
\beq
\h{T}^{(loc)}_{\m\n} (\hxi,\chd) = - f(\hxi) \,\, \la \,\, \h{g}_{\m\n} (\hxi,\chd) + \h{\mscrT}_{\m\n} (\hxi,\chd) - l(\hxi) \,\, \Mfs \, G^{(4)}_{\m\n}(\chd;\hxi)
\eeq
where the matter energy-momentum tensor is now indicated by $\h{\mscrT}_{\m\n}$. Here $l(\hxi)$ is a realization of the Dirac delta function (which may be equal to $f(\hxi)$) which localizes the 4D induced gravity term, and $G^{(4)}_{\m\n}(\chd;\hxi)$ is the Einstein tensor built from the 4D metric $g^{(4)}_{\m\n}(\chd;\hxi)$ which is induced by $\ti{g}_{ab}$ on the $\hxi$-constant 4D slices of the cod-1 brane. It is easy to see that the pure tension solutions derived in section \ref{Pure tension solutions} are solutions of the nested branes realization of the 6D Cascading DGP scenario as well, since for these background solutions we have
\beq
\bar{g}^{(4)}_{\m\n}(\chd;\hxi) = \emn
\eeq
and so the Einstein tensor $G^{(4)}_{\m\n}(\chd;\hxi)$ vanish identically.

On the other hand, in our analysis of the perturbations in the nested branes with induced gravity set-up, the only properties of $\h{\mcalT}_{\m\n}$ we really used are the fact that it vanishes for $\abs{\hxi} > l_2$, the fact that it decays at 4D infinity (which is necessary to use the scalar vector tensor decomposition) and the fact that its pillbox integration satisfies the continuity equation. These properties are satisfied also by $\h{\mscrT}_{\m\n}$ and $l(\hxi) \,\, \Mfs \, G^{(4)}_{\m\n}$, and so the analysis we performed in the previous sections applies also to the Cascading DGP case. In particular, this means that the thin limit of the equations for the first order perturbations around the pure tension solutions is well defined in the nested branes realization of the 6D Cascading DGP scenario. Furthermore, the thin limit equations for the latter model can be obtained from the thin limit equations of the previous section performing the substitution
\beq
\mcalT^{(4)}_{\m\n}(\chd) \rightarrow \mscrT^{(4)}_{\m\n}(\chd) - \Mfs \, G^{(4)}_{\m\n}(\chd) \quad ,
\eeq
where $G^{(4)}_{\m\n}(\chd)$ is the Einstein tensor built from the metric induced on the cod-2 brane, and $\mscrT^{(4)}_{\m\n}$ is obtained from $\h{\mscrT}_{\m\n}$ by performing the pillbox integration across the cod-2 brane, in analogy with (\ref{Kristin}). Unless explicitly said otherwise, in the rest of this section the fields $\pi$, $\hp$, $\pi^{(4)}$, $\hdvfn$ and $\dvf_0$ represent the thin limit configurations $\pi_{\infty}$, $\hp_{\infty}$, $\pi^{(4)}_{\infty}$, $\hdvfn^{\infty}$ and $\dvf_0^{\infty}$, but we omit the $\infty$ symbol.  

\subsubsection{Ef\mbox{}fective master equations for $\pi$ and the critical tension}

We want now to understand why in the nested branes realization of the 6D Cascading DGP scenario there is a critical tension which marks the transition between the pure tension configurations which are plagued by ghosts and the ones which are ghost-free. To do that, we concentrate on the $\pi$ master variable: in particular, we look for approximated (effective) descriptions which allows to obtain master equations which contain the $\pi$ field alone.

Note that, if the background tension $\bla$ is non-vanishing, the derivative part of the $\mu\nu$ components of the cod-2 junction conditions (the Cascading version of equation (\ref{Exagerate1})) links the value of the normal component of the bending $\hdvfn$ on the side of the cod-2 brane with the value of the $\pi$ field on the brane and the trace of the matter energy-momentum tensor on the cod-2 brane
\beq
\label{think}
2 \, M_5^3 \, \tan \bigg( \frac{\bla}{4 M_6^4} \bigg) \, \, \boxf \, \hdvfn\Big\rvert_{+} = M_4^2 \, \boxf \, \pi^{(4)} - \frac{1}{3} \, \mathscr{T} \quad .
\eeq
On the other hand, the $\xi\xi$ component of the pure cod-1 junction conditions (equation (\ref{Gingerxixi})) links the normal component of the bending $\hdvfn$ to the $\pi$ field on the cod-1 brane and the derivative of $\pi$ normally to the cod-1 brane
\beq
2 \, M_6^4 \, \Big( 2 \, \de_{\bar{\mathbf{n}}} \, \pi\evbh - \boxf \, \hdvfn \Big) + \frac{3}{2} \, M_5^3 \, \boxf \, \hp = 0 \quad .
\eeq
Evaluating the latter equation on the side of the cod-2 brane (\emph{i.e.} considering the $\hxi \rightarrow 0^+$ limit of the previous equation), we obtain by continuity a relationship between the value of $\hdvfn$ on the side of the cod-2 brane, the value of $\pi$ on the cod-2 brane and the derivative of $\pi$ normally to the cod-1 brane on the side of the cod-2 brane
\beq
\label{freedomfreedom}
4 \, M_6^4 \, \de_{\bar{\mathbf{n}}} \, \pi\Big\rvert_{+} - 2 \, M_6^4 \, \boxf \, \hdvfn\Big\rvert_{+} + \frac{3}{2} \, M_5^3 \, \boxf \, \hp\Big\rvert_{+} = 0 \quad ,
\eeq
where the latter equation contains function of the 4D variables $\chd$ only, and we introduced the notation
\beq
\de_{\bar{\mathbf{n}}} \, \pi\Big\rvert_{+} = \de_{\bar{\mathbf{n}}} \, \pi\Big\rvert_{\Xd = \bvfd(\hxi = 0^+)} \quad .
\eeq
Therefore, we can then use the two equations (\ref{think}) and (\ref{freedomfreedom}) to obtain a decoupled equation for the field $\pi$, using the fact that by continuity of the $\pi$ field we have $\boxf \, \pi^{(4)} = \boxf \, \hp\big\rvert_{+}$. In fact, expressing $\boxf \, \hdvfn\big\rvert_{+}$ in terms of $\pi^{(4)}$ and $\mscrT$ using the equation (\ref{think}), and inserting the resulting relation in the equation (\ref{freedomfreedom}), we get
\beq
\label{Brigitte}
4 \, M_6^4 \, \tan \bigg( \frac{\bla}{4 M_6^4} \bigg) \, \de_{\bar{\mathbf{n}}} \, \pi\Big\rvert_{+} + \bigg[ \frac{3}{2} \, M_5^3 \, \tan \bigg( \frac{\bla}{4 M_6^4} \bigg) - m_6 \, M_4^2 \bigg] \, \boxf \, \hp\Big\rvert_{+} = - \frac{m_6}{3} \, \mathscr{T} \quad .
\eeq
Note that this equation is exact (at first order in perturbations); considering the thin limit on the cod-2 brane allowed us to find a master equation for the field $\pi$ and its derivatives at the cod-2 brane.

Despite the equation (\ref{Brigitte}) involves only the value of the field $\pi$ near the cod-2 brane, the presence of the normal derivative $\de_{\bar{\mathbf{n}}} \, \pi$ implies that to find a solution of (\ref{Brigitte}) we have to solve the bulk equations and the cod-1 junction conditions, or in other words we still need to solve the complete system of differential equations for $\pi$ and $\hdvfn$. However, it is possible to look for an approximate description which ``decouples'' the dynamics on the cod-2 brane from the dynamics in the bulk and on the cod-1 brane, with the hope to find a master equation which describes the behavior of $\pi$ on the cod-2 brane. We consider in fact the following ``4D limit''
\begin{align}
\label{decouplinglimit1}
\lvert m_6 \, \de_{\bar{\mathbf{n}}} \rvert &\ll \lvert \boxf \rvert \\[2mm]
\label{decouplinglimit2}
\lvert m_5 \, \de_{\bar{\mathbf{n}}} \rvert &\ll \lvert \boxf \rvert
\end{align}
which implies that, in the left hand side of equation (\ref{Brigitte}), we can neglect the first term compared to the second term and to the third term: dividing by $m_6 M_4^2/3$ we then obtain
\beq
\label{critical tension equation}
3 M_4^2 \, \bigg[ \, 1 - \frac{3}{2} \, \frac{m_5}{m_6} \, \tan \bigg( \frac{\bla}{4 M_6^4} \bigg) \bigg] \,\, \boxf \, \pi^{(4)} = \mathscr{T} \quad .
\eeq
This equation contains only the field $\pi$ evaluated on the cod-2 brane, and does not contain derivatives of $\pi$ normally to the cod-2 brane; therefore, it is the effective master equation which describes the behavior of $\pi$ field on the cod-2 brane in the selected range of 4D length scales. Crucially, in this equation the numerical coefficient which multiplies $\boxf \, \pi^{(4)}$ changes sign when $\bla$ becomes equal to the \emph{critical tension}
\beq
\label{critical tension}
\bla_c = 4 M_6^4 \, \arctan \bigg( \frac{2}{3} \, \frac{m_6}{m_5} \bigg) \quad .
\eeq

\subsection{Ghosts and geometrical interpretation}

\subsubsection{Ef\mbox{}fective action and the presence of ghosts}

The sign of the coefficient multiplying $\boxf \, \pi^{(4)}$ in equation (\ref{critical tension equation}) is closely related to the fact that the field $\pi$ is a ghost or not. To see this connection more clearly, we have to describe the dynamics of the field $\pi$ using the Lagrangian (or Hamiltonian) formalism. The action which describes the dynamics of the system is of course the general action (\ref{CascadingDGP6D}); to get the effective action which describes the dynamics of $\pi^{(4)}$ in the 4D limit, we should integrate out of the (quadratic approximation of the) general action all the other fields using the bulk equations and the junction conditions, and impose the conditions (\ref{decouplinglimit1})-(\ref{decouplinglimit2}). However, indicating
\beq
K \equiv 3 M_4^2 \, \bigg[ \, 1 - \frac{3}{2} \, \frac{m_5}{m_6} \, \tan \bigg( \frac{\bla}{4 M_6^4} \bigg) \bigg] \quad ,
\eeq
the equation (\ref{critical tension equation}) tells us that the 4D effective action has to be proportional to the following action
\beq
\label{Alessandra}
S^{(2)}_{\pi^{(4)}} = \int \! d^4 \ch \,\, \bigg[ \, \frac{K}{2} \, \de_{\m} \pi^{(4)} \de^{\m} \pi^{(4)} + \pi^{(4)} \mscr{T} \, \bigg] \quad :
\eeq
therefore, the only thing that we need to determine is the value of the proportionality constant. To do that, it is not necessary to integrate out explicitly the fields in the original action but it is sufficient to look at the coupling with matter: in fact, since we are considering matter sources $\mscrT_{\m\n}$ which are confined on the cod-2 brane, the coupling of gravity with matter in the general action involves only the metric perturbations evaluated on the cod-2 brane. Expanding at quadratic order around the Minkowski spacetime the term in the general action which expresses the gravity-matter coupling, we get
\beq
\int \! d^4 \ch \, \sqrt{- g^{(4)}} \, \mscr{L}_m \simeq \int \! d^4 \ch \, h_{\m\n} \mscr{T}^{\m\n} = \int \! d^4 \ch \, \Big( \pi^{(4)} \mscr{T} + h^{TT}_{\m\n} \mscr{T}_{TT}^{\m\n} \Big) \quad ,
\eeq
which indicates that the action (\ref{Alessandra}) is indeed the correct 4D effective action for $\pi^{(4)}$. We can then say that, integrating out the other fields in the scalar sector and imposing the 4D limit, we generate a $\bla$-dependent kinetic contribution
\beq
S^{(2)}_{\bla} = \int \! d^4 \ch \,\, \bigg[ \, - \frac{9}{4} \, M_4^2 \, \frac{m_5}{m_6} \, \tan \bigg( \frac{\bla}{4 M_6^4} \bigg) \, \de_{\m} \pi^{(4)} \de^{\m} \pi^{(4)} \, \bigg]
\eeq
to the 4D part of the general Lagrangian 
\beq
S_{4}^{(2)} = \int \! d^4 \ch \,\, \bigg[ \, \frac{3 M_4^2}{2} \, \de_{\m} \pi^{(4)} \de^{\m} \pi^{(4)} + \pi^{(4)} \mscr{T} \, \bigg] \quad .
\eeq

It is now easy to check from (\ref{Alessandra}) if $\pi^{(4)}$ is a ghost or not; in fact, a field whose dynamics is described by an action of the form (\ref{Alessandra}) is a ghost if $K > 0$ while it is a healthy field if $K < 0$ (with our choice of the metric signature). Therefore, we can conclude that the field $\pi^{(4)}$ in the nested branes realization of the 6D Cascading DGP model is a ghost if the background tension is smaller than the critical tension $\blac$ introduced in (\ref{critical tension}), while it is a healthy perturbation field if the background tension is bigger than $\blac$. It is important to remember that we obtained this result in the 4D limit (\ref{decouplinglimit1})-(\ref{decouplinglimit2}) and at first order in perturbations, so we cannot say if the presence/absence of the ghost is indeed a feature of the full theory. This result is to be compared with the findings of \cite{deRham:2010rw}, which use a brane-based approach; working directly at the level of the action, they find that the kinetic term for $\pi^{(4)}$ on the cod-2 brane after integrating out the other fields becomes
\beq
\mscr{L}_4^{kin} = \frac{3 M_4^2}{4} \bigg( \frac{3 \bla}{2 m_6 M_4^2} - 1 \bigg) \, \pi^{(4)} \boxf \pi^{(4)} \quad .
\eeq
Such a kinetic term signals the presence (respectively, absence) of a ghost if the background tension $\bla$ is smaller (respectively, bigger) than the critical tension $\blac^{\textup{dRKT}}$, whose value (\ref{critical tension theirs}) is however different from the value (\ref{critical tension}) we find in our analysis. We will comment about this difference in sections \ref{Discussion} and \ref{Numerical check}.

Note that, in the limit $\bla \rightarrow 0$, the action (\ref{Alessandra}) reduces to the action for the scalar sector of (4D) GR; this naively would suggest that GR itself has a ghost\label{observation}. Nonetheless, a careful Hamiltonian analysis of GR permits to show that the constrained structure of the theory renders the $\pi$ field non-propagating. However, this argument is not valid in the case of the 6D Cascading DGP model, since in the latter case the field $\pi$ is the trace of the 4D part of a 6D massless graviton and not the trace part of a 4D one.

\subsubsection{Geometrical interpretation of the critical tension}

We can now try to understand geometrically what is the role of the background tension concerning the dynamics of $\pi^{(4)}$ and the sign of its kinetic term, and in particular why a critical tension emerges at all. First of all, note that the 4D limit equation (\ref{critical tension equation}) for the $\pi^{(4)}$ field can be obtained directly from the the equations (\ref{think}) and (\ref{freedomfreedom}) if we neglect the term $\Msf \de_{\bar{\mathbf{n}}} \pi$ in (\ref{freedomfreedom}), so we can consider the following system of equations
\begin{align}
\label{thinkdeclimit}
2 \, M_5^3 \, \tan \bigg( \frac{\bla}{4 M_6^4} \bigg) \, \, \boxf \, \hdvfn\Big\rvert_{+} &= M_4^2 \, \boxf \, \pi^{(4)} - \frac{1}{3} \, \mathscr{T} \\[2mm]
\label{freedomfreedomdeclimit}
2 \, M_6^4 \, \boxf \, \hdvfn\Big\rvert_{+} &= \frac{3}{2} \, M_5^3 \, \boxf \, \hp\Big\rvert_{+}
\end{align}
as the 4D limit of the system (\ref{think})-(\ref{freedomfreedom}). Furthermore, it is convenient to express these equations in terms of objects which have a clear geometrical meaning also in the thin limit, and in particular it is useful to write the derivative part of the $\mu\nu$ components of the cod-2 junction conditions \eqref{thinkdeclimit} in terms of the mode $\dvf_0 = \dvf^z_{gi}(0)$ which describes the movement of the cod-2 brane in the bulk (remember that, because of the $Z_2$ symmetry inside the cod-1 brane, this movement can be only happen in the $z$ direction). The equations (\ref{thinkdeclimit})-(\ref{freedomfreedomdeclimit}) then read
\begin{align}
\label{thinkdeclimitgeom}
6 \, M_5^3 \, \sin \bigg( \frac{\bla}{4 M_6^4} \bigg) \, \, \boxf \, \dvf_0 - 3 M_4^2 \, \boxf \, \pi^{(4)} &= - \mathscr{T} \\[2mm]
\label{freedomfreedomdeclimitgeom}
2 \, M_6^4 \, \boxf \, \hdvfn\Big\rvert_{+} &= \frac{3}{2} \, M_5^3 \, \boxf \, \hp\Big\rvert_{+} \quad ,
\end{align}
and need to be completed with the continuity conditions
\begin{align}
\label{bendinglink}
\hdvfn\Big\rvert_{+} &= \cos \bigg( \frac{\bla}{4 \Msf} \bigg) \, \dvf_0 \\[2mm]
\label{pilink}
\boxf \, \hp\Big\rvert_{+} &= \boxf \, \pi^{(4)} \quad ,
\end{align}
where the equation (\ref{bendinglink}) expresses the fact that, since the components of the embedding function $\dvf^A$ are continuous (the brane ``does not break''), the movement of the cod-2 brane and the movement of the cod-1 brane near the cod-2 brane are linked. However, since $\hdvfn\big\rvert_{+}$ is constructed from the cod-1 embedding by projecting on the normal vector, and the background normal vector depends on the background tension, $\hdvfn\big\rvert_{+}$ and $\dvf_0$ are linked in a $\bla$-dependent way.

We can then interpret the system of equations (\ref{thinkdeclimitgeom})-(\ref{pilink}) in the following way. The equation (\ref{thinkdeclimitgeom}) tells us that the presence of matter on the cod-2 brane (represented by $\mscrT$) has two effects: on one hand, it excites the metric perturbations on the cod-2 brane (represented by $\pi^{(4)}$) via the 4D induced gravity term, and in a ghostly way (exactly as in GR, however remember the observation at page \pageref{observation}). On the other hand, since the 4D brane is actually part of a 6D set-up and in fact embedded into a 5D cod-1 brane, $\mscrT$ excites also the movement of the cod-2 brane in the bulk (represented by $\dvf_0$), this time in a healthy way. However, it does so in a $\bla$-dependent way, and this excitation mechanism is the more efficient the larger the background tension, while it is completely inefficient when $\bla$ is very small. As we already mentioned, the equation (\ref{bendinglink}) instead tells us that, since the cod-2 brane is embedded inside the cod-1 brane, the movement of the cod-2 brane ``drags'' the cod-1 brane as well; therefore matter on the cod-2 brane indirectly excites $\hdvfn\big\rvert_{+}$. Passing from $\dvf_{0}$ to $\hdvfn\big\rvert_{+}$ we gain an additional $\bla$-dependence, but the sign does not change and so $\mscrT$ excites $\hdvfn\big\rvert_{+}$ in a healthy way. In turn, considering now the equation (\ref{freedomfreedomdeclimitgeom}), $\hdvfn\big\rvert_{+}$ excites the metric perturbations (expressed by the field $\hp$) on the cod-1 brane via the 5D induced gravity term, still in a healthy way; by continuity of the $\hp$ field (equation (\ref{pilink})), we get finally that $\hdvfn\big\rvert_{+}$ excites the metric perturbations on the cod-2 brane $\pi^{(4)}$, in a healthy way.

To sum up, the presence of matter on the cod-2 brane excites the field $\pi^{(4)}$ via two separate channels: it does so directly, because of the 4D induced gravity term, and indirectly via the bending of the cod-1 brane, because of the 5D induced gravity term. Furthermore, we saw above that the first channel excites $\pi^{(4)}$ in a ghostly and $\bla$-independent way, while the second channel excites $\pi^{(4)}$ in a healthy and $\bla$-dependent way. The fact that the field $\pi^{(4)}$ in total is a ghost or not is decided by the fact that the first or the second channel is more efficient than the other. In particular, the existence of the critical tension is due to the competition between these two channels, and its value corresponds to the tension where the two channels are equally efficient. Note finally that the existence of the second channel is entirely due to the higher dimensional structure of the theory. This is seen in the ``action'' approach as the fact that the healthy part of the effective 4D kinetic term (which cures the presence of the ghost for $\bla > \blac$) is created by integrating out the other fields in the 6D and 5D parts of the total action.

\subsection{Discussion}
\label{Discussion}

We want now to understand why the value for the critical tension that we find
\beq
\label{critical tension discussion}
\bla_c = 4 M_6^4 \, \arctan \bigg( \frac{2}{3} \, \frac{m_6}{m_5} \bigg)
\eeq
is different from the the value found in \cite{deRham:2010rw}
\beq
\label{critical tension theirs discussion}
\blac^{\textup{dRKT}} \equiv \frac{2}{3} \, m_6^2 \, \Mfs \quad ,
\eeq
and discuss the consequences of this difference for the viability of the theory.

\subsubsection{Ghost-free regions in the parameter space}

As we already mentioned, the action of the nested brane realization of the 6D Cascading DGP scenario contains two free parameters: $M_6$ and $M_5$, or equivalently $m_6$ and $m_5$. From a phenomenological point of view, it is important to establish which are the constraints on the free parameters coming from the request that there exist an interval of values for the background tension such that there are no ghosts among the perturbations around the pure tension background solutions.

This request is non-trivial because there is an upper limit $\bla_M = 2 \pi \Msf$ for the value of the background tension that we can place on the cod-2 brane, which corresponds to the situation where the deficit angle is $2 \pi$ and the background geometry is pathologic. If the critical tension is bigger than $\bla_M$, then every physically acceptable pure tension configurations is plagued by ghosts (at least at first order in perturbations and in the 4D limit). From the equation (\ref{critical tension theirs discussion}) it is easy to see that the result of \cite{deRham:2010rw} implies
\beq
\frac{\blac^{\textup{dRKT}}}{\bla_M} = \frac{1}{3 \pi} \, \frac{m_6}{m_5} \quad ,
\eeq
and so only if $m_6 < 3 \pi \, m_5$ the critical tension is smaller than the maximum tension: therefore, this implies that the cases $m_6 > 3 \pi \, m_5$ are phenomenologically ruled out. On the other hand, our result (\ref{critical tension discussion}) implies
\beq
\frac{\blac}{\bla_M} = \frac{2}{\pi} \, \arctan \bigg( \frac{2}{3} \, \frac{m_6}{m_5} \bigg) \quad ,
\eeq
which is always smaller than one; this implies that also the theories with $m_6 > m_5$ are phenomenologically viable, and therefore our findings are in stark contrast with the suggestion of \cite{deRham:2007xp} that in order to avoid the ghost the behavior of gravity should cascade from 6D down to 4D ``step by step'' (which happens in the case $m_6 < m_5$). It is interesting to note that, for $m_6 \ll m_5$, our result reads
\beq
\bla_c \simeq 4 M_6^4 \, \bigg( \frac{2}{3} \, \frac{m_6}{m_5} \bigg) = \frac{8}{3} \, m_6^2 \, M_4^2
\eeq
which is actually equal to the result of \cite{deRham:2010rw} apart from a numerical factor of 4: the qualitative difference between our result and the one of \cite{deRham:2010rw} becomes really relevant only when $m_6 \gtrsim m_5$.

\subsubsection{The critical tension and the pillbox integration}

Since the difference between the results (\ref{critical tension discussion}) and (\ref{critical tension theirs discussion}) is highly relevant concerning the phenomenological viability of the theory, it is important to understand where the difference comes from. Remember that to obtain the value of the critical tension we used the $\xi\xi$ component of the cod-1 junction conditions, and the derivative part of the $\m\n$ components of the cod-2 junction conditions. In particular, indicating explicitly the dependence on the index $n$ which realizes the thin limit, the latter is obtained by performing the following pillbox integration
\beq
\label{Palinka}
\lim_{n \rightarrow + \infty} M_5^3 \int_{-}^{+} \! d \hxi \,\, \bvf_{i\,[n]}^{\p \p} \, \boxf \, \hdvf^{i\,[n]}_{gi} = M_4^2 \, \boxf \, \pi^{(4)} - \frac{1}{3} \, \mathscr{T} \quad .
\eeq
In our approach (which we will call route A), to perform the limit of the integral in the left hand side of the previous equation\footnote{We take the $\boxf$ out of the integral, since the functions are smooth in the 4D directions also in the thin limit.}
\beq
\label{Vittoria}
\mscr{I} = \lim_{n \rightarrow + \infty} \int_{-}^{+} \! d \hxi \,\, \bvf_{i\,[n]}^{\p \p} \, \hdvf^{i\,[n]}_{gi}
\eeq
we used the exact relation
\beq
\bvf_{i\,[n]}^{\p \p} \,  \hdvf^{i\,[n]}_{gi} = \d \! \h{\vf}_{\shortparallel}^{[n] \, \p} - \d \h{v}^{[n]}_{\shortparallel}
\eeq
and the fact that (at least in cod-1 GNC) $\d \h{v}^{[n]}_{\shortparallel}$ does not diverge in the thin limit, so its pillbox integration tends to zero when $n \rightarrow + \infty$. Furthermore, the integral of $\d \! \h{\vf}_{\shortparallel}^{[n] \, \p}$ gives $\d \! \h{\vf}_{\shortparallel}^{[n]}\big\rvert_+$, and the continuity of the embedding function, together with the $Z_2$ symmetry inside the cod-1 brane, implies that in the thin limit the latter can be related to $\d \! \h{\vf}_{\perp}\big\rvert_+$, to obtain 
\beq
\label{RiminiRimini1}
\lim_{n \rightarrow + \infty} M_5^3 \int_{-}^{+} \! d \hxi \,\, \bvf_{i\,[n]}^{\p \p} \, \boxf \, \hdvf^{i\,[n]}_{gi} = 2 M_5^3 \, \tan \bigg( \frac{\bla}{4 M_6^4} \bigg) \, \boxf \, \hdvfn \Big\rvert_+ \quad .
\eeq

However, we may take a different route (which we will call route B): in fact, using
\beq
\bvf_{i\,[n]}^{\p\p} = \frac{\Zpp_{[n]}}{\sqrt{1 - \Zpq_{[n]}}} \,\, \bar{n}_{i \,[n]} \quad ,
\eeq
we can write the integrand in tems of the normal component of the bending 
\beq
\bvf_{i\,[n]}^{\p \p} \, \hdvf^{i\,[n]}_{gi} = \frac{\Zpp_{[n]}}{\sqrt{1 - \Zpq_{[n]}}} \,\, \hdvfn^{[n]} \quad .
\eeq
Furthermore, the background junction condition (\ref{Botswana}) implies that $\Zpp_{[n]}/\sqrt{1 - \Zpq_{[n]}}$ is proportional to a realization of the Dirac delta
\beq
\frac{\Zpp_{[n]}}{\sqrt{1 - \Zpq_{[n]}}} = \frac{\bla}{2 M_6^4} \, f_{[n]}(\hxi) \quad ,
\eeq
and so we may be tempted to use the defining property of the Dirac delta
\beq
\label{classicdeltarep}
\lim_{n \rightarrow + \infty} \, \int_{-}^{+} \! d \hxi \,\, f_{[n]}(\hxi) \, \mcal{F}(\hxi) = \mcal{F}(0)
\eeq
to evaluate the integral $\mscr{I}$ as follows
\beq
\mscr{I} = \frac{\bla}{2 M_6^4} \, \lim_{n \rightarrow + \infty} \int_{-}^{+} \! d \hxi \,\, f_{[n]}(\hxi) \,\, \hdvfn^{[n]} = \frac{\bla}{2 M_6^4} \, \hdvfn\Big\rvert_+
\eeq
and obtain

\beq
\label{RiminiRimini2}
\lim_{n \rightarrow + \infty} M_5^3 \int_{-}^{+} \! d \hxi \,\, \bvf_{i\,[n]}^{\p \p} \, \boxf \, \hdvf^{i\,[n]}_{gi} = \frac{\bla}{2 m_6} \, \boxf \, \hdvfn\Big\rvert_+ \quad .
\eeq

Comparing (\ref{RiminiRimini1}) to (\ref{RiminiRimini2}), it is evident that, using the route B to perform the integral $\mscrI$ instead of the route A, the term $\tan \big( \bla/4 M_6^4 \big)$ is substituted by $\bla/4 M_6^4$. If we indeed use the route B to obtain the derivative part of the $\m\n$ components of cod-2 junction conditions, and perform an analysis analogue to the one which from (\ref{think}) and (\ref{freedomfreedom}) leads to the evaluation of the critical tension (\ref{critical tension}), we obtain
\beq
\bla_c = \frac{8}{3} \, m_6^2 \, M_4^2 \quad ,
\eeq
which is exactly the result of \cite{deRham:2010rw} apart from the multiplicative numerical factor of 4. Therefore, we propose that, concerning the critical tension, the difference between our result and the result of \cite{deRham:2010rw} lies in the way the pillbox integration across the cod-2 brane is executed, and more in general in how the singular structure of the perturbation fields at the cod-2 brane is taken care of. More precisely, to reproduce the result of \cite{deRham:2010rw} in our framework we need to assume that the normal component of the bending $\hdvfn^{[n]}$ converges to a continuous function, while to obtain our result we need to assume that the components of the perturbation of the bending $\hdvf^{i \, [n]}_{gi}$ converge to continuous functions; as we show below, these two conditions cannot be both satisfied at the same time.

\subsection{Numerical check}
\label{Numerical check}

To see clearly which of the two ways of performing the pillbox integration is correct, the most straightforward way is to consider a case in which the solution is known exactly, and perform the integration numerically. This is indeed possible in our case since the exact solution is known in the case of pure tension perturbations; we perform below this numerical check. However, before discussing the numerical integration, it is worthwhile to try to understand where the problem of the route B may originate.

\subsubsection{Subtleties in the pillbox integration}

The central point in route B derivation of the pillbox integration is the use of the property of the Dirac delta (\ref{classicdeltarep}) where the function $\mcal{F}(\hxi)$ is, in our specific case,
\beq
\mcal{F}(\hxi) = \hdvfn^{[n]} \quad .
\eeq
The use of the formula (\ref{classicdeltarep}) with this identification of $\mcal{F}$ involves a subtlety, since $\hdvfn^{[n]}$ is a sequence of functions; in fact, the formula (\ref{classicdeltarep}) which holds for a realization $f_{[n]}$ of the Dirac delta is true only is $\mcal{F}$ is a continuous function which is independent of $n$. The idea behind the formula (\ref{classicdeltarep}) is that, since $f_{[n]}$ is peaked around $\hxi = 0$, it probes the function $\mcal{F}$ only around $\hxi = 0$; if $\mcal{F}$ is continuous, in the $n \rightarrow + \infty$ limit it can be considered nearly constant in the $\hxi$-interval where $f_{[n]}$ is peaked, and so it can be taken out of the integral. If $\mcal{F}$ is a sequence of functions $\mcal{F}_{[n]}$, it may develop a non-trivial behavior (for example, a peak) around $\hxi = 0$ in the $n \rightarrow + \infty$ limit, as much as $f_{[n]}$ does: in this case, by no means it can be considered constant and taken out of the integral, since its singular behavior will contribute in a non-trivial way to the integral even in the $n \rightarrow + \infty$ limit. On the other hand, if the sequence of functions $\mcal{F}_{[n]}$ converge uniformly to a continuous function $\mcal{F}_\infty$, the formula (\ref{classicdeltarep}) holds anyway provided we substitute $\mcal{F}(0)$ with $\mcal{F}_{\infty}(0)$ in the right hand side, since in this case the behavior of $\mcal{F}_{[n]}$ is under control (see appendix \ref{Pillbox integration of nested branes}).

From this point of view, the crucial observation is that the function $\hdvfn^{[n]}$ converges to a \emph{discontinuous} function; this implies that, since $\hdvfn^{[n]}$ is smooth for every finite value of $n$, it cannot converge uniformly to its limiting function. Remember in fact the definition of $\hdvfn^{[n]}$
\beq
\hdvfn^{[n]} = \Yp_{[n]} \, \hdvf^{z \, [n]}_{gi} - \Zp_{[n]} \, \hdvf^{y \, [n]}_{gi} \quad ,
\eeq
and remember from the section \ref{Thin limit of the background} that $\Zp_{[n]}$ converges to the function
\beq
\Zp_{[n]} \xrightarrow[n \rightarrow +\infty]{}
\begin{cases}
\sin \Big( \bla/4 \Msf \Big) & \text{for $\hxi > 0$} \\
0 & \text{for $\hxi = 0$} \\
- \sin \Big( \bla/4 \Msf \Big) & \text{for $\hxi < 0$} \quad ,
\end{cases}
\eeq
while $\Yp_{[n]}$ converges to the function
\beq
\Yp_{[n]} \xrightarrow[n \rightarrow +\infty]{}
\begin{cases}
\cos \Big( \bla/4 \Msf \Big) & \text{for $\hxi \neq 0$} \\
1 & \text{for $\hxi = 0$} \quad .
\end{cases}
\eeq
Since the thin limit is well defined, the functions $\hdvf^{z \, [n]}_{gi}$ and $\hdvf^{y \, [n]}_{gi}$ converge to continuous functions $\hdvf^z_{\infty}$ and $\hdvf^y_{\infty}$ which, because of the $Z_2$ symmetry that holds inside the cod-1 brane, are respectively even ($\hdvf^z_{\infty}$) and odd ($\hdvf^y_{\infty}$). This implies that, indicating $\hdvf_0 = \hdvf^z_{\infty}(0)$, we have
\beq
\label{entusiasmo}
\lim_{\hxi \rightarrow 0} \, \hdvfn^{\infty} (\hxi) = \cos \bigg( \frac{\bla}{4 \Msf} \bigg) \, \hdvf_0 \neq \hdvf_0 = \hdvfn^{\infty} (0) \quad ,
\eeq
and this proves that $\hdvfn^{\infty}$ indeed is discontinuous. The non-trivial (peaked) behavior of $\hdvfn^{[n]}$ in the neighborhood of $\hxi = 0$ is confirmed by the numerical plot (figure \ref{normalbending}) obtained in the case of a pure tension perturbation. This is the reason for the mismatch between the predicted value and the output of the numerical integration of the integral $\mscr{I}$ which we discuss below. \label{saggezza}

\subsubsection{The pure tension perturbation case}
\label{The pure tension perturbation case}

We now turn to the case of a pure tension perturbation, where the localized cod-1 energy-momentum tensor reads
\beq
\label{Gigio pure tension}
\hat{T}^{(loc)}_{\m\n \, [n]}(\hxi, \chd) = - f_{[n]}(\hxi) \, \big( \bla + \dla \big) \,\h{g}^{[n]}_{\m\n}(\hxi, \chd) \quad .
\eeq
Analogously to the analysis performed in section \ref{The pure tension case}, we know that the exact solution is of the form
\begin{align}
g^{[n]}_{AB} &= \e_{AB} & \h{g}^{[n]}_{ab} &= \e_{ab} & g^{(4) \, [n]}_{\m\n} &= \e_{\m\n}
\end{align}
\beq
\vf^A(\hxi) = \Big( \mZ_{[n]}(\hxi), \mY_{[n]}(\hxi), 0,0,0,0 \Big) \quad ,
\eeq
since in this case the 4D Einstein tensor vanishes and so the 4D induced gravity term does not play a role. The $z$ and $y$ components of the embedding function can be expressed exactly in terms of the regulating function $\ep_{[n]} (\hxi)$
\beq
\ep_{[n]} (\hxi) \equiv \int_{0}^{\hxi} \! d\zeta \, f_{[n]} (\zeta) \quad ,
\eeq
which is a (even) realization\footnote{For the sake of precision we have $\ep_{[n]}\big\rvert_{\pm} = \pm 1/2$.} of the Heavyside theta function: in fact integrating the junction conditions we get
\beq
\mscr{Z}_{[n]}^{\prime}(\hxi) = \sin \bigg( \frac{\bla + \dla}{2 M_6^4} \, \ep_{[n]} (\hxi) \bigg) \quad ,
\eeq
and at first order in $\dla$ we have
\begin{align}
\mscr{Z}_{[n]}^{\prime}(\hxi) &\simeq \sin \bigg( \frac{\bla}{2 M_6^4} \, \ep_{[n]} (\hxi) \bigg) + \cos \bigg( \frac{\bla}{2 M_6^4} \, \ep_{[n]} (\hxi) \bigg) \, \frac{\dla}{2 M_6^4} \, \ep_{[n]} (\hxi) \\[2mm]
\mscr{Y}_{[n]}^{\prime}(\hxi) &\simeq \cos \bigg( \frac{\bla}{2 M_6^4} \, \ep_{[n]} (\hxi) \bigg) - \sin \bigg( \frac{\bla}{2 M_6^4} \, \ep_{[n]} (\hxi) \bigg)\, \frac{\dla}{2 M_6^4} \, \ep_{[n]} (\hxi) \quad .
\end{align}
On the other hand, the $z$ and $y$ components of the background embedding read
\begin{align}
Z_{[n]}^{\prime}(\hxi) &= \sin \bigg( \frac{\bla}{2 M_6^4} \, \ep_{[n]} (\hxi) \bigg) \\[2mm]
Y_{[n]}^{\prime}(\hxi) &= \cos \bigg( \frac{\bla}{2 M_6^4} \, \ep_{[n]} (\hxi) \bigg) \quad ,
\end{align}
so we can obtain explicitly the perturbations of the $z$ and $y$ components of the embedding $\hdvf_{[n]}^{z} = \mscr{Z}_{[n]} - Z_{[n]}$ and $\hdvf_{[n]}^{y} = \mscr{Y}_{[n]} - Y_{[n]}$
\beq
\hdvf_{[n]}^{z}(\hxi) \simeq \hdvf^{[n]}_0 + \frac{\dla}{2 M_6^4} \, \int_{0}^{\hxi} \! d \z \, \, \ep_{[n]} (\z) \, \cos \bigg( \frac{\bla}{2 M_6^4} \, \ep_{[n]} (\z) \bigg)
\eeq
and
\beq
\hdvf^{y}_{[n]}(\hxi) \simeq - \frac{\dla}{2 M_6^4} \, \int_{0}^{\hxi} \! d \z \, \, \ep_{[n]} (\z) \, \sin \bigg( \frac{\bla}{2 M_6^4} \, \ep_{[n]} (\z) \bigg)
\eeq
where we defined $\hdvf^{[n]}_0 \equiv \hdvf_{[n]}^{z}(0)$. Finally, we can construct the normal component of the bending $\hdvfn^{[n]} = \Yp_{[n]} \, \hdvf_{[n]}^{z} - \Zp_{[n]} \, \hdvf_{[n]}^{y}$ to get
\begin{align}
\label{heyhey}
\hdvfn^{[n]}(\hxi) &\simeq \cos \bigg( \frac{\bla}{2 M_6^4} \, \ep_{[n]} (\hxi) \bigg) \, \hdvf^{[n]}_0 + \frac{\dla}{2 M_6^4} \, \cos \bigg( \frac{\bla}{2 M_6^4} \, \ep_{[n]} (\hxi) \bigg) \, \int_{0}^{\hxi} \! d \z \, \, \ep_{[n]} (\z) \, \cos \bigg( \frac{\bla}{2 M_6^4} \, \ep_{[n]} (\z) \bigg) + \nn \\[2mm]
& + \frac{\dla}{2 M_6^4} \, \sin \bigg( \frac{\bla}{2 M_6^4} \, \ep_{[n]} (\hxi) \bigg) \, \int_{0}^{\hxi} \! d \z \, \, \ep_{[n]} (\z) \, \sin \bigg( \frac{\bla}{2 M_6^4} \, \ep_{[n]} (\z) \bigg) \quad ,
\end{align}
and in particular its value on the side of the cod-2 brane reads
\begin{align}
\label{enoughisenough}
\hdvfn^{[n]}\Big\rvert_{+} &\simeq \cos \bigg( \frac{\bla}{4 M_6^4} \bigg) \, \hdvf^{[n]}_0 + \frac{\dla}{2 M_6^4} \, \cos \bigg( \frac{\bla}{4 M_6^4} \bigg) \, \int_{0}^{+} \! d \z \, \, \ep_{[n]} (\z) \, \cos \bigg( \frac{\bla}{2 M_6^4} \, \ep_{[n]} (\z) \bigg) + \nn \\[2mm]
& + \frac{\dla}{2 M_6^4} \, \sin \bigg( \frac{\bla}{4 M_6^4} \bigg) \, \int_{0}^{+} \! d \z \, \, \ep_{[n]} (\z) \, \sin \bigg( \frac{\bla}{2 M_6^4} \, \ep_{[n]} (\z) \bigg) \quad .
\end{align}

Note that we have expressed the $z$, $y$ and normal components of the bending perturbation in terms of three quantities: $\dla$, $\ep_{[n]}(\hxi)$ and $\hdvf^{[n]}_0$. The first one fixes the amplitude of the tension perturbation, and is therefore a free parameter apart from the fact that it has to satisfy the condition $\dla / \bla \ll 1$. The regulating function $\ep_{[n]}(\hxi)$, instead, expresses the details of the internal structure of the cod-2 brane and is therefore fixed once we choose the system we are working with. For the purpose of checking numerically the validity of route A and B, it is enough to choose a particular realization of $\ep_{[n]}$ and $f_{[n]}$: we use the following realization of the Dirac delta
\beq
\label{fnexplicit}
f_{[n]}(\hxi) = 
\begin{cases}
\dfrac{n}{2 \pi} \, \Big( 1 + \cos\big( n \, \hxi \big) \Big) & \text{for $\abs{\hxi} \leq \dfrac{\pi}{n}$}
\vspace{3mm} \\
0 & \text{for $\abs{\hxi} > \dfrac{\pi}{n}$}
\end{cases}
\eeq
and the associated regulating function
\beq
\label{epnexplicit}
\ep_{[n]}(\hxi) =
\begin{cases}
\dfrac{1}{2 \pi} \, \Big( n \, \hxi + \sin \big( n \, \hxi \big) \Big) & \text{for $\abs{\hxi} \leq \dfrac{\pi}{n}$} \vspace{3mm} \\
\pm \dfrac{1}{2} & \text{for $\hxi \gtrless \pm \, \dfrac{\pi}{n}$} \quad ,
\end{cases}
\eeq
\begin{figure}[htp!]
\begin{center}
\includegraphics{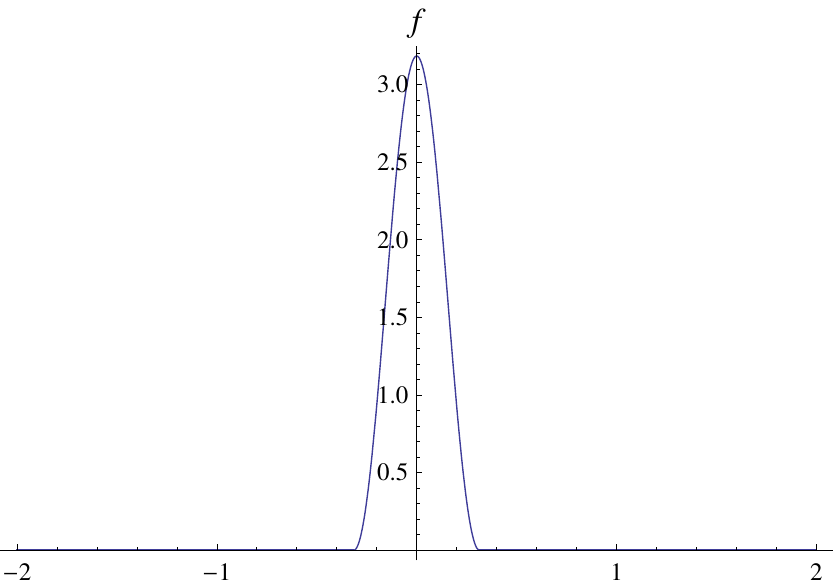}
\caption[The realization $f$ of the Dirac delta]{The realization $f$ of the Dirac delta}
\label{fdelta}
\end{center}
\end{figure}
\begin{figure}[htp!]
\begin{center}
\includegraphics{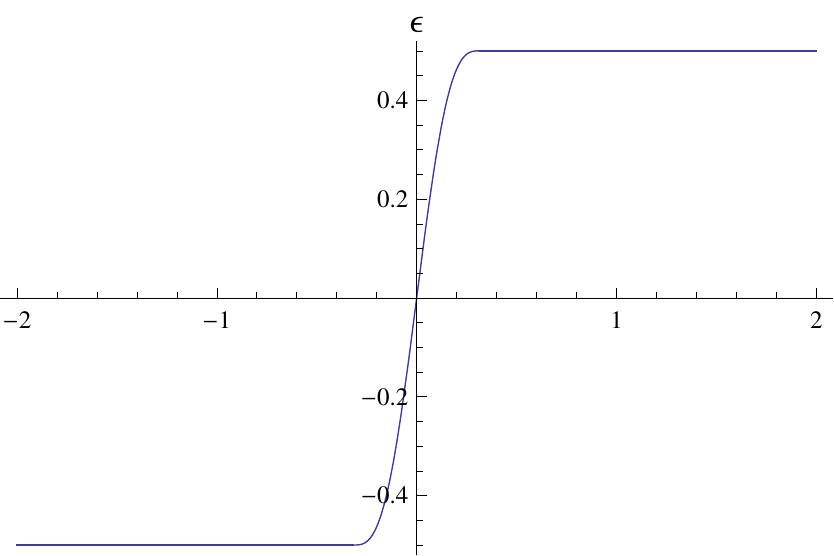}
\caption[The regulating function $\ep$]{The regulating function $\ep$}
\label{epsilon}
\end{center}
\end{figure}
whose plots for $n = 10$ are shown respectively in figure \ref{fdelta} and in figure \ref{epsilon}. Note that in this case the thickness of the (physical) cod-2 brane is $l_2 = \pi/n$, and indeed the thin limit $l_2 \rightarrow 0^+$ mathematically corresponds to the limit $n \rightarrow + \infty$. It is worthwhile to point out that the explicit form (\ref{fnexplicit}) for the function $f_{[n]}$ is of class $\mscr{C}^1$ on all the real axis, but its second derivative does not exist in $\hxi = \pm \pi/n$; however, this is not a problem for what concerns the numerical check since the latter does not involve the derivation of the function $f_{[n]}$ but only its integration.

The 4D field $\hdvf^{[n]}_0$, instead, is in general a dependent quantity, in the sense that its value is determined by the equations of motion once we specify the source configuration. However, in the pure tension case, the equations of motion do not fix its value since a rigid translation of the cod-1 and cod-2 branes is a symmetry of the system: this is expressed by the fact that the left hand side of the equation (\ref{Palinka}) vanishes identically (and of course the right hand side as well), since the integral $\mscr{I}$ is independent from the 4D coordinates, and therefore $\boxf \mscr{I}$ vanishes. However, our aim here is to understand which route (A or B, or none of the two) to evaluate the integral $\mscr{I}$ is correct, independently of the fact that the integral itself does or does not contribute to the equations of motion; therefore, in the particular case we are considering, $\hdvf^{[n]}_0$ can be considered a free parameter as well.

\subsubsection{Numerical pillbox integration}

Having fixed the details of the internal structure of the cod-2 brane ($f_{[n]}$ and $\ep_{[n]}$, equations (\ref{fnexplicit}) and (\ref{epnexplicit})), and obtained the explicit relations for the normal component of the bending (equations (\ref{heyhey}) and (\ref{enoughisenough})), we can numerically compute the integral
\beq
\label{Alkistis}
\mscr{I}_{[n]} = \int_{-}^{+} \! d \hxi \,\, \bvf_{i\,[n]}^{\p \p} \, \hdvf^{i\,[n]}_{gi} = \frac{\bla}{2 M_6^4} \int_{-}^{+} \! d \hxi \,\, f_{[n]}(\hxi) \,\, \hdvfn^{[n]} \quad ,
\eeq
whose limit for $n \rightarrow + \infty$ is the integral $\mscr{I}$ defined in (\ref{Vittoria}). Before doing that, it is worthwhile to plot numerically the function $\hdvfn^{[n]}$ to see explicitly that indeed it has a non-trivial behavior around $\hxi = 0$. For definiteness, we can choose the background tension and the tension perturbation to be
\begin{align}
\label{FraMarche}
\bla &= \frac{3}{4} \, \bla_M & \frac{\dla}{2 \Msf} &= 0.1 \quad ,
\end{align}
which is consistent with the hypothesis that the tension perturbation is small since with this choice $\dla/\bla \simeq 0.04$; furthermore, for the free parameter $\hdvf^{[n]}_0$ we choose the value $\hdvf^{[n]}_0 = 5$.
\begin{figure}[htp!]
\begin{center}
\includegraphics{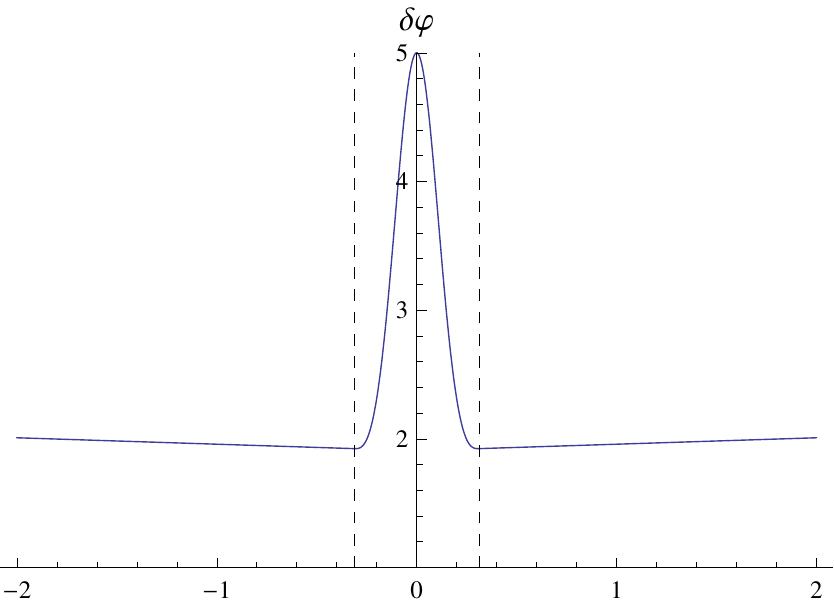}
\caption[The normal component of the bending]{The normal component of the bending $\hdvfn^{[n]}$ in the case of a pure tension perturbation}
\label{normalbending}
\end{center}
\end{figure}
The plot of the normal component of the bending perturbation $\hdvfn^{[n]}(\hxi)$ for $n = 10$ is shown in figure \ref{normalbending}, together with the boundaries of the physical cod-2 brane (represented by the vertical dashed lines): it is evident that indeed $\hdvfn^{[n]}(\hxi)$ is peaked around $\hxi = 0$, consistently with the discussion of page \pageref{saggezza}. Moreover, the non-trivial behavior near $\hxi = 0$ is concentrated only inside the cod-2 brane: the value of the field $\hdvfn^{[n]}$ on the side of the cod-2 brane ($\hdvfn^{[n]}\big\rvert_{+}$) is in fact very different from its value at $\hxi = 0$. This feature does not disappear if we send $n$ to infinity, but remains true for $n$ arbitrarily large: the width of the peak tends to zero, but the difference between $\hdvfn^{[n]}\big\rvert_{+}$ and $\hdvfn^{[n]}(0)$ remains finite. Therefore, the numerical plots confirm the fact that $\hdvfn^{[n]}(\hxi)$ converges to a discontinuous function, as expressed by the equation (\ref{entusiasmo}).

\begin{figure}[htp!]
\begin{center}
\includegraphics{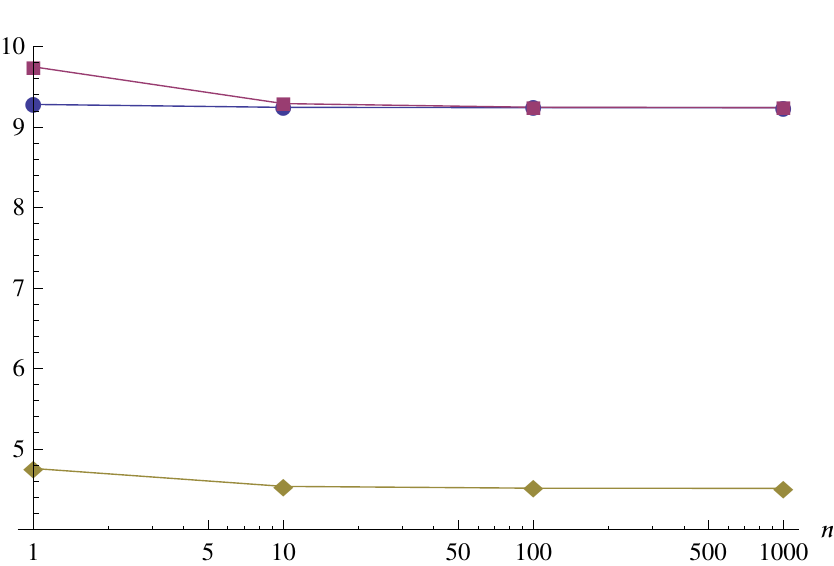}
\caption[Numerical results of the pillbox integration]{Plot of the numerical results of the pillbox integration}
\label{PillboxIntegrationfigure}
\end{center}
\end{figure}

Coming now to the integral $\mscr{I}$, remember that performing the pillbox integration following the route A we obtain
\beq
\label{routeA}
\lim_{n \rightarrow + \infty} \mscr{I}_{[n]} = \lim_{n \rightarrow + \infty} \mscr{A}_{[n]} \quad ,
\eeq
while following the route B we obtain
\beq
\label{routeB}
\lim_{n \rightarrow + \infty} \mscr{I}_{[n]} = \lim_{n \rightarrow + \infty} \mscr{B}_{[n]} \quad ,
\eeq
where we defined
\begin{align}
\mscr{A}_{[n]} &= 2 \, \tan \bigg( \frac{\bla}{4 M_6^4} \bigg) \, \hdvfn^{[n]}\Big\rvert_+ \\[2mm]
\mscr{B}_{[n]} &= \frac{\bla}{2 M_6^4} \,\, \hdvfn^{[n]}\Big\rvert_+ \quad .
\end{align}
To test the validity of the routes A and B, we want to compute independently the values of $\mscr{I}_{[n]}$, $\mscr{A}_{[n]}$ and $\mscr{B}_{[n]}$ by numerical means for several values of $n$, and then see if $\mscr{A}_{[n]}$ or $\mscr{B}_{[n]}$ (or none of the two) converges to $\mscr{I}_{[n]}$ for $n$ large. Note that $\mscr{A}_{[n]}$ and $\mscr{B}_{[n]}$ are approximately equal for $\bla/4 M_6^4 \ll 1$, while they differ significantly when $\bla/4 M_6^4 \sim 1$; therefore, for the purpose of deciding which route is correct, it is useful to choose the background tension close to the maximum tension. Therefore, we stick to the choice (\ref{FraMarche}) for $\bla$ and $\dla$, and also to the choice $\hdvf^{[n]}_0 = 5$. The results of the numerical integration are given in table \ref{pillboxintegrationtable} with $5$ significant digits, and for clarity the same results are plotted in figure \ref{PillboxIntegrationfigure} (note that the plot is semi-logaritmic). It is evident that the points corresponding to $\mscr{A}_{[n]}$ (squares) converge to the points corresponding to $\mscr{I}_{[n]}$ (circles), while the points corresponding to $\mscr{B}_{[n]}$ (diamonds) are significantly distant from the former ones.
\begin{table}[htb!]
\centering
\[
\begin{array}{ccccc}
\toprule
n & 1 & 10 & 10^2 & 10^3 \\
\midrule
\mscr{I}_{[n]} & 9.2794 & 9.2429 & 9.2392 & 9.2388 \\
\mscr{A}_{[n]} & 9.7466 & 9.2896 & 9.2439 & 9.2393 \\
\mscr{B}_{[n]} & 4.7562 & 4.5332 & 4.5109 & 4.5086 \\
\bottomrule
\end{array}
\]
\caption{Numerical results of the pillbox integration}
\label{pillboxintegrationtable}
\end{table}
This implies that the route B to perform the pillbox integration is in general wrong, while the route A (at least in the pure tension perturbation case) is correct; in particular, the pillbox integration performed following the route B gives a lower value compared to the pillbox integration performed following the route A because the route B completely misses the peak of $\hdvfn^{[n]}(\hxi)$ inside the cod-2 brane. These results strongly suggest that the route A is the correct way to perform the pillbox integration across the cod-2 brane in the general case.

The same conclusion can be reached in a slightly different way, by exploiting the fact that $\hdvf^{[n]}_0$ and $\dla$ are independent parameters. In fact, both $\mscr{I}_{[n]}$ and $\hdvfn^{[n]}\Big\rvert_+$ are the sum of a piece multiplied by $\hdvf^{[n]}_0$ (which we call the ``bending piece'') and a piece multiplied by $\dla$ (which we call the ``tension piece''); since these two parameters are independent, if one of the two equations (\ref{routeA}) and (\ref{routeB}) is true then it has to be true also separately for the bending piece and for the tension piece. Note that the bending piece of $\hdvfn^{[n]}\Big\rvert_+$ reads
\beq
\label{bendingN}
\text{bending} \bigg[ \hdvfn^{[n]}\Big\rvert_+ \bigg] = \cos \bigg( \frac{\bla}{4 M_6^4} \bigg) \quad ,
\eeq
while the bending piece of $\mscr{I}_{[n]}$ reads
\beq
\text{bending} \Big[ \mscr{I}_{[n]} \Big] = \frac{\bla}{2 M_6^4} \int_{-}^{+} \! d \hxi \,\, f_{[n]}(\hxi) \,\, \cos \bigg( \frac{\bla}{2 M_6^4} \, \ep_{[n]} (\hxi) \bigg) \quad :
\eeq
the latter integral can be performed exactly changing the integration variable to $\z = \ep_{[n]} (\hxi)$, to obtain
\beq
\text{bending} \Big[ \mscr{I}_{[n]} \Big] = 2 \, \sin \bigg( \frac{\bla}{4 M_6^4} \bigg) \quad .
\eeq
Remembering (\ref{bendingN}), the latter formula reproduces exactly the result of route A
\beq
\text{bending} \Big[ \mscr{I}_{[n]} \Big] = 2 \, \tan \bigg( \frac{\bla}{4 M_6^4} \bigg) \,\, \text{bending} \bigg[ \hdvfn^{[n]}\Big\rvert_+ \bigg] \quad .
\eeq

\subsubsection{Final remarks}

The numerical results obtained for the case of a pure tension perturbation put on firm footing our derivation of the thin limit equations for the nested branes with induced gravity set-up, and also our results for the nested brane realization of the 6D Cascading DGP scenario. In particular, it strongly supports our claim that the correct value of the critical tension is (\ref{critical tension discussion}), and that the models characterized by $m_6 > m_5$, where gravity cascades directly from 6D to 4D, are not phenomenologically ruled out, at least as far as we consider an analysis at first order in perturbations and in the 4D limit.

The subtlety of the pillbox integration, which we discussed, warns us that in this class of models we cannot perform the pillbox integration (in the action or in the equation of motion) by assuming that the singular structure of the perturbed configurations is encoded solely in the Dirac delta functions associated to the background. In fact, in the equations for the perturbations we get terms where the background delta functions are multiplied by the perturbation fields; due to the non-smooth structure of the perturbation fields at the cod-2 brane, they cannot be taken out of the integral but instead they contribute in a non-trivial way to the pillbox integration. Even if the thin limit of the cod-2 brane is well defined in the nested brane with induced gravity set-up, the presence of such subtleties in the pillbox integration can be seen as an indirect confirmation of the fact that the singular structure of branes of codimension higher than one is more complex than the singular structure of codimension-1 branes.

\clearemptydoublepage
\chapter{dRGT massive gravity}
\label{Ghost Free Massive Gravity}

We have seen in the previous chapters that a way to try to explain the apparent late time acceleration of the universe is to modify gravity in the infrared, \ie at large distances. In particular, we have seen that the DGP model provides an interesting way to do that, and in that model gravitational potentials behave like $1/r$ below a crossover scale $r_c$ and like $1/r^2$ above it. However, in particle physics it is not unusual to have a theory which behaves like $1/r$ below a scale and decays much faster above it: Yukawa long ago proposed a model, which ought to describe the pion, in which a scalar field has exactly this property. This is linked with the idea that the mass of a particle fixes the range of the interaction it mediates: massive particles mediate finite range forces, while massless particles mediate infinite range forces. Considering a scalar field, the relativistic field equation for a massless field is the D'Alembert equation
\beq
\Box \phi = T
\eeq
where $T$ is the source. Considering a static, spherically symmetric source, the solution outside the source is
\beq
\phi \propto \frac{1}{r} \quad .
\eeq
However, giving a mass to the particle one obtains the equation of motion
\beq
(\Box - m^2) \phi = T
\eeq
which is the Klein-Gordon equation, and admits a static, spherically symmetric vacuum solution
\beq
\phi \propto \frac{e^{-mr}}{r} \quad .
\eeq
This is known as the \emph{Yukawa potential}, and we can see that it behaves like $\sim 1/r$ for $r \ll r_c$ while it decays exponentially for $r \gg r_c$, where $r_c = 1/m$ is called the \emph{Compton radius}. We then see that the interaction mediated by a massive scalar field has a finite length, set by the Compton radius or equivalently by the inverse mass.

It is quite natural to wonder if we could use this simple idea to modify gravity in the infrared, ``giving a mass'' to the graviton. This relies on the fact that GR can be considered as a theory of a massless field: we will see in fact that GR can be thought as an interacting theory of a massless helicity-2 field, which is consistent with the fact that gravitational interaction in GR have infinite range. More precisely we could try to formulate an interacting theory of a massive spin-2 field, and set its Compton radius of the order of the Hubble radius today $r_c \sim H_0^{-1}$. The hope is that we could construct in this way a theory which accurately reproduces GR below $r_c$, while behaves differently above that radius. Having done that, we could investigate if this modified gravity theory is able to explain the late time acceleration as an effect of gravity behaving differently when the Hubble's radius becomes comparable to the Compton radius.

The idea of formulating a theory of a massive spin-2 field which reduces to GR below the Compton radius is actually quite old, and can be traced back to the works of Fierz and Pauli (FP) in 1939 \cite{FP}. They formulated a theory of a free massive spin-2 field, whose action reduces to the one of linearized GR in the $m \rightarrow 0$ limit. However, the program we sketched above proved to be very difficult to implement. On one hand, it was argued that any non-linear extension of the FP theory leads to the appearing of an additional ``sixth'' degree of freedom and the reintroduction of ghosts \cite{BD}, and therefore the is no sensible way to formulate an interacting theory of a massive spin-2 graviton (apart from considering Lorentz violating theories \cite{Rubakov:2004eb}). On the other hand, it was shown that at linear level the FP theory does not reproduce GR, even below the Compton radius \cite{Iwasaki70,VDV70,Zakharov70}. A possible way out of the latter problem has been suggested by Vainshtein \cite{Vainshtein72}, who proposed that non-linearities could be crucial in restoring the agreement with GR, a mechanism which is known as \emph{Vainshtein mechanism}. Recently, a class of non-linear completions of the Fierz-Pauli theory which are Lorentz invariant and propagate exactly five degrees of freedom has been proposed \cite{deRham:2010ik,deRham:2010kj}. Even before considering cosmological solutions, it is crucial to establish if this class of theories reproduces GR in a suitable range of length scales, and therefore if the Vainshtein mechanism is effective or not.

The main aim of the second part of this thesis is to investigate the effectiveness of the Vainshtein mechanism in the class of theories known as dRGT Massive Gravity \cite{deRham:2010ik,deRham:2010kj}. In this chapter we therefore introduce the theory in its generality, while in the next chapter we focus on static, spherically symmetric solutions and on the Vainshtein mechanism. This chapter is largely based on the recent review \cite{HinterbichlerReview}.

\section{GR as an interacting massless helicity-2 f\mbox{}ield}

Let's consider the action of GR
\beq
\label{GRaction}
S_{GR}[g_{\m\n}, \psi_{(i)}] = \frac{M^2_P}{2} \int \! d^4 x \, \sqrt{-g} \, R \, + S_{M}[g_{\m\n}, \psi_{(i)}]
\eeq
where the $\psi_{(i)}$ are matter fields while the matter action is
\beq
S_{M} = \int \! d^4 x \, \sqrt{-g} \, \mathscr{L}_M \quad .
\eeq
The energy momentum tensor is defined as
\beq
T_{\m\n} \equiv - \frac{2}{\sqrt{-g}} \frac{\d}{\d g^{\m\n}} S_M
\eeq
so the equations of motion are the Einstein equations
\beq
G_{\m\n} = \frac{1}{M_P^2} \, T_{\m\n}
\eeq
where $M_P^2 = 1/8 \pi G$.

\subsection{Linearized GR as a free massless helicity-2 f\mbox{}ield} 

Let's study perturbations around the Minkowski solution
\beq
\label{Minkowskiperturbationsdef}
g_{\m\n} = \eta_{\m\n} + h_{\m\n} \quad .
\eeq
The linearized equations of motion can be deduced by expanding the equations of motions, or equivalently by varying the quadratic part of the action obtained by the expanding (\ref{GRaction}) in terms of $h_{\m\n}$, which reads
\beq
\label{GRquadraticaction}
S^{(2)}_{GR} = \int \! d^4 x \, \frac{M^2_P}{2} \bigg( -\frac{1}{2}\partial_\lambda h_{\mu\nu}\partial^\lambda h^{\mu\nu}+\partial_\mu h_{\nu\lambda}\partial^\nu h^{\mu\lambda}-\partial_\mu h^{\mu\nu}\partial_\nu h+\frac{1}{2}\partial_\lambda h\partial^\lambda h \, \bigg) + h_{\mu\n} T^{\m\n}
\eeq
where indices has been raised using $\eta^{\m\n}$. To study the vacuum dynamics of perturbation from Minkowski spacetime, we can set to zero the energy momentum tensor in the action above: the vacuum equations of motion for $h_{\mu\nu}$ can be then deduced from the action
\beq
\label{GRquadraticactionvacuum}
S^{(2)} = \int \! d^4 x \, \bigg( -\frac{1}{2}\partial_\lambda h_{\mu\nu}\partial^\lambda h^{\mu\nu}+\partial_\mu h_{\nu\lambda}\partial^\nu h^{\mu\lambda}-\partial_\mu h^{\mu\nu}\partial_\nu h+\frac{1}{2}\partial_\lambda h\partial^\lambda h \, \bigg) \quad .
\eeq
We could pretend to forget for a moment where this action comes from, and just study its properties. In general, fields living in Minkowski spacetime can be categorized regarding their transformation properties with respect to Lorentz transformations: in particular, they can be decomposed in components of fixed mass and spin. It can be shown that action (\ref{GRquadraticactionvacuum}) describes exactly a massless helicity-2 field \cite{HinterbichlerReview}. As a consistency check, we can show that a field whose dynamic is described by (\ref{GRquadraticactionvacuum}) propagates two degrees of freedom (d.o.f.), as a massless helicity-2 field should. 

\subsubsection{Degrees of freedom counting} 

To count the number of degrees of freedom which the action (\ref{GRquadraticactionvacuum}) propagates, it is useful to recast the theory in Hamiltonian form. In this case, the dynamical variables are the field components $h_{\m\n}$, but it can be seen that it is impossible to perform the Legendre transform with respect to all of the velocities $\dot{h}_{\m\n}$ since $\dot{h}_{00}$ and $\dot{h}_{0i}$ appear linearly in the Lagrangian. However, since total derivatives in the Lagrangian don't have any effect on the physics of the system, it is possible to integrate by parts in the action: using this freedom, we end up with an action where $\dot{h}_{00}$ and $\dot{h}_{0i}$ do not appear, and instead $h_{00}$ and $h_{0i}$ appear linearly. We can do the Legendre transform of the new action with respect just to the spatial components $\dot{h}_{ij}$: the conjugate momenta are then \cite{HinterbichlerReview}
\beq
\pi_{ij}= \frac{\de \mcalL}{\de \dot{h}_{ij}}= \dot{h}_{ij} - \dot{h}_{kk} \delta_{ij} - \de_{(i}h_{j)0} + 2\de_k h_{0k} \delta_{ij} \quad ,
\eeq
and we can invert this relation to get
\beq
\dot{h}_{ij} = \pi_{ij} - \half \pi_{kk} \delta_{ij} + \de_{(i}h_{j)0} \quad .
\eeq
Note that, since we are splitting space and time, it makes sense to perform purely spatial transformations and so the Kronecker delta $\d_{ij}$ is indeed a tensor. Moreover, note that we are using the convention of implicit sum on repeated indices, but now the indices do not need to be ``up and down'', so for example $\dot{h}_{kk}$ means $\sum_{k=1}^3 \dot{h}_{kk}$. We can then write the Lagrangian as \cite{HinterbichlerReview}
\beq
\mcalL (h, \pi, h_{00}, h_{0i}) = \pi_{ij} \, \dot{h}_{ij} - \mcalH + 2 h_{0i} \big( \de_j \pi_{ij} \big) + h_{00} \big( \triangle h_{ii} - \de_i \de_j h_{ij} \big)
\eeq
where $\mcalH$ depends only on $h_{ij}$, $\pi_{ij}$ and their spatial derivatives. Note that $h_{00}$ and $h_{0i}$ indeed appear linearly, and they are multiplied by terms with no time derivatives: we can interpret $h_{00}$ and $h_{0i}$ as Lagrange multipliers which enforce the (primary) constraints
\beq
\de_j \pi_{ij} = 0 \qquad \qquad  \triangle h_{ii} - \de_i \de_j h_{ij} = 0 \quad ,
\eeq 
and so consider the system described by (\ref{GRquadraticactionvacuum}) as a constrained Hamiltonian system. It can be checked that the matrix whose elements are the Poisson brackets of the constraints between themselves is vanishing when the fields satisfy the constraints, so each of the four constraints generate a gauge transformation, and that the Poisson bracket of the constraints with the Hamiltonian vanishes, so the constraints are conserved by the time evolution. To count the number of degrees of freedom, the $h_{ij}$ and $\pi_{ij}$ are $3 \times 3$ symmetric matrices, so have 6 independent components each. Of these 12 degrees of freedom, 4 can be eliminated using the constraints, and other 4 can be fixed using the gauge transformations. So in the end we are left with 4 phase space degrees of freedom, which correspond to 2 physical degrees of freedom.

\subsubsection{Massless helicity-2 and gauge invariance}

It is remarkable that, even if we didn't start from the complete GR action, we could have arrived at the action (\ref{GRquadraticactionvacuum}) following other paths. As we just said, the requirement that the action describes a massless, helicity-2 field singles out (apart from a multiplicative constant) the action above. Even if we just ask that the action describes a massless field which, upon decomposition in helicity-2, helicity-1 and helicity-0 components, contains a helicity-2 part, then the requirement of absence of ghost instabilities fixes the action to be (\ref{GRquadraticactionvacuum}) \cite{VanNieuwenhuizen73}. Therefore, if we started from a more field theoretical perspective, we would have singled out this action just asking that a massless helicity-2 field plays a role in the gravitational interaction. Note that there is yet another way of deriving this action, this time from the point of view of symmetries. The action (\ref{GRquadraticactionvacuum}) is invariant with respect to the (gauge) transformation
\beq
\label{gaugesymmlin}
h_{\mu\nu} \rightarrow h_{\mu\nu} + \mathcal{L}_{\xi}(\eta)_{\mu\n} = h_{\mu\nu} + \partial_{\mu}\xi_\nu+\partial_{\nu}\xi_\mu
\eeq
where $\xi_{\m}(x)$ is an arbitrary 1-form field. From the perspective of GR, this is just a consequence of diffeomorphism invariance of the full theory, and the transformation above is the linearized form of an infinitesimal coordinate transformation. On the other hand, considering the most general quadratic, local and Lorentz invariant action for a symmetric field $h_{\m\n}$ on Minkowski spacetime, with no more than two derivatives, the requirement of invariance with respect to the transformation (\ref{gaugesymmlin}) fixes the action to be (\ref{GRquadraticactionvacuum}) \cite{HinterbichlerReview,OhanianRuffinibook}, again up to a multiplicative constant. Once again, we may have found the action above just by asking reasonable physical properties plus gauge invariance, without knowing anything about GR. It is tempting to wonder if it is not just a chance that the action which describes linear GR has these properties, and if they may be considered instead the core of GR as a field theory of gravitation.

\subsection{GR as an interacting massless helicity-2 theory}
\label{GR as an interacting massless spin-2 theory}

It can in fact be seen that locality, Lorentz invariance, no higher derivatives and gauge invariance actually fix the theory also at non-linear level. Let's start again from the complete action of GR (\ref{GRaction}): the theory is invariant with respect with general coordinate transformations, which for infinitesimal transformations read
\begin{align}
X^{\m} &\rightarrow X^{\m} - \xi^{\m}(X) \\[1mm]
h_{\m\n} &\rightarrow h_{\m\n} + \dem \xi_{\n} + \den \xi_{\m} + \mathcal{L}_{\xi}(h)_{\, \mu\n} \quad .
\end{align}
Here, the full metric is $g_{\m\n} = \e_{\m\n} + h_{\m\n}$, $\xi^{\m}$ is an infinitesimal vector field and indices are lowered/raised with the flat metric $\e_{\m\n}/\e^{\m\n}$. However, $h_{\m\n}$ is not necessarily small. Expanding around Minkowski space, we can write the full action in terms of powers of $h_{\m\n}$: the quadratic piece gives the action (\ref{GRquadraticaction}), while higher powers of $h_{\m\n}$ can be interpreted as self-interaction pieces. The full action in vacuum schematically will be of the form
\beq
\label{GRexpandvacuum}
S = \int \! d^4 x \, \Big[ \de^2 h^2 + \de^2 h^3 + \cdots + \de^2 h^{n} + \cdots \Big] + 
\eeq
where $\de^2 h^{n}$ means that this piece contains two derivatives and $n$ factors coming from $h_{\m\n}$ (not that there is a second derivative of $h$ to the $n$-th power). The fact that this is an expansion of GR around Minkowski spacetime is encoded in the precise form of the terms which enter at every order, and in the values of the numerical coefficients which stand in front of each term.

\subsubsection{GR as a resummed theory}

However, we may take the opposite perspective: we may start with the action (\ref{GRquadraticactionvacuum}) for a free massless helicity-2 graviton, and ask what higher power interaction terms can be added. The possible terms can be arranged in powers of the perturbations $h$ and their derivatives, so the general non-linear extension of (\ref{GRquadraticactionvacuum}) will contain the type of terms present in (\ref{GRexpandvacuum}) as well as many others. We may ask that the full action resulting from such an operation enjoys gauge invariance: the gauge transformations should reduce to (\ref{gaugesymmlin}) at linear order, but may have higher order corrections. It can be shown \cite{HinterbichlerReview,OhanianRuffinibook} that these requirements are strong enough to force the interaction terms to be exactly the ones of full non-linear GR. Therefore, we may equivalently see the full action of GR not as the starting point, but as the result of the summation of all the terms allowed by gauge invariance for an interacting theory of a massless helicity-2 field.

A note of caution is in order: this ``bottom-up'' construction which allows to see GR as an interacting theory of a massless helicity-2 field relies on the fact that we chose Minkowski space as the starting point. However, from this perspective a ``miracle'' happens when we add up all the interaction terms: despite the fact that we explicitly started from a definite background ($\e_{\m\n}$), which is not dynamical (it is not determined by the theory itself), the field redefinition $h_{\mu\nu} \rightarrow g_{\mu\nu} - \eta_{\mu\nu}$ in the resummed theory completely eliminates the background metric $\e_{\m\n}$ from the action, leaving only the physical metric $g_{\m\n}$. Therefore, the fully interacting action turns out to be background independent, or in other words there is not a prior geometry in the theory.

\subsection{Propagator and relevant scales} 
\label{Propagator and relevant scales}

\subsubsection{Propagator}

Let's study the linear approximation of GR in presence of sources. As we already said, the theory is defined by the action (\ref{GRquadraticaction}) which gives the equations of motion
\beq
\label{lineareqmotionGR}
\mcal{E}_{\m\n}^{\,\,\,\,\,\r\s} h_{\r \s} = M^{-2}_P \, T_{\m\n}
\eeq
where
\beq
\label{Epsilon operator}
\mcal{E}_{\m\n}^{\,\,\,\,\,\r\s} = 
\frac{1}{2} \Big[ \, \d_{(\m}^{\,\,\s} \e^{\la \r} \de_{\la} \de_{\n)} - \e^{\r \s} \de_{\m} \de_{\n} - \d_{\m}^{\,\,\r} \d_{\n}^{\,\,\s} \square - \eta_{\m \n} \left( \e^{\la \r} \e^{\a \s} \de_{\la} \de_{\a} - \e^{\r \s} \square \right) \Big] \quad .
\eeq
We would like to find the propagator of the (linear) theory, which roughly speaking is the solution of the equation above when the source is perfectly localized. However, the gauge invariance enjoyed by the theory implies that, for every configuration of the source term, there are an infinite number of solutions of the equation above and therefore the operator $\mcal{E}_{\m\n}^{\,\,\,\,\,\r\s}$ is not invertible. To find the propagator, we have to fix the gauge and render the differential operator invertible: once we have found the propagator in a particular gauge, the solution of (\ref{lineareqmotionGR}) will be given by the sum of the gauge fixed solution and a pure gauge contribution. We choose to impose the harmonic gauge condition
\beq
\label{lorenzgaugecon}
\de^\mu h_{\mu\nu} -\half \partial_\nu h = 0 \quad ,
\eeq
and using this condition the equation of motion (\ref{lineareqmotionGR}) can be simplified to give
\beq
\label{lineareqmotionGRLorentzgauge}
\mcal{O}_{\m\n}^{\,\,\,\,\,\r\s} h_{\r \s} = M^{-2}_P \, T_{\m\n}
\eeq
where
\beq
\mcal{O}_{\m\n}^{\,\,\,\,\,\r\s} = - \frac{1}{2} \Big[ \d_{\m}^{\,\,\r} \d_{\n}^{\,\,\s} \square - \half \eta_{\m \n} \e^{\r \s} \square \Big] \quad .
\eeq
The propagator $D_{\m\n}^{\,\,\,\,\,\,\a\b}(x; x^{\prime})$ is then defined as the solution to the equation
\beq
\label{defpropagatorGRLorentzgauge}
\mcal{O}_{\m\n}^{\,\,\,\,\,\r\s} D_{\r \s}^{\,\,\,\,\,\,\a\b}(x; x^{\prime}) = \half \Big( \d_{\m}^{\,\,\a} \d_{\n}^{\,\,\b} + \d_{\n}^{\,\,\a} \d_{\m}^{\,\,\b} \Big) \, \d^{(4)}(x-x^{\p}) \quad ,
\eeq
and as in the previous chapter we can factorize a scalar part $D_{S}(x;x^{\p})$ and a purely numerical part which carries the tensor structure $S_{\m\n}^{\,\,\,\,\,\,\a\b}$ (note that the propagator depends only on the difference $(x-x^{\p})$ because of translational symmetry). One has then
\begin{align}
S_{\m\n}^{\,\,\,\,\,\,\a\b} &= \half \Big( \d_{\m}^{\,\,\a} \d_{\n}^{\,\,\b} + \d_{\n}^{\,\,\a} \d_{\m}^{\,\,\b} \Big) - \half \, \e_{\m\n} \e^{\a\b} \\[2mm]
- \half \square \, D_{S}(x-x^{\p}) &= \d^{(4)}(x-x^{\p}) \quad ,
\end{align}
which confirms the formula (\ref{GRtensorstructure}) of the previous chapter.

\subsubsection{Static spherically symmetric solutions and non-linearity scales}

Considering now a static, spherically symmetric source point source of mass $M$: $T_{\a\b}(\vec{x}^{\, \p}) = M \, \d_{\a}^{\,\, 0} \d_{\b}^{\,\, 0} \, \d^{(3)}(\vec{x}^{\, \p})$, we get analogously to section (\ref{Weak GR gravity vs weak DGP gravity})
\beq
h_{\m\n}(r) = S_{\m\n}^{\,\,\,\,\,\,00} \, \frac{M}{M_P^2} \, V_{GR}(r)
\eeq
and so we have that $h_{\m\n}$ is diagonal and 
\begin{align}
\label{GRweakfield1}
h_{00}(r) &= \frac{M}{M_P^2} \, \frac{1}{4 \pi r} = \frac{2GM}{r}\\[2mm]
\label{GRweakfield2}
h_{ii}(r) &= \frac{M}{M_P^2} \, \frac{1}{4 \pi r} \, \d_{ij}  = \frac{2GM}{r} \, \d_{ij} \quad .
\end{align}
Remembering the definition of gravitational potentials
\begin{align}
\label{gravpotentialdef1}
h_{00}(r) &= - 2 \, \Phi(r) \, \\[2mm]
\label{gravpotentialdef2}
h_{ii}(r) &=  - 2 \, \Psi(r) \, \d_{ij} \quad ,
\end{align}
we have that in GR
\begin{align}
\label{GRpotential1}
\Phi(r) &= - \frac{M}{M_P^2} \, \frac{1}{8 \pi r} = - \frac{GM}{r}\\[2mm]
\label{GRpotential2}
\Psi(r) &= - \frac{M}{M_P^2} \, \frac{1}{8 \pi r} = - \frac{GM}{r}
\end{align}
which gives (\ref{gravitationalpotentialsGR}). Note that this solution indeed satisfies the harmonic gauge condition (\ref{lorenzgaugecon}).

To find the scale where non-linearities become important in GR, we should insert the linear solution (\ref{GRpotential1})-(\ref{GRpotential2}) in the full action (\ref{GRexpandvacuum}), and see at what radius(es) the non-linear terms become comparable with the linear ones. Due to the dependence $\propto 1/r$ of the components of $h_{\m\n}$, any term $\de^2 h^{n}$ will be, apart from numerical factors, $\de^2 h^{n} \sim h^{n}/ r^2$ and so become comparable to $\de^2 h^2 \sim h^2 /r^2$ at $r \sim M/M_P^{2}$. We see that all the non-linear terms become comparable to the linear ones at the same scale
\beq
r_g \sim GM \sim \frac{M}{M_P^2}
\eeq
which is therefore the only scale where non-linearities become important in presence of a spherical body of mass $M$.

\section{The Fierz-Pauli theory}

\subsection{The Fierz-Pauli action}

Having seen that GR can be considered in some sense as an interacting theory of a massless helicity-2 field on Minkowski spacetime, the first step in building a non-linear theory of massive gravity is to find the action which describes the dynamics of a free \emph{massive} spin-2 field on Minkowski spacetime. In the perturbative approach to construct the full theory, once we have found this free action we should add interaction terms which extend the theory at full non-linear level. The problem of finding the action which describes a free massive spin-2 field on Minkowski spacetime has been solved already in 1939 by Fierz and Pauli \cite{FP} who proposed the following action for a symmetric tensor $h_{\mu\nu}$
\begin{multline}
\label{FierzPauliaction}
S^{(2)}_{FP} = \int \! d^4 x \, \Big[ - \frac{1}{2} \de_{\la} h_{\mu\nu} \de^{\la} h^{\mu\nu} + \de_{\m} h_{\n \la} \de^{\n} h^{\m \la} - \de_\mu h^{\mu\nu}\partial_\nu h +\\[2mm]
+\frac{1}{2} \partial_\lambda h\partial^\lambda h - \frac{m^2}{2} \big( h_{\mu\nu} h^{\mu\nu} -h^2 \big) \Big]
\end{multline}
which is therefore called the \emph{Fierz-Pauli action}. Analogously to the quadratic action for GR, there are several ways to look at it. We may notice in fact that it is a linear combination of all the possible contractions of two powers of $h_{\m\n}$ with up to two derivatives, which are the terms appearing in (\ref{GRquadraticactionvacuum}) plus two non-derivative terms. The coefficients of this linear combination are such that the derivative part exactly reproduces the quadratic GR action (\ref{GRquadraticactionvacuum}), while the relative coefficient between the two non-derivative terms is fixed to be $-1$: this is known as the \emph{Fierz-Pauli tuning}. However, the most distinctive property of this action is seen from the point of view of the representations of the Lorentz group: this is exactly the action which describes the dynamics of a free massive spin-2 field. Any change in this action would either introduce other degrees of freedom along with the massive spin-2 field, or disrupt the fact that there is a massive spin-2 in the theory. The overall coefficient of the non-derivative terms plays the role of mass of the field, and the part $m^2 ( h_{\mu\nu} h^{\mu\nu} -h^2 )/2$ is then called the mass term. As we did for the quadratic GR action, we can count the degrees of freedom as consistency check of the fact that the Fierz-Pauli action propagates the 5 degrees of freedom of a massive spin-2 field.

\subsubsection{Degrees of freedom counting} 

Analogously to the case of linear GR, the fields $\dot{h}_{00}$ and $\dot{h}_{0i}$ appear linearly in the action, and it is impossible to perform the Legendre transform with respect to all the velocities $\dot{h}_{\m\n}$. Also in this case we integrate by parts to have an action where $\dot{h}_{00}$ and $\dot{h}_{0i}$ do not appear at all. However, in this case the fields $h_{0i}$ do not appear linearly in the action obtained after integrating by parts, since the mass term produces quadratic terms in $h_{0i}$, while $h_{00}$ still appears linearly, despite the mass term. We can do the Legendre transform of the (integrated by parts) action with respect just to the spatial components, and the conjugate momenta have the same form as in the $m = 0$ case \cite{HinterbichlerReview}
\beq
\pi_{ij}= \frac{\de \mcalL}{\de \dot{h}_{ij}}= \dot{h}_{ij} - \dot{h}_{kk} \delta_{ij} - \de_{(i}h_{j)0} + 2\de_k h_{0k} \delta_{ij}
\eeq
and inverting this relation we get as in the $m = 0$ case
\beq
\dot{h}_{ij} = \pi_{ij} - \half \pi_{kk} \delta_{ij} + \de_{(i}h_{j)0} \quad .
\eeq
The contributions from the mass term show up in the Lagrangian, which can be written as \cite{HinterbichlerReview}
\beq
\label{Hamiltonianactionfierzpauli1}
\mcalL (h, \pi, h_{00}, h_{0i}) = \pi_{ij} \, \dot{h}_{ij} - \mcalH + 2 h_{0i} \big( \de_j \pi_{ij} \big) + m^2 h^2_{0i} + h_{00} \big( \triangle h_{ii} - \de_i \de_j h_{ij} - m^2 h_{ii} \big)
\eeq
where again $\mcalH$ depends only on $h_{ij}$, $\pi_{ij}$ and their spatial derivatives, and $h^2_{0i}$ is a shorthand for $\sum_{i} h^2_{0i}$. It is apparent that, as we said, $h_{00}$ still appears linearly, and still multiply a term with no time derivatives, but now the fields $h_{0i}$ appear quadratically. They can be interpreted as auxiliary variables: in this case they don't enforce any constraint, and their equations of motion give
\beq
h_{0i} = - \frac{1}{m^2} \, \de_j \pi_{ij}
\eeq
which can be plugged back into the action (\ref{Hamiltonianactionfierzpauli1}) to give \cite{HinterbichlerReview}
\beq
\label{Hamiltonianactionfierzpauli12}
S = \int \! d^4x \, \Big[ \pi_{ij} \dot{h}_{ij} - \mathscr{H} + h_{00} \Big( \triangle h_{ii}- \de_i \de_j h_{ij} - m^2 h_{ii} \Big) \Big]
\eeq
where
\beq
\mathscr{H} = \mcalH + \frac{1}{m^2} \big( \de_j \pi_{ij} \big)^2 \quad .
\eeq
The field $h_{00}$ instead enforces the (primary) constraint $\mathcal{C}_1 = \triangle h_{ii}- \de_i \de_j h_{ij} - m^2 h_{ii} = 0$. However, this constraint is not automatically preserved by the time evolution of the system, since its Poisson bracket with the Hamiltonian $\mathcal{C}_{2} \equiv \{ \mathcal{C}_1 , \mathscr{H} \}_P$ is neither zero nor proportional to $\mathcal{C}_1$. Therefore, we have to impose also the (secondary) constraint $\mathcal{C}_2 = 0$. The Poisson bracket of $\mathcal{C}_2$ with $H$ is instead linearly dependent with $\mathcal{C}_1$ and $\mathcal{C}_2$, so we don't need to impose any more constraints: in total the number of constraints we need to impose is therefore two. Since the Poisson bracket of the two constraints does not vanish, they don't generate any gauge symmetry. The degrees of freedom are then the $6 +6 = 12$ of $h_{ij}$ and $\pi_{ij}$ minus one for each constraint: we have in total $12-2= 10$ phase space degrees of freedom which correspond to 5 physical degrees of freedom.

\subsubsection{Massive spin-2 and absence of gauge invariance}

Using the Hamiltonian formalism, it is actually quite easy to see why the Fierz-Pauli tuning is necessary: a generic mass term $a \, h_{\mu\nu} h^{\mu\nu} + b \, h^2$ contains $h^2_{00}$ in the form $(a +b ) h^2_{00}$, so only if $a = -b$ we have that $h_{00}$ appears linearly. Explicitly
\beq
a h_{\mu\nu} h^{\mu\nu} + b h^2 = (a + b ) h^2_{00} - 2 a h^2_{0i} - 2 b h_{00} h_{ii} + a h_{ij} h_{ij} + b h_{ii}^2 \quad .
\eeq
We see that if $a = 0$ then $h_{0i}$ appear linearly in the action (due the derivative part), so there are at least 3 constraints and it is impossible to have 10 phase space degrees of freedom. Therefore, if $a = 0$ the action can never propagate the 5 physical degrees of freedom of a massive spin-2 graviton. However, if $a \neq 0$  the $h_{0i}$ become auxiliary variables: if $b \neq -a$ then $h^2_{00}$ appears in the action, and so it is a auxiliary variable as well meaning that there are no constraints at all. Therefore, in the latter case the number of physical degrees of freedom is 6. Only if $a \neq 0$ and $a = -b$ there can be 5 degrees of freedom, which describe the massive spin-2 field.

Note that this action is \emph{not} invariant with respect to the gauge transformation (\ref{gaugesymmlin}): the gauge symmetry is broken by the mass term. Therefore, we cannot construct a non-linear extension by enforcing a non-linear version of the gauge symmetry, as can be done to construct (non-linear) GR from the linear approximation. However, it can be shown that every modification at linearized level of (\ref{FierzPauliaction}) which still propagates a massive spin-2 field, has ghost instabilities \cite{VanNieuwenhuizen73}: necessarily the additional (sixth) degree of freedom turned on by the modification is a ghost. The Fierz-Pauli action is therefore the only quadratic action for a symmetric tensor on Minkowski spacetime which contains a massive spin-2 field and is ghost-free. This is a property we may hope to use as a criterion to build a non-linear extension of the Fierz-Pauli action.

\subsection{The vDVZ discontinuity and Vainshtein mechanism}

We would like now to derive the weak field solution corresponding to a static, point-like mass in the Fierz-Pauli theory. The full Fierz-Pauli action including the source is
\begin{multline}
\label{FierzPauliactionsourced}
S^{(2)}_{FP} = \int \! d^4 x \, \frac{M^2_P}{2} \bigg[ - \frac{1}{2} \de_{\la} h_{\mu\nu} \de^{\la} h^{\mu\nu} + \de_{\m} h_{\n \la} \de^{\n} h^{\m \la} - \de_\mu h^{\mu\nu}\partial_\nu h \, + \\[2mm]
+\frac{1}{2} \partial_\lambda h\partial^\lambda h - \frac{m^2}{2} \big( h_{\mu\nu} h^{\mu\nu} -h^2 \big) \bigg] + h_{\m\n} T^{\m\n}
\end{multline}
and performing the variation with respect to $h^{\m\n}$ we obtain the equation of motion
\begin{multline}
\label{FPfieldequationsourced}
- \half \Big( \square h_{\mu\nu} - \de_{\la} \de_{(\mu} h^\lambda_{\,\, \nu)} + \eta_{\mu\nu} \partial_\lambda \partial_\sigma h^{\lambda\sigma} + \partial_\mu \partial_\nu h - \\[2mm]
- \eta_{\mu\nu} \square h - m^2 ( h_{\mu\nu} -\eta_{\mu\nu} h ) \Big)= M^{-2}_P T_{\mu\nu} \quad .
\end{multline}
We consider conserved sources, for which $\de_{\m} T^{\m\n} = 0$. Note that, differently from GR where the continuity equation is a consequence of the invariance of the theory with respect to reparametrizations, in this case there is no symmetry which guarantees that the energy momentum tensor is covariantly conserved (since the mass term breaks the reparametrization invariance). Therefore the validity of the continuity equation is in this case an assumption, which is nevertheless quite natural at classical level, but is likely to be broken at quantum level by loop corrections. Acting on the equations of motion (\ref{FPfieldequationsourced}) with $\partial^\mu$, we find
\begin{equation}
\label{lorentzg}
\partial^\mu h_{\mu\nu} - \partial_\nu h = 0
\end{equation}
and, plugging this back into (\ref{FPfieldequationsourced}) and taking the trace, we find
\begin{equation}
\label{traceless}
- \frac{3}{2} \, m^2 h = \frac{1}{M^{2}_P} \, T \quad .
\end{equation}
Using the last two relations, we can show that the equations of motion (\ref{FPfieldequationsourced}) are equivalent to the following system of differential equations
\begin{align}
\label{massiveequationbox}
- \half \, \big( \square - m^2 \big) h_{\mu\nu} &= \frac{1}{M^{2}_P} \, \bigg[ \, T_{\mu\nu} - \third \bigg( \eta_{\mu\nu} - \frac{\demden}{m^2} \bigg) \, T \, \bigg] \\[2mm]
\label{massiveequationtransverse}
\de^{\mu} h_{\mu\nu} &= - \frac{2}{3} \frac{1}{M^{2}_P m^2} \, \de_{\nu} T \\[2mm]
\label{nonmelodiredai}
h &= - \frac{2}{3} \frac{1}{M^{2}_P m^2} \, T \quad .
\end{align}

The general solution to (\ref{FPfieldequationsourced}) can be expressed in general as the sum of the general solution of the homogeneous equation plus a particular solution. The former is therefore the general solution of the system
\begin{align}
\big( \square - m^2 \big) h_{\mu\nu} &= 0 \\[2mm]
\de^{\mu} h_{\mu\nu} &= 0 \\[2mm]
h &= 0
\end{align}
and so is a transverse-traceless field. For the particular solution of the sourced equation, we impose boundary conditions which imply that the operator $\big( \square - m^2 \big)$ is invertible, and so the second and third equations (\ref{massiveequationtransverse}) and (\ref{nonmelodiredai}) are implied by the first one (\ref{massiveequationbox}). Therefore, in order to find a particular solution of the sourced field equations (\ref{FPfieldequationsourced}), it is sufficient to find a solution of
\beq
\label{massiveequationbox2}
- \half \, \big( \square - m^2 \big) h_{\mu\nu} = \frac{1}{M^{2}_P} \, \bigg[ \, T_{\mu\nu} - \third \bigg( \eta_{\mu\nu} - \frac{\demden}{m^2} \bigg) \, T \, \bigg] \quad .
\eeq
To solve this equation, it is useful to go to momentum space. We express $h_{\mu\nu}(x)$ and $T_{\mu\nu}(x)$ via their Fourier transforms
\begin{align}
h_{\mu\nu}(x) &= \int \! d^4 p \,\, e^{ip_{\a} x^{\a}} \, h_{\mu\nu}(p)\\[1mm]
T_{\mu\nu}(x) &= \int \! d^4 p \,\, e^{ip_{\a} x^{\a}} \, T_{\mu\nu}(p)
\end{align}
and so we obtain
\beq
\label{sourcegen1}
h_{\mu\nu}(p) = \frac{2}{M^{2}_P} \, \frac{1}{p_{\a} p^{\a} + m^2} \, \bigg[ \, \half \, \, \d_{(\m}^{\,\,\,\, \r} \, \d_{\n)}^{\,\,\,\, \s} - \third \, \Big( \eta_{\mu\nu} + \frac{p_\mu p_\nu}{m^2} \Big) \, \eta^{\r\s} \, \bigg] \, T_{\r \s}(p) \quad .
\eeq
Note that a static source $T_{\mu\nu}(x) = T_{\mu\nu}(\vecx)$ has a Fourier transform of the form
\beq
T_{\mu\nu}(p) = \d(p^0) \, T^{(3)}_{\mu\nu}(\vecp) \quad ,
\eeq
and in particular for a point-like source of mass $M$ we have
\beq
T_{\mu\nu}(\vecx) = M \, \d_{\m}^{\,\, 0} \, \d_{\n}^{\,\, 0} \, \d^{(3)}(\vecx) \qquad \longrightarrow \qquad T_{\mu\nu}(p) = \frac{\d(p^0)}{(2 \pi)^3} \, M \, \d_{\m}^{\,\, 0} \, \d_{\n}^{\,\, 0} \quad .
\eeq
Indicating $r \equiv \sqrt{\vecx^{\,\, 2}}$ and using the formulas
\begin{align}
\int \! \frac{d^3 \vecp}{(2\pi)^3} \, e^{i \vecp \cdot \vecx} \, \frac{1}{\vecp^{\,\, 2} + m^2} &= \frac{1}{4\pi} \, \frac{e^{-mr}}{r} \\[2mm]
\int \! \frac{d^3 \vecp}{(2\pi)^3} \, e^{i \vecp \cdot \vecx} \, \frac{p_i p_j}{\vecp^{\,\, 2} + m^2} &=
- \de_{i} \de_{j} \, \int \! \frac{d^3 \vecp}{(2\pi)^3} \, e^{i \vecp \cdot \vecx} \, \frac{1}{\vecp^{\,\, 2} + m^2} \quad ,
\end{align}
we have \cite{HinterbichlerReview}
\begin{align}
\label{FPweakfieldsolutiontrue00}
h_{00}(x) &= \frac{4}{3} \, \frac{M}{M^2_P} \, \frac{e^{-mr}}{4 \pi r}\\[2mm]
\label{FPweakfieldsolutiontrue0i}
h_{0i}(x) &= 0 \\[2mm]
\label{FPweakfieldsolutiontrueij}
h_{ij}(x) &= \frac{2}{3} \, \frac{M}{M^2_P} \, \frac{e^{-mr}}{4 \pi r} \left[ \frac{1 + mr + m^2 r^2}{m^2 r^2} \delta_{ij} -\frac{1}{m^2 r^4} ( 3 + 3mr + m^2 r^2 ) x_i x_j \right]
\end{align}
where $x_i = \d_{ik} x^{k}$.

\subsubsection{The vDVZ discontinuity}
\label{The VDVZ discontinuity in FP}

Note that, neglecting the term $\demden T$ in (\ref{massiveequationbox2}) and therefore the term $p_\mu p_\nu \, T(p)$ in (\ref{sourcegen1}), we would obtain the following solution
\begin{align}
\label{FPweakfieldsolutionfalse00}
h_{00}(x) &= \frac{4}{3} \, \frac{M}{M^2_P} \, \frac{e^{-mr}}{4 \pi r}\\[1mm]
\label{FPweakfieldsolutionfalse0i}
h_{0i}(x) &= 0 \\[1mm]
\label{FPweakfieldsolutionfalseij}
h_{ij}(x) &= \frac{2}{3} \, \frac{M}{M^2_P} \, \frac{e^{-mr}}{4 \pi r} \, \d_{ij} \quad .
\end{align}
The term $p_\mu p_\nu \, T(p)$ produces a contribution to the metric field which has no observable consequences on a test body whose energy-momentum tensor obeys the conservation equation: in fact, the interaction amplitude $ \int d^4 x \, h_{\m\n} T_{tb}^{\m\n}$ between such a contribution to the metric and the conserved energy-momentum tensor of a test body vanishes. Therefore, regarding measurements like light deflection, planets orbits and so on, the metric (\ref{FPweakfieldsolutiontrue00})-(\ref{FPweakfieldsolutiontrueij}) give the same predictions as the metric (\ref{FPweakfieldsolutionfalse00})-(\ref{FPweakfieldsolutionfalseij}). Let's consider then the metric (\ref{FPweakfieldsolutionfalse00})-(\ref{FPweakfieldsolutionfalseij}): the gravitational potentials reads
\beq
\Phi(r) = - \frac{2}{3} \, \frac{M}{M^2_P} \, \frac{e^{-mr}}{4 \pi r} \qquad \qquad \Psi(r) = \half \, \Phi(r) \quad .
\eeq
For distances larger than the Compton length $r_c \equiv 1/m$, the potentials decay exponentially, with the typical (Yukawa) behavior of massive fields $e^{-mr}/r$. On the other hand, for distances smaller than the Compton wavelength $r \ll r_c$, both of the gravitational potentials have the $1/r$ dependence of GR, but their ratio $\Phi(r)/\Psi(r)$ is twice the GR value. The situation is completely equivalent to the weak field solution of the DGP model inside the crossover scale: this mismatch is responsible for a $25 \%$ relative error in light deflection or planet orbits predictions compared to the GR ones. Note that this conclusion is not affected by taking $m$ as small as we like, since this will only make the Compton radius bigger and bigger without altering what happens well inside the Compton radius itself. However, if we set $m$ to be exactly zero, then the theory is exactly GR and trivially the predictions agree with the GR ones: therefore, there seems to be a discontinuity in the physical predictions of the theory when $m \rightarrow 0$. This has been noted and pointed out independently by Iwasaki \cite{Iwasaki70}, van Dam and Veltman \cite{VDV70} and Zakharov \cite{Zakharov70}, and is known as the \emph{vDVZ discontinuity}. This is \emph{a priori} unexpected, since there seems to be no discontinuity in the $m \rightarrow 0$ limit in the action (\ref{FierzPauliactionsourced}), and it usually assumed that if a theory is continuous in a parameter, then its physical predictions should be continuous in that parameter as well. However, the key point here is that taking the limit $m \rightarrow 0$ in the action is not the correct way to perform the $m \rightarrow 0$ limit in the theory: for example, the action (\ref{FierzPauliactionsourced}) for every $m \neq 0$ propagates 5 degrees of freedom, as we saw, while the $m = 0$ action propagates only two degrees of freedom. Also, the $m = 0$ theory enjoys a gauge invariance which does not hold as soon as $m$ becomes different from zero. Therefore, the number of degrees of freedom and the symmetry properties of the action (\ref{FierzPauliactionsourced}) are not continuous in the $m \rightarrow 0$ limit. We may conclude that the $m \rightarrow 0$ limit of the Fierz-Pauli theory is \emph{not} described by the $m \rightarrow 0$ limit of the Fierz-Pauli action, and in particular the $m \rightarrow 0$ limit of the Fierz-Pauli theory is not GR. To elucidate this, it is useful to construct a different action which enjoys gauge invariance even in the $m \neq 0$ case and gives the same physical predictions of the FP one: this is achieved using the St\"uckelberg language, as we shall see in section \ref{The Fierz-Pauli theory in the Stuckelberg language}.

\subsubsection{The Vainshtein mechanism}
\label{The Vainshtein mechanism}

The conclusion that the $m \rightarrow 0$ limit of the Fierz-Pauli theory is not GR seems to put an end to our hope to use a very small mass for the graviton as a way to explain the cosmological observations which indicate a late time acceleration: it seems that massive gravity is not a modified gravity theory in the sense of section (\ref{Backreaction, dark energy and modified gravity}). However, from the modified gravity perspective the FP theory is just the starting point: since the FP theory is linear (at the level of the field equations), it can never reproduce the strong field behavior of GR. The hope was that the FP theory reproduces the weak field limit of GR for distances smaller than the Compton length, and that a suitable non-linear completion of the FP theory is able to reproduce also the strong field behavior of GR in the same range of length scales. Instead, we found that the FP theory does not reproduce GR either inside or outside the Compton radius. However, it has been proposed by Vainshtein \cite{Vainshtein72} that interaction terms added to the FP action may be effective to restore agreement with GR also at length scales where the weak field approximation in GR is valid. This idea relies on the fact that non-linear terms in the non-linear extension of the FP theory may become relevant at a scale which is much larger than the scale $r_g = GM \sim M/M_P^2$ where non-linear terms become relevant in GR, somewhat similarly to what happens in the DGP model. 

Vainshtein considered a specific non-linear extension of the Fierz-Pauli theory, namely the one obtained adding the mass term $m^2 ( h_{\mu\nu} h^{\mu\nu} -h^2 )/2$ to the \emph{full} non-linear GR action expressed in terms of $\e_{\m\n}$ and $h_{\m\n} = g_{\m\n} - \e_{\m\n}$. Considering a static and spherically symmetric source, he used the following \emph{ansatz} for the metric
\beq
ds^2 = - B(r) dt^2 + C(r) dr^2 + A(r) r^2 d\Omega^2
\eeq
which at linear order (\emph{i.e.} keeping only the quadratic terms in the action) have the vacuum solutions \cite{HinterbichlerReview}
\begin{align}
B_1(r) &= - \frac{8GM}{3} \, \frac{e^{-mr}}{r} \\[1mm]
C_1(r) &= - \frac{8GM}{3} \, \frac{e^{-mr}}{r} \, \frac{1 + mr}{m^2 r^2} \\[1mm]
A_1(r) &= \frac{4GM}{3} \, \frac{e^{-mr}}{r} \, \frac{1 + mr + m^2 r^2}{m^2 r^2}
\end{align}
which are equivalent to (\ref{FPweakfieldsolutiontrue00})-(\ref{FPweakfieldsolutiontrueij}). We can ask how these solutions are modified if we keep also the non-linear terms in the equations of motion (or equivalently the interaction terms in the action). We can write
\begin{align}
B(r) &= B_0(r) + \epsilon B_1(r) + \epsilon^2 B_2(r) + \cdots \\ \nn
C(r) &= C_0(r) + \epsilon C_1(r) + \epsilon^2 C_2(r) + \cdots \\ \nn
A(r) &= A_0(r) + \epsilon A_1(r) + \epsilon^2 A_2(r) + \cdots 
\end{align}
where $A_0 = B_0 = C_0 = 1$ and $\epsilon$ is a parameter that keeps track of which order in non-linearities we are working at. Solving recursively the vacuum equations at each order in $\epsilon$ shows \cite{Vainshtein72,HinterbichlerReview} that the expansion in powers of non-linearities shows up in the solutions for $A$, $B$ and $C$ as an expansion in the parameter $r_V/r$, where 
\beq
\label{VainshteinRadius5}
r_V \equiv \sqrt[5]{\frac{GM}{m^4}} = \sqrt[5]{r_g r_c^4} 
\eeq
is called the \textit{Vainshtein radius}. It follows that non-linearities become important (\emph{i.e.} comparable to the linear terms) when $r \approx r_V$, so the Vainshtein radius is the scale around a mass $M$ below which the linear approximation cannot be trusted. Note that, since we assume that $r_c/r_g \gg 1$, it follows that the Vainshtein radius is much bigger than the Schwarzschild radius $r_V/r_g \gg 1$ and so the scale where non-linearities become important around a spherical object for the Fierz-Pauli theory is indeed much bigger than the scale where this happens in GR. In fact, setting $m = H_0$, for an object like the sun the Vainshtein radius (\ref{VainshteinRadius5}) is $r_V \sim 10^5 \, \textrm{pc}$, which is bigger than the diameter of the Milky Way\footnote{The diameter of the Milky Way is approximately $3 \times 10^{4} \, \textrm{pc}$.}: therefore the linear solution cannot be used to calculate the light bending and the planets' orbits in the solar system. Note also that the definition (\ref{VainshteinRadius5}) for the Vainshtein radius in this non-linear extension of Fierz-Pauli is different from the definition (\ref{VainshteinRadius3}) of the Vainshtein radius in the DGP model: this is not strange, since the Vainshtein radius of a theory depends on the structure of the interaction terms, and theories which have different non-linear structures are likely to have different Vainshtein radii.

To understand if this non-linear extension of the Fierz-Pauli theory reproduces or not the GR predictions inside the Vainshtein radius, we should then solve the full equations (with all the non-linear terms). Note that we have to solve necessarily for three unknown functions, we cannot reduce to just two unknown functions as we do in GR. In fact, reparametrising the radial coordinate according to  
\beq
r \rightarrow \rho(r) = r \sqrt{A(r)}
\eeq
we can eliminate the function $A$ from the metric and write the line element in terms of just two functions
\beq
ds^2 = - \tilde{B}(\r) dt^2 + \tilde{C}(\r) d\r^2 + \r^2 d\Omega^2 \quad .
\eeq
In GR, performing this change of variables in the equations of motion results in the function $A$ disappearing also from them, as a consequence of the fact that the theory is invariant with respect of reparametrisations, and so indeed we can reduce the problem to solving for just two functions. However, the Fierz-Pauli theory is \emph{not} invariant with respect to reparametrisations; as a consequence of this, the function $A \rightarrow \tilde{A}(\r)$, despite disappearing from the metric, remains present in the equations of motion along with $\tilde{B}$ and $\tilde{C}$ when we reparametrise the radial coordinate. Of course, nothing prevents us from performing the change of coordinate and work with the unknown functions $\tilde{A}$, $\tilde{B}$ and $\tilde{C}$ instead of $A$, $B$ and $C$. In fact, Vainshtein suggests that this is a convenient thing to do to study the $m \rightarrow 0$ limit, since, regarding the functions $\tilde{B}$ and $\tilde{C}$, he suggests that the effects of non-linearities inside the Vainshtein radius is just to rescale the numerical factors so that $\tilde{B}/\tilde{C} = 1$, while preserving the $\propto 1/r$ dependence. Instead, the effect of non-linearities changes $\tilde{A}$ quite dramatically. He then concludes \cite{Vainshtein72} that for $m \ll 1$ the functions $\tilde{B}$ and $\tilde{C}$ coincide to a very good approximation with their GR ($m = 0$) values inside the Vainshtein radius, and have a smooth $m \rightarrow 0$ limit. In other words, the non-linear terms in the equation of motion modify all the three functions $A$, $B$ and $C$, but in such a way that, redefining the radial coordinate to get rid of $A$ in the metric, the non-linear solutions for $\tilde{B}$ and $\tilde{C}$ inside the Vainshtein radius agree with the GR solutions, and so the non-linear interaction terms restore the agreement with GR. Further studies on the recovery of GR results in the same non-linear extension of the Fierz-Pauli theory considered by Vainshtein can be found in \cite{Damour:2002gp, Babichev:2009us, Deffayet:2008zz, Babichev:2009jt, Babichev:2010jd}. The mechanism of restoring agreement with GR via non-linear interactions is named after Vainshtein and is known as the \emph{Vainshtein mechanism}.

Note finally that, as the mass $m$ approaches 0, the Vainshtein radius grows and tends to infinity: in the limit $m \rightarrow 0$ the predictions of GR are recovered everywhere, and so the $m \rightarrow 0$ limit is indeed smooth for the theory. Therefore, while the linear Fierz-Pauli theory does not reduce to (linear) GR in the $m \rightarrow 0$ limit, it is possible that a non-linear extension of the Fierz-Pauli theory does reduce to (non-linear) GR in the same limit, and therefore that there is no vDVZ discontinuity at non-linear level.

\subsection{The Fierz-Pauli theory in the St\"uckelberg language}
\label{The Fierz-Pauli theory in the Stuckelberg language}

We mentioned that the weak field predictions of the FP theory are significantly different from the ones of linearized GR, no matter how small is the mass of the FP graviton. Therefore, while the $m \rightarrow 0$ limit of the FP action is smooth and gives the GR action, the physical predictions of the FP theory seem not to be continuous in the $m \rightarrow 0$ limit. This is very surprising, since it usually assumed that if a theory is continuous in a parameter, then its physical predictions should be continuous in that parameter as well. However, as we already mentioned, a deeper look at the structure of the FP theory and of GR casts doubts on the fact that $m \rightarrow 0$ limit of the FP theory is given by the $m \rightarrow 0$ limit of the Fierz-Pauli action: in fact, the FP theory propagates five degrees of freedom, while GR propagates just two degrees of freedom; conversely, GR enjoys gauge invariance, which is instead broken in the Fierz-Pauli theory. Therefore, regarding the symmetry properties and the number of degrees of freedom, the $m \rightarrow 0$ limit of the Fierz-Pauli action is not continuous. It is tempting to conjecture that the vDVZ discontinuity and the discontinuity in symmetry properties and degrees of freedom are linked, and that the $m \rightarrow 0$ limit of the Fierz-Pauli theory is not described by the $m \rightarrow 0$ limit of the Fierz-Pauli action.

To understand the relation between the $m \rightarrow 0$ limit of the Fierz-Pauli theory and GR, thereby possibly sheding light on the origin of the vDVZ discontinuity, we would like to formulate a new theory which gives the same physical predictions of the Fierz-Pauli theory, but whose action in the $m \rightarrow 0$ limit still has the same symmetry properties and number of degrees of freedom of the $m \neq 0$ action.  This task is achieved using the St\"uckelberg formalism. 

\subsubsection{The St\"uckelberg formalism}

Starting from the Fierz-Pauli action \eqref{FierzPauliactionsourced}, we want to formulate a \emph{different} theory which is invariant under gauge transformations, yet gives the same physical predictions of the FP action. This is achieved introducing auxiliary fields, called \emph{St\"uckelberg fields}, whose transformation properties are defined exactly to render the action invariant. Let's in fact perform in the action \eqref{FierzPauliactionsourced} the substitution
\beq
\label{htoHlinear}
h_{\m\n}(x) \rightarrow H_{\m\n}(x) = h_{\m\n}(x) + \de_{(\m} Z_{\n)} \quad :
\eeq
if we impose that the field $Z_{\m}$ \emph{shifts} under gauge transformations
\begin{align}
x^{\p \m} &= x^{\m} - \xi^{\m}(x)\\[1mm]
Z^{\p}_{\m} &= Z_{\m} - \xi^{\m}\\[1mm]
h^{\p}_{\m\n} &= h_{\m\n} + \de_{(\m} \, \xi_{\n)} \quad ,
\end{align}
we have that the resulting action is invariant. Note that, given any field configuration $(h_{\m\n}$, $Z_{\m})$ of the new theory, we can always perform a gauge transformation with parameter $\xi_\mu = Z_{\m}$ in the new action and reobtain the original FP action. Therefore, despite the fact that the new action contains more fields that the original one, the physical prediction of the original action and of the ``covariantized'' one are precisely the same. On the other hand, the field $Z_{\m}$ does \emph{not} transform as a 1-form, but has an unusual transformation property.

Performing the substitution \eqref{htoHlinear} inside the FP action, the kinetic part of the action does not change since the substitution \eqref{htoHlinear} has the same form of a gauge transformation, and that part of the action is invariant (it is the action for linearized GR in fact). The only thing that changes is the mass term (the interaction between $Z_{\m}$ and $T_{\m\n}$ produces a total derivative when the latter is covariantly conserved), and modulo total derivatives we get
\begin{multline}
\label{Zstukelberglinear}
S = \int \! d^4x \, M^2_P \bigg[ \, h^{\m\n} \mcal{E}_{\m\n}^{\,\,\,\,\,\r\s} h_{\r\s} - \frac{m^2}{2} \big( h_{\mu\nu} h^{\mu\nu} -h^2 \big) - \\[2mm]
- \frac{m^2}{2} F_{\mu\nu} F^{\mu\nu} - 2 m^2 \big( h_{\mu\nu}\partial^\mu Z^\nu-h\partial_\mu Z^\mu \big) \bigg] + h_{\mu\nu} T^{\mu\nu}
\end{multline}
where $F_{\m\n} = \de_{[\m} Z_{\n]}$ and we raise/lower indices with the Minkowski metric $\e^{\m\n}$/$\e_{\m\n}$. We can redefine the field $Z_\mu \rightarrow \frac{1}{m} Z_\mu$ to render canonical its kinetic term: if we take the $m\rightarrow 0$ limit, we obtain an action for a massless graviton and a massless vector, which in total have four degrees of freedom. So at this point, we still lose one degree of freedom in the $m\rightarrow 0$ limit.

We can remedy to this problem by introducing an additional substructure in $Z_{\m}$ by singling out explicitly a derivative part: we then write
\beq
\label{helicitydecZlinear}
Z_{\m} = A_{\m} + \de_{\m} \phi
\eeq
and in terms of $A_{\m}$ and $\phi$ the tensor $H_{\m\n}$ reads
\beq
\label{HasAandfilinear}
H_{\mu\nu} = h_{\m\n} + \de_{(\m} A_{\n)} + 2 \, \demden \phi \quad .
\eeq
Note that the decomposition (\ref{helicitydecZlinear}) is invariant with respect to the additional internal symmetry
\begin{align}
\phi(x) &\rightarrow \phi(x) - \La(x)\\[1mm]
A_{\m}(x) &\rightarrow A_{\m}(x) + \de_{\m} \La(x) 
\end{align}
and so there are now two gauge transformation under which the action is invariant
\begin{align}
x^{\p \m} &= x^{\m} - \xi^{\m}(x)\\[1mm]
h^{\p}_{\m\n} &= h_{\m\n} + \de_{(\m} \, \xi_{\n)}\\[1mm]
A^{\p}_{\m} &= A_{\m} - \xi_{\m} + \de_{\m} \La\\[1mm]
\label{ultimalinear}
\phi^{\p} &= \phi - \La \quad .
\end{align}
In terms of the fields $A_\m$ and $\f$, the action \eqref{Zstukelberglinear} takes the form
\begin{multline}
\label{Afistukelberglinear}
S = \int \! d^4x \, M^2_P \bigg[ \, h^{\m\n} \mcal{E}_{\m\n}^{\,\,\,\,\,\r\s} h_{\r\s} - \frac{m^2}{2} \big( h_{\mu\nu} h^{\mu\nu} -h^2 \big) - \frac{m^2}{2} F_{\mu\nu} F^{\mu\nu} - \\[2mm]
- 2 m^2 \big( h_{\mu\nu} \partial^\mu A^\nu -h \partial_\mu A^\mu \big) -2 m^2 \big( h_{\mu\nu} \de^\mu \de^\nu \phi - h \de_{\m} \f \de^{\m} \f \big) \bigg] + h_{\mu\nu} T^{\mu\nu}
\end{multline}
where now $F_{\m\n} = \de_{[\m} A_{\n]}$ and again we have discarded total derivatives (which include the interaction of $A_{\m}$ and $\f$ with $T_{\m\n}$ when the latter is covariantly conserved). Note that the quadratic piece in $\de \de f$ and the mixed term $\de A \de \de f$ does not appear precisely for this reason: these terms rearrange in total derivatives, and therefore have no effect on the dynamic. As we will mention later, this is a consequence of the Fierz-Pauli tuning, since any other choice for the mass term of $h_{\m\n}$ in the starting action produces a quadratic piece in $\de \de f$ and a mixed piece $\de A \de \de f$ which do not arrange themselves into total derivatives.

\subsubsection{The vDVZ discontinuity in the St\"uckelberg language}

Note that, in the action \eqref{Afistukelberglinear}, the field $\f$ does not have a kinetic term on its own, but is kinetically mixed with $h_{\m\n}$. To be able to see more clearly the physical meaning of this action, it is useful to perform a field redefinition which de-mix kinetically the fields $h_{\m\n}$ and $\f$, and at the same time creates a proper kinetic term for the latter field. The redefinition
\begin{align}
\bar{h}_{\m\n} &= h_{\m\n} - m^2 \f \, \e_{\m\n}  \\[1mm]
\bar{A}_{\m} &= A_{\m} \\[1mm]
\bar{\f} &= \f
\end{align}
has precisely this effect, and creates a coupling between $\bar{\f}$ and the trace of the energy-momentum tensor as well. It is convenient to further redefine the fields to render the kinetic terms canonical
\begin{align}
\hat{h}_{\m\n} &= M_P \, \bar{h}_{\m\n} \\[1mm]
\hat{A}_{\m} &= M_P \, m \, \bar{A}_{\m} \\[1mm]
\hat{\f} &= M_P \, m^2 \, \bar{\f} \quad ,
\end{align}
and in terms of the ``hatted'' fields the action \eqref{Afistukelberglinear} reads
\beq
\label{hattedlinearStuckelbergaction}
S = \int \! d^4x \, \bigg[ \, \hat{h}^{\m\n} \mcal{E}_{\m\n}^{\,\,\,\,\,\r\s} \hat{h}_{\r\s} - \half \, \hat{F}_{\mu\nu} \hat{F}^{\mu\nu} - 3 \, \de_{\m} \hat{\f} \de^{\m} \hat{\f} + \frac{1}{M_P} \, \hat{h}_{\m\n} T^{\m\n} + \frac{1}{M_P} \, \hat{\f} \, T + \ldots \, \bigg]
\eeq
where the dots stand for terms which are multiplied by $m$ or $m^2$. The $m \rightarrow 0$ limit of this action describes a theory of a massless graviton, a massless vector and a massless scalar, and so propagates five degrees of freedom exactly as the $m \neq 0$ theory. 

Note that, in the $m \rightarrow 0$ limit, the action for the field $\hat{h}_{\m\n}$ is exactly the GR action (apart the $1/M_P$ rescaling); furthermore, the coupling of the field $\hat{\f}$ with the trace of the energy-momentum tensor remains finite in the limit. Going back to the field $h_{\m\n}$ (whose dynamics is described by the action \eqref{Afistukelberglinear}) we can express it in terms of $\hat{h}_{\m\n}$ and $\hat{\f}$ as
\beq
h_{\m\n} = \frac{\hat{h}_{\m\n}}{M_P} + \frac{\hat{\f}}{M_P} \, \e_{\m\n} \quad :
\eeq
in the $m \rightarrow 0$ limit, it receives contributions both from a tensor field which satisfies the GR equations and a scalar field which couples with $T$ with finite strength. Since by construction the action \eqref{Afistukelberglinear} gives the same physical prediction of the Fierz-Pauli theory, we can conclude that indeed the $m \rightarrow 0$ limit of the FP theory is not equivalent to GR, but rather to a scalar-tensor theory.

Note finally that it is possible to impose gauge conditions which eliminate all the terms in the action \eqref{hattedlinearStuckelbergaction} which are linear in $m$ \cite{HinterbichlerReview}. This gauge transformation completely diagonalizes the action, and in the resulting action all the fields have a canonical kinetic term and a mass term, while only $\hat{h}_{\m\n}$ and $\hat{\f}$ couple to the energy-momentum tensor. Therefore, if we consider a static and spherically symmetric source of mass $M$, the profile (in this gauge) for the fields $\hat{h}_{\m\n}$ and $\hat{\f}$ inside the Compton wavelength $r_c = 1/m$ reads 
\beq
\hat{h}_{\m\n} \sim \frac{M}{M_P} \, \frac{1}{r} \qquad \qquad \hat{\f} \sim \frac{M}{M_P} \, \frac{1}{r} \quad ,
\eeq
apart from numerical factors.

\section{Nonlinear extensions of the Fierz-Pauli theory}

Having discussed the linear theory of a massive graviton, we would like to formulate now a non-linear theory of massive gravity which reproduces the predictions of GR in a suitable range of length scales. To be more precise, we are looking for a theory which can be seen as an interacting theory of a massive graviton: therefore we ask that it reduces to the Fierz-Pauli theory in the weak field approximation, and that it propagates the same number of degrees of freedom (five) as the Fierz-Pauli theory. In the linear case, we can formulate the theory of a massive graviton by starting from the action of a massless graviton (linearized GR), and adding a suitable term (the mass term) which is weighted by a parameter which sets the range of the interaction, and does not contain derivatives of the field: we want to do the same also at non-linear level. Therefore, we consider the full (non-linear) GR Lagrangian and add a ``mass'' term, which in general we take to be non-linear as well: this is to be a term which is weighted by a mass parameter, and contains no derivatives of the metric.

\subsection{Generic non-linear extension}

In full generality, considering a local and Lorentz-invariant theory, such a mass term cannot be built from one metric tensor alone \cite{BD}: in fact, the identity $g_{\m\la} g^{\la\n} = \d_{\m}^{\,\, \n}$ implies that it is impossible to construct a nontrivial scalar function out of $g_{\m\n}$ and $g^{\m\n}$ without using derivatives. Therefore, the theory will contain (at least) two metric tensors: there will be a \emph{physical} metric $\mbfg$, which is the metric test bodies feel and which determine in general the causal structure of the spacetime, and an \emph{absolute} background metric $\mbfg^{(0)}$, which is necessary to create nontrivial traces and contractions. To respect the equivalence principle, we postulate that matter fields couple only to the physical metric. Therefore the action will have the following structure
\beq
\label{actionmassivegravitygeneral}
S = \int \! d^4 x \, \sqrt{-g} \, \bigg[ \, \frac{\MPq}{2} \Big( R[\mbfg] - \frac{m^2}{2} \mathcal{U}[\mbfg,\mbfg^{(0)}] \Big) + \mscr{L}_{M}[\mbfg,\psi_{(i)}] \, \bigg]
\eeq
where the $\psi_{(i)}$ are matter fields. Note that the mass term can equivalently be written as a function of the absolute metric $\mbfg^{(0)}$ and of the physical metric $\mbfg$, or as a function of the absolute metric $\mbfg^{(0)}$ and of the difference between the two metrics $\mbf{h} \equiv \mbfg - \mbfg^{(0)}$, or as a function of the physical metric $\mbfg$ and of the difference $\mbf{h}$. Despite the fact that we may use any absolute metric $\mbfg^{(0)}$, a natural choice is to use the Minkowski metric as the absolute metric, and so in the following we assume $g^{(0)}_{\m\n} = \eta_{\m\n}$. Assuming that the function $\mathcal{U}$ is analytic, we can therefore write the mass term as an (\emph{a priori}) infinite sum of terms where each term contains a fixed number of powers of $h_{\m\n}$, and therefore we can write
\beq
\label{fasVk}
\sqrt{-g} \, \mathcal{U}[\mbfg,\mbfg^{(0)}] = \sqrt{-det(\e)} \, \sum_{k = 2}^{+\infty} V_{k}[\eta,\mbfh]
\eeq
where each term $V_{k}[\eta,\mbfh]$ is a linear combination of all the possible contractions of $k$ factors $h_{\m\n}$ with $k$ factors $\eta^{\a\b}$
\beq
\label{Vketah}
V_{k}[\eta,\mbfh] = \sum_{p \in P_k} c_p^{(k)} \, \e^{\m_1 p(\n_1)}\, \cdots \, \e^{\m_k p(\n_k)} \, h_{\m_{1} \n_{1}} \cdots h_{\m_{k} \n_{k}}
\eeq
where $P_k$ is the group of permutations of $k$ elements, and the sum runs on all the permutations $p$ belonging to $P_k$. Introducing the notation
\beq
\big[ h^n \big] \equiv \e^{\m \a_{1}} \, h_{\a_{1} \b_{1}} \, \e^{\b_{1} \a_{2}} \, h_{\a_{2} \b_{2}} \cdots \e^{\b_{n - 1} \a_{n}} \, h_{\a_{n} \m}
\eeq
for the cyclic contraction of $n$ tensors $h_{\m\n}$, we can write the terms in the following more compact way
\begin{align}
V_2 [\eta,\mbfh] &= B_1 \big[ h^2 \big] + B_2 \big[ h \big]^2 \\[1mm]
V_3 [\eta,\mbfh] &= C_1 \big[ h^3 \big] + C_2 \big[ h^2 \big] \big[ h \big] + C_3 \big[ h \big]^3 \\[1mm]
V_4 [\eta,\mbfh] &= D_1 \big[ h^4 \big] + D_2 \big[ h^3 \big] \big[ h \big] + D_3 \big[ h^2 \big]^2 + D_4 \big[ h^2 \big] \big[ h \big]^2 + D_5 \big[ h \big]^4 \\[1mm]
V_5 [\eta,\mbfh] &= F_1 \big[ h^5 \big] + F_2 \big[ h^4 \big] \big[ h \big] + F_3 \big[ h^3 \big] \big[ h \big]^2 + F_4 \big[ h^3 \big] \big[ h^2 \big] + F_5 \big[ h^2 \big]^2 \big[ h \big] + \nn \\
\phantom{V} &+ F_6 \big[ h^2 \big] \big[ h \big]^3 + F_7 \big[ h \big]^5\label{V5equation}\\[1mm]
\phantom{V} &\vdots \nn 
\end{align}
and the requirement that the weak field limit should reproduce the Fierz-Pauli action implies that $B_2 = - B_1$. Inserting this expression in (\ref{actionmassivegravitygeneral}) and expanding also $\sqrt{-g} \, R[\eta,\mbfh]$ in powers of $h_{\m\n}$, we can see that the resulting action is the one we would obtain in a perturbative approach adding interaction terms to the Fierz-Pauli action (\ref{FierzPauliaction}), with the condition that the derivative interaction terms are exactly the same as in (interacting) GR.

\subsubsection{Degrees of freedom and the Boulware-Deser ghost}

The values of the numerical coefficients $C_i$, $D_i$, $F_i$, \ldots (or at least consistency conditions on their values) are to be found imposing the condition that the theory be a viable theory of an interacting massive spin-2 field. This condition translates in several requirements, both of theoretical and phenomenological nature: from the theoretical point of view, we ask that the theory does not have ghost instabilities and that it propagates exactly 5 degrees of freedom, which match the degrees of freedom of the Fierz-Pauli theory. From the phenomenological point of view, we ask that GR predictions are reproduced in the range of length scales where GR is well tested. Note that, since the FP theory is not gauge invariant, we cannot use the requirement of gauge invariance as a guide to build the non-linear theory: unlike in GR, whose non-linear structure is completely fixed by this requirement, we have to implement directly the conditions relating to the absence of ghosts and the number of degrees of freedom. These are in fact quite strong requirements, and it has been actually claimed that any non-linear extension of the Fierz-Pauli theory necessarily propagates six degrees of freedom and the Hamiltonian is not bounded from below \cite{BD}, meaning that the ``sixth'' degree of freedom is a ghost (usually called the \emph{Boulware-Deser ghost}). Although this conclusion is premature, it has been shown explicitly that any non-linear completion of FP where the (non-linear) mass term is of the form
\beq
\mathcal{U}[\mbfg,\mbfg^{(0)}] = \mathcal{U} \big( \big( \e^{\m\a} \e^{\n\b} - \e^{\m\n} \e^{\a\b} \big) h_{\m\n} h_{\a\b} \big)
\eeq
with\footnote{This condition enforces the fact that the weak field limit is the Fierz-Pauli theory.} $\mathcal{U}^{\p}(0) = 1$, propagates six degrees of freedom and has an Hamiltonian which is unbounded from below. The non-linear completion originally considered by Vainshtein in \cite{Vainshtein72} (see section (\ref{The Vainshtein mechanism})) falls in this category, and is therefore plagued by ghost instabilities.

We could try to tackle the problem in full generality using the Hamiltonian formalism, and try to find consistency relations between the numerical coefficients $C_i$, $D_i$, $F_i$, \ldots above imposing that the theory does not propagate a sixth degree of freedom. However, this approach turns out to be very difficult to implement. Another approach is to first use appropriate limits and approximations of the theory to try to guess what a reasonable non-linear extension could be, and restrict the domain of possible values for the coefficients $C_i$, $D_i$, $F_i$, \ldots: only in a second moment would we use the Hamiltonian formalism, with the hope that the analysis of the selected class of actions turns out to be less cumbersome than the general analysis. We follow the latter approach: the tools we use to simplify the analysis of the non-linear massive actions are provided by the St\"uckelberg language in its full non-linear form, and the use of a ``decoupling'' limit which select relevant subsets of non-linear operators and focus on specific aspects/scales of the non-linear dynamics. To apply the St\"uckelberg formalism to interacting massive gravity, it will be more useful to write the action (\ref{actionmassivegravitygeneral}) in terms of the physical metric $\mbfg$ and the difference between the physical and absolute metric $\mbf{h} \equiv \mbfg - \mbfg^{(0)}$. In complete analogy with what has been done above, we can write 
\beq
\label{fasUk}
\sqrt{-g} \, \mathcal{U}[\mbfg,\mbfg^{(0)}] = \sqrt{-g} \, \sum_{k = 2}^{+\infty} U_{k}[\mbfg,\mbfh]
\eeq
where each term $U_{k}[\mbfg,\mbfh]$ has exactly the same structure of (\ref{Vketah}) with the only difference that each index raised factor $\eta^{\a \b}$ is now substituted with $g^{\a \b}$. Also, introducing the notation
\beq
\langle h^n \rangle \equiv g^{\m \a_{1}} \, h_{\a_{1} \b_{1}} \, g^{\b_{1} \a_{2}} \, h_{\a_{2} \b_{2}} \cdots g^{\b_{n - 1} \a_{n}} \, h_{\a_{n} \m}
\eeq
we can write the terms $U_{k}$ in the more compact way
\begin{align}
\label{U2equation}
U_2 [\mbfg,\mbfh] &= b_1 \langle h^2 \rangle + b_2 \langle h \rangle^2 \\[1mm]
U_3 [\mbfg,\mbfh] &= c_1 \langle h^3 \rangle + c_2 \langle h^2 \rangle \langle h \rangle + c_3 \langle h \rangle^3 \\[1mm]
U_4 [\mbfg,\mbfh] &= d_1 \langle h^4 \rangle + d_2 \langle h^3 \rangle \langle h \rangle + d_3 \langle h^2 \rangle^2 + d_4 \langle h^2 \rangle \langle h \rangle^2 + d_5 \langle h \rangle^4 \\[1mm]
U_5 [\mbfg,\mbfh] &= f_1 \langle h^5 \rangle + f_2 \langle h^4 \rangle \langle h \rangle + f_3 \langle h^3 \rangle \langle h \rangle^2 + f_4 \langle h^3 \rangle \langle h^2 \rangle + f_5 \langle h^2 \rangle^2 \langle h \rangle + \nn \\
\phantom{U} &+ f_6 \langle h^2 \rangle \langle h \rangle^3 + f_7 \langle h \rangle^5 \label{U5equation}\\[1mm]
\phantom{U} &\vdots \nn 
\end{align}
where again the requirement that the weak field limit should reproduce the Fierz-Pauli action implies that $b_2 = - b_1$. These two formulations (\emph{i.e.} in terms of $\eta$ and $\mbfh$ or $\mbfg$ and $\mbfh$) are completely equivalent, and the upper case numerical coefficients $C_i$, $D_i$, $F_i$, \ldots are biunivocally related to the lower case numerical coefficients $c_i$, $d_i$, $f_i$, \ldots: it is possible to see this explicitly expressing the inverse and the determinant of the full metric in terms of the inverse and determinant of the absolute metric
\begin{align}
g^{\mu\nu} &= \e^{\mu\nu} - \e^{\mu\a} \e^{\n\b} \Big( h_{\a\b} - \e^{\la \r} h_{\a\lambda} h_{\r \b} + \e^{\la \r} \e^{\s \t} h_{\a \lambda} h_{\r \s} h_{\t \b} + \cdots \Big) \\[1mm]
\sqrt{-g} &= 1 + \half \, \e^{\mu\nu} h_{\m\n} - \fourth \Big( \e^{\mu\nu} \e^{\a \b} - \half \, \e^{\mu\a} \e^{\nu \b} \Big) h_{\m \a} h_{\n \b} + \cdots 
\end{align}
and substituting in (\ref{fasUk})-(\ref{U5equation}) and finally comparing with (\ref{fasVk})-(\ref{V5equation}).

\subsection{The non-linear St\"uckelberg formalism}

We have seen in section (\ref{The Fierz-Pauli theory in the Stuckelberg language}) that the introduction of auxiliary fields which restore gauge invariance is a powerful tool in studying the Fierz-Pauli theory, since it elucidates the origin of the vDVZ discontinuity and allows to perform the $m \rightarrow 0$ limit of the theory without losing degrees of freedom. We would like to apply the same formalism to the full non-linear massive gravity, as first proposed by \cite{ArkaniHamedGeorgiSchwartz}. As we already mentioned, the theory contains two metrics, the physical metric $\mbfg$ which transforms \emph{covariantly} with respect to general coordinate transformations
\begin{align}
x^{\p\, \m} &= (f^{-1})^{\, \m} (x)\\[1mm]
g^{\p}_{\m\n} (x^{\p}) &= \frac{\de f^{\a}(x^{\p})}{\de x^{\p\, \m}} \, \frac{\de f^{\b}(x^{\p})}{\de x^{\p\, \n}} \, g_{\m\n} (f(x^{\p})) \quad ,
\end{align}
and the absolute metric $\mbfg^{(0)}$ (which we choose to be the Minkowski metric) which transform \emph{invariantly}
\beq
g^{(0) \, \p}_{\m\n} = g^{(0)}_{\m\n} = \e_{\m\n} \quad .
\eeq
To construct a new action which is physically equivalent to (\ref{actionmassivegravitygeneral}) and enjoys invariance with respect to general coordinate transformations, we first promote the absolute metric to a covariant tensor
\beq
\label{Nikki}
\e_{\m\n} \rightarrow \S_{\m\n}(x) \equiv \e_{\a\b} \, \frac{\de \phi^{\a}(x)}{\de x^{\m}} \, \frac{\de \phi^{\b}(x)}{\de x^{\n}}
\eeq
using four \emph{scalar} fields $\phi^{\a}(x)$ which are called the \emph{St\"uckelberg fields}. It can be checked that the chain rule for the derivative of composite functions gives the correct tensorial transformation law for $\S_{\m\n}(x)$. We then define the covariantisation of the difference between the physical and the absolute metric $h_{\m\n}(x) = g_{\m\n}(x) - \e_{\m\n}$ as
\beq
H_{\m\n}(x) \equiv g_{\m\n}(x) - \S_{\m\n}(x) \quad .
\eeq 
Now, remembering the expression
\beq
\label{actionmassivegravitygeneral2}
S = \int \! d^4 x \, \sqrt{-g} \, \bigg[ \, \frac{\MPq}{2} \Big( R[\mbfg] - \frac{m^2}{2} \, \mathcal{U}[\mbfg,h] \Big) + \mscr{L}_{M}[\mbfg,\psi_{(i)}] \, \bigg]
\eeq
where $\mathcal{U}[\mbfg,h] = \sum_{k = 2}^{+\infty} U_{k}[\mbfg,\mbfh]$ has the structure (\ref{U2equation})-(\ref{U5equation}), we can construct a theory which is diffeomorphism invariant by replacing
\beq
\label{htoH}
h_{\m\n}(x) \rightarrow H_{\m\n}(x) \quad .
\eeq
By construction, for every configuration of the St\"uckelberg fields $\phi^\a$ we can perform a suitable coordinate change such that the covariantized absolute metric $\S_{\m\n}(x)$ becomes the Minkowski metric: in this reference system, the covariantized theory and the original theory are equal, and so the two descriptions are physically equivalent.

\subsubsection{Perturbative expansion}
\label{Perturbative expansion}

In order to perform a perturbative analysis, it is useful to define a new object $Z^\a$ which can be considered the perturbation in the St\"uckelberg fields
\beq
\phi^{\a} = x^{\a} - Z^{\a}
\eeq
and so we can express $H_{\m\n}$ in terms of $h_{\m\n}$ and $Z^{\m}$ (we raise/lower indices with the Minkowski metric, so  $Z_{\n} = \e_{\n\a} Z^\a$)
\beq
\label{HmunuZ}
H_{\mu\nu} = h_{\mu\nu} + \de_{(\mu} Z_{\n)} - \e_{\a\b} \de_{\mu} Z^{\a} \de_{\nu} Z^{\b} \quad .
\eeq
Note that $Z^{\m}$ does \emph{not} transform as a vector with respect to general coordinate transformation: under infinitesimal coordinate transformations with gauge parameter $\xi^{\a}$ we have
\begin{align}
x^{\p \m} &= x^{\m} - \xi^{\m}(x)\\[1mm]
Z^{\p \m} &= Z^{\m} - \xi^{\m} + \xi^{\la} \, \de_{\la} Z^{\m} \\[1mm]
h^{\p}_{\m\n} &= h_{\m\n} + \de_{(\m} \, \xi_{\n)} + \mcalL_{\xi}(h)_{\m\n}
\end{align}
and we can see that at linear order $Z^{\m}$ simply shifts. As we did in the linear case, it is useful to introduce an additional substructure in $Z^{\m}$ singling out explicitly a derivative part and writing
\beq
\label{helicitydecZ}
Z_{\m} = A_{\m} + \de_{\m} \phi \quad ,
\eeq
and in terms of $A^{\a}$ and $\phi$ the tensor $H_{\m\n}$ reads
\begin{multline}
\label{stukelbergsecondreplace}
H_{\mu\nu} = h_{\m\n} + \de_{(\m} A_{\n)} + 2 \, \demden \phi - \de_{\m} A^{\a} \de_{\n} A_{\a} - \\[1mm]
- \de_{(\m} A^{\a} \de_{\n)} \de_{\a} \phi - \de_{\m} \de^{\a} \phi \, \de_{\n} \de_{\a} \phi \quad .
\end{multline}
Note that the decomposition (\ref{helicitydecZ}) is invariant with respect to the internal symmetry
\begin{align}
\phi(x) &\rightarrow \phi(x) - \La(x)\\[1mm]
A_{\a}(x) &\rightarrow A_{\a}(x) + \de_{\a} \La(x) \quad ,
\end{align}
and so the fields transform under the joint action of the two symmetries in the following way
\begin{align}
x^{\p \m} &= x^{\m} - \xi^{\m}(x)\\[1mm]
h^{\p}_{\m\n} &= h_{\m\n} + \de_{(\m} \, \xi_{\n)} + \mcalL_{\xi}(h)_{\m\n}\\[1mm]
A^{\p}_{\m} &= A_{\m} - \xi_{\m} + \xi^{\la} \, \de_{\la} A_{\m} + \de_{\m} \La\\[1mm]
\label{ultima}
\phi^{\p} &= \phi + \xi^{\la} \, \de_{\la} \phi - \La \quad .
\end{align}
At linear order, the relations (\ref{HmunuZ})-(\ref{ultima}) reduce to the analogous relations introduced in section (\ref{The Fierz-Pauli theory in the Stuckelberg language}) to study the Fierz-Pauli theory with the St\"uckelberg language. Note finally that $A_{\m}$ and $\phi$ does \emph{not} transform respectively as a vector and as a scalar with respect to general coordinate transformation, as a consequence of the fact that $Z^{\m}$ does not transform as a vector.
We will use in the following the notation
\beq
\label{Pidefinition}
\Pi_{\m\n} \equiv \demden \phi \quad .
\eeq

\section{St\"uckelberg analysis of non-linear massive gravity}

We want now to study the theory defined by the action (\ref{actionmassivegravitygeneral})
\beq
\label{actionmassivegravitygeneral3}
S = \int \! d^4 x \, \sqrt{-g} \, \bigg[ \, \frac{\MPq}{2} \Big( R[\mbfg] - \frac{m^2}{2} \mathcal{U}[\mbfg,\mbfg^{(0)}] \Big) + \mscr{L}_{M}[\mbfg,\psi_{(i)}] \, \bigg]
\eeq
from a perturbative point of view, similarly to what we did in section (\ref{GR as an interacting massless spin-2 theory}) when we interpreted the full theory of GR as a resummation of an infinite expansion in powers of perturbations of the metric around Minkowski spacetime. Expanding the action (\ref{actionmassivegravitygeneral3}) around the vacuum solution $g_{\m\n} = g^{(0)}_{\m\n} = \e_{\m\n}$, we would indeed obtain an interacting theory of the field $h_{\m\n}$. However, since we want to work with a gauge invariant formulation, we first introduce the St\"uckelberg fields by expressing the potential part $\sqrt{-g} \, \mathcal{U}[\mbfg,\mbfg^{(0)}]$ as in (\ref{fasUk}) and performing the replacement (\ref{htoH}). Expanding also the inverse physical metric $g^{\m\n}$ in terms of $h^{\m\n}$, we then obtain an interacting action expressed in terms of the fields $h_{\m\n}$, $A_{\m}$, $\phi$, where the interaction terms are expressed as linear combinations of powers of $h_{\m\n}$, $A_{\m}$, $\phi$ and their derivatives. In the following, we raise/lower indices on perturbation fields with the Minkowski metric.

\subsection{Interaction terms}

Note first of all that the introduction of the St\"uckelberg fields have no effect on the ``Einstein-Hilbert'' part of the action, since it has the same form of a gauge transformation and the Einstein-Hilbert term is gauge invariant. Therefore, the non-linear terms coming from this piece of the action do not contain the St\"uckelberg fields $A$ and $\phi$ and are exactly the same as in GR
\beq
\frac{\MPq}{2} \, \sqrt{-g} \, R[\mbfg] \sim \MPq \, \sum_{k=2}^{+ \infty} \de^2 h^k \sim \sum_{k=2}^{+ \infty} M_P^{2-k} \, \de^2 \tilde{h}^k
\eeq
where $\tilde{h}_{\m\n} = M_P h_{\m\n}$. On the other hand, the mass term is not gauge invariant: since $A_\m$ appears always derived once in the St\"uckelberg formalism and $\f$ appears always derived twice, the interaction terms coming from the mass term will be of the form
\beq
\label{interactionterms}
\frac{\MPq \, m^2}{4} \, \sqrt{-g} \, \mathcal{U} \supset \MPq \, m^2 \, h^i \, (\de A)^j \, (\de\de \f)^r \sim M_P^{2-i-j-r} \, m^{2-j-2r} \, \tilde{h}^i \, (\de \tilde{A})^j \, (\de\de \tilde{\f})^r
\eeq
with $i,j,r \geq 2$ and the tilde fields are defined as follows
\begin{align}
\tilde{h}_{\m\n} &= M_P \, h_{\m\n} \\[1mm]
\tilde{A}_{\m} &= M_P \, m \, A_{\m} \\[1mm]
\tilde{\f} &= M_P \, m^2 \, \f \quad .
\end{align}
To be more precise, note that every $U_k$ for $k \geq 2$ contains a piece $\big[ H^{k} \big]$ which contains all the combinations of the form $h^i \, (\de A)^j \, (\de\de \f)^r$ with $i + j +r = k$. Therefore, if we don't assume the Fierz-Pauli tuning, the most general mass term actually contains \emph{all} the possible combinations of terms of the type (\ref{interactionterms}) with $i +j +r \geq 2$ and $i$, $j$, $r$ non-negative. If we assume the Fierz-Pauli tuning, the quadratic part have a special form, while the interaction part (terms which are cubic or higher in the fields) contains all the possible combinations of terms of the type (\ref{interactionterms}) with $i +j +r \geq 3$ and $i$, $j$, $r$ non-negative.

\subsubsection{Quadratic term}

Let us look at the quadratic terms first, assuming the Fierz-Pauli tuning. They can be obtained using only the part of $H_{\m\n}$ which is linear in $h_{\m\n}$, $\de A_{\m}$, $\de \de \f$ (which we indicate with $\bar{H}_{\m\n}$) and replacing $g^{\m\n}$ with $\e_{\m\n}$, so it reads
\beq
- \frac{\MPq \, m^2}{4} \Big( \big[ \bar{H}^{2} \big] - \big[ \bar{H} \big]^{2} \Big)
\eeq
and is therefore equivalent to the mass term obtained in the St\"uckelberg analysis of the Fierz-Pauli action. Using the tilde fields, it contains (modulo total derivatives) a canonic kinetic term for $\tilde{A}_\m$, the FP mass term for $\tilde{h}_{\m\n}$, a mixing term $m \, \tilde{h} \de \tilde{A}$ and a kinetic mixing between $\tilde{h}$ and $\tilde{\f}$. Note that the quadratic terms in $\tilde{\f}$ appear in the combination
\beq
 \big[ \tilde{\Pi}^{2} \big] - \big[ \tilde{\Pi} \big]^{2}
\eeq
which is indeed a total derivative, however if we don't assume the Fierz-Pauli tuning we would get the term
\beq
b_1 \, \big[ \tilde{\Pi}^{2} \big] + b_2 \, \big[ \tilde{\Pi} \big]^{2}
\eeq
instead. This term is not a total derivative if $b_1 \neq - b_2$, and would give rise to higher derivative terms (\emph{i.e.} terms with derivatives of order three or higher) in the equation of motion for $\tilde{\f}$. Higher derivative terms in the equation of motion are usually associated with ghost instabilities, by the Ostrogradski theorem \cite{Ostrogradski1850,Woodard:2006nt}. This is consistent with the already mentioned result that any violation of the Fierz-Pauli tuning imply that the theory propagates also a sixth degree of freedom, which is a ghost \cite{VanNieuwenhuizen73}. The Fierz-Pauli mass term can therefore be uniquely identified in the St\"uckelberg language at quadratic order by the requirement that the scalar mode $\phi$ does not have higher derivative terms in the equations of motion.

\subsection{Strong coupling scales and decoupling limit}
\label{Strong coupling scales and decoupling limit}

Let us now turn to the interaction terms. As we already mentioned, a general non-linear extension of the Fierz-Pauli theory contains all the possible combinations of terms
\beq
\label{interactionterms2}
M_P^{2-i-j-r} \, m^{2-j-2r} \, \tilde{h}^i \, (\de \tilde{A})^j \, (\de\de \tilde{\f})^r
\eeq
with $i$, $j$, $r$ non-negative and $i+j+r \geq 3$. Note that each of the terms $\tilde{h}^i (\de \tilde{A})^j (\de\de \tilde{\f})^r$ is suppressed by a dimensionful factor
\beq
M_P^{i+j+r-2} \, m^{j+2r-2}
\eeq
where $M_P$ appears with positive power since $i+j+r \geq 3$. This factor sets a (mass) scale $\Lambda_{(ijr)}$
\beq
\label{Lambda}
\Lambda_{(ijr)}^{i+2j+3r-4} = M_P^{i+j+r-2} \, m^{j+2r-2}
\eeq
and, since the kinetic terms are in canonical form, the lowest of these mass scales is the strong coupling scale of the system, which is the scale where quantum corrections become non-negligible and need to be taken into account. Note that, despite $M_P$ appears always with positive power in the suppressing factor, $m$ appears with negative or zero power if $0 \leq j+2r \leq 2\,$: in these cases (which comprise the non-derivative self interaction of $\tilde{h}$ for example) the associated scale $\La$ is bigger than $M_P$. For the other cases (for which $j+2r > 2$) the associated scale $\La$ is smaller than $M_P$, and to see more clearly which is the lowest of these mass scales it is useful to write them in the following way 
\beq
\label{scalesformula}
\Lambda_{\lambda} = \sqrt[\lambda]{M_P \, m^{\lambda-1}} \quad ,
\eeq
where (as it follows from (\ref{Lambda})) we have
\beq
\lambda = \lambda(i,j,r) = \frac{i+2j+3r-4}{i+j+r-2} \quad .
\eeq
Since we assume $m \ll M_P$, we have that the bigger $\la$ the lower the scale $\Lambda_{\lambda}$. Note that in general $\la$ is a rational number: $\la \in \mathbb{Q}$. The strong coupling scale of the system is therefore set by the biggest allowed $\la$, which we call $\la_{max}$: once found $\la_{max}$, we can immediately read the strong coupling scale $\La_{sc} = \La_{\la_{max}}$ from (\ref{scalesformula}).

\subsubsection{Strong coupling scales}

To see which are the allowed values for $\la$, we note that at fixed $i,j$ the function $\lambda(i,j,r)$ becomes a function of $r$ only which is a hyperbola
\beq
\lambda_{i,j}(r) = \frac{ 3r -( 4 -i -2j)}{r -(2 -i-j)} \quad ,
\eeq
apart from the cases $(i,j) = (1,0)$ and $(i,j) = (0,2)$ where $\lambda_{i,j}(r) = 3$ and is independent of $r$. For the other cases, the hyperbola $\lambda_{i,j}(r)$ has the horizontal asymptote $\la = 3$ and the vertical asymptote $\la = 2 -i-j$. Since we have $i+j+r \geq 3$, at fixed $(i,j)$ (which must be positive) only the values $r \geq 3 - i -j$ are allowed, and since they are bigger that the position of the vertical asymptote, it follows that the allowed points $(r,\la(r))$ lie on the branch of the hyperbola which extends to $r \rightarrow +\infty$. It is easy to see that this branch is a decreasing function for the cases $(i,j) = (0,0)$ and $(i,j) = (0,1)$, while is an increasing function in the other cases (apart the particular cases $(i,j) = (1,0)$, $(i,j) = (0,2)$ as mentioned above). Furthermore, in the cases $(i,j) = (0,0)$, $(i,j) = (0,1)$ for which $\lambda_{i,j}(r)$ is a decreasing function, the biggest value for $\la$ is set by the lowest possible value for $r$, which is respectively $r = 3$ and $r = 2$. Therefore we conclude that the allowed values for $\la$ in the case $(i,j) = (0,0)$ lie in the range
\beq
(i,j) = (0,0) \qquad \Rightarrow \qquad 3 < \lambda_{0,0}(r) \leq 5 \quad , \quad r \geq 3 \quad ,
\eeq
and in particular we have
\begin{center}
\begin{tabular}{|l|l|c|c|c|c|c|c|}
\hline
$(i,j) = (0,0)$ & $r$ & $\rightarrow$ & 3 & 4 & 5 & 6 & $\cdots$ \\
\hline
\phantom{$(i,j) = (0,0)$} & $\la(r)$ & $\rightarrow$ & 5 & 4 & 11/3 & 7/2 & $\cdots$ \\
\hline
\end{tabular}
\end{center}
while for $(i,j) = (0,1)$ the allowed values for $\la$ lie in the range
\beq
(i,j) = (0,1) \qquad \Rightarrow \qquad 3 < \lambda_{0,1}(r) \leq 4 \quad , \quad r \geq 2
\eeq
and in particular we have
\begin{center}
\begin{tabular}{|l|l|c|c|c|c|c|c|}
\hline
$(i,j) = (0,1)$ & $r$ & $\rightarrow$ & 2 & 3 & 4 & 5 & $\cdots$ \\
\hline
\phantom{$(i,j) = (0,1)$} & $\la(r)$ & $\rightarrow$ & 4 & 7/2 & 10/3 & 13/4 & $\cdots$ \\
\hline
\end{tabular}
\end{center}
\phantom{a}

\begin{figure}[htp!]
\begin{center}
\includegraphics{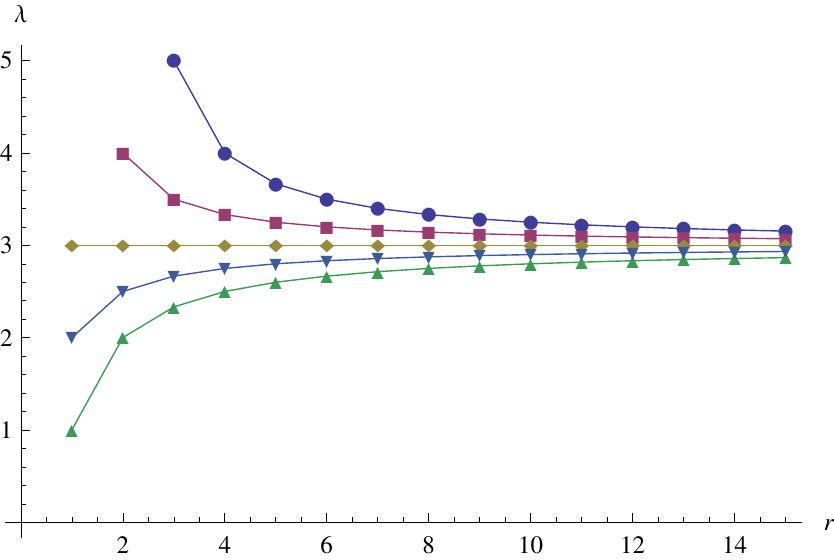}
\caption[Strong coupling scales and the hyperbolae $\big(r, \la_{i,j}(r)\big)$]{The points $\big(r, \la_{i,j}(r)\big)$ for the cases (top to bottom) $(i,j) = (0,0)$, $(i,j) = (0,1)$, $(i,j) = (0,2)$, $(i,j) = (1,1)$, $(i,j) = (2,0)$}
\label{MassiveHyperbolae}
\end{center}
\end{figure}

\noindent As already mentioned, for the cases $(i,j) = (1,0)$ $(i,j) = (0,2)$ we have
\beq
(i,j) = (1,0) \,\, \textrm{or} \,\, (0,2) \qquad \Rightarrow \qquad \lambda_{i,j}(r) = 3 = \textrm{constant} \quad ,
\eeq
while for the other cases we have
\beq
(i,j) \neq \{(0,0),(0,1),(1,0),(0,2)\} \qquad \Rightarrow \qquad \lambda_{i,j}(r) < 3
\eeq
since in the latter cases the relevant branch of the hyperbola is a monotonically increasing function and asymptotes the value $\la = 3$. For clarity, we plot the points $\big(r, \la_{i,j}(r)\big)$ for several choices of $(i,j)$ in figure \ref{MassiveHyperbolae}.

We then conclude that for a generic non-linear mass term (or equivalently for a generic choice of the coefficients $c_i$, $d_i$, $f_i$, $\ldots$) we have $\la_{max} = 5$ and the strong coupling scale of the system is
\beq
\La_{5} = \sqrt[5]{M_P \, m^{4}} \quad ,
\eeq
which is carried only by the cubic self-interaction term of $\tilde{\f}$
\beq
\frac{1}{\La_{5}^{5}} \, \big(\de^2 \tilde{\f}\big)^3 \quad .
\eeq
The second lowest scale is instead 
\beq
\La_{4} = \sqrt[4]{M_P \, m^{3}} \quad ,
\eeq
which is carried by the quartic self-interaction term of $\tilde{\f}$ and by the interaction term which is quadratic in $\tilde{\f}$ and linear in $\tilde{A}$
\beq
\frac{1}{\La_{4}^{8}} \, \big(\de^2 \tilde{\f}\big)^4 \qquad \qquad \frac{1}{\La_{4}^{4}} \, \de \tilde{A} \, \big(\de^2 \tilde{\f}\big)^2 \quad .
\eeq
We then have the higher order self-interaction terms of $\tilde{\f}$ with or without a term which linear in $\tilde{A}$
\beq
\propto \big(\de^2 \tilde{\f}\big)^n \qquad \qquad \propto \de \tilde{A} \, \big(\de^2 \tilde{\f}\big)^l
\eeq
with $n \geq 5$ and $l \geq 3$, which carry scales $\La_{\la}$ such that $3 < \la < 4$, and finally terms of the type
\beq
\frac{1}{\La_{3}^{3(s-1)}} \, \tilde{h} \, (\de\de \tilde{\f})^s \qquad \qquad \frac{1}{\La_{3}^{3p}} \, (\de \tilde{A})^2 \, (\de\de \tilde{\f})^p
\eeq
with $s \geq 2$ and $p \geq 1$, which carry the scale
\beq
\La_{3} = \sqrt[3]{M_P \, m^{2}} \quad .
\eeq
All the remaining terms carry scales $\La_{\la}$ such that $\la < 3$.

\subsubsection{The Vainshtein radius}

Having found the scale where quantum correction become important, we turn now to the scale where classical non-linearities become important. Let's consider a static spherically symmetric source of mass $M$: as we saw in section (\ref{The Fierz-Pauli theory in the Stuckelberg language}), in terms of the redefined fields $\hat{h}_{\m\n}$, $\hat{A}_{\m}$ and $\hat{\f}$ the kinetic terms are in canonical form, and a gauge can be chosen so that there are no mixed terms at quadratic order. Therefore, the fields profile at linear order are ($\sim$ here means ``apart from dimensionless factors'')
\beq
\hat{h}_{\m\n} \sim \frac{M}{M_P} \, \frac{1}{r} \qquad \qquad \hat{\f} \sim \frac{M}{M_P} \, \frac{1}{r}
\eeq
which is to be expected since the ``hatted'' fields, as well as the tilded ``fields'', has dimension $(\mathrm{length})^{-1}$. In particular this implies that also the tilded fields have the same behavior, modulo a gauge mode which has no effect since the theory is now gauge invariant. Therefore it is quite simple to see at which radius each interaction term becomes comparable to the quadratic terms in the action. An interaction term of the form
\beq
\label{interactionterms3}
M_P^{2-i-r} \, m^{2-2r} \, \tilde{h}^i \, (\de\de \tilde{\f})^r
\eeq
gives a contribution
\beq
\sim M_P^{2-i-r} \, m^{2-2r} \, \bigg(\frac{M}{M_P}\bigg)^{i+r} \, \bigg(\frac{1}{r}\bigg)^{i+3r} \quad ,
\eeq
while the quadratic terms give a contribution
\beq
\sim \bigg(\frac{M}{M_P}\bigg)^{\!2} \, \bigg(\frac{1}{r}\bigg)^{\!4}
\eeq
so an interaction term of the form (\ref{interactionterms3}) becomes comparable to the quadratic ones at the radius
\beq
\label{Keys}
r_{(ir)} \sim \Bigg[ \, m^{2-2r} \, \bigg( \frac{M}{\MPq} \bigg)^{i+r-2} \, \Bigg]^{1/(i+3r-4)} \quad .
\eeq

The largest of these radii is the one where the linear theory breaks down (at a classical level), and is therefore the Vainshtein radius of the theory. The interaction terms which correspond to this radius are the ones which first go non-linear when from spatial infinity we move towards the source: they are the only relevant interaction terms when we consider scales close to the Vainshtein radius. To see more clearly which is the biggest radius $r_{(ir)}$ defined by (\ref{Keys}) when $i+j+r \geq 3$, we write it in the following form
\beq
\label{Mission}
r_{(ir)} = r_{\m} = \sqrt[\m]{r_g r_c^{\m -1}}
\eeq
where we have introduced the Compton radius of the theory $r_c = 1/m$ and the gravitational radius $r_g = M/\MPq$ (which depends on the mass of the source). The hierarchy $M_P \gg m$ implies $r_g \ll r_c$, and so the bigger $\m$ the bigger $r_{\m}$: the Vainshtein radius is set by the maximum allowed value for $\m$, which we indicate with $\m_{max}$. Comparing (\ref{Keys}) with (\ref{Mission}) we find
\beq
\m = \m(i,r) = \frac{i+3r-4}{i+r-2} = \la
\eeq
and so $\m$ is precisely equal to the number $\la$ associated to the interaction term individuated by $(ijr) = (i0r)$ which we have introduced when studying the strong coupling scales. In particular, since both $\la_5$, $\la_4$ and $\la_3$ admits operators with $j = 0$, it follows that $\m_{max} = \la_{max}$. Therefore, the interaction terms which set the strong coupling scale are also the terms which set the Vainshtein radius: for the most general mass term the strong coupling scale is $\La_5$ and the Vainshtein radius is
\beq
\label{VainshteinRadius52}
r_{V} = \sqrt[5]{r_g r_c^{4}}
\eeq
where the only term which goes non-linear at this scale is the cubic self-interaction term for $\tilde{\f}$
\beq
\frac{1}{\La_{5}^{5}} \, \big(\de^2 \tilde{\f}\big)^3 \quad .
\eeq
We recover then the result (\ref{VainshteinRadius5}) obtained in a somewhat different way in section \ref{The Vainshtein mechanism}. In that case we were considering the particular non-linear extension of the Fierz-Pauli theory obtained adding the quadratic Fierz-Pauli term to the full non-linear GR action: this action in fact contains the cubic self-interaction term for $\tilde{\f}$, and so the Vainshtein radius is indeed (\ref{VainshteinRadius52}).

\subsubsection{The decoupling limit}

We have seen that there exists in the theory a special subclass of interaction terms which set both the strong coupling scale and the Vainshtein radius. We would like to define a formal limit of the theory which kills all the other interaction terms, and leaves us with a theory which contains only the kinetic terms and this special class of interaction terms.

We notice that, if we formally send $m \rightarrow 0$ and $M_P \rightarrow + \infty$ while keeping $\La_{sc}$ fixed, all the scales $\La$ bigger than $\La_{sc}$ diverge. Therefore, taking this formal limit in the action, all the interaction terms suppressed by scales larger than the strong coupling scale disappear. However, also the source term disappears since it is suppressed by $M_P$. If we want to construct a theory which contains only the desired interaction terms, but where the fields are still sourced by the energy and momentum of matter fields, we have to ask that also the energy-momentum tensor scales in some way in the limit, in order to compensate the fact that $M_P$ diverges. Therefore, we define the so called \emph{decoupling limit} (first introduced by \cite{LutyPorratiRattazzi} in the context of the DGP model) as
\beq
m \rightarrow 0 \,\,\, , \quad M_P \rightarrow +\infty \,\,\, , \quad T_{\m\n} \rightarrow +\infty \,\,\, , \quad \Lambda_{sc} \,\, \mathrm{and}\,\, \frac{T_{\m\n}}{M_P} \,\,\mathrm{fixed} \quad .
\eeq

By construction, this limit does not change the strong coupling scale of the theory and leaves untouched the Vainshtein radius. Therefore, we could see this formal limit as a way to focus on the behavior of the complete theory at the scales corresponding to the strong coupling and the Vainshtein radius: it seems likely that the decoupling limit should be appropriate to study the effectiveness of the Vainshtein mechanism.

\section{dRGT massive gravity}
\label{dRGT massive gravity}

We have so far introduced a very general class of actions (\ref{actionmassivegravitygeneral}) which can be seen as non-linear extensions of the Fierz-Pauli theory. We have then restored gauge invariance using the St\"uckelberg language, and identified the scales where quantum corrections and non-linearities become important. In this section and in the next chapter, we want to select a subset of actions which ought to describe a phenomenologically viable theory of an interacting massive spin-2 field. As we already mentioned, to be viable these actions have to be meet several requirements: they must propagate exactly five degrees of freedom (as many as the free theory of a massive spin-2 field), they have to be free of ghost instabilities, and they have to reproduce GR in the range of scales where GR is well tested, which practically translates to the requirement that there has to be an efficient screening mechanism at work (the Vainshtein mechanism in this case). In this section we deal with the first two requirements, namely the number of degrees of freedom and absence of ghosts, which are anyway closely related \cite{BD}. We will select a two-parameter class of actions, which are shown to propagate the correct number of degrees of freedom. We dedicate the next chapter, instead, to the study of the effectiveness of the Vainshtein mechanism in this restricted class of theories, with the aim to select the range of parameters for which the corresponding theory is phenomenologically viable.

\subsection{The $\La_3$ theory}

As we already mentioned, it is very difficult to impose the condition of having just five degrees of freedom by performing a Hamiltonian analysis of the general action (\ref{actionmassivegravitygeneral}). We instead try to reach the goal in two steps: first we select a subclass of actions which we expect to be good candidates for propagating five degrees of freedom, and only after that we apply the Hamiltonian formalism to properly count the numer of degrees of freedom.

\subsubsection{Arranging self-interactions in total derivatives}

We saw that, at quadratic level, the Fierz-Pauli action is the only action (apart an overall numerical factor) which has no ghosts and propagates exactly five degrees of freedom. We have also seen, using the St\"uckelberg language, that this requirement is precisely equivalent to the requirement that the scalar component $\f$ of the St\"uckelberg fields have no higher derivative terms in the equations of motion (which is in turn linked to the absence of ghosts by Ostrogradski theorem \cite{Ostrogradski1850, Woodard:2006nt}), which implies that quadratic terms in $\de \de \f$ in the action rearrange themselves to produce a total derivative term. We decide to follow this guideline also at full non-linear level, and therefore we look for actions of the form (\ref{actionmassivegravitygeneral}) where, \emph{at every order}, self-interaction terms in $\de \de \f$ rearrange themselves to produce total derivative terms. This is also consistent with the indications in \cite{ArkaniHamedGeorgiSchwartz,Creminelli:2005qk, Deffayet:2005ys} that the non-linear interactions of the scalar mode are related to sixth degree of freedom at full non-linear level.

Since we are (for the time being) only interested in self-interacting terms in $\f$, we may set
\beq
h_{\m\n} = 0 \qquad \qquad A_{\m} = 0 \qquad \qquad H_{\m\n} = 2 \, \Pi_{\m\n} - \Pi_{\m}^{\,\,\,\a} \, \Pi_{\a\n}
\eeq
where indices are raised/lowered with $\eta^{\m\n}/\eta_{\m\n}$ and $\Pi_{\m\n}$ is defined in (\ref{Pidefinition}). The only terms which survive are the ones belonging to the non-linear mass term, and the action takes the form
\beq
\label{maiocheneso}
S = - \frac{\MPq m^2}{4} \, \int \! d^4 x \, \sum_{k = 2}^{+\infty} U_{k}[\Pi]
\eeq
where
\begin{align}
\label{U2equationvabeh}
U_2 [\Pi] &= \big[ H^2 \big] - \big[ H \big]^2 \\[1mm]
U_3 [\Pi] &= c_1 \big[ H^3 \big] + c_2 \big[ H^2 \big] \big[ H \big] + c_3 \big[ H \big]^3 \\[1mm]
U_4 [\Pi] &= d_1 \big[ H^4 \big] + d_2 \big[ H^3 \big] \big[ H \big] + d_3 \big[ H^2 \big]^2 + d_4 \big[ H^2 \big] \big[ H \big]^2 + d_5 \big[ H \big]^4 \\[1mm]
U_5 [\Pi] &= f_1 \big[ H^5 \big] + f_2 \big[ H^4 \big] \big[ H \big] + f_3 \big[ H^3 \big] \big[ H \big]^2 + f_4 \big[ H^3 \big] \big[ H^2 \big] +\nn \\
\phantom{U} & \phantom{=} + f_5 \big[ H^2 \big]^2 \big[ H \big] + f_6 \big[ H^2 \big] \big[ H \big]^3 + f_7 \big[ H \big]^5\label{U5equationvabeh}\\[1mm]
\phantom{U} &\vdots \nn 
\end{align}

The idea is now to work perturbatively order by order, starting at order 3 and choosing (if possible) the coefficients $c_1$, $c_2$, $c_3$ such that the cubic piece in $\Pi$ contained in $U_2[\Pi] + U_3[\Pi]$ is a total derivative, then going to order 4 and choosing (if possible) the coefficients $d_1$, $d_2$, $d_3$, $d_4$, $d_5$ such that the quartic piece in $\Pi$ contained in $U_2[\Pi] + U_3[\Pi] + U_4[\Pi]$ is a total derivative, and so on. The first attempt to realize this program has been done in \cite{Creminelli:2005qk}, where it is was mistakenly concluded that there is no way to tune the free coefficients in (\ref{U2equationvabeh})-(\ref{U5equationvabeh}) in order to produce total derivatives at fourth order and above. Later, it has been proved in \cite{deRham:2010ik} (building on previous works \cite{Gabadadze:2009ja,deRham:2009rm,deRham:2010gu}) that 
it is indeed possible to carry on successfully this procedure at every order. It can be shown \cite{NicolisRattazziTrincherini} that, at every order in $\Pi$, there is essentially only one linear combination of contractions of $\Pi$ which is a total derivative, which at order $n$ is explicitly
\beq
\label{TDtermn}
\mathcal{L}^{TD}_{n}(\Pi) = \sum_{p \in P_n} (-1)^p \, \e^{\m_1 p(\n_1)}\, \cdots \, \e^{\m_n p(\n_n)} \, \Pi_{\m_1 \n_1} \, \cdots \, \Pi_{\m_n \n_n}
\eeq
where the sum runs on all the permutations $p$ of $n$ elements. ``Essentially'' means that all the other linear combination of contractions of $\Pi$ at order $n$ which are total derivatives, are actually proportional to $\mathcal{L}^{TD}_{n}(\Pi)$. Note that, for $n \geq 5$, the sum in (\ref{TDtermn}) vanishes identically by symmetry reasons: therefore, at each order $n$ there is a one-dimensional variety of total derivative terms if $n = 2, 3$ and $4$, while for $n \geq 5$ the variety is zero-dimensional: the total derivative structures have in total three free parameters.

It is actually not difficult to see that it is always possible to tune the coefficients in (\ref{U2equationvabeh})-(\ref{U5equationvabeh}) to rearrange the terms in total derivatives at all orders. If we fix $n$ and insert in $U_n(\Pi)$ only the part of $H_{\m\n}$ which is \emph{linear} in $\Pi$, we generate the most general linear combination of contraction of $n$ tensors $\Pi_{\m\n}$ with $n$ inverse metrics $\eta^{\a\b}$. Therefore we can always use the free coefficients in $U_n(\Pi)$ to compensate exactly for the terms of order $n$ in $\Pi$ which come from the lower orders of the potential, and create the total derivative combination (\ref{TDtermn}) at each order. Furthermore, since there are three free parameters in the total derivatives combinations which correspond to the orders $n = 2,3$ and $4$, there will be a three-parameter class of Lagrangians where the $\f$ self-interactions are removed at all orders. The parameter coming from order two is reabsorbed in the overall mass parameter $m$ in the action, so we end up with a genuinely two-parameters class of massive actions. Explicitly, the values of the tuned coefficients in (\ref{U2equationvabeh})-(\ref{U5equationvabeh}) are \cite{deRham:2010ik} to fourth order
\beq
c_1 = 2 c_3 + \half \quad \quad c_2 = -3 c_3 - \half
\eeq
\begin{align}
d_1 &= -6 d_5 +\frac{1}{16} (24 c_3 +5) & d_2 &= 8 d_5 -\frac{1}{4} ( 6 c_3 +1) \\[1mm]
d_3 &= 3 d_5 - \frac{1}{16} (12 c_3 +1) & d_4 &= - 6 d_5 + \frac{3}{4} c_3 \quad .
\end{align}

\subsubsection{The ef\mbox{}fect on the strong coupling scale}

Considering now the strong coupling scale of the theory, from what we said in section \ref{Strong coupling scales and decoupling limit} we can immediately conclude that the removal of all $\f$ self-interaction terms raises the strong coupling scale to $\La_4$, which is carried by the term
\beq
\frac{1}{\La_{4}^{4}} \, \de \tilde{A} \, \big(\de^2 \tilde{\f}\big)^2 \quad .
\eeq
However, it can be shown \cite{ArkaniHamedGeorgiSchwartz,deRham:2010ik} that the choice of coefficients in the non-linear mass term which remove the self-interaction terms in $\f$, automatically remove also the terms of the form
\beq
M_P^{1-l} \, m^{1-2l} \, \de \tilde{A} \, (\de\de \tilde{\f})^l
\eeq
with $l \geq 2$, which carry the strong coupling scales $\La_\la$ with $4 \geq \la > 3$. Therefore, removing the scalar self-interactions actually raises the strong coupling scale to
\beq
\La_{3} = \sqrt[3]{M_P \, m^{2}}
\eeq
which is carried by terms of the form
\beq
\frac{1}{\La_{3}^{3(s-1)}} \, \tilde{h} \, (\de\de \tilde{\f})^s \qquad \qquad \frac{1}{\La_{3}^{3p}} \, (\de \tilde{A})^2 \, (\de\de \tilde{\f})^p
\eeq
with $s \geq 2$ and $p \geq 1$. Note that these terms are the only terms which survive in the decoupling limit, since we proved in section \ref{Strong coupling scales and decoupling limit} that all the other interaction terms are suppressed by scales $\La_\la$ with $\la < 3$. The two-parameters theory defined by tuning the interaction terms so as to remove the $\f$ self-interactions is usually called the $\La_{3}$ theory.

Note that the vector field $A_{\m}$ does not couple directly to $T^{\m\n}$, and therefore setting it to zero and solving for $h_{\m\n}$ and $\f$ always gives consistent solutions of the theory. This however does not mean that $A_{\m}$ does not play any role. Since $A_{\m}$ couples to $h_{\m\n}$ and $\f$, the most general solution of the theory contains also the $A_{\m}$ field, and in fact the vector sector may contain ghost instabilities (at least around some backgrounds) \cite{Koyama:2011wx}. Setting anyway $A_{\m}$ to zero for the time being, the decoupling limit Lagrangian up to total derivatives is given by the kinetic term for $\tilde{h}_{\m\n}$ plus the part of the mass term which is linear in $\tilde{h}_{\m\n}$. As shown in \cite{deRham:2010ik}, it has at most quartic couplings in $\tilde{h}_{\m\n}$ and $\tilde{\f}$ and explicitly reads
\begin{multline}
\label{lambda3decoupling}
S = \int \! d^4x \, \Bigg[ \tilde{h}^{\mu\nu} {\cal E}_{\mu\nu}^{\,\,\,\,\,\r\s} \tilde{h}_{\r\s} - \half \, \tilde{h}^{\mu\nu} \bigg( -4 \tilde{X}^{(1)}_{\mu\nu}(\tilde{\f}) + \frac{4 (6 c_3 - 1)}{\Lambda_3^3} \, \tilde{X}^{(2)}_{\mu\nu}(\tilde{\f}) + \\[2mm]
\shoveright{+ \frac{16 (8 d_5 +c_3)}{\La_3^6} \, \tilde{X}^{(3)}_{\mu\nu}(\tilde{\f}) \bigg) + \frac{1}{M_P} \, \tilde{h}_{\mu\nu} T^{\mu\nu} \Bigg]}\\
\phantom{a}
\end{multline}
where the operator ${\cal E}_{\mu\nu}^{\,\,\,\,\,\r\s}$ has been defined in (\ref{Epsilon operator}) and the tensors $\tilde{X}^{(n)}_{\mu\nu}$ are of order $n$ in $\tilde{\Pi}$ and are defined in the Appendix \ref{Appendix Total derivative combinations}. Note finally that, in the decoupling limit, the Lagrangian has a finite number of interaction terms between $\tilde{h}_{\mu\nu}$ and $\tilde{\f}$, while it has an infinite number of interaction terms between $\tilde{h}_{\mu\nu}$ and $\tilde{A}_{\mu}$.

\subsubsection{De-mixing in the decoupling limit and galileons} 
\label{Demixing in the decoupling limit and galileons}

In the decoupling limit Lagrangian (\ref{lambda3decoupling}), the scalar mode $\tilde{\f}$ does not have a kinetic term on its own but is kinetically mixed to $\tilde{h}_{\mu\nu}$: furthermore, all the interaction terms are in mixed form. To make more transparent the physical meaning of this action, we would like to disentangle as much as we can the dynamics of $\tilde{h}_{\mu\nu}$ and that of $\tilde{\f}$.

First of all, we kinetically de-mix $\tilde{h}_{\mu\nu}$ and $\tilde{\f}$ by redefining the fields, as we did in section (\ref{The Fierz-Pauli theory in the Stuckelberg language}), and going to the ``hatted'' fields: this transformation creates a canonical kinetic term for $\hat{\f}$, as well as coupling $\hat{\f}$ to the trace of the energy-momentum tensor $T$. 
At this point there are still couplings $\hat{h} \, \hat{\Pi}^2$ and $\hat{h} \, \hat{\Pi}^3$ between $\hat{h}$ and $\hat{\f}$, while derivative self-interaction terms for $\hat{\f}$ have appeared. It is possible to further de-mix the action and remove the cubic $\hat{h} \, \hat{\Pi}^2$ coupling, performing the field redefinition
\beq
\check{h}_{\mu\nu} = \hat h_{\mu\nu} + \frac{2 (6 c_3 -1)}{\Lambda_3^3} \, \partial_\mu \hat\phi \, \partial_\nu \hat \phi \quad .
\eeq
After this operation the Lagrangian reads
\begin{multline}
\label{lambda3decouplingdiag2}
S = \int \! d^4x \, \Bigg[ \, \check{h}^{\mu\nu} {\cal E}_{\mu\nu}^{\,\,\,\,\,\r\s} \check{h}_{\r\s} + \frac{\mathcal{C}_1}{\Lambda_3^6} \, \check{h}^{\mu\nu} \check{X}^{(3)}_{\mu\nu} + \frac{1}{M_P} \, \check{h}_{\mu\nu} T^{\mu\nu} - \\
- 3 \, (\de \check{\f} \cdot \de \check{\f}) + \frac{\mathcal{C}_2}{\Lambda_3^3} \, (\de \check{\f} \cdot \de \check{\f}) \, \square \check{\f} + \frac{\mathcal{C}_3}{\Lambda_3^6}\, (\de \check{\f} \cdot \de \check{\f}) \, \Big( [\check{\Pi}]^2 - [\check{\Pi}^2] \Big) + \\
+ \frac{\mathcal{C}_4}{\Lambda_3^9} \, (\de \check{\f} \cdot \de \check{\f}) \, \Big( [\check{\Pi}]^3 -3 [\check{\Pi}^2] [\check{\Pi}] +2 [\check{\Pi}^3] \Big) + \\
+ \frac{1}{M_P} \, \check{\f}\, T + \frac{\mathcal{C}_5}{\Lambda_3^3 M_P} \, \de_\mu \check{\f} \, \de_\nu \check{\f} \, T^{\mu\nu} \, \Bigg]
\end{multline}
while it is instead not possible to de-mix further the action and remove the quartic mixing $\check{h} \, \check{\Pi}^3$ keeping the action local, since only a nonlocal field redefinition could remove that mixing term. The notation $\big( \de \check{\f} \cdot \de \check{\f} \big)$ here stands for $\big( \de_{\a}\!\check{\f} \, \de^{\a}\!\check{\f} \big)$, while the numerical coefficients $\mathcal{C}_1$, $\mathcal{C}_2$, $\mathcal{C}_3$, $\mathcal{C}_4$ and $\mathcal{C}_5$ depend only on $c_3$ and $d_5$, and their explicit expression can be found for example in \cite{HinterbichlerReview}
\begin{align}
\mathcal{C}_1 &= -8 (8 d_5 + c_3) \\[2mm]
\mathcal{C}_2 &= 6 (6 c_3 - 1) \\[2mm]
\mathcal{C}_3 &= -4 \big( (6 c_3 - 1)^2 -4 (8 d_5 + c_3) \big) \\[2mm]
\mathcal{C}_4 &= -40 (6 c_3 - 1) (8 d_5 + c_3) \\[2mm]
\mathcal{C}_5 &= 2 (6 c_3 - 1) \qquad \qquad \qquad \qquad \qquad .
\end{align}
Note that they are all written in terms of the combinations $6 c_3 - 1$ and $8 d_5 + c_3$, so they all disappear from the action when both these combinations vanish. More precisely, the coupling $\check{h}^{\mu\nu} \check{X}^{(3)}_{\mu\nu}$ disappears when $8 d_5 + c_3 = 0$, irrespectively of whether $6 c_3 - 1 = 0$ vanishes or not, while the coupling $\de_\mu \check{\f} \, \de_\nu \check{\f} \, T^{\mu\nu}$ disappear when $6 c_3 - 1 = 0$, irrespectively of the value of $d_5$.

The action (\ref{lambda3decouplingdiag2}) has several interesting features. First, note that, beside the coupling $\check{\f}\, T$ of the scalar mode with the trace $T$ of the energy-momentum tensor, there is a new form of coupling between $\check{\f}$ and the energy-momentum tensor which involves the derivatives $\de \check{\f}$ and \emph{not} the trace $T$. This implies in particular that the scalar mode $\check{\f}$ couples also to the electromagnetic field, whose energy-momentum tensor is traceless. Second, turning to the interaction terms, apart from the mixed term $\check{h}^{\mu\nu} \check{X}^{(3)}_{\mu\nu} \sim \check{h} \, \check{\Pi}^{3}$ (which disappear from the action when $8 d_5 + c_3 = 0$), the scalar mode has now three self-interaction terms, respectively at order 3, 4 and 5. Dropping the symbol $\check{\phantom{\f}}$ for clarity, the kinetic and the self-interaction terms have the structure
\begin{align}
\mathcal{L}_{2} &= - \half \, \big( \de \f \cdot \de \f \big) \\
\mathcal{L}_{3} &= - \half \, (\de \f \cdot \de \f) \big[ \Pi \big] \\
\mathcal{L}_{4} &= - \half \, (\de \f \cdot \de \f) \big( \big[ \Pi \big]^2 - \big[ \Pi^2 \big] \big) \\
\mathcal{L}_{5} &= - \half \, (\de \f \cdot \de \f) \Big( \big[ \Pi \big]^3 -3 \big[ \Pi \big] \big[ \Pi^2 \big] +2 \big[ \Pi^3 \big] \Big) \quad .
\label{normalGalileons}
\end{align}
These terms are known as \emph{Galileon terms} \cite{NicolisRattazziTrincherini}, and have the defining property that they give rise to equations of motion where the field appears only derived twice, and that they are invariant with respect to the ``galilean'' transformation
\beq
\f \rightarrow \f + b_{\m} x^{\m} + c
\eeq
(for the sake of precision, the Lagrangians are not invariant themselves but the galilean transformation produce a total derivative, therefore the action is invariant). It can be shown \cite{NicolisRattazziTrincherini} that at each order in $\f$ they are the only terms with these properties, up to total derivatives. Historically, apart from the quadratic term, the first of these terms to be studied was the cubic galileon term, which describes the dynamics of the brane bending mode in the decoupling limit of the DGP model (see section \ref{The Vainshtein radius}). It has later been recognized that, in general, an action which produces non-linear equations of motion in which the field appears only through its second derivatives, can be used to modify gravity at large distances since the field may shield itself around a spherical source via the Vainshtein mechanism \cite{NicolisRattazziTrincherini}.

Note that the scalar mode of the St\"uckelberg fields trivially enjoys the galilean symmetry, since by construction it appears only derived twice. Instead, the absence of higher derivatives in the equations of motion (despite the Lagrangian containing second derivatives already) is highly nontrivial. The fact that the decoupling limit of the $\La_3$ theory produces only self-interactions of galileon type, which are ghost free, is a promising signal that the full theory may be indeed free of the BD ghost. Even more, it has been argued in \cite{deRham:2010ik} that the complete decoupling limit Lagrangian (containing also $\check{h}_{\m\n}$ and its coupling with $\check{\f}$) is indeed free of ghosts. Note finally that the galileon interaction terms arise in the decoupling limit only when we de-mix the fields $\tilde{h}_{\m\n}$ and $\tilde{\f}$: in particular, the first transformation $(\tilde{h},\tilde{\f}) \rightarrow (\hat{h},\hat{\f})$ (which de-mixes the kinetic terms) create the cubic and quartic galileon terms, and the second trasformation $(\hat{h},\hat{\f}) \rightarrow (\check{h},\check{\f})$ (which eliminates the $\hat{h} \, \hat{\Pi}^2$ coupling) creates also the fifth galileon term. The de-mixing procedure is on the other hand responsible for the coupling of $\f$ to matter: initially, the field $\tilde{\f}$ in fact does not couple with $T_{\m\n}$; the first redefinition (which removes the kinetic $\tilde{h} \, \Pi$ term) creates the ``trace'' coupling $\hat{\f}\, T$, while the second redefinition (which removes the $\hat{h} \, \hat{\Pi}^2$ term) creates the ``derivative'' coupling $\de_\mu \check{\f} \, \de_\nu \check{\f} \, T^{\mu\nu}$.

\subsection{Resummation of $\La_3$ massive gravity}

In the previous sections we saw that there is a way to tune order by order the coefficients of a generic non-linear extension of the Fierz-Pauli action, in order to avoid the appearence of higher derivatives in the equations of motion for the scalar mode of the St\"uckelberg fields. Although the theory is uniquely defined (once we specify the values of the free parameters), and we could be just satisfied with this perturbative formulation, we may like to reformulate it in a more compact and manageable form.

In fact, if we want to find exact solutions of a theory which is defined as the sum of a perturbative expansion, we have to solve iteratively the equations of motion at each order, obtaining the full solution as an infinite expansion (which quite often we are not able to sum explicitly). If instead we are able to define the theory in an already resummed form, to find exact solutions we have to solve just one equation (although with several components), which is however intrinsically non-linear. This is often more convenient, especially because in general it is easier to perform a Taylor expansion of an object than to resum a perturbative expansion.

\subsubsection{The square root formulation}

We would like then to provide a resummed form of the theory of non-linear massive gravity we defined so far. To do that, we should identify an object which makes it possible to express the full action as the sum of a finite number of terms. Looking back to the problem of rearranging the $\f$ self-interaction terms in total derivatives, we notice that the reason why the tuning of coefficients goes on to an infinite number of orders is that, in the St\"uckelberg language, the generic non-linear mass term is expressed as a power series of $H_{\m\n}$, which is \emph{quadratic} in $\Pi_{\m\n}$. As a consequence, every order $n$ of the potential generates terms in $\Pi_{\m\n}$ which are of order $m > n$, and, as we construct the total derivative at order $n$, we are generating higher order terms which will need to be taken care of. We could try instead to express the generic mass term (\ref{fasUk})-(\ref{U5equation}) of a non-linear extension of FP in terms of an object which is \emph{linear} in $\Pi$, at least when $h_{\m\n}$ and $A_\m$ are vanishing since the condition we want to impose involves $\f$ self-interactions only. 

In fact, this is possible if we define the object \cite{deRham:2010kj}
\beq
\label{Ktensor}
\mathcal{K}^\mu_{\,\,\nu} (g,H) \equiv \delta^\mu_{\,\,\nu} - \sqrt{\delta^\mu_{\,\,\nu} - H^\mu_{\,\,\nu}}
\eeq
where $H^{\m}_{\,\,\n} = g^{\m\la} H_{\la\n}$ and the square root of a matrix $\mathcal{A}^{\m}_{\,\,\n}$ is defined as the matrix $\mathcal{R}^{\m}_{\,\,\n}$ such that $\mathcal{A}^{\m}_{\,\,\n} = \mathcal{R}^{\m}_{\,\,\a} \, \mathcal{R}^{\a}_{\,\,\n}$. Since $\mathcal{K}^\mu_{\,\,\nu}$ can be expressed (at least perturbatively, when its components are small) as power series of $H^\mu_{\,\,\nu}$
\beq
\label{Kassum}
\mathcal{K}^\mu_{\,\,\nu} = \sum_{n=1}^{\infty} \, \tilde{\b}_n \, ( H^n)^\mu_{\,\,\nu} \qquad \qquad  \tilde{\b}_n = - \frac{(2n)!}{(1-2n)(n!)^2 4^n} \quad ,
\eeq 
the most general non-linear extension of the Fierz-Pauli theory (\ref{actionmassivegravitygeneral}) can be expressed as an expansion in powers of the tensor $\mathcal{K}^\mu_{\,\,\nu}$ 
\beq
\label{fasWk}
\sqrt{-g} \, \mathcal{U}[\mbfg,\mbfg^{(0)}] = \sqrt{-g} \, \sum_{k = 2}^{+\infty} W_{k}[\mathcal{K}]
\eeq
where
\begin{align}
\label{W2equation}
W_2 [\mathcal{K}] &= \langle \mathcal{K}^2 \rangle - \langle \mathcal{K} \rangle^2 \\[1mm]
W_3 [\mathcal{K}] &= \tilde{c}_1 \langle \mathcal{K}^3 \rangle + \tilde{c}_2 \langle \mathcal{K}^2 \rangle \langle \mathcal{K} \rangle + \tilde{c}_3 \langle \mathcal{K} \rangle^3 \\[1mm]
W_4 [\mathcal{K}] &= \tilde{d}_1 \langle \mathcal{K}^4 \rangle + \tilde{d}_2 \langle \mathcal{K}^3 \rangle \langle \mathcal{K} \rangle + \tilde{d}_3 \langle \mathcal{K}^2 \rangle^2 + \tilde{d}_4 \langle \mathcal{K}^2 \rangle \langle \mathcal{K} \rangle^2 + \tilde{d}_5 \langle \mathcal{K} \rangle^4 \label{W4equation}\\[1mm]
W_5 [\mathcal{K}] &= \tilde{f}_1 \langle \mathcal{K}^5 \rangle + \ldots \label{W5equation}\\[1mm]
\phantom{W} &\vdots \nn 
\end{align}
and where the angled brackets here mean
\beq
\langle \mathcal{K}^n \rangle = \mathcal{K}^{\m}_{\,\,\a_2} \, \mathcal{K}^{\a_2}_{\,\,\a_3} \, \cdots \, \mathcal{K}^{\a_n}_{\,\,\m} \quad .
\eeq
On the other hand, if we set $h_{\m\n}=0$ and $A_{\m}=0$, remarkably the powers of the linear and the quadratic pieces in $\Pi$ which constitute $H_{\m\n}$ nearly cancel out, when the power expansion of the square root (\ref{Kassum}) is performed, leaving only the linear term
\beq
\mathcal{K}^{\mu}_{\,\,\nu}\Big\rvert_{h=0,A=0} = \delta^\mu_{\,\,\nu} - \sqrt{\delta^\mu_{\,\,\nu} - \big( \Pi^\mu_{\,\,\nu} - \Pi^\mu_{\,\,\a} \, \Pi^\a_{\,\,\nu} \big)} = \Pi^{\mu}_{\,\,\nu}
\eeq
and so $\mathcal{K}^\mu_{\,\,\nu}$ is precisely equal to $\Pi^{\mu}_{\,\,\nu}$ when $h_{\m\n}=0$ and $A_{\m}=0$. Therefore, it is much simpler to impose the condition that the self-interaction terms of $\f$ rearrange in total derivatives when we express the non-linear mass term in terms of $\mathcal{K}^\mu_{\,\,\nu}$, since it reduces to the conditions
\begin{align}
W_3 [\Pi] &= \a_3 \, {\cal L}_3^{\rm TD}(\Pi)\\[1mm]
W_4 [\Pi] &= \a_4 \, {\cal L}_4^{\rm TD}(\Pi)\\[1mm]
W_5 [\Pi] &= 0 \\[1mm]
W_6 [\Pi] &= 0 \\[1mm]
\phantom{W} &\vdots \nn
\end{align}
without any higher order tuning. Comparing with (\ref{galileon3})-(\ref{galileon4}), we deduce
\begin{align}
\tilde{c}_1 &= 2 \, \a_3 & \tilde{c}_2 &= -3 \, \a_3 & \tilde{c}_3 &= \a_3 \\[1mm]
\tilde{d}_1 &= -6 \, \a_4 & \tilde{d}_2 &= 8 \, \a_4 & \tilde{d}_3 &= 3 \, \a_4 & \tilde{d}_4 &= -6 \, \a_4 & \tilde{d}_5 &= \a_4
\end{align}
while $\tilde{f}_i$ and all the coefficients of the orders of $W_k$ higher than four vanish. The coefficients $\a_3$ and $\a_4$ are free parameters, and correspond to the free parameters $c_3$ and $d_5$ in the other formulation.

\subsubsection{The resummed action}

To get the complete action of non-linear massive gravity, we have to reintroduce in some way the fields $h_{\m\n}$ and $A_{\m}$. Since the tensor $\mathcal{K}^\mu_{\,\,\nu}$ naturally contains them, we can \emph{define} the complete action of non-linear massive gravity to be expressed in terms of $\mathcal{K}^\mu_{\,\,\nu}$ precisely in the same way as it is in the case $h_{\m\n}=0$ and $A_{\m}=0$: the action in the resummed form then reads

\beq
\label{resummedactionfinal}
S = \int \! d^4 x \, \sqrt{-g} \, \bigg[ \, \frac{\MPq}{2} \Big( R[\mbfg] - \frac{m^2}{2} \mathcal{U}[\mathcal{K}] \Big) + \mscr{L}_{M}[\mbfg,\psi_{(i)}] \, \bigg]
\eeq
where

\beq\label{resummedpotentialfinal}
\mathcal{U}[\mbfg,\mathcal{K}] = \mathcal{U}_2[\mathcal{K}] + \a_3 \, \mathcal{U}_3[\mathcal{K}] + \a_4 \, \mathcal{U}_4[\mathcal{K}]
\eeq
and
\begin{align}
\mathcal{U}_2 &= (\mathrm{tr} \mathcal{K})^2-\mathrm{tr} (\mathcal{K}^2) \label{Ithilien}\\[1mm]
\mathcal{U}_3 &= (\mathrm{tr} \mathcal{K})^3 - 3 (\mathrm{tr} \mathcal{K})(\mathrm{tr} \mathcal{K}^2) + 2 \, \mathrm{tr} \mathcal{K}^3 \\[1mm]
\mathcal{U}_4 &= (\mathrm{tr} \mathcal{K})^4 - 6 (\mathrm{tr} \mathcal{K})^2 (\mathrm{tr} \mathcal{K}^2) + 8 (\mathrm{tr} \mathcal{K})(\mathrm{tr} \mathcal{K}^3) + 3 (\mathrm{tr} \mathcal{K}^2)^2 - 6 \, \mathrm{tr} \mathcal{K}^4 \quad . \label{Silmaril}
\end{align}
The infinite series of terms which made up the mass term in the previous formulation is expressed, in the resummed form, with just three terms. Note that in \eqref{Ktensor} we have defined the tensor $\mathcal{K}$ in terms of $H^\mu_{\,\,\nu} = g^{\m\a} \, H_{\a\n}$, where $H_{\m\n}$ is the ``covariantization'' of the difference $h_{\m\n}$ between the physical metric $g_{\m\n}$ and the absolute metric $g^{(0)}_{\m\n}$. To construct the theory, we found more convenient to express the theory in terms of $h_{\m\n}$ and $g_{\m\n}$, but now we want to express the full resummed action in terms of the absolute and physical metrics themselves. Remembering that $H_{\m\n}$ is defined as
\beq
H_{\m\n} = g_{\m\n} - \S_{\m\n} \quad ,
\eeq
where the $\boldsymbol{\S}$ tensor is the ``covariantization'' of the absolute metric $g^{(0)}_{\m\n} = \e_{\m\n}$ and is defined as
\beq
\label{Ryuken}
\S_{\m\n}(x) = g^{(0)}_{\a\b} \, \frac{\de \phi^{\a}(x)}{\de x^{\m}} \, \frac{\de \phi^{\b}(x)}{\de x^{\n}} \quad ,
\eeq
we have that
\beq
\delta^\mu_{\,\,\nu} - H^\mu_{\,\,\nu} = g^{\m\a} \S_{\a\n} \quad .
\eeq
We can therefore express the $\mathcal{K}$ tensor in terms of the physical metric $\mbfg$, the absolute metric $\mbfg^{(0)}$ and the St\"uckelberg fields $\phi^{\a}$ as

\beq
\label{Eriador}
\mathcal{K}^\mu_{\,\,\nu} = \d^{\mu}_{\,\,\nu} - \Big[ \sqrt{\mathbf{g^{-1}} \cdot \boldsymbol{\S}\,}\, \Big]^{\mu}_{\,\,\,\,\nu}
\eeq
where the dot stands for the matrix multiplication operation. 

The last expression, together with \eqref{resummedactionfinal} - \eqref{Silmaril}, defines the theory in the resummed form. Note that, by construction, the theory is reparametrization-invariant, by means of the St\"uckelberg fields $\phi^{\a}$. The introduction of the St\"uckelberg fields and the restoration of gauge invariance proved in fact to be very helpful in clarifying the analysis of a general non-linear extension of the Fierz-Pauli theory. However, as we stressed above, a theory with gauge invariance restored by means of St\"uckelberg fields is completely equivalent from a physical point of view to a theory without St\"uckelberg fields where gauge invariance is broken. Without using the St\"uckelberg formalism, the non-linear theory of massive gravity we obtained is described by the action

\beq
\label{resummedactionfinalnonStuckelberg}
S = \int \! d^4 x \, \sqrt{-g} \, \bigg[ \, \frac{\MPq}{2} \Big( R[\mbfg] - \frac{m^2}{2} \mathcal{U} \big[ \mbfg , \mbfg^{(0)} \big] \Big) + \mscr{L}_{M}[\mbfg,\psi_{(i)}] \, \bigg]
\eeq
where

\beq\label{resummedpotentialfinalnonStuckelberg}
\mathcal{U} \big[ \mbfg , \mbfg^{(0)} \big] = \mathcal{U}_2 \Big[ \sqrt{\mathbf{g^{-1}} \cdot \mbfg^{(0)} \,} \,\Big] + \a_3 \, \mathcal{U}_3 \Big[ \sqrt{\mathbf{g^{-1}} \cdot \mbfg^{(0)} \,} \,\Big] + \a_4 \, \mathcal{U}_4 \Big[ \sqrt{\mathbf{g^{-1}} \cdot \mbfg^{(0)} \,} \,\Big]
\eeq
and the explicit form of the potentials can be obtained plugging in \eqref{Ithilien}-\eqref{Silmaril} the expression

\beq
\label{EriadornonStuckelberg}
\mathcal{K}^\mu_{\,\,\nu} = \d^{\mu}_{\,\,\nu} - \Big[ \sqrt{\mathbf{g^{-1}} \cdot \mbfg^{(0)} \,}\, \Big]^{\mu}_{\,\,\,\,\nu} \quad .
\eeq

\subsubsection{Absence of the Boulware-Deser mode and prior geometry}

We go back now to the problem of the number of degrees of freedom. As we already mentioned, a legitimate interacting theory of a massive graviton has to propagate five degrees of freedom, as many as a massive spin-2 field propagates. The absence of a sixth degree of freedom is also important from the point of view of the stability of the theory, since the additional degree of freedom is usually associated with ghost instabilities (Boulware-Deser ghost). The number of degrees of freedom can in principle be established recasting the theory in Hamiltonian form, however (as we said above) performing a full Hamiltonian analysis on the most general non-linear extension of Fierz-Pauli action is very hard. By restoring gauge invariance and asking that the scalar component of the St\"uckelberg fields does not have higher derivatives in the equations of motion, it has been possible to single out a two-parameters class of non-linear extensions of the Fierz-Pauli theory. The hope is that the Hamiltonian analysis of this restricted class of theories turns out to be easier to perform.

A full Hamiltonian analysis on this restricted class of actions has indeed been performed in \cite{Hassan:2011ea, Mirbabayi:2011aa, Kluson:2012wf, Hinterbichler:2012cn}, with the result that it has been confirmed that these actions propagate exactly five degrees of freedom. Therefore, the theories defined by \eqref{Ithilien} - \eqref{Silmaril} and \eqref{resummedactionfinalnonStuckelberg} - \eqref{EriadornonStuckelberg} are legitimate interacting theories of a massive graviton, and are known as \emph{dRGT Massive Gravity} (from the name of the authors de Rham, Gabadadze and Tolley) or also \emph{Ghost-Free Massive Gravity}. The latter denomination is due to the fact that in these theories the Boulware-Deser ghost is absent. However, it is fair to say that the absence of the BD ghost does not imply that the theory is ghost-free, since some of the five degrees of freedom may still be a ghost, at least on some backgrounds \cite{Koyama:2011wx}. Leaving aside this issue, a necessary condition for these theories to be phenomenologically viable is that they reproduce GR results on length scales/configurations where these results are experimentally tested. This implies that they have to admit static spherically symmetric solutions where the Vainshtein mechanism is effective. In the next chapter, we will systematically study static and spherically symmetric solutions in the dRGT massive gravity theories, to characterise in which part of the phase space of theories spanned by ($\a_3$, $\a_4$) we can find solutions which display the Vainshtein mechanism. This is a crucial step in establishing the phenomenological viability of non-linear massive gravity.

Note that the absolute metric $\mbfg^{(0)}$ is explicitly present in the resummed action \eqref{resummedactionfinalnonStuckelberg} - \eqref{EriadornonStuckelberg}, therefore the dRGT Massive Gravity has a prior geometry, which is set by the absolute metric. This is in stark constrast with GR, where the absolute metric disappears from the resummed action when we substitute $h_{\m\n}$ with $g_{\m\n} - g^{(0)}_{\m\n}$, and so there is no prior geometry. It follows in particular that each choice for the absolute geometry generates a different theory of non-linear massive gravity. On the other hand, we can see that the theory really depends on the absolute \emph{geometry}, and not on the coordinates chosen to express the absolute metric. In fact, let's consider two absolute metrics $g^{(0)}_{\m\n}$ and $g^{(0) \p}_{\m\n}$ which describe the same absolute geometry, and so are linked by a change of coordinates: we may introduce an absolute metric manifold $\mathscr{M}_{(0)}$, and two system of references $y^\m$ and $y^{\p \m}$ on $\mathscr{M}_{(0)}$, so that
\beq
g^{(0) \p}_{\m\n} = \frac{\de y^{\a}}{\de y^{\p \m}} \frac{\de y^{\a}}{\de y^{\p \n}} \, g^{(0)}_{\a\b} \quad .
\eeq
The physical metric in general is determined by the absolute metric and the energy-momentum tensor. Let's consider on one side the theory associated with the absolute metric $g^{(0)}_{\m\n}$, and consider a source term $T_{\m\n}$ in this theory, and on the other side the theory associated with the absolute metric $g^{(0) \p}_{\m\n}$, and consider in this second theory a source term $T^{\p}_{\m\n}$ which is linked to $T_{\m\n}$ by the same relation which links $g^{(0)}_{\m\n}$ and $g^{(0) \p}_{\m\n}$
\beq
T^{\p}_{\m\n} = \frac{\de y^{\a}}{\de y^{\p \m}} \frac{\de y^{\a}}{\de y^{\p \n}} \, T_{\a\b} \quad .
\eeq
Let's call $g_{\m\n}$ the solution for the physical metric in the first theory and $g^{\p}_{\m\n}$ the solution for the physical metric in the second theory. If $g_{\m\n}$ and $g^{\p}_{\m\n}$ are \emph{not} linked by the same relation which links the absolute metrics and the source terms, then we may say that the dRGT massive gravity depends not only on the absolute geometry, but also on the coordinate system chosen to express the absolute metric. Conversely, if $g_{\m\n}$ and $g^{\p}_{\m\n}$ are indeed linked by the relation
\beq
g^{\p}_{\m\n} = \frac{\de y^{\a}}{\de y^{\p \m}} \frac{\de y^{\a}}{\de y^{\p \n}} \, g_{\a\b} \quad ,
\eeq
then we may say that the dRGT massive gravity depends only on the absolute geometry, and not on the coordinate system chosen to express the absolute metric.

It is in fact not difficult to see that the latter case is the correct one. In fact, despite the fact that the action \eqref{resummedactionfinalnonStuckelberg} is not invariant with respect to coordinate changes (which change the physical metric and the energy-momentum tensor but leaves untouched the absolute metric), the action is invariant with respect to the \emph{formal} transformation
\beq
\label{formaltransf}
g_{\m\n} \rightarrow g^{\p}_{\m\n} = \frac{\de y^{\a}}{\de y^{\p \m}} \frac{\de y^{\a}}{\de y^{\p \n}} \, g_{\a\b} \quad g^{(0)}_{\m\n} \rightarrow g^{(0) \p}_{\m\n} = \frac{\de y^{\a}}{\de y^{\p \m}} \frac{\de y^{\a}}{\de y^{\p \n}} \, g^{(0)}_{\a\b} \quad T_{\m\n} \rightarrow T^{\p}_{\m\n} = \frac{\de y^{\a}}{\de y^{\p \m}} \frac{\de y^{\a}}{\de y^{\p \n}} \, T_{\a\b}
\eeq
as a consequence of the structure $\sqrt{\mathbf{g^{-1}} \cdot \mbfg^{(0)}}$ in the potential. This is more in general a consequence of the fact that we started from the general action \eqref{actionmassivegravitygeneral} whose potential term is written in terms of contractions of the inverse of the physical metric $g^{\m\n}$ and of the difference between the physical and absolute metric $h_{\m\n} = g_{\m\n} - g^{(0)}_{\m\n}$.

\subsection{Cosmology in dRGT massive gravity}
\label{Cosmology in dRGT massive gravity}

Before turning to the study of spherically symmetric solutions and the Vainshtein mechanism, it is worthwhile to discuss briefly the cosmology of dRGT Massive Gravity, since our interest in modified gravity theories was motivated by the idea to address the cosmological late time acceleration problem.

As we explained in chapter \ref{Introduction}, the study of cosmology is usually performed by modeling the physical spacetime $\mscr{M}$ with a homogeneous and isotropic spacetime $\bar{\mscr{M}}$, whose evolution is assumed to trace the large scale behavior of the real universe. In GR, to find the evolution of the homogeneous and isotropic metric $\bar{\mathbf{g}}$ it is enough to go the the reference system where it assumes the Robertson-Walker form, and solve the Einstein equations where the source is obtained by spatially averaging the ``real'' energy-momentum tensor of the universe. The situation in dRGT massive gravity is more complicated, since there is a prior geometry (which we still assume to be flat): this implies that it is not sufficient to say that the physical metric $\bar{g}_{\m\n}$ is homogeneous and isotropic, but we need to give more information. Consider in fact the formulation of the dRGT massive gravity without St\"uckelberg fields, defined by the equations \eqref{Ithilien}-\eqref{Silmaril} and  (\ref{resummedactionfinalnonStuckelberg})-(\ref{EriadornonStuckelberg}). A homogeneous and isotropic metric, when written in a generic coordinate system, contains more unknown functions than a Robertson-Walker metric, which contains only the scale factor (we assume here that we fixed the sign of the spatial curvature). In GR, all these unknown functions disappear from the equations of motion when we go to the comoving reference system, apart from the scale factor: however, in dRGT massive gravity without Stuckelberg fields, these functions do not disappear from the equations of motion, since the theory is not reparametrization-invariant, and the evolution of the scale factor depends on their expression. If we instead use the formulation defined by \eqref{resummedactionfinal} - \eqref{Silmaril} and (\ref{Eriador}), where the St\"uckelberg fields have been introduced and diffeomorphism invariance has been restored, the change of coordinates from the reference system where the fiducial metric is Minkowski to the reference which is comoving with the isotropic observers in general excites the St\"uckelberg fields, and so we get different equations for the scale factor depending on the form which the physical metric had in the reference system where the fiducial metric was Minkowski.

\subsubsection{The mass term as a cosmological constant}

We want now to understand if the dRGT massive gravity admits cosmological solutions and find what type of expansion histories we may obtain. To do that, it is convenient to restore diffeomorphism invariance by means of the St\"uckelberg fields, and look for solutions where the physical metric is of the Robertson-Walker form. Following \cite{Gratia:2012wt}, it is convenient to consider the following line element for the physical metric
\beq
\label{RWmetricisotropicmg}
ds_{\textup{phys}}^2 = - b^2(t,\r) \, dt^2 + a^2(t,\r) \bigg[ d \r^2 + \r^2 \big( d\theta^2 + \sin^2 \!\theta \, d\phi^2  \big) \bigg]
\eeq
and consider a spherically symmetric ans\"atz for the St\"uckelberg fields
\begin{equation}
\left\{
  \begin{aligned}
   \f^{0} &= f(t, \r) \\[1mm]
   \f^{i} &= g(t, \r) \, \frac{x^i}{r} \quad .
  \end{aligned}
\right.
\label{Forlimpopoli}
\end{equation}
Note that the Robertson-Walker metric is a particular case of the line element (\ref{RWmetricisotropicmg}): in fact, redefining the radial coordinate $r \rightarrow \r$ in the line element (\ref{RWmetric2}) according to $r(\r) = 4\r/(4 + k \r^2)$ (isotropic coordinates\footnote{Note that for $k = 0$ $\r$ is defined on $[0 , + \infty[$ while in the $k = \pm 1$ cases $\r$ is defined on $[0 , + 2[$.}), we obtain a line element of the form (\ref{RWmetricisotropicmg}) where $b(t,\r) = 1$ and
\beq
a(t,\r) = \frac{a(t)}{1+ k \, \r^2/4}
\eeq
where $a(t)$ is the scale factor in the $(t,r,\theta,\f)$ coordinate system. Note furthermore that the configurations \eqref{Forlimpopoli} for the St\"uckelberg fields are isotropic with respect to the origin $\r = 0$ but in general inhomogeneous; the homogeneous and isotropic configurations correspond to the case $\f^{0} = \f^{0}(t)$ and $\f^{i} = x^i$, which are included as the particular case $f(t,\r) = f(t)$ and $g(t, \r) = \r$. From (\ref{RWmetricisotropicmg}) and (\ref{Forlimpopoli}) we can now construct the fiducial metric $\boldsymbol{\S}$ using (\ref{Ryuken}) and the tensor $\mcalK$ using (\ref{Eriador}): the equations of motion for the physical metric and the St\"uckelberg fields are obtained by varying the action (\ref{resummedactionfinal}) with respect to the fields $a$, $b$, $f$ and $g$. Note in particular that we can express the equation of motion for the ``isotropic scale factor'' $a(t, \r)$ and for the lapse function $b(t ,\r)$ in the form of modified Einstein equations
\beq
\bar{G}_{\m\n} = m^2 \, T^{(\mcal{K})}_{\m\n} + \frac{1}{M_P^2} \, T_{\m\n}
\eeq
by defining the effective energy-momentum tensor
\beq
T^{(\mcal{K})}_{\m\n} \equiv \frac{1}{\sqrt{-g}} \, \frac{\d}{\d g^{\m\n}} \int \! d^4 x \, \sqrt{-g} \, \, \frac{\mathcal{U}[\mathcal{K}]}{2}
\eeq
which encodes the contribution of the non-linear mass term to the equations of motion for the physical metric.

The equations of motion for the St\"uckelberg fields $f$ and $g$ are a coupled system of non-linear partial differential equations. Remarkably, there exist \cite{Gratia:2012wt} two branches of solutions of this system of equations where $g$ is independent of $f$ and is linked to the isotropic scale factor $a$ by the simple relation
\beq
\label{Johnny Depp}
g(t,\r) = x_0 \, \r \, a(t,\r) \quad ,
\eeq 
where $x_0$ is a number which is fixed by the values of the free parameters $\a_3$ and $\a_4$ of the model according to
\beq
\label{IkeaVillesse}
x_0 = x_0 (\a_3, \a_4) = \frac{1 + 6 \a_3 + 12 \a_4 \pm \sqrt{1 + 3 \a_3 + 9 \a^2_3 - 12 \a_4}}{3 (\a_3 + 4 \a_4)}
\eeq
and the choice of the sign in $\pm$ distinguishes the two branches. The field $f$, on the other hand, in this case obeys a non-linear partial differential equation in which only $a$ and $b$ appear (since $g$ can be expressed in terms of $a$ using the relation (\ref{Johnny Depp})): this equation is extremely difficult to solve in general and admits different solutions, depending on the boundary condition $f(t , 0)$ which we choose. A priori we expect that, in order to find the expansion history of the universe, we need to find the solution for $f$, since $f$ in general appears in the effective energy-momentum tensor $T^{(\mcal{K})}_{\m\n}$. However, surprisingly enough, for the two branches of solutions we are considering it can be proved \cite{Gratia:2012wt} that the field $f$ disappears from $T^{(\mcal{K})}_{\m\n}$ once we use the relation (\ref{Johnny Depp}) and the equation of motion for $f$. Moreover, the effective energy-momentum tensor $T^{(\mcal{K})}_{\m\n}$ has exactly the form of a perfect fluid with $w = -1$ at rest in the comoving reference, and explicitly we have
\beq
T^{(\mcal{K})}_{\m\n} = - \half \, P_0(x_0) \, g_{\m\n}
\eeq
where $P_0$ is the polynomial
\beq
P_0(x) = -12 -2x (x-6) -12 (x-1) (x-2) \, \a_3 -24 (x-1)^2 \, \a_4 \quad .
\eeq
Note that $T^{\m \,(\mcal{K})}_{\,\,\,\n}$ is not only homogeneous and isotropic but in fact constant. This implies that, concerning these branches of solutions, the fields $a$ and $b$ in dRGT massive gravity obey exactly the same equations that the isotropic scale factor and the lapse function obey in GR in presence of a cosmological constant $\La$ equal to
\beq
\label{windyday}
\La = \frac{m^2}{2} \, P_0 \big( x_0 \big) \quad .
\eeq
We remind that both $P_0$ and $x_0$ depend on the parameters $\a_3$ and $\a_4$. Therefore, the cosmological expansion history in these two branches of solutions in dRGT massive gravity is exactly the same that we obtain in GR when there is a cosmological constant $\La$ whose value is related to $m$, $\a_3$ and $\a_4$ by (\ref{windyday}). It is worthwhile to emphasize that this result do not imply only that the mass term in the dRGT massive gravity models can produce a cosmological acceleration acting at late times as an effective cosmological constant: it acts as a cosmological constant during \emph{all} the expansion history of the universe. Therefore, at the level of the background expansion, dRGT massive gravity and $\La$CDM are not distinguishable, and in particular choosing $m$, $\a_3$ and $\a_4$ carefully these branches of background solutions produce a very good fit to the observational data. This is very important from the point of view of the cosmological constant problem, since the observed value for the effective cosmological constant $\La_{\textup{eff}}$, which is very unnatural if it is due to a true cosmological constant (see section \ref{The acceleration problem}), may be more natural if it is due to the mass $m$ of the graviton.

Note that, very unexpectedly, the dRGT mass term produce a homogeneous and isotropic contributions to the Einstein equations even if the configuration of the St\"uckelberg fields is inhomogeneous (since the field $f$ for the solutions above is inhomogeneous for a generic choice of the boundary condition $f(t,0)$). This possibility was overlooked in \cite{D'Amico:2011jj}, where only homogeneous and isotropic configurations for the St\"uckelberg fields were considered: since it turned out that in this case the theory admits only solutions where the universe is static \cite{D'Amico:2011jj}, it was claimed that no viable homogeneous and isotropic solutions exist in the dRGT massive gravity. A subsequent paper \cite{Gumrukcuoglu:2011ew} found explicitly solutions where both the physical and the fiducial metric are homogeneous and isotropic (in particular the physical metric has negative spatial curvature): it can be shown \cite{Tasinato:2013rza} that these solutions belong to the class of self-accelerating solutions we described above, and correspond to a specific choice of $f(t,0)$. Surprisingly, it can be shown \cite{Gratia:2012wt} that the contribution of the dRGT mass term to the effective Einstein equations remains of the form of a cosmological constant even if not only the St\"uckelberg fields but also the physical metric is inhomogeneous (but still isotropic), and therefore also when the source configuration is isotropic and inhomogeneous.

\subsubsection{The non-linear instability}

To decide if the self accelerating cosmological solutions in dRGT massive gravity are able to fit all the observational data, and therefore provide an explanation for the cosmic acceleration alternative to $\La$CDM, it is necessary to go beyond the background homogeneous and isotropic expansion and study the behavior of perturbations. On one hand, this is necessary because (as we already mentioned) we can observationally probe the behavior of cosmological perturbations at linear and also at non-linear level; on the other hand, it is necessary to make sure that the homogeneous and isotropic solutions are stable.

The study of perturbations around self-accelerating backgrounds in dRGT massive gravity (still considering only the case of flat absolute geometry) concentrated on two families of exact solutions, the self-accelerating solutions (KNT) found in \cite{Koyama:2011xz, Koyama:2011yg}, which are a subset of a class of spherically symmetric solutions (the non-diagonal branch, in the language of chapter \ref{Vainshtein Mechanism in Massive Gravity}), and the ``open'' FLRW self-accelerating solutions (GLM) found in \cite{Gumrukcuoglu:2011ew}. Both these families of solutions belong to the class of cosmological solutions discussed above \cite{Gratia:2012wt, Tasinato:2013rza}, and correspond to different choices of the boundary condition $f(t,0)$: in particular, for the KNT solutions the fiducial metric is isotropic but inhomogeneous, while for the GLM solutions the fiducial metric is homogeneous and isotropic. The physical metric is, as we already mentioned, homogeneous and isotropic in both cases.

The study of perturbations around the solutions KNT reveals \cite{Tasinato:2012ze} that vector perturbations have vanishing kinetic terms at quadratic order in perturbations (at the level of the action). This implies that their dynamics is controlled by higher order interactions, since only if their kinetic terms vanish at every order in the perturbative expansion we can conclude that these fields are not dynamical. Going to higher orders in perturbations, it has been shown \cite{Tasinato:2012ze} that the Hamiltonian is unbounded from below, and therefore the KNT solutions are generically unstable. Regarding the GLM solutions, it has been shown \cite{Gumrukcuoglu:2011zh} that at quadratic order in perturbations the scalar and vector degrees of freedom have vanishing kinetic terms, and so also in this case it is necessary to study perturbations at higher order. A subsequent analysis has shown \cite{DeFelice:2012mx} that all the cosmological solutions which respect homogeneity and isotropy (both for the fiducial and the physical metric) are unstable. These results have been confirmed by the analysis of \cite{Khosravi:2013axa}, where it was shown also that the behavior of perturbations is very different depending on the fact that the fiducial metric is homogeneous or not. Note that, in both cases (KNT and GLM), the instability suggests the presence of a ghost, but the ghost field belongs to the five propagating degrees of freedom of the theory and therefore is not the BD ghost.

We conclude that, to find stable cosmological solutions in the dRGT massive gravity, we have to consider cosmological solutions which are either inhomogeneous or anisotropic \cite{DeFelice:2012mx,DeFelice:2013bxa}. It has been proposed in fact that the theory should admit solutions where the physical spacetime is inhomogeneous at scales \emph{larger} that the Hubble horizon, and where the usual Friedmann-Lema\^{\i}tre-Robertson-Walker cosmology is reproduced due to the Vainshtein mechanism \cite{D'Amico:2011jj}. In this case, the universe would feel the effect of mass of the graviton only when the average density drops below a crossover scale, and roughly speaking the Vainshtein radius of the universe becomes smaller than the Hubble radius. On the other hand, anisotropic solutions have been considered in \cite{Gumrukcuoglu:2012aa}, and it has been shown in \cite{DeFelice:2013awa} that these solutions can be ghost-free for a range of parameters and initial conditions.

\clearemptydoublepage

\chapter{The Vainshtein mechanism in dRGT massive gravity}
\label{Vainshtein Mechanism in Massive Gravity}

In the previous chapter we introduced a class of non-linear completions of the Fierz-Pauli action, known as dRGT massive gravity, which are free of the Boulware-Deser ghost and so seem to be potentially phenomenologically viable. To provide a reliable description of the gravitational interaction, they necessarily have to pass stringent experimental constraints, and agree with the predictions of GR which have been tested to a very high accuracy. A necessary condition for this to happen is that the vDVZ discontinuity is cured by non-linear interactions, or in other words that the Vainshtein mechanism is effective. In particular, since this class of actions has two free parameters (the Fierz-Pauli action has already a free parameter, the mass), it is crucial to understand for which values of the free parameters the Vainshtein mechanism works, and so to identify the regions in the phase space of free parameters which correspond to phenomenologically viable theories. The aim of this chapter is to find a precise answer to this problem. Therefore, we study static, spherically symmetric vacuum solutions in the dRGT massive gravity model with flat absolute geometry, and classify the types of solutions that the theory admits. We then determine in which regions of the two parameters phase space the Vainshtein mechanism is effective.

\section{Spherically symmetric solutions}

We consider the theory defined by equations \eqref{Ithilien} - \eqref{Silmaril} and \eqref{resummedactionfinalnonStuckelberg} - \eqref{EriadornonStuckelberg} in the case where the absolute geometry is flat. To study static and spherically symmetric solutions in this case, we start by expressing the absolute metric $\mbfg^{(0)}$ in spherical coordinates, which are more suited to the symmetry of the problem
\beq
\label{Becky}
ds^2 = g^{(0)}_{\m\n} dy^{\m} dy^{\n} =  -dt^2 + dr^2 + r^2 d \Omega^2
\eeq
where $y^{\m} = (t,r,\theta,\varphi)$ indicates collectively the spherical coordinates. The most general form for the physical metric allowed by the requirement that the latter be static and spherical symmetric is
\beq
\label{genmetr}
d s^2 \, = \, - C(r) \, d t^2 + A(r)\, d r^2 + 2 D(r)\, dt dr + B(r) d \Omega^2
\eeq
and, varying the action (\ref{resummedactionfinal}) and considering vacuum regions, we obtain the following equation of motion
\beq
G_{\mu \nu} = \frac{m^2}{2} T^{{\cal U}}_{\mu \nu}
\eeq
where we have defined 
\beq
T^{{\cal U}}_{\mu \nu} = \frac{1}{\sqrt{-g}}\,\frac{ \delta \sqrt{-g}\ {\cal U}}{\delta g^{\mu \nu}} \quad .
\eeq

\subsection{The two branches}

For metrics of the form (\ref{genmetr}), the Einstein tensor $G_{\mu\nu}$ satisf\mbox{}ies the identity
\beq
D(r)\, G_{tt}+C(r)\,G_{tr}\,=\,0
\eeq
which implies the following algebraic constraint on $T^{{\cal U}}_{\mu \nu}$
\beq
D(r)\, T^{{\cal U}}_{tt} + C(r)\,T^{{\cal U}}_{tr} = 0 \quad .
\eeq
This last equation reduces to
\beq
D(r) \left(b_0 r - \sqrt{B(r)} \right) = 0
\eeq
where $b_0$ is a function of $\alpha_3$ and $\alpha_4$ only \cite{Koyama:2011wx}. This constraint is solved in two possible ways, def\mbox{}ining two class of solutions: either the metric is diagonal $D=0$, which defines the \emph{diagonal branch}, or $B=b_0^2 r^2$, which defines the \emph{non-diagonal branch}. Note that it is possible to map a physical metric belonging to the diagonal branch into one of the non-diagonal branch via a change of coordinates, and viceversa. However, in dRGT massive gravity these two branches are physically distinct. To see it, it is convenient to restore gauge invariance by using the St\"uckelberg formalism. Consider, before introducing the St\"uckelberg fields, a configuration where the absolute metric has the form (\ref{Becky}) and indicate with $\bar{g}_{\m\n}$ a solution of the equations of motion belonging to the diagonal branch, while indicate with $\bar{\bar{g}}_{\m\n}$ a solution of the equations of motion belonging to the non-diagonal branch. We then introduce the St\"uckelberg fields $\phi^{\m}$ and form the ``covariantized'' version of the absolute metric
\beq
\S_{\m\n}(x) = g^{(0)}_{\a\b} \, \frac{\de \phi^{\a}(y)}{\de y^{\m}} \, \frac{\de \phi^{\b}(y)}{\de y^{\n}} \quad ,
\eeq
where (analogously to section \ref{Perturbative expansion}) we decompose the St\"uckelberg fields $\phi^{\m}$ in the following way
\begin{align}
\phi^{t} &= t - \ti{Z}^{t} \\[1mm]
\phi^{r} &= r - \ti{Z}^{r} \\[1mm]
\phi^{\theta} &= \theta - \ti{Z}^{\theta} \\[1mm]
\phi^{\varphi} &= \varphi - \ti{Z}^{\varphi} \quad .
\end{align}
Substituting the absolute metric $\mbfg^{(0)}$ with $\boldsymbol{\S}$ in the action restores gauge invariance in the theory, and it is customary to call \emph{unitary gauge} the situation when $\ti{Z}^{\m} = 0$. Therefore, the configurations $\big( \bar{g}_{\m\n},g^{(0)}_{\m\n} \big)$ and $\big( \bar{\bar{g}}_{\m\n},g^{(0)}_{\m\n} \big)$ we introduced above correspond, upon introducing the St\"uckelberg fields, to a situation where the physical metric is respectively $\bar{g}_{\m\n}$ and $\bar{\bar{g}}_{\m\n}$ in the unitary gauge. Suppose we now change coordinates and map $\bar{g}_{\m\n}$ into a metric $\bar{g}^{\p}_{\m\n}$ which belongs to the non-diagonal branch: the change of coordinates excites some components of the St\"uckelberg fields. Both $\bar{g}^{\p}_{\m\n}$ and $\bar{\bar{g}}_{\m\n}$ are non-diagonal metrics, but in the first case the St\"uckelberg fields are non-zero, while in the second case they vanish. Since the St\"uckelberg fields explicitly appear in the equations of motion, we conclude that $\bar{g}^{\p}_{\m\n}$ and $\bar{\bar{g}}_{\m\n}$ obey \emph{different} equations of motion, and therefore are different. This implies that there are indeed two physically distinct branches of static and spherically symmetric solutions. This is in stark contrast with the GR case, where the theory is gauge invariant without the need to introduce the St\"uckelberg fields. In that case, $\bar{g}^{\p}_{\m\n}$ and $\bar{\bar{g}}_{\m\n}$ obey \emph{the same} equations of motion, and so the two branches are physically identical.

As we shall see shortly, the Vainshtein mechanism in the diagonal branch is related to the role of non-linearities for the radial component of the St\"uckelberg fields. However, it has been shown \cite{deRham:2010tw} that, in the non-diagonal branch, the scalar mode of the St\"uckelberg fields does not couple directly to the energy-momentum tensor in the decoupling limit. In fact, the results of GR in this branch are reproduced without the need of the Vainshtein mechanism: the non-diagonal branch is very interesting and it can be shown that in this branch static, spherically symmetric solutions leads to Schwarzschild or Schwarzschild-(anti) de Sitter solutions \cite{Koyama:2011xz, Koyama:2011yg, Nieuwenhuizen:2011sq, Gruzinov:2011mm, Berezhiani:2011mt, Volkov:2012wp, Comelli:2011wq}. Other interesting discussions on the non-diagonal branch can be found for example in \cite{Gumrukcuoglu:2011ew, Gumrukcuoglu:2011zh, Koyama:2011wx}.

Anyway, we conclude that the only branch which is relevant for the Vainshtein mechanism is the diagonal one: therefore, from now on we will consider only the diagonal branch.

\subsection{The diagonal branch}

To study the diagonal branch, let's start from the following ansatz for the physical metric
\beq
ds^2 = - \ti{N}(r)^2 dt^2 + \ti{F}(r)^{-1} dr^2 + r^2 \ti{H}(r)^{-2} d \Omega^2 \quad ,
\label{diagonal1}
\eeq
and the form (\ref{Becky}) for the absolute metric. To derive the equations of motion, we have to compute the form of the potential $\mathcal{U} \big( \mbfg, \mbfg^{(0)} \big)$ in terms of $\ti{N}(r)$, $\ti{F}(r)$ and $\ti{H}(r)$: this amounts to evaluating the trace of $\sqrt{\mathcal{M}}$, $\mathcal{M}$, $\sqrt{\mathcal{M}}^{\,3}$ and $\mathcal{M}^2$, where $\mathcal{M} = \mbfg^{-1} \mbfg^{(0)}$. Note that, if a matrix $\mathcal{D}$ is diagonal, we have
\beq
tr \sqrt{\mathcal{D}}^{\,\, k} = \sum_{i} \sqrt{\la_{i}}^{\,\, k}
\eeq
where $\la_{i}$, $i = 1, \cdots, 4$ are the eigenvalues of $\mathcal{D}$ and $k$ is a natural number. Furthermore, if a matrix $\mathcal{M}$ is diagonalizable (\emph{i.e.} $\mathcal{M} = \mathcal{A} \mathcal{D} \mathcal{A}^{-1}$, for some invertible matrix $\mathcal{A}$), then we have
\beq
tr \mathcal{M} = tr \big( \mathcal{A} \mathcal{D} \mathcal{A}^{-1} \big) = tr \mathcal{D}
\eeq
and using these relations we find
\beq
tr \sqrt{\mathcal{M}}^{\,\, k} = tr \Big( \mathcal{A} \sqrt{\mathcal{D}} \mathcal{A}^{-1} \cdots \mathcal{A} \sqrt{\mathcal{D}} \mathcal{A}^{-1} \Big) = tr \Big( \mathcal{A} \,\, \sqrt{\mathcal{D}}^{\,\, k} \mathcal{A}^{-1} \Big) =  \sum_{i} \sqrt{\la_{i}}^{\,\, k} \quad .
\eeq
Therefore, to compute $\mathcal{U} \big( \mbfg, \mbfg^{(0)} \big)$ one has to find the eigenvalues of the matrix $\mbfg^{-1} \mbfg^{(0)}$ and plug them in \eqref{Ithilien}-\eqref{Silmaril} : this has been done in \cite{Koyama:2011yg}, where it was found that
\begin{multline}
\sqrt{-g} \, \mathcal{U} \big( \mbfg, \mbfg^{(0)} \big) = - \frac{r^2}{\sqrt{\ti{F}} \ti{H}^2} \bigg[ 2 \Big[ \sqrt{\ti{F}} \big( (2 \ti{H} - 3) \ti{N} + 1 \big) + \ti{H}^2 \ti{N} + \ti{H} (2 - 6 \ti{N}) + 6 \ti{N} - 3 \Big] - \\[2mm]
- 6 \a_3 (\ti{H} - 1) \Big[ \sqrt{\ti{F}} \big( (\ti{H} - 3) \ti{N} + 2 \big) - 2 \ti{H} \ti{N} + \ti{H} + 4 \ti{N} - 3 \Big] - \\[2mm]
- 24 \a_4 (1 - \sqrt{\ti{F}}) (1 - \ti{N}) (1 - \ti{H})^2 \bigg] \quad .
\end{multline}
Varying the action with respect to $\ti{N}(r)$, $\ti{F}(r)$ and $\ti{H}(r)$, one obtains the exact equations of motion for static, spherically symmetric solutions in the diagonal branch \cite{Koyama:2011yg}. These equations are however very complicated, and to solve them it will be convenient to do some approximations.

Note that, in order to study the Vainshtein mechanism, we need to compare the solutions of this theory with the ones of GR: it may turn out to be convenient to rescale the radial coordinate $r \rightarrow \r$ to recast the physical metric in a form where the angular components of the metric are just the square of a radial coordinate, since the linearized Schwarzschild solution has this form. It is crucial to notice, however, that it is impossible to eliminate completely the field $\ti{H}$ from the equations. In fact, if we don't use the St\"uckelberg formalism the theory is not invariant with respect to reparametrizations, and if we perform the coordinate change the field $\ti{H}$ disappears from the line element but does not disappear from the equations of motion. Using the St\"uckelberg formalism, instead, the theory is invariant with respect to reparametrizations and the field $\ti{H}$ itself disappears when we rescale the radius; however, the transformation excites a component of the St\"uckelberg fields, which is related to $\ti{H}$ and appears explicitly in the equations of motion. This is analogous to what happens in the non-linear extension of the Fierz-Pauli action considered by Vainshtein in \cite{Vainshtein72}, as explained in section (\ref{The Vainshtein mechanism}).

Vainshtein \cite{Vainshtein72} in fact suggested that the behavior of the system below the Vainshtein radius is in some sense more transparent with the second coordinate choice, in which the angular components of the metric are just the square of the radial coordinate. In particular, he suggested that, inside the Vainshtein radius, the effect of non-linearities on the two remaining components of the physical metric is just to rescale them by a numerical factor, so that they remain small even around and inside the Vainshtein radius. Instead, the St\"uckelberg field is strongly affected by the non-linearities. Therefore, we perform a coordinate change in the radial coordinate $r \rightarrow \r$ so that in the new coordinate system we have
\beq
ds^2 = - N(\r)^2 dt^2 + F(\r)^{-1} d \r^2 + \r^2 d \Omega^2 \quad ,
\label{diagonal2}
\eeq
and we define $\ti{H}\big( r(\r) \big) = 1 + h(\r)$. We also write
\beq
N(\r) = 1 + \frac{n(\r)}{2}  \qquad F(\r) = 1 + f(\r) \quad ,
\eeq
which for the time being is just a field redefinition.

As we said above, this change of coordinates excites the perturbations of the St\"uckelberg fields $Z^{\m}$. Since the St\"uckelberg fields $\phi^{\m}$ transform as scalars, after changing coordinates we have\footnote{We indicate with $y^{\m}$ and $\ti{Z}^{\m}$ the coordinates and St\"uckelberg fields in the $(t,r,\theta,\vf)$ coordinate system, while we indicate with $y^{\p \m}$ and $Z^{\m}$ the coordinates and St\"uckelberg fields in the $(t,\r,\theta,\vf)$ coordinate system.}
\beq
y^{\p \m}(y) - Z^{\m} \big( y^{\p}(y) \big) = y^{\m} - \ti{Z}^{\m}(y)
\eeq
and since, before changing coordinates, we were in the unitary gauge, we have $\ti{Z}^{\m} = 0$. The fact that only the radial coordinate is involved in the transformation implies then
\begin{align}
Z^{t} &= 0 \\[1mm] 
Z^{\r}(\r) &= \r - r(\r) \\[1mm]
Z^{\theta} &= 0 \\[1mm]
Z^{\vf} &= 0
\end{align}
and, remembering the internal decomposition $Z_{\m} = A_{\m} + \de_{\m} \phi$ and the fact that $\r^2 = r^2/\ti{H}^2$, we have that $A_{\m}$ vanishes and the only non-zero component of $\de_{\m} \phi$ is
\beq
\de_{\r} \phi = - \r \, h(\r) \quad .
\eeq
We conclude that the field $h$ and the scalar component of the St\"uckelberg fields $\phi$ play exactly the same role in this case: we can then work equivalently with the fields $n$, $f$ and $h$, or with $n$, $f$ and $\dot{\phi} \equiv \de_{\r} \phi$. It will turn out to be more convenient to work with $h$ instead of $\dot{\phi}$, so from now on we will work with the fields $n$, $f$ and $h$.

\subsection{Focusing on the Vainshtein mechanism}

Let's first study the behavior around and above the Compton radius $r_c = 1/m$ of solutions which decay at infinity. At linear order in the fields $n$, $f$ and $h$, the physical line element reads
\beq
ds^2 = - (1 + n) \, dt^2 + (1 - f) \, d\rho^2 + \rho^2 d \Omega^2
\eeq
and the equations of motion read \cite{Koyama:2011yg}
\begin{align}
\label{Neq}
0 &= \left( m^2 \rho^2 + 2 \right)f +2 \rho \left( \dot{f} +m^2 \rho^2 \dot{h} +3 \, m^2 \rho h \right) \\[2mm]
0 &= \frac{1}{2} \, m^2 \rho^2 (n -4 h) -\rho \, \dot{n} -f \label{feq} \\[2mm]
0 &= f +\frac{1}{2} \, \rho \, \dot{n} \label{const}
\end{align}
where we have indicated derivatives with respect to $\r$ with an overdot $\dot{\phantom{a}}$. The solutions for $n$ and $f$ are
\begin{align}
n &= - \frac{8 G M}{3 \rho} \, e^{- m \rho} \label{linsoln} \\[2mm]
f &= - \frac{4 G M}{3 \rho} \, (1 + m \rho) \, e^{- m \rho} \label{linsolf} 
\end{align}
where we f\mbox{}ixed the integration constant so that $M$ is the mass of a point particle at the origin, and $8 \pi G = M_{pl}^{-2}$. It is apparent that the solutions display the Yukawa exponential suppression for scales larger than the Compton radius, and for scales smaller than the Compton radius exhibit the vDVZ discontinuity, since the ratio between $n$ and $f$ is 2 in the massless limit $m \rightarrow 0$. This result agrees with the spherically symmetric solutions in the Fierz-Pauli model we found in section (\ref{The VDVZ discontinuity in FP}), and are exactly what we expected: since the dRGT massive gravity is a non-linear completion of the Fierz-Pauli theory, the linearized solution of the equations of motion in the former theory should reproduce the solutions of the latter.

We now want to focus on the Vainshtein mechanism. As we already mentioned, the findings of Vainshtein \cite{Vainshtein72} suggest that, when we focus on scales around and below the Vainshtein radius $r_v$, the effects of non-linearities show up mostly in the St\"uckelberg field, while the gravitational potentials $n$ and $f$ remain small. Therefore, to study the Vainshtein mechanism we decide to treat the gravitational potentials as first order perturbations, and instead keep all the non-linearities in the field $h$. It can be shown \cite{Koyama:2011yg} that in this approximation the equations of motion reduce to the following system of equations
\begin{align}
f &= - \frac{2 G M}{\rho} - (m \rho)^2 \Big[ h - (1+ 3\alpha_3)h^2+(\alpha_3+4\alpha_4) h^3 \Big] \label{solf}\\[4mm]
\dot{n} &= \frac{2 G M}{\rho^2} - m^2 \rho \Big[ h - (\alpha_3 +4\alpha_4) h^3 \Big] \label{soln}
\end{align}
\begin{multline}
\frac{G M}{\rho} \Big[ 1 - 3 (\alpha_3 +4 \alpha_4) h^2 \Big] = - (m \rho)^2 \bigg[ \, \frac{3}{2} h -  3 (1 + 3\alpha_3)h^2 +  \Big. \\
\Big. + \Big( (1 + 3\alpha_3)^2 + 2(\alpha_3 +4\alpha_4) \Big) h^3 - \frac{3}{2}(\alpha_3 +4\alpha_4)^2 h^5 \bigg] \quad . \label{solh}
\end{multline}
Note that the field $h$ obeys a decoupled equation, since the gravitational potentials are not present in \eqref{solh}: this equation is in fact an \emph{algebraic} equation, and for the sake of precision is a polynomial of fifth degree in $h$.

There is another way to derive the system of equations above, starting from the decoupling limit Lagrangian \eqref{lambda3decoupling} \cite{Koyama:2011yg}. As we mentioned in the previous chapter, the decoupling limit leaves the Vainshtein radius fixed and sends the Compton radius to infinity, while sending the gravitational radius to zero: in some sense, this limit focuses on the scales above the gravitational length and below the Compton wavelength. Also, the decoupling limit selects a subclass of the interaction terms which appear in the action, and sends all the others to zero: these terms can be thought to be the ones which are more relevant regarding the effect of non-linear interactions on the linearized solutions when we focus on scales comparable to the Vainshtein radius. We then expect that there should be a connection between the equations for static, spherically symmetric solutions obtained from the decoupling limit Lagrangian and the equations obtained above.

To see it, it is actually more convenient to work with the fields $\check{h}_{\m\n}$ and $\check{\phi}$ , because their dynamics are coupled by just one interaction term, as is apparent in the Lagrangian \eqref{lambda3decouplingdiag2}. Apart from the interaction term $\propto \check{h}^{\mu\nu} \check{X}^{(3)}_{\mu\nu}$, the dynamics of the field $\check{\phi}$ is described by a Galileon Lagrangian: as shown in \cite{NicolisRattazziTrincherini}, for static and spherically symmetric configurations the equations of motion for a Galileon field can be integrated exactly, obtaining an algebraic equation for $\de_{\r} \check{\phi}/\r$
\beq
a_1 \bigg( \frac{\de_{\r} \check{\phi}}{\r} \bigg) + a_2 \bigg( \frac{\de_{\r} \check{\phi}}{\r} \bigg)^2 + a_3 \bigg( \frac{\de_{\r} \check{\phi}}{\r} \bigg)^3 \propto \frac{M}{4 \pi r^3} \quad .
\eeq
The coefficients $a_1$, $a_2$ and $a_3$ depend on the coefficients of the Galileon terms in the Lagrangian \eqref{lambda3decouplingdiag2}: therefore, if we neglect the interaction term $\propto \check{h}^{\mu\nu} \check{X}^{(3)}_{\mu\nu}$, the equation for $\check{\phi}$ is polynomial in $\de_{\r} \check{\phi}/\r$ and it is at most a cubic. As shown in \cite{Koyama:2011yg}, the effect of the interaction term $\propto \check{h}^{\mu\nu} \check{X}^{(3)}_{\mu\nu}$ is to add to the left hand side of the cubic equation above a contribution proportional to
\beq
\label{additionalterm}
\big( 8 d_5 + c_3 \big) \, \bigg( \frac{\de_{\r} \check{n}}{\r} \bigg) \, \bigg( \frac{\de_{\r} \check{\phi}}{\r} \bigg)^2
\eeq
where $\check{n} = \check{h}_{tt}$, and $8 d_5 + c_3$ is proportional to $\a_3 + 4 \a_4$. Varying the action with respect to $\check{h}_{\m\n}$, instead, one obtains that the equations of motion for $\check{n}$ and $\check{f}$: these equations imply that $\de_{\r} \check{n}/\r$ can be expressed as a linear combination of a Newtonian term $GM/\r^3$ and of a term $\propto \big( \a_3 + 4 \a_4 \big) \big( \de_{\r} \check{\phi}/\r \big)^3$, which again comes from the interaction term $\check{h}^{\mu\nu} \check{X}^{(3)}_{\mu\nu}$ in the Lagrangian. Substituting this expression for $\de_{\r} \check{n}/\r$ in the equation for $\check{\phi}$, one obtains the quintic equation \eqref{solh} for $h = \de_{\r} \check{\phi}/\r$: in particular, the $h^5$ term in the quintic is generated by substituting this expression for $\de_{\r} \check{n}/\r$ in \eqref{additionalterm}. Therefore, the interaction term $\check{h}^{\mu\nu} \check{X}^{(3)}_{\mu\nu}$ (which is the only one which cannot be removed from the action by a local field redefinition) is responsible for the fact that the degree of the polynomial equation which $\de_{\r} \check{\phi}/\r$ obey changes from three (as it is in a general Galileon theory) to five. Note however that when $(8 d_5 + c_3) \propto (\a_3 + 4 \a_4) = 0$ this coupling vanishes, and the polynomial equation becomes a cubic as in a Galileon theory. It is possible to verify \cite{Koyama:2011yg} that also the equations \eqref{solf} - \eqref{soln} can be derived from the decoupling limit Lagrangian: this strongly supports the idea that the system of equations \eqref{solf} - \eqref{solh} is a good description of the full theory when we focus on scales comparable to the Vainshtein radius, and therefore this system is the starting point for our analysis of the Vainshtein mechanism in dRGT massive gravity.

\section{The quintic equation}
\label{The quintic equation}

For notational convenience, it is useful to def\mbox{}ine the parameters $\a \equiv 1 + 3 \, \a_3$ and $\b \equiv \a_3 + 4 \, \a_4 \,$: in terms of these new parameters, the system (\ref{solf})-(\ref{solh}) takes the form
\begin{align}
\label{solfab}
f &= - 2 \, \frac{G M}{\rho} - (m \rho)^2 \Big(
h - \alpha h^2 + \beta h^3 \Big) \\[2mm]
\label{solnab}
\dot{n} &= 2 \, \frac{G M}{\rho^2} - m^2 \rho \, \Big( h - \b h^3 \Big)
\end{align}
\beq
\label{solhab}
\frac{3}{2} \, \bq \, \ha{5}(\r) - \Big( \aq + 2 \b \Big) \, \ha{3}(\r) + 3 \, \Big( \a + \b A(\r) \Big) \, \ha{2}(\r)
- \frac{3}{2} \, h(\r) - A(\r) = 0
\eeq \\
\noi where $A(\r) = \big( \r_{v} / \r \big)^{3}$ and $\r_{v}$ is the Vainshtein radius def\mbox{}ined as $\r_{v} \equiv \big( G M / m^{2} \big)^{\! 1/3}$. The new parameters have a clear physical interpretation: in fact, the two combinations of the parameters $c_3$ and $d_5$ which appear in the decoupling limit action (\ref{lambda3decouplingdiag2}) are easily expressed in terms of $\a$ and $\b$
\beq
\a \propto 6 c_3 - 1 \qquad \qquad \b \propto 8 d_5 + c_3 \quad .
\eeq
In particular, the case $\b = 0$ corresponds to a situation where the coupling $\check{h}^{\mu\nu} \check{X}^{(3)}_{\mu\nu}$ is absent and so the field $\check{\f}$ is exactly a Galileon, while the case $\a = 0$ corresponds to a situation where the derivative coupling $\de_\mu \check{\f} \, \de_\nu \check{\f} \, T^{\mu\nu}$ is absent and so the field $\check{\f}$ does not couple to the electromagnetic field. In the case $\a = \b = 0$ all the Galileon self-interaction terms vanish, and in the decoupling limit we are left with a Lagrangian for a free tensor field $\check{h}_{\mu\nu}$ and a free scalar $\check{\f}$ both of which interact with the energy-momentum tensor via non-derivative couplings.

As we already mentioned, the equation \eqref{solhab} does not contain the gravitational potentials $n$ and $f$, so $h$ obeys a decoupled equation: furthermore, if we know the solution for $h$, the fields $f$, $n$ are uniquely determined (up to an integration constant) by the other two equations (\ref{solfab}) and (\ref{solnab}) in terms of $h$. Therefore, our aim has been to study all the solutions which the equation \eqref{solhab} admits, for every value of the parameters $\a$ and $\b$, and characterize their geometrical properties using the equations \eqref{solfab}-\eqref{solnab}. Note that in the particular case of $\beta=0$, the equation for $h$ becomes a cubic equation and it is possible to obtain solutions for $h$ and the metric perturbations exactly. These solutions were studied in \cite{Koyama:2011xz, Koyama:2011yg} and it was shown that the solutions exhibit the Vainshtein mechanism. Therefore, in what follows, we assume $\beta \neq 0$. A systematic approach to Vainshtein ef\mbox{}fects in theories which have connections with massive gravity have been performed in \cite{Kaloper:2011qc}, regarding covariant Galileon theory, and in \cite{Kimura:2011dc, DeFelice:2011th}, regarding general scalar-tensor theories.

\subsection{The quintic equation}

The equation of motion for $h$, which we rewrite here

\beq
\label{quintic}
\frac{3}{2} \, \bq \, \ha{5}(\r) - \Big( \aq + 2 \b \Big) \, \ha{3}(\r) + 3 \, \Big( \a + \b A(\r) \Big) \, \ha{2}(\r)
- \frac{3}{2} \, h(\r) - A(\r) = 0
\eeq

\noi is an algebraic equation for $h$, $A$, $\a$ and $\b$; at f\mbox{}ixed  $\r$, $\a$ and $\b$ it is a polynomial equation of f\mbox{}ifth degree in $h$ (except, as we already mentioned, in the special case $\b = 0$). In the following, we will refer to it as \emph{the quintic equation}. To study the Vainshtein mechanism in this theory, the most convenient thing to do would be to find exact solutions of the quintic equation, derive their physical predictions inside the Vainshtein radius, and determine if they agree with the ones of GR. However, finding exact solutions of this equation is almost impossible: a general theorem of algebra, the \emph{Abel-Ruffini theorem} (see, for example, \cite{AbelRuffini}), states that is impossible to express the general solution of a polynomial equation of degree five or higher in terms of radicals (while it possible for quadratic, cubic and quartic equations). Even if the quintic equation \eqref{quintic} lacks of the $h^4$ term, and so it is not the most general quintic equation, it seems arduous to find explicit solutions as a function of $\r$. 

However, it is indeed possible to find explicitly the number and properties of solutions which the quintic equation admits in a neighborhood of $\r \rightarrow + \infty$, which we call the \emph{asymptotic solutions}, and the number and properties of solutions which the quintic equation admits in a neighborhood of $\r \rightarrow 0^{+}$, which we call the \emph{inner solutions}. This fact offers the possibility to study the Vainshtein mechanism without finding the complete solutions of \eqref{quintic}. In fact, suppose for example that we are able to show that (for some $\a$ and $\b$) there exists a global solution of \eqref{quintic} (\emph{i.e.} a solution which is defined on the domain $\r \in (0, + \infty)$) which interpolates between an inner solution which reproduces GR results, and an asymptotic solution which displays the vDVZ discontinuity. We can then conclude that the Vainshtein mechanism is working for the theory defined by this choice of parameters. More generally, we can make a precise statement on the effectiveness of the Vainshtein mechanism just by characterizing the properties of asymptotic and inner solutions in all the phase space of parameters, and by determining if there are global solutions which interpolate between each couple of asymptotic/inner solutions. In the following, when there is a global solution which interpolates between an inner and an asymptotic solution, we say that there is matching between the two solutions.

This is precisely the approach we take in studying the Vainshtein mechanism in dRGT massive gravity: in sections \ref{Asymptotic and inner solutions} and \ref{Characterization of the asymptotic and inner solutions} we find exactly the number and properties of asymptotic and inner solutions in every point of the phase space, and in the section \ref{Phase space diagram for solutions matching} we discuss the details of the matching between asymptotic and inner solutions. We will not restrict ourselves to asymptotically decaying solutions and to inner solutions which reproduce GR, but we will study the matching properties of all kinds of asymptotic and inner solutions.

It is worthwhile to point out that our starting equations (\ref{solfab})-(\ref{solhab}) were constructed assuming $GM < \rho < 1/m$, but in the following analysis we use the whole radial domain $0 < \r < + \infty$ . On one hand, this allows us to characterize exactly the number and properties of solutions on large and small scales. On the other hand, the picture we have in mind is that the Compton wavelength of the gravitational f\mbox{}ield $\r_c = 1/m$ is of the same order of the Hubble radius today, and that there is a huge hierarchy between $\r_c$ and the gravitational radius\footnote{We are using units where the speed of light speed has unitary value.} $\r_g = GM$, \textit{i.e.} $\r_c / \r_g \ggg 1$. Therefore, we expect that extending the analysis to the whole radial domain captures the correct physical results. Nonetheless, when considering a specific configuration of sources, it is necessary to verify explicitly that extending the domain is indeed a harmless approximation.

\subsection{Symmetry of the quintic and dual formulation}

\subsubsection{Symmetry of the quintic}

To be able to describe how the matching works in all the phase space, in principle we should study
separately every point ($\a$, $\b$). However, this is not necessary since equation (\ref{quintic}) obeys a
remarkable symmetry: def\mbox{}ining the quintic function as
\beq
\label{quinticFunction}
q \, \big( h, A; \a, \b \big) \equiv \frac{3}{2} \, \bq \, \ha{5} - \big( \aq + 2 \b \big) \, \ha{3} + 3 \, \big( \a + \b A \big) \, \ha{2}
- \frac{3}{2} \, h - A \quad ,
\eeq
it is simple to see that
\beq
\label{symmetry}
q \, \Big( \frac{h}{k} , \frac{A}{k} ; k \, \a, k^2 \b \Big) = \frac{1}{k} \, q \, \big( h, A; \a, \b \big) \quad .
\eeq
Therefore if a local solution of (\ref{quintic}) exists for a given $(\alpha,\beta)$ within a certain radial interval, it would also be present for $(k \alpha,k^2 \beta)$, for $k>0$, with $h$ being replaced by $h/k$ and the radial interval rescaled by $1/\sqrt[3]{k}$. As a result, each point belonging to the $\a>0$ part of the parabola $\b = c \, \aq$ of the phase space (with $c$ any non-vanishing constant) shares the same physics, hence having the same number of global solutions and matching properties. The same is true for the points belonging the $\a<0$ part of the parabola. So, to understand the global structure of the phase space, it is suf\mbox{}f\mbox{}icient to analyze one point for each of the half-parabolas present in the phase space.

\subsubsection{Dual formulation}

In order to find the asymptotic and the inner solutions, we need to study the quintic equation in the limits $\r \rightarrow + \infty$ and $\r \rightarrow 0^+$. In particular, we will consider both decaying and diverging solutions. To do this, it is very useful to formulate the theory in terms of quantities which remain finite in the limit.

Note that the radial coordinate $\r$ is defined for $\r \in (0, + \infty)$: this implies that the function $A(\r)$ is always non-zero, and the map $\r \rightarrow A(\r)$ is a diffeomorphism\footnote{By diffeomorphism we mean a smooth and invertible function whose inverse is smooth.} of $(0, + \infty)$ into itself. In particular, this means that we can use equivalently $\r$ and $A$ as radial coordinates: the latter choice is more convenient to study asymptotic solutions, since the limit $\r \rightarrow + \infty$ is expressed as the limit $A \rightarrow 0^+$. Furthermore, it will be useful to work with dimensionless radial coordinates, at least as far as only the solutions of the quintic are concerned, so instead of $\r$ we will often use the coordinate $x \equiv \r/\r_v$ and, as we mentioned, $A = 1/x^3$.

The fact that $A$ is always different from zero implies that a solution $h$ of (\ref{quintic}) never vanishes in the domain of definition, since the quintic function \eqref{quinticFunction} for $h = 0$ is equal to $A$. Therefore, we can divide the quintic equation by $h^5$ obtaining the following quintic equation for $v \equiv 1/h$
\beq
\label{vedremovedremo}
d \big( v, A; \a, \b \big) \equiv A \, \va{5} + \frac{3}{2} \, \va{4} - 3 \, \big( \a + \b A \big) \, \va{3} + \big( \aq + 2 \b \big) \, \va{2} - \frac{3}{2} \, \bq  = 0 \quad .
\eeq
Since we are considering the $\b \neq 0$ case, every solution to the new quintic \eqref{vedremovedremo} is again never vanishing. It follows that, if we find a solution $h$ of the ``original'' quintic equation \eqref{quintic}, then its reciprocal $1/h$ is a solution of the ``new'' quintic \eqref{vedremovedremo}, and conversely the reciprocal of every solution of \eqref{vedremovedremo} is a solution of \eqref{quintic}. This implies that it is completely equivalent to work with the field $h$ or with the field $v$: the quintic equation \eqref{vedremovedremo}, together with the equations which we obtain substituting $h = 1/v$ in the equations \eqref{solfab}-\eqref{solnab}, provides a completely equivalent formulation of the (decoupling limit) theory defined by the equations \eqref{solfab}-\eqref{solhab}. We will refer to the formulation in terms of $v$ as the dual formulation.

It will be useful, especially when studying inner solutions, to work with the $x$ coordinate: to derive the quintic equations in terms of $x$, we can divide the quintic equation \eqref{quintic} by $A$ obtaining the following quintic equation
\beq
\label{hxequation}
b \big( h, x; \a, \b \big) \equiv x^3 \, \bigg( \frac{3}{2} \, \bq \, \ha{5} - \big( \aq + 2 \b \big) \, \ha{3} + 3 \, \a \, \ha{2} - \frac{3}{2} \, h \bigg) + 3 \, \b \, \ha{2} - 1 = 0 \quad .
\eeq
Furthermore, dividing the equation above by $\ha{5}$ we obtain the quintic in the dual formulation in terms of the radial coordinate $x$
\beq
\label{quinticv}
g \big( v, x; \a, \b \big) \equiv \va{5} + \frac{3}{2} \, x^3 \, \va{4} - 3 \, \big( \b + \a \, x^3 \big) \, \va{3} + \big( \aq + 2 \b \big) \, x^3 \, \va{2} - \frac{3}{2} \, \bq \, x^3 = 0 \quad .
\eeq
These four quintic equations provide equivalent descriptions of the same problem, when $\b \neq 0$. Note that the dual formulation is more suited to discuss the $\b \rightarrow 0$ limit of our results and the connection with exact results of the $\b = 0$ case \cite{Koyama:2011yg}, since the quintic equations in the dual formulation remain of degree five even in the $\b \rightarrow 0$ limit.

\section{Asymptotic and inner solutions}
\label{Asymptotic and inner solutions}

We turn now to the study of asymptotic and inner solutions of the quintic equation (\ref{quintic}), in the $\b \neq 0$ case. Interesting results about asymptotic and inner solutions of the quintic equation have been obtained in \cite{Koyama:2011yg} and\footnote{Note that \cite{Chkareuli:2011te} uses convention different from ours, in particular their $\a$ and $\b$ have opposite sign with respect to ours.} \cite{Chkareuli:2011te}, however the existence of the solution was not proved there. Furthermore, an exact characterization of the number of asymptotic and inner solutions in the phase space is missing in these papers. See also \cite{Sjors:2011iv, Cai:2012db} for related studies on the phenomenology of solutions in this branch of massive gravity.

\subsection{Asymptotic solutions}

Let's suppose that a solution $h(\r)$ of the quintic equation (\ref{quintic}) exists in a neighborhood of $\rho = + \infty \,$, and that it has a well def\mbox{}ined limit as $\r \rightarrow + \infty$. We can immediately conclude that this solution cannot be divergent. In fact, suppose that indeed the solution is divergent $|\lim_{\r \rightarrow + \infty} h(\r)| = + \infty \,$: in the dual formulation, this corresponds to the case $\lim_{A \rightarrow 0} v(A) = 0$. Performing the limit $A \rightarrow 0$ in the quintic \eqref{vedremovedremo} one obtains $\b = 0$, which is precisely against our initial assumption. Therefore, asymptotic solutions of the quintic equation (\ref{quintic}) have to be finite.

Suppose now that $\lim_{\r \rightarrow + \infty} h(\r)$ is f\mbox{}inite, and let's call it $C$. Then both of the sides of the quintic equation (\ref{quintic}) have a f\mbox{}inite limit when $\rho \rightarrow + \infty \,$, and taking this limit one gets
\beq
\frac{3}{2} \, \bq \, C^{5} - \big( \aq + 2 \b \big) \, C^{3} + 3 \, \a \, C^{2} - \frac{3}{2} \, C = 0 \quad .
\eeq
It follows then that the allowed asymptotic values at inf\mbox{}inity for $h(\r)$ are the roots of the following equation, which we call the \emph{asymptotic equation}
\beq
\label{asymptotic equation}
\mathscr{A}(y) \equiv \frac{3}{2} \, \bq \, y^{5} - \big( \aq + 2 \b \big) \, y^{3} + 3 \, \a \, y^{2}
- \frac{3}{2} \, y = 0 \quad .
\eeq
Note that $y = 0$ is always a root of this equation, and in fact a simple root (\textit{i.e.} a root of multiplicity one) since $\frac{d}{dy} \mathscr{A}(0) = -3/2 \neq 0 \,$. Dividing by $y$, one obtains that the other asymptotic values for $h(\r)$ are the roots of the \emph{reduced asymptotic equation}
\beq
\label{red asymptotic equation}
\mathscr{A}_{r}(y) \equiv \frac{3}{2} \, \bq \, y^{4} - \big( \aq + 2 \b \big) \, y^{2}
+ 3 \, \a \, y - \frac{3}{2} = 0 \quad .
\eeq
This last equation is a quartic, so it can have up to 4 (real) roots, depending on the specif\mbox{}ic values of $\a$ and $\b$. Since
\beq
\lim_{y \rightarrow -\infty} \mathscr{A}_{r}(y) = +\infty \qquad \mathscr{A}_{r}(0) = - \frac{3}{2} < 0 \qquad \lim_{y \rightarrow +\infty} \mathscr{A}_{r}(y) = +\infty \quad ,
\eeq
we have, by the intermediate value theorem (see, for example, \cite{Clarke}), that the reduced asymptotic equation has always at least two roots, one positive and one negative. For the same reason, it cannot have two positive and two negative roots, since at each simple root the quartic function changes sign.

\begin{figure}[htp!]
\begin{center}
\includegraphics[width=11cm]{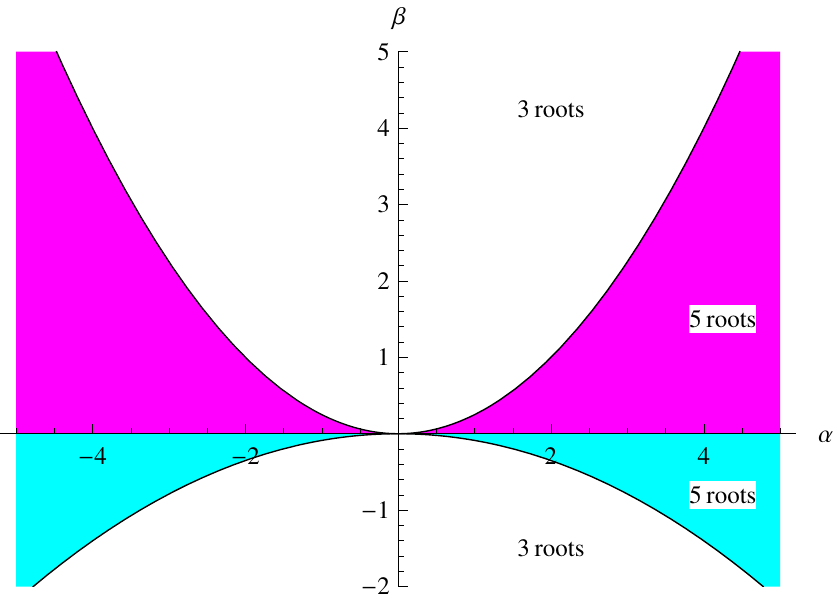}
\caption[Phase space diagram for asymptotic solutions]{phase space diagram for the number of asymptotic solutions}
\label{five roots}
\end{center}
\end{figure}

As we show in the appendix \ref{Roots at infinity}, in the regions of the phase space below the parabola $\b = c_{-} \, \a^2$ and above the parabola $\b = c_{+} \, \a^2$ the asymptotic equation has three real roots, which are simple roots, while in the regions $c_{-} \, \a^2 < \b <0$ and $0 < \b < c_{+} \, \a^2$ the asymptotic equation has five real roots, which are again simple roots. Note that $c_{+} = 1/4$ and $c_{-}$ is the only real root of the equation $8 + 48 \, y - 435 \, y^2 + 676 \, y^3 = 0$. On the two parabolas $\b = c_{\pm} \, \a^2$ (which we call the five-roots-at-infinity parabolas) there are four roots, one of which is a root of multiplicity two. This is summarized in f\mbox{}igure \ref{five roots}.

We name the roots in the following way: the $y=0$ root is denoted as $\textbf{L}$. For the phase space points where there are just three roots, the positive root is denoted as $\textbf{C}_{+}$ and the negative one as $\textbf{C}_{-}$ . For points in the f\mbox{}ive-roots regions, we adopt the following convention. Be $(\a_5, \b_5)$ a point where there are f\mbox{}ive roots. In the same quadrant of the phase space, take another point $(\a_3, \b_3)$ where there are three roots, and a path $\mathscr{C}$ which connects the two points. Following the path $\mathscr{C}$, two of the four non-zero roots of $(\a_5, \b_5)$ smoothly f\mbox{}low to the non-zero roots of $(\a_3, \b_3)$, and are denoted as $\textbf{C}_{+}$ and $\textbf{C}_{-}$ themselves. The other two non-zero roots of $(\a_5, \b_5)$, instead, disappear when (following $\mathscr{C}$) the boundary of the f\mbox{}ive-roots region is crossed, and are denoted as $\textbf{P}_{1}$ and $\textbf{P}_{2}$. We adopt the convention that $\abs{\textbf{P}_{1}} \leq \abs{\textbf{P}_{2}}$. The def\mbox{}inition is independent of the particular choice of the point $(\a_3, \b_3)$ and of the path $\mathscr{C}$ used. A careful study of the asymptotic equation and of its derivatives permits to show that we have $\textbf{C}_{-} < \textbf{C}_{+} < \textbf{P}_{1} < \textbf{P}_{2}$ for $\a > 0$ and $\textbf{P}_{2} < \textbf{P}_{1} < \textbf{C}_{-} < \textbf{C}_{+}$ for $\a < 0$. On the boundaries $\b = c_{\pm} \, \a^2$ we have $\textbf{P}_{1} = \textbf{P}_{2} \equiv \textbf{P}$.

\subsection{Inner solutions}

Suppose now that a solution of the quintic equation exists in a neighborhood of $\r = 0^+$ (possibly not def\mbox{}ined in $\r = 0$), and that it has a well def\mbox{}ined limit when $\r \rightarrow 0^+$. We can immediately see that such a solution cannot tend to zero as $\r \rightarrow 0^+$. In fact, suppose that indeed the solution tends to zero $\lim_{x \rightarrow 0^+} h(x) = 0 \,$: taking the limit in the quintic equation \eqref{hxequation}, we get $-1=0$ which contradicts our assumption. Therefore, if $h(\r)$ is an inner solution then $\lim_{\r \rightarrow 0^+} h (\r) \neq 0$.

This means that, in the dual formulation, all the inner solutions $v(x)$ have a finite limit for $x \rightarrow 0^+$. Considering the quintic in the dual formulation (\ref{quinticv}), the permitted limiting values for a inner solution $v$ are then the roots of the equation obtained performing the limit $x \rightarrow 0^+$ in the quintic (\ref{quinticv}), namely
\beq
\label{inner equation}
\va{5} - 3 \, \b \, \va{3} = 0 \quad .
\eeq
For $\b > 0$ there are three roots, namely $v_{0} = 0$, $v_{+} = + \sqrt{3 \, \b}$ and $v_{-} = - \sqrt{3 \, \b}$ ; for $\b < 0$, instead, there is only the root $v = 0$. Therefore, the permitted limiting behaviors for $h$ when $\r \rightarrow 0^+$ are
\beq
\abs{h(\r)} \rightarrow + \infty
\eeq
for $\b \neq 0$, and
\beq
h \rightarrow \mathbf{F}_{\pm} \equiv \pm \sqrt{\frac{1}{3 \, \b}}
\eeq
only for $\b > 0$.

\subsection{Existence of the asymptotic and inner solutions}

Note that so far we have not proved that inner and asymptotic solutions exist, but just found the values that have to be the limit of these solutions if they exist. The existence and uniqueness of solutions can be proved applying the \emph{implicit function theorem} (known also as Dini's theorem) which we enunciate in appendix \ref{The implicit function theorem}. Regarding asymptotic solutions, to apply the implicit function theorem we can artificially extend the domain of definition of the equation (\ref{quintic}) to $A < 0$ as well: apart from the five-roots-at-infinity boundaries, all the asymptotic roots are simple roots. Therefore we can apply the implicit function theorem, which tells us that there exists a local solution of (\ref{quintic}) associated to every root of the asymptotic equation: restricting now the domain of definition of these local solutions to $A>0$, we obtain the desired asymptotic solutions to the quintic equation. It follows that to each of the asymptotic roots $\textbf{L}$, $\textbf{C}_{+}$, $\textbf{C}_{-}$, $\textbf{P}_{1}$ and $\textbf{P}_{2}$ we can associate a local solution of the quintic equation in a neighborhood of $\r \rightarrow + \infty$, and we indicate the root and the associate local solution with the same letter. 

On the five-roots-at-infinity boundaries, a separate analysis is needed for the double root $\textbf{P}_{1} = \textbf{P}_{2} \equiv \textbf{P}$. It can be shown that for $\a > 0$ and $\b = c_{+}\, \aq$ there are no local solutions of (\ref{quintic}) which tend to $\textbf{P}$ when $\r \rightarrow + \infty$, and the same holds for $\a < 0$ and $\b = c_{-}\, \aq$. On the other hand, for $\a > 0$ and $\b = c_{-}\, \aq$ there are \emph{two} dif\mbox{}ferent local solutions of (\ref{quintic}) which tend to $\textbf{P}$ when $\r \rightarrow + \infty$, and the same holds for $\a < 0$ and $\b = c_{+}\, \aq$. Despite having the same limit for  $\r \rightarrow + \infty$, these two local solutions are dif\mbox{}ferent when $A \neq 0$: we then call $\textbf{P}_{1}$ the solution which in absolute value is smaller, and $\textbf{P}_{2}$ the solution which in absolute value is bigger. Therefore, on the boundaries between the three-roots-at-inf\mbox{}inity regions and the f\mbox{}ive-roots-at-inf\mbox{}inity regions, for $\a \gtrless 0$, $\b = c_{\pm}\, \aq$ there are three asymptotic solutions of (\ref{quintic}), while for $\a \gtrless 0$, $\b = c_{\mp}\, \aq$ there are f\mbox{}ive asymptotic solutions of (\ref{quintic}).

Regarding the inner solutions, the existence of local solutions in a neighborhood of $\r = 0^+$ associated to the limiting values $\textbf{F}_{+}$ and $\textbf{F}_{-}$ can be proved extending the validity of (\ref{quinticv}) to $x < 0$ and applying the implicit function theorem at $(v = \pm \sqrt{3 \b}, x = 0)$. Restricting then to $x > 0$ the domain of definition of the solutions obtained this way, we get two local solutions $v_{\pm}(x)$ of (\ref{quinticv}) which tend to $\pm \sqrt{3 \b}$ as $\r \rightarrow 0^+$: the reciprocal $h_{\pm}(\r) = 1/v_{\pm}(x(\r))$ of these solutions are local solutions of the quintic (\ref{quintic}) in a neighborhood of $\r \rightarrow 0^+$, and are the inner solutions associated to $\textbf{F}_{\pm}$. We will use $\textbf{F}_{\pm}$ to denote both the limiting values and the inner solutions associated to the limiting values. For the solution associated to the limiting value $v = 0$, we cannot apply the implicit function theorem straightaway, because the function $g \big( v, x; \a, \b \big)$ is such that $\frac{\de g}{\de v} = 0$ in $(v,x)=(0,0)$. However, using the results of appendix \ref{Useful properties of the quintic function}, it can be shown that, for $\b > 0$, there always exists a neighborhood of $A \rightarrow + \infty$ where there is a simple root of the quintic (\ref{quintic}) which is $< \textbf{F}_{-}$ and decreases when $A$ increases. Applying the implicit function theorem to (\ref{quintic}) in this neighborhood of $A \rightarrow + \infty$, we obtain a local solution of (\ref{quintic}) which corresponds to the limiting value $v = 0$, which will be denoted by \textbf{D}. For $\b < 0$, instead, there always exists a neighborhood of $A \rightarrow + \infty$ where there is a simple root of the quintic (\ref{quintic}) which is $> \textbf{F}_{+}$ and increases when $A$ increases. Analogously to the $\b > 0$ case, applying the implicit function theorem to (\ref{quintic}) in this neighborhood we obtain a local solution of (\ref{quintic}) which corresponds to the limiting value $v = 0$, which will be denoted as well by \textbf{D}.

\section{Characterization of the asymptotic and inner solutions}
\label{Characterization of the asymptotic and inner solutions}

We sum up here the results obtained in the previous section on the existence and properties of asymptotic and inner solutions of eq.~(\ref{quintic}), together with their leading behaviors and geometrical meaning. We refer to the appendix \ref{Leading behaviors} for the derivation of the leading behaviors.

\subsection{Asymptotic solutions}

In a neighborhood of $\r \rightarrow +\infty$ there are, depending on the value of $(\a, \b)$, three or f\mbox{}ive solutions
to eq.~(\ref{quintic}). In particular:
\begin{itemize}
 \item[-] There is always a decaying solution, which we indicate with $\textbf{L}$. Its asymptotic behavior is
          \beq
          h(\r) = - \frac{2}{3} \left( \frac{\rho_{v}}{\r} \right)^{\! 3} + \, R(\r)
          \eeq
          where $\lim_{\r \rightarrow +\infty} \, \r^3 R(\r) = 0$. This solution corresponds to a spacetime which is
          asymptotically f\mbox{}lat, as one can see from eqs.~(\ref{solfab})-(\ref{solnab}).
 \item[-] Additionally, there are two or four solutions to eq.~(\ref{quintic}) which tend to a f\mbox{}inite,
          non-zero value as $\r \rightarrow +\infty$. We name these solutions with $\textbf{C}_{+}$, $\textbf{C}_{-}$,
          $\textbf{P}_{1}$ and $\textbf{P}_{2}$. Their asymptotic behavior is
          \beq
          h(\r) = C + \, R(\r)
          \eeq
          where $\lim_{\r \rightarrow +\infty} \, R(\r) = 0$ and $C$ is a root of the reduced asymptotic equation
	  (\ref{red asymptotic equation}). From eqs.~(\ref{solfab})-(\ref{solnab}), one can get convinced that these solutions correspond to spacetimes which are asymptotically non-f\mbox{}lat. Interestingly, the leading term in the gravitational potentials scales as $\rho^2$ for large radii, the same scaling which we f\mbox{}ind in a de Sitter spacetime. It is worthwhile to point out that, since we are working on scales below the Compton wavelength of the gravitational f\mbox{}ield, ``asymptotically non-f\mbox{}lat'' really means that (from the point of view of the full and non-approximated theory) the spacetime corresponding to this solution tends to a non-f\mbox{}lat spacetime when the Compton wavelength is approached. To understand the ``true'' asymptotic behavior of this solution, one should use the non-approximated equations. Note that, even if $C$ (and so $h$) is much smaller than one, the gravitational potentials $n$ and $f$ can be very large (as they behave like $\propto \rho^2$ far from the origin in this case): therefore, the linear approximation (for the gravitational potentials) we used to obtain eqs.~(\ref{linsoln})-(\ref{linsolf}) is not valid. Instead, the asymptotic fate of the solution is dictated by the non-linear behavior of the non-approximated equations. This seems not easy to predict without a separate analysis, and we don't attempt to address this interesting problem.
\end{itemize}

\subsection{Inner solutions}

In a neighborhood of $\r \rightarrow 0^+$ there are either one or three solutions to eq.~(\ref{quintic}).
For $\b > 0$ there are exactly three inner solutions, while for $\b < 0$ there is only one inner solution.
In particular:
\begin{itemize}
 \item[-] There is always a diverging solution, which we denote by $\textbf{D}$. Its leading behavior is
          \beq
          h(\r) = - \, \sqrt[3]{\frac{2}{\b}} \, \frac{\r_v}{\r} + R(\r)
          \eeq
          where $\lim_{\r \rightarrow 0^+} \, (R(\r)/\r)$ is f\mbox{}inite. This solution exists for both
          $\b > 0$ and $\b < 0$, with opposite signs for each case.
          Using this solution in eqs.~(\ref{solfab})-(\ref{solnab}), one realizes that the $h^3$ term cancels the $GM/\rho$ term, so the gravitational f\mbox{}ield is self-shielded and does not diverge as $\r \rightarrow 0^+$. This solution is in strong disagreement with gravitational observations.

 \item[-] For $\b > 0$, there are two additional solutions to eq.~(\ref{quintic}), which tend to a f\mbox{}inite, non-zero value as $\r \rightarrow 0^+$.
          We indicate these solutions by $\textbf{F}_{+}$ and $\textbf{F}_{-} \,$. Their leading behavior is
          \beq
          h(\r) = \pm \sqrt{\frac{1}{3 \, \b}} + \, R(\r)
          \eeq
          where $\lim_{\r \rightarrow 0^+} \, R = 0$. Notice that for $\b < 0$ there are no solutions to eq.~(\ref{quintic}) which tend to a f\mbox{}inite value as $\r \rightarrow 0^+$.

          The expressions (\ref{solfab})-(\ref{solnab}) for the gravitational potentials imply that the metric associated to these solutions ($\textbf{F}_{+}$ and $\textbf{F}_{-}$) approximate the linearized Schwarzschild metric as $\r \rightarrow 0^+$.
\end{itemize}
From the behavior of the inner solutions, one concludes that only in the $\b>0$ part of the phase space solutions may exhibit the Vainshtein mechanism, but not necessarily for all values of $\alpha$. In the next subsection we see more in detail how this mechanism works.

\subsection{Vainshtein mechanism and solutions matching}

In order to study where in the phase space the Vainshtein mechanism works, it is useful to compare the
gravitational potentials $f$ and $n$ with their counterparts in the GR case. In the weak f\mbox{}ield limit, the Schwarzschild solution of GR reads
\beq
ds^2 = - \bigg( 1 - \frac{2 G M}{\r} \bigg) \, dt^2 + \bigg( 1 + \frac{2 G M}{\r} \bigg) \, d\rho^2
+ \rho^2 \, d \Omega^2 \quad ,
\eeq	
so by calling $f_{GR} = n_{GR} = - 2 G M / \r$ we obtain
\begin{align}
\label{solfabderx}
\frac{f}{f_{GR}} &= 1 + \frac{1}{2} \, \bigg( \frac{\rho}{\rho_{v}} \bigg)^3 \, \Big( h - \alpha h^2 + \beta h^3 \Big) \\[2mm]
\label{solnabderx}
\frac{n^{\, \p}}{n_{GR}^{\, \p}} &= 1 - \frac{1}{2} \, \bigg( \frac{\rho}{\rho_{v}} \bigg)^3 \, \Big( h - \b h^3 \Big) \quad .
\end{align}
Let us now f\mbox{}irst discuss the asymptotic solutions. For the decaying solution $\textbf{L}$, we have that the linear contribution in $h$ rescales the coef\mbox{}f\mbox{}icients of the Schwarzschild-like terms, so we obtain $f / f_{GR} \rightarrow 2/3$ and $n^{\, \p} / n_{GR}^{\, \p} \rightarrow 4/3$ for $\r \rightarrow +\infty$. For the non-decaying solutions $\textbf{C}_{\pm}$ and  $\textbf{P}_{1,2}$, the leading behavior for $f / f_{GR}$ and $n^{\, \p} / n_{GR}^{\, \p}$ is proportional to $( \r / \r_v )^3$ in both cases, however the proportionality coef\mbox{}f\mbox{}icients generally dif\mbox{}fer since they have a dif\mbox{}ferent functional dependence on $\alpha$ and $\beta$. There are some special cases for $(\a,\b)$ where these asymptotic solutions lead to $f/n\rightarrow 1$ as $\rho\rightarrow+\infty$, and therefore have the same behavior as in a de Sitter spacetime.

Consider instead the inner solutions. For the f\mbox{}inite solutions $\textbf{F}_{\pm}$ we obtain $(f / f_{GR}) \rightarrow 1$ and $(n^{\, \p} / n_{GR}^{\, \p}) \rightarrow 1$ as $\r \rightarrow 0^+$, where the corrections scale like $\r^3$. On the contrary, for the diverging solution $\textbf{D}$, the cubic terms in $h$ cancel out the contribution coming from the Schwarzschild-like terms, as explained above, and so $(f / f_{GR}) \rightarrow 0$ and $(n^{\, \p} / n_{GR}^{\, \p}) \rightarrow 0$ when $\r \rightarrow 0^+$. In this case, corrections are linear in $\r$.

Therefore, any global solution of equation (\ref{quintic}) which interpolates between $\textbf{L}$ and $\textbf{F}_{\pm}$ provides a realization of the Vainshtein mechanism in an asymptotically f\mbox{}lat spacetime, whereas an interpolation between $\textbf{C}_{\pm}$ or $\textbf{P}_{1,2}$ with $\textbf{F}_{\pm}$ exhibits the Vainshtein mechanism in an asymptotically non-f\mbox{}lat spacetime. On the other hand, any asymptotic solution which interpolates with the inner solution $\textbf{D}$ does no lead to the Vainshtein mechanism. These matchings will be explicitly exposed in the next section.

\section{Phase space diagram for solutions matching}
\label{Phase space diagram for solutions matching}

In the previous section, we characterized the number and properties of asymptotic and inner solutions in all the phase space. As we mentioned in section \ref{The quintic equation}, to make precise statements about the effectiveness of the Vainshtein mechanism it is enough to establish (for every point of the phase space) which asymptotic solution is connected to which inner solution by a global solution which interpolates between them. The aim of this section is to study the matching of asymptotic and inner solutions in all the phase space.

\subsection{Local solutions and the shape of the quintic}
\label{Local solutions and the shape of the quintic}

Since finding exact solutions of the quintic equation is extremely difficult, we need another method to determine, given a fixed asymptotic solution and a fixed inner solution, if there exists a global solution interpolating between them. To explain how this can be done, let's first of all note that we may see the quintic function \eqref{quinticFunction}, which is a function of two variables (when we keep $\a$ and $\b$ fixed), as a collection of functions of $h$ whose shape depend continuously on a parameter $A$. This idea can be formalized introducing the \emph{shape function} $q_{A} \, \big( h; \a, \b \big)$ which is defined as
\beq
\label{shapefunction}
q_{A} \, \big( h; \a, \b \big) = q \, \big( h, A ; \a, \b \big) \quad :
\eeq
the shape function is a function of $h$ only, and essentially, given a value of $A$, it is the quintic in $h$ which one obtains keeping fixed $A$ in the quintic function (\ref{quinticFunction}). At every $A$, the shape function has a certain set of zeros $\{r_i(A)\}_i$, which change continuously when $A$ changes: if $h(A)$ is a solution of the quintic equation, by definition $h(A)$ describes the continuous flow with $A$ of a particular zero of the shape function. Since we study the flow with $A$ at $\a$ and $\b$ fixed, for simplicity from now on we will omit to write the dependence from $\a$ and $\b$.

We would like to follow the opposite path, and infer the existence of a solution of the quintic equation from the study of the flow of the zeros of the shape function. This is indeed possible thanks to the implicit function theorem (see appendix \ref{The implicit function theorem}). In fact, if we start from a fixed $\bar{A}$ and find a simple zero $\bar{h}$ of the shape function, the implicit function theorem tells us that there exists a (local) solution $\bar{h}(A)$ of the quintic equation, which is defined in a neighborhood of $\bar{A}$, and which describes the flow with $A$ of the zero $\bar{h}$ we started with. Moreover, as we explain in the appendix \ref{The implicit function theorem}, there is a criterion which permits to infer the existence of global solutions of the quintic equation: if the flow of a zero $\bar{h}$ is such that the zero remains simple\footnote{We say that a zero $\bar{h}$ of the shape function $q_{A} \,( h)$ is simple if $\bar{h}$ is a simple root of the equation $q_{A} \, ( h ) = 0$.} for every value of $A$, then the local solution $\bar{h}(A)$ can be extended maximally to a global solution. Therefore, we are in principle able to find global solutions to the quintic equation just by studying how the shape of $q_{A} \, \big( h \big)$ evolves with $A$.

\subsection{Creation and annihilation of local solutions}
\label{Creation and annihilation of local solutions}

Let's consider instead what happens when, extending a local solution $h(A)$, we reach a point $\ti{A}$ when $d q_A/d h = 0$ and so the zero of the shape function is not simple. This situation graphically means that the shape function has a stationary point on the $h$ axis. Consider for example the case where the shape function has a local minimum below the $h$ axis, and there are two zeros around the minimum. If this minimum translates upwards when $A$ increases and eventually crosses the $h$ axis at a certain $A = \tilde{A}$, the two zeros join together and disappear at the axis crossing: it follows that the two local solutions $h_{12}(A)$ associated to the zeros stop existing at $A = \tilde{A}$. When this happens, by \eqref{derivativeDini} the derivative $dh_{12}/dA$ diverges at $A = \tilde{A}$, but the functions $h_{12}(A)$ remain bounded. The same happens when a local maximum of the shape function crosses the $h$ axis translating downwards. We will say in these cases that two local solution ``annihilate'' at $A = \tilde{A}$. If instead a local minimum of the shape function translates downwards when $A$ increases and crosses the $h$ axis at a certain $A = \tilde{A}$, two new zeros appear at $A = \tilde{A}$ and therefore two local solutions $h_{12}(A)$ of the quintic equation start existing at $A = \tilde{A}$: again, by \eqref{derivativeDini} the derivative $dh_{12}/dA$ diverges at the point $A = \tilde{A}$, but the values of the functions remain bounded. The same happens if a local maximum of the shape function translates upwards and crosses the $h$ axis. We will say in these cases that two local solution ``are created'' at a certain $A = \tilde{A}$. The creation and annihilation of local solutions and its relation with local maxima and minima of the shape function is well illustrated in figure \ref{Numerical4} and in figure \ref{plot8}.

The phenomenon of creation and annihilation of local solutions is found to be a general feature of the phenomenology of equation (\ref{quintic}). In fact, in most part of the phase space the number of asymptotic solution is different from the number of inner solutions: the reason why some of these solutions cannot be continued to all the radial domain $0 < \r < +\infty$ is always that they annihilate with some other local solution. Note that, in general, the solutions are created and annihilated in pairs, and the pairs of solutions have inf\mbox{}inite slope when they are created or they annihilate. Anyway, a note of caution is in order: the fact that a stationary point appears on the $h$ axis does not necessarily means that a solution disappears or is created. For example, if a horizontal inflection point of the shape function crosses the $h$ axis, then there is a value $A = \tilde{A}$ where there is a stationary point on the $h$ axis, and the implicit function theorem cannot be applied. Nevertheless, in this case the solution continues existing, even if at $A = \tilde{A}$ it has an infinite first derivative.

It is crucial to point out that, since the first derivative of a local solution of the quintic equation diverges at a creation/annihilation point, the gravitational potentials associated with this solution have diverging derivatives themselves at this point. This implies that, when a creation/annihilation point is approached, the approximations we used to derive the system of equations (\ref{solfab})-(\ref{solhab}) does not hold anymore (\emph{i.e.} the linear approximation on the gravitational potentials), and to understand what happens to the spacetime described by this solutions we should study the full theory. We don't attempt to do this, and therefore we cannot say anything about what happens to the spacetimes described by local solutions of the quintic equation which in our analysis cannot be extended to the complete radial domain.

\subsection{Analysis strategy}

Our analysis strategy is therefore the following: for every point of the phase space, we start from the zeros of the shape function at infinity $A = 0$ (\emph{i.e.} from the roots of the asymptotic equation), and we follow the evolution of the shape function when $A$ goes from zero to $+ \infty$. In this way, we determine which asymptotic solutions flow into an inner solution, and we determine which asymptotic solutions matches which inner solution. The study is done in three different ways.

On one hand, we study analytically the evolution of the shape function, in particular focusing on the evolution of the number and position of its inflection points. In many cases, the study of the position of the inflection points is enough to establish that in a certain interval of values for $h$ there always (\emph{i.e.} for every value of $A$) exists one simple zero of the shape function, thereby proving analytically the existence of the global solution of the quintic equation which corresponds to this zero. For this study it is necessary to characterize precisely the properties of the shape function at infinity, and the evolution of its properties when $A$ goes from zero to $+ \infty$: the details of the study of these properties are given in the appendices \ref{Useful properties of the quintic function} and \ref{Asymptotic structure of the quintic function}.

On the other hand, we plot numerically the shape function and continuously change the value of $A$ (of course, since it is a numerical procedure the modulation is not really continuous but procedes by small finite steps). Despite being less rigorous than the former procedure, this allows to visualize in a very efficient way the evolution of the shape function. Note that, as we explain in the appendix \ref{Useful properties of the quintic function}, there is no need to follow the evolution till $A \rightarrow + \infty$ because for every $\a$ and $\b$ there is a critical value $A_{crit}$ (which depends on $\a$ and $\b$) such that for $A > A_{crit}$ there are no more creations/annihilations of solutions, and so from the shape function at $A = A_{crit}$ one can infer unambiguously the matching of the solutions. Note that, since $h$ is defined on $(- \infty, +\infty)$, we don't plot the shape function $q_A(h)$ itself but its composition with the tangent function $q_A\big(tg(h)\big)$: this has the effect of compactifying the real axis into the interval $(-\pi/2, +\pi/2)$, and at the same time does not change the number and the relative order of the zeros.

Finally, we check the results of these two (somehow complementary) methods by solving with the software Mathematica\textcopyright$\,$ for symbolic and numeric calculations\footnote{\texttt{http://www.wolfram.com/mathematica/}} the condition of the presence of a stationary point on the $h$ axis. More precisely, we impose the condition that there exist a couple of values ($h$,$A$) where both the shape function $q_{A}(h;\a,\b)$ and its first derivative $d q_{A}/dh$ vanish: solving this condition gives constraints on the values for $\a$ and $\b$, and identifies the regions of the phase space where solution can annihilate/be created.

These three different approaches permit us to characterize the solution matching in a detailed way, and in the next section we present our results.

\subsection{Phase space diagram}

\begin{figure}[htp!]
\begin{center}
\includegraphics[width=12cm]{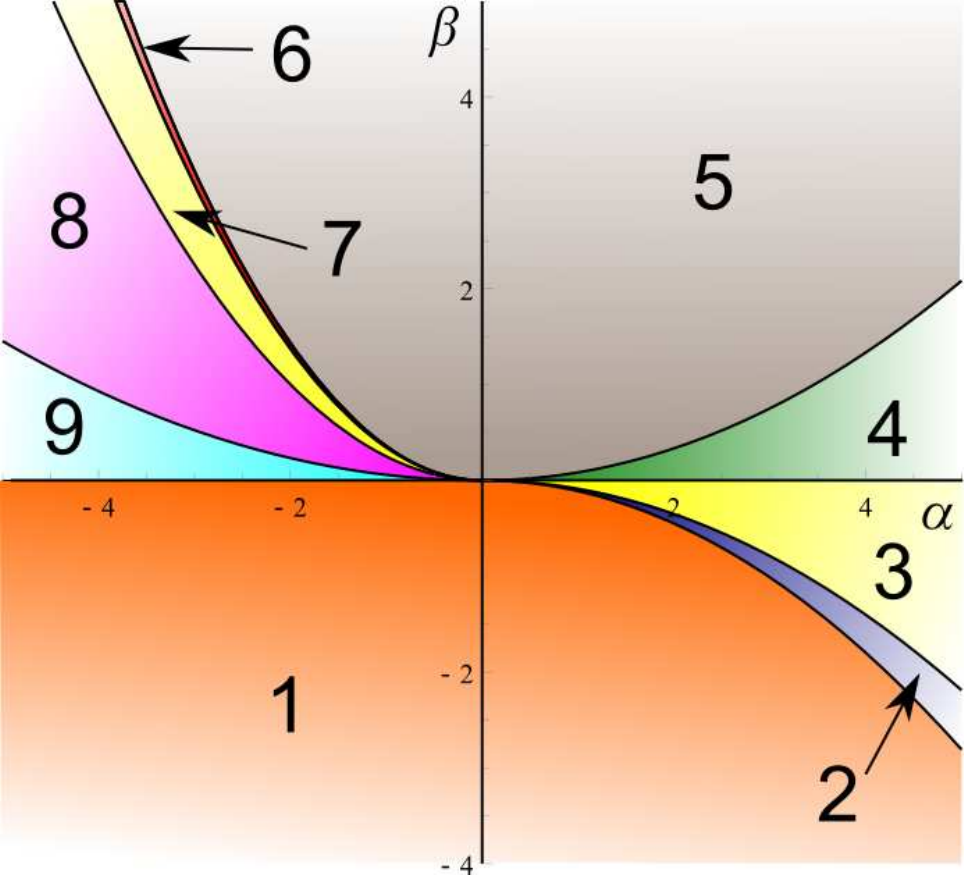}
\caption[Phase space diagram for solutions matching]{Phase space diagram in $(\a,\b)$ for the solutions to the quintic equation
(\ref{quintic}) in $h$, where the dif\mbox{}ferent regions show dif\mbox{}ferent matching of inner
solutions to asymptotic ones. The lines splitting the regions are half parabolas
($\b \propto \aq$, with $\a >0$ or $\a <0$) due to rescaling symmetry of eq.~(\ref{quintic}).}
\label{phase space}
\end{center}
\end{figure}

The phase space diagram which displays our results about solution matching is given in f\mbox{}igure
\ref{phase space}. We discuss separately the $\b > 0$ and $\b < 0$ part of the phase space, and
refer to the f\mbox{}igure for the numbering of the regions. The notation $\textbf{I}
\leftrightarrow \textbf{A}$ means that there is matching between the inner solution $\textbf{I}$
and the asymptotic solution $\textbf{A}$.

\subsubsection{$\b < 0$}

In this part of the phase space, there is only one inner solution, $\textbf{D}$, so there can be at most one global solution to (\ref{quintic}). There are three distinct regions which dif\mbox{}fer in the way the matching works:
\begin{itemize}
 \item[-] region 1: $\textbf{D} \leftrightarrow \textbf{C}_{+}$. In this region, there are three or f\mbox{}ive asymptotic solutions, and only one of them, $\textbf{C}_{+}$, is positive. This solution is the one which connects with the inner solution $\textbf{D}$, which is also positive, leading to the only global solution of eq.~(\ref{quintic}). The boundaries of this region are the line $\b = 0$ for $\a<0$ and the parabola $\b = c_{12} \, \a^2$ for $\a>0$, where $c_{12}$ is the negative\footnote{The equation $-4 - 8 \, y + 88 \, y^2 - 1076 \, y^3 + 2883 \, y^4 = 0$ has only two real roots, one positive and one negative.} root of the equation $-4 - 8 \, y + 88 \, y^2 - 1076 \, y^3 + 2883 \, y^4 = 0$ (approximately, $c_{12} \simeq -0.1124$). On the boundary $\b = c_{12} \, \a^2$ the matching $\textbf{D} \leftrightarrow \textbf{C}_{+}$ still holds, however the solution $h(\r)$ displays an inf\mbox{}lection point with vertical tangent.

 \item[-] region 2: $\textbf{No matching}$. In this region there are three asymptotic solutions. However, none of them can be extended all the way to $\r \rightarrow 0^+$, and so, despite the fact that local solutions exist both at inf\mbox{}inity and near the origin, equation (\ref{quintic}) does not admit any global solution. The boundaries of this region are the parabola $\b = c_{12} \, \a^2$ and the (negative) f\mbox{}ive-roots-at-inf\mbox{}inity parabola $\b = c_{-} \, \a^2$, where $c_{-}$ is the only real root of the equation $8 + 48 \, y - 435 \, y^2 + 676 \, y^3 = 0$ (approximately, $c_{-} \simeq -0.0876$).

 \item[-] region 3: $\textbf{D} \leftrightarrow \textbf{P}_{2}$. This region coincides with the $\a > 0$,
$\b < 0$ part of the f\mbox{}ive roots at inf\mbox{}inity region of the phase space (see f\mbox{}ig.
\ref{five roots}). The largest positive asymptotic solution, $\textbf{P}_{2}$, is the one which connects to $\textbf{D}$, leading to the only global solution of eq.~(\ref{quintic}). On the boundary $\b = c_{-} \, \a^2$ the matching $\textbf{D} \leftrightarrow \textbf{P}_{2}$ still holds, but the solution $h$ seen as a function of $A$ has inf\mbox{}inite derivative in $A=0$.
 \\
 \end{itemize}

\subsubsection{$\b > 0$}

In this part of the phase space, there are three inner solutions, $\textbf{D}$, $\textbf{F}_{+}$ and $\textbf{F}_{-}$, so there can be at most three global solutions to eq.~(\ref{quintic}). There are six distinct regions with dif\mbox{}ferent matching properties:
\begin{itemize}

 \item[-] region 4: $\textbf{F}_{-} \leftrightarrow \textbf{L}$ , $\textbf{D} \leftrightarrow \textbf{C}_{-}$. This region lies inside the $\a > 0$, $\b > 0$ part of the f\mbox{}ive roots at inf\mbox{}inity region of the phase space (see f\mbox{}ig.~\ref{five roots}), so there are f\mbox{}ive asymptotic solutions. Of the f\mbox{}ive asymptotic solution, $\textbf{C}_{-}$ and $\textbf{L}$ can always be extended to $\r \rightarrow 0^+$, while $\textbf{C}_{+}$, $\textbf{P}_{1}$ and  $\textbf{P}_{2}$ cannot. So there are just two global solutions to eq.~(\ref{quintic}). The boundaries of this region are the parabola $\b = c_{45} \, \a^2$, where $c_{45} = 1/12 \simeq 0.0833$, and the line $\b = 0$. On the boundary $\b = c_{45} \, \a^2$ there is the additional matching $\textbf{F}_{+} \leftrightarrow \textbf{C}_{+}$, and the corresponding solution is $h(\r) = const = + \sqrt{1/\,3\,\b}\,$.

 \item[-] region 5: $\textbf{F}_{+} \leftrightarrow \textbf{C}_{+}$ , $\textbf{F}_{-} \leftrightarrow \textbf{L}$, $\textbf{D} \leftrightarrow \textbf{C}_{-}$. In this region there are three or f\mbox{}ive asymptotic solutions; $\textbf{C}_{-}$ , $\textbf{C}_{+}$ and $\textbf{L}$ can always be extended to $\r \rightarrow 0^+$, while $\textbf{P}_{1}$ and $\textbf{P}_{2}$ , where present, cannot. So there are three global solutions to (\ref{quintic}). The boundaries of this region are the parabola $\b = c_{45} \, \a^2$ for $\a > 0$ and the parabola $\b = c_{56} \, \a^2$ for $\a < 0$, where $c_{56} = (5 + \sqrt{13})/24 \simeq 0.3586$. On the $\a < 0$ boundary $\b = c_{56} \, \a^2$ the matching works as in the rest of the region, but the solution $\textbf{F}_{-} \leftrightarrow \textbf{L}$ has an inf\mbox{}lection point with vertical tangent.

 \item[-] region 6: $\textbf{D} \leftrightarrow \textbf{C}_{-}$ , $\textbf{F}_{+} \leftrightarrow \textbf{C}_{+}$. In this region there are three asymptotic solutions, however only two of them can be extended to $\r \rightarrow 0^+$, while $ \textbf{L}$ cannot. Therefore, there are just two global solutions to eq.~(\ref{quintic}). The boundaries of this region are the parabolas $\b = c_{56} \, \a^2$ and $\b = c_{67} \, \a^2$, where $c_{67}$ is the positive root of the equation $-4 - 8 \, y + 88 \, y^2 - 1076 \, y^3 + 2883 \, y^4 = 0$ (approximately, $c_{67} \simeq 0.3423$). On the boundary $\b = c_{67} \, \a^2$ the matching works as in the rest of the region, but the solution $\textbf{D} \leftrightarrow \textbf{C}_{-}$ has an inf\mbox{}lection point with vertical tangent.

 \item[-] region 7: $\textbf{F}_{+} \leftrightarrow \textbf{C}_{+}$. In this region there are three asymptotic solutions, however only one of them can be extended to $\r \rightarrow 0^+$, while $\textbf{L}$ and $\textbf{C}_{-}$ cannot. The boundaries of this region are the parabola $\b = c_{67} \, \a^2$ and the (positive) f\mbox{}ive-roots-at-inf\mbox{}inity parabola $\b = c_{+} \, \a^2$, where $c_{+} = 1/4$. Note that on the ($\a < 0$) part of the parabola $\b = 1/3 \, \aq$ there is the additional matching $\textbf{F}_{-} \leftrightarrow \textbf{C}_{-}$, so for these points there are two global solutions to eq.~(\ref{quintic}). On the boundary $\b = c_{+} \, \a^2$ there are the additional matchings $\textbf{F}_{-} \leftrightarrow \textbf{P}_{1}$ , $\textbf{D} \leftrightarrow \textbf{P}_{2}$, and the solutions corresponding to both these additional matchings, seen as functions of $A$, display an inf\mbox{}inite derivative in $A=0$.

 \item[-] region 8: $\textbf{F}_{+} \leftrightarrow \textbf{C}_{+}$ , $\textbf{F}_{-} \leftrightarrow \textbf{P}_{1}$ , $\textbf{D} \leftrightarrow \textbf{P}_{2}$. This region lies inside the $\a < 0$, $\b > 0$ part of the f\mbox{}ive roots at inf\mbox{}inity region of the phase space (see f\mbox{}ig.~\ref{five roots}), so there are f\mbox{}ive asymptotic solutions. Only three of them can be extended to $\r \rightarrow 0^+$, while $ \textbf{C}_{-}$ and $\textbf{L}$ cannot. The boundaries of this region are the parabolas $\b = c_{+} \, \a^2$ and $\b = c_{89} \, \a^2$, where $c_{89} = (5 - \sqrt{13})/24 \simeq 0.0581$. On the boundary $\b = c_{89} \, \a^2$ the matchings are the same as in the rest of the region, but the solution $h(\r)$ corresponding to the matching $\textbf{F}_{+} \leftrightarrow \textbf{C}_{+}$ has an inf\mbox{}lection point with vertical tangent.

 \item[-] region 9: $\textbf{F}_{-} \leftrightarrow \textbf{P}_{1}$ , $\textbf{D} \leftrightarrow \textbf{P}_{2}$. This region lies inside the $\a < 0$, $\b > 0$ part of the f\mbox{}ive roots at inf\mbox{}inity region of the phase space (see f\mbox{}ig.~\ref{five roots}), so there are again f\mbox{}ive asymptotic solutions. The matching is similar to that of region 8, apart from the fact that $\textbf{C}_{+}$ cannot be extended to $\r \rightarrow 0^+$ anymore; hence there are just two global solutions to eq.~(\ref{quintic}). The boundaries of this region are the parabola $\b = c_{89} \, \a^2$ and line $\b = 0$.
\end{itemize}

\noindent We note that the decaying solution $\textbf{L}$ never connects to the diverging one $\textbf{D}$,
so we cannot have a spacetime which is asymptotically f\mbox{}lat and exhibit the self-shielding of the
gravitational f\mbox{}ield at the origin. On the other hand, f\mbox{}inite non-zero asymptotic solutions ($\textbf{C}_{\pm}$ or $\textbf{P}_{1,2}$) can connect to both f\mbox{}inite and diverging inner solutions.
Therefore, one can have an asymptotically non-f\mbox{}lat spacetime which presents self-shielding at the origin, or an asymptotically non-f\mbox{}lat spacetime which tends to Schwarzschild spacetime for small radii. More precisely, for $\b < 0$ there are only solutions displaying the self-shielding of the gravitational f\mbox{}ield, apart from region 2 where there are no global solutions. Therefore the Vainshtein mechanism never works for $\b < 0$. In contrast, for $\b > 0$ all three kinds of global solutions are present. Solutions with asymptotic f\mbox{}latness and the Vainshtein mechanism are present in regions 4 and 5, while solutions which are asymptotically non-f\mbox{}lat and exhibit the Vainshtein mechanism do exist in all ($\b > 0$) regions but region 4. Finally, solutions which display the self-shielding of the gravitational f\mbox{}ield are present in all ($\b > 0$) regions but region 7.

\section{Numerical solutions}

We said in the previous sections that, having characterized geometrically the asymptotic and inner solutions, to study the Vainshtein mechanism it is enough to know how the matching between asymptotic and inner solutions works. To verify this assertion and corroborate the validity of our results, we solved numerically the system of equations (\ref{solfab}) $-$ (\ref{solhab}) in several points of the phase space and for each of the three different types of matching. We present here the numerical solutions for the $h$ f\mbox{}ield and the gravitational potentials in some representative cases. We choose a specif\mbox{}ic realization for each of the three physically distinct cases, namely asymptotic f\mbox{}latness with Vainshtein mechanism, asymptotically non-f\mbox{}lat spacetime with Vainshtein mechanism, and asymptotically non-f\mbox{}lat spacetime with self-shielded gravitational f\mbox{}ield at the origin. In addition, we consider the case in which there are no global solutions to eq.~(\ref{quintic}). This provides an illustration of what happens, in general, to local solutions of eq.~(\ref{quintic}) which cannot be extended to the whole radial domain, and give an insight on the phenomenology of the equation (\ref{quintic}).

\subsection{Asymptotic f\mbox{}latness with Vainshtein mechanism}

\begin{figure}[htp!]
\begin{center}
\includegraphics[width=11cm]{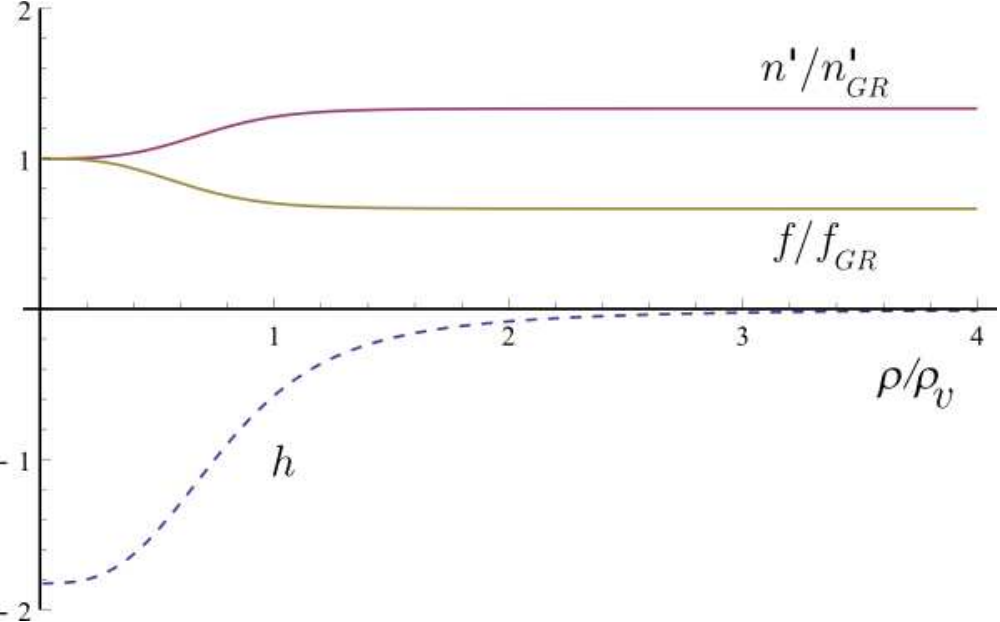}
\caption[Numerical solution for the case $\textbf{F}_{-} \leftrightarrow \textbf{L}$.]{Numerical solutions for the case $\textbf{F}_{-} \leftrightarrow \textbf{L}$.}
\label{Numerical1}
\end{center}
\end{figure}

Let's consider the case in which the solution of eq.~(\ref{quintic}) connects to the decaying solution at inf\mbox{}inity $\textbf{L}$ and to a f\mbox{}inite inner solution (in this case $\textbf{F}_{-}$). In f\mbox{}igure \ref{Numerical1}, the numerical solutions for $h$ (dashed line), $f / f_{GR}$ (bottom continuous line) and $n^{\, \p} / n_{GR}^{\, \p}$ (top continuous line) are plotted as functions of the dimensionless radial coordinate $x \equiv \r / \r_v$. These solutions correspond to the point $(\alpha, \beta) = (0 \, , 0.1)$ of the phase space.

This plot displays very clearly the presence of the vDVZ discontinuity and its resolution via the Vainshtein mechanism. For large scales, $h$ is small and the gravitational potentials behave like the Schwarzschild one, however their ratio is dif\mbox{}ferent from one, unlike the massless case. Note that the ratio of the two potentials for $\r \gg \r_v$ is independent of $m$, so does not approach one as $m \rightarrow 0$ (vDVZ discontinuity). However, on small scales $h$ is strongly coupled, and well inside the Vainshtein radius the two potentials scale again as the Schwarzschild one, but their ratio is now one even if $m \neq 0$. So, the strong coupling of the $h$ f\mbox{}ield on small scales restores the agreement with GR (Vainshtein mechanism).

\subsection{Asymptotically non-f\mbox{}lat spacetime with Vainshtein mechanism}

\begin{figure}[htp!]
\begin{center}
\includegraphics[width=11cm]{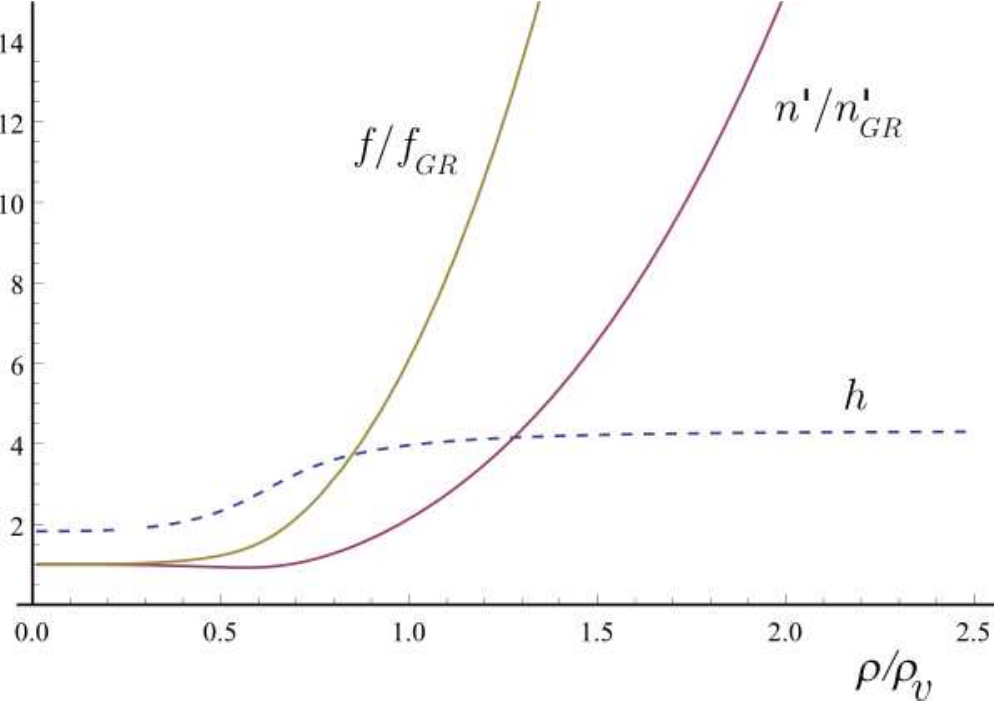}
\caption[Numerical solution for the case $\textbf{F}_{+} \leftrightarrow \textbf{C}_{+}$]{Numerical solutions for the case $\textbf{F}_{+} \leftrightarrow \textbf{C}_{+}$.}
\label{Numerical2}
\end{center}
\end{figure}

Let's consider now the case in which the solution of eq.~(\ref{quintic}) connects to a f\mbox{}inite solution at inf\mbox{}inity and to a f\mbox{}inite inner solution. We consider for def\mbox{}initeness the phase space point $(\alpha, \beta) = (0 \, , 0.1)$. In f\mbox{}igure \ref{Numerical2}, we plot the numerical results for the gravitational potentials (normalized to their GR values) and the global solution of eq.~(\ref{quintic}) which interpolates between the inner solution $\textbf{F}_{+}$ and the asymptotic solution $\textbf{C}_{+}$.

We can see that, on large scales, the gravitational potentials are not only dif\mbox{}ferent one from the other but also behave very dif\mbox{}ferently compared to the GR case. However, on small scales there is a macroscopic region where the two potentials agree, and their ratio with the Schwarzschild potential stays nearly constant and equal to one. Therefore, also in this case the small scale behavior of $h$ guarantees that GR results are recovered, even if the spacetime is not asymptotically f\mbox{}lat. This behavior provides then, in a more general sense, a realization of the Vainshtein mechanism.

\subsection{Asymptotically non-f\mbox{}lat spacetime with self-shielding}

\begin{figure}[htp!]
\begin{center}
\includegraphics[width=11cm]{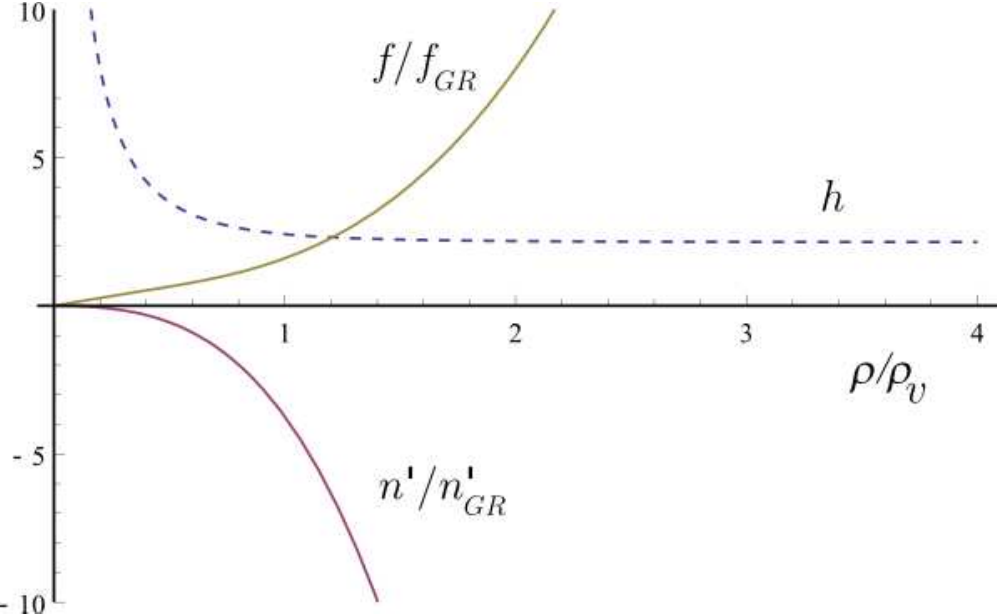}
\caption[Numerical solution for the case  $\textbf{D} \leftrightarrow \textbf{C}_{+}$]{Numerical solutions for the case $\textbf{D} \leftrightarrow \textbf{C}_{+}$.}
\label{numerical3}
\end{center}
\end{figure}

We turn now to the case where the solution of eq.~(\ref{quintic}) connects to a f\mbox{}inite solution at inf\mbox{}inity and to the diverging inner solution. In f\mbox{}igure \ref{numerical3}, we plot the global solution $h$ and the associated gravitational potentials, normalized to their GR values, corresponding to the phase space point $(\alpha, \beta) = (-1 \, , -0.5)$. It is apparent that there are no regions where the solutions behave like in the GR case.

\begin{figure}[htp!]
\begin{center}
\includegraphics[width=11cm]{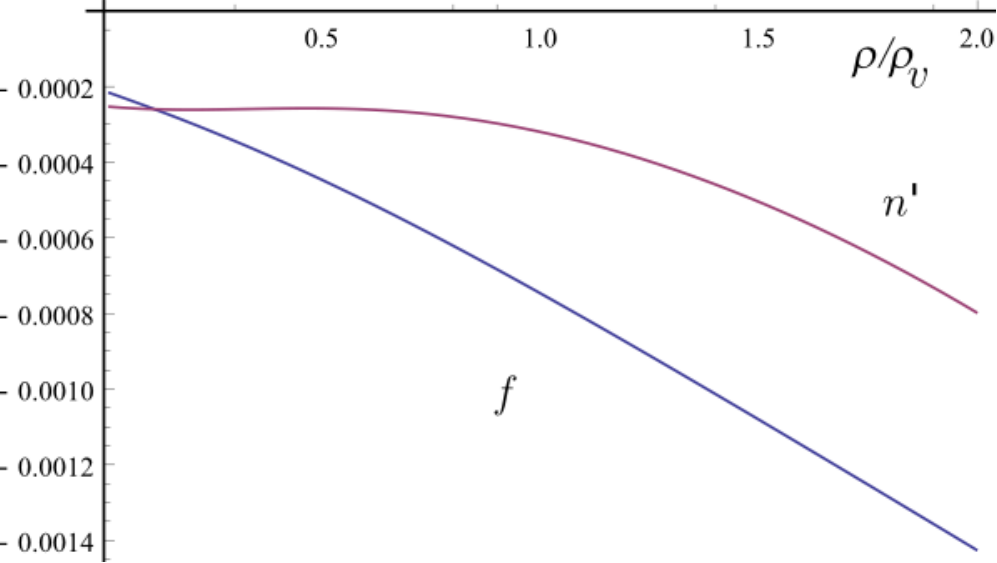}
\caption[Numerical solution for the self-shielding case]{Numerical solutions for the gravitational potentials, for the case
$\textbf{D} \leftrightarrow \textbf{C}_{+}$.}
\label{numerical3nonnorm}
\end{center}
\end{figure}

To see that the gravitational potentials are indeed f\mbox{}inite at the origin, we plot in f\mbox{}igure \ref{numerical3nonnorm} the potentials $f$ and $n'$ themselves, as functions of $\rho/\r_v$. We choose for def\mbox{}initeness the following ratio between the Compton wavelength and the gravitational radius $\r_{c} / \r_{g} = 10^{6}$, and plot the potentials for $0.01 < \rho/\r_v < 2$. Note that, since in this case $\r_{c} / \r_{v} = \sqrt[3]{\r_{c} / \r_{g}} = 10^{2}$, the range where the functions are plotted is well inside the range of validity of our approximations. We can see that the potentials approach a f\mbox{}inite value as $\r \rightarrow 0^+$, and so indeed the gravitational f\mbox{}ield does not diverge at the origin.

\subsection{No matching}

\begin{figure}[htp!]
\begin{center}
\includegraphics[width=11cm]{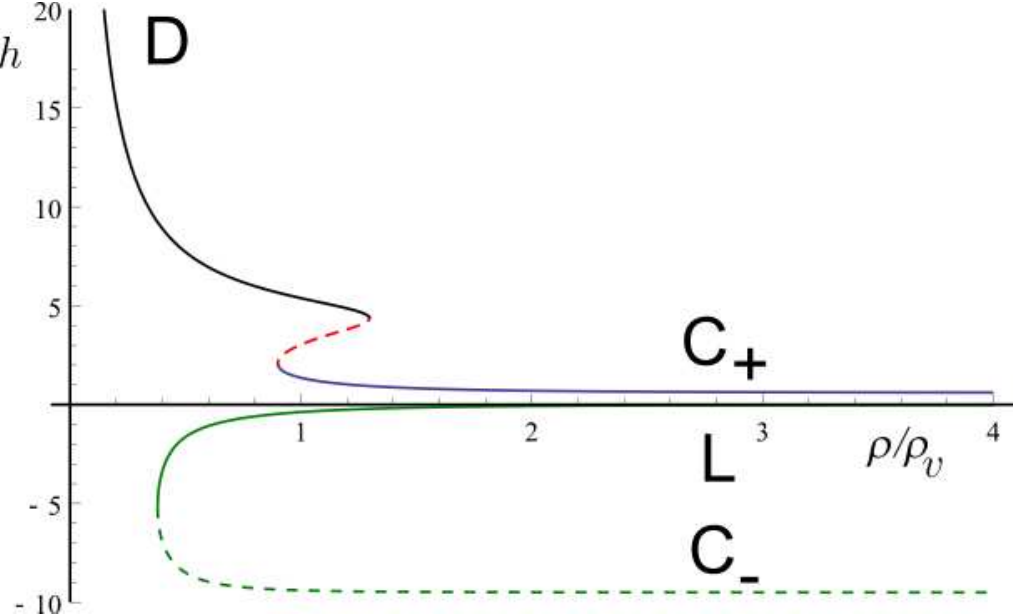}
\caption[Numerical solution for the no-matching case]{Numerical results for all local solutions of eq.~(\ref{quintic}) in the case where there is no
matching.}
\label{Numerical4}
\end{center}
\end{figure}

Finally, we consider the case in which equations (\ref{solfab}) $-$ (\ref{solhab}) do not admit global solutions. We consider for def\mbox{}initeness the phase space point $(\alpha, \beta) = (1 \, , -0.092)$. In f\mbox{}igure \ref{Numerical4} we plot all the local solutions of the quintic equation (\ref{quintic})
as functions of the dimensionless radial coordinate $x \equiv \r / \r_v$.

For $0 < x < 0.38$, there is only one local solution (the top continuous curve), which connects to the diverging inner solution $\textbf{D}$. At $x \simeq 0.38$ a pair of solutions is created (dashed and continuous negative valued curves), and at $x \simeq 0.9$, another pair of solutions is created (positive valued dashed curve and positive valued bottom continuous curve). However, at $x \simeq 1.3$ one of the newly created solutions (the positive valued dashed curve) annihilates with the solution which connects to the inner solution, so for $x > 1.3$ there are three local solutions, which f\mbox{}inally connect with the asymptotic solutions $\textbf{C}_{-}$ , $\textbf{L}$ and $\textbf{C}_{+}$. Therefore, the number of existing local solutions is one for $0 < x < 0.38$, three for $0.38 < x < 0.9$, f\mbox{}ive for $0.9 < x < 1.3$ and three for $x > 1.3$. We can see that, despite the fact that for every $\r$ there is at least one local solution, there does not exist a solution which extends over the whole radial domain. 

\begin{figure}[htp!]%
\begin{center}%
\parbox{4.7cm}{\includegraphics[width=4.7cm]{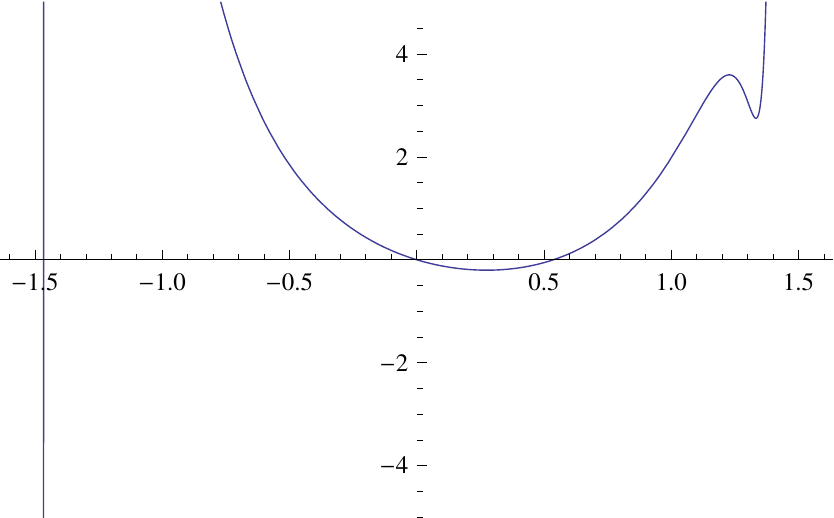}}%
\hspace{2mm}%
\parbox{4.7cm}{\includegraphics[width=4.7cm]{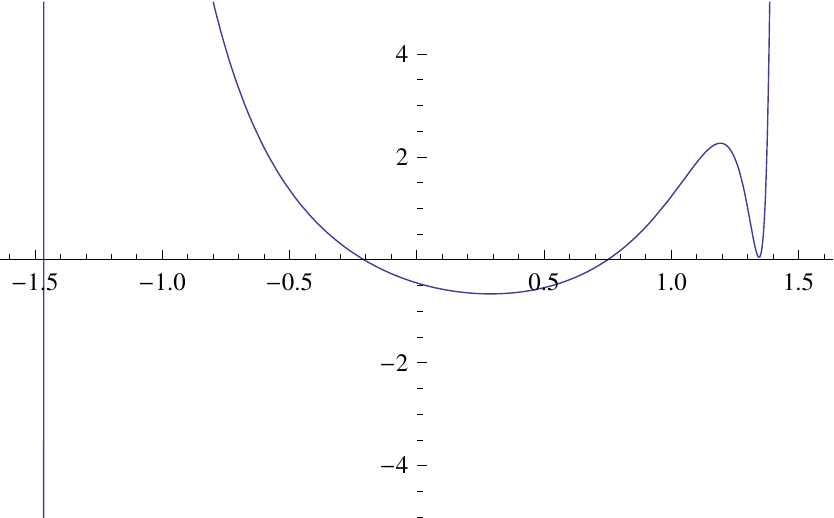}}%
\hspace{2mm}%
\parbox{4.7cm}{\includegraphics[width=4.7cm]{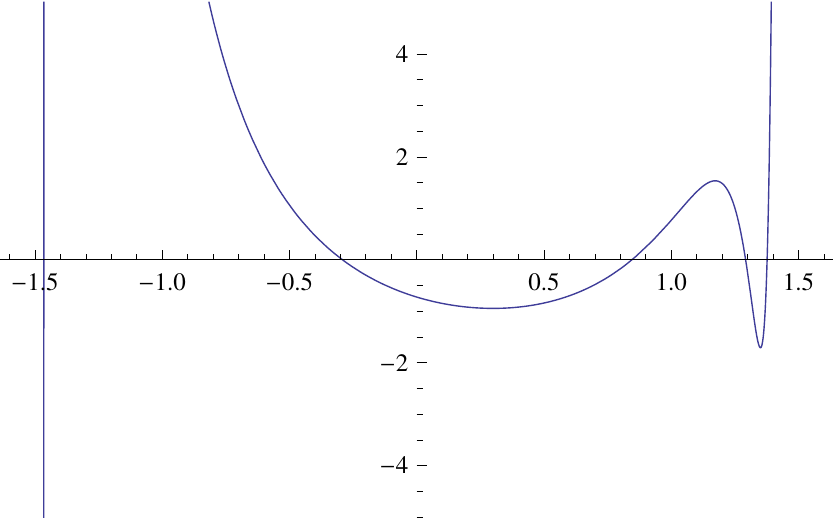}}%
\hspace{2mm}%
\parbox{4.7cm}{\includegraphics[width=4.7cm]{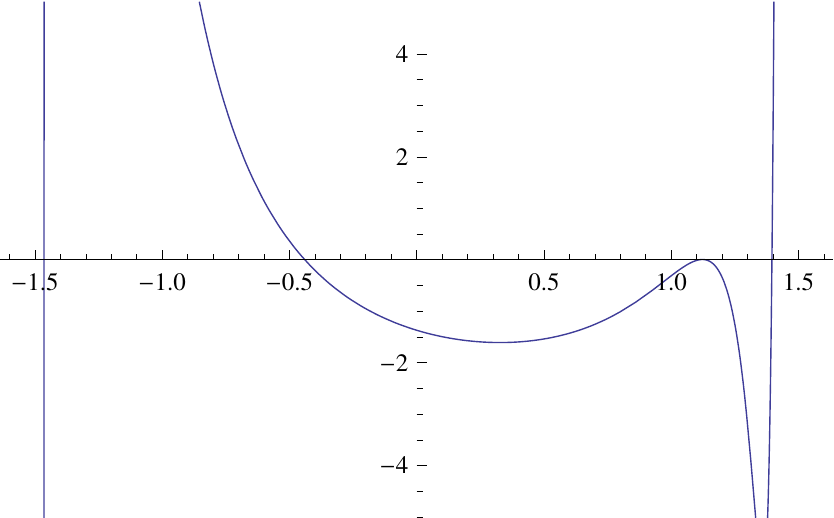}}%
\hspace{2mm}%
\parbox{4.7cm}{\includegraphics[width=4.7cm]{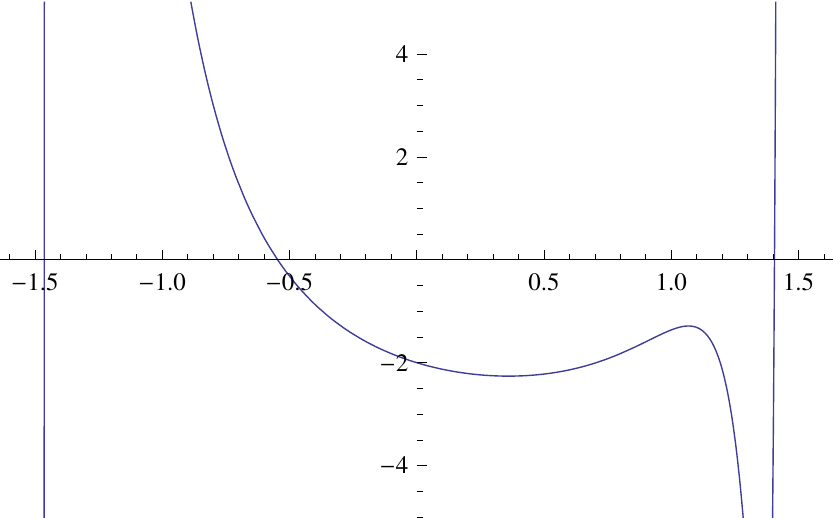}}%
\hspace{2mm}%
\parbox{4.7cm}{\includegraphics[width=4.7cm]{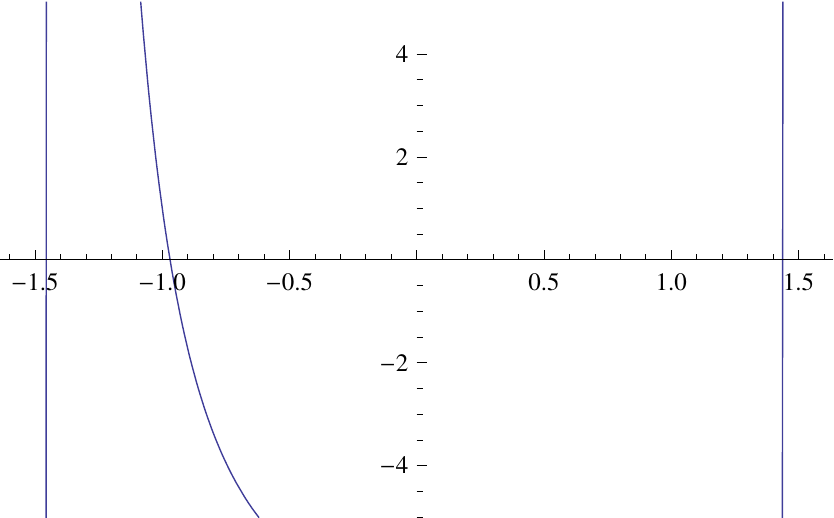}}%
\hspace{2mm}%
\parbox{4.7cm}{\includegraphics[width=4.7cm]{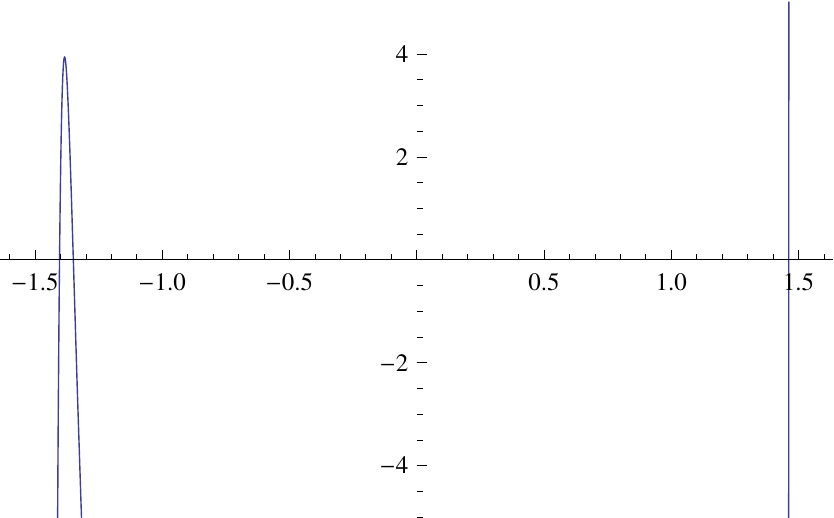}}%
\hspace{2mm}%
\parbox{4.7cm}{\includegraphics[width=4.7cm]{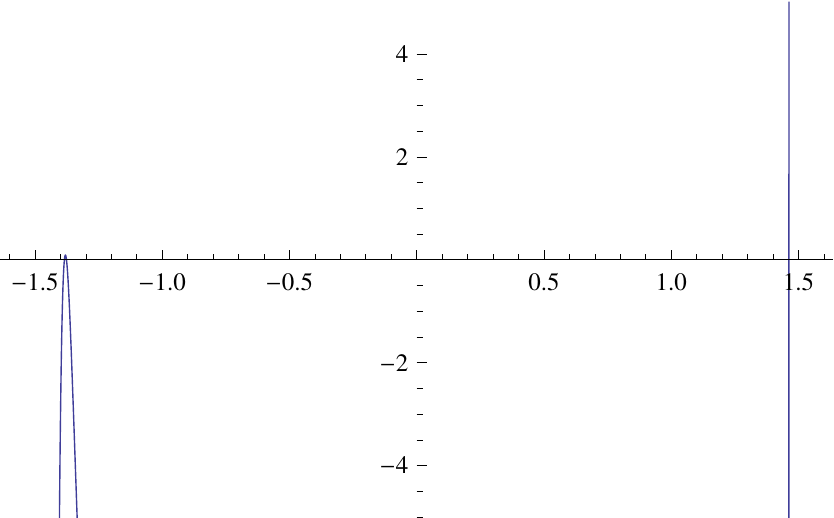}}%
\hspace{2mm}%
\parbox{4.7cm}{\includegraphics[width=4.7cm]{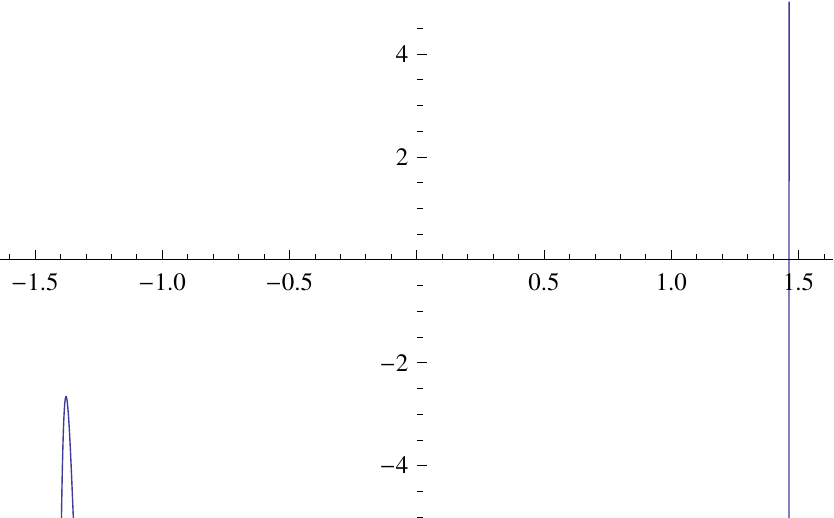}}%
\caption[Quintic function in the no-matching case.]{Quintic function for increasing values of $A$, no-matching case.}%
\label{plot8}
\end{center}%
\end{figure}

To clarify the meaning of figure \ref{Numerical4}, we plot in figure \ref{plot8} several snapshots of the quintic function at different values of $A$, for the same phase space point $(\alpha, \beta) = (1 \, , -0.092)$. Figure \ref{plot8} shows the creation and annihilation of solutions from the point of view of the quintic instead of from the point of view of the implicitly defined functions: note that the quintic is plotted for increasing values of $A= 1/x^3$, while in figure \ref{Numerical4} the local solutions are plotted as functions of $x$. The plots of the quintic correspond to the following values of $A$: $A = 0$, $A = 0.456 \leftrightarrow x = 1.3$, $A = 0.716$, $A = 1.356 \leftrightarrow x = 0.9$, $A = 2$, $A = 6.93$, $A = 17.9$, $A = 18.35 \leftrightarrow x = 0.38$ and $A = 18.68$.

At $A = 0$ there are three roots, one negative, one positive and the zero root, which correspond to the three asymptotic solutions $\textbf{C}_{-}$ , $\textbf{C}_{+}$ and $\textbf{L}$. At $A = 0.456 \leftrightarrow x = 1.3$ a new double root appears, and two local solutions are created: these are the top continuous and dashed curve of figure \ref{Numerical4}. As is apparent in the $A = 0.716$ plot, for $0.456 < A < 1.356$ there are five roots and so five local solutions. At $A = 1.356 \leftrightarrow x = 0.9$ one of the newly created solutions (the top dashed curve of figure \ref{Numerical4}) annihilates with the asymptotic solution $\textbf{C}_{+}$, which ceases existing: for $1.356 < A < 18.35$ there are three roots and therefore three local solutions. At $A = 18.35 \leftrightarrow x = 0.38$ the asymptotic solution $\textbf{C}_{-}$ annihilates with the asymptotic solution $\textbf{L}$, and for $A > 18.35$ only one local solution survives, the one created at $A = 0.456 \leftrightarrow x = 1.3$ which correspond to the top continuous curve in figure \ref{Numerical4}. This solution is the one which connects to the inner solution $\textbf{D}$ when $A \rightarrow + \infty \, \leftrightarrow \, x \rightarrow 0^+$.

Note that, as we discussed in general in section \ref{Creation and annihilation of local solutions} and in appendix \ref{The implicit function theorem}, the solutions are created and annihilated in pairs. Furthermore, the pairs of solutions have inf\mbox{}inite slope when they are created and when they annihilate, while their values remain bounded.

\clearemptydoublepage
\chapter*{Conclusions}

\addcontentsline{toc}{chapter}{Conclusions}

\pagestyle{fancyplain}
\lhead[\fancyplain{}{\sl{Conclusions}}]{\fancyplain{}{}}
\lfoot[\fancyplain{}{}]{\fancyplain{}{}}
\chead[\fancyplain{}{}]{\fancyplain{}{}}
\cfoot[\fancyplain{\rm\thepage}{\rm\thepage}]{\fancyplain{\rm\thepage}{\rm\thepage}}
\rhead[\fancyplain{}{}]{\fancyplain{}{\sl{Conclusions}}}
\rfoot[\fancyplain{}{}]{\fancyplain{}{}}

Recent cosmological observations seem to suggest that the universe is currently undergoing a period of accelerated expansion. Despite being unexpected, this result can be explained assuming the presence of a non-zero and fine-tuned cosmological constant, or the existence of an exotic source of energy which is usually termed dark energy. However, from another point of view, these observations may indicate that General Relativity is not a good description for gravity at very large scales. To test this idea, it is necessary to consider theories whose predictions differ from the ones of General Relativity only at very large scales, and see if they can fit the data in a satisfying way.

In general, theories which modify gravity at large distances involve more degrees of freedom than General Relativity, and for these theories to be phenomenologically viable it is necessary that the extra degrees of freedom do not lead to instabilities and are screened at terrestrial and astrophysical scales. The presence of ghost instabilities is in fact quite a common problem in theories which modify gravity at large distances. Regarding the screening of the extra degrees of freedom, several screening mechanisms have been proposed and among them the Vainshtein mechanism, where derivatives self-interactions of a field are responsible for its screening, is very well known. In this thesis we considered two different classes of theories which modify gravity at large distances, the Cascading DGP and the dRGT massive gravity, and investigated their phenomenological viability. In particular, we investigated the presence of ghosts in the 6D Cascading DGP model, which is the minimal set-up of the Cascading DGP class of models, and we investigated the effectiveness of the Vainshtein mechanism in the dRGT massive gravity theory.

Regarding the 6D Cascading DGP model, we studied perturbations at first order around background configurations where positive tension is localized on the cod-2 brane. To fix the theory uniquely, we chose a particular realization of the set-up where the thickness of the codimension-1 brane is much smaller than the longitudinal thickness of the codimension-2 brane, so that the codimension-2 brane can be considered as a ``ribbon'' lying inside a thin codimension-1 brane. We performed a perturbative analysis in a bulk based approach, where both the metric and the position of the codimension-1 brane are free to fluctuate, and used gauge invariant variables and master variables to deal with the issue of gauge invariance. We showed that, at least for first order perturbations around the background configurations mentioned above, the thin limit of the codimension-2 brane inside the already thin codimension-1 brane is well defined; furthermore, we confirmed that gravity on the codimension-2 brane remains finite even in the codimension-2 thin limit, and therefore a source with a generic (weak) energy-momentum tensor can be localized on the thin codimension-2 brane.

Concerning the presence of ghosts, we confirmed the existence (at least in a specific decoupling limit of the model) of a critical value $\blac$ for the tension of the codimension-2 brane which separates background configurations which possess a ghost in the scalar sector of perturbations ($\bla < \blac$) and background configurations which are ghost-free at first order on perturbations ($\bla > \blac$). However, the expression we found for the critical tension in terms of the free parameters of the model (the mass scales $m_5$ and $m_6$) is different from the expression which appears in the literature: in particular, our result indicate that the critical tension is always smaller than the maximum tension which is possible to put on the codimension-2 brane. This means that, differently from the claims which appear in the literature, the models characterized by $m_6 > m_5$ (where gravity cascades directly from 6D to 4D in a static and spherically symmetric configuration) can be ghost-free if enough tension is put on the codimension-2 brane. To understand why we find a different result, we identified the way to change our hypothesis to reproduce in our framework the literature result, and showed that the two results for the critical tension are obtained by using different procedures to perform the pillbox integration across the codimension-2 brane. We then checked which of the two procedures is correct by performing numerically the pillbox integration in the case of a pure tension perturbation, where the exact solution for the perturbation fields is known. The result obtained by performing the pillbox integration with our procedure converges to the exact result in the thin limit, while the result obtained following the other procedure does not: this supports the claim that our result for the critical tension is the correct one, and that the models characterized by $m_6 > m_5$ are not ruled out by the unavoidable presence of a ghost around these background solutions.

We conclude that the Cascading DGP is a very promising framework to modify gravity, and to address the cosmological constant problem and the cosmological late time acceleration problem. However, several aspects of this framework need to be investigated further. To begin with, it is important to establish if the Vainshtein mechanism is effective and if it restores the agreement with the predictions of GR where the latter is well tested. It would also be necessary to perform a study at higher order in perturbations to confirm the result that putting enough tension on the codimension-2 brane is enough to get rid of ghosts around flat solutions. Furthermore, it would be important to investigate if the Cascading DGP model suffers from a dangerously small strong coupling scale, as its similarity with the DGP model may suggest. Concerning the cosmological constant problem, it would be interesting to see if this model can indeed provide a realization of the self-tuning mechanism, and from another point of view if it provides a realization of the degravitation mechanism at full non-linear level. Concerning cosmology, it is important to derive explicitly solutions of the Friedmann-Lema\^{\i}tre-Robertson-Walker form, which are still missing, and see if the self-accelerating solutions which have been derived are plagued by ghost instabilities or not. Once done that, it would be interesting to see if these cosmological solutions can fit the observational data better than $\La$CDM, and more in general if the agreement with the data is significant or not.

Regarding the dRGT massive gravity, to study the Vainshtein mechanism we considered static and spherically symmetric solutions. Since there are two branches of solutions which satisfy this symmetry requirement, we considered only the branch where the Vainshtein mechanism can be effective (the diagonal branch). We focused on scales smaller than the Compton radius of the gravitational f\mbox{}ield, and considered the weak f\mbox{}ield limit for the gravitational potentials, while keeping all the non-linearities of the scalar mode which is involved in the screening. For every point of the two free-parameters phase space, we characterized completely the number and properties of asymptotic solutions on large scales, and also of inner solutions on small scales. In particular, there are two kinds of asymptotic solutions, one which is asymptotically f\mbox{}lat and another one which is not. There are also two kinds of inner solutions, one which displays the Vainshtein mechanism and the other which exhibits the self-shielding of the gravitational f\mbox{}ield near the origin.

We described under which circumstances the theory admits global solutions interpolating between the asymptotic and inner solutions, and found that the asymptotically f\mbox{}lat solution connects only to inner solutions displaying the Vainshtein mechanism, while solutions which diverge asymptotically can connect to both kinds of inner solutions. Furthermore, we showed that there are some regions in the parameter space where global solutions do not exist, and characterised precisely in which regions of the phase space the Vainshtein mechanism is working. We showed that there is a significant part of the phase space where the Vainshtein mechanism is effective, which correspond to theories which are phenomenologically viable.

Our study embraces all of the phase space spanned by the two parameters of the theory. Notably, we found that, within our approximations, the asymptotic and inner solutions cannot in general be extended to the whole radial domain. In particular, we exhibited  extreme cases in which  global solutions do not exist at all. This happens because at a f\mbox{}inite radius the derivatives of the metric components diverge, while the metric components themselves remain bounded. When the derivatives of the metric cease to be small, the approximations we used to derive the equations under study break down. It would be interesting to study what happens at this radius in the full theory.

In conclusion, the formulation of a consistent interacting theory of a massive spin-2 field, which is explicitly provided by the dRGT massive gravity, is without doubt an important success. On the other hand, some features of the dRGT massive gravity models with flat absolute geometry are not satisfying. Considering the diagonal branch of spherically symmetric solutions, the fact that non-linearities in the helicity-0 mode of the St\"uckelberg fields becomes important at astrophysical scales is associated to a dangerously small strong coupling scale. This is analogous to what happens in the DGP model (see section \ref{Strong coupling problem}), but the problem seem to be even worse in the case of dRGT massive gravity \cite{Burrage:2012ja} (however there is no general consensus on this point, for another opinion see \cite{Berezhiani:2013dw, Berezhiani:2013dca, deRham:2013hsa}). It is also worthwhile to point out that, after our results (chapter \ref{Vainshtein Mechanism in Massive Gravity}) were published, further studies have shown that the solutions with $\b \neq 0$ are unstable in vacuum \cite{Koyama:2013paa, Berezhiani:2013dca}, and also that the solutions with $\b = 0$ and $\a > 0$ (with our conventions) become unstable inside realistic sources \cite{Berezhiani:2013dw} (while the case $\a = \b = 0$ is ruled out by solar system observations \cite{Koyama:2011yg}). The solutions in the non-diagonal branch instead do not suffer from the strong coupling problem, and contains interesting solutions of Schwarzschild-de Sitter form, where the effective cosmological constant is set by the graviton mass. Interestingly, a subclass of these solutions can be mapped by a suitable coordinate transformation in a configuration where the physical metric describes a homogeneous and isotropic self-accelerating universe, while the fiducial metric is inhomogeneous \cite{Koyama:2011xz, Koyama:2011yg}. It was hoped that this solutions may lead to a geometrical realization of the cosmological self-acceleration: however, the study of perturbations around these backgrounds revealed the presence of instabilities \cite{Koyama:2011wx, Tasinato:2012ze}.

Considering more in general self-accelerating cosmological solutions (see section \ref{Cosmology in dRGT massive gravity}), if we assume that the fiducial metric is isotropic we can obtain self-accelerating Friedmann-Lema\^{\i}tre-Robertson-Walker solutions both with homogeneous and inhomogeneous fiducial metric \cite{Gumrukcuoglu:2011ew, Koyama:2011xz, Koyama:2011yg, Gratia:2012wt} (where in the former case only negative spatial curvature is allowed \cite{Gumrukcuoglu:2011ew}, while in the latter case we can have also positive and zero spatial curvature), but all these solutions are unstable \cite{Gumrukcuoglu:2011zh, DeFelice:2012mx, Tasinato:2012ze}. Therefore, to look for stable and self-accelerating cosmological solutions in dRGT massive gravity we need to consider configurations where the fiducial metric is anisotropic \cite{Gumrukcuoglu:2012aa, DeFelice:2013awa}, or where the physical spacetime is inhomogeneous at scales \emph{larger} that the Hubble horizon (in which case we expect the usual Friedmann-Lema\^{\i}tre-Robertson-Walker cosmology to be reproduced due to the Vainshtein mechanism, at least when the energy density of the universe is not too low \cite{D'Amico:2011jj}).

In addition to these unsatisfying aspects, serious issues regarding the general structure of the theory have been uncovered concerning both superluminal propagation, acausalities \cite{Deser:2013eua,Deser:2013qza} and anomalous source effects \cite{Deser:2013rxa}. To overcome the problems listed above, and for example find stable and self-accelerating solutions with an isotropic fiducial metric, it is necessary to extend the theory. Several possibilities have been investigated so far. For example, a natural idea is to consider a non-flat absolute geometry \cite{Hassan:2011tf}, which is often taken to be de Sitter \cite{Langlois:2012hk, deRham:2012kf}, while another natural extension is found by promoting the fiducial metric, which in dRGT massive gravity is non-dynamical, to a dynamical object, obtaining a bigravity theory \cite{Hassan:2011zd, Babichev:2013pfa, Comelli:2011zm, Comelli:2012db, Fasiello:2013woa}. A somehow more elaborate construction is provided by the mass-varying and the quasi-dilaton extensions of the dRGT massive gravity \cite{Huang:2012pe, Saridakis:2012jy, Cai:2012ag, Leon:2013qh, D'Amico:2012zv, Gannouji:2013rwa, Gumrukcuoglu:2013nza, DeFelice:2013tsa}. The study of these extensions of the dRGT massive gravity theory is currently an active area of research.

\clearemptydoublepage
\appendix

\pagestyle{fancyplain}
\lhead[\fancyplain{}{\sl{Appendices}}]{\fancyplain{}{}}
\lfoot[\fancyplain{}{}]{\fancyplain{}{}}
\chead[\fancyplain{}{}]{\fancyplain{}{}}
\cfoot[\fancyplain{\rm\thepage}{\rm\thepage}]{\fancyplain{\rm\thepage}{\rm\thepage}}
\rhead[\fancyplain{}{}]{\fancyplain{}{\sl{Appendices}}}
\rfoot[\fancyplain{}{}]{\fancyplain{}{}}

\chapter{Classical and quantum ghosts}
\label{Classical and quantum ghosts}

In this appendix, we define what a ghost field is and discuss why the presence of a ghost in a (classical or quantum) theory is usually problematic. Let's consider for simplicity the following free Lagrangian density for a relativistic scalar field $\f$ in a Minkowski spacetime (indices are raised/lowered with the flat metric $\e^{\m\n}$/$\e_{\m\n}$)
\beq
\label{ghostLagrangianfreeapp}
\mscr{L} = - \frac{\ep}{2} \, \dem \f \, \de^{\m} \f - \frac{\vep}{2} \, m^2 \f^2
\eeq
where $\ep = \pm 1$ and $\vep = \pm 1$. The momentum conjugated to $\f$ is defined by
\beq
\pi_{\f} \equiv \frac{\de \mscr{L}}{\de \dot{\f}} = \ep \dot{\f} \quad ,
\eeq
and performing the Legendre transform with respect to $\dot{\f}$ (here an overdot indicates a time derivative) we obtain the Hamiltonian density
\beq
\label{ghostHamiltonianfreeapp}
\mscr{H} = \ep \, \Big( \frac{1}{2} \, \dot{\f}^2 + \frac{1}{2} \, \big( \vec{\nabla} \f \big)^{\! 2} \Big) + \frac{\vep}{2} \, m^2 \f^2
\eeq
in terms of which the Hamiltonian is defined as
\beq
H \equiv \int_{\mathbb{R}^3} \! d^3 x \,\, \mscr{H}[\f,\dot{\f}] \quad .
\eeq
As we already mentioned in the main text, in the $\ep =\vep = +1$ case the Hamiltonian is positive semi-definite and therefore bounded from below, while in the $\ep =\vep = -1$ case the Hamiltonian is negative semi-definite and therefore bounded from above. Finally, in the case $\ep = - \vep$, the Hamiltonian is indefinite and so it is not bounded either from below or from above. The field $\f$ is called a \emph{ghost field} if $\ep = -1$, while is called a \emph{tachyon field} if $\vep = -1$. Although these definitions have been given for a relativistic scalar field, it is straightforward to extend them to a more general case: a ghost field is defined as a field which has negative kinetic energy. If the Lagrangian density is not Lorentz-invariant, the part of the kinetic term which decides if the field is a ghost or not is the one which contains the time derivative of the field (the ``velocity'' of the field), or the conjugate momentum in the Hamiltonian formulation

\section{Ghosts at classical level}

A Hamiltonian which is unbounded from below is usually associated with instabilities of the system. However, if a ghost field $\f$ is free, the system is in fact stable since the energy is conserved by time evolution, independently of its sign. In fact, at classical level, an overall sign (or more in general a constant) in front of the complete Lagrangian density of the system has no influence at all, since it does not appear in the equations of motion. Therefore, at classical level, the theory described by the Lagrangian density (\ref{ghostLagrangianfreeapp}) corresponding to $\ep = \vep = +1$ is completely equivalent to theory described by the Lagrangian density corresponding to $\ep = \vep = -1$, and is defined in both cases by the equation of motion (the Klein-Gordon equation)
\beq
\big( \boxf - m^2 \big) \f = 0 \quad .
\eeq
If we consider the following Fourier decomposition
\beq
\f(\vecx, t) = \int_{\mathbb{R}^3} \! \frac{d^{3}p}{(2 \pi)^3} \,\, \ti{\f}_{\vecp} (t) \,\, e^{i \vecp \cdot \vecx} \quad ,
\eeq
we have that every mode is decoupled and obeys the equation
\beq
\ddot{\ti{\f}}_{\vecp} (t) = - (m^2 + \vec{p}^{\,\,2}) \, \ti{\f}_{\vecp} (t)
\eeq
which has only oscillatory solutions of frequency $\o(\vecp) = \sqrt{m^2 + \vec{p}^{\,\,2}}$. Since the plane waves of the Fourier expansion are orthonormal functions, a small perturbation\footnote{We define a perturbation $f(\vecx)$ to be small (respectively, big) if $\int d^3x \, f^2 \ll 1$ (respectively, $\gg 1$).} at $t = t_0$ from the configuration $\f = 0$ has small Fourier coefficients $\ti{\f}_{\vecp} (t_0)$, and the oscillatory behavior ensures that the perturbation remains small at all time. Therefore, the trivial configuration $\f(\vecx , t) = 0$ is stable both in the case $\ep = +1$ and in the case $\ep = -1$. Note instead that, if $\ep = -\vep$, the frequency $\o(\vecp) = \sqrt{\vec{p}^{\,\,2} - m^2}$ becomes imaginary for modes characterized by $\vec{p}^{\,\, 2} < m^2$ and so these mode can grow exponentially, signalling an instability.

However, the situation changes if a (classical) ghost field interacts with a (classical) non-ghost field. Consider in fact the following Lagrangian density for the relativistic scalar fields $\phi$ and $\psi$
\beq
\label{ghostLagrangianinteractingapp}
\mscr{L} = - \frac{\ep}{2} \, \dem \phi \, \de^{\m} \phi  - \frac{\ep}{2} \, m_{\f}^2 \f^2 - \frac{1}{2} \, \dem \psi \, \de^{\m} \psi - \frac{1}{2} \, m_{\psi}^2 \psi^2 - V_{int} \big(\phi, \psi \big)
\eeq
where we assume that the configuration $\f = \ps = 0$ is a local minimum of the potential. Performing the Legendre transformation with respect to $\dot{\f}$ and $\dot{\ps}$, we obtain the Hamiltonian density
\beq
\label{ghostHamiltonianinteractingapp}
\mscr{H} = \frac{\ep}{2} \, \dot{\f}^2 + \frac{\ep}{2} \, \big( \vec{\nabla} \f \big)^{\! 2} + \frac{\ep}{2} \, m_{\f}^2 \f^2 + \frac{1}{2} \, \dot{\psi}^2 + \frac{1}{2} \, \big( \vec{\nabla} \psi \big)^{\! 2} + \frac{1}{2} \, m_{\psi}^2 \psi^2 + V_{int} \big(\phi, \psi \big) \quad .
\eeq
Note first of all that, in the $V_{int} = 0$ case, the state $\f = \ps = 0$ is still stable independently of the sign of $\ep$, as can be established performing an analysis analogous to the one performed in the single field case. However, this does \emph{not} happen because the only states which have energies close to $H = 0$ are small perturbations of the $\f = \ps = 0$ configuration. In fact, while this is true in the $\ep = +1$ case, if $\ep = -1$ there exist an infinite number of different configurations for every value of the energy, and in particular there exist an infinite number of configurations with $H \simeq 0$ which are not small\footnote{We say that two configurations $f_1(\vecx)$ and $f_2(\vecx)$ are very close (respectively, very distant) if their difference $f_1 - f_2$ is small (respectively, big) in the sense of the previous footnote.} perturbations (the are ``highly excited'') of the $\f = \ps = 0$ one\footnote{For example, configurations where $\f$ and $\ps$ are plane waves with zero total 3-momentum and the same (arbitrarily high) amplitude has zero energy (where for simplicity we assumed $m_{\f} = m_{\ps}$, although this is not essential).}. The stability is instead due to the fact that energy is separately conserved for the two fields, and the system cannot reach the infinite region in parameter space where both sectors are indiscriminately excited at fixed total energy. If $V_{int} \neq 0$, it remains true that configuration $\f(\vecx, t) = \ps(\vecx, t) = 0$ is a solution of the equations of motion, so if we prepare the system in the state $\f(\vecx, t_0) = \ps(\vecx, t_0) = 0$ at an initial time $t_0$, the fields $\f$ and $\ps$ will remain in the ``zero'' configuration forever. However, if we perturb this state of a small amount of energy, in the case $\ep = -1$ the interaction may drive energy exchange between the two sectors and the system may indeed evolve towards a highly excited state, while this is not possible in the $\ep = +1$ case. The fact that in the $\ep = -1$ case the instability develops or not (and in case the velocity with which this happens) depends both on the details of the interaction potential and on the initial conditions of the system (roughly speaking, the initial perturbation has to have enough power in the modes which are prone to the instability).

\section{Ghosts at quantum level}

The presence of a ghost, already problematic at classical level, is even more so at quantum level. If we want to define the quantum theory of a field described by the Lagrangian density
\beq
\mscr{L}_{\f} = \frac{1}{2} \, \dem \f \, \de^{\m} \f + \frac{1}{2} \, m^2 \f^2 \quad ,
\eeq
we have two options: either the states which describe the quantum configuration of the field $\hat{\f}$ are assigned negative norm, or they are assigned positive norm (as usual in a quantum theory). The first choice implies that the energy spectrum is bounded from below, so the theory is stable, but the probabilistic interpretation of the theory is lost, and the theory is not predictive. To have a well-defined probabilistic interpretation, we have to choose the second option, which however implies that the energy spectrum is unbounded from below and so if the field interacts with other (non-ghost) quantum fields the theory is prone to instabilities. As we shall see, in this case the instability associated with the presence of the ghost field is much more severe at quantum level than it is at classical level.

Let's consider a relativistic ghost field $\f$ coupled to the Standard Model fields (collectively indicated with $\psi_{(j)}$) described schematically by the following Lagrangian density
\beq
\label{quantumghostSMapp}
\mscr{L} = \mscr{L}_{\f}[\f,\de \f \,] + \mscr{L}_{SM}[\psi_{(j)}, \de \psi_{(j)}] + \mscr{L}_{int}[\f,\psi_{(j)}]
\eeq
where $\mscr{L}_{int}$ is local and describes the interaction of the ghost with the SM fields (the ghost and the SM fields always couple at least gravitationally \cite{Cline:2003gs}, so there is always an effective interaction term, the interaction being graviton mediated, direct or both). We want to see that the system is intrinsically unstable, in the sense that even the vacuum state (\ie the quantum state devoid of particles) is unstable. If we consider a decay channel for the vacuum in which the final configuration, which we indicate with $\mcal{F}$, is made of $n$ particles, the decay rate takes the form \cite{PeskinSchroeder}
\beq
\label{decayratenbodyapp}
\G_{vac \rightarrow \mcal{F}} = \int \bigg( \prod_{i = 1}^{n} \frac{d^3 p_i}{(2 \pi)^3} \, \frac{1}{2 E_i} \bigg) \, \abs{\mcal{M}(vac \rightarrow \{ p_i \}_i)}^2 \, (2 \pi)^4 \, \d^{(4)} \bigg(\sum_{i = 1}^{n} p_i \bigg)
\eeq
where the $\{ p_i \}_i$ are the 4-momenta of the emitted particles and $p_i^0$ is the energy of the particle of mass $m_i$ and 3-momentum $\vecp_i$: the total decay rate is then the sum over all the possible decay channels
\beq
\G_{vac} = \sum_{\mathcal{F}} \, \G_{vac \rightarrow \mcal{F}} \quad .
\eeq
The integral
\beq
\label{relinvnbodyphasespace}
\int \bigg( \prod_{i = 1}^{n} \frac{d^3 p_i}{(2 \pi)^3} \, \frac{1}{2 E_i} \bigg) \, (2 \pi)^4 \, \d^{(4)} \bigg(\sum_{i = 1}^{n} p_i \bigg)
\eeq
is called the \emph{relativistically invariant n-body phase space}, while $\mcal{M}(vac \rightarrow \{ p_i \}_i)$ is called the \emph{relativistically invariant transition matrix element}: as we see in (\ref{decayratenbodyapp}), the decay rate depends both on the phase space available and on the modulation due to the dependence of the matrix element on the momenta. We want now to see that the phase space for the vacuum decay into two ghost $\f$ and two non-ghost particles $\ps$ is infinite, which implies that, if the transition matrix element do not decay steeply with the momenta\footnote{If $\mscr{L}_{int} = \frac{\la}{4} \f^2 \psi^2$, then $\abs{\mcal{M}}^2 \geq \la^2$ and does not decay to zero.}, the decay rate is infinite and the vacuum is subject to a catastrophic instability.

Let's focus on the decay channel where the final state $\mathcal{F}$ is a four-particle state made of a ghost-anti ghost couple and an ordinary particle-anti particle couple. For simplicity, we describe schematically this situation assuming that both the ghost ($\f$) and the ordinary particle ($\ps$) coincide with their anti-particle. Indicating with $\vecp_1$ and $\vecp_2$ the 3-momenta of the ghost particles, and with $\veck_1$ and $\veck_1$ the 3-momenta of the ordinary particles, the relativistically invariant 4-body phase space reads
\beq
\label{relinv4bodyphasespaceapp}
\mcal{I} = \int \frac{d^3 p_1}{(2 \pi)^3 2 E_1} \, \frac{d^3 p_2}{(2 \pi)^3 2 E_2} \, \frac{d^3 k_1}{(2 \pi)^3 2 \o_1} \, \frac{d^3 k_2}{(2 \pi)^3 2 \o_2} \, \frac{(2 \pi)^4}{2! \, 2!} \, \d^{(4)} \big( p_1 + p_2 + k_1 + k_2 \big)
\eeq
where $p_i^0 = - \sqrt{m_{\f}^2 + \vecp_i^{\, 2}}$, $\o_i = \sqrt{m_{\ps}^2 + \veck_i^{\, 2}}$ and the factors $2!$ take into account that the particles $\f$ as well as the particles $\ps$ are identical. The four dimensional delta function enforces the conservation of the total energy and momentum: it selects a volume $\mcal{V}$ in the 12-dimensional momentum space which contains the final momenta configurations which are compatible with energy-momentum conservation, whose measure is the integral $\mcal{I}$. Note that, if $\f$ were an ordinary particle as well, then in the massless case only the state $\vecp_i = \veck_i = (0,0,0)$ would be compatible with the conservation of energy (and in the massive case no states at all): this implies that $\mcal{V}$ would have zero measure, and the vacuum would be stable (since $\G$ would vanish). The presence of quantum ghosts destabilize the vacuum because there exist ``excited'' states at the same energy of the vacuum, and so $\mcal{V}$ has non-zero measure: differently from the classical case, at quantum level we don't even need an initial perturbation to be able to reach these states, the vacuum decays spontaneously.

To evaluate the integral $\mcal{I}$, we may integrate on $\veck_1$ and $\veck_2$ on the sections at fixed $\vecp_1$ and $\vecp_2$, and then integrate on the projection for $\vecp_1$ and $\vecp_2$. However, following \cite{Kaplan:2005rr,Jaccard:2013gla}, it is more convenient to embed $\mcal{V}$ into a 20-dimensional space, where the 8 extra dimensions are the components of the total ``ghost'' 4-momentum $P \equiv p_1 + p_2$ and the components of the total ``ordinary'' 4-momentum $K \equiv k_1 + k_2$, and calculate its area. We can in fact rewrite the total energy-momentum conservation as
\beq
\d^{(4)} ( p_1 + p_2 + k_1 + k_2 ) = \d^{(4)} ( P + K ) \,\, \d^{(4)} ( P - p_1 - p_2 ) \,\, \d^{(4)} ( K - k_1 - k_2 ) 
\eeq
and then integrate over $\vecp_1$, $\vecp_2$ at fixed $P$, and independently integrate over $\veck_1$, $\veck_2$ at fixed $K$. The integration over $\vecp_1$, $\vecp_2$ generates the two-body phase space $\Phi_{\f}^{(2)}(-P^2)$ (which is defined by the general formula (\ref{relinvnbodyphasespace}) in the particular case where $n = 2$) for two identical particles of mass $m_\f$ whose center of mass energy is $- P^2$, while the integration over $\veck_1$, $\veck_2$ generates the two-body phase space $\Phi_{\ps}^{(2)}(-K^2)$ for two identical particles of mass $m_\ps$ whose center of mass energy\footnote{Note that, with our choice of signature for the metric, $-P^2$ and $-K^2$ are non-negative numbers} is $- K^2$: we have then
\begin{align}
\mcal{I} &= \frac{1}{(2 \pi)^4} \int d^4 P \, d^4 K \, \d^{(4)} ( P + K ) \, \Phi_{\f}^{(2)}(-P^2) \, \Phi_{\ps}^{(2)}(-K^2) = \nn \\[2mm]
&= \frac{1}{(2 \pi)^4} \int d^4 P \, \Phi_{\f}^{(2)}(-P^2) \, \Phi_{\ps}^{(2)}(-P^2) \quad . \label{divergentapp}
\end{align}
The relativistically invariant two-body phase space for two identical particles of mass $m$ can be calculated explicitly, and reads \cite{Jaccard:2013gla}
\beq
\Phi^{(2)}(s) = \th(s - 4 m^2) \, \frac{1}{16 \pi} \, \sqrt{1 - \frac{4 m^2}{s}}
\eeq
where $\th(x)$ is the Heavyside theta function: it is easy to see that $\Phi^{(2)}(s)$ tends to a non-zero constant when $s \rightarrow + \infty$. Therefore, the integral $\mcal{I}$ is badly divergent, as can be deduced from (\ref{divergentapp}). To understand better where the divergence comes from, we can rewrite (\ref{divergentapp}) as an integral over the modulus $s$ of the sections $(P^0, \vecP)$ at $s = -P^2$ fixed. Adding a fifth dimension $s$ in the integration and inserting a delta function $\d(s + P^2)$, we get
\beq
\mcal{I} = \frac{1}{(2 \pi)^4} \int_{0}^{+\infty} ds \, \Phi_{\f}^{(2)}(s) \, \Phi_{\ps}^{(2)}(s) \, \int_{\mathbb{R}^4} d P^0 \, d \vecP \,\, \d \big(s - (P^0)^2 + \vecP^2 \big) \quad ,
\eeq
and using the property of the Dirac delta function
\beq
\d \big( f(x) \big) = \frac{1}{\abs{f^{\p}(x)}} \,\, \d(x)
\eeq
we obtain
\beq
\mcal{I} = \frac{1}{(2 \pi)^4} \int_{0}^{+\infty} \!\! ds \, \Phi_{\f}^{(2)}(s) \, \Phi_{\ps}^{(2)}(s) \, \int_{\mathbb{R}^3} d \vecP \,\, \frac{1}{2 \sqrt{s + {\vec{P}}^{2}}} \quad .
\eeq
We note that not only the integral in $s$ is divergent, since the three dimensional integral in $\vecP$ does not decay for $s \rightarrow + \infty$, but the three-dimensional integral in $\vecP$ itself is divergent, since written in spherical coordinates it becomes
\beq
\label{messieursandmadames}
\int_{\mathbb{R}^3} d \vecP \,\, \frac{1}{2 \sqrt{s + {\vec{P}}^{2}}} = 2 \pi \int_{0}^{+\infty} \!\! d\z \, \frac{\z^2}{\sqrt{s + \z^2}}
\eeq
which diverges indeed.

Therefore, under the assumption that the relativistically invariant transition matrix element $\mcal{M}$ for the decay $vac \rightarrow \f \, \f \, \ps \, \ps$ does not decay steeply as the modulus of the momenta go to infinity, the decay rate for this process is diverging. This implies that the total decay rate $\G$ of the vacuum is diverging as well: therefore, in a system described by the Lorentz-invariant Lagrangian density (\ref{quantumghostSMapp}), where $\mscr{L}_{int}$ is local, the vacuum is catastrophically unstable.

\section{Ghosts in effective theories}

It is worthwhile to spend few words on the physical meaning of the decay rate of the vacuum being divergent. The decay rate is a probability density (in space and time) of decay: in other words, if $V$ is a 3-dimensional volume and $T$ is a time interval, the quantity
\beq
N = V \, T \, \G_{vac \rightarrow \f \f \ps \ps}
\eeq
gives the average number of quadruples $\f \f \ps \ps$ emitted in the volume $V$ in the time interval $T$ by the decay of the vacuum (regardless of the momenta of the emitted particles). If in the integral (\ref{decayratenbodyapp}) we integrate over a specific interval of energy for the emitted particles, the result multiplied by $V$ and $T$ gives the average number of quadruplets $\f \f \ps \ps$ emitted in the volume $V$ in the time interval $T$ which have energies in the selected interval. Therefore, if $\G_{vac \rightarrow \f \f \ps \ps}$ is divergent, we expect to detect an infinite number of particles emitted by the vacuum decay in every volume and time interval, no matter how small: furthermore, despite the total emitted energy is zero by energy conservation, the total energy emitted by the vacuum decay in each separate sector (ordinary and ghost) is infinite, independently of the volume and time interval we consider.

It is important to point out that, of course, the quantum state which describes our universe is not the vacuum, since the universe is not empty: this observation may cast doubts on the real significance of the vacuum instability. However, the instability due to spontaneous creation of couples of ghosts and ordinary particles with zero total energy production is not peculiar to the vacuum, but is common to every quantum state. Furthermore, the density of the outer space is very low and we can to a first approximation consider it empty: therefore, it seems reasonable to estimate the fluxes of particles coming from space (produced by the decay of a realistic quantum state) with the fluxes that the decay of the vacuum would produce. This estimate is clearly in contrast with the observations, because in reality we don't observe streams of particles of infinite flux coming from space: the observed properties of the cosmological backgrounds of each kind of detected particle therefore can be used to put an upper limit on the decay rate of the vacuum for the decay channels in which the same type of particle is produced. In particular, the stringent constraint on the vacuum decay rate comes from the diffuse gamma ray background \cite{Cline:2003gs}, which put bounds on the decay rate for the channel $vac \rightarrow \f \f \g \g$.

It seems then that observations rule out the possibility of considering systems described by a Lagrangian density if the form (\ref{quantumghostSMapp}). However, it is usually assumed that such a Lagrangian density is \emph{not} an exact description of reality, \emph{i.e.} valid at arbitrarily high energies and arbitrarily small distances, but is rather believed to be an effective description which is trustable only in a definite range of energy/length scales. In particular, it is customary to assume that we can trust it only when the momenta of the emitted particles stay below an appropriate cut-off: this restriction limits the volume of phase space available for the decay, and we may hope that this limitation renders the phase space integral (and the decay rate) finite. However, if the cut-off on the momenta is imposed in a Lorentz-invariant way, the phase space integral remains divergent. In fact, the only Lorentz-invariant way to put a cut-off on the total ghost momentum $P$ is to restrict the domain of $s = P^2$: however, such a constraint does not put an upper bound to the allowed value of $\vecP^2$, and does not influence the integral on $\vecP$ at $s$ fixed (\ref{messieursandmadames}). Therefore, to render the decay rate finite we need to cut-off the momenta in a Lorentz non-invariant way: both $s$ and $\vecP^2$ has to be cut-off.

\subsection*{Momentum cut-off, Lorentz violation and non-locality}

The presence of a Lorentz non-invariant momentum cut-off can in principle be ascribed to two reasons: it may be that above the cut-off the correct description of nature is Lorentz-violating, or it may be that the Lorentz symmetry is spontaneously broken, for example by the fact that the system is created in the vacuum state at a finite initial time $t_i$ in a specific reference system. Note that the Lorentz symmetry has to be spontaneously broken somehow, as a consequence of the fact that, as we have shown above, the vacuum of a system containing a ghost can be at best metastable. The quantum state of the system described by (\ref{quantumghostSMapp}) cannot have been the vacuum state for an infinite amount of time in the past: otherwise, independently of the fact that the decay rate is finite or infinite, it would have already decayed anyway, producing a infinite amount of radiation. If we assume that the Big Bang (or more precisely, the time when the energy of the universe becomes smaller than the energy cut-off) marks the beginning of the validity of the effective description (\ref{quantumghostSMapp}), the Lorentz invariance is spontaneously broken by the fact that the system can be created in the vacuum only for $t \gtrsim 0$ in the comoving reference: this introduces an effective cut-off on the $\vecP$ integration (\ref{messieursandmadames}) \cite{Cline:2003gs,Garriga:2012pk}, and renders the decay rate finite if there is at the same time a Lorentz-preserving cut-off on $s$.

However, it has been shown that the Lorentz non-invariant cut-off on the 3-momenta induced by the finite age of the universe, together with a Lorentz-preserving cut-off on $s$, is not enough to make the theory consistent with observations \cite{Cline:2003gs}: in fact, current bounds on the flux of the diffuse gamma rays background put the following constraint on the value of the Lorentz preserving cut-off $\La$
\beq
\La \lesssim 10^{-3} \, \textrm{eV}
\eeq
which would imply modification to the Newton's law of gravitational attraction at distances $\gtrsim 0.2$ mm, in contradiction with experiments \cite{Kapner:2006si}.

Interestingly, there is a claim \cite{Garriga:2012pk} that if the theory becomes \emph{non-local} above the energy cut-off, the decay rate of the vacuum can be suppressed. As an example, consider a ghost field $\f$ and a standard field $\ps$ whose interaction is described by the action
\begin{multline}
\label{non local interaction}
S_{I} = \frac{\la}{4} \int d^{4} x \, d^{4} z \, d^{4} y_{1} \, d^{4} y_{2} \,\, \f \big(x +z +y_1 \big) \, \f \big(x +z - y_1 \big) \\
g\big(z, y_1, y_2 \big) \, \ps \big(x -z +y_2 \big) \, \ps \big(x -z - y_2 \big) \quad ,
\end{multline}
which is the non-local generalization of a quartic $\la \, \f^2 \ps^2$ coupling. Here $g(z, y_1, y_2)$ is the non-local form factor, while $y_1$ (respectively, $y_2$) is the coordinate distance between the points at which the two ghosts (respectively, standard) fields interact, and $z$ is the coordinate distance between the interaction points of the ghost couple and the standard couple. The non-local properties of the interaction, and in particular the fact that the interaction is non-local above or below a cut-off, is linked to the properties of the Fourier transform $G(q^{\m}, q_1^{\m}, q_2^{\m})$ of the form factor: in particular, the interaction is local when $G$ is constant, and so $g$ is 12-dimensional Dirac delta. If the theory is Lorentz-invariant, then $G$ can only depend on the square moduli of $q^{\m}$, $q_1^{\m}$ and $q_2^{\m}$ and on the scalar products $\pi_1 \equiv q_\m \, q_1^{\m}$, $\pi_2 \equiv q_\m \, q_2^{\m}$ and $\pi_{12} \equiv q_{1 \m} \, q_2^{\m}$; a Lorentz-invariant cut-off on the theory constrains the values of these 6 Lorentz-invariant quantities, and restricts the domain of integration in momentum space involved in the calculation of the decay amplitude.

The authors of \cite{Garriga:2012pk} claim that, if the theory (\ref{non local interaction}) is Lorentz-invariant and non-local above a cut-off $\La$, the Lorentz-violating cut-off due to the finite age of the universe can be sufficient to produce a decay rate consistent with the observations, since the Lorentz-preserving cut-off can be slightly higher than the previous bound
\beq
\La \lesssim (1.8 - 5.6) \times 10^{-3} \, \textrm{eV} \quad ,
\eeq
which is marginally consistent with the experimental data on small distances modifications of GR \cite{Garriga:2012pk}. Note that non-locality does not necessarily imply a lack of causality in the theory \cite{Jaccard:2013gla}, and that even though a generic non-local theory may violate Lorentz-invariance, there exists a class of non-local theories which does not violate it \cite{Garriga:2012pk}.

If we nevertheless demand that the theory is local also above the cut-off, then compatibility with observational constraints implies that the (fundamental) description of nature has to be Lorentz-violating, and has to be such that, above the cut-off, the ghosts and the ordinary fields are completely decoupled. In particular, there cannot be direct couplings and even gravity cannot mediate interactions between the two sectors: the Lorentz invariance in this case cannot be a fundamental property but just an effective symmetry which emerges at low energy. More precisely, it can be shown \cite{Cline:2003gs,Kaplan:2005rr} that the new Lorentz-violating gravitational physics which serves to decouple ghost and ordinary matter has to become relevant at momenta $\m$ which are smaller than the weak scale energies which we probe in colliders, in particular the observations of the diffuse gamma ray background imply\footnote{1 MeV $= 10^3$ eV $\sim 1.602 \times 10^{-13}$ J}
\beq
\m \lesssim 1 \, \textrm{MeV} \quad .
\eeq

Therefore, we conclude that we could accommodate ghosts (without violating observational constraints) in a (local and Lorentz-invariant) effective description of an underlying theory, provided the latter theory is either non-local or Lorentz violating above a momentum cut-off scale $\m$. In both cases, observational constraints imply that this scale has to be lower than the energy scales we probe in particle colliders: although the possibility that General Relativity breaks down at these energies is unorthodox, strictly speaking it is not forbidden by experiments since the new gravitational physics not necessarily has to produce relevant signatures in the colliders experimental set-ups. The requirements of locality and Lorentz-invariance are at the core of our current conceptual understanding of nature: we can say that, to accommodate low energy effective ghosts, we have to accept a very unorthodox situation, in which violations of these basic assumptions happen (in a relative sense) quite close to our experimental reach.

Note that so far we have assumed that the fundamental theory possesses two sectors of fields, a ghost sector and an ordinary one, precisely as the effective theory. There is in principle another possibility, namely the fact that the very existence of the ghost sector is just an effective property and that the fundamental theory is ghost free. In this case, the contribution of the high-energy modes is automatically zero, because above the cut-off the ghost sector does not exist. However, for the same reasons outlined above, also in this case the fundamental theory has to be Lorentz-violating above the cut-off, and the momentum cut-off has to be lower than the energy scales we probe in particle colliders.

\chapter{Conical geometry}
\label{Conical geometry}

In this appendix we discuss the geometry of a right circular two-dimensional conical surface, which for simplicity we call simply a cone. To do that, we first derive its metric by embedding the 2-cone in the three-dimensional Euclidean space $\mathbb{E}^3$, and after that we define the 2-cone as an intrinsic object. We then discuss the properties of the vertex of the cone, and we conclude considering regularized versions of the cone.

\section{The cone as an extrinsic object}
\label{The cone as an extrinsic object}

We define an embedded two-dimensional cone as the set of points $(x,y,z) \in \mathbb{E}^3$ which satisfy
\begin{align}
\label{embeddedcone}
\sqrt{x^2 + y^2} &= z \, \tan \theta & z &\geq 0 \quad ,
\end{align}
where the angle $\theta \in ( 0, \pi /2)$ is called the \emph{opening} of the cone and the point $(x,y,z) = (0,0,0)$ is called the \emph{vertex} of the cone. Note that the limiting case $\theta = 0$ corresponds to a degenerate cone, \emph{i.e.} a half-line, while the limiting case $\theta = \pi /2$ corresponds to the configuration where the cone is actually a plane. To obtain the metric on the cone, we need to equip it with a coordinate system and find the explicit form for the embedding function: the metric structure of $\mathbb{E}^3$ induces a metric structure on the cone via its embedding. Note that, since the embedded cone has a cusp at the vertex, we cannot obtain the induced metric there by pulling back the 3D Euclidean metric, since at the vertex the embedding is not derivable and so the definition (\ref{Induced metric general}) of the induced metric cannot be used.

Because of the circular symmetry of the system, it is convenient to derive the induced metric using coordinates which respect this symmetry. Using the cylindrical coordinates $(r, \vartheta, z)$ in the 3D Euclidean space and using $r$ and $\vartheta$ to parametrize the cone (apart from a half-line which has the vertex as the initial point), the embedding function reads

\beq
\vfd (r, \vartheta) = \Big( r, \vartheta, \frac{r}{\tan \theta} \Big)
\eeq
and using the general formula (\ref{Induced metric general}) the metric induced on the cone by the Euclidean metric is

\begin{align}
\label{conicalmetricr}
\ti{\g}_{rr}(r, \vartheta) &= \frac{1}{\sin^2 \theta} & \ti{\g}_{r\vartheta}(r, \vartheta) &= 0 & \ti{\g}_{\vartheta\vartheta}(r, \vartheta) &= r^2 \quad .
\end{align}
Redefining the radial coordinate according to

\beq
r \rightarrow \r(r) = \frac{r}{\sin \theta}
\eeq
we arrive at the line element

\beq
\label{Susanna}
d s^2 = d \r^2 + \b^2 \, \r^2 \, d \vartheta^2
\eeq
where

\beq
\label{Zoppas}
\b = \sin \theta \quad .
\eeq

\noi Note that $0 < \b < 1$ and that $\b \rightarrow 1$ when $\theta \rightarrow \pi/2$, which is expected since in this limit the cone tends to a bidimensional flat space. As we already mentioned, this reference system does not cover the whole cone, but we can cover the whole cone (apart the vertex) using two coordinate charts, for example one where the angular coordinate $\vartheta$ is defined on $(0, 2 \pi)$ and one where $\vartheta$ is defined on $(-\pi, \pi)$.

\section{The cone as an intrinsic object}

Using the results obtained above, we can define the notion of a cone as an intrinsic object. We define the cone as a two dimensional Riemannian manifold which is covered by two coordinate systems $(\r, \vartheta)$, $(\r', \vartheta')$ and in which the line element takes the form
\beq
\label{cono}
ds^2 = d\r^2 + \b^2 \, \r^2 \, d\vartheta^2
\eeq
with $\b \in (0, 1)$, while an analogue expression holds in the ``primed'' reference system. In both the reference systems the radial coordinate ($\r$ or $\r'$) is defined on $ (0,+ \infty)$, while the angular coordinate $\vartheta$ is defined on $(0, 2 \pi )$ and the angular coordinate $\vartheta'$ is defined on $(- \pi, \pi )$: the transition function between the two reference systems is given by $\r' = \r$, $\vartheta' = \vartheta - \pi$. The behavior of the geometry in the proximity of $\r = 0$, and therefore the properties of the vertex from the intrinsic point of view, deserve a separate discussion which is presented in the next subsection. 

Note that the meaning of the parameter $\b$, which is linked to the opening of the cone in the extrinsic description, is not clear in the intrinsic description. However, we can change coordinates $(\r, \vartheta) \rightarrow (\r, \Theta)$ redefining the angular coordinate according to
\beq
\vartheta \rightarrow \Theta(\vartheta) = \b \, \vartheta \quad ,
\eeq
so that the line element in the new coordinate reads
\beq
\label{conodeficit}
ds^2 = d\r^2 + \r^2 \, d\Theta^2
\eeq
and we have $\r \in (0, + \infty)$ and $\Theta \in (0, 2 \pi \b)$. It is apparent that, removing a half-line from the cone, we obtain a manifold which is isomorphic to the flat two dimensional space $\mathbb{E}^2$ with a slice of angular opening $2 \pi (1 - \b)$ removed. Therefore, we can think of the (intrinsic) cone as a two dimensional flat space with a slice removed and the two boundaries identified. The angle $\a \equiv 2 \pi (1 - \b)$ is called the \emph{deficit angle} of the cone.

Using this result we can reconstruct the extrinsic description of the cone from the intrinsic one: first of all, we can embed the 2D flat space with a slice removed in the 3D Euclidean space, keeping it flat in a extrinsic sense. Then we can bend it without deforming it (\emph{i.e.} without changing the intrinsic distance between its points) and join the boundaries of the removed slice: ``gluing'' this two boundaries, we obtain the full cone embedded in $\mathbb{E}^3$. The opening of the embedding obtained this way is determined by $\b$ again via (\ref{Zoppas}).

\subsection{The vertex of the cone}

Despite the fact that the extrinsic geometry is singular at the vertex, the Riemann tensor built with the metric \eqref{cono} remains well behaved as we approach the vertex (it actually vanishes identically), and the metric components of (\ref{cono}) themselves do not diverge at the origin. Furthermore, we can reach the vertex starting from any point of the cone and following a (radial) geodesic, within a finite range of values of the affine parameter. This may suggest that, from the intrinsic point of view, the vertex of the cone is not a singular point. However, the fact that the vertex is a special point remains imprinted in the affine properties of the geometry defined by (\ref{cono}). In fact, the parallel transport of a vector along a closed loop which surrounds the vertex (more precisely, a non-contractible loop) rotates the vector by an amount which does not depend on the shape of the loop, while parallel transport along a closed loop which does not surround the vertex (a contractible loop) leaves the vector invariant.

To see this explicitly, consider for simplicity a circular loop $\mscr{L}_R$ of coordinate radius $R$, and choose the parametrization
\beq
\s : \t \rightarrow (R,\t) \qquad , \qquad \t \in (0, 2 \pi) \quad .
\eeq
The differential equation\footnote{We indicate derivatives with respect to the parameter $\t$ with an overdot.} for the parallel transport of a vector $\vec{v}$
\beq
\dot{v}^i + \G^{i}_{jk} \Big\rvert_{\s(\t)} \dot{\s}^j \, v^k = 0
\eeq
reads in our case
\beq
\dot{v}^i = \mathcal{M}^{i}_{\,\,j} \, v^j \quad ,
\eeq
where
\begin{displaymath}
\mathcal{M} = \begin{pmatrix}
               0 & \bq R \\
               - 1/R  & 0 \quad .
              \end{pmatrix}
\end{displaymath}
Supposing that the radial component of the vector does not vanish when parallel transported, it is useful to define the ratio between the angular and radial component of the vector
\beq
y(\t) \equiv \frac{v^{\vartheta}(\t)}{v^{\r}(\t)}
\eeq
in terms of which the equation for parallel transport becomes
\beq
\dot{y} = - \frac{1 + \bq R^2 y^2}{R}
\eeq
which can be integrated to give
\beq
\label{solymar}
\arctan y(0) - \arctan y (2 \pi) = 2 \pi \b \quad .
\eeq
The angle $\o$ between a vector $\vec{v}_0$ and the vector $\vec{v}_{2 \pi}$ obtained by parallel transporting $\vec{v}_0$ along the loop $\mscr{L}_R$ satisfies
\beq
\label{marysol}
\cos \o = \frac{1 + \bq R^2 \, y(2 \pi) \, y(0)}{\sqrt{1 + \bq R^2 \, y^2(2 \pi)} \, \sqrt{1 + \bq R^2 \, y^2(0)}}
\eeq
and inserting \eqref{solymar} into \eqref{marysol} we obtain
\beq
\abs{\o} = 2 \pi (1 - \b) = \a \quad .
\eeq
We can conclude that the parallel transport of a vector along the loop $\mscr{L}_R$ rotates the vector by an angle which is independent of the radius $R$ of the loop and is equal to the deficit angle of the cone.

On the contrary, if we consider a loop $\mscr{L}$ which does not surround the vertex, the parallel transport along $\mscr{L}$ can be decomposed into parallel transports along loops which are contained in the coordinate chart where the metric (\ref{conodeficit}) is defined. Since the metric (\ref{conodeficit}) is flat, the parallel transport along these loops leaves any vector invariant. This result in turn implies that, if $\mscr{L}_{\star}$ is a loop which surrounds the vertex and can be obtained as a continuous deformation of a circular loop $\mscr{L}_R$, then the parallel transport of any vector along $\mscr{L}_{\star}$ rotates the vector by the same angle $\a$ as the parallel transport along $\mscr{L}_{R}$ does.

These results imply that the holonomy of a loop changes discontinuously when the loop crosses the vertex: for the sake of precision, it changes from a rotation of angle $\a$ to the identity or the other way around. This is consistent with the naive idea that the vertex possesses a (localized) infinite curvature.

\section{The regularized cone}

We want now to consider a regularized version of the cone, in which the tip is smoothed out to give a 2D surface which is regular everywhere. We then define an embedded regularized cone as the set of points $(x,y,z) \in \mathbb{E}^3$ which satisfy
\beq
z = \vf(x,y)
\eeq
where
\begin{equation}
 \vf(x,y) \quad \left\{
  \begin{aligned}
   \phantom{i} &= \frac{\sqrt{x^2 + y^2}}{\tan \theta} \qquad & \qquad \textrm{for} \, \sqrt{x^2 + y^2} &\geq R \\[2mm]
   \phantom{i} &= \f(x,y) \qquad & \qquad \textrm{for} \, \sqrt{x^2 + y^2} &< R \quad ,
  \end{aligned}
\right.
\end{equation}
and the \emph{regularization function} $\f(x,y)$ is of class $\mscr{C}^{\infty}$ for $r = \sqrt{x^2 + y^2} < R$ and such that the function $\vf(x,y)$ is (at least) of class $\mscr{C}^{1}$ in $r = R$. It is easy to see that, outside the (coordinate) \emph{regularization radius} $R$, the regularized cone coincides with the ``sharp'' cone (\ref{embeddedcone}) and in particular its opening is still the angle $\theta \in ( 0, \pi /2)$.

We may assume that the regulating functions is circularly symmetric, and define the function $f(r)$ such that
\beq
f \big( \sqrt{x^2 + y^2} \big) = \vf (x,y)
\eeq
which in particular implies
\beq
f(r) = \frac{r}{\tan \th} \qquad \textrm{for} \qquad r \geq R \quad .
\eeq
Note that we have the following relationship between $f$ and $\vf$
\beq
f(r) = \vf \Big\rvert_{(x,y) = (r,0)}
\eeq
which implies that $f$ is of class $\mscr{C}^\infty$ for $0 < r < R$ and it is of class $\mscr{C}^1$ in $r = R$; furthermore we have
\beq
\lim_{r \rightarrow 0^+} f^{(n)} (r) = \de_{x}^n \vf \Big\rvert_{(x,y) = (0,0)}
\eeq
where $f^{(n)}$ is the $n$-th derivative of $f$. Using the same procedure as in section \ref{The cone as an extrinsic object}, we can parametrize the regularized cone with the radial and angular coordinates $(r, \vth)$ of the 3D cylindrical coordinates and calculate the induced metric. We obtain that the induced metric is
\begin{align}
\ti{\g}_{rr}(r, \vth) &= 1 + \dot{f}^{\, 2}(r) & \ti{\g}_{r\vartheta}(r, \vth) &= 0 & \ti{\g}_{\vartheta\vartheta}(r, \vth) &= r^2
\end{align}
where we indicated derivatives with respect to $r$ with an overdot $\dot{\phantom{a}}$. Also in this case we redefine the radial coordinate $r \rightarrow \r$ to have the radial-radial component of the metric equal to one, which amounts to require
\beq
\label{Michelaicogl}
\frac{d \r(r)}{dr} = \sqrt{1 + \dot{f}^{\, 2}(r)} \quad ,
\eeq
and so the induced metric in the $(\r , \vth)$ reference system is characterized by the line element

\beq
\label{regularizedcone}
ds^2 = d\r^2 + b^2(\r) \, d\vartheta^2
\eeq
where $b(\r) \equiv r(\r)$. Note that (\ref{Michelaicogl}) implies that $\r(r)$ is a monotonically increasing function, which is of class $\mscr{C}^1$ in $r = R$ and of class $\mscr{C}^\infty$ in $0 < r < R$ and $r > R$: in particular this implies that $\r(r)$ is invertible, so $b(\r)$ is well defined, and it is natural to choose the integration constant so that $\r(0) = 0$ (since $\r$ is a radial coordinate) which implies
\beq
\r_R \equiv \r(R) = \int_{0}^{R} \sqrt{1 + \dot{f}^{\, 2}(\t)} \,\, d\t \quad .
\eeq
Furthermore, (\ref{Michelaicogl}) implies that $\lim_{r \rightarrow 0^+} \dot{\r}(r) = 1$, and so we have that
\beq
\lim_{\r \rightarrow 0^+} b^{\p}(\r) = 1
\eeq
where we indicated derivatives with respect to $\r$ with a prime $\phantom{i}^{\p}$. Defining $\r_0 \equiv \r_R - R/\b$, where $\b \in (0,1)$ is defined as in (\ref{Zoppas}), we then have that for $r \geq R$

\beq
\r(r) = \frac{r}{\b} + \r_0
\eeq
and finally

\beq
\label{vag31}
b(\r) = \b (\r - \r_0) \qquad \textrm{for} \qquad \r \geq \r_R \quad .
\eeq
We then reproduce outside the regularization radius the metric (\ref{cono}) of the ``sharp'' cone, apart from an additive constant.

\subsection{The geometry of the regularized cone}

We now want to understand the geometrical meaning of the metric
\beq
\label{cononew}
ds^2 = d\r^2 + b^2(\r) \, d\vartheta^2
\eeq
where $b(\r)$ is such that: it has the form (\ref{vag31}) for $\r \geq \r_R$, is of class $\mscr{C}^1$ in $\r = \r_R$ and of class $\mscr{C}^\infty$ in $0 < r < \r_R$; furthermore, its derivatives have a finite limit for $\r \rightarrow 0^+$ and in particular $b^{\p}(0^+) = 1$. It can be shown that the only non-zero connection coefficients are
\begin{align}
\G^{\r}_{\vartheta \vartheta} &= - b \, b^{\p} & \G^{\vartheta}_{\r \vartheta} &= \frac{b^{\p}}{b}
\end{align}
while the Ricci tensor reads

\begin{align}
R_{\r\r} &= - \frac{b^{\p\p}}{b} & R_{\r \vartheta} &= 0 & R_{\vartheta \vartheta} &= - b \, b^{\p\p}
\end{align}
and the Ricci scalar reads

\beq
R = - 2 \, \frac{b^{\p\p}}{b}
\eeq
It can be checked explicitly that the Einstein tensor vanishes identically, which is true for every 2D Riemannian space.

Therefore, the metric (\ref{cononew}) with the properties outlined above describes a two-dimensional Riemannian space which is flat for $\r > \r_R$, while the curvature is concentrated inside the (physical) regularization radius $\r_R$ . Note that the total curvature concentrated inside the ball $B_{R}$ of radius $\r_R$ and center $\r = 0$
\beq
\mcal{R} \equiv \int_{B_{R}} \!\! d\r \, d\vth \,\, \sqrt{\g} \,\, R(\r) = 4 \pi \, \Big( b^{\p}(0^+) - b^{\p}(\r_R) \Big) = 2 \, \a
\eeq
is independent on the details of the regularization: it is rigidly fixed by the opening of the ``outer'' part of the cone (which is expressed by $\b$), and is in fact twice the deficit angle. This result justifies the previous assertion that we may see the vertex of the sharp cone as a point where an infinite amount of curvature is localized. In fact, we may see a sharp cone with a given opening as the limit of a sequence of regularized cones $\mathscr{R}_n$ with the same fixed opening and regularization radiuses $R_n$ which tends to zero as $n \rightarrow +\infty$. The result obtained above implies that the total curvature contained inside the regularization radius is independent of $n$, while the area of the part of the cone where the curvature is present tends to zero in the limit. Therefore, the density of curvature inside the smoothed tip of the cone diverges as $n \rightarrow +\infty$: in the limit, we may think that the total curvature $\mcal{R} = 2 \, \a$ is localized on the vertex, which in this sense is characterized by an infinite Ricci curvature.

\chapter{Conical space in brane-based coordinates}
\label{Conical space in brane-based coordinates}

In this appendix we summarize the physical properties of the configurations (\ref{backgroundthemline}) introduced by \cite{deRham:2010rw}, and show that they are actually equivalent to the bulk-based configurations that we derived in section \ref{Pure tension solutions}.

\section{The geometry of the brane-based solution}

Let's consider a 6D spacetime covered by a coordinate chart $(z,y,\xd)$, where $z$ is defined on $(0, +\infty)$ and $(y,\xd)$ are defined on $\mathbb{R}^5$, whose geometry is defined by the metric (\ref{backgroundthemline})
\beq
\label{backgroundthemlineapp}
ds^2 = \big(1 + \b^2 \big) \, dz^2 + 2 \b \, \ep(y) \, dz dy + dy^2 + \e_{\m \n} dx^{\m} dx^{\n}
\eeq
where $\b$ is a real parameter and $\ep(y)$ is a smooth function which is a regularized version of the step function\footnote{Here $\theta$ is the Heavyside theta function.} $\s (y) \equiv 2 \, \theta (y) - 1$. More precisely, $\ep(y)$ is monotonically increasing and odd with respect to the reflection $y \rightarrow -y$, satisfies the condition
\beq
\label{HeyJo}
\lim_{y \rightarrow \pm \infty} \ep(y) = \pm 1
\eeq
and its first derivative is peaked around $y = 0$. Let's suppose that a (thin) cod-1 brane is placed at $z = 0$, and let's choose to parametrize it with the bulk coordinates $(y,\xd)$, and that a (thin) cod-2 brane is placed at $z = y = 0$, and let's choose to parametrize it with the bulk coordinates $(\xd)$. Note that the reference system is not Gaussian Normal to the cod-1 brane, since the metric (\ref{backgroundthemlineapp}) has non-zero $zy$ components. It is not difficult to see that the 6D Riemann tensor built from the metric (\ref{backgroundthemlineapp}) vanishes identically (independently of the form of $\ep$), and that the induced metrics on the cod-1 and cod-2 branes are respectively the 5D and the 4D Minkowski metrics (again independently of the form of $\ep$), so the intrinsic geometries of the bulk and of the branes are flat. However, the extrinsic geometry of the cod-1 brane is non-trivial, since we have
\begin{align}
\bar{K}_{\m \n} &= 0 & \bar{K}_{\m y} &= 0 & \bar{K}_{y y} &= - \, \beta \, \dfrac{\epsilon^{\prime}(y)}{\sqrt{ 1+ \beta^2 \big( 1 - \epsilon(y)^2 \big)}} 
\end{align}
and so we conclude that $\ep$ influences just the extrinsic geometry of the cod-1 brane (in this appendix a prime $\phantom{i}^{\p}$ indicates a derivative with respect to $y$). Suppose now that the cod-1 brane contains a pure tension source localized around the cod-2 brane, so that the energy-momentum on the cod-1 brane is of the form
\beq
\bar{T}_{ab}^{(loc)} = - \bla \, f(y) \,\, \d_{a}^{\,\,\,\m} \, \d_{b}^{\,\,\,\n} \,\, \bar{g}^{(4)}_{\m\n}
\eeq
where $f(y)$ is a positive, even and normalized function (so it is a regularized version of the Dirac delta function) which describes the details of the distribution of the tension inside a thick cod-2 brane whose boundaries are $y = \pm l_2$. The only non-trivial component of the junction conditions reads
\beq
\label{Kelly Nock}
\frac{\beta \, \epsilon^{\prime}(y)}{\sqrt{ 1+ \beta^2 \big( 1- \epsilon^2(y) \big)}} = \frac{\bla}{2 M_6^4} \, f(y)
\eeq
and is to be supplemented with the initial condition $\ep(0) = 0$ which has to be satisfied by symmetry reasons: the equation above in particular implies that $\ep$ is constant for $\abs{y} > l_2$, and the condition (\ref{HeyJo}) then implies that
\beq
\ep(y) = \pm 1 \qquad \qquad \textrm{for} \qquad \qquad y \gtrless l_2 \quad .
\eeq
The equation (\ref{Kelly Nock}) can be integrated exactly, and integrating it over the interval $(-l_2 , l_2)$ we obtain
\beq
\label{Julie}
\arctan \b = \frac{\bla}{4 M_6^4} \quad :
\eeq
this implies that for $\abs{y} > l_2$ the metric (\ref{backgroundthemlineapp}) is rigidly fixed by the total amount of tension $\bla$ present inside the thick cod-2 brane, while for $\abs{y} < l_2$ the shape of $\ep(y)$ explicitly depends on the details of how the tension is distributed inside the thick cod-2 brane (expressed by $f$). This implies in particular that the thin limit of these configurations exists: if we consider the limit in which $l_2 \rightarrow 0^+$ while $\bla$ remains constant (so $f$ tends to a Dirac delta), the result obtained above implies that $\b$ remains constant and $\ep$ tends to the step function.

\section{Equivalence with the bulk-based description}

The geometry of the bulk-branes system corresponding to the metric (\ref{backgroundthemlineapp}) is however not evident. The fact that the Riemann tensor is identically vanishing in the bulk implies that (\ref{backgroundthemlineapp}) is equivalent to a portion of a 6D Minkowski space written in a non-trivial coordinate system: to have a transparent idea of the geometry of the configuration (\ref{backgroundthemlineapp}), we can try to find a coordinate transformation which maps it into the 6D Minkowski space: the geometrical meaning of the configuration will then be encoded in the embedding of the cod-1 brane, which after the coordinate change will be non-trivial. Note that in the following part of this appendix we don't use the conventions which we use in the rest of the thesis, and in particular a tilde does not necessarily identify cod-1 quantities.

\subsection{The change of coordinates}

Let's start from the configuration (\ref{backgroundthemlineapp})
\begin{equation*}
\overline{g}^{(6)}_{zz} = 1 + \b^2 \quad \qquad \overline{g}^{(6)}_{zy} = \b \, \ep (y) \quad \qquad \overline{g}^{(6)}_{yy} = 1 \quad \qquad \overline{g}^{(6)}_{z \m} = \overline{g}^{(6)}_{y \m} = 0 \quad \qquad \overline{g}^{(6)}_{\m\n} = \e_{\m\n} 
\end{equation*}
and consider the following coordinate transformation 
\begin{equation*}
(\star) \quad \left\{
\begin{aligned}
  z(\tilde{z}, \tilde{y}, \tilde{x}^{\cdot}) &= (1+\b^{2})^{-1/2} \,\, \tilde{z}\\
  y(\tilde{z}, \tilde{y}, \tilde{x}^{\cdot}) &= \tilde{y}\\
  x^{\mu}(\tilde{z}, \tilde{y}, \tilde{x}^{\cdot}) &= \tilde{x}^{\mu}
\end{aligned}
\right.
\end{equation*}
which brings the metric into the form
\beqnn
\tilde{g}^{(6)}_{zz} = 1 \quad \qquad \tilde{g}^{(6)}_{zy} = \frac{\b \, \ep (\tilde{y})}{\sqrt{1+ \b^2}} \quad \qquad \tilde{g}^{(6)}_{yy} = 1 \quad \qquad \tilde{g}^{(6)}_{z \m} = \tilde{g}^{(6)}_{y \m} = 0 \quad \qquad \tilde{g}^{(6)}_{\m\n} = \e_{\m\n} \quad .
\eeqnn
Consider next the following coordinate transformation
\begin{equation*}
(\star \star) \quad \left\{
\begin{aligned}
  \tilde{z}(\hat{z}, \hat{y}, \hat{x}^{\cdot}) &= \hat{z} - \mathscr{F}(\hat{y})\\
  \tilde{y}(\hat{z}, \hat{y}, \hat{x}^{\cdot}) &= \hat{y}\\
  \tilde{x}^{\mu}(\hat{z}, \hat{y}, \hat{x}^{\cdot}) &= \hat{x}^{\mu}
\end{aligned}
\right.
\end{equation*}
which brings the metric into the form
\begin{gather*}
\hat{g}^{(6)}_{zz} = 1 \qquad \qquad \hat{g}^{(6)}_{zy} = - \frac{d\mathscr{F}}{d \hat{y}} \, + \, \frac{\b \, \ep (\hat{y})}{\sqrt{1+ \b^2}} \qquad \qquad \hat{g}^{(6)}_{yy} = \Big( \frac{d\mathscr{F}}{d \hat{y}} \Big)^{\!2} - 2 \, \frac{d\mathscr{F}}{d \hat{y}} \, \frac{\b \, \ep (\hat{y})}{\sqrt{1+ \b^2}} + 1 \notag \\
\\
\hat{g}^{(6)}_{z \m} = \hat{g}^{(6)}_{y \m} = 0 \qquad \qquad \qquad \qquad \qquad \qquad \hat{g}^{(6)}_{\m\n} = \e_{\m\n} \quad . \notag
\end{gather*}
Asking that $\hat{g}^{(6)}_{zy} = 0$ amounts to impose
\beq
\label{Rapidita}
\frac{d\mathscr{F}}{d \hat{y}} (\hat{y}) = \frac{\b \, \ep (\hat{y})}{\sqrt{1+ \b^2}}
\eeq
which in turn implies
\begin{gather*}
\hat{g}^{(6)}_{zz} = 1 \quad \qquad \hat{g}^{(6)}_{zy} = 0 \quad \qquad \hat{g}^{(6)}_{yy} = 1 - \, \Bigg( \!\frac{\b \, \ep (\hat{y})}{\sqrt{1+ \b^2}} \! \Bigg)^{\!\!2} \quad \qquad \hat{g}^{(6)}_{z \m} = \hat{g}^{(6)}_{y \m} = 0 \quad \qquad \hat{g}^{(6)}_{\m\n} = \e_{\m\n} \quad .
\end{gather*}
Finally consider the following coordinate transformation
\begin{equation*}
(\star \! \star \! \star) \quad \left\{
\begin{aligned}
  \hat{z}(Z, Y, X^{\cdot}) &= Z\\
  \hat{y}(Z, Y, X^{\cdot}) &= \mathscr{G}(Y)\\
  \hat{x}^{\mu}(Z, Y, X^{\cdot}) &= X^{\mu}
\end{aligned}
\right.
\end{equation*}
which brings the metric into the form:
\begin{gather*}
g^{(6)}_{zz} = 1 \qquad \qquad g^{(6)}_{zy} = 0 \qquad \qquad g^{(6)}_{yy} = \, \Bigg( \frac{d\mathscr{G}}{d Y} \Bigg)^{\!\!2} \, \Bigg[ 1 - \Bigg( \!\frac{\b \, \ep (\mathscr{G}(Y))}{\sqrt{1+ \b^2}} \! \Bigg)^{\!\!2} \, \Bigg] \notag \\
\\
g^{(6)}_{z \m} = g^{(6)}_{y \m} = 0 \qquad \qquad \qquad \qquad \qquad \qquad g^{(6)}_{\m\n} = \e_{\m\n} \quad . \notag
\end{gather*}
Asking that $g^{(6)}_{yy} = 1$ amounts to
\beq
\label{sunnysun}
\Bigg( \frac{d\mathscr{G}}{d Y} \Bigg)^{\!\!2} =  \frac{1+ \b^2}{1+ \b^{2} \, \Big( 1 - \ep^{2} \big( \mathscr{G}(Y) \big) \Big)}
\eeq
which implies
\beqnn
g^{(6)}_{AB} = \e_{AB} \quad .
\eeqnn
Therefore, provided that the functions $\mathscr{F}$ and $\mathscr{G}$ exist, the composition of the three coordinates changes ($\star$), ($\star\star$) and ($\star\star\star$) transforms the initial metric (\ref{backgroundthemlineapp}) into the 6D Minkowski metric. The existence of solutions of the differential equation (\ref{Rapidita}) is ensured by the fact that the function $\ep$, being continuous, is primitivable; concerning the existence of the function $\mscr{G}$, note first of all that the right hand side of (\ref{sunnysun}) never vanishes, so there are two classes of solutions characterised by the fact that $d\mathscr{G}/dY$ is positive or negative. These two choices for the sign of $d\mathscr{G}/dY$ correspond to the fact that the new ``$y$'' coordinate ($Y$) has the same or the opposite orientation with respect to the old ``$y$'' coordinate ($\hat{y}$): we choose to impose that $d\mathscr{G}/dY$ is positive, which means that the $Y$ coordinate has the same orientation as $\hat{y}$. Therefore, we can rewrite the equation (\ref{sunnysun}) as
\beq
\label{sunnysun2}
\frac{d\mathscr{G}}{d Y} =  \mathcal{D} \big( \mathscr{G}(Y) \big)
\eeq
where
\beq
\label{sunnysun3}
\mathcal{D} \big( \mathscr{G} \big) = \sqrt{\frac{1+ \b^2}{1+ \b^{2} \, \Big( 1 - \ep^{2} \big( \mathscr{G} \big) \Big)}} \quad .
\eeq
Since both $\ep$ and $\ep^\p$ are smooth and bounded by hypothesis, the function $\mcal{D}$ is (globally) Lipschitzian: therefore, the Picard-Lindel\"of theorem (see for example \cite{Hale}) ensures that, for each choice of the initial condition, there exists a unique local solution to the equation (\ref{sunnysun2}). Furthermore, the fact that $\mcal{D}$ is smooth and bounded both from below and from above (we have in fact $1 \leq \mathcal{D} ( \mathscr{G} ) \leq \sqrt{1 + \bq}\,$) implies that the local solution can be extended to a global solution. Note finally that, since the right hand side of the equation (\ref{sunnysun2}) never vanishes, it follows that $\mathscr{G}(Y)$ is a diffeomorphism and therefore invertible.

Therefore, we can indeed find a change of coordinates which maps the metric (\ref{backgroundthemlineapp}) into the 6D Minkowski metric: in the new reference system, the geometrical meaning of the configuration is encoded in the trajectory of the cod-1 brane, which is defined by $\mscr{F}$ and $\mscr{G}$. In synthesis, we have passed from a trivial embedding and a non-trivial metric to a non-trivial embedding and a trivial metric.

\subsection{The new embedding of the cod-1 brane}

To find the embedding of the cod-1 brane in the new bulk reference system, note first of all that we can still parametrize the cod-1 brane and the cod-2 brane with the ``old'' coordinates $(y, \xd)$ and $\xd$. Furthermore, as a consequence of the three coordinate changes, a point $(z,y,\xd) = (0,y,\xd)$ on the cod-1 brane is mapped into the point $(Z,Y,\Xd) = (\mathscr{F}(y),\mathscr{G}^{-1}(y),\xd)$, and in particular a point $(z,y,\xd) = (0,0,\xd)$ on the cod-2 brane is mapped into the point $(Z,Y,\Xd) = (\mathscr{F}(0),\mathscr{G}^{-1}(0),\xd)$. Therefore, the embedding of the cod-1 brane into the 6D Minkowski space is then \beq
\label{Summertime}
\vf^A(y,\xd) = \big( \mcal{Z}(y),\mcal{Y}(y), \xd \big)
\eeq
where $\mcal{Z}(y) \equiv \mscr{F}(y)$ and $\mcal{Y}(y) \equiv \mscr{G}^{-1}(y)$: note that, as a consequence of (\ref{Rapidita}) and (\ref{sunnysun}), the components of the embedding function $\mcal{Z}$ and $\mcal{Y}$ satisfy
\beq
\mcalZpq (y) + \mcalYpq (y) = 1 \quad .
\eeq
This was actually to be expected, since, using the embedding which corresponds to the new bulk coordinates, the $yy$ component of the metric induced on the cod-1 brane reads
\beq
\ti{g}_{yy}(y) = \mcalZpq (y) + \mcalYpq (y)
\eeq
while using the old bulk coordinates we had $\ti{g}_{ab} = \e_{ab}$, and we know that induced metric on the brane is not influenced by a change of the bulk coordinates. 

The components of the embedding function (\ref{Summertime}) are not uniquely determined by the differential equations (\ref{Rapidita}) and (\ref{sunnysun}), since to determine them we need to add some initial conditions. We choose to impose that the position of the cod-2 brane have the same bulk coordinates before and after the coordinate changes, which means to ask that $\mscr{F}(0) = 0$ and $\mscr{G}(0) = 0$: the non-trivial components of the embedding function of the cod-1 brane are then determined by the following Cauchy problems
\beq
\label{mcalZ Cauchy problem}
\left\{
\begin{aligned}
\mcal{Z}^{\p}(y) &= \frac{\b \, \ep (y)}{\sqrt{1+ \b^2}} \\[2mm]
\mcal{Z}(0) &= 0 
\end{aligned}
\right.
\eeq
and
\beq
\label{mcalY Cauchy problem}
\left\{
\begin{aligned}
\mcal{Y}^{\p}(y) &= \sqrt{\frac{1 + \bq \big( 1 - \ep^2(y) \big)}{1 + \bq}} \\[2mm]
\mcal{Y}(0) &= 0 \quad .
\end{aligned}
\right.
\eeq

\noi The Cauchy problem for $\mcal{Z}$ is implicitly solved by
\beq
\mcal{Z} (y) = \frac{\b}{\sqrt{1+ \b^2}} \, \int_{0}^{y} \! d \z \, \ep (\z)
\eeq
and we remember that $\ep(\z)$ is an odd, monotonically increasing function which is constant and equal to one (minus one) for $\z \geq l_2$ ($\z \leq -l_2$). This implies that the solution $\mcal{Z}(y)$ is even, and that, for $\abs{y} \geq l_2$ (\emph{i.e.} outside the thick cod-2 brane), $\mcal{Z}$ reads
\beq
\label{BarbieGirl 1}
\mcal{Z} (y) = \frac{\b}{\sqrt{1+ \b^2}} \, \abs{y} + \mcal{Z}_0
\eeq
where $\mcal{Z}_0$ is an integration constant which is responsible for the smooth matching of the internal and the external solutions, and explicitly reads
\beq
\mcal{Z}_0 = \frac{\b}{\sqrt{1+ \b^2}} \, \int_{0}^{l_2} \! d \z \, \ep (\z) - \frac{\b}{\sqrt{1+ \b^2}} \, l_2 \quad .
\eeq
Note that, as we perform the thin limit on the thick cod-2 brane and $l_2 \rightarrow 0^+$, as we already mentioned $\b$ remains constant while $\ep (\z)$ remains bounded, so $\mcal{Z}_0$ tends to zero: therefore, the thin limit of $\mcal{Z}$ exists and is given by
\beq
\label{DoctorJones 1}
\mcal{Z} (y) = \frac{\b}{\sqrt{1+ \b^2}} \, \abs{y} \quad .
\eeq
Concerning the Cauchy problem for $\mcal{Y}$, also in this case the function on the right hand side of (\ref{mcalY Cauchy problem}) is continuous (since $-1 \leq \ep(y) \leq 1$) and therefore it is primitivable, so there exists a unique solution of (\ref{mcalY Cauchy problem}) which is defined on all the real axis and has the implicit form
\beq
\label{mscrY primitive}
\mcal{Y}(y) = \int_{0}^{y} \! d \z \, \sqrt{\frac{1 + \bq \big( 1 - \ep^2(\z) \big)}{1 + \bq}} \quad .
\eeq
Furthermore, the fact that $\ep(y)$ is odd implies that the solution for $\mcal{Y}$ is odd, and since
\beq
\label{Justina}
\sqrt{\frac{1}{1 + \bq}} \leq \mcalYp(y) \leq 1
\eeq
we have that $\mcal{Y}$ is a diffeomorphism. The existence of the solution for $\mcal{Y}$ and the fact that it is a diffeomorphism are also directly implied by the fact that $\mcal{Y}$ is the inverse function of $\mscr{G}$. In particular, the relation (\ref{mscrY primitive}) implies that for $y \geq l_2$ the solution for $\mcal{Y}$ has the form
\beq
\label{BarbieGirl 2}
\mcal{Y} (y) = \frac{1}{\sqrt{1+ \b^2}} \,\, y + \mcal{Y}_0
\eeq
where the integration constant $\mcal{Y}_0$ reads
\beq
\label{mscrY0}
\mcal{Y}_0 = \int_{0}^{l_2} \! d \z \, \sqrt{\frac{1 + \bq \big( 1 - \ep^2(\z) \big)}{1 + \bq}} - \frac{l_2}{\sqrt{1 + \bq}} \quad .
\eeq
Analogously to the $\mcal{Z}$ case, $\mcal{Y}_0$ tends to zero when we perform the thin limit on the thick cod-2 brane and $l_2 \rightarrow 0^+$: therefore, the thin limit of $\mcal{Y}$ exists and is given by
\beq
\label{DoctorJones 2}
\mcal{Y} (y) = \frac{1}{\sqrt{1+ \b^2}} \,\, y \quad .
\eeq
Note finally that (\ref{Julie}) implies
\begin{align}
\label{SaintEmilion}
\frac{\b}{\sqrt{1+ \b^2}} &= \sin \bigg( \frac{\bla}{4 \Msf} \bigg) & \frac{1}{\sqrt{1+ \b^2}} &= \cos \bigg( \frac{\bla}{4 \Msf} \bigg) \quad .
\end{align}

Therefore, the configuration defined by the bulk metric (\ref{backgroundthemlineapp}) where the cod-1 brane and cod-2 brane are placed respectively at $z = 0$ and $z = y = 0$, is equivalent to a configuration where the bulk metric is the 6D Minkowski metric and the embedding of the cod-1 metric is of the form (\ref{Summertime}) while the cod-2 brane is placed at $y = 0$. The latter configuration has the same form of the pure tension solutions we found in section \ref{Pure tension solutions} using the bulk-based approach. Taking into account the relation (\ref{SaintEmilion}), and comparing the explicit expressions (\ref{BarbieGirl 1}) and (\ref{BarbieGirl 2}) (and the thin limit versions (\ref{DoctorJones 1}) and (\ref{DoctorJones 2})) for the components of the embedding function with the expressions (\ref{Youpiyou}) and (\ref{Annabelle}) (and the thin limit versions (\ref{CestLaVie})) for the analogous components in the bulk-based approach, we conclude that the brane-based configurations (\ref{backgroundthemline}) introduced in \cite{deRham:2010rw} are actually equivalent to the bulk-based configurations we derived in section \ref{Pure tension solutions}.

\chapter{Pillbox integration of nested branes}
\label{Pillbox integration of nested branes}

In this appendix, we perform explicitly the pillbox integrations which appear in the left hand side of equations (\ref{kindness der}) and (\ref{kindness trace2}), namely
\begin{align}
\label{GoPuteolanaGo}
\mscr{I} &= \lim_{n \rightarrow + \infty} \int_{-}^{+} \! d \hxi \,\, \bvf_{i \, [n]}^{\p \p} \, \hdvf^{i \, [n]}_{gi} \\[2mm]
\label{TakTak}
\mcalI &= \lim_{n \rightarrow + \infty} \int_{-}^{+} \! d \hxi \, \bigg( \bn_{[n]}^i \bn_{[n]}^j \, \Big( \bn^{[n]}_k \bvf_{[n]}^{k \, \p \p} \Big) \, \h{h}^{gi \, [n]}_{ij} + 2 \, \bn^{[n]}_i \, \hdvf^{i \, [n] \, \p \p}_{gi} \bigg) \quad .
\end{align}
Note that these expressions are not numbers, but functions of the 4D coordinates $\chd$: we omit to indicate this dependence in the following.

Before performing the integrations, it is useful to sum up some properties of the background tangent and normal vectors, and to discuss how the perturbation of the parallel and normal components of the embedding are linked to the analogous components of the perturbation of the parallel vector $\d \mathbf{v}_{(\xi)}$. The background tangent vector $\bvf_i^{\p}$ and the background normal vector $\bn^i$ are constructed from the first $\xi$-derivative of the $z$ and $y$ components of the background embedding; they are orthonormal
\begin{align}
\bvf_i^{\p} \, \bvf^{i \, \p} &= 1 & \bvf_i^{\p} \, \bn^{i} &= 0 & \bn_i \, \bn^{i} &= 1
\end{align}
and their derivative with respect to $\xi$ satisfies
\begin{align}
\bvf^{i \, \p\p} &= \frac{\Zpp}{\sqrt{1 - \Zpq}} \,\, \bn^i & \bn^{i \, \p} &= - \frac{\Zpp}{\sqrt{1 - \Zpq}} \,\, \bvf^{i \, \p} \quad .
\end{align}
Remembering the definition of the normal and parallel component of the perturbation of the embedding
\begin{align}
\label{CostaCoffee}
\dvfn &\equiv \bn_i \, \dvf^i_{gi} & \dvfp &\equiv \bvf_i^{\p} \, \dvf^i_{gi}
\end{align}
and the definition of the normal and parallel component of the perturbation of the tangent vector
\begin{align}
\label{Ellie}
\dvn &\equiv \bn_i \, \dvf^{i \, \p}_{gi} & \dvsp &\equiv \bvf_i^{\p} \, \dvf^{i \, \p}_{gi} \quad ,
\end{align}
we can relate $\dvn$ and $\dvsp$ to $\dvfn$ and $\dvfp$ in the following way
\begin{align}
\dvn &= \dvfn^{\p} + \frac{\Zpp}{\sqrt{1 - \Zpq}} \,\, \dvfp \\[2mm]
\dvsp &= \dvfp^{\p} - \frac{\Zpp}{\sqrt{1 - \Zpq}} \,\, \dvfn \quad .
\end{align}
Using the relations above, we can express the quantities $\bn_i \, \dvf^{i \, \p\p}_{gi}$ and $\bvf_i^\p \, \dvf^{i \, \p\p}_{gi}$, where the perturbation of the embedding is derived twice with respect to $\xi$, as follows
\begin{align}
\label{Riverside}
\bn_i \, \dvf^{i \, \p\p}_{gi} &= \dvn^{\p} + \frac{\Zpp}{\sqrt{1 - \Zpq}} \,\, \dvsp \\[2mm]
\bvf_i^\p \, \dvf^{i \, \p\p}_{gi} &= \dvsp^{\p} - \frac{\Zpp}{\sqrt{1 - \Zpq}} \,\, \dvn
\end{align}
and we can express the quantities $\bvf_i^{\p\p} \, \dvf^{i}_{gi}$ and $\bn_i^{\p} \, \dvf^{i}_{gi}$, where the background embedding is derived twice with respect to $\xi$, as follows
\begin{align}
\label{GoSpursGo}
\bvf_i^{\p\p} \, \dvf^{i}_{gi} &= \dvfp^{\p} - \dvsp \\[2mm]
\bn_i^{\p} \, \dvf^{i}_{gi} &= \dvfn^{\p} - \dvn \quad .
\end{align}

\section{The induced gravity part}

We turn now to the evaluation of the pillbox integral (\ref{GoPuteolanaGo}). Indicating explicitly the dependence on $n$ of the domain of integration, the integral reads
\beq
\mscr{I} = \lim_{n \rightarrow + \infty} \int_{- l_{2}^{[n]}}^{+ l_{2}^{[n]}} \!\! d \hxi \,\,\, \bvf_{i \, [n]}^{\p \p} \, \hdvf^{i \, [n]}_{gi}
\eeq
and, using the relation (\ref{GoSpursGo}), we get
\beq
\mscr{I} = \lim_{n \rightarrow + \infty} \int_{- l_{2}^{[n]}}^{+ l_{2}^{[n]}} \!\! d \hxi \,\,\, \hdvfp^{[n] \, \p} - \lim_{n \rightarrow + \infty} \int_{- l_{2}^{[n]}}^{+ l_{2}^{[n]}} \!\! d \hxi \,\,\, \dhvsp^{[n]} \quad .
\eeq
Note that, from the definition (\ref{Ellie}), $\dhvsp^{[n]}$ is constructed from the first derivative of the embedding functions only (no second derivatives), so (using our ansatz) it remains bounded even in the $n \rightarrow + \infty$ limit. Since in this limit the domain of integration shrinks to a domain of zero measure, we obtain
\beq
\lim_{n \rightarrow + \infty} \int_{- l_{2}^{[n]}}^{+ l_{2}^{[n]}} \!\! d \hxi \,\,\, \dhvsp^{[n]} = 0
\eeq
and so we conclude that
\beq
\mscr{I} = 2 \lim_{n \rightarrow + \infty} \hdvfp^{[n]}\Big\rvert_{l_{2}^{[n]}} = 2 \,\,\hdvfp^{\infty}\Big\rvert_{0^{+}} \quad .
\eeq

The parallel component of the bending, despite being non-zero outside the cod-2 brane, does not appear in the pure cod-1 junction conditions. It is then useful to express the integral $\mscr{I}$ in terms of quantities which have a more direct geometrical interpretation. The embedding of the cod-2 brane in the bulk $\a^{\cdot}(\chd)$ is obtained by composing the cod-2 embedding into the cod-1 brane $\ta^{\cdot}(\chd)$ and the cod-1 embedding $\vfd(\hxid)$
\beq
\a^A(\chd) = \vf^A \big( \ta^{\cdot}(\chd) \big) \quad .
\eeq
Since the embedding of the cod-2 brane into the cod-1 brane is trivial both at background and at perturbative level (equation (\ref{Joan})), we have 
\begin{align}
\a^z(\chd) &= \vf^z \big( 0, \chd \big) & \a^y(\chd) &= 0 & \a^{\m}(\chd) &= \ch^{\m}
\end{align}
where $\a^y$ vanishes as a consequence of the $Z_2$ symmetry present inside the cod-1 brane. Therefore the movement of the cod-2 brane in the bulk is described by the functions $\d \!\a^z \big( \chd \big) = \dvf^z \big( 0, \chd \big)$ and  $\d \!\a^\m \big( \chd \big) = \dvf^{\m} \big( 0, \chd \big)$. In particular, we introduce the gauge invariant description of the movement of the brane in the $z$ direction
\beq
\dvf_0 \big( \chd \big) \equiv \dvf^z_{gi}\big( 0, \chd \big) \quad .
\eeq
Using the definition (\ref{CostaCoffee}) of $\dvfp$, and the fact that the components $\hdvf^{i \, [n]}_{gi}$ are continuous also in the thin limit, we get
\beq
\hdvfp^{\infty}\Big\rvert_{0^{+}} = \lim_{n \rightarrow + \infty} \bvf_{i\, [n]}^{\p}\Big\rvert_{l_{2}^{[n]}} \, \hdvf^{i \, [n]}_{gi}\Big\rvert_{l_{2}^{[n]}} = \Zp_{\infty}\Big\rvert_{0^+} \, \dvf_0^{\infty} \quad ,
\eeq 
where we used the fact that $\hdvf^{y \, \infty}_{gi}\big\rvert_{0}$ vanishes for symmetry reasons. Therefore, we can express the integral $\mscr{I}$ in terms of the movement of the cod-2 brane in the bulk as follows
\beq
\mscr{I} = 2 \, \sin \bigg( \frac{\bla}{4 \Msf} \bigg) \, \dvf_0^{\infty} \quad .
\eeq
Alternatively, we can use the definition (\ref{CostaCoffee}) of $\dvfn$ to express the integral $\mscr{I}$ in terms of $\dvfn^{\infty}$. In fact, analogously to what we did above for $\dvfp\big\rvert_{0^{+}}$, we can express $\dvfn^{\infty}\big\rvert_{0^{+}}$ in terms of $\dvf_0^{\infty}$ as follows
\beq
\hdvfn^{\infty}\Big\rvert_{0^{+}} = \lim_{n \rightarrow + \infty} \bn_{i}^{[n]}\Big\rvert_{l_{2}^{[n]}} \, \hdvf^{i \, [n]}_{gi}\Big\rvert_{l_{2}^{[n]}} = \Yp_{\infty}\Big\rvert_{0^+} \, \dvf_0^{\infty}
\eeq 
and therefore we get
\beq
\mscr{I} = 2 \, \tan \bigg( \frac{\bla}{4 \Msf} \bigg) \, \hdvfn^{\infty}\Big\rvert_{0^{+}} \quad .
\eeq

\section{The extrinsic curvature part}

We consider now the pillbox integration (\ref{TakTak})
\beq
\mcalI = \lim_{n \rightarrow + \infty} \int_{- l_{2}^{[n]}}^{+l_{2}^{[n]}} \!\! d \hxi \, \bigg( \bn_{[n]}^i \bn_{[n]}^j \, \Big( \bn^{[n]}_k \bvf_{[n]}^{k \, \p \p} \Big) \, \h{h}^{gi \, [n]}_{ij} + 2 \, \bn^{[n]}_i \, \hdvf^{i \, [n] \, \p \p}_{gi} \bigg) \quad .
\eeq
To perform this integration, it is useful to recast the integrand in a more convenient form. First of all, we can express the quantity $\bn^{[n]}_i \, \hdvf^{i \, [n] \, \p \p}_{gi}$ using (\ref{Riverside}) and the relation
\beq
\frac{\Zpp_{[n]}}{\sqrt{1 - \Zpq_{[n]}}} = \bn^{[n]}_k \bvf_{[n]}^{k \, \p \p} \quad ,
\eeq
to obtain
\beq
\bn^{[n]}_i \, \hdvf^{i \, [n] \, \p \p}_{gi} = \dhvn^{[n] \, \p} + \dhvsp^{[n]} \, \Big( \bn^{[n]}_k \bvf_{[n]}^{k \, \p \p} \Big) \quad .
\eeq
Secondly, we remember from (\ref{LeJeuxSonFait}) that the $\xi\xi$ component of the perturbation of the induced metric reads
\beq
\label{Caravan}
\ti{h}^{[n]}_{\xi\xi} = \bvf_{[n]}^{i \, \p} \bvf_{[n]}^{j \, \p} \, \ti{h}_{ij}^{gi \, [n]} + 2 \, \dvsp^{[n]} \quad ,
\eeq
and therefore in cod-1 GNC we have
\beq
\label{Maretta}
\dhvsp^{[n]} = - \half \, \bvf_{[n]}^{i \, \p} \bvf_{[n]}^{j \, \p} \, \h{h}_{ij}^{gi \, [n]} \quad .
\eeq
Using the relations (\ref{Caravan}) and (\ref{Maretta}) we can express the integral $\mcalI$ as follows
\beq
\label{ponk}
\mcalI = 4 \, \dhvn^{\infty}\Big\rvert_{0^+} + \lim_{n \rightarrow + \infty} \int_{- l_{2}^{[n]}}^{+ l_{2}^{[n]}} \!\! d \hxi \,\,\, \Big( \bn_{[n]}^i \bn_{[n]}^j - \bvf_{[n]}^{i \, \p} \bvf_{[n]}^{j \, \p} \Big) \, \Big( \bn^{[n]}_k \bvf_{[n]}^{k \, \p \p} \Big) \, \h{h}^{gi \, [n]}_{ij} \quad .
\eeq

Since $\Yp_{[n]}$ can be written as $\sqrt{1 - \Zpq_{[n]}}$, the integral in the right hand side of the equation (\ref{ponk}) can be expressed as a linear combination of integrals of the form
\beq
\label{cibocibo}
\lim_{n \rightarrow + \infty} \int_{- l_{2}^{[n]}}^{+ l_{2}^{[n]}} \!\! d \hxi \,\,\, \Zpp_{[n]}\big( \hxi \big) \, \mcal{A} \Big( \Zp_{[n]} \big( \hxi \big) \Big) \, \mcal{G}_{[n]} \big( \hxi \big) \quad ,
\eeq
where $\mcal{G}_{[n]} \big( \hxi \big) = \h{h}^{gi \, [n]}_{ij}$ for a specific choice of $ij$. For example, the term
\beq
\bn_{[n]}^z \bn_{[n]}^z \, \Big( \bn^{[n]}_k \bvf_{[n]}^{k \, \p \p} \Big) \, \h{h}^{gi \, [n]}_{zz}
\eeq
gives rise to an integral of the form (\ref{cibocibo}) with
\begin{align}
\mcal{A} \big( \z \big) &=\sqrt{1 - \z^2} & \mcal{G}_{[n]} \big( \hxi \big) &= \h{h}^{gi \, [n]}_{zz} \big( \hxi \big) \quad .
\end{align}
We want now to develop a general method to evaluate pillbox integrals of the form (\ref{cibocibo}) when $\mcal{G}_{[n]}$ is a sequence of smooth functions which converges uniformly to a continuous function $\mcal{G}_{\infty}$, as indeed happens for the functions $\h{h}^{gi \, [n]}_{ij}$.

\subsection{General pillbox integration method}

Note that the functions $\mcal{G}_{[n]}$ and $\mcal{A}$ have always a definite parity, and they are either both even or both odd: we consider first the case in which $\mcal{A}$ is even. Let's suppose that the function $\mcal{A}$ is continuous and integrable on every closed interval $[0, x]$ where $0< x <1$ (this is always true for the integrals which appear in (\ref{ponk})). Then it follows that the sequence of functions
\beq
\mcal{F}_{[n]}\big( \hxi \big) \equiv \Zpp_{[n]}\big( \hxi \big) \, \mcal{A} \Big( \Zp_{[n]} \big( \hxi \big) \Big)
\eeq
is proportional to a realization of the Dirac delta. In fact, $\mcal{F}_{[n]}\big( \hxi \big)$ vanishes identically for $\abs{\hxi} > l_2^{[n]}$ (as can be deduced from the relation (\ref{Balotelli})), and its integral on the interval $[-l_2^{[n]} , +l_2^{[n]}]$ is independent from $n$, despite the fact that $l_2^{[n]} \rightarrow 0$ in the $n \rightarrow + \infty$ limit. This can be explicitly seen by changing variable $\hxi \rightarrow \z = \Zp_{[n]} \big( \hxi \big)$ in the integral (we indicate here $x \equiv  \sin ( \bla/4 \Msf )$)
\beq
\int_{-l_2^{[n]}}^{+l_2^{[n]}} \!\! d\hxi \,\, \mcal{F}_{[n]}\big( \hxi \big) = \int_{-x}^{+x} \!\! d\z \,\, \mcal{A}\big( \z \big) \equiv 2 \, \mcalI_{\mcal{A}}
\eeq
where we defined $\mcalI_{\mcal{A}}$ as the integral of $\mcal{A}$ on the interval $[0,x]$, to be consistent with the case where $\mcal{A}$ is odd. Therefore, we conclude that
\beq
\mcal{F}_{[n]}\big( \hxi \big) \xrightarrow[n \rightarrow + \infty]{} 2 \, \mcalI_{\mcal{A}} \,\, \d \big( \hxi \big) \quad .
\eeq
We want to show that, provided $\mcal{G}_{[n]}$ has the properties mentioned above, the integrals of the type (\ref{cibocibo}) can be evaluated using (a generalized version of) the properties of the Dirac delta, namely the following relation holds
\beq
\label{Palacinka}
\lim_{n \rightarrow + \infty} \int_{- l_{2}^{[n]}}^{+ l_{2}^{[n]}} \!\! d \hxi \,\,\, \mcal{F}_{[n]}\big( \hxi \big) \, \mcal{G}_{[n]} \big( \hxi \big) = 2 \, \mcalI_{\mcal{A}} \,\, \mcal{G}_{\infty} \big( 0 \big) \quad .
\eeq
To begin with, note that to each $n \in \mathbb{N}$ we can associate a positive number $\vep_n$ which is the upper bound of the absolute difference between $\mcal{G}_{[n]} \big( \hxi \big)$ and $\mcal{G}_{\infty} \big( \hxi \big)$ where $\hxi$ belongs to the interval $I_{[n]} \equiv [- l_{2}^{[n]} , l_{2}^{[n]}]$
\beq
\vep_n \equiv \textrm{sup}_{\hxi \in I_{[n]}} \Big\{ \big\lvert \mcal{G}_{[n]} \big( \hxi \big) - \mcal{G}_{\infty} \big( \hxi \big) \big\rvert \Big\} \quad .
\eeq
Since the functions $\mcal{G}_{[n]} \big( \hxi \big)$ are continuous and the intervals $I_{[n]}$ are compact, $\vep_n$ is finite for every value of $n$. Crucially, the fact that the sequence of functions $\mcal{G}_{[n]}$ converges \emph{uniformly} to $\mcal{G}_{\infty}$ implies that
\beq
\label{Kuguluf}
\lim_{n \rightarrow + \infty} \vep_n = 0 \quad ,
\eeq
while if the convergence is pointwise but not uniform then the numerical sequence $\vep_n$ may even diverge. Now, the fact that $\vep_n$ is positive implies that
\beq
\label{Boulevards}
\int_{- l_{2}^{[n]}}^{+ l_{2}^{[n]}} \!\! d \hxi \,\,\, \mcal{F}_{[n]} \, \Big( \mcal{G}_{\infty}  - \vep_n \Big) \leq \int_{- l_{2}^{[n]}}^{+ l_{2}^{[n]}} \!\! d \hxi \,\,\, \mcal{F}_{[n]} \, \mcal{G}_{[n]}  \leq \int_{- l_{2}^{[n]}}^{+ l_{2}^{[n]}} \!\! d \hxi \,\,\, \mcal{F}_{[n]} \, \Big( \mcal{G}_{\infty}  + \vep_n \Big)
\eeq
and, since $\mcal{G}_{\infty} \big( \hxi \big)$ is continuous, using the properties of the Dirac delta we have
\beq
\label{Monmartre}
\lim_{n \rightarrow + \infty} \int_{- l_{2}^{[n]}}^{+ l_{2}^{[n]}} \!\! d \hxi \,\,\, \mcal{F}_{[n]} \big( \hxi \big) \, \Big( \mcal{G}_{\infty} \big( \hxi \big) \pm \vep_k \Big) = 2 \, \mcalI_{\mcal{A}} \, \Big( \mcal{G}_{\infty} \big( 0 \big) \pm \vep_k \Big)
\eeq
where in the last relation $k$ and $n$ are considered as independent parameters. The relations (\ref{Kuguluf}), (\ref{Boulevards}) and (\ref{Monmartre}) together imply that the formula (\ref{Palacinka}) holds, and therefore we conclude that
\beq
\label{Place Vendome}
\lim_{n \rightarrow + \infty} \int_{- l_{2}^{[n]}}^{+ l_{2}^{[n]}} \!\! d \hxi \,\,\, \Zpp_{[n]}\big( \hxi \big) \, \mcal{A} \Big( \Zp_{[n]} \big( \hxi \big) \Big) \, \mcal{G}_{[n]} \big( \hxi \big) = 2 \, \mcalI_{\mcal{A}} \,\, \mcal{G}_{\infty} \big( 0 \big) \quad ,
\eeq
where
\beq
\label{quay}
\mcalI_{\mcal{A}} = \int_{0}^{x} \!\! d\z \,\, \mcal{A}\big( \z \big)
\eeq
and $x = \sin \big( \bla/4 \Msf \big)$.

In the case where $\mcal{A}$ is odd (and $\mcal{G}_{[n]}$ as well), its integral on the interval $[-x,x]$ vanishes by symmetry reasons, although the integral (\ref{cibocibo}) doesn't. Moreover, the integral $\mcalI_{\mcal{A}}$ defined in (\ref{quay}) is different from zero: it is then useful to express the integral (\ref{cibocibo}) as twice the integral on the interval $[0,l_{2}^{[n]}]$ of the same integrand, and work only with positive values of $\hxi$. With this proviso, the relations (\ref{Kuguluf}) and (\ref{Boulevards}) hold also in this case (substituting $- l_2^{[n]}$ with $0$), and the relation (\ref{Monmartre}) becomes
\beq
\label{Monmartre2}
\lim_{n \rightarrow + \infty} \int_{0}^{+ l_{2}^{[n]}} \!\! d \hxi \,\,\, \mcal{F}_{[n]} \big( \hxi \big) \, \Big( \mcal{G}_{\infty} \big( \hxi \big) \pm \vep_k \Big) = \mcalI_{\mcal{A}} \, \Big( \mcal{G}_{\infty} \big( 0 \big) \pm \vep_k \Big) \quad .
\eeq
Therefore, the formula (\ref{Place Vendome}) holds also for $\mcal{A}$ odd. However, if $\mcal{G}_{[n]}$ is odd then $\mcal{G}_{\infty}\big( 0 \big) = 0$, and so we conclude that in this case the integral (\ref{cibocibo}) vanishes.

\subsection{Final result}

We can finally use the formula (\ref{Place Vendome}) to compute the pillbox integration (\ref{ponk}). Note that $\h{h}^{gi \, [n]}_{zz}$ and $\h{h}^{gi \, [n]}_{yy}$ are even functions of $\hxi$ while $\h{h}^{gi \, [n]}_{zy}$ is odd: it follows that the integration of the terms involving $\h{h}^{gi \, [n]}_{zy}$ vanish, and (\ref{ponk}) can be rewritten as
\beq
\label{ponk2}
\mcalI = 4 \, \dhvn^{\infty}\Big\rvert_{0^+} + \lim_{n \rightarrow + \infty} \int_{- l_{2}^{[n]}}^{+ l_{2}^{[n]}} \!\! d \hxi \,\,\, \frac{\Zpp_{[n]}}{\sqrt{1 - \Zpq_{[n]}}} \, \Big( \Ypq_{[n]} - \Zpq_{[n]} \Big) \, \Big( \h{h}^{gi \, [n]}_{zz} - \h{h}^{gi \, [n]}_{yy} \Big) \quad .
\eeq
Defining the integrals
\begin{align}
\mcalI_1 &\equiv \int_{0}^{x} \sqrt{1 - \z^2} \,\, d \z \\[2mm]
\mcalI_2 &\equiv \int_{0}^{x} \frac{\z^2}{\sqrt{1 - \z^2}} \,\, d \z
\end{align}
and using (\ref{Place Vendome}) and (\ref{quay}), we obtain
\beq
\label{ponk3}
\mcalI = 4 \, \dhvn^{\infty}\Big\rvert_{0^+} + 2 \, \Big( \mcalI_1 - \mcalI_2 \Big) \, \Big( \h{h}^{gi \, \infty}_{zz} (0) - \h{h}^{gi \, \infty}_{yy} (0) \Big) \quad .
\eeq
Indicating $\h{h}^{gi \, \infty}_{zz}$ and $\h{h}^{gi \, \infty}_{yy}$ respectively with $\h{h}^{\infty}_{zz}$ and $\h{h}^{\infty}_{yy}$, and evaluating explicitly the integrals
\begin{align}
\mcalI_1 &= \frac{1}{2} \, \Big( \arcsin x \, + x \, \sqrt{1 - x^2} \, \Big) \\[2mm]
\mcalI_2 &= \frac{1}{2} \, \Big( \arcsin x \, - x \, \sqrt{1 - x^2} \, \Big) \quad ,
\end{align}
we obtain
\beq
\label{ponk4}
\mcalI = 4 \, \dhvn^{\infty}\Big\rvert_{0^+} + \sin \bigg( \frac{\bla}{2 \Msf} \bigg) \, \Big( \h{h}^{\infty}_{zz} (0) - \h{h}^{\infty}_{yy} (0) \Big) \quad .
\eeq

Also in this case, it is useful to express the integral $\mcal{I}$ in terms of $\dvfn^{\infty}$, which appears in the pure cod-1 junction conditions. To do that, we remember that outside the cod-2 brane we have $\dhvn^{[n]} = \hdvfn^{\p \, [n]}$, which implies that for $\hxi \neq 0$ we have $\dhvn^{\infty}\big( \hxi \big) = \hdvfn^{\infty \, \p}\big( \hxi \big)$; this implies that
\beq
\dhvn^{\infty}\big\rvert_{0^+} = \hdvfn^{\infty \, \p}\big\rvert_{0^+} \quad ,
\eeq
and therefore we conclude that
\beq
\label{ponk5}
\mcalI = 4 \, \hdvfn^{\infty \, \p}\Big\rvert_{0^+} + \sin \bigg( \frac{\bla}{2 \Msf} \bigg) \, \Big( \h{h}^{\infty}_{zz} (0) - \h{h}^{\infty}_{yy} (0) \Big) \quad .
\eeq

\chapter{Total derivative combinations}
\label{Appendix Total derivative combinations}

We review here the main definitions and properties of total derivative combinations of the field $\f$ (and related objects) considered in section \ref{dRGT massive gravity}.

\section{Total derivative combinations of $\Pi_{\m\n}$}

Let's remind the definition of the object $\Pi$ constructed from the second derivatives of the field $\f$
\beq
\Pi_{\mu\nu} = \demden \phi \quad .
\eeq
As already mentioned in the main text, at every order in $\Pi$ (or equivalently in $\f$) there is a unique (up to an overall constant) contraction of $\Pi$ factors (we raise/lower indices with the Minkowski metric $\e^{\m\n}/\e_{\m\n}$ ) which is in the form of a total derivative. Explicitly, at order $n$ it takes the form \cite{NicolisRattazziTrincherini}
\beq
\label{TDtermnappendix}
\mathcal{L}^{TD}_{n}(\Pi) = \sum_p (-1)^p \, \e^{\m_1 p(\n_1)}\, \cdots \, \e^{\m_n p(\n_n)} \, \Pi_{\m_1 \n_1} \, \cdots \, \Pi_{\m_n \n_n} \quad ,
\eeq
where the sum runs on all the permutations $p$ of $n$ elements. To facilitate the comparison with the $\Pi$ structures coming from the non-linear mass term, we can group together some of the contractions in (\ref{TDtermnappendix}) using the fact that $\e^{\mu\nu}$ and $\Pi_{\mu\nu}$ are symmetric, and using the notation
\beq
\big[ \Pi^n \big] \equiv \e^{\m \a_{1}} \, \Pi_{\a_{1} \b_{1}} \, \e^{\b_{1} \a_{2}} \, \Pi_{\a_{2} \b_{2}} \cdots \e^{\b_{n - 1} \a_{n}} \, \Pi_{\a_{n} \m}
\eeq
we obtain
\begin{align}
\label{galileon1}
{\cal L}_1^{\rm TD}(\Pi) &= [\Pi] \\[1mm]
\label{galileon2}
{\cal L}_2^{\rm TD}(\Pi) &= [\Pi]^2 - [\Pi ^2] \\[1mm]
\label{galileon3}
{\cal L}_3^{\rm TD}(\Pi) &= [\Pi]^3 -3 [\Pi][\Pi ^2] +2 [\Pi^3] \\[1mm]
\label{galileon4}
{\cal L}_4^{\rm TD}(\Pi) &= [\Pi]^4 -6 [\Pi ^2][\Pi]^2 +8 [\Pi ^3][\Pi] +3 [\Pi^2]^2 -6 [\Pi ^4] \quad .
\end{align}

\noi Note that the terms ${\cal L}_n^{\rm TD}(\Pi)$ vanish identically for $n \geq 5$ (in general, they vanish for $n > D$, where $D$ is the spacetime dimension), and ${\cal L}_2^{\rm TD} (h)$ is the Fierz-Pauli term. Furthermore, they satisfy a recursion relation
\beq
{\cal L}_n^{\rm TD}(\Pi) = - \sum_{m=1}^n (-1)^m \frac{(n-1)!}{(n-m)!} \left[ \Pi^m \right] {\cal L}_{n-m}^{\rm TD}(\Pi)
\eeq
with $ {\cal L}_0^{\rm TD}(\Pi)=1$.

\section{The $X^{(n)}_{\mu\nu}$ tensors}

From the total derivative Lagrangians ${\cal L}_n^{\rm TD}(\Pi)$, we can construct the tensors $X^{(n)}_{\mu\nu}$ by deriving with respect to $\Pi^{\m\n}$
\beq
X^{(n)}_{\mu\nu} = \frac{1}{n+1} \, \frac{\de}{\de \Pi^{\mu\nu}} \, {\cal L}_{n+1}^{\rm TD}(\Pi) \quad ,
\eeq
obtaining in general
\beq
X^{(n)}_{\mu\nu} = \sum_{m=0}^n (-1)^m \frac{n!}{(n-m)!} \, \Pi^m_{\mu\nu} \, {\cal L}_{n-m}^{\rm TD}(\Pi) \quad .
\eeq
The tensors $X^{(n)}_{\mu\nu}$ satisfy the recursion relation

\beq
X^{(n)}_{\mu\nu} = -n \, \Pi_\mu^{\,\,\a} X^{(n-1)}_{\alpha\nu} + \Pi^{\alpha\beta} X^{(n-1)}_{\alpha\beta} \eta_{\mu\nu}
\eeq
and, since ${\cal L}_n^{\rm TD}(\Pi)$ vanishes for $n > 4$, they vanish for $n \geq 4$ ($n \geq D$ in a spacetime of dimension $D$). Explicitly they read
\begin{align*}
X^{(0)}_{\mu\nu} &= \eta_{\mu\nu} \\[1mm]
X^{(1)}_{\mu\nu} &= \left[ \Pi \right] \eta_{\mu\nu} - \Pi_{\mu\nu} \\[1mm]
X^{(2)}_{\mu\nu} &= \left( \left[ \Pi \right]^2 -\left[ \Pi^2 \right] \right) \eta_{\mu\nu} -2 \left[ \Pi \right] \Pi_{\mu\nu} + 2 \Pi^2_{\mu\nu} \\[1mm]
X^{(3)}_{\mu\nu} &= \left( \left[ \Pi \right]^3 -3 \left[ \Pi \right] \left[ \Pi^2 \right] +2 \left[ \Pi^3 \right] \right) \eta_{\mu\nu} -3 \left( \left[ \Pi \right]^2 - \left[ \Pi^2 \right] \right) \Pi_{\mu\nu} +6 \left[ \Pi \right] \Pi^2_{\mu\nu} -6 \Pi^3_{\mu\nu} \quad .
\end{align*}

The following relations involving the massless kinetic operator (\ref{Epsilon operator}) make clear which is the form of transformations we can perform on $h_{\m\n}$ to remove the mixing terms $h^{\m\n} X^{(j)}_{\mu\nu}$ from the $\La_3$ action in the decoupling limit
\bea &&{\cal E}_{\mu\nu}^{\ \ \alpha\beta} \left(\phi \, \eta_{\alpha\beta}\right) = - (D-2) \, X^{(1)}_{\mu\nu} \\[1mm]
&& {\cal E}_{\mu\nu}^{\ \ \alpha\beta} \left( \partial_\alpha \phi \, \partial_\beta \phi \right) = X^{(2)}_{\mu\nu} \quad .
\eea

Finally, it can be shown that the $X^{(n)}_{\mu\nu}$ tensors are symmetric and identically conserved
\begin{align}
X^{(n)}_{\mu\nu} &= X^{(n)}_{\nu\mu} \\[2mm]
\partial^\mu X^{(n)}_{\mu\nu} &= 0 \qquad .
\end{align}

\chapter{The implicit function theorem}
\label{The implicit function theorem}

The implicit function theorem, also known as Dini's theorem, is used repeatedly throughout the text. Although the theorem is more general, we give here its formulation in the specific case of a function of two (real) variables. For the proof, see \cite{Rudin} for the general case and \cite{Giusti} for the particular case treated here.

\section{Formulation of the theorem}

\newtheorem{theorem}{Theorem}
\begin{theorem}[Implicit function theorem, or Dini's theorem]
Let $F(x,y)$ be a function defined in an open set $A \subset \mathbb{R}^2$, and let $F$ be derivable with continuous partial derivatives. Be $(x_0,y_0) \in A$ such that
\beq
\label{conditionsDini}
F(x_0,y_0) = 0 \qquad, \qquad \frac{\de F}{\de y}(x_0,y_0) \neq 0 \quad .
\eeq
Then there exist:
\begin{description}
 \item[-] An open neighborhood $U$ of $x_0$ and an open neighborhood $V$ of $y_0$, such that $U \times V \subset A$ ;
 \item[-] A function $f: U \rightarrow V$ such that, for all $(x,y) \in U \times V$, we have
\beq
F(x,y) = 0 \qquad \Leftrightarrow \qquad y = f(x) \quad .
\eeq
 \end{description}
Furthermore, the function $x \rightarrow f(x)$ is derivable with continuous derivative, and we have
\beq
\label{derivativeDini}
f^{\p} (x) = - \frac{\de_x F \big( x , f(x) \big)}{\de_y F \big(x , f(x) \big)} \quad .
\eeq
\end{theorem}

Roughly speaking, the implicit function theorem states that, provided the conditions (\ref{conditionsDini}) are satisfied, a zero of a function of two real variables defines implicitly a functional relation between the two variables, at least locally. Furthermore, it says that this functional relation is regular, and gives an expression for the derivative of the function which links the two variables. Note that the conditions (\ref{conditionsDini}) are sufficient but not necessary for the existence of the ``implicit'' solution.

\subsection{The quintic equation and implicit functions}

The implicit function theorem is crucial for our analysis of the Vainshtein mechanism in massive gravity, since (at $\a$, $\b$ fixed) the equation which the field $h(\r)$ obeys (the quintic equation) is of the form $F(h(\r),\r) = 0$. Note that it is equivalent to work with $\r$ as a radial coordinate or with $x = \r/\r_v$, or $A = 1/x^3$, since all these coordinates are related by diffeomorphisms. If we work with the coordinate $A$, the solutions for the field $h(A)$ are then implicitly defined by the equation $q \, \big( h(A), A; \a, \b \big) = 0$, where the quintic function $q$ is defined in (\ref{quinticFunction}).

At $\a$ and $\b$ fixed, the function $q \, \big( h, A \big)$ is defined on $\mathbb{R} \times (0, + \infty)$ and is derivable an arbitrary number of times with continuous partial derivatives. Suppose that we find, at a certain $A = \bar{A}$ (\emph{i.e.} at a certain radius $\bar{\r} = \r_v/\sqrt[3]{\bar{A}}$ ), a root $\bar{h}$ of the equation $q_{\bar{A}} \, \big( h \big) = 0$, where $q_{A} \, \big( h \big)$ is the shape function \eqref{shapefunction}: the condition $\frac{\de F}{\de y}(x_0,y_0) \neq 0$ translates in this case to the fact that $\bar{h}$ is a \emph{simple root} of the equation $q_{\bar{A}} \, \big( h \big) = 0$. Therefore, if we find at a certain $A = \bar{A}$ a simple root $\bar{h}$ of the equation $q_{\bar{A}} \, \big( h \big) = 0$, then the conditions (\ref{conditionsDini}) are satisfied, and the implicit function theorem assures us that there exist a neighborhood of $\bar{A}$ (\emph{i.e.} a neighborhood of $\bar{\r}$) where there exists a solution $h(A)$ of the quintic equation such that $h(\bar{A}) = \bar{h}$.

\subsection{Maximal extension of implicitly def\mbox{}ined solutions}

Our aim in the end is to find \emph{global} solutions of the quintic equation, that is solutions $h(A)$ of the quintic equation which are defined for $A \in (0, + \infty)$. Therefore, it is important to establish when a local solution can be extended to the whole radial domain. Suppose we have a local solution $h(A)$ of the quintic equation defined on $(A_{i}, A_{f}) \subset (0, + \infty)$. If the conditions (\ref{conditionsDini}) are satisfied also at $A = A_{i}$ and $A = A_{f}$, we can extend the solution to an interval $(A^{(2)}_{i}, A^{(2)}_{f}) \supset (A_{i}, A_{f})$, and we can iterate this procedure. Therefore, we can extend the local solution until we reach a point $\tilde{A}$ where the conditions (\ref{conditionsDini}) are not both satisfied: this can happen only if one of the following conditions are true
\begin{enumerate}
 \item $\lim_{A \rightarrow \tilde{A}} \,\, \abs{h(A)} = + \infty$
 \item $\dfrac{\de q}{\de h}(\ti{h},\tilde{A}) = 0$
\end{enumerate}
where in the second case $\ti{h} \equiv \lim_{A \rightarrow \tilde{A}} \, h(A)$. However, it is possible to see that the first case cannot happen. In fact, suppose hypothetically that there exists a solution $h(A)$ of the quintic equation such that $\lim_{A \rightarrow \tilde{A}} \,\, \abs{h(A)} = + \infty$ with $\tilde{A}$ finite and non-zero. This means that, in the dual formulation, there is a solution $v(A)$ of the equation (\ref{vedremovedremo}) such that $\lim_{A \rightarrow \tilde{A}} \, v(A) = 0$, with $\tilde{A}$ finite and non-zero: this implies that $\lim_{A \rightarrow \tilde{A}} \, d \big( v(A), A; \a, \b \big) = \frac{3}{2} \, \bq \neq 0$, since we are considering the $\b \neq 0$ case. But, by the continuity of the function $d \big( v, A; \a, \b \big)$ and the fact that $d \big( v(A), A; \a, \b \big) = 0$ identically since $v(A)$ is a solution of (\ref{vedremovedremo}), we have that $\lim_{A \rightarrow \tilde{A}} \, d \big( v, A; \a, \b \big) = 0$. The hypothesis led us to a contradiction, so it follows that there cannot exist solutions $h(A)$ of the quintic equation such that $\lim_{A \rightarrow \tilde{A}} \,\, \abs{h(A)} = + \infty$ with $\tilde{A}$ finite and non-zero.

Therefore, a local solution $h(A)$ of the quintic equation can be extended until we meet a finite and non-zero $\tilde{A}$ where $\frac{\de q}{\de h}(\ti{h},\tilde{A}) = 0$ (with $\ti{h} \equiv \lim_{A \rightarrow \tilde{A}} \, h(A)$), or equivalently until we meet a finite and non-zero $\tilde{A}$ where the function $q_{A} \, \big( h; \a, \b \big)$ has a stationary point on the horizontal axis. Note that, when this happens, the derivative $\pr{h}(A)$ of the solution diverges as $A \rightarrow \tilde{A}$, as can be deduced from \eqref{derivativeDini}, while the solution $h(A)$ itself remains bounded.

\chapter{Useful properties of the quintic function}
\label{Useful properties of the quintic function}

We discuss here some important properties of the quintic function, which are useful for the analytic study of the solutions matching in chapter \ref{Vainshtein Mechanism in Massive Gravity}. Despite we study the quintic equation (\ref{quintic}) in the domain of definition $h \in (- \infty, + \infty)$, $A \in (0, + \infty)$, $\a \in (- \infty, + \infty)$ and $\b \in (- \infty, 0) \cup (0, + \infty)$, it is very useful to extend the domain of definition of $A$ to $A = 0$ as well, which corresponds to the asymptotic limit $\r \rightarrow + \infty$.

\section{General properties}

The quintic function and its derivatives reads explicitly

\begin{align}
q \, \big( h, A; \a, \b \big) &= \frac{3}{2} \, \bq \, \ha{5} - \big( \aq + 2 \b \big) \, \ha{3} + 3 \, \big( \a + \b A \big) \, \ha{2} - \frac{3}{2} \, h - A \label{quinticzeroderivative}\\[4mm]
q^{\p} \, \big( h, A; \a, \b \big) &= \frac{15}{2} \, \bq \, \ha{4} - 3 \, \big( \aq + 2 \b \big) \, \ha{2} + 6 \, \big( \a + \b A \big) \, h - \frac{3}{2} \label{quinticfirstderivative}\\[5mm]
q^{\p\p} \, \big( h, A; \a, \b \big) &= 30 \, \bq \, \ha{3} - 6 \, \big( \aq + 2 \b \big) \, h + 6 \, \big( \a + \b A \big) \label{quinticsecondderivative}\\[6mm]
q^{\p\p\p} \, \big( h, A; \a, \b \big) &= 90 \, \bq \, \ha{2} - 6 \, \big( \aq + 2 \b \big) \label{quinticthirdderivative}
\end{align} \\

\noindent where we indicated the derivatives with respect to $h$ with a prime $\phantom{a}^{\p}$. Note first of all that
\beq
\lim_{h \rightarrow + \infty} q \, \big( h, A; \a, \b \big) = + \infty \qquad , \qquad \lim_{h \rightarrow - \infty} q \, \big( h, A; \a, \b \big) = - \infty \quad ,
\eeq
and that
\begin{align}
q \, \big( 0, A; \a, \b \big) &= - A \leq 0 \\[3mm]
q^{\p} \, \big( 0, A; \a, \b \big) &= - \frac{3}{2} < 0 \\[3mm]
q^{\p\p} \, \big( 0, A; \a, \b \big) &= 6 \, \big( \a + \b A \big) \quad .
\end{align}

\noi Therefore, for the intermediate value theorem, there is always (for every value of $A$) a root of the quintic for $h \in (0, + \infty)$. In particular, if we take into account the multiplicity of the roots, there is always an odd number of real roots. Note that $q^{\p} \, \big( 0, A; \a, \b \big)$ is indipendent of $A$, while $q \, \big( 0, A; \a, \b \big)$ is linear and decreasing with respect to $A$. We may see the evolution with $A$ of the quintic as the sum of an overall rigid translation due to the constant term of the polynomial, and of a change of shape due to the contribution $3 \b A \, \ha{2}$ to the quadratic piece of the polynomial. 

\section{Evolution with $A$}

\subsection{The quintic function}

To study how the quintic function evolves with $A$, let's consider its partial derivative with respect to $A$. It is easy to verify that
\beq
\frac{\de q}{\de A}\big( h, A; \a, \b \big) = 3 \, \b \ha{2} - 1 \quad ,
\eeq
and this relation implies that, if $\b < 0$, we have
\beq
\b < 0 \qquad \Rightarrow \qquad \frac{\de q}{\de A}\big( h, A; \a, \b \big) < 0
\eeq
for every $h$, $A$ and $\a$. Therefore, if $\b < 0$, at every $h$ the value of the quintic function decreases monotonically when $A$ goes from $0$ to $+ \infty$. On the other hand, if  $\b > 0$ we have

\begin{equation}
\b > 0  \qquad \Rightarrow \qquad \left\{ \begin{aligned}
\quad \frac{\de q}{\de A}\big( h, A; \a, \b \big) &< 0 & \text{for} \quad \abs{h} &< \frac{1}{\sqrt{3 \b}} \\[2mm]
\quad \frac{\de q}{\de A}\big( h, A; \a, \b \big) &> 0 & \text{for} \quad \abs{h} &> \frac{1}{\sqrt{3 \b}} \\[2mm]
\quad \frac{\de q}{\de A}\big( h, A; \a, \b \big) &= 0 & \text{for} \quad \abs{h} &= \frac{1}{\sqrt{3 \b}} \quad ,
\end{aligned}
\right.
\end{equation}
and we conclude that, at every $h$ such that $-1/\sqrt{3 \b} < h < 1/\sqrt{3 \b}$, the value of the quintic function decreases monotonically when $A$ goes from $0$ to $+ \infty$, while it increases monotonically at every $h$ such that $h < -1/\sqrt{3 \b}$ or $h > 1/\sqrt{3 \b}$. Finally, there are two fixed points of the evolution of the quintic with $A$, which correspond to the following values for $h$
\beq
h = \pm \frac{1}{\sqrt{3 \b}} = \mathbf{F}_{\pm}
\eeq
which (as already indicated above) are precisely the limiting values of the finite inner solutions $\textbf{F}_{\pm}$.

\subsection{The f\mbox{}irst derivative}

Consider now the first derivative of the quintic $q^{\p} \, \big( h, A; \a, \b \big)$. We have 
\beq
\frac{\de q^{\p}}{\de A}\big( h, A; \a, \b \big) = 6 \, \b h \quad ,
\eeq
which implies that the only fixed point of the evolution of $q^{\p}$ corresponds to the value $h = 0$, and (as already mentioned) we have
\beq
q^{\p} \, \big( 0, A; \a, \b \big) = - \frac{3}{2}
\eeq
independently of $\a$ and $\b$. Furthermore, we have that
\begin{equation}
\b < 0  \qquad \Rightarrow \qquad \left\{ \begin{aligned}
\quad \frac{\de q^{\p}}{\de A}\big( h, A; \a, \b \big) &< 0 & \text{for} \quad h &> 0 \\[2mm]
\quad \frac{\de q^{\p}}{\de A}\big( h, A; \a, \b \big) &> 0 & \text{for} \quad h &< 0
\end{aligned}
\right.
\end{equation}
so, for $\b < 0$, at every fixed $h > 0$ the first derivative of the quintic decreases when $A$ goes from $0$ to $+ \infty$, while it increases at every fixed $h < 0$. Conversely, we have that
\begin{equation}
\b > 0  \qquad \Rightarrow \qquad \left\{ \begin{aligned}
\quad \frac{\de q^{\p}}{\de A}\big( h, A; \a, \b \big) &< 0 & \text{for} \quad h &< 0 \\[2mm]
\quad \frac{\de q^{\p}}{\de A}\big( h, A; \a, \b \big) &> 0 & \text{for} \quad h &> 0
\end{aligned}
\right.
\end{equation}
and so, for $\b < 0$, at every fixed $h > 0$ the first derivative of the quintic increases when $A$ goes from $0$ to $+ \infty$, while it decreases at every fixed $h < 0$.

\subsection{The second derivative}

For what concerns the second derivative of the quintic $q^{\p\p} \, \big( h, A; \a, \b \big)$, we have
\beq
\frac{\de q^{\p\p}}{\de A}\big( h, A; \a, \b \big) = 6 \, \b
\eeq
and this implies that there are no fixed points in the evolution with $A$ of $q^{\p\p}$. In fact, from \eqref{quinticsecondderivative} it is evident that $q^{\p\p} \, \big( h, A; \a, \b \big)$ translates rigidly when $A$ changes, and in particular translates towards $h \rightarrow + \infty$ when $\b > 0$ while  translates towards $h \rightarrow - \infty$ when $\b < 0$. Note that the value of $\a$ sets the value of the second derivative in $h = 0$ at $A = 0$
\beq
q^{\p\p} \, \big( 0, 0; \a, \b \big) = 6 \, \a \quad ,
\eeq
and that
\beq
\lim_{h \rightarrow + \infty} q^{\p\p} \, \big( h, A; \a, \b \big) = + \infty \qquad , \qquad \lim_{h \rightarrow - \infty} q^{\p\p} \, \big( h, A; \a, \b \big) = - \infty \quad .
\eeq
This implies that, for every value of $\a$ and $\b$ (still with $\b \neq 0$), there is always a critical value $A_{crit}(\a,\b)$ such that the second derivative $q^{\p\p} \, \big( h, A; \a, \b \big)$ has one and only one root for $A > A_{crit}(\a,\b)$. This root is negative when $\b$ is positive, and conversely is positive when $\b$ is negative. Therefore, for $A > A_{crit}(\a,\b)$, the quintic has zero inflection points for $h > 0$ and one inflection point for $h < 0$ in the case $\b > 0$, while has one inflection point for $h > 0$ and zero inflection points for $h < 0$ in the case $\b < 0$. Roughly speaking, this critical value for $A$ can be regarded as the value after which there cannot be anymore creations and annihilations of local solutions.

Note that, since the second derivative $q^{\p\p} \, \big( h, A; \a, \b \big)$ translates rigidly when $A$ changes, it is very useful to characterize completely its shape at infinity (\emph{i.e.} at $A = 0$) for every value of $\a$ and $\b$ in the phase space.

\chapter{Asymptotic structure of the quintic function}
\label{Asymptotic structure of the quintic function}

In this and in the next appendix, we summarize the main properties of the quintic function \eqref{quinticFunction} when $A = 0$, which corresponds to the asymptotic limit $\r \rightarrow + \infty$. In these appendices, when we say that a function has some property at infinity we mean at radial infinity, \emph{i.e.} at $A= 0$.

As we mentioned above, for $A = 0$ the quintic function reduces to the asymptotic function
\beq
\label{asymptotic function}
\mathscr{A} \big( h ; \a, \b \big) = \frac{3}{2} \, \bq \, h^{5} - \big( \aq + 2 \b \big) \, h^{3} + 3 \, \a \, h^{2}
- \frac{3}{2} \, h
\eeq
which can be factorized as
\beq
\label{factorization}
\mathscr{A}\big( h ; \a, \b \big) = h \,\, \mathscr{A}_{r}\big( h ; \a, \b \big)
\eeq
where the function $\mathscr{A}_{r}\big( h ; \a, \b \big)$ is called the reduced asymptotic function and reads
\beq
\label{red asymptotic function}
\mathscr{A}_{r}\big( h ; \a, \b \big) = \frac{3}{2} \, \bq \, h^{4} - \big( \aq + 2 \b \big) \, h^{2}
+ 3 \, \a \, h - \frac{3}{2} \quad .
\eeq
Note that, as a consequence of the symmetry (\ref{symmetry}) of the quintic function, the asymptotic function has the following symmetry
\beq
\label{asymptoticsymmetry}
\mathscr{A} \, \Big( \frac{h}{k} ; k \, \a, k^2 \b \Big) = \frac{1}{k} \, \mathscr{A} \, \big( h; \a, \b \big)
\eeq
which, differently from the symmetry (\ref{symmetry}), holds also for $k < 0$. Therefore, we may restrict the study of the asymptotic function only to the semi-plane $\a > 0$.

\section{Study of the second derivative}

In order to study analytically the matching of solutions, it is very important to establish how many inflection points the quintic function has at infinity, and where they are located in relation to the fixed points of the quintic.

\subsection{Inf\mbox{}lection points at inf\mbox{}inity}

The second derivative of the quintic at $A = 0$ is equal to the second derivative of the asymptotic function which reads
\beq
\label{asymptoticfunctionsecondderivative}
\mathscr{A}^{\p\p} \big( h ; \a, \b \big) = 30 \, \bq \, \ha{3} - 6 \, \big( \aq + 2 \b \big) \, h + 6 \, \a \quad .
\eeq
To find the number of roots of $\mathscr{A}^{\p\p}$, it is enough to study just the case $\a > 0$, since the symmetry \eqref{asymptoticsymmetry} implies that the number of roots at $(-\a, \b)$ and at $(\a, \b)$ are equal. Considering then the case $\a > 0$, the function $\mathscr{A}^{\p\p} \big( h ; \a, \b \big)$ has the following properties
\beq
\label{asymptoticfunctionsecondderivativeproperties}
\lim_{h \rightarrow - \infty} \mathscr{A}^{\p\p} \big( h ; \a, \b \big) = - \infty \qquad
\mathscr{A}^{\p\p} \big( 0 ; \a, \b \big) > 0 \qquad \lim_{h \rightarrow + \infty} \mathscr{A}^{\p\p} \big( h ; \a, \b \big) = + \infty \quad ,
\eeq
so for the intermediate value theorem there is always a root of $\mathscr{A}^{\p\p} \big( h ; \a, \b \big)$ for $h < 0$, which we call $r_0$. To understand if there are other roots, it is useful to study its first derivative
\beq
\label{asymptoticfunctionthirdderivative}
\mathscr{A}^{\p\p\p} \big( h ; \a, \b \big) = 90 \, \bq \, \ha{2} - 6 \, \big( \aq + 2 \b \big) \quad :
\eeq
it is easy to check that the quadratic equation $\mathscr{A}^{\p\p\p} \big( h ; \a, \b \big) = 0$ admits solutions only if
\beq
\b \geq - \half \, \aq \quad ,
\eeq
in which case the roots are
\beq
\label{rootsthirdderivativeasymp}
h_{\pm} = \pm \frac{\sqrt{\aq + 2 \b}}{\sqrt{15} \, \abs{\b}} \quad .
\eeq
Therefore, for $\b \leq - (1/2) \, \aq$ the function $\mathscr{A}^{\p\p\p} \big( h ; \a, \b \big)$ is positive for all values of $h$, and the function $\mathscr{A}^{\p\p} \big( h ; \a, \b \big)$ is monotonically increasing. On the other hand, for $\b > - (1/2) \aq$ the function $\mathscr{A}^{\p\p} \big( h ; \a, \b \big)$ has a relative minimum at $h = h_+$ and a relative maximum at $h = h_-$. The number of roots of the equation $\mathscr{A}^{\p\p} \big( h ; \a, \b \big) = 0$ is determined by the fact that $\mathscr{A}^{\p\p} \big( h_+ ; \a, \b \big)$ is positive or negative: if it is positive, then the equation $\mathscr{A}^{\p\p} = 0$ has only one root (which has negative value), while if it is negative the equation $\mathscr{A}^{\p\p} = 0$ has three roots (one root which has negative value and two roots, $r_1$ and $r_2$, which have positive values). The phase space boundaries between the regions where $\mathscr{A}^{\p\p} = 0$ has three roots and the regions where $\mathscr{A}^{\p\p} = 0$ has one root are defined by the condition $\mathscr{A}^{\p\p} \big( h_+ ; \a, \b \big) = 0$: in this case, the equation $\mathscr{A}^{\p\p} = 0$ has two roots, one simple root and one double root. The condition $\mathscr{A}^{\p\p} \big( h_+ ; \a, \b \big) = 0$ is equivalent to the following condition on $y = \b/\aq$
\beq
8 \, y^3 - \frac{87}{4} \, y^2 + 6 \, y + 1 = 0 \quad :
\eeq
this equation is a cubic and has positive discriminant, therefore has three real roots whose approximated values are $y_1 = in_1 \simeq -0.115898$,  $y_2 = in_2 \simeq 0.452816$ and $y_3 = in_3 \simeq 2.38183$. It can be checked that for $- 0.5 \, \aq < \b < in_1 \, \aq$ and for $in_2 \, \aq < \b < in_3 \, \aq$ we have $\mathscr{A}^{\p\p} \big( h_+ ; \a, \b \big) > 0$, while for $in_1 \, \aq < \b < 0$, $0 < \b < in_2 \, \aq$ and $\b > in_3 \, \aq$ we have $\mathscr{A}^{\p\p} \big( h_+ ; \a, \b \big) < 0$.

\begin{figure}[htp!]
\begin{center}
\includegraphics[width=10cm]{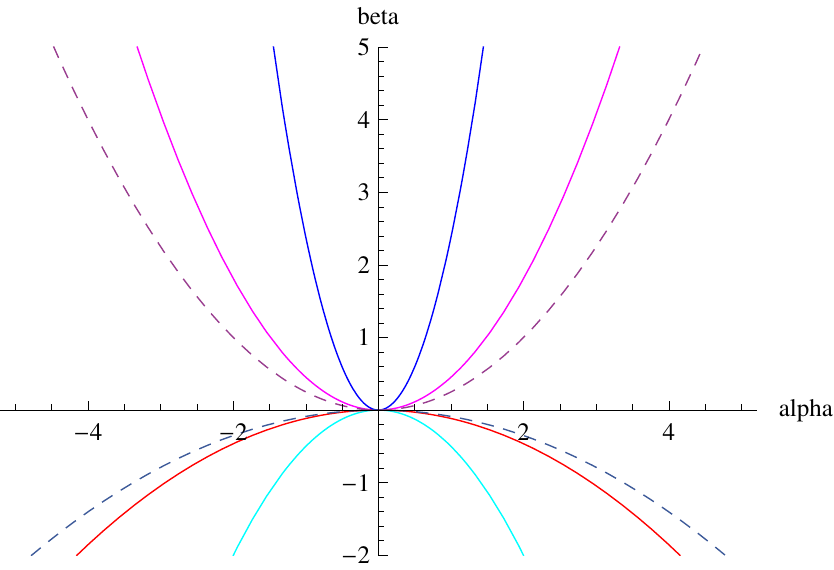}
\caption[Inflection points at infinity parabolas]{Inflection points at infinity and five roots parabolas}
\label{inflection points at infinity}
\end{center}
\end{figure}

Therefore, for $\b < \aq$ the function $\mathscr{A}^{\p\p} \big( h ; \a, \b \big)$ is monotonic and the quintic function has one inflection point at infinity. For $\b > \aq$ the function $\mathscr{A}^{\p\p} \big( h ; \a, \b \big)$ is not monotonic, and:
\begin{itemize}
 \item for $- 0.5 \, \aq < \b < in_1 \, \aq$ the quintic function has one inflection point at infinity;
 \item for $in_1 \, \aq < \b < 0$ and for $0 < \b < in_2 \, \aq$ the quintic function has three inflection points at infinity;
 \item for $in_2 \, \aq < \b < in_3 \, \aq$ the quintic function has one inflection point at infinity;
 \item for $\b > in_3 \, \aq$ the quintic function has three inflection points at infinity.
\end{itemize}
This is summarized in figure \ref{inflection points at infinity}, where the parabolas $\b = -0.5 \, \aq$, $\b = in_1 \, \aq$, $\b = in_2 \, \aq$ and $\b = in_3 \, \aq$ are displayed together with the five-roots-at-infinity parabolas (which are the dashed curves).

\subsection{Inf\mbox{}lection points and f\mbox{}ixed points}

To study analytically the matching of solutions, it is useful to know if the inflection points of the asymptotic function are located at a value of $h$ which is larger or smaller than the fixed points $h = \mathbf{F}_{\pm}$. We consider here only the case $\a > 0$ and $\b > 0$, since $\mathbf{F}_{\pm}$ are defined only for $\b$ positive. 

Let's consider first the negative root $r_0$. The properties \eqref{asymptoticfunctionsecondderivativeproperties} imply that $\mathscr{A}^{\p\p}$ is negative for $\b < r_0$, while is positive for $r_0 < \b < 0$: therefore, we have that if $\mathscr{A}^{\p\p} \big( \mathbf{F}_{-} ; \a, \b \big) < 0$, then we have $\mathbf{F}_{-} < r_0$, while if $\mathscr{A}^{\p\p} \big( \mathbf{F}_{-} ; \a, \b \big) > 0$ we have $\mathbf{F}_{-} > r_0$. Indicating $z = \sqrt{\b}/\a$, we have explicitly
\beq
\mathscr{A}^{\p\p} \big( \mathbf{F}_{-} ; \a, \b \big) = 2 \frac{\aq}{\sqrt{3 \b}} \, \big( z^2 +3 \sqrt{3} z + 3 \big) \quad ,
\eeq
and the roots of the quadratic equation $z^2 +3 \sqrt{3} z + 3 = 0$ are both negative. Therefore, for $\a > 0$ we have $\mathscr{A}^{\p\p} \big( \mathbf{F}_{-} ; \a, \b \big) > 0$, which implies that $r_0 < \mathbf{F}_{-}$.

Let's consider now the positive roots $r_1$ and $r_2$, and let's introduce the convention $r_1 < r_2$. The properties \eqref{asymptoticfunctionsecondderivativeproperties} imply that $\mathscr{A}^{\p\p}$ is positive for $0 < \b < r_1$ and $\b > r_2$, while is negative for $r_1 < \b < r_2$: therefore, we have that if $\mathscr{A}^{\p\p} \big( \mathbf{F}_{+} ; \a, \b \big) < 0$, then $r_1 < \mathbf{F}_{+} < r_2$. On the other hand, if $\mathscr{A}^{\p\p} \big( \mathbf{F}_{+} ; \a, \b \big) > 0$ it follows that either $\mathbf{F}_{+} < r_1 < r_2$ or $r_1 < r_2 < \mathbf{F}_{+}$: in particular, we have that if $\mathscr{A}^{\p\p\p} \big( \mathbf{F}_{+} ; \a, \b \big) < 0$ then $\mathbf{F}_{+} < r_1 < r_2$, while if $\mathscr{A}^{\p\p\p} \big( \mathbf{F}_{+} ; \a, \b \big) > 0$ we have $r_1 < r_2 < \mathbf{F}_{+}$. Still indicating $z = \sqrt{\b}/\a$, we have explicitly
\beq
\mathscr{A}^{\p\p} \big( \mathbf{F}_{+} ; \a, \b \big) = -2 \frac{\aq}{\sqrt{3 \b}} \, \big( z^2 -3 \sqrt{3} z + 3 \big) \quad ,
\eeq
and the roots of the quadratic equation $z^2 -3 \sqrt{3} z + 3 = 0$ are
\beq
z_{12} = \frac{\sqrt{3}}{2} \, \big( 3 \pm \sqrt{5} \big) \quad .
\eeq
Defining $k_1 = (3/4) \, \big( 3 - \sqrt{5} \big)^2$ and $k_2 = (3/4) \, \big( 3 + \sqrt{5} \big)^2$, we have

\begin{equation}
\mathscr{A}^{\p\p} \big( \mathbf{F}_{+} ; \a, \b \big) \,: \qquad \left\{ \begin{aligned}
\phantom{B} &< 0 & \text{for}& \quad 0 < \b < k_1 \aq \,\,\, \text{and} \,\,\, \b > k_2 \aq \\[2mm]
\phantom{B} &> 0 & \text{for}& \quad k_1 \aq < \b < k_2 \aq \quad ,
\end{aligned}
\right.
\end{equation}
where $k_1$ and $k_2$ have the approximate values $k_1 \simeq 0.437694$ and $k_2 \simeq 20.5623$. Furthermore, we have
\beq
\mathscr{A}^{\p\p\p} \big( \mathbf{F}_{+} ; \a, \b \big) = 6 \, \big( 3 \b - \aq \big) \quad ,
\eeq
and so
\begin{equation}
\mathscr{A}^{\p\p\p} \big( \mathbf{F}_{+} ; \a, \b \big) \qquad : \qquad \left\{ \begin{aligned}
\phantom{B} &< 0 & \text{for} \quad 0 &< \b < \frac{1}{3}\, \aq \\[2mm]
\phantom{B} &> 0 & \text{for} \quad \b &> \frac{1}{3}\, \aq \quad .
\end{aligned}
\right.
\end{equation}
We can then conclude that
\begin{itemize}
 \item for $0 < \b < k_1 \, \aq$ we have the ordering $r_1 < \mathbf{F}_{+} < r_2$;
 \item for $k_1 \, \aq < \b < in_2 \, \aq$ we have the ordering $r_1 < r_2 < \mathbf{F}_{+}$;
 \item for $in_2 \, \aq < \b < in_3 \, \aq$ there are no inflection points for $h >0$;
 \item for $in_3 \, \aq < \b < k_2 \, \aq$ we have the ordering $r_1 < r_2 < \mathbf{F}_{+}$;
 \item for $\b > k_2 \, \aq$ we have the ordering $r_1 < \mathbf{F}_{+} < r_2$.
\end{itemize}

\chapter{Roots at inf\mbox{}inity}
\label{Roots at infinity}

We continue the summary started in the previous appendix about the main properties of the quintic function \eqref{quinticFunction} when $A = 0$, which corresponds to the asymptotic limit $\r \rightarrow + \infty$. We want to study here how many zeros the asymptotic function has, in relation to the value of $\a$ and $\b$.

\section{Zeros of the asymptotic function}

The asymptotic function \eqref{asymptotic function} is a quintic, and therefore can have at most five real zeros. As we explained in section \ref{Asymptotic and inner solutions}, $h = 0$ is always a zero, and in fact a simple one\footnote{As we mentioned in section \ref{Local solutions and the shape of the quintic}, we say that $y$ is a simple/double zero of a function $f$ if $y$ is a simple/double root of the equation $f = 0$}. From the factorization (\ref{factorization}) it follows that, to find the other zeros of the asymptotic function, we can study the zeros of the reduced asymptotic function $\mathscr{A}_{r}\big( h ; \a, \b \big)$
\beq
\label{reducedasymptoticfunction}
\mathscr{A}_{r}\big( h ; \a, \b \big) = \frac{3}{2} \, \bq \, h^{4} - \big( \aq + 2 \b \big) \, h^{2}
+ 3 \, \a \, h - \frac{3}{2} \quad .
\eeq
This function (see section \ref{Asymptotic and inner solutions}) has always two zeros, one positive and one negative, and can have up to 4 real zeros, depending on the specif\mbox{}ic values of $\a$ and $\b$.

\subsection{Five-roots-at-inf\mbox{}inity boundaries}

The regions where the asymptotic functions has five zeros, if they exist, have to be inside the regions where there are three inflection points at infinity, since it is impossible to have five zeros and just one or two inflection points. Since the function $\mathscr{A}\big( h ; \a, \b \big)$ changes smoothly with $\a$ and $\b$, the boundaries between regions where there are f\mbox{}ive zeros and regions where there are three zeros are found enforcing that $\mathscr{A}\big( h ; \a, \b \big) = 0$ has a multiple root. In this case the asymptotic function has to have a stationary point on the horizontal axis, and so if $h$ is the multiple root then we have $\mathscr{A}\big( h ; \a, \b \big) = \mathscr{A}^{\p}\big( h ; \a, \b \big) = 0$. Asking that this condition is satisfied for some $h$ and solving this condition with the software Mathematica, we get that the asymptotic function has a stationary point on the horizontal axis only if $\b = c_{+} \, \a^2$ and $\b = c_{-} \, \a^2$, where $c_{+} = 1/4$ and $c_{-}$ is the only real root of the equation $8 + 48 \, y - 435 \, y^2 + 676 \, y^3 = 0$ which has the approximate value $c_{-} \simeq - 0.0876193$. 
%
%
The regions above the positive parabola and below the negative one have only three zeros, which are simple zeros, while the regions between the two parabolas (except $\b = 0$) have f\mbox{}ive zeros, which are again simple zeros. On the boundaries $\b = c_{\pm} \, \a^2$ between the three-zeros regions and the f\mbox{}ive-zeros regions there are four zeros, one of which is a zero of multiplicity two. Note that this result implies that for $\b > in_3 \, \aq$, where in principle there could be five zeros (since there are three inflection points), there are nevertheless only three zeros. This is summarized in figure \ref{five roots}.

\begin{figure}[htp!]%
\begin{center}%
\parbox{4.7cm}{\includegraphics[width=4.7cm]{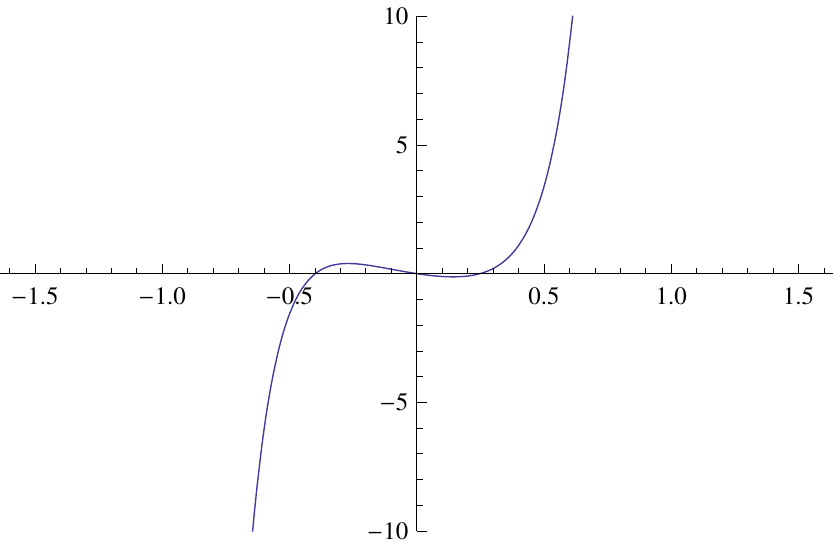}}%
\hspace{2mm}%
\parbox{4.7cm}{\includegraphics[width=4.7cm]{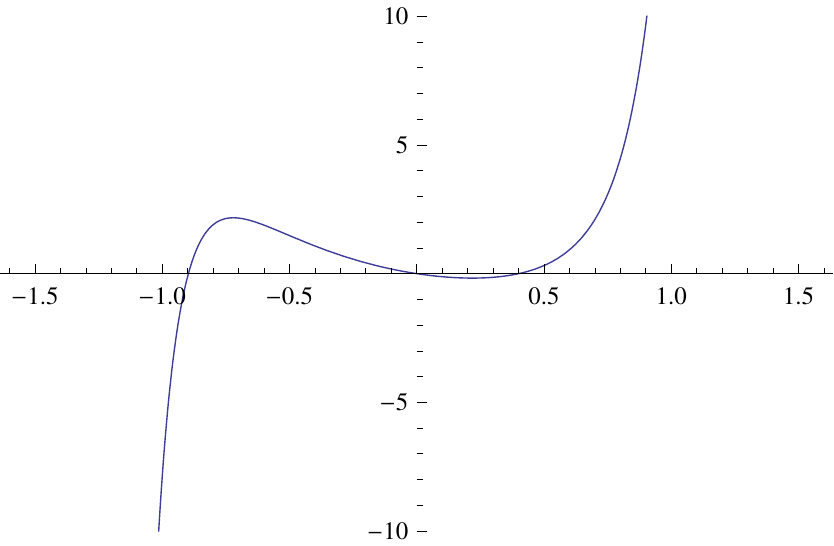}}%
\hspace{2mm}%
\parbox{4.7cm}{\includegraphics[width=4.7cm]{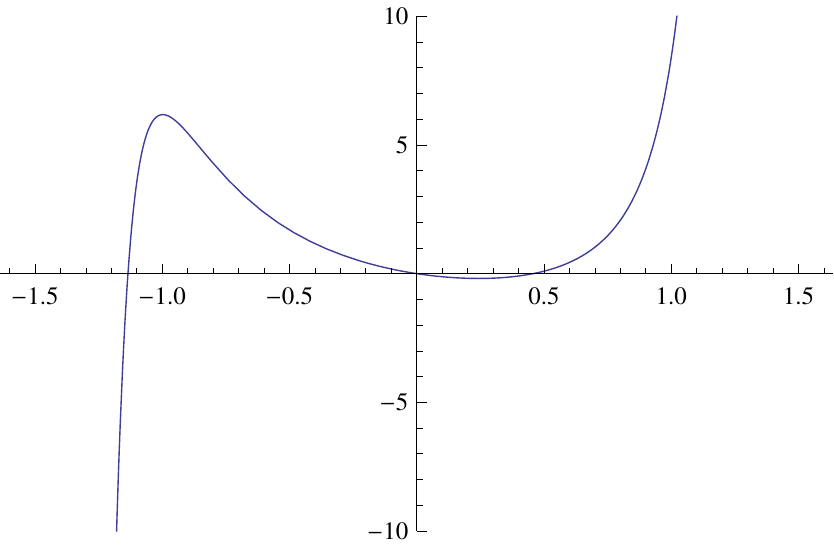}}%
\hspace{2mm}%
\parbox{4.7cm}{\includegraphics[width=4.7cm]{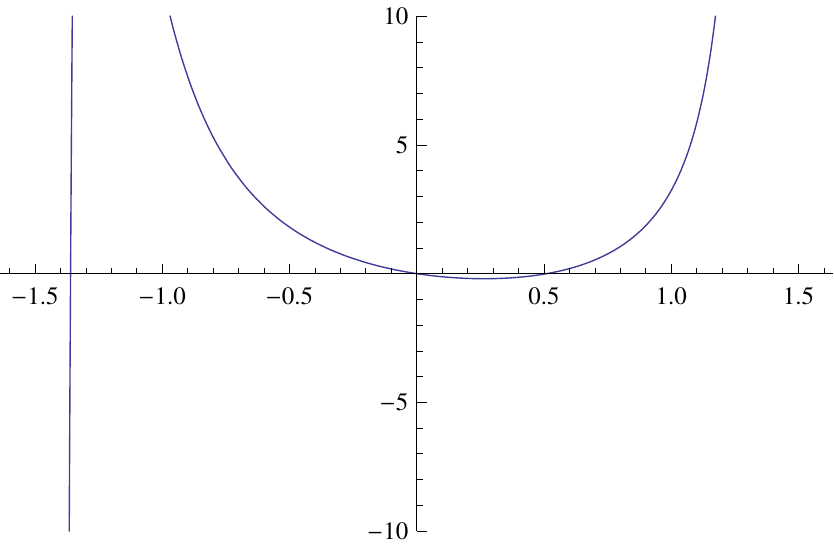}}%
\hspace{2mm}%
\parbox{4.7cm}{\includegraphics[width=4.7cm]{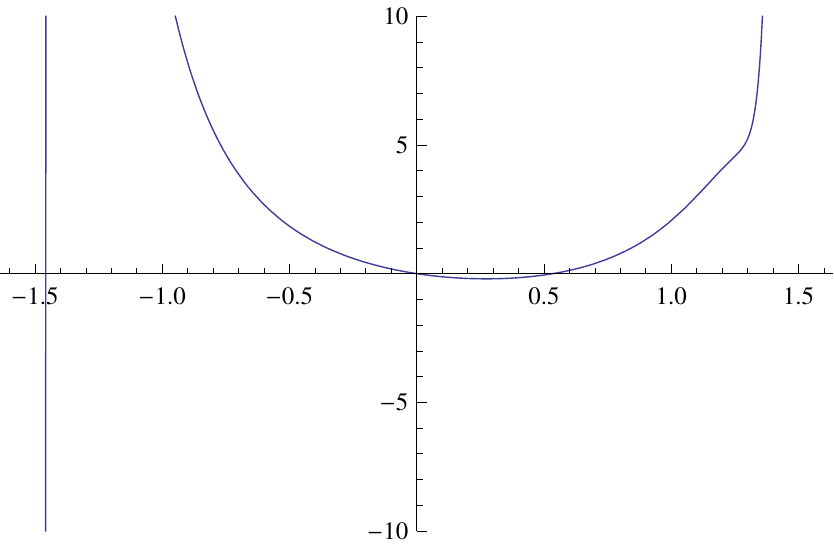}}%
\hspace{2mm}%
\parbox{4.7cm}{\includegraphics[width=4.7cm]{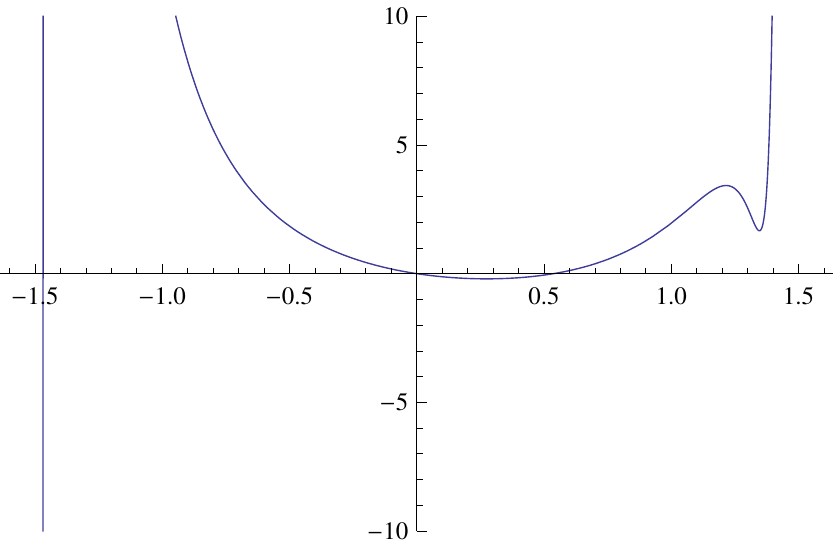}}%
\hspace{2mm}%
\parbox{4.7cm}{\includegraphics[width=4.7cm]{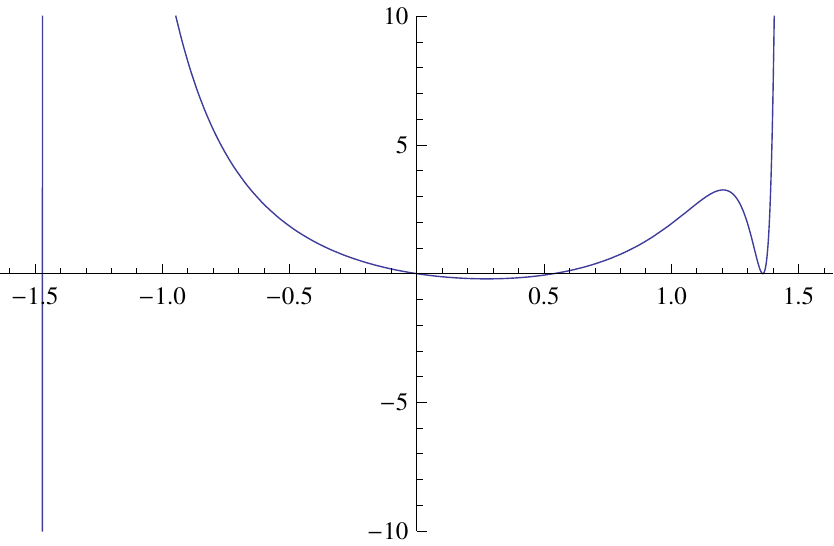}}%
\hspace{2mm}%
\parbox{4.7cm}{\includegraphics[width=4.7cm]{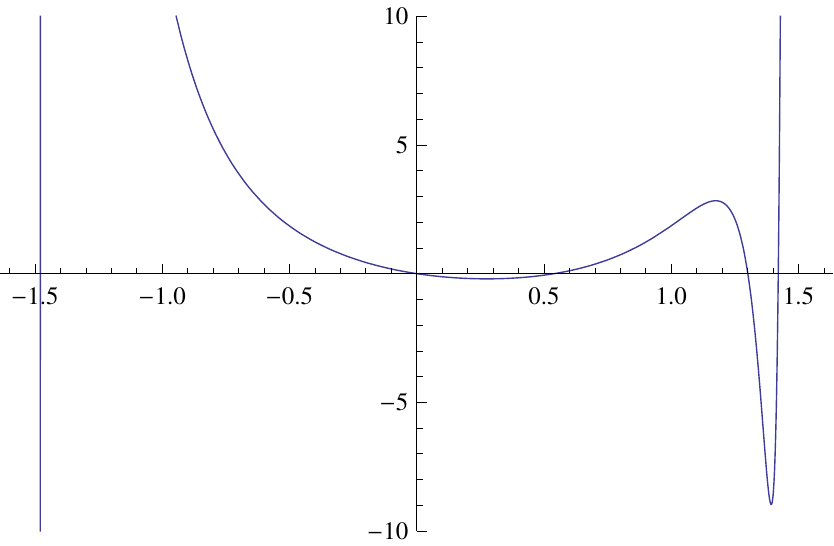}}%
\hspace{2mm}%
\parbox{4.7cm}{\includegraphics[width=4.7cm]{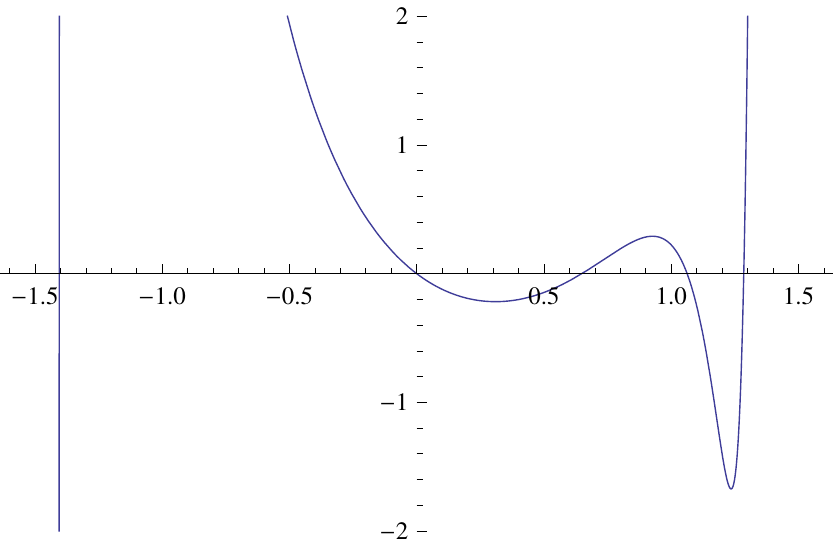}}%
\hspace{2mm}%
\parbox{4.7cm}{\includegraphics[width=4.7cm]{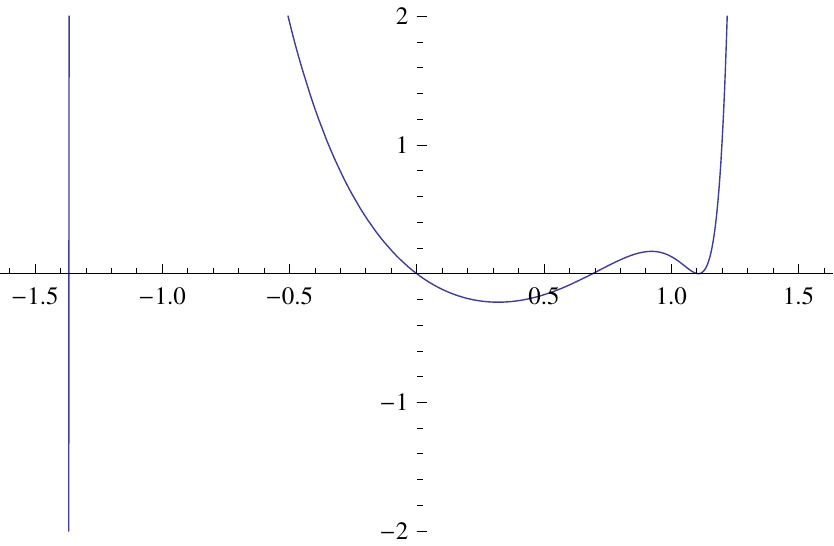}}%
\hspace{2mm}%
\parbox{4.7cm}{\includegraphics[width=4.7cm]{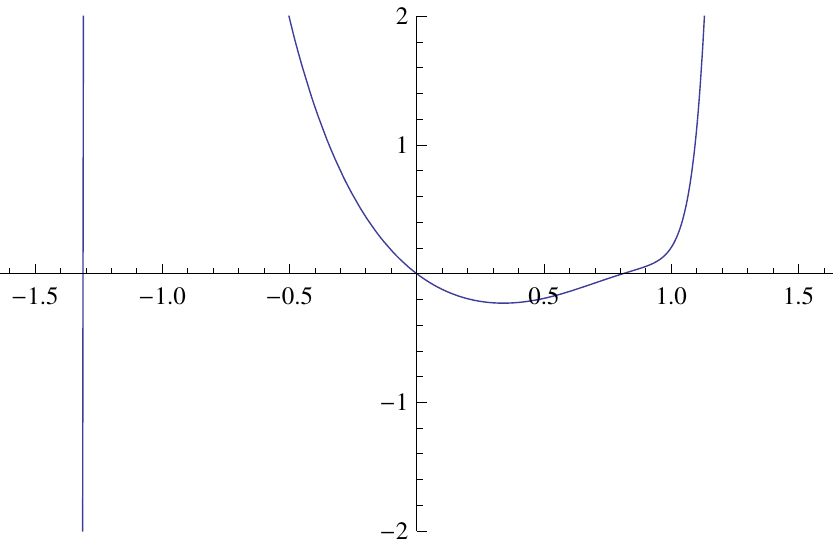}}%
\hspace{2mm}%
\parbox{4.7cm}{\includegraphics[width=4.7cm]{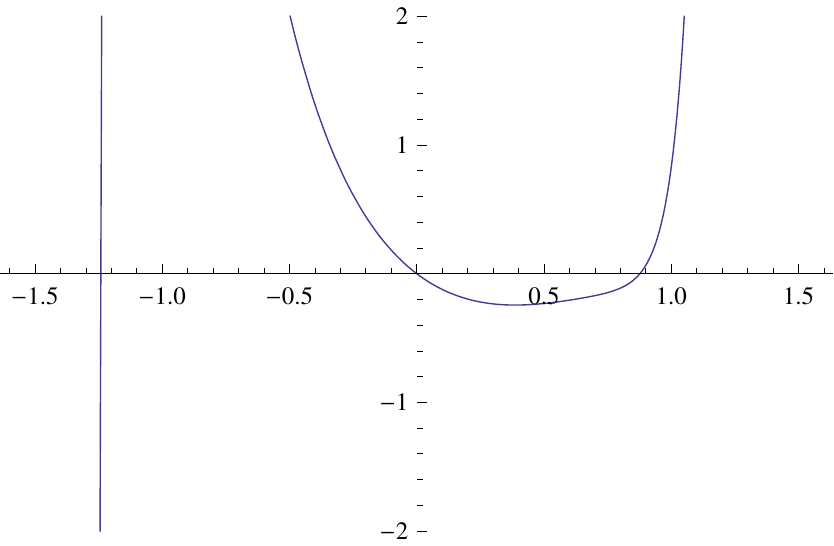}}%
\hspace{2mm}%
\parbox{4.7cm}{\includegraphics[width=4.7cm]{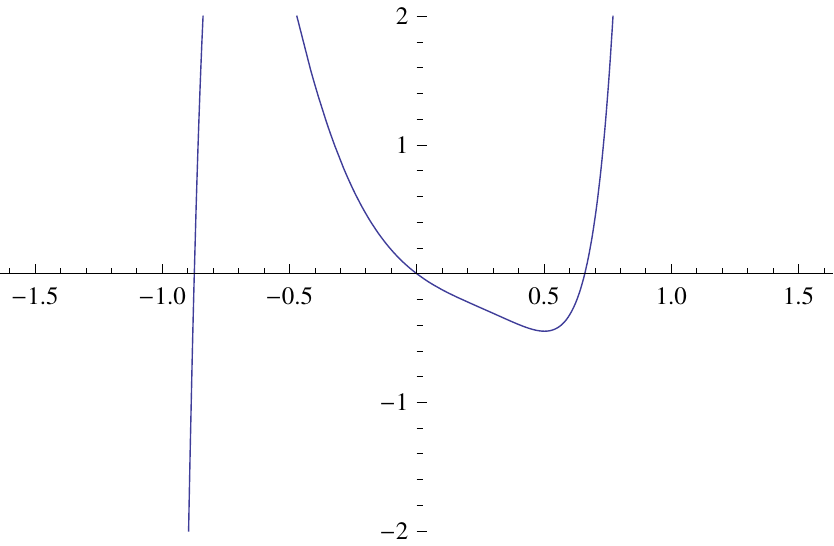}}%
\hspace{2mm}%
\parbox{4.7cm}{\includegraphics[width=4.7cm]{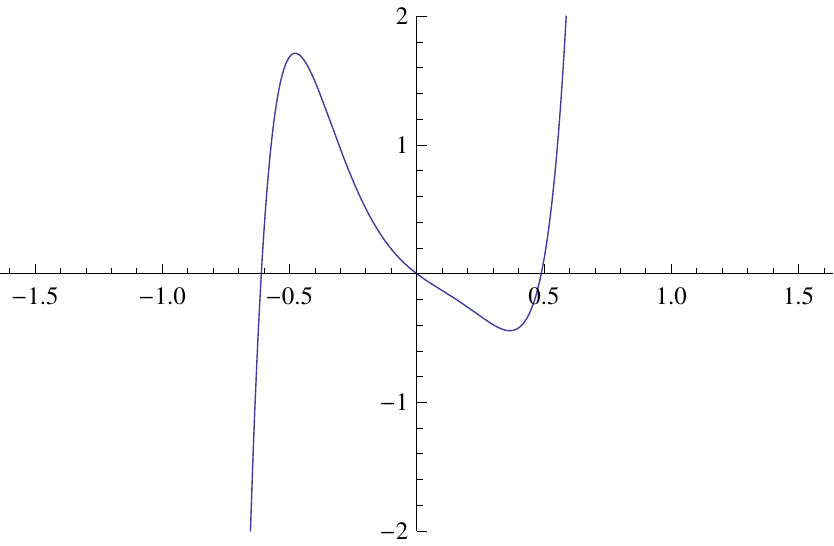}}%
\hspace{2mm}%
\parbox{4.7cm}{\includegraphics[width=4.7cm]{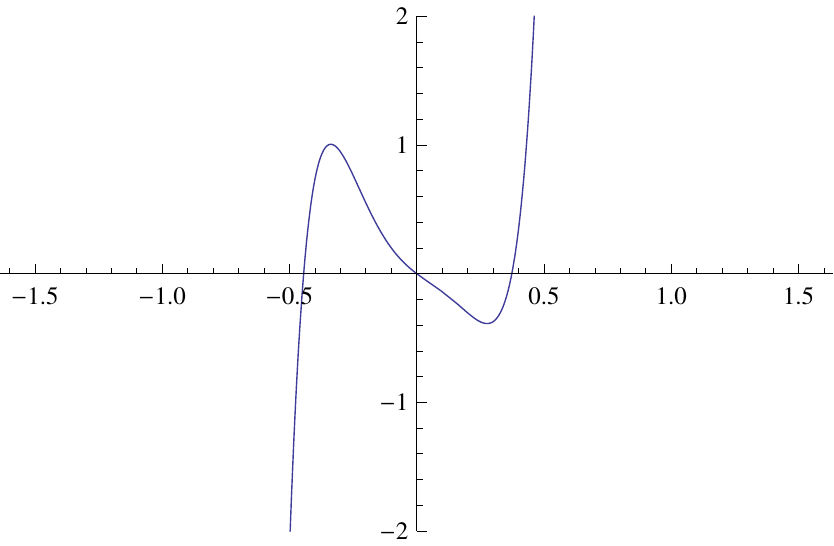}}%
\end{center}%
\label{asymptoticfunctionsaposbvar}
\caption{Asymptotic function at $\a = 1$ for increasing values of $\b$.}%
\end{figure}

These findings have been verified plotting the asymptotic function for many values of $\a$ and $\b$. Note that, because of the symmetry (\ref{asymptoticsymmetry}), we can set $\a = 1$ and vary only the parameter $\b$. In figure (\ref{asymptoticfunctionsaposbvar}) we plot the asymptotic function for $\a = 1$ and increasing values of this parameter: because of space constraints, we plot the function only for fifteen values of $\b$, and precisely for $\b = -5 \, , \, \b = -1 \, , \, \b = -0.5 \, , \, \b = -0.2 \, , \, \b = -0.1 \, , \, \b = -0.09 \, , \, \b = c_{-} \, , \, \b = -0.08 \, , \, \b = 0.19 \, , \, \b = c_{+} \, , \, \b = 0.38 \, , \, \b = 0.5 \, , \, \b = 2 \, , \, \b = 5 \, , \, \b = 10 \, $. For the sake of precision, as already mentioned we don't plot the function itself but its composition with the tangent function, since this compactifies the real axis into the interval $(-\pi/2, +\pi/2)$ and at the same time does not change the number and the relative position of the zeros.

\chapter{Leading behaviors}
\label{Leading behaviors}

In this appendix we study the leading behaviors of the inner and asymptotic solutions. As previously mentioned we consider only the $\b \neq 0$ case.

\section{Finite asymptotic and inner solutions}

For the f\mbox{}inite inner solutions $\textbf{F}_{\pm}$ and f\mbox{}inite non-zero asymptotic solutions $\textbf{C}_{\pm}$ and $\textbf{P}_{1,2}$, the behavior is
\beq
h(\r) = C + R(\r)
\eeq
where $C \neq 0$ is their limiting value, and $R$ is respectively such that $\lim_{\r \rightarrow 0^+}
R = 0$ (inner solutions) and $\lim_{\r \rightarrow + \infty} R = 0$ (asymptotic solutions).

\section{Asymptotic decaying solution $\mathbf{L}$}

Let's consider the solution $\mathbf{L}$, which satisf\mbox{}ies $\lim_{\r \rightarrow +\infty} h(\r) = 0$. Dividing the quintic equation (\ref{quintic}) by $h$, we get
\beq
\frac{3}{2} \, \bq \, \ha{4} - \big( \aq + 2 \b \big) \, \ha{2}
+ 3 \, \a \, h - \frac{3}{2}  = \bigg( \frac{\r_v}{\r} \bigg)^{\! 3} \, \bigg( \frac{1}{h} - 3 \, \b \, h \bigg) \quad .
\eeq
The left hand side has a f\mbox{}inite limit when $\r \rightarrow +\infty$, so the same has to hold for the right hand
side: taking this limit in the equation above gives
\beq
\lim_{\r \rightarrow +\infty} \bigg( \frac{\r_v}{\r} \bigg)^{\! 3} \, \frac{1}{h} = - \frac{3}{2} \quad ,
\eeq
which implies that
\beq
h(\r) = - \frac{2}{3} \, \bigg( \frac{\r_v}{\r} \bigg)^{\! 3} + R(\r)
\eeq
with $\lim_{\r \rightarrow +\infty} \r^3 R(\r) = 0$.

\section{Inner diverging solution $\mathbf{D}$}

Let's consider now the solution $\mathbf{D}$, which satisf\mbox{}ies $\lim_{\r \rightarrow 0^+} \abs{h(\r)} = +\infty$. Dividing the equation (\ref{quinticv}) by $v^3$, one f\mbox{}inds that
\beq
\label{divergingv}
\va{2} - 3 \, \b = \bigg( \frac{\r}{\r_v} \bigg)^{\! 3} \, \frac{1}{\va{3}} \, \Big( - \frac{3}{2} \, \va{4}
+ 3 \, \a \, \va{3} - \big( \aq + 2 \b \big) \, \va{2} + \frac{3}{2} \, \bq \Big) \quad .
\eeq
One more time, the left hand side has a f\mbox{}inite limit when $\r \rightarrow 0^+$, so the same should hold for the right hand side. Therefore, the $\rho \rightarrow 0^+$ limit in the equation above gives
\beq
\lim_{\r \rightarrow 0^+} \bigg( \frac{\r}{\r_v} \bigg)^{\! 3} \, \frac{1}{\va{3}} = - \frac{2}{\b} \quad ,
\eeq
and so
\beq
\label{vrest}
v(\r) = - \sqrt[3]{\frac{\b}{2}} \, \frac{\r}{\r_v} + \mathrm{R}(\r)
\eeq
with $\lim_{\r \rightarrow 0^+} \mathrm{R}(\r)/\r = 0$. To understand the behavior of the gravitational potentials (\ref{solfab})-(\ref{solnab}) in this case, it is useful to calculate the next to leading order behavior. In fact, it turns out that, after going back to $h = 1/ v$, the leading behavior precisely cancels the Schwarzschild-like contribution, so to understand if the gravitational potentials are f\mbox{}inite at the origin it is essential to know how $\mathrm{R}$ behaves for very small radii. Inserting (\ref{vrest}) into (\ref{quinticv}) and dividing by $x^5$, one obtains taking the limit $\r \rightarrow 0^+$ that
\beq
\lim_{\r \rightarrow 0^+} \frac{\mathrm{R}}{x^3} = \frac{1}{9 \, \b} \bigg( \aq + \frac{3}{2} \, \b \bigg) \quad ,
\eeq
where $x = \r / \r_v$. We have then
\beq
v(\r) = - \sqrt[3]{\frac{\b}{2}} \, \frac{\r}{\r_v} + \mathcal{N} \, \Big( \frac{\r}{\r_v} \Big)^{\! 3} + \mathcal{R}(\r) \quad ,
\eeq
where
\beq
\mathcal{N} = \frac{1}{9 \, \b} \, \Big( \aq + \frac{3}{2} \, \b \Big)
\eeq
and $\lim_{\r \rightarrow 0^+} ( \mathcal{R}(\r) / \r^3 ) = 0$. Finally, going back to the function $h$ we get
\beq
h(\r) =  - \sqrt[3]{\frac{2}{\b}} \, \frac{\r_v}{\r} - \mathcal{M} \, \frac{\r}{\r_v} + \mathscr{R}(\r) \quad ,
\eeq
where
\beq
\mathcal{M} = \frac{1}{9} \, \sqrt[3]{\frac{4}{\b^5}} \, \bigg( \aq + \frac{3}{2} \, \b \bigg)
\eeq
and $\lim_{\r \rightarrow 0^+} (\mathscr{R}(\r)/\r) = 0$. It can be shown that in the special case $\aq + 3 \, \b / 2 = 0$, the next to leading order term scales as $\r^2$ instead of $\r$, and that $\lim_{\r \rightarrow 0^+} (\mathscr{R}(\r)/\r^2) = 0$.

Therefore, we can conclude that in general the diverging inner solution $\textbf{D}$ is such that
\beq
h(\r) =  - \sqrt[3]{\frac{2}{\b}} \, \frac{\r_v}{\r} + R(\r) \quad ,
\eeq
where $\lim_{\r \rightarrow 0^+} (R(\r)/\r)$ is f\mbox{}inite (zero in the special case $\aq + 3 \, \b / 2 = 0$).

\clearemptydoublepage

\pagestyle{fancyplain}
\renewcommand{\chaptermark}[1]{\markboth{#1}{}}
\renewcommand{\sectionmark}[1]{\markright{\thesection\ #1}}
\lhead[\fancyplain{}{\rm\thepage} \mbox{  } \mbox{  } \mbox{  } \sl{Bibliography} ]{\fancyplain{}{}}
\lfoot[\fancyplain{}{}]{\fancyplain{}{}}
\chead[\fancyplain{}{}]{\fancyplain{}{}}
\cfoot[\fancyplain{\rm\thepage}{}]{\fancyplain{\rm\thepage}{}}
\rhead[\fancyplain{}{}]{\fancyplain{}{\sl{Bibliography} \mbox{  } \mbox{  } \mbox{  } \rm\thepage}}
\rfoot[\fancyplain{}{}]{\fancyplain{}{}}

\clearemptydoublepage

\end{document}